\documentclass{article}
\usepackage{makeidx}
\usepackage{amsfonts}
\usepackage{amssymb}
\usepackage{graphicx}
\usepackage{amsmath}
\usepackage{geometry}
\usepackage{appendix}

\newcommand{\func}[1]{\operatorname{#1}}
\let\dprod\prod
\input{tcilatex}
\begin{document}

\title{Statistical Field Theory and Networks of Spiking Neurons }
\author{Pierre Gosselin\thanks{%
Pierre Gosselin: Institut Fourier, UMR 5582 CNRS-UGA, Universit\'{e}
Grenoble Alpes, BP 74, 38402 Saint Martin d'H\`{e}res, France\ E-Mail:
Pierre.Gosselin@univ-grenoble-alpes.fr} \and A\"{\i}leen Lotz\thanks{%
A\"{\i}leen Lotz: Cerca Trova, BP 114, 38001 Grenoble Cedex 1, France.\
E-mail: a.lotz@cercatrova.eu} \and Marc Wambst\thanks{%
Marc Wambst : Marc Wambst: IRMA, UMR 7501, CNRS, Universit\'{e} de
Strasbourg, 7 rue Ren\'{e} Descartes, 67084 Strasbourg Cedex, France.\
E-Mail: wambst@math.unistra.fr}}
\date{May 2022}
\maketitle

\begin{abstract}
This paper models the dynamics of a large set of interacting neurons within
the framework of statistical field theory. We use a method initially
developed in the context of statistical field theory \cite{Kl} and later
adapted to complex systems in interaction \cite{GL1}\cite{GL2}.\ Our model
keeps track of individual interacting neurons' dynamics but also preserves
some of the features and goals of neural field dynamics, such as indexing a
large number of neurons by a space variable. This paper thus bridges the
scale of individual interacting neurons and the macro-scale modelling of
neural field theory.
\end{abstract}

\section{Introduction}

Bridging micro and macro behaviors remains largely problematic for systems
with large number of degrees of freedom. When studying neural activity, we
either directly start from a macro description of the system, or from a
micro description that is then treated numerically.

At the macroscopic scale, mean - or neural - fields, that model large
populations of neurons as homogeneous structures and index individual
neurons by some spatial coordinates can describe several patterns of brain
activity. Following Wilson, Cowan and Amari (\cite{R1}\cite{R2}\cite{R3}\cite%
{R4}\cite{R5}\cite{R7}\cite{R8}\cite{R9}\cite{R10}), neural fields dynamics
is usually studied in the continuum limit\textbf{\ }and neural activity is
represented by a macroscopic variable, the population-averaged firing rate.
The Mean Field approach is an effective theory in which degrees of freedom
of some underlying processes are aggregated.

Mean field theory has been extended along various lines and has a wide range
of applications.

It allows for travelling wave solutions (see \cite{Wv1}\cite{Wv2} and
references therein). Stochastic effects in firing rates may be introduced 
\cite{S1}\cite{S2}\cite{S3}\cite{S4}\cite{S5} to model perturbations and
diffusion patterns in the pulse waves dynamics and account for noisy
transitions between different mean field regimes (see \cite{Tl}). Besides,
mean fields can be extended to study the impact of neural network topology
on spatial configurations of neural activity\textbf{\ }(see \cite{N1},
developments in \cite{N2}, and references therein).

Last but not least, the Mean field approach has been extended using the
tools of statistical field theory \cite{P1}\cite{P2}\cite{P3}\cite{P4}\cite%
{P5}\cite{P6}\cite{P7}. Statistical fields stand for the neural activity -
or spike counts - at each point of the network. Because it keeps track of
covariances between neural activity at different points, the perturbation
expansion of the effective action goes beyond mean field approximation.%
\textbf{\ }However, since the fields considered represent densities of
activity, this extension of mean field theory remains at the collective
level rather than deriving from the microscopic features of the network.

Despite the convenience and applications of the mean field formalim, it uses
simplifying assumptions to account for the microscopic level, such as delays
in interactions or variations of neurons connectivity. Besides, they cannot
account for emerging behaviors.

At the microscopic scale, a vast litterature at the intersection of
dynamical systems, complex systems and neural networks focuses on\ neurons
dynamics and interactions (see \cite{P8}\cite{P9}\cite{P10} and references
therein).

In strands of litterature such as cognitive neurodynamics or computational
neurosciences, neural processes result from the interactions of assemblies
of individual neurons. The lower scale allows for a finer account of the
interrelation between neurons' connectivity and firing rates than the one of
neural fields. Usually, no spatial indices are assumed:\ neurons are not
positioned in a spatial structure, and the resolution of the model relies on
numerical studies. This approach accounts for neurons' cyclical dynamics,
changes in oscillation regimes (for an account, see \cite{P8} and references
therein) and, more important to us, for the emergence of local connectivity
and higher scale phenomena, such as binding problem or polychronization (%
\cite{V1}\cite{V2}\cite{V3}\cite{V4}\cite{V5}\cite{V6}\cite{V7}\cite{V8}\cite%
{V9}\cite{V10}\cite{V11}). However, unlike mean fields, these models lack an
analytical treatment of collective effect.

The present method bridges the gap between the macro-scale modelling of
neural field theory and the assembly of interacting neurons. It is based on
a method initially developed in \cite{Kl} and later adapted to complex
systems in interaction \cite{GL1}\cite{GL2}\cite{GL3}\cite{GL4}. Our model
of statistical field theory keeps track of individual interacting neurons
dynamics while preserving some features and goals of neural field dynamics.
For instance, neurons are indexed by a space variable, to find continuous
dynamic equations for the whole system. But, unlike Mean Field Theory and
its extensions, our fields do not directly describe any neural activity.\ As
in Statistical Field Theory (see \cite{Kl}), they are rather abstract
complex valued functionals bearing microscopic information to a larger
collective scale. Closer to our approach would be (\cite{FC1}) and (\cite%
{FC2}), that use partition functions for the whole system of neurons, or (%
\cite{FC3}), that works with an effective action. Yet these approaches use
either simplified assumptions at the micro-level, or \textit{a priori }%
assumptions about the effective action.

Our approach recovers features studied in some extensions of mean field
theory. Our results are inherently stochastic: the field describes
interactions of neurons subject to dynamics' uncertainty. We recover some
traveling waves patterns. Beyond that, our formalism has several advantages.
It highlights the influence of some internal variables on the dynamics of
firing rates. It may provide a direct approach to phase transitions
phenomena, i.e. the impact of collective patterns on individual ones, by
studying the effective action of the system. Besides, it allows for a wide
range of extensions.

The present field theory results from a two-step process. In a first step,
the standard formalism of the dynamic equations of a large set of
interacting neurons ( \cite{V11}) is modified to account for the dynamic
nature of neurons connectivity (see \cite{IFR}). From this, we deduce the
firing frequencies' dynamic equations of a large set of neurons.

In a second step, this set of dynamic equations is transformed into a
second-quantized Euclidean field theory (see \cite{GL1}\cite{GL2}\cite{GL3}
for the method). This field description includes both collective and
individual aspects of the system. The dynamics of the whole system is
encompassed both in the action functional for the field and its associated
partition function. To understand the role of collective effects in the
system, we compute the effective action of the system using standard
techniques of field theory. The minimum of this effective action is the
vacuum of the theory, i.e. the background field, in which the system
evolves. It depends on some internal parameters and external currents and
impacts individual neurons dynamics.

We then compute a local approximation for the individual frequencies'
dynamic equations that depends on internal and external parameters and on
the background field. Depending on the connectivity between neurons, the
wave equation may display some non-linear aspects, such as
position-dependent coefficients. In the linear approximation, the dynamics
reduces to some wave equation, that is either dissipative, stable or
explosive. The presence of the background field stabilizes the frequency
equation.\ Some traveling waves are solutions of the system, which shows the
importance of the phase of the system for the wave dynamics. Moreover, the
field formalism allows to derive frequencies dynamic equations beyond the
local approximation. This stresses the importance of the interdependence of
the system of frequencies.

The successive derivatives of the effective action with respect to the field
yield the correlation functions of the system. These correlation functions
compute the joint probabilities of transition for an arbitrary number of
neurons and depend directly on the form of the vacuum. These correlation
functions for an arbitrary number of points yield an alternative and
complementary description of the system frequencies' equations. They compute
a joint probability density for the frequencies at different point. These
probabilities depend on time, and this linear dependency reflects the
ondulatory behaviour of frequencies, and on the background field. Its
presence ensures coordination to some extent between frequencies.

The field formalism is then generalized to include several extensions. We
show for instance how to include dynamic equations for the connectivity
functions. Although these equations are "classical" differential equations,
they could also be described by a field formalism. We also show that two
types of neurons, inhibitory and excitatory, may be included and their
interactions described by the inclusion of two interacting fields in the
model.

This paper is organised as follows.\ Section 2 describes the individual
dynamics of neurons in interaction. Section 3 describes the field theoretic
formulation of the model and section 4 computes the effective action of the
system. In section 5, we derive the minimum of the effective action. In
section 6 we find the general form of the frequencies' equation. Section 7
computes the static equilibrium. In section 8, we derive the differential
equation for frequencies in the local approximation and show the existence
of traveling waves solutions. We also present some extensions of our model
and discuss the implication of these extensions. Section 9 derives the
frequencies' equation beyond the local approximation. In section 10, we
derive a general form for the correlation functions in presence of strong or
weak background field and present their interpretation in term of joined
probabilities for frequencies at different points. Section 11 is a conclusion

\section{Individual dynamics and probability density of the system}

Following \cite{GL1}\cite{GL2}\cite{GL3}, we describe a system of a large
number of neurons ($N>>1$). We define their individual equations. Then, we
write a probability density for the configurations of the whole system over
time.

\subsection{Individual dynamics}

We follow the description of \cite{V11} for coupled quadratic
integrate-and-fire (QIF) neurons, but use the additional hypothesis that
each neuron is characterized by its position in some spatial range.

Each neuron's potential $X_{i}\left( t\right) $ satisfies the differential
equation:%
\begin{equation}
\dot{X}_{i}\left( t\right) =\gamma X_{i}^{2}\left( t\right) +J_{i}\left(
t\right)  \label{ptn}
\end{equation}%
for $X_{i}\left( t\right) <X_{p}$, where $X_{p}$ denotes the potential level
of a spike. When $X=X_{p}$, the potential is reset to its resting value $%
X_{i}\left( t\right) =X_{r}<X_{p}$. For the sake of simplicity, following (%
\cite{V11}) we have chosen the squared form $\gamma X_{i}^{2}\left( t\right) 
$ in (\ref{ptn}). However any form $f\left( X_{i}\left( t\right) \right) $\
could be used. The current of signals reaching cell $i$ at time $t$ is
written $J_{i}\left( t\right) $.

Our purpose is to find the system dynamics in terms of the spikes'
frequencies. First, we consider the time for the $n$-th spike of cell $i$, $%
\theta _{n}^{\left( i\right) }$. This is written as a function of $n$, $%
\theta ^{\left( i\right) }\left( n\right) $. Then, a continuous
approximation $n\rightarrow t$ allows to write the spike time variable as $%
\theta ^{\left( i\right) }\left( t\right) $. We thus have replaced:

\begin{equation*}
\theta _{n}^{\left( i\right) }\rightarrow \theta ^{\left( i\right) }\left(
n\right) \rightarrow \theta ^{\left( i\right) }\left( t\right)
\end{equation*}%
The continuous approximation could be removed, but is convenient and
simplifies the notations and computations. We assume now that the timespans
between two spikes are relatively small. The time between two spikes for
cell $i$ is obtained by writing (\ref{ptn}) as:%
\begin{equation*}
\frac{dX_{i}\left( t\right) }{dt}=\gamma X_{i}^{2}\left( t\right)
+J_{i}\left( t\right)
\end{equation*}%
and by inverting this relation to write:%
\begin{equation*}
dt=\frac{dX_{i}}{\gamma X_{i}^{2}+J^{\left( i\right) }\left( \theta ^{\left(
i\right) }\left( n-1\right) \right) }
\end{equation*}%
Integrating the potential between two spikes thus yields:%
\begin{equation*}
\theta ^{\left( i\right) }\left( n\right) -\theta ^{\left( i\right) }\left(
n-1\right) \simeq \int_{X_{r}}^{X_{p}}\frac{dX}{\gamma X^{2}+J^{\left(
i\right) }\left( \theta ^{\left( i\right) }\left( n-1\right) \right) }
\end{equation*}%
Replacing $J^{\left( i\right) }\left( \theta ^{\left( i\right) }\left(
n-1\right) \right) $ by its average value during the small time period $%
\theta ^{\left( i\right) }\left( n\right) -\theta ^{\left( i\right) }\left(
n-1\right) $, we can consider $J^{\left( i\right) }\left( \theta ^{\left(
i\right) }\left( n-1\right) \right) $ as constant in first approximation, so
that:

\begin{eqnarray*}
\theta ^{\left( i\right) }\left( n\right) -\theta ^{\left( i\right) }\left(
n-1\right) &\simeq &\frac{\left[ \arctan \left( \sqrt{\frac{\gamma }{%
J^{\left( i\right) }\left( \theta ^{\left( i\right) }\left( n-1\right)
\right) }}X\right) \right] _{X_{r}}^{X_{p}}}{\sqrt{\gamma J^{\left( i\right)
}\left( \theta ^{\left( i\right) }\left( n-1\right) \right) }} \\
&=&\frac{\left[ \arctan \left( \frac{1}{X}\sqrt{\frac{J^{\left( i\right)
}\left( \theta ^{\left( i\right) }\left( n-1\right) \right) }{\gamma }}%
\right) \right] _{X_{p}}^{X_{r}}}{\sqrt{\gamma J^{\left( i\right) }\left(
\theta ^{\left( i\right) }\left( n-1\right) \right) }}=\frac{\arctan \left( 
\frac{\left( \frac{1}{X_{r}}-\frac{1}{X_{p}}\right) \sqrt{\frac{J^{\left(
i\right) }\left( \theta ^{\left( i\right) }\left( n-1\right) \right) }{%
\gamma }}}{1+\frac{J^{\left( n\right) }\left( \theta ^{\left( n-1\right)
}\right) }{\gamma X_{r}X_{p}}}\right) }{\sqrt{\gamma J^{\left( i\right)
}\left( \theta ^{\left( i\right) }\left( n-1\right) \right) }}
\end{eqnarray*}%
For $\gamma $ normalized to $1$ and $\frac{J^{\left( n\right) }\left( \theta
^{\left( n-1\right) }\right) }{X_{r}X_{p}}<<1$, this is: 
\begin{equation}
\theta ^{\left( i\right) }\left( n\right) -\theta ^{\left( i\right) }\left(
n-1\right) \equiv G\left( \theta ^{\left( i\right) }\left( n-1\right)
\right) =\frac{\arctan \left( \left( \frac{1}{X_{r}}-\frac{1}{X_{p}}\right) 
\sqrt{J^{\left( i\right) }\left( \theta ^{\left( i\right) }\left( n-1\right)
\right) }\right) }{\sqrt{J^{\left( i\right) }\left( \theta ^{\left( i\right)
}\left( n-1\right) \right) }}  \label{spt}
\end{equation}%
The frequency or firing rate at $t$, $\omega _{i}\left( t\right) $, is
defined by the inverse time span (\ref{spt}) between two spikes:%
\begin{eqnarray*}
\omega _{i}\left( t\right) &=&\frac{1}{G\left( \theta ^{\left( i\right)
}\left( n-1\right) \right) } \\
&\equiv &F\left( \theta ^{\left( i\right) }\left( n-1\right) \right) =\frac{%
\sqrt{J^{\left( i\right) }\left( \theta ^{\left( i\right) }\left( n-1\right)
\right) }}{\arctan \left( \left( \frac{1}{X_{r}}-\frac{1}{X_{p}}\right) 
\sqrt{J^{\left( i\right) }\left( \theta ^{\left( i\right) }\left( n-1\right)
\right) }\right) }
\end{eqnarray*}%
Since we consider small time intervals between two spikes, we can write:%
\begin{equation}
\theta ^{\left( i\right) }\left( n\right) -\theta ^{\left( i\right) }\left(
n-1\right) \simeq \frac{d}{dt}\theta ^{\left( i\right) }\left( t\right)
-\omega _{i}^{-1}\left( t\right) =\varepsilon _{i}\left( t\right)
\label{dnm}
\end{equation}%
where the white noise perturbation $\varepsilon _{i}\left( t\right) $ for
each period was added to account for any internal uncertainty in the time
span $\theta ^{\left( i\right) }\left( n\right) -\theta ^{\left( i\right)
}\left( n-1\right) $. This white noise is independent from the instantaneous
inverse frequency $\omega _{i}^{-1}\left( t\right) $. We assume these $%
\varepsilon _{i}\left( t\right) $ to have variance $\sigma ^{2}$, so that
equation (\ref{dnm}) writes: 
\begin{equation}
\frac{d}{dt}\theta ^{\left( i\right) }\left( t\right) -G\left( \theta
^{\left( i\right) }\left( t\right) ,J^{\left( i\right) }\left( \theta
^{\left( i\right) }\left( t\right) \right) \right) =\varepsilon _{i}\left(
t\right)  \label{dnq}
\end{equation}%
The $\omega _{i}\left( t\right) $ are computed by considering the overall
current which, using the discrete time notation, is given by:%
\begin{equation}
\hat{J}^{\left( i\right) }\left( \left( n-1\right) \right) =J^{\left(
i\right) }\left( \left( n-1\right) \right) +\frac{\kappa }{N}\sum_{j,m}\frac{%
\omega _{j}\left( m\right) }{\omega _{i}\left( n-1\right) }\delta \left(
\theta ^{\left( i\right) }\left( n-1\right) -\theta ^{\left( j\right)
}\left( m\right) -\frac{\left\vert Z_{i}-Z_{j}\right\vert }{c}\right)
T_{ij}\left( \left( n-1,Z_{i}\right) ,\left( m,Z_{j}\right) \right)
\label{crt}
\end{equation}%
The quantity $J^{\left( i\right) }\left( \left( n-1\right) \right) $ denotes
an external current. The term inside the sum is the average current sent to $%
i$ by neuron $j$ during the short time span $\theta ^{\left( i\right)
}\left( n\right) -\theta ^{\left( i\right) }\left( n-1\right) $. The
function $T_{ij}\left( \left( n-1,Z_{i}\right) ,\left( m,Z_{j}\right)
\right) $ is the transfer function between cells $j$ and $i$. It measures
the level of connectivity between $i$ and $j$. We assume that: 
\begin{equation*}
T_{ij}\left( \left( n-1,Z_{i}\right) ,\left( m,Z_{j}\right) \right) =T\left(
\left( n-1,Z_{i}\right) ,\left( m,Z_{j}\right) \right)
\end{equation*}%
The transfer function of $Z_{j}$ on $Z_{i}$ only depends on positions and
times. It models the transfer function as an average transfer between local
zones of the thread. This transfer function is typically considered as
gaussian or decreasing exponentially with the distance between neurons, so
that the closer the cells, the more connected they are.

We can justify the other terms arising in (\ref{crt}): given the distance $%
\left\vert Z_{i}-Z_{j}\right\vert $ between the two cells and the signals'
velocity $c$, signals arrive with a delay $\frac{\left\vert
Z_{i}-Z_{j}\right\vert }{c}$. The spike emitted by cell $j$ at time $\theta
^{\left( j\right) }\left( m\right) $ has thus to satisfy: 
\begin{equation*}
\theta ^{\left( i\right) }\left( n-1\right) <\theta ^{\left( j\right)
}\left( m\right) +\frac{\left\vert Z_{i}-Z_{j}\right\vert }{c}<\theta
^{\left( i\right) }\left( n\right)
\end{equation*}%
to reach cell $i$ during the timespan $\left[ \theta ^{\left( i\right)
}\left( n-1\right) ,\theta ^{\left( i\right) }\left( n\right) \right] $.
This relation must be represented by a step function in the current formula.
However given our approximations, this can be replaced by: 
\begin{equation*}
\delta \left( \theta ^{\left( i\right) }\left( n-1\right) -\theta ^{\left(
j\right) }\left( m\right) -\frac{\left\vert Z_{i}-Z_{j}\right\vert }{c}%
\right)
\end{equation*}%
as in (\ref{crt}). However, this Dirac function must be weighted by the
number of spikes emitted during the rise of the potential. This number is
the ratio $\frac{\omega _{j}\left( m\right) }{\omega _{i}\left( n-1\right) }$
that counts the number of spikes emitted by neuron $j$ towards neuron $i$
between the spikes $n-1$ and $n$ of neuron $i$. Again, this is valid for
relatively small timespans between two spikes. For larger timespans, the
frequencies should be replaced by their average over this period of time.

The sum over $m$ and $i$ is the overall contribution to the current from any
possible spike of the thread, provided it arrives at $i$ during the interval 
$\theta ^{\left( i\right) }\left( n\right) -\theta ^{\left( i\right) }\left(
n-1\right) $ considered. Note that the current (\ref{crt}) is partly an
endogenous variable. It depends on signals external to $i$, but depends also
on $i$ through $\omega _{i}\left( n-1\right) $. This is a consequence of the
intrication between the system's elements.

In the sequel, we will work in the continuous approximation, so that formula
(\ref{crt}) is replaced by:%
\begin{equation}
\hat{J}^{\left( i\right) }\left( t\right) =J^{\left( i\right) }\left(
t\right) +\frac{\kappa }{N}\int \sum_{j}\frac{\omega _{j}\left( s\right) }{%
\omega _{i}\left( t\right) }\delta \left( \theta ^{\left( i\right) }\left(
t\right) -\theta ^{\left( j\right) }\left( s\right) -\frac{\left\vert
Z_{i}-Z_{j}\right\vert }{c}\right) T_{ij}\left( \left( t,Z_{i}\right)
,\left( s,Z_{j}\right) \right) ds  \label{crT}
\end{equation}

Formula (\ref{crT}) shows that the dynamic equation (\ref{dnm}) has to be
coupled with the frequency equation:%
\begin{eqnarray}
\omega _{i}\left( t\right) &=&G\left( \theta ^{\left( i\right) }\left(
t\right) ,\hat{J}\left( \theta ^{\left( i\right) }\left( t\right) \right)
\right) +\upsilon _{i}\left( t\right)  \label{cstrt} \\
&=&\frac{\sqrt{\hat{J}^{\left( i\right) }\left( t\right) }}{\arctan \left(
\left( \frac{1}{X_{r}}-\frac{1}{X_{p}}\right) \sqrt{\hat{J}^{\left( i\right)
}\left( t\right) }\right) }+\upsilon _{i}\left( t\right)  \notag
\end{eqnarray}%
and $J^{\left( i\right) }\left( t\right) $ is defined by (\ref{crT}). A
white noise $\upsilon _{i}\left( t\right) $ accounts for the possible
deviations from this relation, due to some internal or external causes for
each cell. We assume that the variances of $\upsilon _{i}\left( t\right) $
are constant, and equal to $\eta ^{2}$, such that $\eta ^{2}<<\sigma ^{2}$.
\ 

\subsection{Probability density for the system}

Due to the stochastic nature of equations (\ref{dnq}) and (\ref{cstrt}), the
dynamics of a single neuron can be described by the probability density $%
P\left( \theta ^{\left( i\right) }\left( t\right) ,\omega _{i}^{-1}\left(
t\right) \right) $ for a path $\left( \theta ^{\left( i\right) }\left(
t\right) ,\omega _{i}^{-1}\left( t\right) \right) $ which is given by, up to
a normalization factor:

\begin{equation}
P\left( \theta ^{\left( i\right) }\left( t\right) ,\omega _{i}^{-1}\left(
t\right) \right) =\exp \left( -A_{i}\right)  \label{dnmcs}
\end{equation}%
where:%
\begin{equation}
A_{i}=\frac{1}{\sigma ^{2}}\int \left( \frac{d}{dt}\theta ^{\left( i\right)
}\left( t\right) -\omega _{i}^{-1}\left( t\right) \right) ^{2}dt+\int \frac{%
\left( \omega _{i}^{-1}\left( t\right) -G\left( \theta ^{\left( i\right)
}\left( t\right) ,\hat{J}\left( \theta ^{\left( i\right) }\left( t\right)
\right) \right) \right) ^{2}}{\eta ^{2}}dt  \label{dnmcszz}
\end{equation}%
(see \cite{GL1} and \cite{GL2}). The integral is taken over a time period
that depends on the time scale of the interactions. Actually, the
minimization of (\ref{dnmcszz})\ imposes both (\ref{dnm}) and (\ref{cstrt}),
so that the probability density is, as expected, centered around these two
conditions, i.e. (\ref{dnm}) and (\ref{cstrt}) are satisfied in mean. A
probability density for the whole system is obtained by summing $S_{i}$ over
all agents. We thus define:%
\begin{equation}
P\left( \left( \theta ^{\left( i\right) }\left( t\right) ,\omega
_{i}^{-1}\left( t\right) \right) _{i=1...N}\right) =\exp \left( -A\right)
\label{Prdn}
\end{equation}%
with:%
\begin{equation}
A=\sum_{i}A_{i}=\sum_{i}\frac{1}{\sigma ^{2}}\int \left( \frac{d}{dt}\theta
^{\left( i\right) }\left( t\right) -\omega _{i}^{-1}\left( t\right) \right)
^{2}dt+\int \frac{\left( \omega _{i}^{-1}\left( t\right) -G\left( \theta
^{\left( i\right) }\left( t\right) ,\hat{J}\left( \theta ^{\left( i\right)
}\left( t\right) \right) \right) \right) ^{2}}{\eta ^{2}}dt  \label{ctnt}
\end{equation}

\section{Field theoretic description of the system}

\subsection{translation of Equation (\protect\ref{ctnt}) in terms of field
theory}

We have shown in \cite{GL1}\cite{GL2}\cite{GL3} that the probabilistic
description of the system (\ref{Prdn}) is equivalent to a statistical field
formalism. In such a formalism, the system is collectively described by a
field that is an element of the Hilbert space of complex functions. The
arguments of\ these functions are the same as those describing an individual
neuron. A shortcut of the translation of systems similar to (\ref{ctnt}) in
terms of field, is given in \cite{GL4} . The next paragraph gives an account
of this method.

\subsubsection{Principle\textbf{\ }}

\paragraph{General form of the statistical weight}

In general, we assume a system in which individual agents (here cells) are
described by vectors $\mathbf{X}_{i}\left( t\right) $\ of arbitrary
dimension, and such that the exponent of the statistical weight of the
system has the form:%
\begin{eqnarray}
&&\sum_{i}\int \left( \frac{d\mathbf{X}_{i}^{\left( \alpha \right) }\left(
t\right) }{dt}-\int \sum_{j,k,l...}f^{\left( \alpha \right) }\left( \mathbf{X%
}_{i}\left( t\right) ,\mathbf{X}_{j}\left( t_{j}\right) ,\mathbf{X}%
_{k}\left( t_{k}\right) ...\right) dt_{i}dt_{j}dt_{k}...\right) ^{2}dt
\label{mNZ} \\
&&+\sum_{i}\int \sum_{j,k,l...}g\left( \mathbf{X}_{i}\left( t_{i}\right) ,%
\mathbf{X}_{j}\left( t_{j}\right) ,\mathbf{X}_{k}\left( t_{k}\right)
...\right) dt_{i}dt_{j}dt_{k}...  \notag
\end{eqnarray}%
for some functions $f^{\left( \alpha \right) }$ and $g$. The introduction of
index $\alpha $ represents the dynamics of the $\alpha $-th coordinate of a
variable $\mathbf{X}_{i}\left( t\right) $ as a function of the other agents.
If we replace $\hat{J}\left( \theta ^{\left( i\right) }\left( t\right)
\right) $ in (\ref{ctnt}) by its expression (\ref{crT}), we can check that (%
\ref{ctnt}) has the form (\ref{mNZ}). This point is detailed below.

\paragraph{Translation in terms of fields}

The translation itself can be divided into two relatively simple processes,
but varies slightly depending on the type of terms that appear in the
various minimization functions.

\subparagraph{Term without temporal derivative}

The terms in (\ref{mNZ}) that include indexed variables but no temporal
derivative terms are the easiest to translate.\ They are of the form:%
\begin{equation*}
\sum_{i}\int \sum_{j,k,l...}g\left( \mathbf{X}_{i}\left( t_{i}\right) ,%
\mathbf{X}_{j}\left( t_{j}\right) ,\mathbf{X}_{k}\left( t_{k}\right)
...\right) dt_{i}dt_{j}dt_{k}...
\end{equation*}%
These terms describe the whole set of interactions between agents
characterized by their variables $\mathbf{X}_{i}\left( t\right) ,\mathbf{X}%
_{j}\left( t\right) ,\mathbf{X}_{k}\left( t\right) $...

In the field translation, agents are described by a field $\Psi \left( 
\mathbf{X}\right) $ where $\mathbf{X}$ is a vector of the same dimension as
the $\mathbf{X}_{i}$.

In a first step, the variables indexed $i$ such as $\mathbf{X}_{i}\left(
t\right) $ are replaced by variables $\mathbf{X}$ in the expression of $g$.
The variables indexed $j$,$k$,$l$,$m...$, such as $\mathbf{X}_{j}\left(
t\right) $, $\mathbf{X}_{k}\left( t\right) $... are replaced by $\mathbf{X}%
^{\prime },\mathbf{X}^{\prime \prime }$, and so on for all the indices in
the function. This yields the expression:

\begin{equation*}
\sum_{i}\sum_{j,k,l,m...}g\left( \mathbf{X},\mathbf{X}^{\prime },\mathbf{X}%
^{\prime \prime }...\right)
\end{equation*}%
In a second step, each sum is replaced by a weighted integration symbol: 
\begin{equation*}
\sum_{i}\rightarrow \int \left\vert \Psi \left( \mathbf{X}\right)
\right\vert ^{2}d\mathbf{X}\text{, }\sum_{j}\rightarrow \int \left\vert \Psi
\left( \mathbf{X}^{\prime }\right) \right\vert ^{2}d\mathbf{X}^{\prime }%
\text{, }\sum_{k}\rightarrow \int \left\vert \Psi \left( \mathbf{X}^{\prime
\prime }\right) \right\vert ^{2}d\mathbf{X}^{\prime \prime }
\end{equation*}%
which leads to the translation:%
\begin{eqnarray}
&&\sum_{i}\int \sum_{j,k,l...}g\left( \mathbf{X}_{i}\left( t_{i}\right) ,%
\mathbf{X}_{j}\left( t_{j}\right) ,\mathbf{X}_{k}\left( t_{k}\right)
...\right) dt_{i}dt_{j}dt_{k}...  \notag \\
&\rightarrow &\int g\left( \mathbf{X},\mathbf{X}^{\prime },\mathbf{X}%
^{\prime \prime }...\right) \left\vert \Psi \left( \mathbf{X}\right)
\right\vert ^{2}\left\vert \Psi \left( \mathbf{X}^{\prime }\right)
\right\vert ^{2}\left\vert \Psi \left( \mathbf{X}^{\prime \prime }\right)
\right\vert ^{2}d\mathbf{X}d\mathbf{X}^{\prime }d\mathbf{X}^{\prime \prime
}...  \label{tln}
\end{eqnarray}%
Note that this formula can be generalized if we compose the previous formula
by any function $G$. The translation of an expression of the form:%
\begin{equation}
\sum_{i}\int G\left( \int \sum_{j,k,l...}g\left( \mathbf{X}_{i}\left(
t_{i}\right) ,\mathbf{X}_{j}\left( t_{j}\right) ,\mathbf{X}_{k}\left(
t_{k}\right) ...\right) dt_{j}dt_{k}...\right) dt_{i}  \label{cmG}
\end{equation}%
is obtained by expanding (\ref{cmG}) in powers of $\int
\sum_{j,k,l...}g\left( \mathbf{X}_{i}\left( t_{i}\right) ,\mathbf{X}%
_{j}\left( t_{j}\right) ,\mathbf{X}_{k}\left( t_{k}\right) ...\right)
dt_{j}dt_{k}...$ and using (\ref{tln}). We find:%
\begin{eqnarray}
&&\sum_{i}\int G\left( \int \sum_{j,k,l...}g\left( \mathbf{X}_{i}\left(
t_{i}\right) ,\mathbf{X}_{j}\left( t_{j}\right) ,\mathbf{X}_{k}\left(
t_{k}\right) ...\right) dt_{j}dt_{k}...\right) dt_{i}  \notag \\
&\rightarrow &\int \left\vert \Psi \left( \mathbf{X}\right) \right\vert
^{2}G\left( \int g\left( \mathbf{X},\mathbf{X}^{\prime },\mathbf{X}^{\prime
\prime }...\right) \left\vert \Psi \left( \mathbf{X}^{\prime }\right)
\right\vert ^{2}\left\vert \Psi \left( \mathbf{X}^{\prime \prime }\right)
\right\vert ^{2}d\mathbf{X}^{\prime }d\mathbf{X}^{\prime \prime }...\right) d%
\mathbf{X}  \label{tnL}
\end{eqnarray}

\subparagraph{Term with temporal derivative}

The terms in (\ref{mNZ}) that imply a variable temporal derivative are of
the form:%
\begin{equation}
\sum_{i}\int \left( \frac{d\mathbf{X}_{i}^{\left( \alpha \right) }\left(
t\right) }{dt}-\int \sum_{j,k,l...}f^{\left( \alpha \right) }\left( \mathbf{X%
}_{i}\left( t\right) ,\mathbf{X}_{j}\left( t_{j}\right) ,\mathbf{X}%
_{k}\left( t_{k}\right) ...\right) dt_{i}dt_{j}dt_{k}\right) ^{2}dt
\label{edr}
\end{equation}%
The method of translation is similar to the above, but the time derivative
adds an additional operation.

In a first step, we translate the terms without derivative inside the
parenthesis:%
\begin{equation}
\int \sum_{j,k,l...}f^{\left( \alpha \right) }\left( \mathbf{X}_{i}\left(
t\right) ,\mathbf{X}_{j}\left( t_{j}\right) ,\mathbf{X}_{k}\left(
t_{k}\right) ...\right) dt_{i}dt_{j}dt_{k}  \label{ntr}
\end{equation}%
The translation of this type of term has already been presented in the
previous paragraph.\ Note however that, in (\ref{ntr}), there is no sum over 
$i$,\ so that the translation includes neither the integral over $X$, nor
the factor $\left\vert \Psi \left( \mathbf{X}\right) \right\vert ^{2}$.

The translation of (\ref{ntr}) is therefore, as before:%
\begin{equation}
\int f^{\left( \alpha \right) }\left( \mathbf{X},\mathbf{X}^{\prime },%
\mathbf{X}^{\prime \prime }...\right) \left\vert \Psi \left( \mathbf{X}%
^{\prime }\right) \right\vert ^{2}\left\vert \Psi \left( \mathbf{X}^{\prime
\prime }\right) \right\vert ^{2}d\mathbf{X}^{\prime }d\mathbf{X}^{\prime
\prime }  \label{trn}
\end{equation}%
A free variable $\mathbf{X}$ remains, which will be integrated later, when
we account for the external sum $\sum_{i}$. We will call $\Lambda (\mathbf{X}%
)$ the expression obtained in (\ref{trn}):%
\begin{equation}
\Lambda (\mathbf{X})=\int f^{\left( \alpha \right) }\left( \mathbf{X},%
\mathbf{X}^{\prime },\mathbf{X}^{\prime \prime }...\right) \left\vert \Psi
\left( \mathbf{X}^{\prime }\right) \right\vert ^{2}\left\vert \Psi \left( 
\mathbf{X}^{\prime \prime }\right) \right\vert ^{2}d\mathbf{X}^{\prime }d%
\mathbf{X}^{\prime \prime }  \label{bdt}
\end{equation}%
In a second step, we account for the derivative in time by using field
gradients. To do so, and as a rule, we replace :%
\begin{equation}
\sum_{i}\left( \frac{d\mathbf{X}_{i}^{\left( \alpha \right) }\left( t\right) 
}{dt}-\int \sum_{j,k,l...}f^{\left( \alpha \right) }\left( \mathbf{X}%
_{i}\left( t\right) ,\mathbf{X}_{j}\left( t_{j}\right) ,\mathbf{X}_{k}\left(
t_{k}\right) ...\right) dt_{i}dt_{j}dt_{k}\right) ^{2}  \label{inco}
\end{equation}%
by:%
\begin{equation}
\int \Psi ^{\dag }\left( \mathbf{X}\right) \left( -\nabla _{\mathbf{X}%
^{\left( \alpha \right) }}\left( \frac{\sigma _{\mathbf{X}^{\left( \alpha
\right) }}^{2}}{2}\nabla _{\mathbf{X}^{\left( \alpha \right) }}+\Lambda (%
\mathbf{X})\right) \right) \Psi \left( \mathbf{X}\right) d\mathbf{X}
\label{Trl}
\end{equation}%
The variance $\sigma _{\mathbf{X}^{\left( \alpha \right) }}^{2}$ reflects
the probabilistic nature of the model which is hidden behind the field
formalism. This variance represents the characteristic level of uncertainty
of the system's dynamics. It is a parameter of the model. Note also that in (%
\ref{Trl}), the integral over $\mathbf{X}$ reappears at the end, along with
the square of the field $\left\vert \Psi \left( \mathbf{X}\right)
\right\vert ^{2}$.\ This square is split into two terms, $\Psi ^{\dag
}\left( \mathbf{X}\right) $ and $\Psi \left( \mathbf{X}\right) $, with a
gradient operator inserted in between.\bigskip

\subsubsection{Translation of (\protect\ref{ctnt})}

In our context, the field depends on the three variables $\left( \theta
,Z,\omega \right) $,\ and writes $\Psi \left( \theta ,Z,\omega \right) $.
The field dynamics is described by an action functional for the field and
its associated partition function. This partition function reflects both
collective and individual aspects of the system, and allows to recover
correlation functions for an arbitrary number of neurons.

The field theoretic version of (\ref{dnmcszz}) is obtained using (\ref{ctnt}%
): a correspondence detailed in \cite{GL1}\cite{GL2}) yields an action $%
S\left( \Psi \right) $ for a field $\Psi \left( \theta ,Z,\omega \right) $
and a statistical weight $\exp \left( -\left( S\left( \Psi \right) \right)
\right) $ for each configuration $\Psi \left( \theta ,Z,\omega \right) $ of
this field. The functional $S\left( \Psi \right) $ is decomposed in two
parts corresponding to the two contributions in (\ref{ctnt}).

The first term of (\ref{ctnt}):%
\begin{equation}
\frac{1}{\sigma ^{2}}\int \left( \frac{d}{dt}\theta ^{\left( i\right)
}\left( t\right) -\omega _{i}^{-1}\left( t\right) \right) ^{2}dt  \label{fsT}
\end{equation}%
is a term with temporal derivative. Its form is simple since the function $%
f^{\left( \alpha \right) }$ in (\ref{inco}) depends only on the variable $%
\mathbf{X}_{i}\left( t\right) =\left( \theta ^{\left( i\right) }\left(
t\right) ,\omega _{i}^{-1}\left( t\right) ,Z_{i}\right) $. Actually $%
f^{\left( \theta \right) }\left( \mathbf{X}_{i}\left( t\right) \right)
=\omega _{i}^{-1}\left( t\right) $. Using (\ref{Trl}), the term (\ref{fsT})
is thus replaced by the corresponding quadratic functional in field theory :%
\begin{equation}
-\frac{1}{2}\Psi ^{\dagger }\left( \theta ,Z,\omega \right) \nabla \left( 
\frac{\sigma ^{2}}{2}\nabla -\omega ^{-1}\right) \Psi \left( \theta
,Z,\omega \right)  \label{thtdnmcs}
\end{equation}%
where $\sigma ^{2}$ is the variance of the errors $\varepsilon _{i}$.

The field functional that corresponds to the second term of (\ref{dnmcszz}):%
\begin{equation*}
V=\int \frac{\left( \omega _{i}^{-1}\left( t\right) -G\left( \theta ^{\left(
i\right) }\left( t\right) ,\hat{J}\left( \theta ^{\left( i\right) }\left(
t\right) \right) \right) \right) ^{2}}{\eta ^{2}}dt
\end{equation*}
is obtained by expanding the formula (\ref{crT}) for the current induced by
all $j$:

\begin{eqnarray}
V &=&\frac{1}{2\eta ^{2}}\int dt\sum_{i}\left( \omega _{i}^{-1}\left(
t\right) \right.  \label{Cpnt} \\
&&-\left. G\left( J\left( \theta ^{\left( i\right) }\left( t\right)
,Z_{i}\right) +\frac{\kappa }{N}\int ds\sum_{j}\frac{\omega _{j}\left(
s\right) T_{ij}\left( \left( t,Z_{i}\right) ,s,Z_{j}\right) }{\omega
_{i}\left( t\right) }\delta \left( \theta ^{\left( i\right) }\left( t\right)
-\theta ^{\left( j\right) }\left( s\right) -\frac{\left\vert
Z_{i}-Z_{j}\right\vert }{c}\right) \right) \right) ^{2}  \notag
\end{eqnarray}%
with $\eta <<1$, which is the constraint (\ref{cstrt}) imposed
stochastically. Its corresponding potential in field theory is obtained
straightforwardly by using the translation (\ref{tnL}):%
\begin{equation}
\frac{1}{2\eta ^{2}}\int \left\vert \Psi \left( \theta ,Z,\omega \right)
\right\vert ^{2}\left( \omega ^{-1}-G\left( J\left( \theta ,Z\right) +\int 
\frac{\kappa }{N}\frac{\omega _{1}T\left( Z,\theta ,Z_{1},\theta -\frac{%
\left\vert Z-Z_{1}\right\vert }{c}\right) }{\omega }\left\vert \Psi \left(
\theta -\frac{\left\vert Z-Z_{1}\right\vert }{c},Z_{1},\omega _{1}\right)
\right\vert ^{2}dZ_{1}d\omega _{1}\right) \right) ^{2}  \label{ptntl}
\end{equation}%
To simplify, we will write in the sequel:%
\begin{equation*}
T\left( Z,\theta ,Z_{1},\theta -\frac{\left\vert Z-Z_{1}\right\vert }{c}%
\right) \equiv T\left( Z,\theta ,Z_{1}\right)
\end{equation*}%
The field action is then the sum of (\ref{thtdnmcs}) and (\ref{ptntl}):%
\begin{eqnarray}
S &=&-\frac{1}{2}\Psi ^{\dagger }\left( \theta ,Z,\omega \right) \nabla
\left( \frac{\sigma _{\theta }^{2}}{2}\nabla -\omega ^{-1}\right) \Psi
\left( \theta ,Z,\omega \right)  \label{lfS} \\
&&+\frac{1}{2\eta ^{2}}\int \left\vert \Psi \left( \theta ,Z,\omega \right)
\right\vert ^{2}\left( \omega ^{-1}-G\left( J\left( \theta ,Z\right) +\int 
\frac{\kappa }{N}\frac{\omega _{1}}{\omega }\left\vert \Psi \left( \theta -%
\frac{\left\vert Z-Z_{1}\right\vert }{c},Z_{1},\omega _{1}\right)
\right\vert ^{2}T\left( Z,\theta ,Z_{1}\right) dZ_{1}d\omega _{1}\right)
\right) ^{2}  \notag
\end{eqnarray}%
$\allowbreak $

\subsection{Projection on dependent frequency states:}

Using the fact that $\eta ^{2}<<1$, and noting that in this case, field
configurations $\Psi \left( \theta ,Z,\omega \right) $ such that:%
\begin{equation*}
\omega ^{-1}-G\left( J\left( \theta ,Z\right) +\int \frac{\kappa }{N}\frac{%
\omega _{1}}{\omega }\left\vert \Psi \left( \theta -\frac{\left\vert
Z-Z_{1}\right\vert }{c},Z_{1},\omega _{1}\right) \right\vert ^{2}T\left(
Z,\theta ,Z_{1}\right) dZ_{1}d\omega _{1}\right) \neq 0
\end{equation*}%
have negligible statistical weight, we can simplify (\ref{lfS}) and restrict
the fields to those of the form: 
\begin{equation}
\Psi \left( \theta ,Z\right) \delta \left( \omega ^{-1}-\omega ^{-1}\left(
J,\theta ,Z,\left\vert \Psi \right\vert ^{2}\right) \right)  \label{prt}
\end{equation}%
where $\omega ^{-1}\left( J,\theta ,Z,\Psi \right) $ satisfies:%
\begin{eqnarray*}
\omega ^{-1}\left( J,\theta ,Z,\left\vert \Psi \right\vert ^{2}\right)
&=&G\left( J\left( \theta ,Z\right) +\int \frac{\kappa }{N}\frac{\omega
_{1}T\left( Z,\theta ,Z_{1},\theta -\frac{\left\vert Z-Z_{1}\right\vert }{c}%
\right) }{\omega \left( J,\theta ,Z,\left\vert \Psi \right\vert ^{2}\right) }%
\left\vert \Psi \left( \theta -\frac{\left\vert Z-Z_{1}\right\vert }{c}%
,Z_{1},\omega _{1}\right) \right\vert ^{2}dZ_{1}d\omega _{1}\right) \\
&=&G\left( J\left( \theta ,Z\right) +\int \frac{\kappa }{N}\frac{\omega
_{1}T\left( Z,\theta ,Z_{1},\theta -\frac{\left\vert Z-Z_{1}\right\vert }{c}%
\right) }{\omega \left( J,\theta ,Z,\left\vert \Psi \right\vert ^{2}\right) }%
\left\vert \Psi \left( \theta -\frac{\left\vert Z-Z_{1}\right\vert }{c}%
,Z_{1}\right) \right\vert ^{2}\right. \\
&&\times \left. \delta \left( \omega _{1}^{-1}-\omega ^{-1}\left( J,\theta -%
\frac{\left\vert Z-Z_{1}\right\vert }{c},Z_{1},\left\vert \Psi \right\vert
^{2}\right) \right) dZ_{1}d\omega _{1}\right)
\end{eqnarray*}%
The last equation simplifies to yield:%
\begin{equation}
\omega ^{-1}\left( J,\theta ,Z,\left\vert \Psi \right\vert ^{2}\right)
=G\left( J\left( \theta ,Z\right) +\int \frac{\kappa }{N}\frac{\omega \left(
J,\theta -\frac{\left\vert Z-Z_{1}\right\vert }{c},Z_{1},\Psi \right)
T\left( Z,\theta ,Z_{1},\theta -\frac{\left\vert Z-Z_{1}\right\vert }{c}%
\right) }{\omega \left( J,\theta ,Z,\left\vert \Psi \right\vert ^{2}\right) }%
\left\vert \Psi \left( \theta -\frac{\left\vert Z-Z_{1}\right\vert }{c}%
,Z_{1}\right) \right\vert ^{2}dZ_{1}\right)  \label{qf}
\end{equation}%
The configurations $\Psi \left( \theta ,Z,\omega \right) $ that minimize the
potential (\ref{ptntl}) can now be considered: the field $\Psi \left( \theta
,Z,\omega \right) $ is projected on the subspace (\ref{prt}) of functions of
two variables, and we can therefore replace in (\ref{ptntl}):%
\begin{equation*}
\omega ^{-1}\rightarrow \omega ^{-1}\left( J,\theta ,Z,\left\vert \Psi
\right\vert ^{2}\right)
\end{equation*}%
The "classical" effective action becomes (see appendix 0):%
\begin{equation}
-\frac{1}{2}\Psi ^{\dagger }\left( \theta ,Z\right) \left( \nabla _{\theta
}\left( \frac{\sigma ^{2}}{2}\nabla _{\theta }-\omega ^{-1}\left( J,\theta
,Z,\left\vert \Psi \right\vert ^{2}\right) \right) \right) \Psi \left(
\theta ,Z\right)  \label{nmR}
\end{equation}%
with $\omega ^{-1}\left( J,\theta ,Z,\left\vert \Psi \right\vert ^{2}\right) 
$ given by equation (\ref{qf}).

The form of the transfer function $T\left( Z,\theta ,Z_{1}\right) $ can
ultimately be refined. Using a simplified version of \cite{IFR}, appendix 6
shows that, at the individual level and in first approximation, the transfer
functions are modelled by a product between a spatial factor $T\left(
Z,Z_{1}\right) $ and a function $W$ of the frequencies $\omega \equiv \omega
\left( J,\theta ,Z,\left\vert \Psi \right\vert ^{2}\right) $, and $\omega
_{1}\equiv \omega \left( J,\theta -\frac{\left\vert Z-Z_{1}\right\vert }{c}%
,Z_{1},\left\vert \Psi \right\vert ^{2}\right) $. The function $W$ is
increasing in $\omega $ and decreasing in $\omega _{1}$. Without loss of
generality, we will consider $W$ as an increasing function of $\left( \frac{%
\omega }{\omega _{1}}\right) $, so that: 
\begin{equation}
T\left( Z,\theta ,Z_{1}\right) =T\left( Z,Z_{1}\right) W\left( \frac{\omega 
}{\omega _{1}}\right)  \label{Wqn}
\end{equation}

\subsection{Inclusion of collective stabilization potential}

To stabilize the number of active connexions, we ultimately modify (\ref{nmR}%
) by including collective terms. These terms correspond to some overall
regulatory processes that are not accounted for in the model in its actual
form. When there is no "competition" between inhibitory and excitatory
mechanisms, such a potential models the ability for a system to return to
some minimal equilibrium activity.

To do so, we add to the action functional (\ref{nmR}) a potential $V\left(
\Psi \right) $ that maintains and activates new connections. The system's
action thus becomes:%
\begin{equation}
S=-\frac{1}{2}\Psi ^{\dagger }\left( \theta ,Z\right) \left( \nabla _{\theta
}\left( \frac{\sigma ^{2}}{2}\nabla _{\theta }-\omega ^{-1}\left( J,\theta
,Z,\left\vert \Psi \right\vert ^{2}\right) \right) \right) \Psi \left(
\theta ,Z\right) +V\left( \Psi \right)  \label{nMR}
\end{equation}
We choose the following form for $V\left( \Psi \right) $: 
\begin{equation}
V\left( \Psi \right) =\int \left\vert \Psi \left( \theta ,Z\right)
\right\vert ^{2}U_{0}\left( \int \left\vert \Psi \left( \theta -\frac{%
\left\vert Z-Z^{\prime }\right\vert }{c},Z^{\prime }\right) \right\vert
^{2}\right)  \label{pp}
\end{equation}
where $U_{0}$ is a $U$ shaped potential, with $U_{0}\left( 0\right) =0$ so
that $U_{0}$ has a minimum for some positive value of $\left\vert \Psi
\left( \theta -\frac{\left\vert Z-Z^{\prime }\right\vert }{c},Z^{\prime
}\right) \right\vert ^{2}$. At this minimum, the potential $U_{0}$ is
negative.

Note that, expression (\ref{pp}) models the interactions between activity at
time $\theta $ of cells located at $Z$, and activity at time $\theta -\frac{%
\left\vert Z-Z^{\prime }\right\vert }{c}$ of those located at any point $%
Z^{\prime }$. The delay $\frac{\left\vert Z-Z^{\prime }\right\vert }{c}$ in
time is induced by travelling time of signals between $Z$ and $Z^{\prime }$.

The potential (\ref{pp}) can be written as a series expansion. We choose the
following decomposition: 
\begin{equation}
V\left( \Psi \right) =-\zeta _{1}\int \left( \left\vert \Psi \left( \theta
,Z\right) \right\vert ^{2}\left\vert \Psi \left( \theta -\frac{\left\vert
Z-Z^{\prime }\right\vert }{c},Z^{\prime }\right) \right\vert ^{2}\right)
+\sum_{n=2}^{\infty }\zeta _{n}\int \left\vert \Psi \left( \theta ,Z\right)
\right\vert ^{2}\left( \dprod\limits_{i=1}^{n-1}\left\vert \Psi \left(
\theta -\frac{\left\vert Z-Z_{i}\right\vert }{c},Z_{i}\right) \right\vert
^{2}\right)  \label{ps}
\end{equation}%
The first term in (\ref{ps}):%
\begin{equation*}
-\zeta _{1}\int \left( \left\vert \Psi \left( \theta ,Z\right) \right\vert
^{2}\left\vert \Psi \left( \theta -\frac{\left\vert Z-Z^{\prime }\right\vert 
}{c},Z^{\prime }\right) \right\vert ^{2}\right)
\end{equation*}
with $\zeta _{1}>0$, accounts for a minimal number of connections that are
permanently maintained. The magnitude of this factor depends on the external
activity $J$.

The second term in (\ref{ps}) models a global limitation mechanism. It
increases with the overall number of connections and currents, so that we
assume for $n\geqslant 2$:%
\begin{equation*}
\sum_{k=2}^{n}\zeta _{k}\left\langle \dprod\limits_{i=1}^{k-1}\left\vert
\Psi \left( \theta -\frac{\left\vert Z-Z_{i}\right\vert }{c},Z_{i}\right)
\right\vert ^{2}\right\rangle >0
\end{equation*}%
where the bracket $\left\langle {}\right\rangle $ denotes the expectation
value of the product of fields.

The coefficients $\zeta _{n}$ can be set to $0$ for $n>N$, where $N$ is an
arbitrary threshold. The term proportional to $-\zeta _{1}$ \ and the
contribution for $n=2$ can be gathered to rewrite the collective potential
as:%
\begin{equation}
V\left( \Psi \right) =\sum_{n=2}^{\infty }\zeta ^{\left( n\right) }\int
\left\vert \Psi \left( \theta ,Z\right) \right\vert ^{2}\left(
\dprod\limits_{i=1}^{n-1}\left\vert \Psi \left( \theta -\frac{\left\vert
Z-Z_{i}\right\vert }{c},Z_{i}\right) \right\vert ^{2}\right)  \label{sp}
\end{equation}%
where $\zeta ^{\left( n\right) }=\zeta _{n}$ for $n>2$, and $\zeta ^{\left(
2\right) }=\zeta _{2}-\zeta _{1}$. We assume that $\zeta ^{\left( 2\right)
}<0$, so that a nontrivial minimal collective state exists. The classical
action is thus:%
\begin{eqnarray}
&&-\frac{1}{2}\Psi ^{\dagger }\left( \theta ,Z\right) \left( \nabla _{\theta
}\left( \frac{\sigma ^{2}}{2}\nabla _{\theta }-\omega ^{-1}\left( J,\theta
,Z,\left\vert \Psi \right\vert ^{2}\right) \right) \right) \Psi \left(
\theta ,Z\right) +V\left( \Psi \right)  \label{lcn} \\
&\equiv &-\frac{1}{2}\Psi ^{\dagger }\left( \theta ,Z\right) \left( \nabla
_{\theta }\left( \frac{\sigma ^{2}}{2}\nabla _{\theta }-\omega ^{-1}\left(
J,\theta ,Z,\left\vert \Psi \right\vert ^{2}\right) \right) \right) \Psi
\left( \theta ,Z\right) +\sum_{n=2}^{\infty }\zeta ^{\left( n\right) }\int
\left\vert \Psi \left( \theta ,Z\right) \right\vert ^{2}\left(
\dprod\limits_{i=1}^{n-1}\left\vert \Psi \left( \theta -\frac{\left\vert
Z-Z_{i}\right\vert }{c},Z_{i}\right) \right\vert ^{2}\right)  \notag
\end{eqnarray}%
Since we are merely interested in the relative magnitudes of the
coefficients $\sigma _{\theta }^{2}$ and quantities such as $\omega ^{-1}$,
we can impose a constraint on the coefficients $\zeta ^{\left( n\right) }$.
As a relative benchmark, we choose: 
\begin{eqnarray*}
&&\sum_{n=2}^{\infty }\zeta ^{\left( n\right) }\int \left\vert \Psi \left(
\theta ,Z\right) \right\vert ^{2}\left(
\dprod\limits_{i=1}^{n-1}\left\langle \left\vert \Psi \left( \theta -\frac{%
\left\vert Z-Z_{i}\right\vert }{c},Z_{i}\right) \right\vert
^{2}\right\rangle \right) \\
&=&\int \left\vert \Psi \left( \theta ,Z\right) \right\vert ^{2}U_{0}\left(
\int \left\langle \left\vert \Psi \left( \theta -\frac{\left\vert
Z-Z^{\prime }\right\vert }{c},Z^{\prime }\right) \right\vert
^{2}\right\rangle \right) \simeq 1
\end{eqnarray*}%
Several extensions of the formalism will be considered.\ The details are
left for further research.

\subsection{Including excitatory vs inhibitory interactions}

The previous formalism can be extended to include inhibitory currents. To do
so, two different types of cells, each defined by a different field, are
introduced. We write $\Psi _{1}\left( \theta ,Z,\omega \right) $ and $\Psi
_{2}\left( \theta ,\tilde{Z},\tilde{\omega}\right) $ for excitatory and
inhibitory neurons respectively. A straightforward generalization to an
arbitrary number of types of cells is presented at the end of the section.

\subsubsection{Asymetric interaction between two types of cells}

We consider two types of cells. One set is composed of interacting cells, as
described by the previous formalism, while the other acts as inhibitor or
regulator on the first set. The influence of each type of cell on the other
translates through the actions of the induced currents. Assuming identical
transfer functions for both types of fields, the corresponding potential
terms for the frequencies are, for the first type of cells:%
\begin{eqnarray}
&&\frac{1}{2\eta ^{2}}\int \left\vert \Psi _{1}\left( \theta ,Z,\omega
\right) \right\vert ^{2}\left( \omega ^{-1}-G\left( J\left( \theta ,Z\right)
+\int \frac{\kappa }{N}\frac{\omega _{1}}{\omega }\left\vert \Psi _{1}\left(
\theta -\frac{\left\vert Z-Z_{1}\right\vert }{c},Z_{1},\omega _{1}\right)
\right\vert ^{2}T\left( Z,\theta ,Z_{1}\right) dZ_{1}d\omega _{1}\right.
\right.  \notag \\
&&\left. \left. -\int \frac{\kappa }{N}\frac{\tilde{\omega}_{1}}{\omega }%
\left\vert \Psi _{2}\left( \theta -\frac{\left\vert Z-\tilde{Z}%
_{1}\right\vert }{c},\tilde{Z}_{1},\tilde{\omega}_{1}\right) \right\vert
^{2}T\left( Z,\theta ,\tilde{Z}_{1}\right) d\tilde{Z}_{1}d\tilde{\omega}%
_{1}\right) \right) ^{2}  \label{tpn}
\end{eqnarray}%
and, for the second type:%
\begin{equation}
\frac{1}{2\eta ^{2}}\int \left\vert \Psi _{2}\left( \theta ,\tilde{Z},\tilde{%
\omega}\right) \right\vert ^{2}\left( \tilde{\omega}^{-1}-G\left( J\left(
\theta ,Z\right) +\int \frac{\kappa }{N}\frac{\omega _{1}}{\tilde{\omega}}%
\left\vert \Psi _{1}\left( \theta -\frac{\left\vert Z-Z_{1}\right\vert }{c}%
,Z_{1},\omega _{1}\right) \right\vert ^{2}T\left( Z,\theta ,Z_{1}\right)
dZ_{1}d\omega _{1}\right) \right) ^{2}  \label{tpd}
\end{equation}%
This models the fact that the second system merely inhibits the first one.

As in section 3.2, we project the fields on the frequency-dependent states
defined by (\ref{tpn}) and (\ref{tpd}).\ The resulting action for the system
is:%
\begin{eqnarray}
S &=&-\frac{1}{2}\Psi _{1}^{\dagger }\left( \theta ,Z\right) \nabla _{\theta
}\left( \frac{\sigma _{\theta }^{2}}{2}\nabla _{\theta }-\omega ^{-1}\left(
J,\theta ,Z,\Psi _{1},\Psi _{2}\right) \right) \Psi _{1}\left( \theta
,Z\right)  \label{cnt} \\
&&-\frac{1}{2}\Psi _{2}^{\dagger }\left( \theta ,\tilde{Z}\right) \nabla
_{\theta }\left( \frac{\sigma _{\theta }^{2}}{2}\nabla _{\theta }-\tilde{%
\omega}^{-1}\left( J,\theta ,Z,\Psi _{1},\Psi _{2}\right) \right) \Psi
_{2}\left( \theta ,\tilde{Z}\right)  \notag
\end{eqnarray}%
where the frequencies satisfy:%
\begin{eqnarray}
\omega ^{-1}\left( J,\theta ,Z,\Psi _{1},\Psi _{2}\right) &=&G\left( J\left(
\theta ,Z\right) +\int \frac{\kappa }{N}\frac{\omega \left( J,\theta -\frac{%
\left\vert Z-Z_{1}\right\vert }{c},Z_{1},\Psi \right) }{\omega }\left\vert
\Psi _{1}\left( \theta -\frac{\left\vert Z-Z_{1}\right\vert }{c}%
,Z_{1}\right) \right\vert ^{2}T\left( Z,\theta ,Z_{1}\right) dZ_{1}\right. 
\notag \\
&&\left. -\int \frac{\kappa }{N}\frac{\tilde{\omega}\left( J,\theta -\frac{%
\left\vert Z-Z_{1}\right\vert }{c},\tilde{Z}_{1},\Psi \right) }{\omega }%
\left\vert \Psi _{2}\left( \theta -\frac{\left\vert Z-\tilde{Z}%
_{1}\right\vert }{c},\tilde{Z}_{1}\right) \right\vert ^{2}T\left( Z,\theta ,%
\tilde{Z}_{1}\right) d\tilde{Z}_{1}\right)  \label{bnt} \\
\tilde{\omega}^{-1}\left( J,\theta ,Z,\Psi _{1},\Psi _{2}\right) &=&G\left(
J\left( \theta ,Z\right) +\int \frac{\kappa }{N}\frac{\omega \left( J,\theta
-\frac{\left\vert Z-Z_{1}\right\vert }{c},Z_{1},\Psi \right) T\left(
Z,\theta ,Z_{1},\theta -\frac{\left\vert Z-Z_{1}\right\vert }{c}\right) }{%
\tilde{\omega}}\right.  \notag \\
&&\times \left. \left\vert \Psi _{2}\left( \theta -\frac{\left\vert
Z-Z_{1}\right\vert }{c},Z_{1}\right) \right\vert ^{2}dZ_{1}\right)
\label{btn}
\end{eqnarray}%
Ultimately, a collective potential can be added, as in section 3.3:%
\begin{equation}
\sum_{n=2}^{\infty }\zeta ^{\left( n\right) }\int \left\vert \Psi \left(
\theta ,Z\right) \right\vert ^{2}\left( \dprod\limits_{i=1}^{n-1}\left\vert
\Psi \left( \theta -\frac{\left\vert Z-Z_{i}\right\vert }{c},Z_{i}\right)
\right\vert ^{2}\right)  \label{pcn}
\end{equation}%
where we define:%
\begin{equation*}
\left\vert \Psi \left( \theta ,Z\right) \right\vert ^{2}=\left\vert \Psi
_{1}\left( \theta ,Z\right) \right\vert ^{2}+\left\vert \Psi _{2}\left(
\theta ,Z\right) \right\vert ^{2}
\end{equation*}

Potential (\ref{pcn}) models an equilibrium that results from both
excitatory and inhibitory activities.

\subsubsection{$n$ interacting fields}

The previous $2$-fields description may be generalized to describe $n$
interacting types of cells, with arbitrary interactions. Each type of cells
is caracterized by its frequency $i=1,...,n$, and interacts either
positively or negatively with each other. Each type is defined by a field $%
\Psi _{i}$ and frequencies $\omega _{i}\left( \theta ,Z\right) $. The
general version of (\ref{cnt}), that includes (\ref{pcn}), becomes:%
\begin{eqnarray}
S &=&-\frac{1}{2}\sum_{i}\Psi _{i}^{\dagger }\left( \theta ,Z\right) \nabla
_{\theta }\left( \frac{\sigma _{\theta }^{2}}{2}\nabla _{\theta }-\omega
_{i}^{-1}\left( J,\theta ,Z,\Psi _{1},\Psi _{2}\right) \right) \Psi
_{i}\left( \theta ,Z\right)  \label{tcn} \\
&&+\sum_{i}\sum_{n=2}^{\infty }\zeta _{i}^{\left( n\right) }\int \left\vert
\Psi _{i}\left( \theta ,Z\right) \right\vert ^{2}\left(
\dprod\limits_{i=1}^{n-1}\left\vert \Psi _{i}\left( \theta -\frac{\left\vert
Z-Z_{i}\right\vert }{c},Z_{i}\right) \right\vert ^{2}\right)  \notag
\end{eqnarray}%
and equations for frequencies are defined by:

\begin{eqnarray}
\omega _{i}\left( \theta ,Z\right) &=&F_{i}\left( J\left( \theta \right) +%
\frac{\kappa }{N}\int T\left( Z,Z_{1}\right) \frac{\omega _{j}\left( \theta -%
\frac{\left\vert Z-Z_{1}\right\vert }{c},Z_{1}\right) }{\omega _{i}\left(
\theta ,Z\right) }G^{ij}\right.  \label{ftm} \\
&&\times \left. W\left( \frac{\omega _{i}\left( \theta ,Z\right) }{\omega
_{j}\left( \theta -\frac{\left\vert Z-Z_{1}\right\vert }{c},Z_{1}\right) }%
\right) \left( \mathcal{\bar{G}}_{0j}\left( 0,Z_{1}\right) +\left\vert \Psi
_{j}\left( \theta -\frac{\left\vert Z-Z_{1}\right\vert }{c},Z_{1}\right)
\right\vert ^{2}\right) dZ_{1}\right)  \notag
\end{eqnarray}

The $n\times n$ matrix $G$ has coefficients in the interval $\left[ -1,1%
\right] $. In the sequel, the sum over index $j$ is implicit. For instance,
if $n=2$, the matrix $g$:%
\begin{equation*}
G=\left( 
\begin{array}{cc}
1 & -g \\ 
-g & 0%
\end{array}%
\right)
\end{equation*}%
represents inhibitory interactions between the two populations defined in (%
\ref{tpn}) and (\ref{tpd}).

In the sequel, the computations will focus on the one-field basic model. The
implications for several fields model will be discussed at the end of the
paper.

\section{Effective action}

\subsection{Principle}

Appendices 1, 2 and 3 present the computation of the effective action
associated to the action functional (\ref{lcn}). To do so, appendix 1
computes the two-points Green functions of the system from which the lowest
order expansion in power of field of the effective action is derived, while
appendix 2 considers the whole series of graphs. Appendix 3 then provides a
compact expression for the effective action.

\subsection{Effective action at the tree order}

The effective action at the lowest order in powers of fields is computed
through the two-points Green function, that are computed by a graphs
expansion with free propagator.

\subsubsection{Propagator for the free action}

Appendix 1.1.2 and 1.1.3 compute the two-points Green functions associated
to (\ref{lcn}) by a graph expansion using the propagator associated to the
"free action":%
\begin{equation}
-\Psi ^{\dagger }\left( \theta ,Z\right) \nabla _{\theta }\left( \frac{%
\sigma ^{2}}{2}\nabla _{\theta }-\omega ^{-1}\left( J,\theta ,Z,0\right)
\right) \Psi \left( \theta ,Z\right)  \label{cw}
\end{equation}%
where $\omega ^{-1}\left( J,\theta ,Z,0\right) $ is the inverse frequency
given by (\ref{qf}) for $\Psi \equiv 0$, i.e. \ $\omega ^{-1}\left( J,\theta
,Z,\Psi \right) =G\left( J\left( \theta ,Z\right) \right) $. Action (\ref%
{lcn}) thus decomposes as:%
\begin{equation}
-\frac{1}{2}\Psi ^{\dagger }\left( \theta ,Z\right) \left( \nabla _{\theta
}\left( \frac{\sigma ^{2}}{2}\nabla _{\theta }-\omega ^{-1}\left( J,\theta
,Z,0\right) \right) \right) +\frac{1}{2}\Psi ^{\dagger }\left( \theta
,Z\right) \nabla _{\theta }\omega ^{-1}\left( J,\theta ,Z,\left\vert \Psi
\right\vert ^{2}\right) \Psi \left( \theta ,Z\right) +V\left( \Psi \right)
\label{dcp}
\end{equation}%
and the two last terms in the previous expression are considered
perturbatively in the computation of the graphs.

We find in appendix 1.1.1 that for an external current decomposed in a
static and dynamic part $\bar{J}\left( Z\right) +J\left( Z,\theta \right) $,
the expression for $G\left( J\left( \theta ,Z\right) \right) $ can be
approximated by $G\left( \bar{J}\left( Z\right) \right) $, so that: 
\begin{equation*}
\omega ^{-1}\left( J,\theta ,Z,0\right) =G\left( \bar{J}\left( Z\right)
+J\left( \theta \right) \right) \simeq G\left( \bar{J}\left( Z\right) \right)
\end{equation*}%
Given our choice for the function $G$, we find:%
\begin{equation}
\omega ^{-1}\left( J,\theta ,Z,0\right) \simeq \frac{\arctan \left( \left( 
\frac{1}{X_{r}}-\frac{1}{X_{p}}\right) \sqrt{\bar{J}\left( Z\right) }\right) 
}{\sqrt{\bar{J}\left( Z\right) }}=\frac{1}{\bar{X}_{r}\left( Z\right) }
\label{frz}
\end{equation}%
Using (\ref{frz}), the propagator associated to (\ref{cw}) $\mathcal{G}%
_{0}\left( \theta ,\theta ^{\prime },Z,Z^{\prime }\right) $ can be directly
computed. It satifies:%
\begin{equation*}
-\nabla _{\theta }\left( \frac{\sigma ^{2}}{2}\nabla _{\theta }-\omega
^{-1}\left( J,\theta ,Z,0\right) \right) \mathcal{G}_{0}\left( \theta
,\theta ^{\prime },Z,Z^{\prime }\right) =\delta \left( Z-Z^{\prime }\right)
\delta \left( \theta -\theta ^{\prime }\right)
\end{equation*}%
and we find:%
\begin{equation}
\mathcal{G}_{0}\left( \theta ,\theta ^{\prime },Z,Z^{\prime }\right) =\delta
\left( Z-Z^{\prime }\right) \frac{\exp \left( -\Lambda _{1}\left( Z\right)
\left( \theta -\theta ^{\prime }\right) \right) }{\Lambda \left( Z\right) }%
H\left( \theta -\theta ^{\prime }\right)  \label{rg}
\end{equation}%
where: 
\begin{eqnarray*}
\Lambda \left( Z\right) &=&\sqrt{\frac{\pi }{2}}\sqrt{\left( \frac{1}{\sigma
^{2}\bar{X}_{r}\left( Z\right) }\right) ^{2}+\frac{2\alpha }{\sigma ^{2}}} \\
\Lambda _{1}\left( Z\right) &=&\sqrt{\left( \frac{1}{\sigma ^{2}\bar{X}%
_{r}\left( Z\right) }\right) ^{2}+\frac{2\alpha }{\sigma ^{2}}}-\frac{1}{%
\sigma ^{2}\bar{X}_{r}\left( Z\right) }
\end{eqnarray*}%
and $H$ is the heaviside function:%
\begin{eqnarray*}
H\left( \theta -\theta ^{\prime }\right) &=&0\text{ for }\theta -\theta
^{\prime }<0 \\
&=&1\text{ for }\theta -\theta ^{\prime }>0
\end{eqnarray*}%
For the sake of simplicity, we often discard the factor $\delta \left(
Z-Z^{\prime }\right) $ and write $\mathcal{G}_{0}\left( \theta ,\theta
^{\prime },Z\right) $ for $\mathcal{G}_{0}\left( \theta ,\theta ^{\prime
},Z,Z^{\prime }\right) $. In the sequel, for the sake of simplicity, the
dependency in $Z$ of $\bar{X}_{r}\left( Z\right) $, $\Lambda \left( Z\right) 
$, $\Lambda _{1}\left( Z\right) $ will be implicit, so that we will write: 
\begin{equation*}
\bar{X}_{r}\left( Z\right) \equiv \bar{X}_{r}\text{, }\Lambda \left(
Z\right) \equiv \Lambda \text{, }\Lambda _{1}\left( Z\right) \equiv \Lambda
_{1}
\end{equation*}

\subsubsection{Graphs expansion and two point Green functions}

Having found the propagator $\mathcal{G}_{0}\left( \theta ,\theta ^{\prime
},Z\right) $ associated to (\ref{cw}), appendix 1.1.2 computes the graphs
associated to the decomposition (\ref{dcp}). The two points Green function
is shown to be equal to be the inverse of the operator:%
\begin{equation}
-\frac{1}{2}\nabla _{\theta }\frac{\sigma _{\theta }^{2}}{2}\nabla _{\theta
}+\frac{1}{2}\left[ \frac{\delta \left[ \Psi ^{\dagger }\left( \theta
^{\prime },Z\right) \nabla _{\theta }\left( \omega ^{-1}\left( J,\theta
,Z,\left\vert \Psi \right\vert ^{2}\right) \Psi \left( \theta ,Z\right)
\right) \right] }{\delta \left\vert \Psi \right\vert ^{2}}\right] 
_{\substack{ \theta ^{\prime }=\theta  \\ \left\vert \Psi \left( \theta
,Z\right) \right\vert ^{2}=\mathcal{G}_{0}\left( 0,Z\right) }}+\left[ \frac{%
\delta \left[ V\left( \Psi \right) \right] }{\delta \left\vert \Psi
\right\vert ^{2}}\right] _{\substack{ \left\vert \Psi \left( \theta
,Z\right) \right\vert ^{2}  \\ =\mathcal{G}_{0}\left( 0,Z\right) }}
\label{grt}
\end{equation}%
where:%
\begin{equation*}
\mathcal{G}_{0}\left( 0,Z\right) =\mathcal{G}_{0}\left( \theta ,\theta
,Z\right) =\frac{\exp \left( -\Lambda _{1}\left( Z\right) \left( \theta
-\theta ^{\prime }\right) \right) }{\Lambda \left( Z\right) }H\left( \theta
-\theta ^{\prime }\right)
\end{equation*}
and $\left\vert \Psi \right\vert ^{2}\left[ \frac{\delta }{\delta \left\vert
\Psi \right\vert ^{2}}\right] $ is a shorthand for:%
\begin{equation}
\int dZ^{\prime }\left\vert \Psi \left( \theta -\frac{\left\vert Z-Z^{\prime
}\right\vert }{c},Z^{\prime }\right) \right\vert ^{2}\times \frac{\delta }{%
\delta \left( \left\vert \Psi \left( \theta -\frac{\left\vert Z-Z^{\prime
}\right\vert }{c},Z^{\prime }\right) \right\vert ^{2}\right) }  \label{sht}
\end{equation}

\subsubsection{Effective action at the lowest order}

The effective order at the lowest order is derived directly from (\ref{grt}%
). Appendix 3 shows that:%
\begin{equation}
\Gamma _{0}\left( \Psi ^{\dagger },\Psi \right) =\Psi ^{\dagger }\left(
\theta ,Z\right) \left[ \frac{\delta \left[ S_{cl}\left( \Psi ^{\dagger
},\Psi \right) \right] }{\delta \left\vert \Psi \right\vert ^{2}}\right] 
_{\substack{ \left\vert \Psi \left( \theta ,Z\right) \right\vert ^{2}  \\ =%
\mathcal{G}_{0}\left( 0,Z\right) }}\Psi \left( \theta ,Z\right)
\label{fctnbs}
\end{equation}%
with:%
\begin{eqnarray}
&&S_{cl}\left( \Psi ^{\dagger },\Psi \right) =-\frac{1}{2}\Psi ^{\dagger
}\left( \theta ,Z\right) \left( \nabla _{\theta }\left( \frac{\sigma
_{\theta }^{2}}{2}\nabla _{\theta }-\omega ^{-1}\left( J,\theta
,Z,\left\vert \Psi \right\vert ^{2}\right) \right) \right) \Psi \left(
\theta ,Z\right)  \label{scL} \\
&&+\alpha \int \left\vert \Psi \left( \theta ^{\left( i\right)
},Z_{i}\right) \right\vert ^{2}+V\left( \Psi \right)  \notag
\end{eqnarray}%
and the brackets notation given in equation (\ref{sht}). Alternatively,
formula (\ref{fctnbs}) can also be written:

\begin{eqnarray}
\Gamma _{0}\left( \Psi ^{\dagger },\Psi \right) &=&-\frac{1}{2}\Psi
^{\dagger }\left( \theta ,Z\right) \left( \nabla _{\theta }\frac{\sigma ^{2}%
}{2}\left( \nabla _{\theta }-\left( \omega ^{-1}\left( \bar{J},Z,\mathcal{G}%
_{0}\right) +\frac{\delta \left[ \omega ^{-1}\left( \bar{J},Z,\mathcal{G}%
_{0}\right) \right] }{\delta \mathcal{G}_{0}\left( 0,Z\right) }\mathcal{G}%
_{0}\left( \theta ^{\prime },\theta ,Z\right) \right) \right) \right)
_{\theta ^{\prime }=\theta }\Psi \left( \theta ,Z\right)  \notag \\
&&+\alpha \int \left\vert \Psi \left( \theta ,Z\right) \right\vert ^{2}+\Psi
^{\dagger }\left( \theta ,Z\right) \left[ \frac{\delta \left[ V\left( \Psi
\right) \right] }{\delta \left\vert \Psi \right\vert ^{2}}\right] 
_{\substack{ \left\vert \Psi \left( \theta ,Z\right) \right\vert ^{2}  \\ =%
\mathcal{G}_{0}\left( 0,Z\right) }}\Psi \left( \theta ,Z\right)  \label{Rf}
\end{eqnarray}%
where $\omega ^{-1}\left( \bar{J},Z,\mathcal{G}_{0}\right) $ is the static
inversed frequency defined as the solution of the equation: 
\begin{equation}
\omega ^{-1}\left( \bar{J}\left( Z\right) ,Z,\mathcal{G}_{0}\right) =G\left( 
\bar{J}\left( Z\right) +\int \frac{\kappa }{N}\frac{\omega \left( \bar{J}%
,Z_{1},\mathcal{G}_{0}\right) }{\omega \left( \bar{J},Z,\mathcal{G}%
_{0}\right) }\mathcal{G}_{0}\left( 0,Z_{1}\right) T\left( Z,\theta
,Z_{1}\right) dZ_{1}\right)  \label{cp}
\end{equation}%
In formula (\ref{cp}), $\bar{J}\left( Z\right) $ is the average over time of 
$J\left( \theta ,Z\right) $. As a consequence, $\omega ^{-1}\left( \bar{J}%
\left( Z\right) ,Z,\mathcal{G}_{0}\right) $ solves:%
\begin{equation}
\omega ^{-1}\left( \bar{J}\left( Z\right) ,Z,\mathcal{G}_{0}\right) =G\left( 
\bar{J}\left( Z\right) +\int \frac{\kappa }{N}\frac{\omega ^{-1}\left( \bar{J%
}\left( Z_{1}\right) ,Z_{1},\mathcal{G}_{0}\right) }{\omega ^{-1}\left( \bar{%
J}\left( Z\right) ,Z,\mathcal{G}_{0}\right) }T\left( Z,\theta ,Z_{1}\right) 
\mathcal{G}_{0}\left( 0,Z_{1}\right) dZ_{1}\right)  \label{Cp}
\end{equation}%
Once $\omega ^{-1}\left( Z,\mathcal{G}_{0}\right) $ is known, (\ref{Rf})
implies that the effective action at the tree-order is given by:where $%
V\left( \Psi \right) $ is defined in (\ref{pp}).

\subsection{Effective action at higher orders}

\subsubsection{General formula}

Appendix $2$ shows that the effective action is a series of corrections to
the classical effective action (\ref{fctnbs}):%
\begin{equation*}
\Gamma \left( \Psi ^{\dagger },\Psi \right) =\sum_{n=2}^{\infty }\int \left(
\dprod\limits_{l=1}^{n}\Psi ^{\dagger }\left( \theta _{f}^{\left( l\right)
},Z_{l}\right) \right) S_{n}\left( \left( \theta _{f}^{\left( l\right)
},\theta _{i}^{\left( l\right) },Z_{l}\right) \right) \left(
\dprod\limits_{l=1}^{n}\Psi \left( \theta _{i}^{\left( l\right)
},Z_{l}\right) \right)
\end{equation*}%
The contribution $S_{n}\left( \left( \theta _{f}^{\left( l\right) },\theta
_{i}^{\left( l\right) },Z_{l}\right) \right) $ for a given $n$ is the $n$
points effective vertex. It is the sum of one-particle irreducible graphs ($%
1 $PI) with $n$ lines labelled by their position $Z_{l}$, $l=1,...,n$ \ plus
their starting and ending points $\left( \theta _{f}^{\left( l\right)
},\theta _{i}^{\left( l\right) }\right) $.

To build the series of graphs, we first consider the $l$ points vertices:%
\begin{eqnarray*}
\hat{V}_{2l}\left( \left\{ \left( \theta ^{\left( k_{i}\right)
},Z_{k_{i}}\right) \right\} _{i=1,...,l}\right) &=&\frac{1}{l!}\left[ \frac{%
\delta ^{l}\left[ \int \Psi ^{\dagger }\left( \theta ,Z\right) \nabla
_{\theta }\omega ^{-1}\left( J,\theta ,Z\right) \Psi \left( \theta ,Z\right)
dZd\theta +V\left( \Psi \right) \right] }{\dprod\limits_{i=1}^{l}\delta
\left\vert \Psi \left( \theta ^{\left( k_{i}\right) },Z_{k_{i}}\right)
\right\vert ^{2}}\right] _{\left\vert \Psi \left( \theta ,Z\right)
\right\vert ^{2}=\mathcal{G}_{0}\left( 0,Z\right) } \\
&=&\frac{1}{l!}\left[ \frac{\delta ^{l}\left[ S_{cl}\left( \Psi ^{\dagger
},\Psi \right) \right] }{\dprod\limits_{i=1}^{l}\delta \left\vert \Psi
\left( \theta ^{\left( k_{i}\right) },Z_{k_{i}}\right) \right\vert ^{2}}%
\right] _{\left\vert \Psi \left( \theta ,Z\right) \right\vert ^{2}=\mathcal{G%
}_{0}\left( 0,Z\right) }
\end{eqnarray*}%
for $l=2,...,n$ and $\left( \theta ^{\left( k_{i}\right) },Z_{k_{i}}\right)
\in \left\{ \left( \theta _{i},Z_{i}\right) \right\} _{i=1,...,n}$ with $%
\theta _{i}\in \left[ \theta _{i}^{\left( i\right) },\theta _{f}^{\left(
i\right) }\right] $ and $\left( \theta ^{\left( k_{i}\right)
},Z_{k_{i}}\right) \neq \left( \theta ^{\left( k_{j}\right)
},Z_{k_{j}}\right) $ for $i\neq j$.

These vertices are represented graphically by associating a point $\left(
\theta ,Z\right) _{\hat{V}}$ to each vertex $\hat{V}$, which differs from
all the $\left( \theta ^{\left( i\right) },Z_{i}\right) $ and $\left( \theta
,Z\right) _{\hat{V}}\neq \left( \theta ,Z\right) _{\hat{V}^{\prime }}$ when $%
\hat{V}\neq \hat{V}^{\prime }$. We draw $l$ lines from $\left( \theta
,Z\right) _{\hat{V}}$ ending at the points $\left( \theta ^{\left(
k_{i}\right) },Z_{k_{i}}\right) $.

We then consider the series of graphs $l=2,...,n$, with an arbitrary number
of vertices $\hat{V}_{2l}\left( \left\{ \left( \theta ^{\left( k_{i}\right)
},Z_{k_{i}}\right) \right\} _{i=1,...,l}\right) $, joining the points $%
\left( \theta ^{\left( k_{i}\right) },Z_{k_{i}}\right) $. To each internal
segment between $\theta $ and $\theta ^{\prime }$ at position $Z$, we
associate a propagator $\mathcal{G}_{0}\left( \theta ,\theta ^{\prime
},Z\right) $ defined in (\ref{rg}).\ The effective vertex $S_{n}\left(
\left( \theta _{f}^{\left( l\right) },\theta _{i}^{\left( l\right)
},Z_{l}\right) \right) $\ is obtained by summing the $n$ lines-$1$PI graphs.

Defining:%
\begin{eqnarray}
\hat{S}_{cl}\left( \Psi ^{\dagger },\Psi \right) &\equiv &S_{cl}\left( 
\mathcal{G}_{0}\left( 0,Z\right) +\left\vert \Psi \right\vert ^{2}\right)
\label{fth} \\
&\equiv &-\frac{1}{2}\left( \left( \nabla _{\theta }\left( \frac{\sigma
_{\theta }^{2}}{2}\nabla _{\theta }-\omega ^{-1}\left( \mathcal{G}_{0}\left(
0,Z\right) +\left\vert \Psi \left( \theta ,Z\right) \right\vert ^{2}\right)
\right) \right) \left( \mathcal{G}_{0}\left( \theta ^{\prime },\theta
,Z\right) +\Psi ^{\dagger }\left( \theta ^{\prime },Z\right) \Psi \left(
\theta ,Z\right) \right) \right) _{\theta ^{\prime }=\theta }  \notag \\
&&+\alpha \int \left( \mathcal{G}_{0}\left( 0,Z\right) \mathcal{+}\left\vert
\Psi \left( \theta ,Z\right) \right\vert ^{2}\right) +V\left( \left( 
\mathcal{G}_{0}\left( 0,Z\right) \mathcal{+}\left\vert \Psi \left( \theta
,Z\right) \right\vert ^{2}\right) \right)  \notag
\end{eqnarray}%
with $V\left( \left( \mathcal{G}_{0}\left( 0,Z\right) \mathcal{+}\left\vert
\Psi \left( \theta ,Z\right) \right\vert ^{2}\right) \right) $ given by (\ref%
{pp}):%
\begin{equation}
V\left( \mathcal{G}_{0}\left( 0,Z\right) \mathcal{+}\left\vert \Psi \left(
\theta ,Z\right) \right\vert ^{2}\right) =\int \left( \mathcal{G}_{0}\left(
0,Z\right) \mathcal{+}\left\vert \Psi \left( \theta ,Z\right) \right\vert
^{2}\right) U_{0}\left( \int \mathcal{G}_{0}\left( 0,Z^{\prime }\right)
+\left\vert \Psi \left( \theta -\frac{\left\vert Z-Z^{\prime }\right\vert }{c%
},Z^{\prime }\right) \right\vert ^{2}\right)  \label{ptd}
\end{equation}%
the effective action writes as a series expansion (see appendix 3):

\begin{eqnarray}
\Gamma \left( \Psi ^{\dagger },\Psi \right) &=&\hat{S}_{cl}\left( \Psi
^{\dagger },\Psi \right) +\sum_{\substack{ j\geqslant 2  \\ m\geqslant 2}}%
\sum _{\substack{ \left( p_{l}^{i}\right) _{m\times j}  \\ %
\sum_{i}p_{l}^{i}\geqslant 2}}\int \left( \dprod\limits_{l=1}^{j}\Psi
^{\dagger }\left( \theta _{f}^{\left( l\right) },Z_{l}\right) \right)  \notag
\\
&&\times \dprod\limits_{i=1}^{m}\left[ \underset{\prod\limits_{l}\left[
\theta _{i}^{\left( l\right) },\theta _{f}^{\left( l\right) }\right]
^{p_{l}^{i}}}{\int }\frac{\delta ^{\sum_{l}p_{l}^{i}}\left[ \hat{S}%
_{cl}\left( \Psi ^{\dagger },\Psi \right) \right] }{\dprod\limits_{l=1}^{j}%
\dprod\limits_{k_{l}^{i}=1}^{p_{l}^{i}}\delta \left\vert \Psi \left( \theta
^{\left( k_{l}^{i}\right) },Z_{_{l}}\right) \right\vert ^{2}}%
\dprod\limits_{l=1}^{j}\dprod\limits_{k_{l}^{i}=1}^{p_{l}^{i}}d\theta
^{\left( k_{l}^{i}\right) }\right]  \notag \\
&&\times \frac{\dprod\limits_{l=1}^{j}\exp \left( -\Lambda _{1}\left( \theta
_{f}^{\left( l\right) }-\theta _{i}^{\left( l\right) }\right) \right) }{%
m!\dprod\limits_{k}\left( \sharp _{j,m,k}\left( \left( p_{l}^{i}\right)
\right) \right) !\Lambda ^{\sum_{i,l}p_{l}^{i}}}\left(
\dprod\limits_{l=1}^{j}\Psi \left( \theta _{i}^{\left( l\right)
},Z_{l}\right) \right)  \label{ffc}
\end{eqnarray}%
with $\left( \sharp _{j,m,k}\left( \left( p_{l}^{i}\right) \right) \right) !$
standing for the number of external lines with multiple lines of valence $%
k\geqslant 2$:%
\begin{equation}
\sharp _{j,m,k}\left( \left( p_{l}^{i}\right) \right) =\sum_{l=1}^{j}\delta
_{k,\sum_{i=1}^{m}p_{l}^{i}}  \label{dsM}
\end{equation}%
and where:%
\begin{equation*}
\theta _{i}^{\left( l\right) }<\theta ^{\left( k_{l}^{i}\right) }<\theta
_{f}^{\left( l\right) }
\end{equation*}%
The notation $\left( p_{l}^{i}\right) $ in (\ref{dsM}) stands for the
dependency of $\sharp _{j,m,k}\left( \left( p_{l}^{i}\right) \right) $ in
the whole set of indices $\left( p_{l}^{i}\right) $ with $i=1...m$ and $%
l=1...j$.

In the local approximation, when $\Lambda _{1}>1$, $\exp \left( -\Lambda
_{1}\left( \theta _{f}^{\left( l\right) }-\theta _{i}^{\left( l\right)
}\right) \right) $ can be replaced by $\frac{1}{\Lambda _{1}}\delta \left(
\theta _{f}^{\left( l\right) }-\theta _{i}^{\left( l\right) }\right) $ and
the effective action writes:%
\begin{equation}
\hat{S}_{cl}\left( \Psi ^{\dagger },\Psi \right) +\sum_{j\geqslant 2}\int %
\left[ \left( \dprod\limits_{l=1}^{j}\Psi ^{\dagger }\left( \theta ^{\left(
l\right) },Z_{l}\right) \right) \sum_{\substack{ m\geqslant 1,\left(
p_{l}^{i}\right) _{m\times j}  \\ \sum_{i}p_{l}^{i}\geqslant 2}}\frac{%
\dprod\limits_{i=1}^{m}\left[ \frac{\delta ^{\sum_{l}p_{l}^{i}}\left[ \hat{S}%
_{cl}\left( \Psi ^{\dagger },\Psi \right) \right] }{\dprod\limits_{l=1}^{j}%
\delta ^{\sum_{l}p_{l}^{i}}\left\vert \Psi \left( \theta ^{\left( l\right)
},Z_{_{l}}\right) \right\vert ^{2}}\right] }{m!\dprod\limits_{k}\left(
\sharp _{j,m,k}\left( \left( p_{l}^{i}\right) \right) \right) !\Lambda
_{1}^{j}\Lambda ^{\sum_{i,l}p_{l}^{i}}}\left( \dprod\limits_{l=1}^{j}\Psi
\left( \theta ^{\left( l\right) },Z_{l}\right) \right) \right]  \label{lcf}
\end{equation}

\paragraph{Estimation of the effective action's series expansion}

A series expansion for (\ref{lcf}) can be derived. For strong fields, we
have:%
\begin{equation*}
\frac{\delta ^{\sum_{l}p_{l}^{i}}\left[ \hat{S}_{cl}\left( \Psi ^{\dagger
},\Psi \right) \right] }{\dprod\limits_{l=1}^{j}\dprod%
\limits_{k_{l}^{i}=1}^{p_{l}^{i}}\delta \left\vert \Psi \left( \theta
^{\left( l\right) },Z_{_{l}}\right) \right\vert ^{2}}\simeq \frac{1}{2}\int 
\frac{\delta ^{\sum_{l}p_{l}^{i}}\nabla _{\theta }\omega ^{-1}\left(
\left\vert \Psi \left( \theta ,Z\right) \right\vert ^{2}\right) }{%
\dprod\limits_{l=1}^{j}\dprod\limits_{k_{l}^{i}=1}^{p_{l}^{i}}\delta
\left\vert \Psi \left( \theta ^{\left( l\right) },Z_{_{l}}\right)
\right\vert ^{2}}\left\vert \Psi \left( \theta ,Z\right) \right\vert ^{2}
\end{equation*}%
The derivatives of $\omega ^{-1}\left( J,\theta ,Z,\mathcal{G}_{0}\left(
0,Z\right) +\left\vert \Psi \right\vert ^{2}\right) $ can be estimated as:

\begin{eqnarray}
\frac{\delta ^{n}\omega ^{-1}\left( J,\theta ,Z\right) }{\dprod%
\limits_{i=1}^{n}\delta \left\vert \Psi \left( \theta -l_{i},Z_{i}\right)
\right\vert ^{2}} &\simeq &\frac{\exp \left( -cl_{n}-\alpha \left(
\sum_{r=1}^{n-1}\left( \left( c\left( l_{r}-l_{r+1}\right) \right)
^{2}-\left\vert Z_{r}-Z_{r+1}\right\vert ^{2}\right) \right) \right) }{D^{n}}
\label{nps} \\
&&\times H\left( cl_{n}-\sum_{i=1}^{n-1}\left\vert Z_{i}-Z_{i+1}\right\vert
\right) \dprod\limits_{i=1}^{n}\frac{\omega ^{-1}\left( J,\theta
-l_{i},Z_{i}\right) }{\left\vert \Psi \left( \theta -l_{i},Z_{l}\right)
\right\vert ^{2}}  \notag
\end{eqnarray}%
where $D$ is a constant (see appendix 6).

As a consequence, ordering the $\theta ^{\left( l\right) }$ by $\theta
^{\left( l\right) }<...<\theta ^{\left( 1\right) }$, we define $\theta
_{i,j}=\min_{l,p_{l}^{i}\neq 0}\left( \theta _{l}\right) $. 
\begin{eqnarray}
&&\frac{\delta ^{\sum_{l}p_{l}^{i}}\left[ \hat{S}_{cl}\left( \Psi ^{\dagger
},\Psi \right) \right] }{\dprod\limits_{l=1}^{j}\delta
^{p_{l}^{i}}\left\vert \Psi \left( \theta ^{\left( l\right)
},Z_{_{l}}\right) \right\vert ^{2}}  \label{spn} \\
&\simeq &-\frac{c}{2}\int \frac{\exp \left( -c\left( \theta _{i}-\theta
_{i,j}\right) -\alpha \left( \sum_{l=1,p_{l}^{i}\neq 0}^{j}\left( \left(
c\left( \theta ^{\left( l-1,i\right) }-\theta ^{\left( l,i\right) }\right)
\right) ^{2}-\left\vert Z_{l-1}^{\left( i\right) }-Z_{l}^{\left( i\right)
}\right\vert ^{2}\right) \right) \right) }{D^{\sum_{l}p_{l}^{i}}}  \notag \\
&&\times H\left( c\left( \theta _{i}-\theta _{i,j}\right)
-\sum_{l=1,p_{l}^{i}\neq 0}^{j}\left\vert Z_{l-1}^{\left( i\right)
}-Z_{l}^{\left( i\right) }\right\vert \right) \dprod\limits_{l=1}^{j}\left( 
\frac{\omega ^{-1}\left( J,\theta ^{\left( l\right) },Z_{_{l}}\right) }{%
\left\vert \Psi \left( \theta ^{\left( l\right) },Z_{_{l}}\right)
\right\vert ^{2}}\right) ^{\sum_{i}p_{l}^{i}}\left\vert \Psi \left( \theta
_{i},Z_{i}\right) \right\vert ^{2}d\theta _{i}dZ_{i}  \notag
\end{eqnarray}%
We use the convention $\theta ^{\left( l,i\right) }=\theta ^{\left( l\right)
}$, $Z_{l}^{\left( i\right) }=Z_{l}$ for $l>0$ and $\theta ^{\left(
l,0\right) }=\theta _{i}$, $Z_{l}^{\left( 0\right) }=Z_{i}$. We can thus
write (\ref{lcf}) as:%
\begin{eqnarray}
&&\hat{S}_{cl}\left( \Psi ^{\dagger },\Psi \right) -\sum_{j\geqslant 1}\int
\sum_{\substack{ m\geqslant 2,\left( p_{l}^{i}\right) _{m\times j}  \\ %
\sum_{i}p_{l}^{i}\geqslant 2}}\dprod\limits_{l=1}^{j}d\theta ^{\left(
l\right) }dZ_{l}\left( \frac{\omega ^{-1}\left( J,\theta ^{\left( l\right)
},Z_{_{l}}\right) }{\left\vert \Psi \left( \theta ^{\left( l\right)
},Z_{_{l}}\right) \right\vert ^{2}}\right) ^{\sum_{i}p_{l}^{i}}\left\vert
\Psi \left( \theta ^{\left( l\right) },Z_{l}\right) \right\vert ^{2}
\label{cls} \\
&&\times \frac{\dprod\limits_{i=1}^{m}\int c\exp \left( -c\left( \theta
_{i}-\theta _{i,j}\right) -\alpha \left( \sum_{l=1,p_{l}^{i}\neq
0}^{j}\left( \left( c\left( \theta ^{\left( l-1,i\right) }-\theta ^{\left(
l,i\right) }\right) \right) ^{2}-\left\vert Z_{l-1}^{\left( i\right)
}-Z_{l}^{\left( i\right) }\right\vert ^{2}\right) \right) \right) \left\vert
\Psi \left( \theta _{i},Z_{i}\right) \right\vert ^{2}d\theta _{i}dZ_{i}}{%
\left( -2\right) ^{m}D^{\sum_{i,l}p_{l}^{i}}m!\dprod\limits_{k}\left( \sharp
_{k}\right) !\Lambda _{1}^{j}\Lambda ^{\sum_{i,l}p_{l}^{i}}}  \notag
\end{eqnarray}%
where the expressions in the sum include an implicit Heaviside function: 
\begin{equation*}
H\left( -c\left( \theta _{i}-\theta _{i,j}\right) -\alpha \left(
\sum_{l=1,p_{l}^{i}\neq 0}^{j}\left( \left( c\left( \theta ^{\left(
l-1,i\right) }-\theta ^{\left( l,i\right) }\right) \right) ^{2}-\left\vert
Z_{l-1}^{\left( i\right) }-Z_{l}^{\left( i\right) }\right\vert ^{2}\right)
\right) \right)
\end{equation*}

\subsubsection{Alternative form for the effective action}

An alternative form for the effective action can be derived for
computational reasons (see appendix 3.3). We give its full form, and its
local approximation, which is more tractable.

\paragraph{Full form}

We rewrite $\hat{S}_{cl}\left( \Psi ^{\dagger },\Psi \right) $, up to the
constant $\alpha \int \mathcal{G}_{0}\left( 0,Z_{i}\right) $:%
\begin{eqnarray*}
\hat{S}_{cl}\left( \Psi ^{\dagger },\Psi \right) &\simeq &\int \Psi
^{\dagger }\left( \theta ,Z\right) \left( -\frac{1}{2}\left( \nabla _{\theta
}\left( \frac{\sigma _{\theta }^{2}}{2}\nabla _{\theta }-\omega ^{-1}\left(
\left\vert \Psi \left( \theta ,Z\right) \right\vert ^{2}\right) \right)
\right) +\alpha +U\left( \int \left\vert \Psi \left( \theta -\frac{%
\left\vert Z-Z^{\prime }\right\vert }{c},Z^{\prime }\right) \right\vert
^{2}\right) \right) \Psi \left( \theta ,Z\right) \\
&\equiv &\int \Psi ^{\dagger }\left( \theta ,Z\right) L\left( \Psi ^{\dagger
}\left( \theta ,Z\right) ,\Psi \left( \theta ,Z\right) \right) \Psi \left(
\theta ,Z\right)
\end{eqnarray*}%
In the above:

\begin{equation*}
U\left( \int \left\vert \Psi \left( \theta -\frac{\left\vert Z-Z^{\prime
}\right\vert }{c},Z^{\prime }\right) \right\vert ^{2}\right)
\end{equation*}
is obtained by the series expansion of: 
\begin{equation*}
V\left( \mathcal{G}_{0}\left( 0,Z\right) \mathcal{+}\left\vert \Psi \left(
\theta ,Z\right) \right\vert ^{2}\right)
\end{equation*}
defined in (\ref{ptd}), whose terms of degree $2$ and higher in fields have
been collected.

The effective action is given by:%
\begin{eqnarray}
&&\Gamma \left( \Psi ^{\dagger },\Psi \right) =\hat{S}_{cl}\left( \Psi
^{\dagger },\Psi \right) +\sum_{\substack{ j\geqslant 1  \\ m\geqslant 1}}%
\sum _{\substack{ p_{l},\left( p_{l}^{i}\right) _{m\times j}  \\ %
p_{l}+\sum_{i}p_{l}^{i}\geqslant 2}}\int \Psi ^{\dagger }\left( \theta
,Z\right) \frac{\delta ^{\sum_{l}p_{l}}\left[ L\left( \Psi ^{\dagger }\left(
\theta ,Z\right) ,\Psi \left( \theta ,Z\right) \right) \right] }{%
\dprod\limits_{l=1}^{j}\dprod\limits_{k_{l}^{i}=1}^{p_{l}}\delta \left\vert
\Psi \left( \theta ^{\left( k_{l}^{i}\right) },Z_{_{l}}\right) \right\vert
^{2}}a_{j,m}\left( \theta ,\theta _{i}\right) \Psi \left( \theta
_{i},Z\right)  \notag \\
&&\times \int \left( \dprod\limits_{l=1}^{j}\Psi ^{\dagger }\left( \theta
_{f}^{\left( l\right) },Z_{l}\right) \right) \dprod\limits_{i=1}^{m}\left[
\int_{\dprod\limits_{l=1}^{j}\left[ \theta _{i}^{\left( l\right) },\theta
_{f}^{\left( l\right) }\right] ^{p_{l}^{i}}}\frac{\delta ^{\sum_{l}p_{l}^{i}}%
\left[ \hat{S}_{cl,\theta }\left( \Psi ^{\dagger },\Psi \right) \right] }{%
\dprod\limits_{l=1}^{j}\dprod\limits_{k_{l}^{i}=1}^{p_{l}^{i}}\delta
\left\vert \Psi \left( \theta ^{\left( k_{l}^{i}\right) },Z_{_{l}}\right)
\right\vert ^{2}}\dprod\limits_{l=1}^{j}\dprod%
\limits_{k_{l}^{i}=1}^{p_{l}^{i}}d\theta ^{\left( k_{l}^{i}\right) }\right] 
\notag \\
&&\times \frac{\exp \left( -\Lambda _{1}\left( \theta _{f}^{\left( l\right)
}-\theta _{i}^{\left( l\right) }\right) \right) }{m!\sharp _{j,m}\left(
\left( p_{l}^{i}\right) \right) \Lambda ^{\sum_{i}p_{l}^{i}}}\left(
\dprod\limits_{l=1}^{j}\Psi \left( \theta _{i}^{\left( l\right)
},Z_{l}\right) \right)  \label{ltn}
\end{eqnarray}%
where the kernel $a_{j,m}\left( \theta ,\theta _{i}\right) $ is defined as: 
\begin{equation}
a_{1}\left( \theta ,\theta _{i}\right) =\frac{\exp \left( -\Lambda
_{1}\left( \theta -\theta _{i}^{\left( l\right) }\right) \right) }{\Lambda
^{\sum_{i}p_{l}^{i}}}\prod\limits_{i=1}^{m}\int_{\left[ \theta _{i}^{\left(
l\right) },\theta _{f}^{\left( l\right) }\right] ^{p^{i}}}\frac{\delta
^{p^{i}}}{\dprod\limits_{k_{l}^{i}=1}^{p^{i}}\delta \left\vert \Psi \left(
\theta ^{\left( k^{i}\right) },Z_{m}\right) \right\vert ^{2}}d\theta
^{\left( k^{i}\right) }  \label{krt}
\end{equation}%
\ and 
\begin{equation}
a_{j,m}\left( \theta ,\theta _{i}\right) =\delta \left( \theta -\theta
_{i}\right) +\frac{\sharp _{j+1,m+1}\left( \left( p_{l}\right) ,\left(
p_{l}^{i}\right) \right) }{\bar{\sharp}_{j+1,m+1}\left( \left( p_{l}\right)
,\left( p_{l}^{i}\right) \right) }\frac{\exp \left( -\Lambda _{1}\left(
\theta -\theta _{i}^{\left( l\right) }\right) \right) }{\Lambda
^{\sum_{i}p_{l}^{i}}}\prod\limits_{i=1}^{m}\int_{\left[ \theta _{i}^{\left(
l\right) },\theta _{f}^{\left( l\right) }\right] ^{p^{i}}}\frac{\delta
^{p^{i}}}{\dprod\limits_{k_{l}^{i}=1}^{p^{i}}\delta \left\vert \Psi \left(
\theta ^{\left( k^{i}\right) },Z_{m}\right) \right\vert ^{2}}d\theta
^{\left( k^{i}\right) }  \label{krT}
\end{equation}%
for $j>1$. The factors $\sharp _{j,m}$ and $\frac{1}{\bar{\sharp}_{j,m}}$\
are given by:%
\begin{eqnarray}
\sharp _{j,m}\left( \left( p_{l}^{i}\right) \right)
&=&\dprod\limits_{k}\left( \bar{\sharp}_{j,m,k}\left( p_{l}^{i}\right)
\right) !=\left( \sum_{l=1}^{j}\delta _{k,\sum_{i=1}^{m}p_{l}^{i}}\right) !
\label{dsL} \\
\frac{1}{\bar{\sharp}_{j,m}\left( \left( p_{l}^{i}\right) \right) } &=&\frac{%
1}{\dprod\limits_{k}\left( \bar{\sharp}_{j,m,k}\left( p_{l}^{i}\right)
\right) !}=\sum_{p=1}^{j}\frac{1}{\dprod\limits_{k}\left(
\sum_{l=1}^{j}\delta _{k,\sum_{i=1}^{m}p_{l}^{i}+\delta _{l,p}}\right) !}
\label{dSM}
\end{eqnarray}%
The expression $\sharp _{j,m,k}\left( \left( p_{l}^{i}\right) \right) $ has
been defined in (\ref{dsM}). The notations $\sharp _{j+1,m+1}\left( \left(
p_{l}\right) ,\left( p_{l}^{i}\right) \right) $ and $\bar{\sharp}%
_{j+1,m+1}\left( \left( p_{l}\right) ,\left( p_{l}^{i}\right) \right) $ in (%
\ref{krT}) are defined by (\ref{dsL}) and (\ref{dSM}) in which the
multi-indices $\left( p_{l}^{i}\right) _{l=1,...l}^{i=1,...m}$ are replaced
by the collection obtained by gathering $\left( p_{l}\right) _{i=1...m}$ and 
$\left( p_{l}^{i}\right) _{l=1,...l}^{i=1,...m}$.

The derivatives $i=1,,,,m$ implicitly act independently on each factor:%
\begin{equation*}
\int_{\dprod\limits_{l=1}^{j}\left[ \theta _{i}^{\left( l\right) },\theta
_{f}^{\left( l\right) }\right] ^{p_{l}^{i}}}\frac{\delta ^{\sum_{l}p_{l}^{i}}%
\left[ \hat{S}_{cl,\theta }\left( \Psi ^{\dagger },\Psi \right) \right] }{%
\dprod\limits_{l=1}^{j}\dprod\limits_{k_{l}^{i}=1}^{p_{l}^{i}}\delta
\left\vert \Psi \left( \theta ^{\left( k_{l}^{i}\right) },Z_{_{l}}\right)
\right\vert ^{2}}\dprod\limits_{l=1}^{j}\dprod%
\limits_{k_{l}^{i}=1}^{p_{l}^{i}}d\theta ^{\left( k_{l}^{i}\right) }
\end{equation*}%
in (\ref{ltn}).

\paragraph{Local approximation}

In the local approximation, (\ref{ltn}) simplifies and becomes:%
\begin{eqnarray}
&&\Gamma \left( \Psi ^{\dagger },\Psi \right) =\hat{S}_{cl}\left( \Psi
^{\dagger },\Psi \right) +\sum_{\substack{ j\geqslant 1  \\ m\geqslant 1}}%
\sum _{\substack{ p_{l},\left( p_{l}^{i}\right) _{m\times j}  \\ %
p_{l}+\sum_{i}p_{l}^{i}\geqslant 2}}\int \int \Psi ^{\dagger }\left( \theta
,Z\right) \frac{\delta ^{\sum_{l}p_{l}}\left[ L\left( \Psi ^{\dagger }\left(
\theta ,Z\right) ,\Psi \left( \theta ,Z\right) \right) \right] }{%
\dprod\limits_{l=1}^{j}\dprod\limits_{k_{l}^{i}=1}^{p_{l}}\delta \left\vert
\Psi \left( \theta ^{\left( l\right) },Z_{_{l}}\right) \right\vert ^{2}}\Psi
\left( \theta ,Z\right)  \label{ltN} \\
&&\times \left( \dprod\limits_{l=1}^{j}\Psi ^{\dagger }\left( \theta
^{\left( l\right) },Z_{l}\right) \right) \dprod\limits_{i=1}^{m}\left[ \frac{%
\delta ^{\sum_{l}p_{l}^{i}}\left[ \hat{S}_{cl,\theta }\left( \Psi ^{\dagger
},\Psi \right) \right] }{\dprod\limits_{l=1}^{j}\delta
^{p_{l}^{i}}\left\vert \Psi \left( \theta _{l},Z_{_{l}}\right) \right\vert
^{2}}\right] \frac{\left( \dprod\limits_{l=1}^{j}\Psi \left( \theta ^{\left(
l\right) },Z_{l}\right) \right) }{m!\left( \sharp _{j,m}\left( \left(
p_{l}^{i}\right) \right) \right) \Lambda ^{\sum_{i}p_{l}^{i}}}%
\dprod\limits_{l=1}^{j}dZ_{l}d\theta _{l}  \notag
\end{eqnarray}%
with:%
\begin{equation*}
a_{1,m}=1
\end{equation*}%
\ and 
\begin{equation*}
a_{j,m}=1+\frac{\sharp _{j+1,m+1}\left( \left( p_{l}\right) ,\left(
p_{l}^{i}\right) \right) }{\bar{\sharp}_{j+1,m+1}\left( \left( p_{l}\right)
,\left( p_{l}^{i}\right) \right) }
\end{equation*}%
for $j>1$

\subsubsection{Estimation of the series expansion for the alternative form
of the effective action}

The effective action can be approximated using (\ref{ltn}). We first
estimate the derivatives of $\hat{S}_{cl}\left( \Psi ^{\dagger },\Psi
\right) $. For a slowly varying potential $V\left( \Psi \right) $, we have:%
\begin{equation*}
\frac{\delta ^{\sum_{l}p_{l}^{i}}\left[ \hat{S}_{cl}\left( \Psi ^{\dagger
},\Psi \right) \right] }{\dprod\limits_{l=1}^{j}\delta
^{\sum_{l}p_{l}^{i}}\left\vert \Psi \left( \theta ^{\left( l\right)
},Z_{_{l}}\right) \right\vert ^{2}}\simeq \frac{\delta ^{\sum_{l}p_{l}^{i}}%
\left[ \int \Psi ^{\dagger }\left( \theta ,Z\right) \nabla _{\theta }\omega
^{-1}\left( J,\theta ,Z,\mathcal{G}_{0}\left( 0,Z\right) +\left\vert \Psi
\right\vert ^{2}\right) \Psi \left( \theta ,Z\right) dZd\theta \right] }{%
\dprod\limits_{l=1}^{j}\delta ^{\sum_{l}p_{l}^{i}}\left\vert \Psi \left(
\theta ^{\left( l\right) },Z_{_{l}}\right) \right\vert ^{2}}
\end{equation*}%
This leads in the approximation of slowly varying background fields to the
following effective action:%
\begin{eqnarray}
&&\Gamma \left( \Psi ^{\dagger },\Psi \right)  \label{ftV} \\
&=&\hat{S}_{cl}\left( \Psi ^{\dagger },\Psi \right) +\sum_{\substack{ %
j\geqslant 1  \\ m\geqslant 1}}\sum_{\substack{ p_{l},\left(
p_{l}^{i}\right) _{m\times j}  \\ p_{l}+\sum_{i}p_{l}^{i}\geqslant 2}}\int
\int \left( \dprod\limits_{l=1}^{j}\Psi ^{\dagger }\left( \theta ^{\left(
l\right) },Z_{l}\right) \right) \prod\limits_{i=1}^{m}\left( \left( \frac{1%
}{2}\frac{\delta ^{\sum_{l}p_{l}^{i}}\left( \nabla _{\theta }\omega
^{-1}\left( \theta _{i+1},Z_{i+1},\left\vert \Psi \right\vert ^{2}\right)
\right) }{\dprod\limits_{l=1}^{j}\delta ^{p_{l}^{i}}\left\vert \Psi \left(
\theta ^{\left( l\right) },Z_{_{l}}\right) \right\vert ^{2}}\left\vert \Psi
\left( \theta _{i+1},Z_{i+1}\right) \right\vert ^{2}\right) \right.  \notag
\\
&&\left. \times \left( 1+\sum_{p_{i}}\left( \frac{\sharp _{j+1,i}\left(
\left( \left( p_{i}\right) ,\left( p_{l}^{i}\right) \right) \right) }{\bar{%
\sharp}_{j+1,i}\left( \left( \left( p_{i}\right) ,\left( p_{l}^{i}\right)
\right) \right) }\left( \frac{\delta }{\delta \left\vert \Psi ^{\dagger
}\left( \theta _{i},Z_{i}\right) \right\vert ^{2}}\right) ^{p_{i}}\right)
\right) \right) a_{j,m}\left( \int \Psi ^{\dagger }\left( \theta ,Z\right) 
\frac{1}{2}\frac{\delta ^{\sum_{l}p_{l}}\left( \nabla _{\theta }\omega
^{-1}\left( \theta ,Z,\left\vert \Psi \right\vert ^{2}\right) \right) }{%
\dprod\limits_{l=1}^{j}\delta ^{p_{l}}\left\vert \Psi \left( \theta ^{\left(
l\right) },Z_{_{l}}\right) \right\vert ^{2}}\Psi \left( \theta ,Z\right)
\right)  \notag \\
&&\times \frac{\left( \dprod\limits_{l=1}^{j}\Psi \left( \theta ^{\left(
l\right) },Z_{l}\right) \right) }{\left( m-1\right) !\Lambda
^{\sum_{i}p_{l}^{i}}}\prod\limits_{i=1}^{m}d\theta
_{i}dZ_{_{l}}\dprod\limits_{l=1}^{j}d\theta ^{\left( l\right) }dZ_{l}  \notag
\end{eqnarray}%
with the convention that $\left( \theta _{m+1},Z_{m+1}\right) =\left( \theta
,Z\right) $.

\paragraph{Strong fields approximation}

For strong fields, the derivatives $\frac{\delta }{\delta \left\vert \Psi
^{\dagger }\left( \theta _{i},Z_{i}\right) \right\vert ^{2}}$ are negligible
and (\ref{ftV}) reduces to:%
\begin{eqnarray}
&&\Gamma \left( \Psi ^{\dagger },\Psi \right)  \label{ts} \\
&=&\hat{S}_{cl}\left( \Psi ^{\dagger },\Psi \right) +\sum_{\substack{ %
j\geqslant 1  \\ m\geqslant 1}}\sum_{\substack{ p_{l},\left(
p_{l}^{i}\right) _{m\times j}  \\ p_{l}+\sum_{i}p_{l}^{i}\geqslant 2}}\int
\int \prod\limits_{i=1}^{m}\left( \int \Psi ^{\dagger }\left( \theta
_{i},Z_{i}\right) \left\{ \frac{1}{2}\frac{\delta ^{\sum_{l}p_{l}^{i}}\left(
\nabla _{\theta }\omega ^{-1}\left( \theta _{i},Z_{i},\left\vert \Psi
\right\vert ^{2}\right) \right) }{\dprod\limits_{l=1}^{j}\delta
^{\sum_{l}p_{l}^{i}}\left\vert \Psi \left( \theta ^{\left( l\right)
},Z_{_{l}}\right) \right\vert ^{2}}\right\} \Psi \left( \theta
_{i},Z_{i}\right) \right)  \notag \\
&&\times \left( a_{j,m}\int \Psi ^{\dagger }\left( \theta ,Z\right) \left\{ 
\frac{1}{2}\frac{\delta ^{\sum_{l}p_{l}}\left( \nabla _{\theta }\omega
^{-1}\left( \theta ,Z,\left\vert \Psi \right\vert ^{2}\right) \right) }{%
\dprod\limits_{l=1}^{j}\delta ^{p_{l}}\left\vert \Psi \left( \theta ^{\left(
l\right) },Z_{_{l}}\right) \right\vert ^{2}}\right\} \Psi \left( \theta
,Z\right) \right)  \notag \\
&&\times \left( \dprod\limits_{l=1}^{j}\left\vert \Psi \left( \theta
^{\left( l\right) },Z_{l}\right) \right\vert ^{2}\right)
\prod\limits_{i=1}^{m}d\theta _{i}dZ_{_{l}}\dprod\limits_{l=1}^{j}d\theta
^{\left( l\right) }dZ_{l}  \notag
\end{eqnarray}

The derivatives of $\omega ^{-1}\left( J,\theta ,Z,\mathcal{G}_{0}\left(
0,Z\right) +\left\vert \Psi \right\vert ^{2}\right) $ are computed in
appendix 6, and given in (\ref{nps}). We order the $\theta ^{\left( l\right)
}$ by $\theta ^{\left( l\right) }<...<\theta ^{\left( 1\right) }$, and
define $\theta _{i}=\min_{l,p_{l}^{i}\neq 0}\left( \theta _{l}\right) $. We
thus write (\ref{nps}) as: 
\begin{eqnarray}
&&\frac{\delta ^{\sum_{l}p_{l}^{i}}\omega ^{-1}\left( \left\vert \Psi \left(
\theta ,Z\right) \right\vert ^{2}\right) }{\dprod\limits_{l=1}^{j}\delta
^{p_{l}^{i}}\left\vert \Psi \left( \theta ^{\left( l\right)
},Z_{_{l}}\right) \right\vert ^{2}}  \label{nst} \\
&\simeq &\int \frac{\exp \left( -c\left( \theta -\theta _{i}\right) -\alpha
\left( \sum_{l=1,p_{l}^{i}\neq 0}^{j}\left( \left( c\left( \hat{\theta}%
^{\left( l-1\right) }-\hat{\theta}^{\left( l\right) }\right) \right)
^{2}-\left\vert \hat{Z}_{l-1}-\hat{Z}_{l}\right\vert ^{2}\right) \right)
\right) }{D^{\sum_{l}p_{l}^{i}}}  \notag \\
&&\times H\left( c\left( \theta -\theta _{i}\right) -\sum_{l=1,p_{l}^{i}\neq
0}^{j}\left\vert \hat{Z}_{l-1}-\hat{Z}_{l}\right\vert \right)
\dprod\limits_{l=1}^{j}\left( \frac{\omega ^{-1}\left( J,\theta ^{\left(
l\right) },Z_{_{l}}\right) }{\left\vert \Psi \left( \theta ^{\left( l\right)
},Z_{_{l}}\right) \right\vert ^{2}}\right) ^{\sum_{i}p_{l}^{i}}\left\vert
\Psi \left( \theta ,Z\right) \right\vert ^{2}  \notag
\end{eqnarray}%
where $D$ is a constant. We use the convention $\hat{\theta}^{\left(
l\right) }=\theta ^{\left( l\right) }$, $\hat{Z}_{l}^{\left( i\right)
}=Z_{l} $ for $l>0$ and $\hat{\theta}^{\left( 0\right) }=\theta $, $\hat{Z}%
_{0}=Z$. As a consequence, (\ref{ts}) writes:%
\begin{eqnarray}
&&\hat{S}_{cl}\left( \Psi ^{\dagger },\Psi \right) +\sum_{j\geqslant 1}\sum 
_{\substack{ m\geqslant 2,\left( p_{l}^{i}\right) _{m\times j}  \\ %
\sum_{i}p_{l}^{i}\geqslant 2}}c\left[ \frac{a_{j}}{2}\int
\dprod\limits_{l=1}^{j}dl_{l}dZ_{l}\left( \frac{\omega ^{-1}\left( J,\theta
^{\left( l\right) },Z_{_{l}}\right) }{\left\vert \Psi \left( \theta ^{\left(
l\right) },Z_{_{l}}\right) \right\vert ^{2}}\right)
^{\sum_{i}p_{l}^{i}}\left\vert \Psi \left( \theta ^{\left( l\right)
},Z_{_{l}}\right) \right\vert ^{2}\right.  \label{clS} \\
&&\left. \times a_{j,m}\int \frac{\dprod\limits_{i=1}^{m}\left[ \exp \left(
-c\left( \theta -\theta _{i}\right) -\alpha \left( \sum_{l=1,p_{l}^{i}\neq
0}^{j}\left( \left( c\left( \hat{\theta}^{\left( l-1\right) }-\hat{\theta}%
^{\left( l\right) }\right) \right) ^{2}-\left\vert \hat{Z}_{l-1}-\hat{Z}%
_{l}\right\vert ^{2}\right) \right) \right) \left\vert \Psi \left( \theta
,Z\right) \right\vert ^{2}\right] }{\left( -2\right)
^{m}D^{\sum_{i,l}p_{l}^{i}}m!\left( \sharp _{j,m}\left( \left(
p_{l}^{i}\right) \right) \right) \Lambda _{1}^{j}\Lambda
^{\sum_{i,l}p_{l}^{i}}}d\theta dZ\right]  \notag
\end{eqnarray}

\paragraph{Weak fields approximation}

For weak fields, the main contribution of the derivatives $\left( \frac{%
\delta }{\delta \left\vert \Psi ^{\dagger }\left( \theta _{i},Z_{i}\right)
\right\vert ^{2}}\right) ^{p_{i}}$ is obtained for $p_{i}=1$\ (see appendix
3.3):%
\begin{equation*}
\frac{\delta ^{\sum_{l}p_{l}^{i}}\left[ \hat{S}_{cl,\theta }\left( \Psi
^{\dagger },\Psi \right) \right] }{\dprod\limits_{l=1}^{j}\dprod%
\limits_{k_{l}^{i}=1}^{p_{l}^{i}}\delta \left\vert \Psi \left( \theta
^{\left( k_{l}^{i}\right) },Z_{_{l}}\right) \right\vert ^{2}}\simeq \frac{j}{%
2}\frac{\delta ^{\sum_{l}p_{l}^{i}}\left( \nabla _{\theta _{1}}\omega
^{-1}\left( \left\vert \Psi \left( \theta _{1},Z\right) \right\vert
^{2}\right) \right) }{\dprod\limits_{l=1}^{j}\dprod%
\limits_{k_{l}^{i}=1}^{p_{l}^{i}}\delta \left\vert \Psi \left( \theta
^{\left( k_{l}^{i}\right) },Z_{_{l}}\right) \right\vert ^{2}}
\end{equation*}%
and, in the local approximation, the effective action is:%
\begin{eqnarray}
&&\Gamma \left( \Psi ^{\dagger },\Psi \right)  \label{fw} \\
&=&\hat{S}_{cl}\left( \Psi ^{\dagger },\Psi \right) +\sum_{\substack{ %
j\geqslant 1  \\ m\geqslant 1}}\sum_{\substack{ p_{l},\left(
p_{l}^{i}\right) _{m\times j}  \\ p_{l}+\sum_{i}p_{l}^{i}\geqslant 2}}\left(
\prod\limits_{i=1}^{m}\frac{\sharp _{j+1,m}\left( \left( p_{m},\left(
p_{l}^{m}\right) \right) \right) }{4\bar{\sharp}_{j+1,m}\left( \left(
p_{i},\left( p_{l}^{m}\right) \right) \right) }\right) \frac{a_{j,m}}{2} 
\notag \\
&&\times \int \int \Psi ^{\dagger }\left( \theta ,Z\right) \frac{\delta
^{\sum_{l,i}p_{l}^{i}+p_{l}}}{\dprod\limits_{l=1}^{j}\delta
^{\sum_{i}p_{l}^{i}+p_{l}}\left\vert \Psi \left( \theta ^{\left( l\right)
},Z_{_{l}}\right) \right\vert ^{2}}\left( \frac{\left( \nabla _{\theta
}\omega ^{-1}\left( \theta ,Z,\left\vert \Psi \right\vert ^{2}\right)
\right) }{\dprod\limits_{l=1}^{j}\delta ^{p_{l}^{i}}\left\vert \Psi \left(
\theta ^{\left( l\right) },Z_{_{l}}\right) \right\vert ^{2}}\right)
^{m+1}\Psi \left( \theta ,Z\right) \dprod\limits_{l=1}^{j}\left\vert \Psi
\left( \theta ^{\left( l\right) },Z_{l}\right) \right\vert ^{2}d\theta
^{\left( l\right) }dZ_{l}  \notag
\end{eqnarray}%
which, using the notations of the derivation of (\ref{clS}), leads to an
expression in terms of inverse frequencies:%
\begin{eqnarray}
&&\hat{S}_{cl}\left( \Psi ^{\dagger },\Psi \right) +\sum_{j\geqslant 1}\sum 
_{\substack{ m\geqslant 2,\left( p_{l}^{i}\right) _{m\times j}  \\ %
\sum_{i}p_{l}^{i}\geqslant 2}}c\left[ \frac{ja_{j}}{2}\int
\dprod\limits_{l=1}^{j}dl_{l}dZ_{l}\left( \frac{\omega ^{-1}\left( J,\theta
^{\left( l\right) },Z_{_{l}}\right) }{\left\vert \Psi \left( \theta ^{\left(
l\right) },Z_{_{l}}\right) \right\vert ^{2}}\right)
^{\sum_{i}p_{l}^{i}}\left\vert \Psi \left( \theta ^{\left( l\right)
},Z_{_{l}}\right) \right\vert ^{2}\right. \\
&&\left. \int \times \frac{\dprod\limits_{i=1}^{m}\left[ \exp \left(
-c\left( \theta -\theta _{i}\right) -\alpha \left( \sum_{l=1,p_{l}^{i}\neq
0}^{j}\left( \left( c\left( \hat{\theta}^{\left( l-1\right) }-\hat{\theta}%
^{\left( l\right) }\right) \right) ^{2}-\left\vert \hat{Z}_{l-1}-\hat{Z}%
_{l}\right\vert ^{2}\right) \right) \right) \right] }{\left( -2\right)
^{m}D^{\sum_{i,l}p_{l}^{i}}m!\left( \sharp _{j,m}\left( \left(
p_{l}^{i}\right) \right) \right) \Lambda _{1}^{j}\Lambda
^{\sum_{i,l}p_{l}^{i}}}d\theta dZ\right]  \notag
\end{eqnarray}

\section{Non-trivial minimum}

\subsection{Classical effective action}

The effective action has a minimum for a wide range of parameters (see
appendix 3.4). The corresponding background field decomposes into a constant
part $\Psi _{0}$ and a contribution that depends on the external current. We
show that for slowly varying currents $J\left( \theta ,Z_{i}\right) $, and
for $\left\vert \zeta ^{\left( n\right) }\right\vert >\omega ^{-1}\left(
J\left( \theta \right) ,\theta ,Z,\mathcal{G}_{0}\right) $, the minimum of $%
\Gamma \left( \Psi \right) $ is reached for the fields $\Psi \left( \theta
,Z\right) $ and $\Psi ^{\dagger }\left( \theta ,Z\right) $ that decompose
as: 
\begin{equation}
\Psi \left( \theta ,Z\right) =\Psi _{0}\left( \theta ,Z\right) +\delta \Psi
\left( \theta ,Z\right)  \label{prf}
\end{equation}%
and: 
\begin{equation}
\Psi ^{\dagger }\left( \theta ,Z\right) =\Psi _{0}^{\dagger }\left( \theta
,Z\right) +\delta \Psi ^{\dagger }\left( \theta ,Z\right)  \label{dxf}
\end{equation}%
where $\left\vert \delta \Psi \left( \theta ,Z\right) \right\vert
<<\left\vert \Psi _{0}\left( \theta ,Z\right) \right\vert $ and $\left\vert
\delta \Psi ^{\dagger }\left( \theta ,Z\right) \right\vert <<\left\vert \Psi
_{0}^{\dagger }\left( \theta ,Z\right) \right\vert $. The fields $\Psi
_{0}\left( \theta ,Z\right) $ and $\Psi _{0}^{\dagger }\left( \theta
,Z\right) $\ minimize the potential:%
\begin{equation*}
\alpha \int \left\vert \Psi \left( \theta ^{\left( i\right) },Z_{i}\right)
\right\vert ^{2}dZ_{i}+\sum \frac{\zeta ^{\left( n\right) }}{n!}\left( 
\mathcal{G}_{0}\left( 0,Z_{j}\right) +\int \left\vert \Psi \left( \theta
_{i}^{\left( i\right) }-\frac{\left\vert Z_{i}-Z_{j}\right\vert }{c}%
,Z_{j}\right) \right\vert ^{2}dZ_{j}\right) ^{n}
\end{equation*}%
This minimum exists for $\alpha <<1$ and for $\left\vert \zeta ^{\left(
2\right) }\right\vert $ large. It is reached for a value $X_{0}$ of $\int
\left\vert \Psi \left( \theta ^{\left( i\right) },Z_{i}\right) \right\vert
^{2}dZ_{i}$, and its value is, up to an irrelevant phase:

\begin{equation*}
\Psi _{0}\left( \theta ^{\left( i\right) },Z_{i}\right) =\Psi _{0}^{\dagger
}\left( \theta ,Z\right) =\sqrt{\frac{X_{0}}{V}},
\end{equation*}%
where $V$ is the volume of the thread.

The expression for $\delta \Psi \left( \theta ,Z\right) $ and $\delta \Psi
^{\dagger }\left( \theta ,Z\right) $ arising in equations (\ref{prf}) and (%
\ref{dxf}) are found in appendix 3.4, which shows that the second-order
expansion of the effective action is:%
\begin{eqnarray}
\Gamma \left( \Psi \right) &=&-\frac{1}{2}\delta \Psi ^{\dagger }\left(
\theta ,Z\right) \left( \nabla _{\theta }\left( \frac{\sigma _{\theta }^{2}}{%
2}\nabla _{\theta }-\omega ^{-1}\left( J\left( \theta \right) ,\theta ,Z,%
\mathcal{G}_{0}+X_{0}+\sqrt{X_{0}}\left( \delta \Psi ^{\dagger }+\delta \Psi
\right) \right) \right) \right) \Psi _{0}\left( \theta ,Z\right)  \label{hft}
\\
&&-\frac{1}{2}\delta \Psi ^{\dagger }\left( \theta ,Z\right) \left( \nabla
_{\theta }\left( \frac{\sigma _{\theta }^{2}}{2}\nabla _{\theta }-\omega
^{-1}\left( J\left( \theta \right) ,\theta ,Z,\mathcal{G}_{0}+\left\vert
\Psi \right\vert ^{2}\right) \right) \right) \delta \Psi \left( \theta
,Z\right)  \notag \\
&&+\frac{1}{2}\delta \Psi ^{\dagger }\left( \theta ,Z\right) U^{\prime
\prime }\left( X_{0}\right) \delta \Psi \left( \theta ,Z\right)  \notag
\end{eqnarray}%
and that the first-order equations for $\delta \Psi ^{\dagger }\left( \theta
,Z\right) $ and $\delta \Psi \left( \theta ,Z\right) $ are:%
\begin{eqnarray}
\delta \Psi ^{\dagger } &=&0  \notag \\
\delta \Psi \left( \theta ,Z\right) &=&\left( \frac{\left( \nabla _{\theta
}\left( \frac{\sigma _{\theta }^{2}}{2}\nabla _{\theta }-\omega ^{-1}\left(
J\left( \theta \right) ,\theta ,Z,\mathcal{G}_{0}+\left\vert \Psi
\right\vert ^{2}\right) \right) \right) }{U^{\prime \prime }\left(
X_{0}\right) -\left( \nabla _{\theta }\left( \frac{\sigma _{\theta }^{2}}{2}%
\nabla _{\theta }-\omega ^{-1}\left( J\left( \theta \right) ,\theta ,Z,%
\mathcal{G}_{0}+\left\vert \Psi \right\vert ^{2}\right) \right) \right) }%
\right) \Psi _{0}\left( \theta ,Z\right)  \label{dt}
\end{eqnarray}%
Setting $V=1$ yields, in first approximation:%
\begin{equation}
\delta \Psi \left( \theta ,Z\right) \simeq -\frac{\nabla _{\theta }\omega
^{-1}\left( J\left( \theta \right) ,\theta ,Z,\mathcal{G}_{0}+\left\vert
\Psi \right\vert ^{2}\right) }{U^{\prime \prime }\left( X_{0}\right) }X_{0}
\label{tdd}
\end{equation}%
This relation is sufficient to derive the next section's frequencies
equations, but can however be used to find $\delta \Psi \left( \theta
,Z\right) $, at our order of approximation (see appendix 3.4). In the local
approximation and for slowly varying currents, we show that the minimization
of action (\ref{fth}) yields:%
\begin{eqnarray}
\delta \Psi \left( \theta ,Z\right) &=&\left( G^{-1}\left( -\frac{U^{\prime
\prime }\left( X_{0}\right) }{X_{0}}\exp \left( H^{-1}\left( \frac{\theta }{%
\Gamma \left( \mathcal{G}_{0}\left( Z_{1}\right) +\sqrt{X_{0}}\right) }%
+d\right) \right) \right) -J\left( \theta ,Z\right) \right)  \label{sdt} \\
&&\times \exp \left( H^{-1}\left( \frac{\theta }{\Gamma \left( \mathcal{G}%
_{0}\left( Z_{1}\right) +\sqrt{X_{0}}\right) }+d\right) \right)  \notag
\end{eqnarray}%
with: 
\begin{equation*}
H\left( Y\right) =\int \frac{dY}{G^{-1}\left( -\frac{U^{\prime \prime
}\left( X_{0}\right) }{X_{0}}\exp Y\right) -J\left( \theta ,Z\right) }
\end{equation*}%
and:%
\begin{equation*}
\Gamma =\int \frac{\kappa }{N}\frac{\left\vert Z-Z_{1}\right\vert }{c}%
T\left( Z,\theta ,Z_{1},\theta -\frac{\left\vert Z-Z_{1}\right\vert }{c}%
\right) dZ_{1}
\end{equation*}%
The constant $d$ is chosen so that $\lim_{\theta \rightarrow \infty }\delta
\Psi \left( \theta ,Z\right) =0$.

The field $\Psi \left( \theta ^{\left( j\right) },Z_{j}\right) $ is the -
phase-dependent - background field. It is null in the trivial phase, so that
the effective action is the "classical" one. In a non-trivial phase, $\Psi
\left( \theta ^{\left( j\right) },Z_{j}\right) $ is not null and may be
time-dependent. It describes the accumulation of currents or signals that
shapes the long-term dynamics of frequencies. Incidentally, we note that a
non-trivial minimum that depends on the system parameters should allow for
phase transition in the system of frequencies. This question is left for
further work.

\subsection{Including higher order corrections}

Equation (\ref{clS}) yields the perturbative corrections that modify the
classical effective action and its minimum. These corrections modify
equations (\ref{dt}) and (\ref{tdd}) by shifting $U^{\prime \prime }\left(
X_{0}\right) \rightarrow $ $U^{\prime \prime }\left( X_{0}\right) -C\left(
X_{0}\right) $ (see appendix 3.4 for the expression of $C\left( X_{0}\right) 
$), which in turn modifies the solution (\ref{sdt}):%
\begin{eqnarray}
\delta \Psi \left( \theta ,Z\right) &=&\left( G^{-1}\left( -\frac{U^{\prime
\prime }\left( X_{0}\right) -C\left( X_{0}\right) }{X_{0}}\exp \left(
H^{-1}\left( \frac{\theta }{\Gamma \left( \mathcal{G}_{0}\left( Z_{1}\right)
+\sqrt{X_{0}}\right) }+d\right) \right) \right) -J\left( \theta ,Z\right)
\right) \\
&&\times \exp \left( H^{-1}\left( \frac{\theta }{\Gamma \left( \mathcal{G}%
_{0}\left( Z_{1}\right) +\sqrt{X_{0}}\right) }+d\right) \right)  \notag
\end{eqnarray}%
where the constant $d$ is set to ensure $\lim_{\theta \rightarrow \infty
}\delta \Psi \left( \theta ,Z\right) =0$.

\section{Equation for frequencies: General form}

\subsection{Principle}

To find the frequencies, we use the form (\ref{ltn}) for the effective
action. Writing $\Gamma \left( \Psi ^{\dagger },\Psi \right) $ in the local
approximation as:%
\begin{equation}
\Gamma \left( \Psi ^{\dagger },\Psi \right) \simeq \int \Psi ^{\dagger
}\left( \theta ,Z\right) \left( -\nabla _{\theta }\left( \frac{\sigma
_{\theta }^{2}}{2}\nabla _{\theta }-\omega ^{-1}\left( J\left( \theta
\right) ,\theta ,Z,\mathcal{G}_{0}+\left\vert \Psi \right\vert ^{2}\right)
\right) \delta \left( \theta _{f}-\theta _{i}\right) +\Omega \left( \theta
,Z\right) \right) \Psi \left( \theta ,Z\right)  \label{rfq}
\end{equation}%
with:%
\begin{eqnarray}
\Omega \left( \theta ,Z\right) &=&\int \sum_{\substack{ j\geqslant 1  \\ %
m\geqslant 1}}\sum_{\substack{ \left( p_{l}^{i}\right) _{m\times j}  \\ %
p_{l}+\sum_{i}p_{l}^{i}\geqslant 2}}\frac{\delta ^{\sum_{l}p_{l}}\left[
L\left( \Psi ^{\dagger }\left( \theta ,Z\right) ,\Psi \left( \theta
,Z\right) \right) \right] }{\dprod\limits_{l=1}^{j}\dprod%
\limits_{k_{l}^{i}=1}^{p_{l}}\delta \left\vert \Psi \left( \theta ^{\left(
l\right) },Z_{_{l}}\right) \right\vert ^{2}}\Psi \left( \theta ,Z\right)
\label{mG} \\
&&\times \frac{a_{j}}{j!}\left( \dprod\limits_{l=1}^{j}\Psi ^{\dagger
}\left( \theta _{f}^{\left( l\right) },Z_{l}\right) \right)
\dprod\limits_{i=1}^{m}\left[ \frac{\delta ^{\sum_{l}p_{l}^{i}}\left[ \hat{S}%
_{cl,\theta }\left( \Psi ^{\dagger },\Psi \right) \right] }{%
\dprod\limits_{l=1}^{j}\dprod\limits_{k_{l}^{i}=1}^{p_{l}^{i}}\delta
\left\vert \Psi \left( \theta ^{\left( l\right) },Z_{_{l}}\right)
\right\vert ^{2}}\right] \left( \dprod\limits_{l=1}^{j}\Psi \left( \theta
_{i}^{\left( l\right) },Z_{l}\right) \right)  \notag
\end{eqnarray}%
The effective frequency can be identified as:%
\begin{equation*}
\nabla _{\theta }\omega _{e}^{-1}\left( J\left( \theta \right) ,\theta ,Z,%
\mathcal{G}_{0}+\left\vert \Psi \right\vert ^{2}\right) =\nabla _{\theta
}\omega ^{-1}\left( J\left( \theta \right) ,\theta ,Z,\mathcal{G}%
_{0}+\left\vert \Psi \right\vert ^{2}\right) +\Omega \left( \theta ,Z\right)
\end{equation*}%
that is:%
\begin{equation}
\omega _{e}^{-1}\left( J\left( \theta \right) ,\theta ,Z,\mathcal{G}%
_{0}+\left\vert \Psi \right\vert ^{2}\right) =\omega ^{-1}\left( J\left(
\theta \right) ,\theta ,Z,\mathcal{G}_{0}+\left\vert \Psi \right\vert
^{2}\right) +\int^{\theta }\Omega \left( \theta ,Z\right)  \label{fft}
\end{equation}%
where $\omega \left( J\left( \theta \right) ,\theta ,Z,\mathcal{\bar{G}}%
_{0}+\left\vert \Psi \right\vert ^{2}\right) $ is the solution of: 
\begin{eqnarray}
\omega ^{-1}\left( \theta ,Z\right) &=&G\left( J\left( \theta \right) +\frac{%
\kappa }{N}\int T\left( Z,Z_{1}\right) \frac{\omega \left( \theta -\frac{%
\left\vert Z-Z_{1}\right\vert }{c},Z_{1}\right) }{\omega \left( \theta
,Z\right) }\right.  \label{fqt} \\
&&\times \left. W\left( \frac{\omega \left( \theta ,Z\right) }{\omega \left(
\theta -\frac{\left\vert Z-Z_{1}\right\vert }{c},Z_{1}\right) }\right)
\left( \mathcal{\bar{G}}_{0}\left( 0,Z_{1}\right) +\left\vert \Psi \left(
\theta -\frac{\left\vert Z-Z_{1}\right\vert }{c},Z_{1}\right) \right\vert
^{2}\right) dZ_{1}\right)  \notag
\end{eqnarray}

Using the form of $\Psi \left( \theta ^{\left( j\right) },Z_{j}\right) =\Psi
_{0}\left( \theta ^{\left( j\right) },Z_{j}\right) +\delta \Psi \left(
\theta ,Z\right) $ derived in section 5, we find an expression for $\omega
^{-1}\left( J\left( \theta \right) ,\theta ,Z,\mathcal{G}_{0}+\left\vert
\Psi \right\vert ^{2}\right) $.

The second term $\int^{\theta }\Omega \left( \theta ,Z\right) $ in (\ref{fft}%
) represents corrections due to the interactions. Using (\ref{mG}), we can
find its expression as a series expansion in terms of frequencies and field.
In the next two paragraph, we limit ourselves to the cases of strong and
weak field approximation, respectively.

\subsection{Strong field approximation}

Using (\ref{ts}), the strong field approximation is:%
\begin{eqnarray}
&&\omega _{e}^{-1}\left( J\left( \theta \right) ,\theta ,Z,\mathcal{G}%
_{0}+\left\vert \Psi \right\vert ^{2}\right) =\omega ^{-1}\left( J\left(
\theta \right) ,\theta ,Z,\mathcal{G}_{0}+\left\vert \Psi \right\vert
^{2}\right)  \label{tsf} \\
&&+\int^{\theta }\sum_{\substack{ j\geqslant 1  \\ m\geqslant 1}}\sum 
_{\substack{ p_{l},\left( p_{l}^{i}\right) _{m\times j}  \\ %
p_{l}+\sum_{i}p_{l}^{i}\geqslant 2}}\frac{1}{2}\frac{\delta
^{\sum_{l}p_{l}}\left( \nabla _{\theta }\omega ^{-1}\left( \left\vert \Psi
\left( \theta ,Z\right) \right\vert ^{2}\right) \right) }{%
\dprod\limits_{l=1}^{j}\dprod\limits_{k_{l}^{i}=1}^{p_{l}}\delta \left\vert
\Psi \left( \theta ^{\left( k_{l}^{i}\right) },Z_{_{l}}\right) \right\vert
^{2}}  \notag \\
&&\times \int \frac{a_{j}}{j!}\left( \dprod\limits_{l=1}^{j}\Psi ^{\dagger
}\left( \theta _{f}^{\left( l\right) },Z_{l}\right) \right)
\dprod\limits_{i=1}^{m}\left[ \frac{\delta ^{\sum_{l}p_{l}^{i}}\omega
^{-1}\left( \left\vert \Psi \left( \theta ,Z\right) \right\vert ^{2}\right) 
}{\dprod\limits_{l=1}^{j}\dprod\limits_{k_{l}^{i}=1}^{p_{l}^{i}}\delta
\left\vert \Psi \left( \theta ^{\left( l\right) },Z_{_{l}}\right)
\right\vert ^{2}}\left\vert \Psi \left( \theta ,Z\right) \right\vert ^{2}%
\right] \left( \dprod\limits_{l=1}^{j}\Psi \left( \theta _{i}^{\left(
l\right) },Z_{l}\right) \right)  \notag
\end{eqnarray}%
We will compute in section 8.4 the lowest order terms of the correction
terms in (\ref{tsf}) and inspect their impact on the frequencies dynamics.

\subsection{Weak field approximation}

Using (\ref{fw}), the weak field approximation is:%
\begin{eqnarray}
&&\omega _{e}^{-1}\left( J\left( \theta \right) ,\theta ,Z,\mathcal{G}%
_{0}+\left\vert \Psi \right\vert ^{2}\right) =\omega ^{-1}\left( J\left(
\theta \right) ,\theta ,Z,\mathcal{G}_{0}+\left\vert \Psi \right\vert
^{2}\right)  \label{wff} \\
&&+\int^{\theta }\sum_{\substack{ j\geqslant 1  \\ m\geqslant 1}}\sum 
_{\substack{ p_{l},\left( p_{l}^{i}\right) _{m\times j}  \\ %
p_{l}+\sum_{i}p_{l}^{i}\geqslant 2}}\frac{1}{2}\frac{\delta
^{\sum_{l}p_{l}}\left( \nabla _{\theta }\omega ^{-1}\left( \left\vert \Psi
\left( \theta ,Z\right) \right\vert ^{2}\right) \right) }{%
\dprod\limits_{l=1}^{j}\dprod\limits_{k_{l}^{i}=1}^{p_{l}}\delta \left\vert
\Psi \left( \theta ^{\left( k_{l}^{i}\right) },Z_{_{l}}\right) \right\vert
^{2}}  \notag \\
&&\times \int \frac{a_{j}}{j!}\left( \dprod\limits_{l=1}^{j}\Psi ^{\dagger
}\left( \theta _{f}^{\left( l\right) },Z_{l}\right) \right)
\dprod\limits_{i=1}^{m}\left[ \frac{j}{2}\frac{\delta
^{\sum_{l}p_{l}^{i}}\left( \nabla _{\theta }\omega ^{-1}\left( \left\vert
\Psi \left( \theta ,Z\right) \right\vert ^{2}\right) \right) }{%
\dprod\limits_{l=1}^{j}\dprod\limits_{k_{l}^{i}=1}^{p_{l}^{i}}\delta
\left\vert \Psi \left( \theta ^{\left( k_{l}^{i}\right) },Z_{_{l}}\right)
\right\vert ^{2}}\right] \left( \dprod\limits_{l=1}^{j}\Psi \left( \theta
_{i}^{\left( l\right) },Z_{l}\right) \right)  \notag
\end{eqnarray}

\section{Static equilibrium for frequencies}

Discarding the corrections terms $\int^{\theta }\Omega \left( \theta
,Z\right) $, a static solution of (\ref{fft}) can be found for a constant
background $\Psi \left( \theta ^{\left( 1\right) },Z_{1}\right) \simeq \Psi
_{0}\left( Z_{1}\right) $ and constant current, i.e. $J=\bar{J}$, $\omega
\left( \theta ,Z\right) =\omega \left( Z\right) $. For a static solution, it
implies $\left( \Psi ^{\dag }\mathcal{G}_{0}^{-1}\Psi \right) =0$, or
equivalently: $\delta \Psi \left( \theta ,Z\right) =\nabla _{\theta }\omega
^{-1}\left( J\left( \theta \right) ,\theta ,Z,\mathcal{G}_{0}+X_{0}\right)
=0 $, and:%
\begin{equation*}
\omega _{e}^{-1}\left( J\left( \theta \right) ,\theta ,Z\right) =\omega
^{-1}\left( \bar{J},Z,\mathcal{G}_{0}\left( 0,Z\right) +X_{0}\right)
\end{equation*}%
where $\omega \left( \bar{J}\left( Z\right) ,\mathcal{G}_{0}\left(
0,Z_{i}\right) +X_{0}\right) $ is solution of:%
\begin{equation}
\omega \left( Z\right) =F\left( \bar{J}+\frac{\kappa }{N}\int T\left(
Z,Z_{1}\right) \frac{\omega \left( Z_{1}\right) }{\omega \left( Z\right) }%
W\left( \frac{\omega \left( Z\right) }{\omega \left( Z_{1}\right) }\right) 
\mathcal{\bar{G}}_{0}\left( 0,Z_{i}\right) dZ_{1}\right)  \label{frs}
\end{equation}%
and:%
\begin{equation}
\mathcal{\bar{G}}_{0}\left( 0,Z_{i}\right) \simeq \mathcal{G}_{0}\left(
0,Z_{i}\right) +X_{0}  \label{sg}
\end{equation}%
Moreover, in the absence of external sources, i.e. for $J\left( \theta
,Z\right) =0$, the solution of (\ref{frs}) can be written $\omega _{0}\left(
Z\right) $, which satisfies: 
\begin{equation}
\omega _{0}\left( Z\right) =F\left( \frac{\kappa }{N}\int T\left(
Z,Z_{1}\right) \frac{\omega \left( Z_{1}\right) }{\omega \left( Z\right) }%
W\left( \frac{\omega \left( Z\right) }{\omega \left( Z_{1}\right) }\right) 
\mathcal{\bar{G}}_{0}\left( 0,Z_{i}\right) dZ_{1}\right)  \label{ndpt}
\end{equation}%
where $\frac{1}{N}\int dZ_{1}$ is normalized to $1$.

\section{ Differential equation for frequencies in the local approximation}

A local approximation of (\ref{fft}) under some position-independent static
equilibrium can be derived. It will be later generalized to a
position-dependent equilibrium.

\subsection{Assumptions}

First, assume that the background field:%
\begin{equation*}
\Psi \left( \theta ^{\left( j\right) },Z_{j}\right) =\Psi _{0}\left(
Z_{j}\right) +\delta \Psi \left( \theta ^{\left( j\right) },Z_{j}\right)
\end{equation*}%
satisfies:%
\begin{equation*}
\left\vert \delta \Psi \left( \theta ^{\left( j\right) },Z_{j}\right)
\right\vert <<\left\vert \Psi _{0}\left( Z_{j}\right) \right\vert
\end{equation*}%
Then, considering a translation-independent transfer function, i.e. $T\left(
Z,Z_{1}\right) =T\left( Z-Z_{1}\right) $ with $d>>1$, and neglecting border
effects, equation (\ref{ndpt}) simplifies and yields a constant solution:%
\begin{equation}
\omega _{0}=F\left( \frac{TW\left( 1\right) }{\bar{\Lambda}}\right)
\label{bct}
\end{equation}%
where:%
\begin{eqnarray*}
T &=&\frac{\kappa }{N}\int T\left( Z,Z_{1}\right) dZ_{1} \\
\frac{T}{\bar{\Lambda}} &=&\frac{\kappa }{N}\int T\left( Z,Z_{1}\right) 
\mathcal{G}_{0}\left( 0,Z_{1}\right) dZ_{1}
\end{eqnarray*}%
Ultimately, we assume that the transfer functions are symmetric, that is:%
\begin{equation}
T\left( Z,Z_{1}\right) =T\left( Z_{1},Z\right)  \label{smt}
\end{equation}

\subsection{Local equation for frequencies}

Note that, given (\ref{Wqn}) and (\ref{smt}), we have:%
\begin{eqnarray*}
&&\frac{\partial }{\partial \omega \left( \theta ,Z\right) }\left( \frac{%
\kappa }{N}\int W\left( \frac{\omega }{\omega _{1}}\right) dZ_{1}\right)
_{\omega _{1}\left( \theta _{1},Z_{1}\right) =\omega \left( \theta ,Z\right)
} \\
&&+\frac{\partial }{\partial \omega _{1}\left( \theta ,Z\right) }\left( 
\frac{\kappa }{N}\int W\left( \frac{\omega }{\omega _{1}}\right)
dZ_{1}\right) _{\omega _{1}\left( \theta _{1},Z_{1}\right) =\omega \left(
\theta ,Z\right) }=\frac{\kappa }{N}\int \frac{\left( W^{\prime }\left(
1\right) -W^{\prime }\left( 1\right) \right) }{\omega \left( \theta
,Z\right) }dZ_{1}=0
\end{eqnarray*}%
We can find a local approximation of (\ref{fft}) if we expand $\omega \left(
J\left( \theta \right) ,\theta ,Z,\mathcal{G}_{0}+\left\vert \Psi
\right\vert ^{2}\right) $ to the second-order in $Z-Z_{1}$, and consider the
other terms in the right-hand side of (\ref{fft}) as corrections. The
equation for $\omega \left( J\left( \theta \right) ,\theta ,Z,\mathcal{G}%
_{0}+\left\vert \Psi \right\vert ^{2}\right) $ is:%
\begin{eqnarray}
&&F^{-1}\left( \omega \left( J\left( \theta \right) ,\theta \right) \right)
\label{vnq} \\
&=&\left( J\left( \theta \right) +\int \frac{\kappa T\left( Z,Z_{1}\right) }{%
N}\frac{\omega \left( \theta -\frac{\left\vert Z-Z_{1}\right\vert }{c}%
,Z_{1}\right) }{\omega \left( \theta ,Z\right) }W\left( \frac{\omega \left(
\theta ,Z\right) }{\omega \left( \theta -\frac{\left\vert Z-Z_{1}\right\vert 
}{c},Z_{1}\right) }\right) \right.  \notag \\
&&\times \left. \left( \mathcal{G}_{0}\left( 0,Z_{1}\right) +\left\vert \Psi
_{0}+\delta \Psi \left( \theta -\frac{\left\vert Z-Z_{1}\right\vert }{c}%
,Z_{1}\right) \right\vert ^{2}\right) dZ_{1}\right)  \notag
\end{eqnarray}

We then expand $\omega \left( \theta -\frac{\left\vert Z-Z_{1}\right\vert }{c%
},Z_{1}\right) $ around $\omega \left( \theta ,Z\right) $ to the
second-order in $Z-Z_{1}$\ and compute the integrals, which yields for the
right-hand side of (\ref{vnq}):%
\begin{eqnarray*}
&&J\left( \theta \right) +\int \frac{\kappa T\left( Z,Z_{1}\right) }{N}\frac{%
\omega \left( \theta -\frac{\left\vert Z-Z_{1}\right\vert }{c},Z_{1}\right) 
}{\omega \left( \theta ,Z\right) }W\left( \frac{\omega \left( \theta
,Z\right) }{\omega \left( \theta -\frac{\left\vert Z-Z_{1}\right\vert }{c}%
,Z_{1}\right) }\right) \\
&&\times \left( \mathcal{G}_{0}\left( 0,Z_{1}\right) +\left\vert \Psi
_{0}\left( Z_{1}\right) +\delta \Psi \left( \theta -\frac{\left\vert
Z-Z_{1}\right\vert }{c},Z_{1}\right) \right\vert ^{2}\right) dZ_{1} \\
&\simeq &J\left( \theta \right) +\frac{TW\left( 1\right) }{\bar{\Lambda}}+%
\frac{\hat{f}_{1}\nabla _{\theta }\omega \left( \theta ,Z\right) }{\omega
\left( \theta ,Z\right) }+\frac{\hat{f}_{3}\nabla _{\theta }^{2}\omega
\left( \theta ,Z\right) }{\omega \left( \theta ,Z\right) }+c^{2}\frac{\hat{f}%
_{3}\nabla _{Z}^{2}\omega \left( \theta ,Z\right) }{\omega \left( \theta
,Z\right) }+T\Psi _{0}\delta \Psi \left( \theta ,Z\right)
\end{eqnarray*}%
where we defined:%
\begin{eqnarray}
\hat{f}_{1} &=&\frac{W^{\prime }\left( 1\right) -W\left( 1\right) }{c}\Gamma
_{1}\text{, }\hat{f}_{3}=\frac{\left( W\left( 1\right) -W^{\prime }\left(
1\right) \right) \Gamma _{2}}{c^{2}}  \label{ctf} \\
\Gamma _{1} &=&\frac{\kappa }{NX_{r}}\int \left\vert Z-Z_{1}\right\vert
T\left( Z,Z_{1}\right) \mathcal{\bar{G}}_{0}\left( 0,Z_{1}\right) dZ_{1} 
\notag \\
\Gamma _{2} &=&\frac{\kappa }{2NX_{r}}\int \left( Z-Z_{1}\right) ^{2}T\left(
Z,Z_{1}\right) \mathcal{\bar{G}}_{0}\left( 0,Z_{1}\right) dZ_{1}  \notag
\end{eqnarray}%
and:%
\begin{equation*}
T\Psi _{0}\delta \Psi \left( \theta ,Z\right) =\int \frac{\kappa T\left(
Z,Z_{1}\right) }{N}\Psi _{0}\left( Z_{1}\right) \delta \Psi \left( \theta -%
\frac{\left\vert Z-Z_{1}\right\vert }{c},Z_{1}\right) dZ_{1}
\end{equation*}%
Using (\ref{bct}), equation (\ref{vnq}) then becomes:%
\begin{equation}
F^{-1}\left( \omega \left( J\left( \theta \right) ,\theta \right) \right)
-F^{-1}\left( \omega _{0}\right) =J\left( \theta \right) +\frac{\hat{f}%
_{1}\nabla _{\theta }\omega \left( \theta ,Z\right) }{\omega \left( \theta
,Z\right) }+\frac{\hat{f}_{3}\nabla _{\theta }^{2}\omega \left( \theta
,Z\right) }{\omega \left( \theta ,Z\right) }+c^{2}\hat{f}_{3}\frac{\nabla
_{Z}^{2}\omega \left( \theta ,Z\right) }{\omega \left( \theta ,Z\right) }%
+T\Psi _{0}\delta \Psi \left( \theta ,Z\right)  \label{fsr}
\end{equation}%
Using also the local linear approximation for $\delta \Psi \left( \theta
,Z\right) $ derived in appendix 4.4.2:%
\begin{eqnarray}
\delta \Psi \left( \theta ,Z\right) &\simeq &\frac{\nabla _{\theta }\omega
\left( \theta ,Z,\mathcal{G}_{0}+\left\vert \Psi _{0}\right\vert ^{2}\right) 
}{U^{\prime \prime }\left( X_{0}\right) \omega ^{2}\left( J\left( \theta
\right) ,\theta ,Z,\mathcal{G}_{0}+\left\vert \Psi _{0}\right\vert
^{2}\right) }\Psi _{0}  \label{bgd} \\
&\simeq &\frac{\nabla _{\theta }\omega \left( \theta ,Z,\mathcal{G}%
_{0}\right) }{U^{\prime \prime }\left( X_{0}\right) \omega ^{2}\left(
J\left( \theta \right) ,\theta ,Z,\mathcal{G}_{0}\right) }\Psi _{0}  \notag
\end{eqnarray}%
leads to:%
\begin{eqnarray*}
T\delta \Psi \left( \theta ,Z\right) &\simeq &\delta \Psi \left( \theta
,Z\right) -\Gamma _{1}\nabla _{\theta }\delta \Psi \left( \theta ,Z\right) \\
&\simeq &N_{1}\nabla _{\theta }\omega \left( \theta ,Z,\mathcal{G}%
_{0}\right) -N_{2}\nabla _{\theta }\omega \left( \theta ,Z,\mathcal{G}%
_{0}\right)
\end{eqnarray*}%
with:%
\begin{eqnarray*}
N_{1} &=&\frac{\Psi _{0}\left( Z\right) }{U^{\prime \prime }\left(
X_{0}\right) \omega ^{2}\left( J\left( \theta \right) ,\theta ,Z,\mathcal{G}%
_{0}\right) } \\
N_{2} &=&\frac{\Gamma _{1}\Psi _{0}\left( Z\right) }{U^{\prime \prime
}\left( X_{0}\right) \omega ^{2}\left( J\left( \theta \right) ,\theta ,Z,%
\mathcal{G}_{0}\right) }
\end{eqnarray*}%
We assume that $F^{-1}$ is slowly varying, so that:%
\begin{equation*}
F^{-1}\left( \omega \left( J\left( \theta \right) ,\theta \right) \right)
-F^{-1}\left( \omega _{0}\right) \simeq \Gamma _{0}\left( \omega \left(
J\left( \theta \right) ,\theta \right) -\omega _{0}\right)
\end{equation*}%
with\footnote{%
Given our assumption that $F$ is an increasing function, $f>0$.
\par
{}}:%
\begin{equation*}
f=\left( F^{-1}\right) ^{\prime }\left( \frac{\kappa }{N}\int T\left(
Z,Z_{1}\right) W\left( 1\right) dZ_{1}\mathcal{\bar{G}}_{0}\left(
0,Z_{1}\right) \right)
\end{equation*}%
and define:%
\begin{equation*}
\Omega \left( \theta ,Z\right) =\omega \left( \theta ,Z\right) -\omega _{0}
\end{equation*}%
As a result, the expansion of (\ref{fsr}) for a non-static current is then:%
\begin{equation}
f\Omega \left( \theta ,Z\right) =J\left( \theta \right) +\left( \frac{\hat{f}%
_{1}}{\omega \left( \theta ,Z\right) }+N_{1}\right) \nabla _{\theta }\Omega
\left( \theta ,Z\right) +\left( \frac{\hat{f}_{3}}{\omega \left( \theta
,Z\right) }-N_{2}\right) \nabla _{\theta }^{2}\Omega \left( \theta ,Z\right)
+\frac{c^{2}\hat{f}_{3}}{\omega \left( \theta ,Z\right) }\nabla
_{Z}^{2}\Omega \left( \theta ,Z\right)  \label{pw}
\end{equation}

\subsection{Stable traveling waves solutions of (\protect\ref{pw})}

When $J\left( \theta \right) $ is set to $0$, equation (\ref{pw}) has some
stable non linear oscillatory solutions, for a certain range of the
parameters. As a consequence, equation (\ref{pw}) behaves locally as a wave
equation, provided that $\hat{f}_{3}>0$, $\frac{\hat{f}_{3}}{\omega \left(
\theta ,Z\right) }-N_{2}$ $<0$, and $\omega \left( \theta ,Z\right) \ $%
varies slowly. An approximative solution of the type:

\begin{equation*}
\omega \left( \theta ,Z\right) =\exp \left( ik_{0}\theta -ikZ\right)
\end{equation*}%
can be found by writing: 
\begin{equation}
k_{0}^{2}+i\frac{\frac{\hat{f}_{1}}{\bar{\omega}\left( \theta ,Z\right) }%
+N_{1}}{N_{2}-\frac{\hat{f}_{3}}{\bar{\omega}\left( \theta ,Z\right) }}k_{0}-%
\frac{\frac{c^{2}\hat{f}_{3}}{\bar{\omega}\left( \theta ,Z\right) }k^{2}+f}{%
N_{2}-\frac{\hat{f}_{3}}{\bar{\omega}\left( \theta ,Z\right) }}=0  \label{ds}
\end{equation}%
and:%
\begin{equation*}
k_{0}=-\frac{i}{2}\left( \frac{\hat{f}_{1}}{\bar{\omega}\left( \theta
,Z\right) }+N_{1}\right) \pm \sqrt{\frac{\frac{c^{2}\hat{f}_{3}}{\bar{\omega}%
\left( \theta ,Z\right) }k^{2}+f}{N_{2}-\frac{\hat{f}_{3}}{\bar{\omega}%
\left( \theta ,Z\right) }}-\frac{1}{4}\left( \frac{\frac{\hat{f}_{1}}{\bar{%
\omega}\left( \theta ,Z\right) }+N_{1}}{N_{2}-\frac{\hat{f}_{3}}{\bar{\omega}%
\left( \theta ,Z\right) }}\right) ^{2}}
\end{equation*}%
where $\bar{\omega}\left( \theta ,Z\right) $ is the average of $\omega
\left( \theta ,Z\right) $ in some range of time.

The approximative solution is oscillatory and explosive when the
discriminant of (\ref{ds}) is positive and in a range for $\omega \left(
\theta ,Z\right) $, such that $\frac{\hat{f}_{1}}{\omega \left( \theta
,Z\right) }+N_{1}>0$. The solution is oscillatory and dampening for $\frac{%
\hat{f}_{1}}{\omega \left( \theta ,Z\right) }+N_{1}<0$.

Let define $\omega _{1}$ the value of $\omega \left( \theta ,Z\right) $ such
that $\frac{\hat{f}_{1}}{\omega _{1}}+N_{1}=0$. For $\omega \left( \theta
,Z\right) >\omega _{1}$ the solution of (\ref{pw}) presents an increasing
amplitude, and for $\omega \left( \theta ,Z\right) >\omega _{1}$ the
solution of (\ref{pw}) is decreasing in amplitude.

If $\omega _{1}>\omega _{0}$, then $\omega _{0}$\ is a stable point since it
belongs to the domain in which $\frac{\hat{f}_{1}}{\omega \left( \theta
,Z\right) }+N_{1}<0$. Oscillatory patterns will dampen towards $\omega _{0}$.

If $\omega _{1}<\omega _{0}$, the frequency $\omega _{0}$\ is an unstable
point. However, the change in sign of $\frac{\hat{f}_{1}}{\omega \left(
\theta ,Z\right) }+N_{1}$\ for some ranges in the parameters, induces an
oscillatory\ pattern around $\omega _{0}$. Actually, when the oscillation of 
$A$ is of quite constant amplitute, the time $T_{<\omega _{1}}$ spent by the
system below $\omega _{1}$ is proportional to $\arccos \left( \frac{\omega
_{0}-\omega _{1}}{A}\right) $, and the time $T_{>\omega _{1}}$ spent by the
system above $\omega _{1}$ is proportional to $1-\arccos \left( \frac{\omega
_{0}-\omega _{1}}{A}\right) $. For $\omega _{1}$ large enough and during the
time $T_{<\omega _{1}}$ the system amplitude is multiplied by a term of
order $\exp \left( -\omega _{1}T_{<\omega _{1}}\right) $, whereas during the
time $T_{>\omega _{1}}$ the system amplitude is multiplied by a term of
order $\exp \left( \frac{1}{N_{1}}T_{>\omega _{1}}\right) $. Since $\omega
_{0}-\omega _{1}>0$, the relation $T_{<\omega _{1}}<T_{>\omega _{1}}$ is
always true. However, the overall factor $\exp \left( -\omega _{1}T_{<\omega
_{1}}+\frac{1}{N_{1}}T_{>\omega _{1}}\right) $ may, depending on the
system's parameters and on $T_{<\omega _{1}}$, be lower or grater than $1$.
Moreover its magnitude depends on the dynamics, since the amplitude $A$ that
determines $T_{<\omega _{1}}$ is itself time-dependent.

When $\omega \left( 0,Z\right) >\omega _{1}$, $A$ increases, and the time $%
T_{<\omega _{1}}$ increases with $A$. For some values of the parameters, the
dampening factor $\exp \left( -\omega _{1}T_{<\omega _{1}}\right) $
cumulated during time $T_{<\omega _{1}}$ becomes dominant with respect to
the increasing factor $\exp \left( \frac{1}{N_{1}}T_{>\omega _{1}}\right) $.
As a consequence, $\exp \left( -\omega _{1}T_{<\omega _{1}}+\frac{1}{N_{1}}%
T_{>\omega _{1}}\right) >1$. The dynamic pattern turns from explosive to
dampened. Thus, the average amplitude $A$ decreases, and $T_{<\omega _{1}}$
diminishes. At some point, $\exp \left( -\omega _{1}T_{<\omega _{1}}+\frac{1%
}{N_{1}}T_{>\omega _{1}}\right) $ becomes greater than $1$ and the amplitude
increases again. The resulting dynamics thus presents stable oscillations
that are irregular in amplitude. The system does not converge toward $\omega
_{0}$ which remains unstable, but presents the characteristic of some non
linear travelling wave.

Note that this result is more general than the one obtained in the linear
approximation. In our context, the possibility of travelling stable
oscillation is obtained for a whole range of parameters whereas the linear
approximation implies more restrictive condition. Actually, in the linear
approximation, equation (\ref{pw}) reduces to:%
\begin{equation}
f\Omega \left( \theta ,Z\right) =J\left( \theta \right) +\left( \frac{\hat{f}%
_{1}}{\omega _{0}}+N_{1}\right) \nabla _{\theta }\Omega \left( \theta
,Z\right) +\left( \frac{\hat{f}_{3}}{\omega _{0}}-N_{2}\right) \nabla
_{\theta }^{2}\Omega \left( \theta ,Z\right) +\frac{c^{2}\hat{f}_{3}}{\omega
_{0}}\nabla _{Z}^{2}\Omega \left( \theta ,Z\right)  \label{wp}
\end{equation}%
and the condition for some stable linear travelling wave is obtained only
for one value:%
\begin{equation}
\frac{\hat{f}_{1}}{\omega _{0}}+N_{1}=0  \label{nl}
\end{equation}%
However, once the possibility of travelling wave is understood, one can
replace (\ref{pw}) by its linear version (\ref{wp}) where (\ref{nl}) is
assumed to be satisfied.

\subsection{Interaction corrections to the wave equation (\protect\ref{pw})}

Equation (\ref{fft}) yields the corrective terms to $\omega ^{-1}\left(
J\left( \theta \right) ,\theta ,Z,\mathcal{G}_{0}+\left\vert \Psi
\right\vert ^{2}\right) $. We focus on the weak field approximation (\ref%
{wff}) to ensure corrections of small magnitude.\ Since $p_{l}+%
\sum_{i}p_{l}^{i}\geqslant 2$, the lowest order correction is for $m=1$ and $%
p_{l}+\sum_{i}p_{l}^{i}\geqslant 2=2$, and appendix 4.3 shows that:%
\begin{equation}
\omega _{e}^{-1}\left( J\left( \theta \right) ,\theta ,Z,\mathcal{G}%
_{0}+\left\vert \Psi \right\vert ^{2}\right) =\omega ^{-1}\left( J\left(
\theta \right) ,\theta ,Z,\mathcal{G}_{0}+\left\vert \Psi \right\vert
^{2}\right) +Z
\end{equation}%
where:%
\begin{eqnarray}
Z &=&\int^{\theta }d\theta \sum_{\substack{ j\geqslant 1  \\ m\geqslant 1}}%
\sum_{\substack{ p_{l},\left( p_{l}^{i}\right) _{m\times j}  \\ %
p_{l}+\sum_{i}p_{l}^{i}\geqslant 2}}\left( \prod\limits_{i=1}^{m}\frac{%
\sharp _{j+1,m}\left( \left( p_{m},\left( p_{l}^{m}\right) \right) \right) }{%
4\bar{\sharp}_{j+1,m}\left( \left( p_{i},\left( p_{l}^{m}\right) \right)
\right) }\right) \frac{a_{j,m}}{2} \\
&&\times \int \prod\limits_{i=1}^{m}\left\{ \frac{\delta
^{\sum_{l}p_{l}^{i}}\left( \nabla _{\theta }\omega ^{-1}\left( \theta
,Z,\left\vert \Psi \right\vert ^{2}\right) \right) }{\dprod\limits_{l=1}^{j}%
\delta ^{p_{l}^{i}}\left\vert \Psi \left( \theta ^{\left( l\right)
},Z_{_{l}}\right) \right\vert ^{2}}\right\} \frac{\delta
^{\sum_{l}p_{l}}\left( \nabla _{\theta }\omega ^{-1}\left( \theta
,Z,\left\vert \Psi \right\vert ^{2}\right) \right) }{\dprod\limits_{l=1}^{j}%
\delta ^{p_{l}}\left\vert \Psi \left( \theta ^{\left( l\right)
},Z_{_{l}}\right) \right\vert ^{2}}\dprod\limits_{l=1}^{j}\left\vert \Psi
\left( \theta ^{\left( l\right) },Z_{l}\right) \right\vert ^{2}d\theta
^{\left( l\right) }dZ_{l}  \notag
\end{eqnarray}%
This series take into account the interaction between the frequencies and
the background field. To find detailed results, we limit ourselves to the
lowest order corrections. We show in appendix 3, that these corrections have
the form:%
\begin{eqnarray}
\omega _{e}^{-1}\left( J\left( \theta \right) ,\theta ,Z,\mathcal{G}%
_{0}+\left\vert \Psi \right\vert ^{2}\right) &=&\omega ^{-1}\left( J\left(
\theta \right) ,\theta ,Z,\mathcal{G}_{0}+\left\vert \Psi \right\vert
^{2}\right)  \label{mge} \\
&&-\frac{f}{4\left( \frac{\hat{f}_{3}}{\omega \left( \theta ,Z\right) }%
-N_{2}\right) }\int \int^{\theta }\left( \frac{\delta \left( \omega
^{-1}\left( J\left( \theta \right) ,\theta ,Z,\mathcal{G}_{0}+\left\vert
\Psi \right\vert ^{2}\right) \right) }{\delta \left\vert \Psi \left( \theta
^{\left( l\right) },Z_{_{l}}\right) \right\vert ^{2}}\right) ^{2}  \notag \\
&&+\int \frac{1}{4}\nabla _{\theta }\left( \frac{\delta \left( \omega
^{-1}\left( J\left( \theta \right) ,\theta ,Z,\mathcal{G}_{0}+\left\vert
\Psi \right\vert ^{2}\right) \right) }{\delta \left\vert \Psi \left( \theta
^{\left( l\right) },Z_{_{l}}\right) \right\vert ^{2}}\right) ^{2}\left\vert
\Psi \left( \theta ^{\left( l\right) },Z_{_{l}}\right) \right\vert ^{2} 
\notag \\
&&+\frac{\frac{\hat{f}_{1}}{\omega \left( \theta ,Z\right) }+N_{1}}{4\left( 
\frac{\hat{f}_{3}}{\omega \left( \theta ,Z\right) }-N_{2}\right) }\int
\left( \frac{\delta \left( \omega ^{-1}\left( J\left( \theta \right) ,\theta
,Z,\mathcal{G}_{0}+\left\vert \Psi \right\vert ^{2}\right) \right) }{\delta
\left\vert \Psi \left( \theta ^{\left( l\right) },Z_{_{l}}\right)
\right\vert ^{2}}\right) ^{2}\left\vert \Psi \left( \theta ^{\left( l\right)
},Z_{_{l}}\right) \right\vert ^{2}  \notag
\end{eqnarray}%
In (\ref{mge}), the second and the third contributions:%
\begin{eqnarray}
&&-\frac{f}{4\left( \frac{\hat{f}_{3}}{\omega \left( \theta ,Z\right) }%
-N_{2}\right) }\int \int^{\theta }\left( \frac{\delta \left( \omega
^{-1}\left( J\left( \theta \right) ,\theta ,Z,\mathcal{G}_{0}+\left\vert
\Psi \right\vert ^{2}\right) \right) }{\delta \left\vert \Psi \left( \theta
^{\left( l\right) },Z_{_{l}}\right) \right\vert ^{2}}\right) ^{2}\left\vert
\Psi \left( \theta ^{\left( l\right) },Z_{_{l}}\right) \right\vert ^{2}
\label{ctB} \\
&&+\int \frac{1}{4}\nabla _{\theta }\left( \frac{\delta \left( \omega
^{-1}\left( J\left( \theta \right) ,\theta ,Z,\mathcal{G}_{0}+\left\vert
\Psi \right\vert ^{2}\right) \right) }{\delta \left\vert \Psi \left( \theta
^{\left( l\right) },Z_{_{l}}\right) \right\vert ^{2}}\right) ^{2}\left\vert
\Psi \left( \theta ^{\left( l\right) },Z_{_{l}}\right) \right\vert ^{2} 
\notag
\end{eqnarray}%
\bigskip describe the influence of the collective state defined by $%
\left\vert \Psi \left( \theta ^{\left( l\right) },Z_{_{l}}\right)
\right\vert ^{2}$ on the frequency at position $Z$ and time $\theta $. Under
our previous assumption that $\frac{\hat{f}_{3}}{\omega \left( \theta
,Z\right) }-N_{2}$ $<0$, and given that $f>0$, the first term in (\ref{ctB})
is positive, and thus, this term reduces $\omega \left( J\left( \theta
\right) ,\theta ,Z,\mathcal{G}_{0}+\left\vert \Psi \right\vert ^{2}\right) $%
. The higher the sensitivity $\frac{\delta \left( \omega ^{-1}\left( J\left(
\theta \right) ,\theta ,Z,\mathcal{G}_{0}+\left\vert \Psi \right\vert
^{2}\right) \right) }{\delta \left\vert \Psi \left( \theta ^{\left( l\right)
},Z_{_{l}}\right) \right\vert ^{2}}$ of the frequency to the collective
state, the more the frequency of the wave is reduced. The effect of this
smooting is cumulative in time, as shown by the integral over time arising
in this term. \ The second contribution in (\ref{ctB}) amplifies this
smoothing. Actually, this term is positive when $\frac{\delta \left( \omega
^{-1}\left( J\left( \theta \right) ,\theta ,Z,\mathcal{G}_{0}+\left\vert
\Psi \right\vert ^{2}\right) \right) }{\delta \left\vert \Psi \left( \theta
^{\left( l\right) },Z_{_{l}}\right) \right\vert ^{2}}$, i.e. the sensitivity
of frequency to the background field, increases in absolute value.\ As a
consequence, it reduces the oscillations of $\omega \left( J\left( \theta
\right) ,\theta ,Z,\mathcal{G}_{0}+\left\vert \Psi \right\vert ^{2}\right) $
when the frequency's dependency in the background field increases at
position $Z$ and time $\theta $.

The fourth term in (\ref{mge}):%
\begin{equation*}
\frac{\frac{\hat{f}_{1}}{\omega \left( \theta ,Z\right) }+N_{1}}{4\left( 
\frac{\hat{f}_{3}}{\omega \left( \theta ,Z\right) }-N_{2}\right) }\int
\left( \frac{\delta \left( \omega ^{-1}\left( J\left( \theta \right) ,\theta
,Z,\mathcal{G}_{0}+\left\vert \Psi \right\vert ^{2}\right) \right) }{\delta
\left\vert \Psi \left( \theta ^{\left( l\right) },Z_{_{l}}\right)
\right\vert ^{2}}\right) ^{2}\left\vert \Psi \left( \theta ^{\left( l\right)
},Z_{_{l}}\right) \right\vert ^{2}
\end{equation*}%
reinforces the mechanism of oscillation stabilization described in section
8.3. It has the sign of $-\left( \frac{\hat{f}_{1}}{\omega \left( \theta
,Z\right) }+N_{1}\right) $, given our assumption $\frac{\hat{f}_{3}}{\omega
\left( \theta ,Z\right) }-N_{2}$ $<0$ ensuring oscillatory behavior of $%
\omega \left( J\left( \theta \right) ,\theta ,Z,\mathcal{G}_{0}+\left\vert
\Psi \right\vert ^{2}\right) $. Thus, the correction to $\omega \left(
J\left( \theta \right) ,\theta ,Z,\mathcal{G}_{0}+\left\vert \Psi
\right\vert ^{2}\right) $ induced by this term has the sign of $\left( \frac{%
\hat{f}_{1}}{\omega \left( \theta ,Z\right) }+N_{1}\right) $: for $\frac{%
\hat{f}_{1}}{\omega \left( \theta ,Z\right) }+N_{1}>0$, the approximative
solution is oscillatory and explosive. Thus, the correction amplifies the
oscillations of $\omega \left( \theta ,Z\right) $ and the stabilization
mechanism applies. For $\omega \left( \theta ,Z\right) $ such that $\frac{%
\hat{f}_{1}}{\omega \left( \theta ,Z\right) }+N_{1}<0$, the correction term
turns negative and further decreases $\omega \left( \theta ,Z\right) $.

The series of higher corrections is computed in appendix 4.3. It shows that,
in the local approximation, the frequencies can be described by a wave
equation whose form depends on the stabilization potential and the evolution
of the background itself.

\subsection{Some extensions}

\subsubsection{Multiple components field}

A multiple-components field which describes excitatory vs inhibitory
currents leads to frequencies equations that are similar to (\ref{ftm}) when
interaction corrections are neglected :%
\begin{eqnarray}
\omega _{i}\left( \theta ,Z\right) &=&F_{i}\left( J\left( \theta \right) +%
\frac{\kappa }{N}\int T\left( Z,Z_{1}\right) \frac{\omega _{j}\left( \theta -%
\frac{\left\vert Z-Z_{1}\right\vert }{c},Z_{1}\right) }{\omega _{i}\left(
\theta ,Z\right) }G^{ij}\right.  \label{tmp} \\
&&\times \left. W\left( \frac{\omega _{i}\left( \theta ,Z\right) }{\omega
_{j}\left( \theta -\frac{\left\vert Z-Z_{1}\right\vert }{c},Z_{1}\right) }%
\right) \left( \mathcal{\bar{G}}_{0j}\left( 0,Z_{1}\right) +\left\vert \Psi
_{j}\left( \theta -\frac{\left\vert Z-Z_{1}\right\vert }{c},Z_{1}\right)
\right\vert ^{2}\right) dZ_{1}\right)  \notag
\end{eqnarray}%
Similar computations to those leading to (\ref{pw}) yield:%
\begin{equation}
f\Omega \left( \theta ,Z\right) =J_{i}\left( \theta \right) +\left( \frac{%
\hat{f}_{1ij}}{\omega _{i}\left( \theta ,Z\right) }+N_{1i}\delta
_{ij}\right) \nabla _{\theta }\Omega _{j}\left( \theta ,Z\right) +\left( 
\frac{\hat{f}_{3ij}}{\omega _{i}\left( \theta ,Z\right) }-N_{2i}\right)
\nabla _{\theta }^{2}\Omega _{j}\left( \theta ,Z\right) +\frac{c^{2}\hat{f}%
_{3ij}}{\omega \left( \theta ,Z\right) }\nabla _{Z}^{2}\Omega _{j}\left(
\theta ,Z\right)  \label{wpw}
\end{equation}%
where the sum over $j$ is implicit and with:%
\begin{eqnarray}
N_{1i} &=&\frac{\Psi _{0i}\left( Z\right) }{U^{\prime \prime }\left(
X_{0}\right) \omega _{i}^{2}\left( J\left( \theta \right) ,\theta ,Z,%
\mathcal{G}_{0}\right) }  \label{vctr} \\
N_{2i} &=&\frac{\Gamma _{1i}\Psi _{0i}\left( Z\right) }{U^{\prime \prime
}\left( X_{0}\right) \omega _{i}^{2}\left( J\left( \theta \right) ,\theta ,Z,%
\mathcal{G}_{0}\right) }  \notag
\end{eqnarray}%
\begin{eqnarray}
\hat{f}_{1ij} &=&\frac{W^{\prime }\left( \frac{\omega _{i}\left( \theta
,Z\right) }{\omega _{j}\left( \theta ,Z\right) }\right) -W\left( \frac{%
\omega _{i}\left( \theta ,Z\right) }{\omega _{j}\left( \theta ,Z\right) }%
\right) }{c}\Gamma _{1ij}\text{, }\hat{f}_{3ij}=\frac{\left( W\left( \frac{%
\omega _{i}\left( \theta ,Z\right) }{\omega _{j}\left( \theta ,Z\right) }%
\right) -W^{\prime }\left( \frac{\omega _{i}\left( \theta ,Z\right) }{\omega
_{j}\left( \theta ,Z\right) }\right) \right) \Gamma _{2ij}}{c^{2}}
\label{mtrx} \\
\Gamma _{1ij} &=&\frac{\kappa }{NX_{r}}G^{ij}\int \left\vert
Z-Z_{1}\right\vert T\left( Z,Z_{1}\right) \mathcal{\bar{G}}_{0j}\left(
0,Z_{1}\right) dZ_{1}  \notag \\
\Gamma _{2ij} &=&\frac{\kappa }{2NX_{r}}G^{ij}\int \left( Z-Z_{1}\right)
^{2}T\left( Z,Z_{1}\right) \mathcal{\bar{G}}_{0j}\left( 0,Z_{1}\right) dZ_{1}
\notag
\end{eqnarray}%
Equation (\ref{wpw}) describes the interactions of several non linear
traveling waves.

\subsubsection{Non constant background frequency}

In the above, we considered translation-invariant transfer functions.
Although correct in first approximation, this hypothesis does not hold in
general. For instance, it may be invalidated by finite volume of the system
or border conditions. Moreover, since the whole system depends on the
collective state, one may expect that endogeneizing the transfer functions
induce the emergence of states with position-dependent transfer functions. A
mechanism for this emergence is described in section 8.6.

We will thus consider transfer functions of the form $T\left( Z,Z_{1}\right) 
$. To make things simpler, we dismiss the corrections to the frequencies due
to the potential and the background field and focus on the linear
approximation (\ref{wp}) of (\ref{pw}).

The derivation of the linearized expansion of (\ref{fft}) around $\omega
_{0}\left( Z\right) $ is similar to that of (\ref{pw}), but now yields a
wave equation in an inhomogeneous medium: 
\begin{equation}
\sigma _{_{\theta }}^{2}\nabla _{\theta }^{2}\hat{\Omega}\left( \theta
,Z\right) =g_{0}\left( Z\right) \hat{\Omega}\left( \theta ,Z\right)
-g_{1}\left( Z\right) \nabla _{\theta }\hat{\Omega}\left( \theta ,Z\right)
+g_{2}\left( Z\right) \nabla _{Z}^{2}\hat{\Omega}\left( \theta ,Z\right)
\label{dnqgvr}
\end{equation}%
where we defined:%
\begin{eqnarray*}
\hat{\Omega}\left( \theta ,Z\right) &=&\frac{\Omega \left( \theta ,Z\right) 
}{\omega _{0}\left( Z\right) } \\
g_{1}\left( Z\right) &=&\frac{\Gamma _{1}^{\prime }\left( Z\right) -\Gamma
_{1}\left( Z\right) }{c\sigma _{_{\theta }}^{2}+\frac{\Gamma _{2}^{\prime
}\left( Z\right) -\Gamma _{2}\left( Z\right) }{c}} \\
g_{2}\left( Z\right) &=&\frac{\Gamma _{2}^{\prime }\left( Z\right) -\Gamma
_{2}\left( Z\right) }{\sigma _{_{\theta }}^{2}+\frac{\Gamma _{2}^{\prime
}\left( Z\right) -\Gamma _{2}\left( Z\right) }{c^{2}}} \\
\Gamma _{1}\left( Z\right) &=&\frac{\kappa }{NX_{r}}\int \frac{\left\vert
Z-Z_{1}\right\vert T\left( Z,Z_{1}\right) \frac{\omega _{0}\left(
Z_{1}\right) }{\omega _{0}\left( Z\right) }W\left( \frac{\omega _{0}\left(
Z\right) }{\omega _{0}\left( Z_{1}\right) }\right) dZ_{1}}{\omega _{0}\left(
Z\right) \sqrt{\frac{\pi }{8}\left( \frac{1}{\bar{X}_{r}}\right) ^{2}+\frac{%
\pi }{2}\alpha }}\Gamma _{0}\left( Z\right) \\
\Gamma _{1}^{\prime } &=&\frac{\kappa }{NX_{r}}\int \frac{\left\vert
Z-Z_{1}\right\vert T\left( Z,Z_{1}\right) \frac{\omega _{0}\left(
Z_{1}\right) }{\omega _{0}\left( Z\right) }W^{\prime }\left( \frac{\omega
_{0}\left( Z\right) }{\omega _{0}\left( Z_{1}\right) }\right) dZ_{1}}{\omega
_{0}\left( Z\right) \sqrt{\frac{\pi }{8}\left( \frac{1}{\bar{X}_{r}}\right)
^{2}+\frac{\pi }{2}\alpha }}\Gamma _{0} \\
\Gamma _{2} &=&\frac{\kappa }{2NX_{r}}\frac{\int \left( Z-Z_{1}\right)
^{2}T\left( Z,Z_{1}\right) \frac{\omega _{0}\left( Z_{1}\right) }{\omega
_{0}\left( Z\right) }W\left( \frac{\omega _{0}\left( Z\right) }{\omega
_{0}\left( Z_{1}\right) }\right) dZ_{1}}{\omega _{0}\left( Z\right) \sqrt{%
\frac{\pi }{8}\left( \frac{1}{\bar{X}_{r}}\right) ^{2}+\frac{\pi }{2}\alpha }%
}\Gamma _{0} \\
\Gamma _{2}^{\prime }\left( Z\right) &=&\frac{\kappa }{2NX_{r}}\frac{\int
\left( Z-Z_{1}\right) ^{2}T\left( Z,Z_{1}\right) \frac{\omega _{0}\left(
Z_{1}\right) }{\omega _{0}\left( Z\right) }W^{\prime }\left( \frac{\omega
_{0}\left( Z\right) }{\omega _{0}\left( Z_{1}\right) }\right) dZ_{1}}{\omega
_{0}\left( Z\right) \sqrt{\frac{\pi }{8}\left( \frac{1}{\bar{X}_{r}}\right)
^{2}+\frac{\pi }{2}\alpha }}\Gamma _{0}\left( Z\right) \\
\Gamma _{0}\left( Z\right) &=&G^{\prime }\left( \frac{\kappa }{N}\int \frac{%
T\left( Z,Z_{1}\right) W\left( \frac{\omega _{0}\left( Z\right) }{\omega
_{0}\left( Z_{1}\right) }\right) dZ_{1}}{\sqrt{\frac{\pi }{8}\left( \frac{1}{%
\bar{X}_{r}}\right) ^{2}+\frac{\pi }{2}\alpha }}\right)
\end{eqnarray*}

\subsubsection{Arbitrary transfer functions}

We can derive a straightforward generalization of (\ref{pw}) by considering
anisotropic transfer functions. So far, we have assumed that: 
\begin{equation*}
\int \left( Z-Z_{1}\right) _{i}\left( Z-Z_{1}\right) _{j}T\left(
Z,Z_{1}\right) \frac{\omega _{0}\left( Z_{1}\right) }{\omega _{0}\left(
Z\right) }W\left( \frac{\omega _{0}\left( Z\right) }{\omega _{0}\left(
Z_{1}\right) }\right) dZ_{1}=\delta _{i,j}
\end{equation*}%
where $\delta _{i,j}$ is the Kronecker symbol. Relaxing this condition, we
can replace $f_{2}\left( Z\right) \rightarrow f_{2}^{ij}\left( Z\right) $, $%
g_{2}\left( Z\right) \rightarrow g_{2}^{ij}\left( Z\right) =\frac{%
f_{2}^{ij}\left( Z\right) }{1+f_{3}\left( Z\right) }$. Equation (\ref{pw})
becomes: 
\begin{equation}
\nabla _{\theta }^{2}\Omega \left( \theta ,Z\right) =g_{0}\left( Z\right)
\Omega \left( \theta ,Z\right) +g_{1}\left( Z\right) \nabla _{\theta }\Omega
\left( \theta ,Z\right) +g_{2}^{ij}\left( Z\right) \nabla _{Z_{i}}\nabla
_{Z_{j}}\Omega \left( \theta ,Z\right)  \label{nsT}
\end{equation}%
for distributions:%
\begin{eqnarray*}
f_{2}\left( Z\right) &=&\left( \omega _{0}W^{\prime }\left( 1\right)
-W\left( 1\right) \right) \Gamma _{2}^{ij} \\
\Gamma _{2}^{ij} &=&\frac{\kappa }{2NX_{r}}\frac{\int \left( Z-Z_{1}\right)
_{i}\left( Z-Z_{1}\right) _{j}T\left( Z,Z_{1}\right) dZ_{1}}{\omega _{0}%
\sqrt{\frac{\pi }{8}\left( \frac{1}{\bar{X}_{r}}\right) ^{2}+\frac{\pi }{2}%
\alpha }}\Gamma _{0}
\end{eqnarray*}%
Equation (\ref{nsT}) is a wave equation in an anisotropic medium, the
anisotropy being described by the metric tensor $g_{2}^{ij}\left( Z\right) $.

\subsection{Including transfer functions dynamics}

Until now, we merely considered the dynamics of frequencies, and transfer
functions were considered in first approximation as depending on
frequencies. We will now briefly show how the model can be generalized by
including dynamic oscillations for the transfer function. To simplify the
formula, we consider a constant background frequency and restrain to the
linear approximation defined by equation (\ref{wp}), but the computations
can be generalized to a position dependent background (see appendix 7), and
the idea can be translated to the non-local context presented in the next
section.

To account for the dynamic nature of the transfer functions $T\left(
Z,Z_{1},\omega ,\omega _{1}\right) $, we associate to equation (\ref{wp}) an
evolution equation for $T\left( Z,Z_{1},\omega ,\omega _{1}\right) $. Using (%
\ref{rsf}), we replace $T\left( Z,Z_{1},\omega ,\omega _{1}\right) $ by a
general function $T\left( Z,Z_{1},\theta \right) $ that is a priori
independent from frequencies. Thus, around the equilibrium defined by the
background frequency $\omega _{0}$, the function $T\left( Z,Z_{1},\theta
\right) $ writes:%
\begin{equation*}
T\left( Z,Z_{1},\theta \right) =T_{0}\left( Z,Z_{1}\right) +h\left(
Z,Z_{1}\right) \hat{T}\left( Z,\theta ,Z_{1}\right)
\end{equation*}%
where $T_{0}\left( Z,Z_{1}\right) $ is the transfer function in this
equilibrium. The function $\hat{T}\left( Z,\theta ,Z_{1}\right) $\
represents the fluctuations around this equilibrium. The expansion of $G$\
around $\omega _{0}$ becomes:

\begin{eqnarray}
&&G\left( \frac{\kappa }{N}\int \frac{\omega _{0}+\Omega \left( \theta -%
\frac{\left\vert Z-Z_{1}\right\vert }{c},Z_{1}\right) -\Omega \left( \theta
,Z\right) }{\omega _{0}\sqrt{\frac{\pi }{8}\left( \frac{1}{\bar{X}_{r}}%
\right) ^{2}+\frac{\pi }{2}\alpha }}T\left( Z,Z_{1},\theta \right)
dZ_{1}\right)  \label{vrg} \\
&\simeq &\omega _{0}+\Gamma _{0}\left( \int \left( \frac{\Omega \left(
\theta -\frac{\left\vert Z-Z_{1}\right\vert }{c},Z_{1}\right) -\Omega \left(
\theta ,Z\right) }{\omega _{0}\sqrt{\frac{\pi }{8}\left( \frac{1}{\bar{X}_{r}%
}\right) ^{2}+\frac{\pi }{2}\alpha }}T_{0}\left( Z,Z_{1}\right) +\frac{%
h\left( Z,Z_{1}\right) \hat{T}\left( Z,\theta ,Z_{1}\right) }{\sqrt{\frac{%
\pi }{8}\left( \frac{1}{\bar{X}_{r}}\right) ^{2}+\frac{\pi }{2}\alpha }}%
\right) \right)  \notag
\end{eqnarray}%
As a consequence, equation (\ref{wp}) is replaced by:

\begin{equation*}
\sigma _{_{\theta }}^{2}\nabla _{\theta }^{2}\Omega \left( \theta ,Z\right)
=\Omega \left( \theta ,Z\right) +\frac{\Gamma _{1}}{c}\nabla _{\theta
}\Omega \left( \theta ,Z\right) -\Gamma _{2}\nabla _{Z}^{2}\Omega \left(
\theta ,Z\right) -\frac{\Gamma _{2}}{c^{2}}\nabla _{\theta }^{2}\Omega
\left( \theta ,Z\right) -\Gamma _{0}\hat{T}\left( Z,\theta \right)
\end{equation*}%
where we define:%
\begin{equation*}
\hat{T}\left( Z,\theta \right) =\int \frac{h\left( Z,Z_{1}\right) \hat{T}%
\left( Z,\theta ,Z_{1}\right) }{\sqrt{\frac{\pi }{8}\left( \frac{1}{\bar{X}%
_{r}}\right) ^{2}+\frac{\pi }{2}\alpha }}
\end{equation*}%
and:%
\begin{eqnarray*}
\Gamma _{1} &=&\frac{\kappa }{NX_{r}}\int \frac{\left\vert
Z-Z_{1}\right\vert T_{0}\left( Z,Z_{1}\right) dZ_{1}}{\omega _{0}\sqrt{\frac{%
\pi }{8}\left( \frac{1}{\bar{X}_{r}}\right) ^{2}+\frac{\pi }{2}\alpha }}%
\Gamma _{0} \\
\Gamma _{2} &=&\frac{\kappa }{2NX_{r}}\frac{\int \left( Z-Z_{1}\right)
^{2}T_{0}\left( Z,Z_{1}\right) dZ_{1}}{\omega _{0}\sqrt{\frac{\pi }{8}\left( 
\frac{1}{\bar{X}_{r}}\right) ^{2}+\frac{\pi }{2}\alpha }}\Gamma _{0} \\
\Gamma _{0} &=&G^{\prime }\left( \frac{\kappa }{N}\int \frac{T_{0}\left(
Z,Z_{1}\right) dZ_{1}}{\sqrt{\frac{\pi }{8}\left( \frac{1}{\bar{X}_{r}}%
\right) ^{2}+\frac{\pi }{2}\alpha }}\right)
\end{eqnarray*}%
The dynamics for $\hat{T}\left( Z,\theta \right) $ derived in appendix 2
yields a system of dynamic equations for $\left( \Omega \left( \theta
,Z\right) ,\hat{T}\left( Z,\theta \right) \right) $ that are similar to (\ref%
{wp}) for a slowly varying $\hat{T}\left( Z,\theta \right) $: 
\begin{eqnarray}
0 &=&\sigma _{_{\theta }}^{2}\nabla _{\theta }^{2}\Omega \left( \theta
,Z\right) -\left( \Omega \left( \theta ,Z\right) +\frac{\Gamma _{1}}{c}%
\nabla _{\theta }\Omega \left( \theta ,Z\right) -\Gamma _{2}\nabla
_{Z}^{2}\Omega \left( \theta ,Z\right) -\frac{\Gamma _{2}}{c^{2}}\nabla
_{\theta }^{2}\Omega \left( \theta ,Z\right) -\Gamma _{0}\hat{T}\left(
Z,\theta \right) \right)  \label{dng} \\
0 &=&\frac{\nabla _{\theta }^{2}\hat{T}\left( Z,\theta \right) }{\lambda }%
+U_{1}\left( \omega _{0}\right) \nabla _{\theta }\hat{T}\left( Z,\theta
\right) +U_{2}\left( \omega \right) \hat{T}\left( Z,\theta \right)
\label{dnt} \\
&&-\left( \rho \bar{C}\left( Z\right) h_{C}^{\prime }\left( \omega
_{0}\right) -\frac{\rho \left( D\left( Z\right) \hat{T}_{0}\left( Z\right)
h_{D}^{\prime }\left( \omega _{0}\right) +\bar{C}_{0}\left( Z\right)
h_{C}^{\prime }\left( \omega _{0}\right) \right) }{\lambda \tau }\right)
\Omega \left( Z,\theta \right)  \notag \\
&&-\frac{\rho D\left( Z\right) h_{D}^{\prime }\left( \omega _{0}\right)
\left( \Gamma _{1}\nabla _{\theta }\Omega \left( Z,\theta \right) -\left(
\Gamma _{1}\nabla _{\theta }^{2}\Omega \left( Z,\theta \right) +\Gamma
_{2}\nabla _{Z}^{2}\Omega \left( Z,\theta \right) \right) \right) }{\lambda
\tau }  \notag
\end{eqnarray}%
with:%
\begin{eqnarray*}
\bar{C}\left( Z\right) &=&\frac{1}{\sqrt{\frac{\pi }{8}\left( \frac{1}{\bar{X%
}_{r}}\right) ^{2}+\frac{\pi }{2}\alpha }}\int h\left( Z,Z_{1}\right)
C\left( Z_{1}\right) \\
\bar{C}_{0}\left( Z\right) &=&\frac{1}{\sqrt{\frac{\pi }{8}\left( \frac{1}{%
\bar{X}_{r}}\right) ^{2}+\frac{\pi }{2}\alpha }}\int h\left( Z,Z_{1}\right)
C\left( Z_{1}\right) \hat{T}_{0}\left( Z,Z_{1}\right) \\
\hat{T}_{0}\left( Z\right) &=&\frac{1}{\sqrt{\frac{\pi }{8}\left( \frac{1}{%
\bar{X}_{r}}\right) ^{2}+\frac{\pi }{2}\alpha }}\int h\left( Z,Z_{1}\right) 
\hat{T}_{0}\left( Z,Z_{1}\right)
\end{eqnarray*}%
Solving the system (\ref{dng}) and (\ref{dnt}) implies, depending on the
parameters, the existence of oscillatory solutions, both for frequencies of
activity and transfer functions. These oscillatory solutions illustrate the
constant interaction between cells' activity and the strength of
connectivity between these cells.

To conclude this section, remark that in the limit of slowly varying
transfer functions, equation (\ref{dng}) has constant coefficients, i.e.
describes wave propagation in an homogeneous medium. However, beyond this
approximation, equation (\ref{dng}) is replaced by:%
\begin{equation*}
\sigma _{_{\theta }}^{2}\nabla _{\theta }^{2}\Omega \left( \theta ,Z\right)
=\Omega \left( \theta ,Z\right) +\frac{\Gamma _{1}\left( \theta ,Z\right) }{c%
}\nabla _{\theta }\Omega \left( \theta ,Z\right) -\Gamma _{2}\left( \theta
,Z\right) \nabla _{Z}^{2}\Omega \left( \theta ,Z\right) -\frac{\Gamma
_{2}\left( \theta ,Z\right) }{c^{2}}\nabla _{\theta }^{2}\Omega \left(
\theta ,Z\right)
\end{equation*}%
with:%
\begin{eqnarray*}
\Gamma _{1}\left( \theta ,Z\right) &=&\frac{\kappa }{NX_{r}}\int \frac{%
\left\vert Z-Z_{1}\right\vert T\left( Z,Z_{1},\theta \right) dZ_{1}}{\omega
_{0}\sqrt{\frac{\pi }{8}\left( \frac{1}{\bar{X}_{r}}\right) ^{2}+\frac{\pi }{%
2}\alpha }}\Gamma _{0} \\
\Gamma _{2}\left( \theta ,Z\right) &=&\frac{\kappa }{2NX_{r}}\frac{\int
\left( Z-Z_{1}\right) ^{2}T\left( Z,Z_{1},\theta \right) dZ_{1}}{\omega _{0}%
\sqrt{\frac{\pi }{8}\left( \frac{1}{\bar{X}_{r}}\right) ^{2}+\frac{\pi }{2}%
\alpha }}\Gamma _{0} \\
\Gamma _{0}\left( \theta ,Z\right) &=&G^{\prime }\left( \frac{\kappa }{N}%
\int \frac{T\left( Z,Z_{1},\theta \right) dZ_{1}}{\sqrt{\frac{\pi }{8}\left( 
\frac{1}{\bar{X}_{r}}\right) ^{2}+\frac{\pi }{2}\alpha }}\right)
\end{eqnarray*}%
The dependency in $\left( \theta ,Z\right) $ is driven by the oscillations (%
\ref{dnt}) of the transfer functions. As a consequence, and as stated in
section 8.5.2, the frequencies propagate as waves in an inhomogeneous
medium, this inhomogeneity being time-dependent.

\subsection{Some implications of the differential equation for frequencies
in the linear approximation}

To assess the implications of the wave equations and find the propagation of
an external signal at some particular points to the whole thread, we must
compute the Green functions associated to the linear approximation equations
(\ref{wp}) and (\ref{dnqgvr}).

\subsubsection{Green functions and external signals}

The Green function of (\ref{pw}) and (\ref{dnqgvr}) are found using the
usual Fourier representation. We focus on the retarded Green functions that
model the wave propagation initiated by a source.

\paragraph{Constant background frequency}

We first consider (\ref{pw}). As explained in section 8.3, once the
existence of stable solutions has been established, we can set: 
\begin{equation*}
\frac{\hat{f}_{1}}{\omega \left( \theta ,Z\right) }+N_{1}=0
\end{equation*}%
and replace (\ref{pw}) with its linear approximation (\ref{wp}) for $J\left(
\theta \right) =0$, that writes: 
\begin{equation}
g_{2}\nabla _{Z}^{2}\Omega \left( \theta ,Z\right) -\nabla _{\theta
}^{2}\Omega \left( \theta ,Z\right) =g_{0}\Omega \left( \theta ,Z\right)
\label{KG}
\end{equation}%
with:%
\begin{equation*}
g_{0}=-\frac{f}{\left( \frac{\hat{f}_{3}}{\omega _{0}}-N_{2}\right) }\text{, 
}g_{2}=-\frac{\frac{c^{2}\hat{f}_{3}}{\omega _{0}}}{\left( \frac{\hat{f}_{3}%
}{\omega _{0}}-N_{2}\right) }
\end{equation*}%
Given our assumptions in section 8.3, both $g_{0}$ and $g_{2}$ are positive.

Equation (\ref{KG}) is of Klein-Gordon type and can be normalized by setting 
$g_{2}=1$ and writing $g_{0}=m^{2}$. Using its Fourier representation, the
retarded Green function of (\ref{pw}) is given by:%
\begin{equation}
\mathcal{G}\left( Z,Z^{\prime },t,t^{\prime }\right) =\int dk\frac{\exp
\left( ik.\left( Z-Z^{\prime }\right) -i\omega _{k}\left( t-t^{\prime
}\right) \right) }{\omega _{k}}H\left( t-t^{\prime }\right)  \label{Grfr}
\end{equation}%
with $\omega _{k}=\sqrt{k^{2}+m^{2}}$. This integral can be computed and
yields:%
\begin{equation}
\mathcal{G}\left( Z,Z^{\prime },t,t^{\prime }\right) =H\left( t-t^{\prime
}\right) \left( \frac{1}{2\pi }\delta \left( t-t^{\prime }\right) -\frac{%
mJ_{1}\left( m\sqrt{\left( t-t^{\prime }\right) ^{2}-\left( Z-Z^{\prime
}\right) ^{2}}\right) }{\sqrt{\left( t-t^{\prime }\right) ^{2}-\left(
Z-Z^{\prime }\right) ^{2}}}\right)  \label{Gr}
\end{equation}%
where $J_{1}$ is the $n=1$ Bessel function.

To inspect the implications of (\ref{Gr}), we merely need to approximate it
for small oscillations. For $g_{0}>>g_{2}$, i.e. $m^{2}>1$, we can expand $%
\sqrt{k^{2}+m^{2}}$ at the lowest order in $\frac{k^{2}}{m^{2}}$, and write (%
\ref{Grfr}), up to terms of order $\frac{1}{m^{3}}$, as:

\begin{equation}
\mathcal{G}\left( Z,Z^{\prime },t,t^{\prime }\right) \simeq \int dk\frac{%
\exp \left( ik.\left( Z-Z^{\prime }\right) -i\left( m+\frac{k^{2}}{m}\right)
\left( t-t^{\prime }\right) \right) }{m}H\left( t-t^{\prime }\right)
\label{nrt}
\end{equation}%
Computing the Fourier transform in (\ref{nrt}), the function $\mathcal{G}%
_{0}\left( Z,Z^{\prime },t,t^{\prime }\right) $ can be approximated by:%
\begin{equation}
\mathcal{G}\left( Z,Z^{\prime },t,t^{\prime }\right) =\exp \left( i\left( 
\frac{m}{2}\frac{\left( Z-Z^{\prime }\right) ^{2}}{\left( t-t^{\prime
}\right) }-m\left( t-t^{\prime }\right) \right) \right) H\left( t-t^{\prime
}\right)  \label{Gcst}
\end{equation}%
which shows that the Green function $\mathcal{G}\left( Z,Z^{\prime
},t,t^{\prime }\right) $ represents the path integral of a particle under
the constant potential $m$.

\paragraph{Non-constant background frequency}

The Green function of equation (\ref{dnqgvr}) is a generalization of (\ref%
{Gr}) and has been studied in the context of covariant quantum field theory.
However, (\ref{Gcst}) shows that a path integral formulation for the Green
function can produced. If $g_{2}\left( Z\right) $\ varies slowly with $Z$,
the analog of (\ref{Gcst}) with non-constant coefficients is:

\begin{equation}
\mathcal{G}\left( Z,Z^{\prime },t,t^{\prime }\right) =\int \exp \left(
i\left( \int_{z\left( t^{\prime }\right) =Z^{\prime }}^{z\left( t\right)
=Z}\left( \frac{\sqrt{\frac{g_{0}\left( z\left( s\right) \right) }{%
g_{2}\left( z\left( s\right) \right) }}}{2}\left( \frac{dz\left( s\right) }{%
ds}\right) ^{2}-\sqrt{g_{0}\left( z\left( s\right) \right) }\right)
ds\right) \right) Dz\left( s\right) H\left( t-t^{\prime }\right)
\label{Gvar}
\end{equation}%
The sum is over the set of paths $z\left( s\right) $ starting from $%
Z^{\prime }$ and ending at $Z$ in a time span of $t-t^{\prime }$. \ The
derivation of (\ref{Gvar}) is straightforward. If we neglect $g_{1}\left(
Z\right) $ as in the derivation of (\ref{KG}), (\ref{dnqgvr}) writes:

\begin{equation*}
\sigma _{_{\theta }}^{2}\nabla _{\theta }^{2}\hat{\Omega}\left( \theta
,Z\right) =g_{0}\left( Z\right) \hat{\Omega}\left( \theta ,Z\right)
+g_{2}\left( Z\right) \nabla _{Z}^{2}\hat{\Omega}\left( \theta ,Z\right)
\end{equation*}%
We then cut the time span $t-t^{\prime }$ into slices $\Delta t$, such that $%
g_{0}\left( Z\right) $ and $g_{2}\left( Z\right) $ can be considered
constant in a domain of radius $c\Delta t$. The Green function for a time
span $\Delta t$ is given by a formula similar to (\ref{Gcst}), except that $%
g_{2}\left( Z\right) \neq 1$:%
\begin{equation}
\mathcal{G}\left( z\left( s+\Delta t\right) ,z\left( s\right) ,\Delta
t\right) =\exp \left( i\left( \frac{\sqrt{\frac{g_{0}\left( z\left( s\right)
\right) }{g_{2}\left( z\left( s\right) \right) }}}{2}\frac{\left( z\left(
s+\Delta t\right) -z\left( s\right) \right) ^{2}}{\Delta t}-g_{0}\left(
z\left( s\right) \right) \Delta t\right) \right)  \label{Grm}
\end{equation}%
Under these assumptions, the convolution of (\ref{Grm}) over the time slices
yields ultimately formula (\ref{Gvar}).

\subsubsection{Propagation of external signals}

\paragraph{Constant coefficients}

With the Green function (\ref{Gcst}), we can compute the diffusion of an
external source along the thread by convolution. We assume an external
source:%
\begin{equation}
J\left( t,Z\right) =\exp \left( -i\omega _{0}t\right) \delta \left(
Z-Z_{0}\right)  \label{sgn}
\end{equation}%
which describes a signal located in $Z_{0}$, with frequency $\omega _{0}$.
Using (\ref{Gcst}), the amplitude $\Omega \left( t,Z\right) $ is:%
\begin{eqnarray*}
\Omega \left( t,Z\right) &=&\int \exp \left( i\left( \frac{m}{2}\frac{\left(
Z-Z_{0}\right) ^{2}}{\left( t-t^{\prime }\right) }-\omega _{0}t-\left(
m-\omega _{0}\right) \left( t-t^{\prime }\right) \right) \right) H\left(
t-t^{\prime }\right) dt^{\prime } \\
&=&\frac{\exp \left( -i\omega _{0}t-i\sqrt{m}\left\vert \left( m-\omega
_{0}\right) \right\vert \left\vert Z-Z_{0}\right\vert +i\pi \right) }{\sqrt{%
\left\vert \left( m-\omega _{0}\right) \right\vert }}
\end{eqnarray*}%
and for a signal including a whole range of frequencies:%
\begin{equation}
\hat{f}\left( t,Z\right) =\int f\left( \omega _{0}\right) \exp \left(
-i\omega _{0}t\right) d\omega _{0}  \label{snl}
\end{equation}%
the corresponding response of the thread is:%
\begin{equation*}
\Omega \left( t,Z\right) =\int \frac{\exp \left( -i\omega _{0}t-i\sqrt{m}%
\left\vert \left( m-\omega _{0}\right) \right\vert \left\vert
Z-Z_{0}\right\vert +i\pi \right) }{\sqrt{\left\vert \left( m-\omega
_{0}\right) \right\vert }}f\left( \omega _{0}\right) d\omega _{0}
\end{equation*}%
We assume that the range of frequencies in (\ref{snl}) is such that $%
m-\omega _{0}>0$, so that: 
\begin{eqnarray*}
\Omega \left( t,Z\right) &=&\int \frac{\exp \left( -i\omega _{0}t-i\sqrt{m}%
\left( m-\omega _{0}\right) \left\vert Z-Z_{0}\right\vert +i\pi \right) }{%
\sqrt{\left\vert \left( m-\omega _{0}\right) \right\vert }}f\left( \omega
_{0}\right) d\omega _{0} \\
&=&\int \frac{\exp \left( -i\omega _{0}\left( t-\sqrt{m}\left\vert
Z-Z_{0}\right\vert \right) \right) }{\sqrt{\left\vert \left( m-\omega
_{0}\right) \right\vert }}f\left( \omega _{0}\right) d\omega _{0}\frac{\exp
\left( -i\left( \sqrt{m}\right) ^{3}\left\vert Z-Z_{0}\right\vert +i\pi
\right) }{\sqrt{\left\vert \left( m-\omega _{0}\right) \right\vert }}
\end{eqnarray*}%
To simplify, we also assume that the frequencies of the signal satisfy $%
\left\vert \omega _{0}\right\vert <<m$, so that: 
\begin{equation}
\Omega \left( t,Z\right) \simeq \hat{f}\left( t-\sqrt{m}\left\vert
Z-Z_{0}\right\vert ,Z_{0}\right) \frac{\exp \left( -i\left( \sqrt{m}\right)
^{3}\left\vert Z-Z_{0}\right\vert +i\pi \right) }{m}  \label{rsp}
\end{equation}%
The whole past history of the signal is present in the frequencies at time $%
t $, and is thus recorded in the system of oscillations. The result (\ref%
{rsp}) can be extended for several independent sources. When these sources
are located in two points $Z_{1}$, $Z_{2}$ that emit signals $\hat{f}%
_{1}\left( t\right) $ and $\hat{f}_{2}\left( t\right) $ respectively, with
frequencies below $m$, the response is:%
\begin{eqnarray}
\Omega \left( t,Z\right) &\simeq &\hat{f}_{1}\left( t-\sqrt{m}\left\vert
Z-Z_{0}\right\vert \right) \frac{\exp \left( -i\left( \sqrt{m}\right)
^{3}\left\vert Z-Z_{0}\right\vert +i\pi \right) }{\sqrt{\left\vert \left(
m-\omega _{0}\right) \right\vert }}  \label{rspp} \\
&&+\hat{f}_{2}\left( t-\sqrt{m}\left\vert Z-Z_{0}\right\vert \right) \frac{%
\exp \left( -i\left( \sqrt{m}\right) ^{3}\left\vert Z-Z_{0}\right\vert +i\pi
\right) }{\sqrt{\left\vert \left( m-\omega _{0}\right) \right\vert }}  \notag
\end{eqnarray}%
The response defined by (\ref{rspp}) may present some interference
phenomena, depending on $\hat{f}_{1}$ and $\hat{f}_{2}$, as usual in waves
dynamics.\ 

\paragraph{Non constant coefficients}

Considering non constant coefficients in (\ref{Gvar}) translates the
hypothesis of position-dependent transfer functions between cells. The
implications of this assumption may be understood using formula (\ref{rspp}%
). Assume a thread divided in two regions, each characterized by constant
coefficients $g_{0}$ and $g_{2}$ and only connected via two "entry points".
This can be modelled by $g_{2}=0$ on the border of the two regions, and $%
g_{2}>>1$ at these two points.

Formula (\ref{Gvar}) implies that paths that do not cross the border at
points $Z_{1}$or $Z_{2}$\ do not contribute to the Green function. Actually,
the factor $\sqrt{\frac{g_{0}\left( z\left( s\right) \right) }{g_{2}\left(
z\left( s\right) \right) }}$ that arises in the weight (\ref{Gvar}) of such
paths induces large oscillations in the vicinity of the border that cancel
the contribution of the paths.

\ As a consequence, the paths contributing to the Green function have to
cross at $Z_{1}$or $Z_{2}$, which induces some interference phenomena (\ref%
{rspp}) on the transmitted signal.

More generally, the dependency of the transfer functions in $Z$\ along the
paths impacts the results, even for simple signals (\ref{sgn}). Actually,
the various paths reaching a point $Z$ of the thread contribute to the Green
function (\ref{rspp}). They each acquire a phase that depends on both the
path and the characteristic of the medium encountered.\ These phases may
create interferences between the paths. The trained networks may present
some particular learned features in the coefficients $g_{0}\left( Z\right) $
and $g_{2}\left( Z\right) $, i.e. their transfer functions, that would
produce either constructive or destructive interferences for the signals.

\paragraph{Non-static equilibrium}

The equations of the previous paragraph may be generalized for a
non-constant and slowly varying background solution. For a non-static
potential and for currents of large magnitude, a slowly varying solution of
the type:%
\begin{eqnarray*}
\omega _{0}\left( \theta ,Z\right) &=&F\left( J\left( \theta \right) +\frac{%
\kappa }{N}\int T\left( Z,Z_{1}\right) \frac{\omega _{0}\left( \theta -\frac{%
\left\vert Z-Z_{1}\right\vert }{c},Z_{1}\right) }{\omega _{0}\left( \theta
,Z\right) }\right. \\
&&\times \left. W\left( \frac{\omega _{0}\left( \theta ,Z\right) }{\omega
_{0}\left( \theta -\frac{\left\vert Z-Z_{1}\right\vert }{c},Z_{1}\right) }%
\right) \left( \mathcal{G}_{0}\left( 0,Z_{1}\right) +\left\vert \Psi
_{0}\left( \theta -\frac{\left\vert Z-Z_{1}\right\vert }{c},Z_{1}\right)
\right\vert ^{2}\right) dZ_{1}\right)
\end{eqnarray*}%
may exist, and we can expand (\ref{fqt}) around $\omega _{0}\left( \theta
,Z\right) $ in series of $\delta \Psi $. Minimizing the effective action
(see appendix 3) yields the values of $\delta \Psi $. Equation (\ref{fsr})
and the definition of the coefficients (\ref{ctf}) are still valid, but now $%
\mathcal{\bar{G}}_{0}\left( 0,Z_{1}\right) =\mathcal{G}_{0}\left(
0,Z_{1}\right) +\left\vert \Psi _{0}\right\vert ^{2}$ has to be replaced by
a time dependent propagator $\mathcal{\bar{G}}_{0}\left( \theta
,Z_{1}\right) =\mathcal{G}_{0}\left( 0,Z_{1}\right) +\left\vert \Psi
_{0}\left( \theta -\frac{\left\vert Z-Z_{1}\right\vert }{c},Z_{1}\right)
\right\vert ^{2}$. The coefficients arising in (\ref{fsr}) thus become time
dependent.

\section{Beyond local approximation}

\subsection{One-field system}

Section 8 focused on the wave equation for frequencies, i.e. the local
solutions of (\ref{fqt}), plus the corrections defined in (\ref{fft}). This
section goes one step further and studies the dynamics for frequencies
without the locality assumption. To do so, we dismiss the correction terms
due to the interaction between the system and the stabilization potential in
(\ref{fft}), and thus study the solutions of\ (\ref{fqt}) by rewriting the
equation:

\begin{eqnarray}
\omega \left( J,\theta ,Z\right) &=&F\left( J\left( \theta \right) +\frac{%
\kappa }{N}\int T\left( Z,Z_{1}\right) \frac{\omega \left( \theta -\frac{%
\left\vert Z-Z_{1}\right\vert }{c},Z_{1}\right) W\left( \frac{\omega \left(
\theta ,Z\right) }{\omega \left( \theta -\frac{\left\vert Z-Z_{1}\right\vert 
}{c},Z_{1}\right) }\right) }{\omega \left( \theta ,Z\right) }\right.
\label{qtF} \\
&&\times \left. \left( \mathcal{\bar{G}}_{0}\left( 0,Z_{1}\right)
+\left\vert \Psi \left( \theta -\frac{\left\vert Z-Z_{1}\right\vert }{c}%
,Z_{1}\right) \right\vert ^{2}\right) dZ_{1}\right)  \notag
\end{eqnarray}

\subsubsection{Series expansion of (\protect\ref{qtF})}

To write a non-local solution of (\ref{qtF}), we use the series expansion in 
$\left\vert \Psi \left( \theta ^{\left( j\right) },Z_{1}\right) \right\vert
^{2}$ of the right-hand side of (\ref{qtf}) and write:%
\begin{eqnarray}
\omega \left( J,\theta ,Z\right) &=&\omega \left( \theta ,Z\right)
_{\left\vert \Psi \right\vert ^{2}=0}  \label{psn} \\
&&+\int \sum_{n=1}^{\infty }\left( \frac{\delta ^{n}\omega \left( J,\theta
,Z\right) }{\dprod\limits_{i=1}^{n}\delta \left\vert \Psi \left( \theta
-l_{i},Z_{i}\right) \right\vert ^{2}}\right) _{\left\vert \Psi \right\vert
^{2}=0}\dprod\limits_{i=1}^{n}\left\vert \Psi \left( \theta
-l_{i},Z_{i}\right) \right\vert ^{2}  \notag
\end{eqnarray}%
The first term in (\ref{psn}), $\omega \left( \theta ^{\left( i\right)
},Z\right) _{\left\vert \Psi \right\vert ^{2}=0}$, is a solution of:%
\begin{equation}
\omega \left( \theta ,Z\right) _{\left\vert \Psi \right\vert ^{2}=0}=F\left(
J+\frac{\kappa }{N}\int T\left( Z,Z_{1}\right) \frac{\omega _{\left\vert
\Psi \right\vert ^{2}=0}\left( \theta -\frac{\left\vert Z-Z_{1}\right\vert }{%
c},Z_{1}\right) }{\omega _{\left\vert \Psi \right\vert ^{2}=0}\left( \theta
,Z\right) }W\left( \frac{\omega _{\left\vert \Psi \right\vert ^{2}=0}\left(
\theta ,Z\right) }{\omega _{\left\vert \Psi \right\vert ^{2}=0}\left( \theta
-\frac{\left\vert Z-Z_{1}\right\vert }{c},Z_{1}\right) }\right) \left( 
\mathcal{\bar{G}}_{0}\left( 0,Z_{1}\right) \right) dZ_{1}\right)  \label{srf}
\end{equation}%
To find the internal dynamics of the system, we will first consider a
constant external current $J\left( \theta \right) =J$, typically $J$ $=0$,
but the results of this section will be valid for a non static current $%
J\left( \theta \right) $. The static solution of (\ref{srf}) satifies:%
\begin{eqnarray*}
\omega \left( J,Z\right) &=&F\left( J+\frac{\kappa }{N}\int T\left(
Z,Z_{1}\right) \frac{\omega \left( Z_{1}\right) }{\omega \left( Z\right) }%
W\left( \frac{\omega \left( Z\right) }{\omega \left( Z_{1}\right) }\right) 
\mathcal{\bar{G}}_{0}\left( 0,Z_{i}\right) dZ_{1}\right) \\
&\equiv &F\left[ J,\omega ,Z\right]
\end{eqnarray*}%
we assume this solution to be known, and we chose to expand $\omega \left(
J,\theta ,Z\right) $ in (\ref{psn}) around this solution, the dynamics being
given by $\left\vert \Psi \left( \theta ^{\left( j\right) },Z_{1}\right)
\right\vert ^{2}$. We thus set:%
\begin{equation*}
\omega \left( \theta ,Z\right) _{\left\vert \Psi \right\vert ^{2}=0}=\omega
\left( J,Z\right)
\end{equation*}%
Appendices 5 and 6 compute the derivatives $\left( \frac{\delta ^{n}\omega
\left( J,\theta ,Z\right) }{\dprod\limits_{i=1}^{n}\delta \left\vert \Psi
\left( \theta -l_{i},Z_{i}\right) \right\vert ^{2}}\right) _{\left\vert \Psi
\right\vert ^{2}=0}$\ in (\ref{psn}).

Defining:%
\begin{eqnarray}
&&\hat{T}\left( \theta ,Z,Z_{1},\omega ,\Psi \right)  \label{vdr} \\
&=&\frac{\frac{\kappa }{N}\omega \left( J,\theta ,Z\right) T\left(
Z,Z_{1}\right) F^{\prime }\left[ J,\omega ,\theta ,Z,\Psi \right] }{\omega
^{2}\left( J,\theta ,Z\right) +\left( \int \frac{\kappa }{N}\omega \left(
J,\theta -\frac{\left\vert Z-Z^{\prime }\right\vert }{c},Z^{\prime }\right)
\left( \mathcal{\bar{G}}_{0}\left( 0,Z^{\prime }\right) +\left\vert \Psi
\left( \theta -\frac{\left\vert Z-Z^{\prime }\right\vert }{c},Z^{\prime
}\right) \right\vert ^{2}\right) T\left( Z,Z^{\prime }\right) dZ^{\prime
}\right) F^{\prime }\left[ J,\omega ,\theta ,Z,\Psi \right] }  \notag
\end{eqnarray}%
and the operator $\hat{T}$\ with kernel:%
\begin{eqnarray}
\hat{T}\left( \left( Z^{\left( l-1\right) },\theta ^{\left( l-1\right)
}\right) ,\left( Z^{\left( l\right) },\theta ^{\left( l\right) }\right)
\right) &=&\hat{T}\left( \theta -\sum_{j=1}^{l-1}\frac{\left\vert Z^{\left(
j-1\right) }-Z^{\left( j\right) }\right\vert }{c},Z^{\left( l-1\right)
},Z^{\left( l\right) },\omega _{0}\right)  \label{rnL} \\
&&\times \delta \left( \left( \theta ^{\left( l\right) }-\theta ^{\left(
l-1\right) }\right) -\frac{\left\vert Z^{\left( l-1\right) }-Z^{\left(
l\right) }\right\vert }{c}\right)  \notag
\end{eqnarray}%
appendix 5 shows that:%
\begin{eqnarray}
\frac{\delta \omega \left( J,\theta ,Z\right) }{\delta \left\vert \Psi
\left( \theta -l_{1},Z_{1}\right) \right\vert ^{2}} &=&\sum_{n=1}^{\infty
}\int \frac{\omega \left( J,\theta -\sum_{l=1}^{n}\frac{\left\vert Z^{\left(
l-1\right) }-Z^{\left( l\right) }\right\vert }{c},Z_{1}\right) }{\left( 
\mathcal{\bar{G}}_{0}\left( 0,Z_{1}\right) +\left\vert \Psi \left( \theta
-l_{1},Z_{1}\right) \right\vert ^{2}\right) }  \label{rdn} \\
&&\times \dprod\limits_{l=1}^{n}\hat{T}\left( \theta -\sum_{j=1}^{l-1}\frac{%
\left\vert Z^{\left( j-1\right) }-Z^{\left( j\right) }\right\vert }{c}%
,Z^{\left( l-1\right) },Z^{\left( l\right) },\omega ,\Psi \right) \delta
\left( l_{1}-\sum_{l=1}^{n}\frac{\left\vert Z^{\left( l-1\right) }-Z^{\left(
l\right) }\right\vert }{c}\right) \dprod\limits_{l=1}^{n-1}dZ^{\left(
l\right) }  \notag
\end{eqnarray}%
Appendix 6 builds on (\ref{rdn}) to compute the derivative arising in the
series expansion (\ref{psn}):%
\begin{equation}
\left( \frac{\delta ^{n}\omega \left( J,\theta ,Z\right) }{%
\dprod\limits_{i=1}^{n}\delta \left\vert \Psi \left( \theta
-l_{i},Z_{i}\right) \right\vert ^{2}}\right) _{\left\vert \Psi \right\vert
^{2}=0}\dprod\limits_{i=1}^{n}\left\vert \Psi \left( \theta
-l_{i},Z_{i}\right) \right\vert ^{2}  \label{rcs}
\end{equation}%
by a graphical representation. We associate the squared field $\left\vert
\Psi \left( \theta -l_{i},Z_{i}\right) \right\vert ^{2}$ to each point $%
Z_{i} $ and draw $m$ lines for $m=1,...,n$. One of them at least is starting
from $Z$. These lines are composed of an arbitrary number of segments and
all the points $Z_{i}$ are crossed by one line. Each line ends at a point $%
Z_{i}$. The starting points of the lines branch either at $Z$ or at some
point of an other line. There are $m$ branching points of valence $k$
including the starting point at $Z$. Apart from $Z$, the branching points
have valence $3,...,n-1$.

To each line $i$ of length $L_{i}$, we associate the factor:%
\begin{eqnarray}
F\left( line_{i}\right) &=&\dprod\limits_{l=1}^{L_{i}}\frac{\frac{\kappa }{N}%
T\left( Z^{\left( l-1\right) },Z^{\left( l\right) }\right) F^{\prime }\left[
J,\omega _{0},\theta -\sum_{j=1}^{l-1}\frac{\left\vert Z^{\left( j-1\right)
}-Z^{\left( j\right) }\right\vert }{c},Z^{\left( l-1\right) }\right] }{%
\omega _{0}\left( J,\theta -\sum_{j=1}^{l-1}\frac{\left\vert Z^{\left(
j-1\right) }-Z^{\left( j\right) }\right\vert }{c},Z^{\left( l-1\right)
}\right) }  \label{flN} \\
&&\times \frac{\omega _{0}\left( J,\theta -\sum_{l=1}^{L_{i}}\frac{%
\left\vert Z^{\left( l-1\right) }-Z^{\left( l\right) }\right\vert }{c}%
,Z_{i}\right) }{\mathcal{\bar{G}}_{0}\left( 0,Z_{i}\right) }  \notag \\
&=&\dprod\limits_{l=1}^{L_{i}}\hat{T}\left( \theta -\sum_{j=1}^{l-1}\frac{%
\left\vert Z^{\left( j-1\right) }-Z^{\left( j\right) }\right\vert }{c}%
,Z^{\left( l-1\right) },Z^{\left( l\right) },\omega _{0},\Psi \right) \frac{%
\omega _{0}\left( J,\theta -\sum_{l=1}^{L_{i}}\frac{\left\vert Z^{\left(
l-1\right) }-Z^{\left( l\right) }\right\vert }{c},Z_{i}\right) }{\mathcal{%
\bar{G}}_{0}\left( 0,Z_{i}\right) }  \notag
\end{eqnarray}%
and to each branching point $\left( X,\theta \right) =B$ of valence $k+2$,
we associate the factor:%
\begin{equation}
F\left( \left( X,\theta \right) \right) =\frac{\delta ^{k}\left( \frac{\frac{%
\kappa }{N}T\left( Z,Z^{\left( l\right) }\right) F^{\prime }\left[ J,\theta
,\omega _{0},Z^{\left( l\right) }\right] \mathcal{\bar{G}}_{0}\left(
0,Z^{\left( l\right) }\right) }{\omega _{0}\left( J,\theta ,Z^{\left(
l\right) }\right) }\right) }{\delta ^{k}\omega _{0}\left( J,\theta
,Z^{\left( l\right) }\right) }  \label{fcT}
\end{equation}%
and (\ref{rcs}) writes as a series of lines contributions connected by the
branching points:%
\begin{eqnarray}
&&\left( \frac{\delta ^{n}\omega \left( J,\theta ,Z\right) }{%
\dprod\limits_{i=1}^{n}\delta \left\vert \Psi \left( \theta
-l_{i},Z_{i}\right) \right\vert ^{2}}\right) _{\left\vert \Psi \right\vert
^{2}=0}\dprod\limits_{i=1}^{n}\left\vert \Psi \left( \theta
-l_{i},Z_{i}\right) \right\vert ^{2}  \label{rgr} \\
&=&\left( \sum_{m=1}^{n}\sum_{i=1}^{m}\sum_{\left(
line_{1},...,line_{m}\right) }\dprod\limits_{i}F\left( line_{i}\right)
\dprod\limits_{B}F\left( B\right) \right) \dprod\limits_{i=1}^{n}\left\vert
\Psi \left( \theta -l_{i},Z_{i}\right) \right\vert ^{2}  \notag
\end{eqnarray}%
The graphical representation is generic. The integration over the set of
lines also accounts for the degenerate case of lines that share some
segments.

\subsubsection{Path integral description}

\paragraph{Formalism}

Appendix 6 uses formula (\ref{rgr})\ to derive a non-local formula for the
successive derivatives of $\omega \left( J,\theta ,Z\right) $ and $\omega
^{-1}\left( J,\theta ,Z\right) $. Moreover, equation (\ref{rgr}) allows to
rewrite the expansion (\ref{snp}) as the sum of graphs for an auxiliary
complex field $\Lambda \left( Z_{i},\theta _{i}\right) $. The idea is to
regroup the graphs in (\ref{rgr}) so that their sum becomes a sum over
graphs drawn between an arbitrary number of branch points, seen as vertices
of arbitrary valence $k$ and associated factor (\ref{fcT}). These vertices
are connected by the edges of the graph with associated Green functions $%
\frac{1}{1-\left( 1+\left\vert \Psi \right\vert ^{2}\right) \hat{T}}$ where $%
\hat{T}$ is the operator whose kernel is defined in (\ref{rnL}). The factor $%
\left\vert \Psi \right\vert ^{2}$ is the operator multiplication by $%
\left\vert \Psi \left( \theta ,Z\right) \right\vert ^{2}$ at point $\left(
\theta ,Z\right) $.

Appendix 6 shows that:

\begin{eqnarray}
\omega \left( \theta ,Z\right) &=&\omega _{0}\left( J,\theta ,Z\right)
+\sum_{n=1}^{\infty }\frac{1}{n!}\frac{\int \hat{T}\Lambda ^{\dag }\left(
Z,\theta \right) \int \dprod\limits_{i=1}^{n}\omega _{0}\left( J,\theta
_{i},Z_{i}\right) \left\vert \Psi \left( J,\theta _{i},Z_{i}\right)
\right\vert ^{2}\Lambda \left( Z_{i},\theta _{i}\right) d\left( Z_{i},\theta
_{i}\right) \exp \left( -S\left( \Lambda \right) \right) \mathcal{D}\Lambda 
}{\exp \left( -S\left( \Lambda \right) \right) \mathcal{D}\Lambda }  \notag
\\
&=&\omega _{0}\left( J,\theta ,Z\right) +\frac{\int \hat{T}\Lambda ^{\dag
}\left( Z,\theta \right) \exp \left( -S\left( \Lambda \right) +\int \Lambda
\left( X,\theta \right) \omega _{0}\left( J,\theta ,Z\right) \left\vert \Psi
\left( J,\theta ,Z\right) \right\vert ^{2}d\left( X,\theta \right) \right) 
\mathcal{D}\Lambda }{\int \exp \left( -S\left( \Lambda \right) \right) 
\mathcal{D}\Lambda }  \label{ft}
\end{eqnarray}%
The action for the fields $\Lambda $ and $\Lambda ^{\dag }$ is: 
\begin{eqnarray*}
S\left( \Lambda \right) &=&\int \Lambda \left( Z,\theta \right) \left(
1-\left\vert \Psi \right\vert ^{2}\hat{T}\right) \Lambda ^{\dag }\left(
Z,\theta \right) d\left( Z,\theta \right) \\
&&-\int \Lambda \left( Z,\theta \right) \hat{T}\left( \theta -\frac{%
\left\vert Z-Z^{\left( 1\right) }\right\vert }{c},Z,Z^{\left( 1\right)
},\omega _{0}+\hat{T}\Lambda ^{\dag }\right) \Lambda ^{\dag }\left(
Z^{\left( 1\right) },\theta -\frac{\left\vert Z-Z^{\left( 1\right)
}\right\vert }{c}\right) dZdZ^{\left( 1\right) }d\theta
\end{eqnarray*}%
with:%
\begin{eqnarray*}
&&\hat{T}\left( \theta -\frac{\left\vert Z^{\left( 1\right) }-Z\right\vert }{%
c},Z^{\left( 1\right) },Z,\omega _{0}+\hat{T}\Lambda ^{\dag }\right) \\
&=&\hat{T}\left( \theta -\frac{\left\vert Z^{\left( 1\right) }-Z\right\vert 
}{c},Z^{\left( 1\right) },Z,\omega _{0}\left( Z,\theta \right) +\int \hat{T}%
\left( \theta -\frac{\left\vert Z-Z^{\left( 1\right) }\right\vert }{c}%
,Z^{\left( 1\right) },Z,\omega _{0}\right) \Lambda ^{\dag }\left( Z^{\left(
1\right) },\theta -\frac{\left\vert Z-Z^{\left( 1\right) }\right\vert }{c}%
\right) dZ^{\left( 1\right) }\right)
\end{eqnarray*}%
We then show that, in the saddle point approximation, the detrended
frequency:%
\begin{equation*}
\Omega \left( \theta ,Z\right) =\omega \left( \theta ,Z\right) -\omega
_{0}\left( J,\theta ,Z\right)
\end{equation*}%
satisfies the following equation: 
\begin{equation}
\Omega -\hat{T}\left( \Omega +\omega _{0}\right) \left\vert \Psi \right\vert
^{2}-\hat{T}\frac{\omega _{0}\Omega }{\omega _{0}+\Omega }=0  \label{mgf}
\end{equation}%
Equation (\ref{mgf}) can be used in two ways.

\paragraph{Series expansion}

A first application of the frequencies equation (\ref{mgf}) considers the
background field as an external field $\left\vert \Psi \right\vert ^{2}$.
This case arises when the system is coupled to an external source $J\left(
Z,\theta \right) $ that shapes the background field. A solution of (\ref{mgf}%
) can be found as a series expansion in the field $\left\vert \Psi
\right\vert ^{2}$ (see appendix 6). The dominant terms of the series is:%
\begin{eqnarray}
\omega \left( Z,\theta \right) &=&\omega _{0}\left( J,\theta ,Z\right) +\int
\sum_{k=0}^{\infty }\frac{\exp \left( -c\sum_{i=0}^{k}l_{i}-\alpha \left(
1+\left\langle \left\vert \Psi \right\vert ^{2}\right\rangle \right) \left(
\sum_{i=0}^{k}\left( cl_{i}\right) ^{2}-\sum_{l=0}^{k-1}\frac{\left\vert
Z_{i}-Z_{i+1}\right\vert }{c}\right) \right) }{B^{k+1}}  \notag \\
&&\times \dprod\limits_{i=1}^{k}\left( \frac{\omega _{0}\left( \theta
-l_{i},Z_{i}\right) }{\omega _{0}\left( \theta -l_{i},Z_{i}\right) +A\omega
_{0}\left\vert \Psi \right\vert ^{2}\left( \theta -l_{i},Z_{i}\right) }%
\right) \frac{\omega _{0}\left( J,\theta -l_{k},Z_{k}\right) }{\left(
1+\left\langle \left\vert \Psi \right\vert ^{2}\right\rangle \right) }%
\left\vert \Psi \left( \theta -l_{k},Z_{k}\right) \right\vert
^{2}dZ_{i}dl_{i}  \label{qrf}
\end{eqnarray}%
where $\left\langle \left\vert \Psi \right\vert ^{2}\right\rangle $ is the
average of $\left\vert \Psi \right\vert ^{2}$ over the thread.

For $\omega ^{-1}\left( Z,\theta \right) $, we obtain:%
\begin{eqnarray}
\omega ^{-1}\left( Z,\theta \right) &=&\omega _{0}^{-1}\left( J,\theta
,Z\right)  \notag \\
&&+\frac{G^{\prime }\left[ J,\omega ,\theta ,Z,\Psi \right] }{F^{\prime }%
\left[ J,\omega ,\theta ,Z,\Psi \right] }\int \sum_{k=0}^{\infty }\frac{\exp
\left( -c\sum_{i=0}^{k}l_{i}-\alpha \left( 1+\left\langle \left\vert \Psi
\right\vert ^{2}\right\rangle \right) \left( \sum_{i=0}^{k}\left(
cl_{i}\right) ^{2}-\sum_{l=0}^{k-1}\frac{\left\vert Z_{i}-Z_{i+1}\right\vert 
}{c}\right) \right) }{D^{k+1}}  \notag \\
&&\times \dprod\limits_{i=1}^{k}\left( \frac{\omega _{0}\left( \theta
-l_{i},Z_{i}\right) }{\omega _{0}\left( \theta -l_{i},Z_{i}\right) +A\omega
_{0}\left\vert \Psi \right\vert ^{2}\left( \theta -l_{i},Z_{i}\right) }%
\right) \frac{\omega _{0}^{-1}\left( J,\theta -l_{k},Z_{k}\right) }{\left(
1+\left\langle \left\vert \Psi \right\vert ^{2}\right\rangle \right) }%
\left\vert \Psi \left( \theta -l_{k},Z_{k}\right) \right\vert
^{2}dZ_{i}dl_{i}  \label{qfr}
\end{eqnarray}%
The full series expansion for $\omega \left( Z,\theta \right) $\ is derived
in appendix 6.

\paragraph{Non local differential equation}

Equation (\ref{mgf}) can also be used to obtain a non-local version of the
frequencies equation (\ref{fqt}). To do so, we replace the background field
by a function of the frequencies:%
\begin{equation*}
\Psi \left( \theta ,Z\right) =\frac{\nabla _{\theta }\omega \left( J\left(
\theta \right) ,\theta ,Z,\mathcal{G}_{0}+\left\vert \Psi _{0}\right\vert
^{2}\right) }{U^{\prime \prime }\left( X_{0}\right) \omega ^{2}\left(
J\left( \theta \right) ,\theta ,Z,\mathcal{G}_{0}+\left\vert \Psi
_{0}\right\vert ^{2}\right) }\Psi _{0}\left( \theta ,Z\right)
\end{equation*}%
that rewrites as:%
\begin{equation*}
\left\vert \Psi \right\vert ^{2}=\frac{X_{0}^{2}}{U^{\prime \prime }\left(
X_{0}\right) }\frac{\nabla _{\theta }\Omega }{\left( \omega _{0}+\Omega
\right) ^{2}}
\end{equation*}%
Equation (\ref{mgf}) thus becomes:%
\begin{equation}
\Omega -\hat{T}\frac{\frac{X_{0}^{2}}{U^{\prime \prime }\left( X_{0}\right) }%
\nabla _{\theta }\Omega +\omega _{0}\Omega }{\omega _{0}+\Omega }=0
\label{mg}
\end{equation}%
This is a non-linear equation that generalizes (\ref{fqt}). The second-order
expansion in derivatives of the right-hand side of (\ref{mg}) yields a
second order linear differential equation similar to the type derived in
sections 8.2 and 8.3\textbf{\ }and confirms the possibility of waves
propagation phenomenom.

Remark that equation (\ref{mg}) is still valid for any background field
related to $\omega $ by a relation of the type:%
\begin{equation}
\left\vert \Psi \right\vert ^{2}=f\left( \omega ,\nabla _{\theta }^{l}\omega
\right)
\end{equation}%
which leads to the following dynamic equation:%
\begin{equation*}
\Omega -\hat{T}\omega f\left( \omega ,\nabla _{\theta }^{l}\omega \right) -%
\hat{T}\frac{\omega _{0}\left( \omega -\omega _{0}\right) }{\omega }=0
\end{equation*}

\paragraph{Corrections to the saddle point}

Ultimately, appendix 6 also yields the corrections to the saddle point
approximation. Replacing equation (\ref{ft}) by: 
\begin{equation}
\omega =\omega _{0}+\hat{T}\Lambda _{0}^{\dag }\left( \det \left( 1+\left( 1+%
\frac{\omega _{0}\left( 2\omega _{0}+\hat{T}\Lambda _{0}^{\dag }\right) }{%
\left( \omega _{0}+\hat{T}\Lambda _{0}^{\dag }\right) ^{2}}\hat{T}\left(
1+\left\vert \Psi \right\vert ^{2}\right) \hat{T}\right) ^{-1}\right)
\right) ^{-1}  \label{ftr}
\end{equation}%
and considering equation (\ref{ftr}) along with the defining equation for $%
\Lambda _{0}^{\dag }$: 
\begin{subequations}
\begin{equation}
\Lambda _{0}^{\dag }-\left( \left\vert \Psi \right\vert ^{2}+\frac{\omega
_{0}}{\omega _{0}+\hat{T}\Lambda _{0}^{\dag }}\right) \hat{T}\Lambda
_{0}^{\dag }-\omega _{0}\left\vert \Psi \right\vert ^{2}=0
\end{equation}%
yields the modified version of (\ref{mg}): 
\end{subequations}
\begin{eqnarray*}
&&\Omega \det \left( 1+\left( 1+\frac{\omega _{0}\left( 2\omega _{0}+\Omega
\right) }{\left( \omega _{0}+\Omega \right) ^{2}}\hat{T}\left( 1+f\left(
\omega ,\nabla _{\theta }^{l}\omega \right) \right) \hat{T}\right)
^{-1}\right) \\
&=&\hat{T}\left( \left( f\left( \omega ,\nabla _{\theta }^{l}\omega \right) +%
\frac{\omega _{0}}{\omega _{0}+\Omega \det \left( 1+\left( 1+\frac{\omega
_{0}\left( 2\omega _{0}+\Omega \right) }{\left( \omega _{0}+\Omega \right)
^{2}}\hat{T}\left( 1+f\left( \omega ,\nabla _{\theta }^{l}\omega \right)
\right) \hat{T}\right) ^{-1}\right) }\right) \right. \\
&&\left. \times \Omega \det \left( 1+\left( 1+\frac{\omega _{0}\left(
2\omega _{0}+\Omega \right) }{\left( \omega _{0}+\Omega \right) ^{2}}\hat{T}%
\left( 1+f\left( \omega ,\nabla _{\theta }^{l}\omega \right) \right) \hat{T}%
\right) ^{-1}\right) +\omega _{0}f\left( \omega ,\nabla _{\theta }^{l}\omega
\right) \right)
\end{eqnarray*}

\subsection{Several interacting fields}

The results of section 9.1 can be extended in the case of two types of
interactions. Consider $n$ populations, each caracterized by their
frequencies $i=1,...,n$, and interacting either positively or negatively.
Each population is defined by a field $\Psi _{i}$ and frequencies $\omega
_{i}\left( \theta ,Z\right) $. Equations for frequencies are defined by (\ref%
{ftm}):

\begin{eqnarray}
\omega _{i}\left( \theta ,Z\right) &=&F_{i}\left( J\left( \theta \right) +%
\frac{\kappa }{N}\int T\left( Z,Z_{1}\right) \frac{\omega _{j}\left( \theta -%
\frac{\left\vert Z-Z_{1}\right\vert }{c},Z_{1}\right) }{\omega _{i}\left(
\theta ,Z\right) }G^{ij}\right.  \label{fML} \\
&&\times \left. W\left( \frac{\omega _{i}\left( \theta ,Z\right) }{\omega
_{j}\left( \theta -\frac{\left\vert Z-Z_{1}\right\vert }{c},Z_{1}\right) }%
\right) \left( \mathcal{\bar{G}}_{0j}\left( 0,Z_{1}\right) +\left\vert \Psi
_{j}\left( \theta -\frac{\left\vert Z-Z_{1}\right\vert }{c},Z_{1}\right)
\right\vert ^{2}\right) dZ_{1}\right)  \notag
\end{eqnarray}%
The coefficients of the $n\times n$ matrix $G$ belong to the interval $\left[
-1,1\right] $. The sum over indices is implicit for $j$.

The resolution of (\ref{fML}) is similar to that of (\ref{fqt}), but with a
vector of frequencies. The series expansion of this vector is:%
\begin{eqnarray}
\omega \left( \theta ,Z\right) &=&\omega \left( \theta ,Z\right)
_{\left\vert \Psi \right\vert ^{2}=0}  \label{psN} \\
&&+\int \sum_{n=1}^{\infty }\left( \frac{\delta ^{n}\omega \left( J,\theta
,Z\right) }{\dprod\limits_{i=1}^{n}\delta \left\vert \Psi \left( \theta
-l_{i},Z_{i}\right) \right\vert ^{2}}\right) _{\left\vert \Psi \right\vert
^{2}=0}\dprod\limits_{i=1}^{n}\left\vert \Psi \left( \theta
-l_{i},Z_{i}\right) \right\vert ^{2}  \notag
\end{eqnarray}%
where the expression of the first order derivative is similar to (\ref{rdn}):%
\begin{eqnarray*}
\left( \frac{\delta \omega \left( J,\theta ,Z\right) }{\delta \left\vert
\Psi \left( \theta -l_{1},Z_{1}\right) \right\vert ^{2}}\right) _{\left\vert
\Psi \right\vert ^{2}=0} &=&\sum_{n=1}^{\infty }\int \dprod\limits_{l=1}^{n}%
\hat{T}\left( \theta -\sum_{j=1}^{l-1}\frac{\left\vert Z^{\left( j-1\right)
}-Z^{\left( j\right) }\right\vert }{c},Z^{\left( l-1\right) },Z^{\left(
l\right) },\omega _{0},0\right) \\
&&\times \Omega _{0}\left( J,\theta -\sum_{l=1}^{n}\frac{\left\vert
Z^{\left( l-1\right) }-Z^{\left( l\right) }\right\vert }{c},Z_{1}\right)
\times \delta \left( l_{1}-\sum_{l=1}^{n}\frac{\left\vert Z^{\left(
l-1\right) }-Z^{\left( l\right) }\right\vert }{c}\right)
\dprod\limits_{l=1}^{n-1}dZ^{\left( l\right) }
\end{eqnarray*}%
where $\omega \left( \theta ^{\left( i\right) },Z\right) $ and $\omega
\left( \theta ^{\left( i\right) },Z\right) _{\left\vert \Psi \right\vert
^{2}=0}=\omega _{0}$ are vectors of frequencies. We define $D\left(
\left\vert \Psi \right\vert ^{2}\right) $ as a diagonal matrix with
components $\left\vert \Psi _{i}\right\vert ^{2}$ on the diagonal. More
generally, for any expression $H\left( \omega _{0i},\left\vert \Psi
_{i}\right\vert ^{2}\right) $, we define $D\left( H\left( \omega
_{0},\left\vert \Psi \right\vert ^{2}\right) \right) $ as the diagonal
matrix with components $H\left( \omega _{0i},\left\vert \Psi _{i}\right\vert
^{2}\right) $. The expressions $\left( \frac{\delta \omega \left( J,\theta
,Z\right) }{\delta \left\vert \Psi \left( \theta -l_{1},Z_{1}\right)
\right\vert ^{2}}\right) _{\left\vert \Psi \right\vert ^{2}=0}$ and $\hat{T}%
\left( \theta -\sum_{j=1}^{l-1}\frac{\left\vert Z^{\left( j-1\right)
}-Z^{\left( j\right) }\right\vert }{c},Z^{\left( l-1\right) },Z^{\left(
l\right) },\omega _{0},0\right) $ are $n\times n$ matrices:%
\begin{equation*}
\left( \left( \frac{\delta \omega \left( J,\theta ,Z\right) }{\delta
\left\vert \Psi \left( \theta -l_{1},Z_{1}\right) \right\vert ^{2}}\right)
_{\left\vert \Psi \right\vert ^{2}=0}\right) _{ij}=\left( \frac{\delta
\omega _{i}\left( J,\theta ,Z\right) }{\delta \left\vert \Psi _{j}\left(
\theta -l_{1},Z_{1}\right) \right\vert ^{2}}\right) _{\left\vert \Psi
\right\vert ^{2}=0}
\end{equation*}%
\begin{eqnarray*}
&&\hat{T}_{ij}\left( \theta ,Z,Z_{1}\omega ,\Psi \right) \\
&=&\frac{G^{ij}\frac{\kappa }{N}\omega _{i}\left( J,\theta ,Z\right) T\left(
Z,Z_{1}\right) F^{\prime }\left[ J,\omega ,\theta ,Z,\Psi \right] }{\omega
_{i}^{2}\left( J,\theta ,Z\right) +G^{ij}\left( \int \frac{\kappa }{N}\omega
_{j}\left( J,\theta -\frac{\left\vert Z-Z^{\prime }\right\vert }{c}%
,Z^{\prime }\right) \left( \mathcal{\bar{G}}_{0j}\left( 0,Z^{\prime }\right)
+\left\vert \Psi _{j}\left( \theta -\frac{\left\vert Z-Z^{\prime
}\right\vert }{c},Z^{\prime }\right) \right\vert ^{2}\right) T\left(
Z,Z^{\prime }\right) dZ^{\prime }\right) F^{\prime }\left[ J,\omega ,\theta
,Z,\Psi \right] }
\end{eqnarray*}%
and the operator $\hat{T}$\ with kernel:%
\begin{eqnarray*}
\hat{T}\left( \left( Z^{\left( l-1\right) },\theta ^{\left( l-1\right)
}\right) ,\left( Z^{\left( l\right) },\theta ^{\left( l\right) }\right)
\right) &=&\hat{T}\left( \theta -\sum_{j=1}^{l-1}\frac{\left\vert Z^{\left(
j-1\right) }-Z^{\left( j\right) }\right\vert }{c},Z^{\left( l-1\right)
},Z^{\left( l\right) },\omega _{0}\right) \\
&&\times \delta \left( \left( \theta ^{\left( l\right) }-\theta ^{\left(
l-1\right) }\right) -\frac{\left\vert Z^{\left( l-1\right) }-Z^{\left(
l\right) }\right\vert }{c}\right)
\end{eqnarray*}%
The successives derivatives in (\ref{psN}) are given by a formula similar to
(\ref{rgr}) 
\begin{equation}
\left( \frac{\delta ^{n}\omega \left( J,\theta ,Z\right) }{%
\dprod\limits_{i=1}^{n}\delta \left\vert \Psi \left( \theta
-l_{i},Z_{i}\right) \right\vert ^{2}}\right) _{\left\vert \Psi \right\vert
^{2}=0}=\left( \sum_{m=1}^{n}\sum_{i=1}^{m}\sum_{\left(
line_{1},...,line_{m}\right) }\dprod\limits_{i}F\left( line_{i}\right)
\dprod\limits_{B}F\left( B\right) \right)  \label{rgR}
\end{equation}%
where the various quantities $\left( \frac{\delta ^{n}\omega \left( J,\theta
,Z\right) }{\dprod\limits_{i=1}^{n}\delta \left\vert \Psi \left( \theta
-l_{i},Z_{i}\right) \right\vert ^{2}}\right) _{\left\vert \Psi \right\vert
^{2}=0}$, $F\left( line_{i}\right) $ and $F\left( B\right) $ are tensors
whose precise form and dimensions are given in appendix 6.3.

The resolution for frequencies follows the single field case, and yields
(see appendix 6.3):%
\begin{equation*}
\Omega -\hat{T}\left( \Omega +\omega _{0}\right) \left\vert \Psi \right\vert
^{2}-\hat{T}\left( \frac{\omega _{0}}{\omega _{0}+\Omega }\right) \Omega =0
\end{equation*}%
where $\left( \Omega +\omega _{0}\right) \left\vert \Psi \right\vert ^{2}$
and $\left( \frac{\omega _{0}}{\omega _{0}+\Omega }\right) \Omega $ are the
vectors with components $\left( \Omega +\omega _{0}\right) _{i}\left\vert
\Psi \right\vert _{i}^{2}$ and $\left( \frac{\omega _{i0}}{\omega
_{i0}+\Omega _{i}}\right) \Omega _{i}$, respectively. The approximate series
expansion for $\omega \left( Z,\theta \right) $ and $\omega ^{-1}\left(
Z,\theta \right) $ are given in appendix 6.3: 
\begin{eqnarray}
\omega \left( Z,\theta \right) &=&\omega _{0}\left( J,\theta ,Z\right) +\int
\sum_{k=0}^{\infty }\dprod\limits_{i=0}^{k-1}\frac{\exp \left(
-cl_{i}-\left( 1+D\left( \left\langle \left\vert \Psi \right\vert
^{2}\right\rangle \right) \right) \Lambda \left( \left( cl_{i}\right) ^{2}-%
\frac{\left\vert Z_{i}-Z_{i+1}\right\vert }{c}\right) \right) }{B} \\
&&\times D\left( \frac{\omega _{0}\left( \theta -l_{i},Z_{i}\right) }{\omega
_{0}\left( \theta -l_{i},Z_{i}\right) +A\omega _{0}\left\vert \Psi
\right\vert ^{2}\left( \theta -l_{i},Z_{i}\right) }\frac{\omega _{0}\left(
J,\theta -l_{k},Z_{k}\right) }{\left( 1+D\left( \left\langle \left\vert \Psi
\right\vert ^{2}\right\rangle \right) \right) }\right)  \notag \\
&&\times \frac{\exp \left( -cl_{k}-\left( 1+D\left( \left\langle \left\vert
\Psi \right\vert ^{2}\right\rangle \right) \right) \Lambda \left( \left(
cl_{i}\right) ^{2}-\frac{\left\vert Z_{k-1}-Z_{k}\right\vert }{c}\right)
\right) }{B}\left\vert \Psi \left( \theta -l_{k},Z_{k}\right) \right\vert
^{2}dZ_{i}dl_{i}
\end{eqnarray}%
\begin{eqnarray}
\omega ^{-1}\left( Z,\theta \right) &=&\omega _{0}^{-1}\left( J,\theta
,Z\right) \\
&&+D\left( \frac{G^{\prime }\left[ J,\omega ,\theta ,Z,\Psi \right] }{%
F^{\prime }\left[ J,\omega ,\theta ,Z,\Psi \right] }\right) \int
\sum_{k=0}^{\infty }\dprod\limits_{i=0}^{k-1}\frac{\exp \left(
-cl_{i}-\left( 1+D\left( \left\langle \left\vert \Psi \right\vert
^{2}\right\rangle \right) \right) \Lambda \left( \left( cl_{i}\right) ^{2}-%
\frac{\left\vert Z_{i}-Z_{i+1}\right\vert }{c}\right) \right) }{B}  \notag \\
&&\times D\left( \frac{\omega _{0}\left( \theta -l_{i},Z_{i}\right) }{\omega
_{0}\left( \theta -l_{i},Z_{i}\right) +A\omega _{0}\left\vert \Psi
\right\vert ^{2}\left( \theta -l_{i},Z_{i}\right) }\frac{\omega _{0}\left(
J,\theta -l_{k},Z_{k}\right) }{\left( 1+D\left( \left\langle \left\vert \Psi
\right\vert ^{2}\right\rangle \right) \right) }\right)  \notag \\
&&\times \frac{\exp \left( -cl_{k}-\left( 1+D\left( \left\langle \left\vert
\Psi \right\vert ^{2}\right\rangle \right) \right) \Lambda \left( \left(
cl_{i}\right) ^{2}-\frac{\left\vert Z_{k-1}-Z_{k}\right\vert }{c}\right)
\right) }{B}\left\vert \Psi \left( \theta -l_{k},Z_{k}\right) \right\vert
^{2}dZ_{i}dl_{i}  \notag
\end{eqnarray}%
The full series expansion for $\omega \left( Z,\theta \right) $ is given in
the same appendix.

\section{Correlation functions and probabilistic interpretation}

The correlation functions of the field theory can be interpreted in terms of
the system's dynamics. They compute the joint probability for a set of
frequencies at different points during a certain interval of time. We first
compute and interpret the two points correlation functions.\ We then
generalize to an arbitrary number of points. It is in this context that the
interdependence of frequencies at different points appear.

\subsection{Two points correlation functions}

The correlation functions are found by computing the derivatives of the
effective action with respect to the classical background field. The two
points Green function is the inverse of the second derivative of the
effective action $\Gamma \left( \Psi ^{\dagger },\Psi \right) $:%
\begin{equation*}
\Gamma _{1,1}\left( \left( \theta _{f},Z_{f}\right) ,\left( \theta
_{i},Z_{i}\right) \right) =\frac{\delta ^{2}\Gamma \left( \Psi ^{\dagger
},\Psi \right) }{\delta \Psi ^{\dagger }\left( \theta _{f},Z_{f}\right)
\delta \Psi \left( \theta _{i},Z_{i}\right) }
\end{equation*}%
In first approximation, we have:%
\begin{equation}
\Gamma _{1,1}\left( \left( \theta _{f},Z_{f}\right) ,\left( \theta
_{i},Z_{i}\right) \right) =-\nabla _{\theta }\left( \frac{\sigma _{\theta
}^{2}}{2}\nabla _{\theta }-\omega ^{-1}\left( J\left( \theta \right) ,\theta
,Z,\mathcal{G}_{0}+\left\vert \Psi \right\vert ^{2}\right) \right) \delta
\left( \theta _{f}-\theta _{i}\right) +\hat{\Gamma}_{1,1}\left( \left(
\theta _{f},Z_{f}\right) ,\left( \theta _{i},Z_{i}\right) ,\Psi ^{\dagger
},\Psi \right)
\end{equation}%
with $\hat{\Gamma}_{1,1}\left( \left( \theta _{f},Z_{f}\right) ,\left(
\theta _{i},Z_{i}\right) ,\Psi ^{\dagger },\Psi \right) $ given by the
second derivative of (\ref{ffc}):%
\begin{eqnarray}
&&\hat{\Gamma}_{1,1}\left( \left( \theta _{f},Z_{f}\right) ,\left( \theta
_{i},Z_{i}\right) ,\Psi ^{\dagger },\Psi \right)  \label{gmm} \\
&=&\sum_{\substack{ j\geqslant 2  \\ m\geqslant 2}}\sum_{\substack{ \left(
p_{l}^{i}\right) _{m\times j}  \\ \sum_{i}p_{l}^{i}\geqslant 2}}%
\sum_{l_{f}^{\prime }=1,l_{i}^{\prime }=1}^{j}\int \frac{\left(
\dprod\limits_{l=1,l\neq l_{f}^{\prime }}^{j}\Psi ^{\dagger }\left( \theta
_{f}^{\left( l\right) },Z_{l}\right) \right) }{m!\dprod\limits_{k}\left(
\sharp _{k}\right) !}\times \frac{\dprod\limits_{l=1}^{j}\exp \left(
-\Lambda _{1}\left( \theta _{f}^{\left( l\right) }-\theta _{i}^{\left(
l\right) }\right) \right) }{\Lambda ^{\sum_{i,l}p_{l}^{i}}}  \notag \\
&&\times \dprod\limits_{i=1}^{m}\left[ \int_{\dprod\limits_{l=1}^{j}\left[
\theta _{i}^{\left( l\right) },\theta _{f}^{\left( l\right) }\right]
^{p_{l}^{i}}}\frac{\delta ^{\sum_{l}p_{l}^{i}}\left[ \hat{S}_{cl}\left( \Psi
^{\dagger },\Psi \right) \right] }{\dprod\limits_{l=1}^{j}\dprod%
\limits_{k_{l}^{i}=1}^{p_{l}^{i}}\delta \left\vert \Psi \left( \theta
^{\left( k_{l}^{i}\right) },Z_{_{l}}\right) \right\vert ^{2}}%
\dprod\limits_{l=1}^{j}\dprod\limits_{k_{l}^{i}=1}^{p_{l}^{i}}d\theta
^{\left( k_{l}^{i}\right) }\right] _{\substack{ \left( \theta ^{\left(
l_{f}^{\prime }\right) },Z_{l_{f}^{\prime }}\right) =\left( \theta
_{f},Z_{f}\right)  \\ \left( \theta ^{\left( l_{i}^{\prime }\right)
},Z_{l_{i}^{\prime }}\right) =\left( \theta _{i},Z_{i}\right) }}\times
\left( \dprod\limits_{l=1,l\neq l_{i}^{\prime }}^{j}\Psi \left( \theta
_{i}^{\left( l\right) },Z_{l}\right) \right)  \notag
\end{eqnarray}%
In these expressions, the derivatives that correspond to the impact of
propagation between $\theta _{i}$ and $\theta _{f}$ of the signal have been
neglected. In the local approximation, equation (\ref{gmm}) writes:%
\begin{eqnarray}
&&\hat{\Gamma}_{1,1}\left( \left( \theta ,Z_{f}\right) ,\left( \theta
,Z_{i}\right) ,\Psi ^{\dagger },\Psi \right)  \label{mmg} \\
&=&\sum_{\substack{ j\geqslant 2  \\ m\geqslant 2}}\sum_{\substack{ \left(
p_{l}^{i}\right) _{m\times j}  \\ \sum_{i}p_{l}^{i}\geqslant 2}}%
\sum_{l_{f}^{\prime }=1,l_{i}^{\prime }=1}^{j}\int \left[ \left(
\dprod\limits_{l=1,l\neq l_{f}^{\prime }}^{j}\Psi ^{\dagger }\left( \theta
^{\left( l\right) },Z_{l}\right) \right) \right.  \notag \\
&&\times \left. \frac{\dprod\limits_{i=1}^{m}\left[ \frac{\delta
^{\sum_{l}p_{l}^{i}}\left[ \hat{S}_{cl}\left( \Psi ^{\dagger },\Psi \right) %
\right] }{\dprod\limits_{l=1}^{j}\delta ^{p_{l}^{i}}\left\vert \Psi \left(
\theta ^{\left( l\right) },Z_{_{l}}\right) \right\vert ^{2}}\right] }{%
m!\dprod\limits_{k}\left( \sharp _{k}\right) !\Lambda _{1}^{j}\Lambda
^{\sum_{i,l}p_{l}^{i}}}\left( \dprod\limits_{l=1,l\neq l_{i}^{\prime
}}^{j}\Psi \left( \theta ^{\left( l\right) },Z_{l}\right) d\theta ^{\left(
l\right) }dZ_{l}\right) \right] _{_{\substack{ \left( \theta ^{\left(
l^{\prime }\right) },Z_{l_{f}^{\prime }}\right) =\left( \theta ,Z_{f}\right) 
\\ \left( \theta ^{\left( l^{\prime }\right) },Z_{l_{i}^{\prime }}\right)
=\left( \theta ,Z_{i}\right) }}}
\end{eqnarray}%
with $p_{l}^{i}=0$ or $1$. The two points correlation function is then:%
\begin{equation}
G_{2}\left( \left( \theta _{f},Z_{f}\right) ,\left( \theta _{i},Z_{i}\right)
\right) =\mathcal{G}\left( \left( \theta _{f},Z_{f}\right) ,\left( \theta
_{i},Z_{i}\right) \right) +\mathcal{G\ast }\sum_{n\geqslant 2}\left(
-1\right) ^{n-1}\left( \hat{\Gamma}_{1,1}\left( \Psi ^{\dagger },\Psi
\right) \ast \mathcal{G}\right) ^{n}  \label{nrg}
\end{equation}%
where $\mathcal{G}\left( \left( \theta _{f},Z_{f}\right) ,\left( \theta
_{i},Z_{i}\right) \right) $ satisfies:

\begin{equation*}
-\nabla _{\theta }\left( \frac{\sigma _{\theta }^{2}}{2}\nabla _{\theta
}-\omega ^{-1}\left( J\left( \theta \right) ,\theta ,Z,\mathcal{G}%
_{0}+\left\vert \Psi \right\vert ^{2}\right) \right) \mathcal{G}\left(
\left( \theta _{f},Z_{f}\right) ,\left( \theta _{i},Z_{i}\right) \right) =%
\mathcal{\delta }\left( \theta _{f}-\theta _{i}\right)
\end{equation*}%
and $\hat{\Gamma}_{1,1}\left( \Psi ^{\dagger },\Psi \right) $ is the
operator with kernel $\hat{\Gamma}_{1,1}\left( \left( \theta
_{f},Z_{f}\right) ,\left( \theta _{i},Z_{i}\right) ,\Psi ^{\dagger },\Psi
\right) $. Appendix 4.1 yields an expression for $\hat{\Gamma}_{1,1}\left(
\left( \theta _{f},Z_{f}\right) ,\left( \theta _{i},Z_{i}\right) ,\Psi
^{\dagger },\Psi \right) $ and $\mathcal{G}\left( \left( \theta
_{f},Z_{f}\right) ,\left( \theta _{i},Z_{i}\right) \right) $ in the
approximation of relatively slow variations of frequencies. \ We find: 
\begin{equation*}
\hat{\Gamma}_{1,1}\left( \left( \theta ,Z_{f}\right) ,\left( \theta
,Z_{i}\right) ,\Psi ^{\dagger },\Psi \right) =\omega ^{-1}\left( \theta
_{i},Z_{i}\right) \omega ^{-1}\left( \theta _{f},Z_{f}\right) \Psi \left(
\theta _{f},Z_{f}\right) \Psi ^{\dag }\left( \theta _{i},Z_{i}\right)
C\left( \bar{\omega},\Psi \right)
\end{equation*}%
where:%
\begin{equation*}
C\left( \bar{\omega},\Psi \right) =\frac{1}{2D\Lambda }\frac{1}{2D\Lambda }%
\exp \left( \int \frac{\omega ^{-1}\left( \bar{\theta},\bar{Z}\right) }{%
\Lambda _{1}}d\bar{\theta}d\bar{Z}\right) \times \exp \left( -\int \frac{c}{%
2D\Lambda }\left\vert \Psi \left( \bar{\theta},\bar{Z}\right) \right\vert
^{2}d\bar{\theta}d\bar{Z}\right)
\end{equation*}%
Moreover, we define the average frequency at $Z$ over a time span $\left[
\theta ,\theta ^{\prime }\right] $ as:%
\begin{equation*}
\bar{\omega}^{-1}\left( J\left( \theta \right) ,\theta ,Z,\mathcal{G}%
_{0}+\left\vert \Psi \right\vert ^{2}\right) \equiv \left\langle \omega
^{-1}\left( J\left( \theta \right) ,\theta ,Z,\mathcal{G}_{0}+\left\vert
\Psi \right\vert ^{2}\right) \right\rangle _{\left[ \theta ,\theta ^{\prime }%
\right] }
\end{equation*}%
so that we obtain:%
\begin{eqnarray}
&&\mathcal{G}\left( \theta _{f},\theta _{i},Z_{f},Z_{i}\right)  \label{grv}
\\
&\simeq &\delta \left( Z_{f}-Z_{i}\right) \frac{1}{\sqrt{\frac{\pi }{2}}}%
\frac{\exp \left( -\left( \sqrt{\left( \frac{\bar{\omega}^{-1}\left( J\left(
\theta \right) ,\theta ,Z,\mathcal{G}_{0}+\left\vert \Psi \right\vert
^{2}\right) }{\sigma ^{2}}\right) ^{2}+\frac{2\alpha }{\sigma ^{2}}}-\frac{%
\bar{\omega}^{-1}\left( J\left( \theta \right) ,\theta ,Z,\mathcal{G}%
_{0}+\left\vert \Psi \right\vert ^{2}\right) }{\sigma ^{2}}\right) \left(
\theta -\theta ^{\prime }\right) \right) }{\sqrt{\left( \frac{\bar{\omega}%
^{-1}\left( J\left( \theta \right) ,\theta ,Z,\mathcal{G}_{0}+\left\vert
\Psi \right\vert ^{2}\right) }{\sigma ^{2}}\right) ^{2}+\frac{2\alpha }{%
\sigma ^{2}}}}H\left( \theta -\theta ^{\prime }\right)  \notag
\end{eqnarray}%
As a consequence, the solution of (\ref{nrg}) is the series expansion:

\begin{eqnarray}
G_{2}\left( \theta _{f},\theta _{i},Z_{f},Z_{i}\right) &\simeq &\mathcal{G}%
\left( \theta _{f},\theta _{i},Z_{f},Z_{i}\right)  \label{grx} \\
&&+\int \left( \prod\limits_{k=1}^{n}d\theta _{k}dZ_{k}\right) \mathcal{G}%
\left( \theta _{f},\theta _{n},Z_{f},Z_{f}\right) \frac{\Psi \left( \theta
_{n},Z_{f}\right) }{\omega \left( \theta _{n},Z_{f}\right) }\times
\sum_{n\geqslant 2}\left( -1\right) ^{n-1}  \notag \\
&&\times \prod\limits_{k=1}^{n-1}\left( \left\vert \Psi \left( \theta
_{k},Z_{k}\right) \right\vert ^{2}\left( \omega ^{-1}\left( \theta
_{k},Z_{k}\right) \right) ^{2}\mathcal{G}\left( \theta _{k+1},\theta
_{k},Z_{k},Z_{k}\right) \right) ^{n-1}\frac{\Psi ^{\dag }\left( \theta
_{1},Z_{i}\right) }{\omega \left( \theta _{1},Z_{i}\right) }\mathcal{G}%
\left( \theta _{1},\theta _{i},Z_{i},Z_{i}\right)  \notag
\end{eqnarray}%
In the following, we study the correlation functions with an arbitrary
number of points.

\subsection{$\left( k,n\right) $ points correlation functions}

The $\left( k,n\right) $ points correlation functions are derived in the
standard way (see appendix 4.2).

The correlation functions are obtained from the $\left( k,n\right) $-th
effective vertex $\Gamma _{k,n}$:%
\begin{equation*}
\Gamma _{k,n}\left( \left( \theta _{f}^{\left( l\right) },Z_{l}\right)
_{l=1,..,k},\left( \theta _{i}^{\left( l\right) },Z_{l}\right)
_{l=1,..,n}\right) =\frac{\delta ^{k+n}\Gamma \left( \Psi ^{\dagger },\Psi
\right) }{\delta ^{k}\left( \Psi ^{\dagger }\left( \theta _{f}^{\left(
l\right) },Z_{l}\right) \right) _{l=1,..,k}\delta ^{n}\left( \Psi \left(
\theta _{i}^{\left( l\right) },Z_{l}\right) \right) _{l=1,..,n}}
\end{equation*}%
through standard techniques. Appendix 4 shows that, in first approximation:

\begin{eqnarray}
&&G_{k,n}\left( \left( \theta _{f}^{\left( l\right) },Z_{l}\right)
_{l=1,..,k},\left( \theta _{i}^{\left( l\right) },Z_{l}\right)
_{l=1,..,n}\right)  \label{gRX} \\
&=&\sum_{\sigma _{k},\sigma _{n}}\sum_{u=0}^{\inf \left( k,n\right)
}\prod\limits_{j=0}^{u}G_{2}\left( \left( \theta _{f}^{\left( j\right)
},Z_{f}\right) ,\left( \theta _{i}^{\left( i\right) },Z_{i}\right) \right) 
\notag \\
&&\times \sum_{i=1,j=1}^{k-u,n-u}\left( -1\right) ^{i+j}\sum_{\substack{ %
P_{i}\left( k-u\right)  \\ P_{j}\left( n-u\right) }}\dprod\limits_{\substack{
r\in P_{i}\left( k-u\right)  \\ s\in P_{j}\left( n-u\right) }}\Gamma
_{k_{r},n_{s}}\left( \left( \theta _{f}^{\left( l,r,u\right)
},Z_{l,r,u}\right) _{l=1,..,k_{r}},\left( \theta _{i}^{\left( l,s,u\right)
},Z_{l,s,u}\right) _{l=1,..,n_{s}}\right)  \notag
\end{eqnarray}%
where $P_{i}\left( k\right) $ and $P_{j}\left( n\right) $ denote the
partitions of $k$ and $n$ in $i$ and $j$ subsets: 
\begin{equation*}
\cup _{r}\left( \theta _{f}^{\left( l,r,u\right) },Z_{l,r,u}\right)
_{l=1,..,k_{r}}=\left( \theta _{f}^{\left( l\right) },Z_{l}\right)
_{l=u+1,..,k}
\end{equation*}%
and: 
\begin{equation*}
\cup _{s}\left( \theta _{i}^{\left( l,s,u\right) },Z_{l,s,u}\right)
_{l=u+1,..,n_{s}}=\left( \theta _{i}^{\left( l\right) },Z_{l}\right)
_{l=u+1,..,n}
\end{equation*}%
as ordered sets. The sum over $\sigma _{k}$ and $\sigma _{n}$ is over all
permutations of the $\left( \theta _{f}^{\left( l\right) },Z_{l}\right)
_{l=1,..,k}$ and $\left( \theta _{i}^{\left( l\right) },Z_{l}\right)
_{l=1,..,n}$, respectively.

\subsection{Interpretation: joined probabilities for frequencies}

Equations (\ref{grx}) and (\ref{gRX}) can be interpreted in terms of joined
probabilities for frequencies at different points of the thread. We first
consider the two points correlation functions.

\subsubsection{Two points functions}

At the perturbative zeroth order, the function $\mathcal{G}_{0}\left( \theta
,\theta ^{\prime },Z\right) $ is the Green function of the operator:%
\begin{equation*}
-\nabla _{\theta }\left( \frac{\sigma _{\theta }^{2}}{2}\nabla _{\theta
}-\omega ^{-1}\left( J\left( \theta \right) ,\theta ,Z,\mathcal{G}%
_{0}\right) \right) +\alpha
\end{equation*}%
and is given by (\ref{rg}):%
\begin{equation*}
\mathcal{G}_{0}\left( \theta ,\theta ^{\prime },Z,Z^{\prime }\right) =\delta
\left( Z-Z^{\prime }\right) \frac{\exp \left( -\left( \sqrt{\left( \frac{1}{%
\sigma ^{2}\bar{X}_{r}}\right) ^{2}+\frac{2\alpha }{\sigma ^{2}}}-\frac{1}{%
\sigma ^{2}\bar{X}_{r}}\right) \left( \theta -\theta ^{\prime }\right)
\right) }{\Lambda }H\left( \theta -\theta ^{\prime }\right)
\end{equation*}%
This function is the Laplace transform of the function $\hat{G}_{0Z}\left(
\theta ,\theta ^{\prime },\Delta n\right) $:%
\begin{equation*}
\mathcal{G}_{0}\left( \theta ,\theta ^{\prime },Z,Z^{\prime }\right) =\int 
\hat{G}_{0Z}\left( \theta ,\theta ^{\prime },\Delta n\right) \exp \left(
-\alpha \Delta n\right) d\alpha
\end{equation*}%
The form of $\hat{G}_{0Z}\left( \theta ,\theta ^{\prime },\Delta n\right) $
is irrelevant here.

The function $\hat{G}_{0Z}$ computes the probability of a time interval $%
\theta -\theta ^{\prime }$ for $\Delta n$ spikes of the potential at point $%
Z $. The Laplace transform $\mathcal{G}_{0}\left( \theta ,\theta ^{\prime
},Z,Z^{\prime }\right) $ computes the probability of a time interval $\theta
-\theta ^{\prime }$ for a random number of spikes $\Delta n$ with average $%
\frac{1}{\alpha }$. Since the spikes' frequency is $\frac{\Delta n}{\theta
-\theta ^{\prime }}$, $\mathcal{G}_{0}\left( \theta ,\theta ^{\prime
},Z,Z^{\prime }\right) $ computes the average probability of a frequency $%
\frac{1}{\alpha \left( \theta -\theta ^{\prime }\right) }$ of spikes.
Computing the average $\left\langle \left( \theta -\theta ^{\prime }\right)
\right\rangle $ confirms this point:%
\begin{eqnarray*}
\mathcal{G}_{0}\left( \theta ,\theta ^{\prime },Z,Z^{\prime }\right)
&=&\delta \left( Z-Z^{\prime }\right) \frac{\exp \left( -\left( \sqrt{\left( 
\frac{1}{\sigma ^{2}\bar{X}_{r}}\right) ^{2}+\frac{2\alpha }{\sigma ^{2}}}-%
\frac{1}{\sigma ^{2}\bar{X}_{r}}\right) \left( \theta -\theta ^{\prime
}\right) \right) }{\Lambda }H\left( \theta -\theta ^{\prime }\right) \\
&\simeq &\delta \left( Z-Z^{\prime }\right) \frac{\exp \left( -\alpha \bar{X}%
_{r}\left( \theta -\theta ^{\prime }\right) \right) }{\Lambda }H\left(
\theta -\theta ^{\prime }\right)
\end{eqnarray*}%
so that $\left\langle \left( \theta -\theta ^{\prime }\right) \right\rangle =%
\frac{1}{\alpha \bar{X}_{r}}$. The average inverse frequency is then $\alpha
\left\langle \left( \theta -\theta ^{\prime }\right) \right\rangle =\frac{1}{%
\bar{X}_{r}}$.

As a consequence, the Green function $\mathcal{G}_{0}\left( \theta ,\theta
^{\prime },Z,Z^{\prime }\right) $ computed at $\alpha =1$ can be interpreted
as the probability, at time $\frac{\theta +\theta ^{\prime }}{2}$ and
position $Z$, of a spikes' frequency equal to $\frac{1}{\theta -\theta
^{\prime }}$. The same applies for higher order correlation functions.

Including the higher order corrections (\ref{grv}) and (\ref{grx}) lead to
the same conclusion, but using (\ref{grx}) for $Z_{f}=Z_{i}$ shows the
interdependencies of frequencies. Actually, this can be rewritten: 
\begin{eqnarray}
G_{2}\left( \theta _{f},\theta _{i},Z_{i},Z_{i}\right) &\simeq &\mathcal{G}%
\left( \theta _{f},\theta _{i},Z_{i},Z_{i}\right) \\
&&+\int \left( \prod\limits_{k=1}^{n}d\theta _{k}dZ_{k}\right) \mathcal{G}%
\left( \theta _{f},\theta _{n},Z_{i},Z_{i}\right) \frac{\Psi \left( \theta
_{n},Z_{f}\right) }{\omega \left( \theta _{n},Z_{f}\right) }\times
\sum_{n\geqslant 2}\left( -1\right) ^{n-1}  \notag \\
&&\times \prod\limits_{k=1}^{n-1}\left( \frac{\Psi ^{\dag }\left( \theta
_{k},Z_{k}\right) }{\omega ^{-1}\left( \theta _{k},Z_{k}\right) }\mathcal{G}%
\left( \theta _{k+1},\theta _{k},Z_{k},Z_{k}\right) \frac{\Psi \left( \theta
_{k},Z_{k}\right) }{\omega \left( \theta _{k},Z_{k}\right) }\right) \frac{%
\Psi ^{\dag }\left( \theta _{1},Z_{i}\right) }{\omega \left( \theta
_{1},Z_{i}\right) }\mathcal{G}\left( \theta _{1},\theta
_{i},Z_{i},Z_{i}\right)  \notag
\end{eqnarray}

The probability $\mathcal{G}\left( \theta _{f},\theta
_{i},Z_{i},Z_{i}\right) $ of frequency $\theta _{f}-\theta _{i}$ on an
interval centered on $\frac{\theta _{f}+\theta _{i}}{2}$ is modified
recursively by probabilities $\mathcal{G}\left( \theta _{k+1},\theta
_{k},Z_{k},Z_{k}\right) $ at other points and times with a factor $\frac{%
\Psi \left( \theta _{k},Z_{k}\right) }{\omega \left( \theta
_{k},Z_{k}\right) }\frac{\Psi ^{\dag }\left( \theta _{1},Z_{i}\right) }{%
\omega \left( \theta _{1},Z_{i}\right) }$. This factor measures the rates of
interaction between different points. The probability $\mathcal{G}\left(
\theta _{1},\theta _{i},Z_{i},Z_{i}\right) $ impacts $\mathcal{G}\left(
\theta _{2},\theta _{1},Z_{1},Z_{1}\right) $, that itself impacts $\mathcal{G%
}\left( \theta _{3},\theta _{2},Z_{2},Z_{2}\right) $ and so on, until $%
\mathcal{G}\left( \theta _{f},\theta _{n},Z_{i},Z_{i}\right) $ closes the
series of successive modifications. The sum over times and space yields the
impact of the whole system on the frequencies at $Z_{i}$.

\subsubsection{$\left( n,n\right) $ Green functions}

The $\left( n,n\right) $ Green function $G_{n,n}\left( \left( \theta
_{f}^{\left( l\right) },Z_{l}\right) _{l=1,..,n},\left( \theta _{i}^{\left(
l\right) },Z_{l}\right) _{l=1,..,n}\right) $ computes the transition
probability of $\left( \theta _{i}^{\left( n\right) },Z_{l}\right)
_{l=1,..,n}$ to $\left( \theta _{f}^{\left( n\right) },Z_{l}\right)
_{l=1,..,n}$ for $i=1...l$ for an average number of spikes of $\frac{1}{%
\alpha }$, so that:%
\begin{eqnarray}
&&\left( G_{n,n}\left( \left( \theta _{f}^{\left( n\right) },Z_{l}\right)
_{l=1,..,n},\left( \theta _{i}^{\left( n\right) },Z_{l}\right)
_{l=1,..,n}\right) \right) _{\alpha =1}  \label{jpn} \\
&=&P\left( \omega \left( Z_{1},\frac{\theta _{f}^{\left( 1\right) }+\theta
_{i}^{\left( 1\right) }}{2}\right) =\frac{1}{\theta _{f}^{\left( 1\right)
}-\theta _{i}^{\left( 1\right) }},...,\omega \left( Z_{l},\frac{\theta
_{f}^{\left( n\right) }+\theta _{i}^{\left( n\right) }}{2}\right) =\frac{1}{%
\theta _{f}^{\left( n\right) }-\theta _{i}^{\left( n\right) }}\right)  \notag
\end{eqnarray}%
computes the joined probability for a set of $n$ frequencies at points $%
Z_{1} $,..., $Z_{n}$ and times $\frac{\theta _{f}^{\left( 1\right) }+\theta
_{i}^{\left( 1\right) }}{2}$,..., $\frac{\theta _{f}^{\left( n\right)
}+\theta _{i}^{\left( n\right) }}{2}$. Equation (\ref{jpn}) can be rewritten
in terms of density for the set of variables $\theta ^{\left( l\right) }=%
\frac{\theta _{f}^{\left( l\right) }+\theta _{i}^{\left( l\right) }}{2}$ and 
$\omega ^{\left( i\right) }=\frac{1}{\theta _{f}^{\left( l\right) }-\theta
_{i}^{\left( l\right) }}$:%
\begin{equation*}
P\left( \left( \omega ^{\left( i\right) },\theta ^{\left( i\right) }\right)
_{1\leqslant i\leqslant n}\right) =\left( \dprod\limits_{l=1}^{n}\frac{1}{%
\left( \omega ^{\left( l\right) }\right) ^{2}}\right) \left( G_{n,n}\left(
\left( \theta _{f}^{\left( n\right) },Z_{l}\right) _{l=1,..,n},\left( \theta
_{i}^{\left( n\right) },Z_{l}\right) _{l=1,..,n}\right) \right) _{\alpha =1}
\end{equation*}%
Using (\ref{jpn}), we can now interpret equations (\ref{grx}) and (\ref{gRX}%
). Writing: 
\begin{eqnarray*}
&&G_{n,n}\left( \left( \theta _{f}^{\left( l\right) },Z_{l}\right)
_{l=1,..,k},\left( \theta _{i}^{\left( l\right) },Z_{l}\right)
_{l=1,..,n}\right) \\
&=&\sum_{\sigma _{n},\sigma
_{n}}\sum_{u=0}^{n}\prod\limits_{j=0}^{u}G_{2}\left( \left( \theta
_{f}^{\left( j\right) },Z_{f}\right) ,\left( \theta _{i}^{\left( i\right)
},Z_{i}\right) \right) \\
&&\times \sum_{i=1,j=1}^{n-u}\left( -1\right) ^{i+j}\sum_{\substack{ %
P_{i}\left( n-u\right)  \\ P_{j}\left( n-u\right) }}\dprod\limits_{\substack{
r\in P_{i}\left( n-u\right)  \\ s\in P_{j}\left( n-u\right) }}\Gamma
_{n_{r},n_{s}}\left( \left( \theta _{f}^{\left( l,r,u\right)
},Z_{l,r,u}\right) _{l=1,..,n_{r}},\left( \theta _{i}^{\left( l,s,u\right)
},Z_{l,s,u}\right) _{l=1,..,n_{s}}\right)
\end{eqnarray*}%
in terms of probabilities:%
\begin{eqnarray*}
&&\left( \dprod\limits_{l=1}^{n}\left( \omega ^{\left( l\right) }\right)
^{2}\right) P\left( \left( \omega ^{\left( i\right) },\theta ^{\left(
i\right) }\right) _{1\leqslant i\leqslant n}\right) \\
&=&\sum_{\sigma _{n},\sigma _{n}}\sum_{u=0}^{n}\prod\limits_{j=0}^{u}\left(
\omega ^{\left( j\right) }\right) ^{2}P\left( \omega ^{\left( j\right)
},\theta ^{\left( j\right) }\right) \sum_{i=1,j=1}^{n-u}\left( -1\right)
^{i+j} \\
&&\times \sum_{\substack{ P_{i}\left( n-u\right) ,  \\ P_{j}\left(
n-u\right) }}\dprod\limits_{\substack{ r\in P_{i}\left( n-u\right)  \\ s\in
P_{j}\left( n-u\right) }}\Gamma _{n_{r},n_{s}}\left( \left( \theta ^{\left(
l,r,u\right) }+\frac{\left( \omega ^{\left( l,r,u\right) }\right) ^{-1}}{2}%
,Z_{l,r,u}\right) _{l=1,..,n_{r}},\left( \theta ^{\left( l,r,u\right) }-%
\frac{\left( \omega ^{\left( l,r,u\right) }\right) ^{-1}}{2}%
,Z_{l,s,u}\right) _{l=1,..,n_{s}}\right)
\end{eqnarray*}%
has an interpretation similar to the $2$-points Green function. The first
terms $\prod\limits_{j=0}^{u}\left( \omega ^{\left( j\right) }\right)
^{2}P\left( \omega ^{\left( j\right) },\theta ^{\left( j\right) }\right) $
represent an independent distribution for the frequencies at different
points, and the corrective terms measure the mutual dependencies due to the
interactions in the background field. Moreover, for $l=m=1$, the
probabilistic interpretation is an alternate description to the frequencies'
local differential equation.

\section{Conclusion}

We have presented a field theoretic framework for a system with a large
number of interacting spiking neurons, and showed its implications on the
dynamics of the system frequencies.

The field framework and the existence of collective or background states
allow for stable traveling wave solutions and correlated frequencies at
different points. These correlations are measured by the $n$ point Green
functions and induces a non-locality in frequencies wave equations. which we
accounted for by deriving non local equations for the frequencies. Besides,
some non-locality also emerges in the impact of the external current on the
background field. An external current may shape the form of the background
field, which in turn conditions the thread in which frequencies waves
propagate.

We have presented several further extensions of our framework. First, we
have extended our formalism to multi-component fields, to include different
types of cells interacting with each others. Second, we have accounted for
the possibility of time and position-dependent transfer functions, where the
dependency results from the strength of cells interactions. This extension
induces frequencies' wave equations with non constant coefficients, that are
waves in a non-homogeneous medium, whose non-homogeneity is described by a
metric.

Our results have been obtained using a dynamic evolution for transfer
functions that depends on the background field for the system of cells. A
straightforward extension would be to design a field formalism for the
transfer functions themselves, in interaction with the cells'
field.\pagebreak

\section*{Appendix 0}

When we restrict the fields to those of the form: 
\begin{equation}
\Psi \left( \theta ,Z\right) \delta \left( \omega ^{-1}-\omega ^{-1}\left(
J,\theta ,Z,\left\vert \Psi \right\vert ^{2}\right) \right)
\end{equation}%
where $\omega ^{-1}\left( J,\theta ,Z,\Psi \right) $ satisfies:%
\begin{eqnarray}
&&\omega ^{-1}\left( J,\theta ,Z,\left\vert \Psi \right\vert ^{2}\right) \\
&=&G\left( J\left( \theta ,Z\right) +\int \frac{\kappa }{N}\frac{\omega
\left( J,\theta -\frac{\left\vert Z-Z_{1}\right\vert }{c},Z_{1},\Psi \right)
T\left( Z,\theta ,Z_{1},\theta -\frac{\left\vert Z-Z_{1}\right\vert }{c}%
\right) }{\omega \left( J,\theta ,Z,\left\vert \Psi \right\vert ^{2}\right) }%
\left\vert \Psi \left( \theta -\frac{\left\vert Z-Z_{1}\right\vert }{c}%
,Z_{1}\right) \right\vert ^{2}dZ_{1}\right)  \notag
\end{eqnarray}%
The classical effective action writes: 
\begin{subequations}
\begin{equation}
-\frac{1}{2}\int \Psi ^{\dagger }\left( \theta ,Z\right) \Psi \left( \theta
,Z\right) \delta \left( \omega ^{-1}-\omega ^{-1}\left( J,\theta
,Z,\left\vert \Psi \right\vert ^{2}\right) \right) \left( \left( \frac{%
\sigma ^{2}}{2}\nabla _{\theta }-\omega ^{-1}\right) \nabla _{\theta
}\right) \Psi \left( \theta ,Z\right) \delta \left( \omega ^{-1}-\omega
^{-1}\left( J,\theta ,Z,\left\vert \Psi \right\vert ^{2}\right) \right)
\label{brt}
\end{equation}

We can replace the first $\delta $ function by $1$ to normalize the
projection on the frequency dependent states.. The action of $\nabla
_{\theta }$ on $\Psi \left( \theta ,Z\right) \delta \left( \omega
^{-1}-\omega ^{-1}\left( J,\theta ,Z,\left\vert \Psi \right\vert ^{2}\right)
\right) $ yields: 
\end{subequations}
\begin{eqnarray}
&&\nabla _{\theta }\left( \Psi \left( \theta ,Z\right) \delta \left( \omega
^{-1}-\omega ^{-1}\left( J,\theta ,Z,\left\vert \Psi \right\vert ^{2}\right)
\right) \right)  \label{rlT} \\
&=&\left( \nabla _{\theta }\Psi \left( \theta ,Z\right) \right) \delta
\left( \omega ^{-1}-\omega ^{-1}\left( J,\theta ,Z,\left\vert \Psi
\right\vert ^{2}\right) \right)  \notag \\
&&-\left( \nabla _{\theta }\omega ^{-1}\left( J,\theta ,Z,\left\vert \Psi
\right\vert ^{2}\right) \right) \Psi \left( \theta ,Z\right) \delta ^{\prime
}\left( \omega ^{-1}-\omega ^{-1}\left( J,\theta ,Z,\left\vert \Psi
\right\vert ^{2}\right) \right)  \notag
\end{eqnarray}%
Inserting the result (\ref{rlT}) in (\ref{brt}) leads to:%
\begin{eqnarray*}
&&-\frac{1}{2}\int \Psi ^{\dagger }\left( \theta ,Z\right) \Psi \left(
\theta ,Z\right) \left( \left( \frac{\sigma ^{2}}{2}\nabla _{\theta }-\omega
^{-1}\right) \right) \left( \nabla _{\theta }\Psi \left( \theta ,Z\right)
\right) \delta \left( \omega ^{-1}-\omega ^{-1}\left( J,\theta ,Z,\left\vert
\Psi \right\vert ^{2}\right) \right) \\
&&+\frac{1}{2}\int \Psi ^{\dagger }\left( \theta ,Z\right) \Psi \left(
\theta ,Z\right) \left( \left( \frac{\sigma ^{2}}{2}\nabla _{\theta }-\omega
^{-1}\right) \right) \Psi \left( \theta ,Z\right) \delta ^{\prime }\left(
\omega ^{-1}-\omega ^{-1}\left( J,\theta ,Z,\left\vert \Psi \right\vert
^{2}\right) \right) \\
&=&-\frac{1}{2}\int \Psi ^{\dagger }\left( \theta ,Z\right) \Psi \left(
\theta ,Z\right) \left( \left( \frac{\sigma ^{2}}{2}\nabla _{\theta }-\omega
^{-1}\left( J,\theta ,Z,\left\vert \Psi \right\vert ^{2}\right) \right)
\right) \nabla _{\theta }\Psi \left( \theta ,Z\right) \\
&&-\frac{1}{2}\int \Psi ^{\dagger }\left( \theta ,Z\right) \Psi \left(
\theta ,Z\right) \left( \left( \frac{\sigma ^{2}}{2}\nabla _{\theta }-\nabla
_{\theta }\omega ^{-1}\left( J,\theta ,Z,\left\vert \Psi \right\vert
^{2}\right) \right) \right) \Psi \left( \theta ,Z\right)
\end{eqnarray*}%
and the sum of the two last terms is, as in the text:%
\begin{equation*}
-\frac{1}{2}\int \Psi ^{\dagger }\left( \theta ,Z\right) \left( \nabla
_{\theta }\left( \frac{\sigma ^{2}}{2}\nabla _{\theta }-\omega ^{-1}\left(
J,\theta ,Z,\left\vert \Psi \right\vert ^{2}\right) \right) \right) \Psi
\left( \theta ,Z\right)
\end{equation*}

\section*{Appendix 1. Vertices of (\protect\ref{lcn}) involved in the
computation of the $2n$ Green functions}

To find the effective action associated to (\ref{lcn}) and the collective
term (\ref{sp}), we proceed in several steps. The first one is to find the
vertices involved in the computation of the Green functions. To do so, we
will expand the action (\ref{lcn}) in series of field. This produces a
series of an infinite series of vertices. However, given that the two points
Green function are not symmetric by time reversal, we will show that only
the $2n$ first terms are involved in the computation of the $2n$ Green
functions. We will then estimate these vertices using the recursive relation
(\ref{prt}) between frequencies depending on field. These results will be
used in the next section to find the graph expansion of the system's
partition function.

\subsection*{1.1 Estimation of the two points Green function}

We start with the two points Green function and prove (\ref{fctnbs}). To do
so, we will expand the action functional in series of the field $\Psi $. The
two points Green functions will be computed by using the "free" action's
propagator, obtained by replacing $\omega ^{-1}\left( J,\theta ,Z,\Psi
\right) $ with $\omega ^{-1}\left( J,\theta ,Z,0\right) $ in (\ref{lcn}).
The free action is:%
\begin{equation}
S_{0}=-\frac{1}{2}\Psi ^{\dagger }\left( \theta ,Z\right) \nabla _{\theta
}\left( \frac{\sigma ^{2}}{2}\nabla _{\theta }-\omega ^{-1}\left( J,\theta
,Z,0\right) \right) \Psi \left( \theta ,Z\right)  \label{zs}
\end{equation}%
and the series in field of (\ref{lcn}) will be considered, as usual, as a
perturbation expansion.

\subsubsection*{1.1.1 "Free" action propagator.}

Now, we compute the propagator associated to (\ref{zs}). We decompose the
external current into a static and a time dependent parts $\bar{J}+J\left(
\theta \right) $ where $\bar{J}$ can be thought as the time average of the
current. We will consider that $\left\vert \bar{J}\left( Z\right)
\right\vert >\left\vert J\left( \theta ,Z\right) \right\vert $. At zeroth
order in current $J\left( \theta \right) $, the function $\omega ^{-1}\left(
J,\theta ,Z,0\right) $ satisfies:%
\begin{eqnarray}
\omega ^{-1}\left( J,\theta ,Z,0\right) &=&G\left( \bar{J}+J\left( \theta
\right) \right)  \label{mgv} \\
&\simeq &G\left( \bar{J}\left( Z\right) \right) =\frac{\arctan \left( \left( 
\frac{1}{X_{r}}-\frac{1}{X_{p}}\right) \sqrt{\bar{J}\left( Z\right) }\right) 
}{\sqrt{\bar{J}\left( Z\right) }}=\frac{1}{\bar{X}_{r}\left( Z\right) }%
\equiv \frac{1}{\bar{X}_{r}}  \notag
\end{eqnarray}%
where the dependence in $Z$ of $\bar{X}_{r}$ will be understood. As a
consequence $\omega \left( \theta ,Z\right) $ is thus approximatively equal
to $\bar{X}_{r}$. Under this approximation:%
\begin{equation*}
S_{0}=-\Psi ^{\dagger }\left( \theta ,Z\right) \nabla _{\theta }\left( \frac{%
\sigma ^{2}}{2}\nabla _{\theta }-\frac{1}{\bar{X}_{r}}\right) \Psi \left(
\theta ,Z\right)
\end{equation*}%
and the Green function of the operator $\nabla _{\theta }\left( \frac{\sigma
^{2}}{2}\nabla _{\theta }-\frac{1}{\bar{X}_{r}}\right) $ is computed as:%
\begin{equation}
\left\langle \Psi ^{\dagger }\left( \theta ,Z\right) \Psi \left( \theta
^{\prime },Z\right) \right\rangle \equiv \mathcal{G}_{0}\left( \left( \theta
,Z\right) ,\left( \theta ^{\prime },Z^{\prime }\right) \right) \equiv 
\mathcal{G}_{0}\left( \theta ,\theta ^{\prime },Z\right) =\delta \left(
Z-Z^{\prime }\right) \int \frac{\exp \left( ik\left( \theta -\theta ^{\prime
}\right) \right) }{\frac{\sigma ^{2}}{2}k^{2}+ik\frac{1}{\bar{X}_{r}}+\alpha 
}dk  \label{Grnrr}
\end{equation}%
The right hand side of (\ref{Grnrr}) can be computed as: 
\begin{eqnarray}
\int \frac{\exp \left( ik\left( \theta -\theta ^{\prime }\right) \right) }{%
\frac{\sigma ^{2}}{2}k^{2}+ik\frac{1}{\bar{X}_{r}}+\alpha }dk &=&\exp \left( 
\frac{\theta -\theta ^{\prime }}{\sigma ^{2}\bar{X}_{r}}\right) \int \frac{%
\exp \left( ik\left( \theta -\theta ^{\prime }\right) \right) }{\frac{\sigma
^{2}}{2}k^{2}+\frac{1}{2}\left( \frac{1}{\sigma \bar{X}_{r}}\right)
^{2}+\alpha }dk  \notag \\
&=&\frac{1}{\sqrt{\frac{\pi }{2}}}\frac{\exp \left( -\sqrt{\left( \frac{1}{%
\sigma ^{2}\bar{X}_{r}}\right) ^{2}+\frac{2\alpha }{\sigma ^{2}}}\left\vert
\theta -\theta ^{\prime }\right\vert \right) }{\sqrt{\left( \frac{1}{\sigma
^{2}\bar{X}_{r}}\right) ^{2}+\frac{2\alpha }{\sigma ^{2}}}}\exp \left( \frac{%
\theta -\theta ^{\prime }}{\sigma ^{2}\bar{X}_{r}}\right)  \label{rdrrz}
\end{eqnarray}%
and this is quickly suppressed for $\theta -\theta ^{\prime }<0$. This is
the direct consequence of non-hermiticity of operator. In the sequel, for $%
\sigma ^{2}\bar{X}_{r}<<1$, we can thus consider that:%
\begin{equation}
\mathcal{G}_{0}\left( \theta ,\theta ^{\prime },Z\right) =\delta \left(
Z-Z^{\prime }\right) \frac{1}{\sqrt{\frac{\pi }{2}}}\frac{\exp \left(
-\left( \sqrt{\left( \frac{1}{\sigma ^{2}\bar{X}_{r}}\right) ^{2}+\frac{%
2\alpha }{\sigma ^{2}}}-\frac{1}{\sigma ^{2}\bar{X}_{r}}\right) \left(
\theta -\theta ^{\prime }\right) \right) }{\sqrt{\left( \frac{1}{\sigma ^{2}%
\bar{X}_{r}}\right) ^{2}+\frac{2\alpha }{\sigma ^{2}}}}H\left( \theta
-\theta ^{\prime }\right)  \label{rdrzr}
\end{equation}%
where $H$ is the Heaviside function:%
\begin{eqnarray*}
H\left( \theta -\theta ^{\prime }\right) &=&0\text{ for }\theta -\theta
^{\prime }<0 \\
&=&1\text{ for }\theta -\theta ^{\prime }>0
\end{eqnarray*}%
Formula (\ref{rdrzr}) for the propagator is sufficient to compute the graphs
expansion in the next paragraphs. We can check that the corrections due to a
non-static current do not modify the result at a good level of
approximation. Considering the following form for $G\left( J\left( \theta
,Z\right) \right) $: 
\begin{equation*}
G\left( J\left( \theta ,Z\right) \right) =\frac{\arctan \left( \left( \frac{1%
}{X_{r}}-\frac{1}{X_{p}}\right) \sqrt{J\left( \theta ,Z\right) }\right) }{%
\sqrt{J\left( \theta ,Z\right) }}
\end{equation*}%
For relatively high frequency firing rates, i.e., small periods of time
between two spikes, we can write in first approximation:%
\begin{eqnarray*}
G\left( \bar{J}+J\left( \theta ,Z\right) \right) &\simeq &G\left( \bar{J}%
\right) +J\left( \theta ,Z\right) G^{\prime }\left( \bar{J}\right) \\
&=&\frac{1}{\bar{X}_{r}}+J\left( \theta ,Z\right) G^{\prime }\left( \bar{J}%
\right)
\end{eqnarray*}%
and replace (\ref{Grnrr}) by the Green function of:%
\begin{equation*}
\nabla _{\theta }\left( \frac{\sigma ^{2}}{2}\nabla _{\theta }-G\left(
J\left( \theta ,Z\right) \right) \right) \simeq \nabla _{\theta }\left( 
\frac{\sigma ^{2}}{2}\nabla _{\theta }-\frac{1}{\bar{X}_{r}}-J\left( \theta
,Z\right) G^{\prime }\left( \bar{J}\right) \right)
\end{equation*}%
As a consequence, the inverse frequency $\mathcal{G}_{0}\left( \theta
,\theta ^{\prime },Z\right) $ defined in (\ref{rdrzr}) is replaced by:%
\begin{eqnarray*}
\mathcal{G}_{0}\left( \left( \theta ,Z\right) ,\left( \theta ^{\prime
},Z^{\prime }\right) \right) &=&\delta \left( Z-Z^{\prime }\right) \frac{1}{%
\sqrt{\frac{\pi }{2}}}\frac{\exp \left( -\left( \sqrt{\left( \frac{1}{\sigma
^{2}\bar{X}_{r}}\right) ^{2}+\frac{2\alpha }{\sigma ^{2}}}-\frac{1}{\sigma
^{2}\bar{X}_{r}}\right) \left( \theta -\theta ^{\prime }\right) \right) }{%
\sqrt{\left( \frac{1}{\sigma ^{2}\bar{X}_{r}}\right) ^{2}+\frac{2\alpha }{%
\sigma ^{2}}}}H\left( \theta -\theta ^{\prime }\right) \\
&&\times \left( 1-\frac{1}{\sqrt{\frac{\pi }{2}}}\frac{G^{\prime }\left( 
\bar{J}\right) }{\sqrt{\left( \frac{1}{\sigma ^{2}\bar{X}_{r}}\right) ^{2}+%
\frac{2\alpha }{\sigma ^{2}}}}\int_{\theta }^{\theta ^{\prime }}J\left(
\theta ^{\prime \prime },Z\right) d\theta ^{\prime \prime }\right)
\end{eqnarray*}%
Since $J\left( \theta ,Z\right) $ is a deviation around the static part $%
\bar{J}$, the corrective term:%
\begin{equation*}
-\frac{1}{\sqrt{\frac{\pi }{2}}}\frac{G^{\prime }\left( \bar{J}\right) }{%
\sqrt{\left( \frac{1}{\sigma ^{2}\bar{X}_{r}}\right) ^{2}+\frac{2\alpha }{%
\sigma ^{2}}}}\int_{\theta }^{\theta ^{\prime }}J\left( \theta ^{\prime
\prime },Z\right) d\theta ^{\prime \prime }
\end{equation*}
vanishes quickly as $\theta -\theta ^{\prime }$ increases, which justifies
approximation (\ref{rdrzr}).

\subsubsection*{1.1.2 perturbation expansion and the two points Green
function}

Formula (\ref{rdrzr}) allows to compute higher order contributions to the
Green function of action (\ref{lcn}) by using a graph expansion. Actually,
writing $\omega ^{-1}\left( \theta ,Z\right) $ for $\omega ^{-1}\left(
J,\theta ,Z,\Psi \right) $ when no ambiguity is possible, the higher order
contribution for the series expansion of $\omega ^{-1}\left( \theta
,Z\right) $\ in fields are obtained by solving recursively:%
\begin{equation}
\omega ^{-1}\left( J,\theta ,Z\right) =G\left( J\left( \theta ,Z\right)
+\int \frac{\kappa }{N}\frac{\omega \left( J,\theta -\frac{\left\vert
Z-Z_{1}\right\vert }{c},Z_{1}\right) }{\omega \left( J,\theta ,Z\right) }%
\left\vert \Psi \left( \theta -\frac{\left\vert Z-Z_{1}\right\vert }{c}%
,Z_{1}\right) \right\vert ^{2}T\left( Z,\theta ,Z_{1}\right) dZ_{1}d\omega
_{1}\right)  \label{gnf}
\end{equation}%
This will be done precisely in the next paragraph. For now, it is enough to
note that given (\ref{gnf}), the recursive expansion in $\omega ^{-1}\left(
J,\theta ,Z\right) $ of the potential term in (\ref{lcn}):

\begin{equation}
\frac{1}{2}\Psi ^{\dagger }\left( \theta ,Z\right) \nabla \left( G\left(
J\left( \theta ,Z\right) +\int \frac{\kappa }{N}\frac{\omega \left( J,\theta
-\frac{\left\vert Z-Z_{1}\right\vert }{c},Z_{1}\right) }{\omega \left(
J,\theta ,Z\right) }\left\vert \Psi \left( \theta -\frac{\left\vert
Z-Z_{1}\right\vert }{c},Z_{1}\right) \right\vert ^{2}T\left( Z,Z_{1}\right)
dZ_{1}\right) \right) \Psi \left( \theta ,Z\right)  \label{ptnlcn}
\end{equation}%
induces the presence of products in the series expansion of the two points
Green function:%
\begin{eqnarray}
&&\dprod\limits_{i=1}^{m}\int \Psi ^{\dagger }\left( \theta ^{\left(
i\right) },Z_{i}\right) \nabla _{\theta ^{\left( i\right)
}}\dprod\limits_{k=1}^{k_{i}}\left(
\dprod\limits_{l=1}^{l_{k}}\dprod\limits_{\alpha \left( l\right)
=1}^{n\left( \alpha \left( l\right) \right) }\int \left\vert \Psi \left(
\theta ^{\left( i\right) }-\frac{\left\vert Z_{i}-Z_{\alpha \left( l\right)
}^{\left( 1\right) }\right\vert +...+\left\vert Z_{\alpha \left( l\right)
}^{\left( l-1\right) }-Z_{\alpha \left( l\right) }^{\left( l\right)
}\right\vert }{c},Z_{\alpha \left( l\right) }^{\left( l\right) }\right)
\right\vert ^{2}\right)  \notag \\
&&\qquad \qquad \qquad \times dZ_{\alpha \left( l\right) }^{\left( 1\right)
}...dZ_{\alpha \left( l\right) }^{\left( l_{k}\right) }\Psi \left( \theta
^{\left( i\right) },Z_{i}\right) d\theta ^{\left( i\right) }dZ_{i}
\label{pdtc}
\end{eqnarray}%
with $n\left( \alpha \left( l\right) \right) \geqslant n\left( \alpha \left(
l^{\prime }\right) \right) $ for $l>l^{\prime }$ and $m\in 
\mathbb{N}
$. The function $\delta \left( Z-Z^{\prime }\right) $ in\ (\ref{Grnrr}) and
the use of Wick's theorem imply that all subgraphs drawn from this product
reduce to a product of\ free Green functions (\ref{rdrzr}) of the following
form (the gradient terms and the indices $\alpha \left( l\right) $ are not
included and do not impact the reasoning):%
\begin{eqnarray}
&&\int \dprod\limits_{i}\mathcal{G}_{0}\left( \theta ^{\left( i\right)
}-\sum_{l\leqslant n_{i}}\frac{\left\vert Z_{i}-Z_{i}^{\left( l\right)
}\right\vert }{c},\theta ^{\left( i+1\right) }-\sum_{k\leqslant n_{i+1}}%
\frac{\left\vert Z_{i+1}-Z_{i+1}^{\left( k\right) }\right\vert }{c}%
,Z_{i}^{\left( n_{i}\right) },Z_{i}^{\left( n_{i+1}\right) }\right)  \notag
\\
&&\times \delta \left( Z_{1}-Z_{i}^{\left( n_{i}\right) }\right) \delta
\left( Z_{1}-Z_{i+1}^{\left( n_{i+1}\right) }\right) dZ_{i}^{\left(
n_{i}\right) }dZ_{i+1}^{\left( n_{i+1}\right) }\dprod\limits_{i}d\theta
^{\left( i\right) }  \notag \\
&=&\int \dprod\limits_{i}\mathcal{G}_{0}\left( \theta ^{\left( i\right)
}-\sum_{l\leqslant n}\frac{\left\vert Z_{i}-Z_{1}^{\left( l\right)
}\right\vert }{c},\theta ^{\left( i+1\right) }-\sum_{k\leqslant m}\frac{%
\left\vert Z_{i+1}-Z_{1}^{\left( k\right) }\right\vert }{c},Z_{1}\right)
\dprod\limits_{i}d\theta ^{\left( i\right) }  \notag \\
&=&\int \dprod\limits_{i}\mathcal{G}_{0}\left( \theta ^{\left( i\right)
},\theta ^{\left( i+1\right) },Z_{1}\right) \dprod\limits_{i}d\theta
^{\left( i\right) }  \label{lp}
\end{eqnarray}%
by change of variable in the successive integrations. Moreover, the
cancelation of $\mathcal{G}_{0}\left( \theta ,\theta ^{\prime },Z\right) $
for $\theta <\theta ^{\prime }$ implies that this product is different from
zero only for $\theta ^{\left( i\right) }<\theta ^{\left( i+1\right) }$. As
a consequence, for all closed loops $\theta _{1}<...<\theta ^{\left(
i\right) }<\theta ^{\left( i+1\right) }<...\theta _{n}=\theta _{1}$, the
contribution (\ref{lp}) for loop graphs reduces to:%
\begin{equation*}
\dprod\limits_{i}\mathcal{G}_{0}\left( \theta _{1},\theta _{1},Z_{1}\right)
=\dprod\limits_{i}\mathcal{G}_{0}\left( 0,Z_{1}\right)
\end{equation*}%
with (see (\ref{rdrzr})):%
\begin{equation*}
\mathcal{G}_{0}\left( 0,Z\right) =\frac{1}{\sqrt{\frac{\pi }{2}\left( \frac{1%
}{\sigma ^{2}\bar{X}_{r}}\right) ^{2}+\frac{2\pi \alpha }{\sigma ^{2}}}}
\end{equation*}%
As a consequence, the contribution of (\ref{pdtc}) to the two points Green
function between an initial and final state:%
\begin{eqnarray}
&&\left\langle \Psi ^{\dagger }\left( \theta _{in},Z_{in}\right) \int
\dprod\limits_{i=1}^{m}\Psi ^{\dagger }\left( \theta ^{\left( i\right)
},Z_{i}\right) \right.  \notag \\
&&\times \nabla _{\theta ^{\left( i\right)
}}\dprod\limits_{k=1}^{k_{i}}\left( \left( \dprod\limits_{l=1}^{l_{k}}\int
\left\vert \Psi \left( \theta ^{\left( i\right) }-\frac{\left\vert
Z_{i}-Z^{\left( 1\right) }\right\vert +...+\left\vert Z^{\left( l-1\right)
}-Z^{\left( l\right) }\right\vert }{c},Z^{\left( l\right) }\right)
\right\vert ^{2}dZ^{\left( 1\right) }...dZ^{\left( l_{k}\right) }\right)
\right)  \notag \\
&&\times \left. \Psi \left( \theta ^{\left( i\right) },Z_{i}\right) d\theta
^{\left( i\right) }dZ_{i}\Psi \left( \theta _{fn},Z_{fn}\right) \right\rangle
\label{prtr}
\end{eqnarray}%
reduces to sums and integrals of the type:%
\begin{eqnarray}
&&\delta \left( Z_{in}-Z_{fn}\right) \sum_{p}\mathcal{G}_{0}\left( \theta
_{in},\theta _{1},Z_{in}\right) \mathcal{G}_{0}\left( \theta _{1},\theta
_{2},Z_{in}\right) ...\mathcal{G}_{0}\left( \theta _{p},\theta
_{fn},Z_{in}\right)  \label{grpn} \\
&&\times \left( \sum_{n}\sum_{\left\{ L_{1}^{\left( p\right)
},...,L_{n}^{\left( p\right) }\right\} }\dprod_{m=1}^{n}\left( \mathcal{G}%
_{0}\left( 0,0,Z_{m}\right) \right) ^{l\left( L_{m}^{\left( p\right)
}\right) }\right)  \notag
\end{eqnarray}%
where $\left\{ L_{1}^{\left( p\right) },...,L_{n}^{\left( p\right) }\right\} 
$ is the set of all $n$-uplet of possible closed loops that can be drawn
from the remaining variables in (\ref{prtr}) once $p$ variables have been
chosen.

The result (\ref{grpn}) is the same as if in (\ref{ptnlcn}) the potential
had been expanded to the second order in $\Psi $ and in all terms of higher
order, $\left\vert \Psi \left( \theta ,Z\right) \right\vert ^{2}$ had been
replaced by $\mathcal{G}_{0}\left( 0,Z\right) $.

Now, writing $\omega \left( J,\theta ,Z,\left\vert \Psi \right\vert
^{2}\right) $ for $\omega $ and $\omega \left( 0\right) =\omega \left(
J,\theta ,Z,0\right) $ (i.e. when we set $\Psi \equiv 0$), this means that
the $2$ points Green functions are computed using the free action:%
\begin{eqnarray}
&&-\frac{1}{2}\Psi ^{\dagger }\left( \theta ,Z\right) \nabla _{\theta
}\left( \frac{\sigma _{\theta }^{2}}{2}\nabla _{\theta }-\omega ^{-1}\left(
0\right) \right) \Psi \left( \theta ,Z\right)  \label{fctc} \\
&&+\frac{1}{2}\Psi ^{\dagger }\left( \theta ,Z\right) \sum_{n>0}\frac{\nabla
_{\theta }\left( \omega ^{-1}\right) ^{\left( \left[ n\right] \right)
}\left( 0\right) }{\left[ n\right] !}\left( \mathcal{G}_{0}\left( 0,Z\right)
\right) ^{n}\Psi \left( \theta ,Z\right)  \notag \\
&&+\sum_{n>0}\left( \nabla _{\theta }\frac{\left( \omega ^{-1}\right)
^{\left( \left[ n-1\right] \right) }\left( 0\right) \left\vert \Psi
\right\vert ^{2}}{\left[ n-1\right] !}\left( \mathcal{G}_{0}\left(
0,Z\right) \right) ^{n-1}\mathcal{G}_{0}\left( \theta ,\theta ^{\prime
},Z\right) \right) _{\theta ^{\prime }=\theta }  \notag \\
&=&-\frac{1}{2}\Psi ^{\dagger }\left( \theta ,Z\right) \nabla _{\theta
}\left( \frac{\sigma _{\theta }^{2}}{2}\nabla _{\theta }-\omega ^{-1}\left(
0\right) \right) \Psi \left( \theta ,Z\right) +\frac{1}{2}\Psi ^{\dagger
}\left( \theta ,Z\right) \sum_{n>0}\nabla _{\theta }\left( \left( \omega
^{-1}\right) \left( \mathcal{G}_{0}\left( 0,Z\right) \right) -\omega
^{-1}\left( 0\right) \right) \Psi \left( \theta ,Z\right)  \notag \\
&&+\Psi ^{\dagger }\left( \theta ,Z\right) \left( \nabla _{\theta ^{\prime
}}\left( \left( \omega ^{-1}\right) ^{\left( \left[ 1\right] \right) }\left( 
\mathcal{G}_{0}\left( 0,Z\right) \right) \Psi \left( \theta ^{\prime
},Z\right) \mathcal{G}_{0}\left( \theta ,\theta ^{\prime },Z\right) \right)
\right) _{\theta ^{\prime }=\theta }  \notag \\
&\equiv &-\frac{1}{2}\Psi ^{\dagger }\left( \theta ,Z\right) \left( \nabla
_{\theta }\frac{\sigma _{\theta }^{2}}{2}\nabla _{\theta }\right) \Psi
\left( \theta ,Z\right) +\frac{1}{2}\left\vert \Psi \right\vert ^{2}\left[ 
\frac{\delta \left[ \Psi ^{\dagger }\left( \theta ^{\prime },Z\right) \nabla
_{\theta }\omega ^{-1}\left( J,\theta ,Z,\left\vert \Psi \right\vert
^{2}\right) \Psi \left( \theta ,Z\right) \right] }{\delta \left\vert \Psi
\right\vert ^{2}}\right] _{\left\vert \Psi \left( \theta ,Z\right)
\right\vert ^{2}=\mathcal{G}_{0}\left( 0,Z\right) }  \notag
\end{eqnarray}%
where $\frac{\left( \omega ^{-1}\right) ^{\left( \left[ n\right] \right)
}\left( 0\right) }{\left[ n\right] !}$ is a short notation for:%
\begin{equation*}
\sum_{l_{i}}\int \dprod\limits_{i=1}^{n}dZ_{l_{i}}^{\left( 1\right)
}...dZ_{l_{i}}^{\left( l_{i}\right) }\left( \frac{\delta ^{n}\left[ \omega
^{-1}\left( J,\theta ,Z,\left\vert \Psi \right\vert ^{2}\right) \right] }{%
\dprod\limits_{i=1}^{n}\delta \left( \left\vert \Psi \left( \theta -\frac{%
\left\vert Z-Z_{l_{i}}^{\left( 1\right) }\right\vert +...+\left\vert
Z^{_{l_{i}}\left( l-1\right) }-Z_{l_{i}}^{\left( l_{i}\right) }\right\vert }{%
c},Z_{l_{i}}^{\left( l_{i}\right) }\right) \right\vert ^{2}\right) }\right)
_{\left\vert \Psi \right\vert =0}
\end{equation*}%
and $\frac{\left( \omega ^{-1}\right) ^{\left( \left[ n-1\right] \right)
}\left( 0\right) \left\vert \Psi \right\vert ^{2}}{\left[ n-1\right] !}$
stands for:%
\begin{eqnarray*}
&&\sum_{l_{i}}\int \dprod\limits_{i=1}^{n-1}dZ_{l_{i}}^{\left( 1\right)
}...dZ_{l_{i}}^{\left( l_{i}\right) }\left( \frac{\delta ^{n-1}\left[ \omega
^{-1}\left( J,\theta ,Z,\left\vert \Psi \right\vert ^{2}\right) \right] }{%
\dprod\limits_{i}\delta \left( \left\vert \Psi \left( \theta -\frac{%
\left\vert Z-Z_{l_{i}}^{\left( 1\right) }\right\vert +...+\left\vert
Z_{l_{i}}^{\left( l-1\right) }-Z_{l_{i}}^{\left( l_{i}\right) }\right\vert }{%
c},Z_{l_{i}}^{\left( l_{i}\right) }\right) \right\vert ^{2}\right)
^{k_{l_{i}}}}\right) _{\left\vert \Psi \right\vert =0} \\
&&\times \sum_{j=1}^{n-1}\left\vert \Psi \left( \theta -\frac{\left\vert
Z-Z_{l_{j}}^{\left( 1\right) }\right\vert +...+\left\vert Z_{l_{i}}^{\left(
l-1\right) }-Z_{l_{j}}^{\left( l_{j}\right) }\right\vert }{c}%
,Z_{l_{j}}^{\left( l_{j}\right) }\right) \right\vert ^{2}
\end{eqnarray*}%
Similar notation is valid for $\frac{\left( \omega ^{-1}\right) ^{\left( %
\left[ n\right] \right) }\left( \mathcal{G}_{0}\left( 0,0,Z\right) \right)
\left\vert \Psi \right\vert ^{2}}{\left[ n-1\right] !}$, the derivatives are
evaluated at $\left\vert \Psi \left( \theta ,Z\right) \right\vert ^{2}=%
\mathcal{G}_{0}\left( 0,0,Z\right) $.

We have also used $\left\vert \Psi \right\vert ^{2}\left[ \frac{\delta }{%
\delta \left\vert \Psi \right\vert ^{2}}\right] $ as a shorthand for:%
\begin{eqnarray}
&&\sum_{l}\int \left( \frac{dZ_{l}^{\left( 1\right) }...dZ_{l}^{\left(
l\right) }}{\left( k_{l}\right) !}\right) \left\vert \Psi \left( \theta -%
\frac{\left\vert Z-Z_{l}^{\left( 1\right) }\right\vert +...+\left\vert
Z_{l}^{\left( l-1\right) }-Z_{l}^{\left( l_{j}\right) }\right\vert }{c}%
,Z_{l}^{\left( l\right) }\right) \right\vert ^{2}  \label{drvf} \\
&&\times \frac{\delta }{\delta \left( \left\vert \Psi \left( \theta -\frac{%
\left\vert Z-Z_{l}^{\left( 1\right) }\right\vert +...+\left\vert
Z_{l}^{\left( l-1\right) }-Z_{l}^{\left( l\right) }\right\vert }{c}%
,Z_{l}^{\left( l\right) }\right) \right\vert ^{2}\right) }  \notag
\end{eqnarray}%
Ultimately, the computation of the Green function involves the series
expansion of the potential $V\left( \Psi \right) $. We have seen above (see
equation(\ref{grpn})) that the graphs generated by this expansion are the
same as if in (\ref{ptnlcn}) the potential had been expanded to the second
order in $\Psi $ and in all terms of higher order, $\left\vert \Psi \left(
\theta ,Z\right) \right\vert ^{2}$ had been replaced by $\mathcal{G}%
_{0}\left( 0,Z\right) $. As a consequence, the second order Green functions
are computed with the action:%
\begin{eqnarray*}
&&-\frac{1}{2}\Psi ^{\dagger }\left( \theta ,Z\right) \left( \nabla _{\theta
}\frac{\sigma _{\theta }^{2}}{2}\nabla _{\theta }\right) \Psi \left( \theta
,Z\right) \\
&&+\frac{1}{2}\left\vert \Psi \right\vert ^{2}\left[ \frac{\delta \left[
\Psi ^{\dagger }\left( \theta ^{\prime },Z\right) \nabla _{\theta }\left(
\omega ^{-1}\left( J,\theta ,Z,\left\vert \Psi \right\vert ^{2}\right) \Psi
\left( \theta ,Z\right) \right) \right] }{\delta \left\vert \Psi \right\vert
^{2}}\right] _{\substack{ \left\vert \Psi \left( \theta ,Z\right)
\right\vert ^{2}  \\ =\mathcal{G}_{0}\left( 0,Z\right) }}+\left\vert \Psi
\right\vert ^{2}\left[ \frac{\delta \left[ V\left( \Psi \right) \right] }{%
\delta \left\vert \Psi \right\vert ^{2}}\right] _{\substack{ \left\vert \Psi
\left( \theta ,Z\right) \right\vert ^{2}  \\ =\mathcal{G}_{0}\left(
0,Z\right) }}
\end{eqnarray*}%
Equivalently, this means that the $2$ points Green functions are the inverse
of the operator:%
\begin{equation*}
-\frac{1}{2}\nabla _{\theta }\frac{\sigma _{\theta }^{2}}{2}\nabla _{\theta
}+\frac{1}{2}\left[ \frac{\delta \left[ \Psi ^{\dagger }\left( \theta
^{\prime },Z\right) \nabla _{\theta }\left( \omega ^{-1}\left( J,\theta
,Z,\left\vert \Psi \right\vert ^{2}\right) \Psi \left( \theta ,Z\right)
\right) \right] }{\delta \left\vert \Psi \right\vert ^{2}}\right] 
_{\substack{ \left\vert \Psi \left( \theta ,Z\right) \right\vert ^{2}  \\ =%
\mathcal{G}_{0}\left( 0,Z\right) }}+\left[ \frac{\delta \left[ V\left( \Psi
\right) \right] }{\delta \left\vert \Psi \right\vert ^{2}}\right] 
_{\substack{ \left\vert \Psi \left( \theta ,Z\right) \right\vert ^{2}  \\ =%
\mathcal{G}_{0}\left( 0,Z\right) }}
\end{equation*}

\subsection*{1.2 Higher order vertices involved in the effective action}

\subsubsection*{1.2.1 General form of the vertices}

To compute the $2n$ points Green functions, we proceed as for the two points
function and consider a series expansion of the potential in powers of $\Psi
\left( \theta ,Z\right) $. In products $\dprod\limits_{i=1}^{n}\left\vert
\Psi \left( \theta _{i},Z_{i}\right) \right\vert ^{2}$, $n-k$ factors $%
\left\vert \Psi \left( \theta _{i},Z_{i}\right) \right\vert ^{2}$ are
replaced by $\mathcal{G}_{0}\left( 0,0,Z_{i}\right) $ at the higher orders.
A derivation similar to (\ref{fctc}) then shows that $2n$ Green functions
are computed by using the expansion of the action:%
\begin{eqnarray}
&&-\frac{1}{2}\Psi ^{\dagger }\left( \theta ,Z\right) \left( \nabla _{\theta
}\frac{\sigma _{\theta }^{2}}{2}\nabla _{\theta }\right) \Psi \left( \theta
,Z\right)  \label{srsp} \\
&&+\frac{1}{2}\sum_{n\geqslant k\geqslant 0}\left\vert \Psi \right\vert
^{2k}\left( \frac{\delta ^{k}}{\left[ k\right] !\delta ^{k}\left\vert \Psi
\right\vert ^{2}}\left[ \Psi ^{\dagger }\left( \theta ^{\prime },Z\right)
\nabla _{\theta }\left( \omega ^{-1}\left( J,\theta ,Z,\left\vert \Psi
\right\vert ^{2}\right) \Psi \left( \theta ,Z\right) \right) \right] \right)
_{\left\vert \Psi \left( \theta ,Z\right) \right\vert ^{2}=\mathcal{G}%
_{0}\left( 0,Z\right) }  \notag
\end{eqnarray}%
where $\left\vert \Psi \right\vert ^{2k}\frac{\delta ^{k}}{\left[ n\right]
!\delta ^{k}\left\vert \Psi \right\vert ^{2}}$ generalizes (\ref{drvf}) and
stands for:%
\begin{eqnarray*}
&&\sum_{l_{i}}\int \dprod\limits_{i=1}^{k}\left( dZ_{l_{i}}^{\left( 1\right)
}...dZ_{l_{i}}^{\left( l_{i}\right) }\right) \left\vert \Psi \left( \theta -%
\frac{\left\vert Z-Z_{l_{j}}^{\left( 1\right) }\right\vert +...+\left\vert
Z_{l_{i}}^{\left( l-1\right) }-Z_{l_{j}}^{\left( l_{j}\right) }\right\vert }{%
c},Z_{l_{i}}^{\left( l_{i}\right) }\right) \right\vert ^{2} \\
&&\times \left( \frac{\delta ^{k}}{\dprod\limits_{i}\delta \left( \left\vert
\Psi \left( \theta -\frac{\left\vert Z-Z_{l_{i}}^{\left( 1\right)
}\right\vert +...+\left\vert Z_{l_{i}}^{\left( l-1\right)
}-Z_{l_{i}}^{\left( l_{i}\right) }\right\vert }{c},Z_{l_{i}}^{\left(
l_{i}\right) }\right) \right\vert ^{2}\right) ^{k_{l_{i}}}}\right)
\end{eqnarray*}%
Equation (\ref{srsp}) can be shown recursively. To compute the $2n$
correlation functions, the subgraphs with $2k$ legs, $k<n$, are given by (%
\ref{srsp}) at order $2k$. For $k=n$, the classical action yields a vertex:%
\begin{equation*}
\frac{1}{2}\left( \frac{\delta ^{n}}{\left[ n\right] !\delta ^{n}\left\vert
\Psi \right\vert ^{2}}\left[ \Psi ^{\dagger }\left( \theta ^{\prime
},Z\right) \nabla _{\theta }\left( \omega ^{-1}\left( J,\theta ,Z,\left\vert
\Psi \right\vert ^{2}\right) \Psi \left( \theta ,Z\right) \right) \right]
\right) _{\left\vert \Psi \left( \theta ,Z\right) \right\vert
^{2}=0}\left\vert \Psi \right\vert ^{2n}
\end{equation*}%
For $k>n$, a similar argument as in paragraph 1.1 in the vertex:%
\begin{equation*}
\frac{1}{2}\left( \frac{\delta ^{k}}{\left[ k\right] !\delta ^{k}\left\vert
\Psi \right\vert ^{2}}\left[ \Psi ^{\dagger }\left( \theta ^{\prime
},Z\right) \nabla _{\theta }\left( \omega ^{-1}\left( J,\theta ,Z,\left\vert
\Psi \right\vert ^{2}\right) \Psi \left( \theta ,Z\right) \right) \right]
\right) _{\left\vert \Psi \left( \theta ,Z\right) \right\vert
^{2}=0}\left\vert \Psi \right\vert ^{2k}
\end{equation*}%
$k-n$ factor $\left\vert \Psi \left( \theta ,Z\right) \right\vert ^{2}$ have
to be replaced by $\mathcal{G}_{0}\left( 0,0,Z\right) $. Summing over $k$,
it means that the $2n$ vertex is computed with:%
\begin{equation*}
\frac{1}{2}\sum_{l=0}^{\infty }\left( \frac{\delta ^{l+n}}{\left[ l+n\right]
!\delta ^{l+n}\left\vert \Psi \right\vert ^{2}}\left[ \Psi ^{\dagger }\left(
\theta ^{\prime },Z\right) \nabla _{\theta }\left( \omega ^{-1}\left(
J,\theta ,Z,\left\vert \Psi \right\vert ^{2}\right) \Psi \left( \theta
,Z\right) \right) \right] \right) _{\left\vert \Psi \left( \theta ,Z\right)
\right\vert ^{2}=0}\left[ C_{l+n}^{l}\right] \left( \mathcal{G}_{0}\left(
0,0,Z\right) \right) ^{l}\left\vert \Psi \right\vert ^{2n}
\end{equation*}%
where the symbol $\left[ C_{l+n}^{l}\right] $ reminds that among the product 
$\left\vert \Psi \left( \theta _{1},Z_{1}\right) \right\vert
^{2}...\left\vert \Psi \left( \theta _{l+n},Z_{l+n}\right) \right\vert ^{2}$
we sum over all the $C_{l+n}^{l}$ possibilities to replace $l$ factor $%
\left\vert \Psi \left( \theta _{j},Z_{j}\right) \right\vert ^{2}$ by $%
\mathcal{G}_{0}\left( 0,0,Z_{j}\right) $. Summing the series, we find for
the $2n$ vertices: 
\begin{eqnarray*}
&&\frac{1}{2}\sum_{l=0}^{\infty }\left( \frac{\delta ^{l+n}}{\left[ l+n%
\right] !\delta ^{l+n}\left\vert \Psi \right\vert ^{2}}\left[ \Psi ^{\dagger
}\left( \theta ^{\prime },Z\right) \nabla _{\theta }\left( \omega
^{-1}\left( J,\theta ,Z,\left\vert \Psi \right\vert ^{2}\right) \Psi \left(
\theta ,Z\right) \right) \right] \right) _{\left\vert \Psi \left( \theta
,Z\right) \right\vert ^{2}=0}\left[ C_{l+n}^{l}\right] \left( \mathcal{G}%
_{0}\left( 0,0,Z\right) \right) ^{l}\left\vert \Psi \right\vert ^{2n} \\
&=&\frac{1}{2}\left( \frac{\delta ^{n}}{\left[ n\right] !\delta
^{n}\left\vert \Psi \right\vert ^{2}}\left[ \Psi ^{\dagger }\left( \theta
^{\prime },Z\right) \nabla _{\theta }\left( \omega ^{-1}\left( J,\theta
,Z,\left\vert \Psi \right\vert ^{2}\right) \Psi \left( \theta ,Z\right)
\right) \right] \right) _{\left\vert \Psi \left( \theta ,Z\right)
\right\vert ^{2}=\mathcal{G}_{0}\left( 0,Z\right) }\left\vert \Psi
\right\vert ^{2n}
\end{eqnarray*}%
as requested.

Below, to compute the higher order corrections to the effective potential,
it will be useful to write (\ref{srsp}) with an other set of variables. We
replace: 
\begin{equation*}
\Psi \left( \theta -\frac{\left\vert Z-Z_{l_{i}}^{\left( 1\right)
}\right\vert +...+\left\vert Z_{l_{i}}^{\left( l-1\right)
}-Z_{l_{i}}^{\left( l_{i}\right) }\right\vert }{c},Z_{l_{i}}^{\left(
l_{i}\right) }\right)
\end{equation*}%
by $\Psi \left( \theta -l_{i},Z_{i}\right) $ where $l_{i}$ represents an
arbitrary delay time. As a consequence, the $2n$-th vertex:%
\begin{equation*}
V_{2n}=\left\vert \Psi \right\vert ^{2n}\left( \frac{\delta ^{n}}{\left[ n%
\right] !\delta ^{n}\left\vert \Psi \right\vert ^{2}}\left[ \Psi ^{\dagger
}\left( \theta ^{\prime },Z\right) \nabla _{\theta }\left( \omega
^{-1}\left( J,\theta ,Z,\left\vert \Psi \right\vert ^{2}\right) \Psi \left(
\theta ,Z\right) \right) \right] \right) _{\left\vert \Psi \left( \theta
,Z\right) \right\vert ^{2}=\mathcal{G}_{0}\left( 0,Z\right) }
\end{equation*}%
becomes (where $\omega ^{-1}\left( J,\theta ,Z\right) $ stands for $\omega
^{-1}\left( J,\theta ,Z,\left\vert \Psi \right\vert ^{2}\right) $ when no
confusion is possible):%
\begin{eqnarray}
V_{2n} &=&\frac{1}{2\left( n\right) !}\left[ \int \Psi ^{\dagger }\left(
\theta ,Z\right) \frac{\delta ^{n}\left[ \int \Psi ^{\dagger }\left( \theta
,Z\right) \nabla _{\theta }\omega ^{-1}\left( J,\theta ,Z\right) \Psi \left(
\theta ,Z\right) dZd\theta \right] }{\dprod\limits_{i=1}^{n}\delta
\left\vert \Psi \left( \theta -l_{i},Z_{i}\right) \right\vert ^{2}}\right.
\label{pvdn} \\
&&\times \left. \dprod\limits_{i=1}^{n-1}\left\vert \Psi \left( \theta
-l_{i},Z_{i}\right) \right\vert ^{2}dl_{i}\Psi \left( \theta ,Z\right)
dZd\theta \right] _{\left\vert \Psi \left( \theta ,Z\right) \right\vert ^{2}=%
\mathcal{G}_{0}\left( 0,Z\right) }  \notag \\
&=&\frac{1}{2\left( n-1\right) !}\int \Psi ^{\dagger }\left( \theta
,Z\right) \nabla _{\theta }\left[ \frac{\delta ^{n-1}\omega ^{-1}\left(
J,\theta ,Z\right) }{\dprod\limits_{i=1}^{n-1}\delta \left\vert \Psi \left(
\theta -l_{i},Z_{i}\right) \right\vert ^{2}}\right] _{\substack{ \left\vert
\Psi \left( \theta ,Z\right) \right\vert ^{2}  \\ =\mathcal{G}_{0}\left(
0,Z\right) }}\dprod\limits_{i=1}^{n-1}\left\vert \Psi \left( \theta
-l_{i},Z_{i}\right) \right\vert ^{2}dZ_{i}\Psi \left( \theta ,Z\right)
dZd\theta dl_{i}  \notag \\
&&+\frac{1}{2n!}\int \nabla _{\theta }\left[ \frac{\delta ^{n}\left( \nabla
_{\theta }\omega ^{-1}\left( J,\theta ,Z\right) \mathcal{G}_{0}\left( \theta
,\theta ^{\prime },Z\right) \right) _{\theta =\theta ^{\prime }}}{%
\dprod\limits_{i=1}^{n}\delta \left\vert \Psi \left( \theta
-l_{i},Z_{i}\right) \right\vert ^{2}}\right] _{\substack{ \left\vert \Psi
\left( \theta ,Z\right) \right\vert ^{2}  \\ =\mathcal{G}_{0}\left(
0,Z\right) }}\dprod\limits_{i=1}^{n}\left\vert \Psi \left( \theta
-l_{i},Z_{i}\right) \right\vert ^{2}\dprod\limits_{i=1}^{n}dZ_{i}dZd\theta
dl_{i}  \notag
\end{eqnarray}

\subsubsection*{1.2.2 Estimation of (\protect\ref{pvdn})}

Expression (\ref{pvdn}) can also be rewritten:%
\begin{eqnarray}
V_{2n} &=&\frac{1}{2}\int \Psi ^{\dagger }\left( \theta ,Z\right) \nabla
_{\theta }\frac{\delta ^{n-1}\omega ^{-1}\left( J,\theta ,Z\right) }{%
\dprod\limits_{i=1}^{n-1}\delta \left\vert \Psi \left( \theta
-l_{i},Z_{i}\right) \right\vert ^{2}}\dprod\limits_{i=1}^{n-1}\left\vert
\Psi \left( \theta -l_{i},Z_{i}\right) \right\vert
^{2}\dprod\limits_{i=1}^{n-1}dZ_{i}\Psi \left( \theta ,Z\right) dZdl_{i}
\label{vrtp} \\
&&+\int \mathcal{G}_{0}^{\prime }\left( Z\right) \frac{\delta ^{n}\omega
^{-1}\left( J,\theta ,Z\right) }{\dprod\limits_{i=1}^{n}\delta \left\vert
\Psi \left( \theta -l_{i},Z_{i}\right) \right\vert ^{2}}\dprod%
\limits_{i=1}^{n}\left\vert \Psi \left( \theta -l_{i},Z_{i}\right)
\right\vert ^{2}\dprod\limits_{i=1}^{n}dZ_{i}dZdl_{i}  \notag \\
&&+\int \mathcal{G}_{0}\left( Z\right) \frac{\delta ^{n}\nabla _{\theta
}\omega ^{-1}\left( J,\theta ,Z\right) }{\dprod\limits_{i=1}^{n}\delta
\left\vert \Psi \left( \theta -l_{i},Z_{i}\right) \right\vert ^{2}}%
\dprod\limits_{i=1}^{n}\left\vert \Psi \left( \theta -l_{i},Z_{i}\right)
\right\vert ^{2}\dprod\limits_{i=1}^{n}dZ_{i}dZdl_{i}  \notag
\end{eqnarray}%
with:%
\begin{equation*}
\mathcal{G}_{0}^{\prime }\left( Z\right) =\left( \frac{\nabla _{\theta }%
\mathcal{G}_{0}\left( \theta ,\theta ^{\prime },Z\right) }{2}\right)
_{\theta =\theta ^{\prime }}
\end{equation*}%
However, the two last terms in (\ref{vrtp}) come from the backreaction of
the $n$ vertices on the whole system, and can be neglected in first
approximation. Actually in a neighborhood of the permanent regime, we have:%
\begin{equation*}
\mathcal{G}_{0}\left( Z\right) \frac{\delta ^{n}\omega ^{-1}\left( J,\theta
,Z\right) }{\dprod\limits_{i=1}^{n}\delta \left\vert \Psi \left( \theta
-l_{i},Z_{i}\right) \right\vert ^{2}}<<\frac{\delta ^{n-1}\omega ^{-1}\left(
J,\theta ,Z\right) }{\dprod\limits_{i=1}^{n-1}\delta \left\vert \Psi \left(
\theta -l_{i},Z_{i}\right) \right\vert ^{2}}
\end{equation*}%
The neglected terms will be reintroduced later. We can thus consider that:%
\begin{equation}
V_{2n}=\frac{1}{2\left( n-1\right) !}\int \Psi ^{\dagger }\left( \theta
,Z\right) \nabla _{\theta }\frac{\delta ^{n-1}\omega ^{-1}\left( J,\theta
,Z\right) }{\dprod\limits_{i=1}^{n-1}\delta \left\vert \Psi \left( \theta
-l_{i},Z_{i}\right) \right\vert ^{2}}\dprod\limits_{i=1}^{n-1}\left\vert
\Psi \left( \theta -l_{i},Z_{i}\right) \right\vert
^{2}\dprod\limits_{i=1}^{n-1}dZ_{i}dl_{i}\Psi \left( \theta ,Z\right)
d\theta dZ  \label{vnd}
\end{equation}%
The neglected contributions will be reintroduced in Appendix 4.

The terms in (\ref{vnd}) are the coefficients obtained by the expansion of $%
\omega ^{-1}\left( J,\theta ,Z\right) $ in powers of $\Psi ^{\dagger }\left(
\theta ,Z\right) \Psi \left( \theta ,Z\right) $. It is valid for $\left\vert
\Psi \left( \theta ,Z\right) \right\vert <1$. For $\left\vert \Psi \left(
\theta ,Z\right) \right\vert >1$, we can expand $\omega ^{-1}\left( J,\theta
,Z\right) $ in powers of $\frac{1}{\left\vert \Psi \left( \theta ,Z\right)
\right\vert }$. Given the form of $F$ and since $\arctan \left( x\right) =%
\frac{\pi }{2}-\arctan \left( \frac{1}{x}\right) $, the expansion is
obtained by replacing the derivatives of $F$ by those of $-x^{2}F$ and by
replacing $\omega $ with $\omega ^{-1}$.

Formula (\ref{vnd}) yields the vertices $V_{2n}$, $n\leqslant N$,
intervening in the computation of the $2N$ correlation functions. We have to
estimate the derivatives arising in (\ref{vnd}), before computing the
effective action.

\section*{Appendix 2 General form of the graphs and convergence of the
graphs expansions}

We compute the sum of graphs involved in the partition function with source
term, i.e. the graphs deduced from the.interaction terms (\ref{pvdn}). This
is done in several step. We first give the general form of these graphs.
Then, we compute the factors arising from the vertices (\ref{pvdn}). This
allows to find the full sum of graphs, and to show its convergence.

\subsection*{2.1 General form of the graphs}

The sum of graphs with $2n$-th external points is obtained by considering
any graph between these points and made of $2$\ points propagators connected
by vertices $V_{2l}$ defined in (\ref{pvdn}), where $l\leqslant n$. The $2l$
vertices are given by:%
\begin{equation}
V_{2l}\left( \left\{ \left( \theta ^{\left( k_{i}\right) },Z_{k_{i}}\right)
\right\} _{i=1,...,l}\right) =\frac{1}{l!}\left[ \frac{\delta ^{n}\left[
\int \Psi ^{\dagger }\left( \theta ,Z\right) \nabla _{\theta }\omega
^{-1}\left( J,\theta ,Z\right) \Psi \left( \theta ,Z\right) dZd\theta \right]
}{\dprod\limits_{i=1}^{l}\delta \left\vert \Psi \left( \theta ^{\left(
k_{i}\right) },Z_{k_{i}}\right) \right\vert ^{2}}\right] _{\left\vert \Psi
\left( \theta ,Z\right) \right\vert ^{2}=\mathcal{G}_{0}\left( 0,Z\right) }
\end{equation}%
To these vertices will be added the contributions of a stabilization
potential. If we write this potential $V\left( \Psi \right) $, the vertex is
modified as:%
\begin{eqnarray}
&&\hat{V}_{2l}\left( \left\{ \left( \theta ^{\left( k_{i}\right)
},Z_{k_{i}}\right) \right\} _{i=1,...,l}\right)  \label{mv} \\
&=&\frac{1}{l!}\left[ \frac{\delta ^{l}\left[ \int \Psi ^{\dagger }\left(
\theta ,Z\right) \nabla _{\theta }\omega ^{-1}\left( J,\theta ,Z\right) \Psi
\left( \theta ,Z\right) dZd\theta +\Psi ^{\dagger }\left( \theta ,Z\right)
V\left( \Psi \right) \Psi \left( \theta ,Z\right) \right] }{%
\dprod\limits_{i=1}^{l}\delta \left\vert \Psi \left( \theta ^{\left(
k_{i}\right) },Z_{k_{i}}\right) \right\vert ^{2}}\right] _{\substack{ %
\left\vert \Psi \left( \theta ,Z\right) \right\vert ^{2}  \\ =\mathcal{G}%
_{0}\left( 0,Z\right) }}  \notag
\end{eqnarray}

The graphs have no loop drawn between two legs of any of the external points
(these contributions are already taken into account by the expansion around $%
\mathcal{G}_{0}\left( 0,Z\right) =\frac{1}{\Lambda }$). The absence of
internal loops implies that the $2n$-th points graphs are made of $n$ lines $%
P_{i}$. Each line $P_{i}$ is associated to a point $Z_{i}$. It is drawn
between an initial time $\theta _{i}^{\left( i\right) }$ and a final time $%
\theta _{f}^{\left( i\right) }$. We can thus write the $2n$-th external
points as $\left( \theta _{i}^{\left( i\right) },\theta _{f}^{\left(
i\right) },Z_{i}\right) _{i=1,...,n}$ with $\theta _{i}^{\left( i\right)
}<\theta _{f}^{\left( i\right) }$. The vertex $\hat{V}_{2l}\left( \left\{
\left( \theta ^{\left( k_{i}\right) },Z_{k_{i}}\right) \right\}
_{i=1,...,l}\right) $ can be represented by a point $\left( Z,\theta \right) 
$ from which is issued $2l$ legs ending at the points $\left( \theta
^{\left( k_{i}\right) },Z_{k_{i}}\right) $. A graph thus consists in lines $%
P_{i}$ that are cut by an arbitrary number of vertices of valence $%
l\leqslant n$ of the form $\hat{V}_{2l}\left( \left\{ \left( \theta ^{\left(
k_{i}\right) },Z_{k_{i}}\right) \right\} _{i=1,...,l}\right) $ with $\theta
_{i}^{\left( k_{i}\right) }<\theta ^{\left( k_{i}\right) }<\theta
_{f}^{\left( k_{i}\right) }$. We associate a propagator $\mathcal{G}%
_{0}\left( \theta ,\theta ^{\prime },Z\right) =\frac{\exp \left( -\Lambda
_{1}\left( \theta -\theta ^{\prime }\right) \right) }{\Lambda }H\left(
\theta -\theta ^{\prime }\right) $ to each segment of the graph between two
vertices connected to the line labelled by $Z$ at $\theta $ and $\theta
^{\prime }$,

As in (\ref{pvdn}), the vertices $\hat{V}_{2l}\left( \left\{ \left( \theta
^{\left( k_{i}\right) },Z_{k_{i}}\right) \right\} _{i=1,...,l}\right) $ can
be decomposed in two terms:%
\begin{eqnarray}
&&\frac{1}{2\left( n-1\right) !}\int \left[ \frac{\delta ^{l-1}\left( \nabla
_{\theta }\omega ^{-1}\left( J,\theta ,Z\right) +V\left( \Psi \right)
\right) }{\dprod\limits_{i=1}^{l-1}\delta \left\vert \Psi \left( \theta
^{\left( k_{i}\right) },Z_{k_{i}}\right) \right\vert ^{2}}\right]
_{\left\vert \Psi \left( \theta ,Z\right) \right\vert ^{2}=\mathcal{G}%
_{0}\left( 0,Z\right) }\delta \left( \theta ^{\left( k_{i}\right) }-\theta
\right) \delta \left( Z_{k_{i}}-Z\right) dZd\theta dl_{i}  \notag \\
&&+\frac{1}{2n!}\int \left[ \frac{\delta ^{l}\left( \nabla _{\theta }\omega
^{-1}\left( J,\theta ,Z\right) \mathcal{G}_{0}\left( \theta ,\theta ^{\prime
},Z\right) +\mathcal{G}_{0}\left( \theta ,\theta ,Z\right) V\left( \Psi
\right) \right) _{\theta =\theta ^{\prime }}}{\dprod\limits_{i=1}^{n}\delta
\left\vert \Psi \left( \theta ^{\left( k_{i}\right) },Z_{k_{i}}\right)
\right\vert ^{2}}\right] _{\left\vert \Psi \left( \theta ,Z\right)
\right\vert ^{2}=\mathcal{G}_{0}\left( 0,Z\right) }dZd\theta  \notag
\end{eqnarray}

The second term computes the impact of the propagator $\mathcal{G}_{0}\left(
\theta ,\theta ^{\prime },Z\right) $ on the background $\Psi $, that is, on
the whole system. As a consequence, its contribution can be neglected. Only
the first term remains in first approximation and it is equivalent to
constrain one of the derivatives in (\ref{mv}) to act on $\Psi ^{\dagger
}\left( \theta ,Z\right) $ and $\Psi \left( \theta ,Z\right) $ in the
integral. We will compute the graphs in this approximation and account
ultimately for the corrections due to the neglected terms.

We can picture picture the vertices (\ref{mv}) as a box cutting the lines $%
P_{k_{i}}$. The contributions associated to the segments between two
vertices are $2$ points Green functions $\mathcal{G}_{0}\left( \theta
^{\left( k_{i}\right) },\theta ^{\left( l_{i}\right) },Z_{k_{i}}\right) $.
We can transform all the vertices of valence $2l\leqslant 2n$ as $2n$ points
vertices. To do so, we define for$\ \left\{ k_{1},...,k_{l}\right\} \equiv
\left\{ k_{i}\right\} _{l}\subset \left\{ 1,...,n\right\} $ the $2n$ vertex: 
\begin{equation*}
\left[ \hat{V}_{2n}^{\left\{ k_{i}\right\} _{l}}\left( \left\{ \left( \theta
^{\left( m\right) },Z_{m}\right) \right\} _{m=1,...,n}\right) \right] =\hat{V%
}_{2l}\left( \left\{ \left( \theta ^{\left( k_{i}\right) },Z_{k_{i}}\right)
\right\} _{i=1,...,l}\right) \otimes \left( \mathcal{G}_{0}^{-1}\right)
^{\otimes \overline{\left\{ k_{i}\right\} _{i=1,...,l}}}
\end{equation*}%
where $\overline{\left\{ k_{i}\right\} _{i=1,...,l}}$ is the complement of $%
\left\{ k_{i}\right\} _{i=1,...,l}$ in $\left\{ 1,...,n\right\} $. The
operators $\mathcal{G}_{0}^{-1}$ are local and depend on two variables:%
\begin{equation*}
\mathcal{G}_{0}^{-1}\equiv \mathcal{G}_{0}^{-1}\left( \theta ^{\left(
l_{i}\right) },Z_{l_{i}}\right)
\end{equation*}%
Then:%
\begin{equation*}
\left( \mathcal{G}_{0}^{-1}\right) ^{\otimes \overline{\left\{ k_{i}\right\}
_{i=1,...,l}}}=\dprod\limits_{l_{i}\subset \overline{\left\{ k_{i}\right\}
_{i=1,...,l}}}\mathcal{G}_{0}^{-1}\left( \theta ^{\left( l_{i}\right)
},Z_{l_{i}}\right)
\end{equation*}

The vertex $\left[ \hat{V}_{2n}^{\left\{ k_{i}\right\} _{l}}\left( \left\{
\left( \theta ^{\left( m\right) },Z_{m}\right) \right\} _{m=1,...,n}\right) %
\right] $ is represented by a box cutting all the lines $P_{i}$. For the
variables $\left\{ \left( \theta ^{\left( k_{i}\right) },Z_{k_{i}}\right)
\right\} _{i=1,...,l}$ the propagators on each side of the box are
convoluted with $\hat{V}_{2l}\left( \left\{ \left( \theta ^{\left(
k_{i}\right) },Z_{k_{i}}\right) \right\} _{i=1,...,l}\right) $. For the
other variables, pairs of propagators are convoluted with their inverse,
producing a single propagator, as needed. In the sequel, we write:%
\begin{equation*}
\left[ \hat{V}_{2n}^{\left\{ k_{i}\right\} _{l}}\left( \left\{ \left( \theta
^{\left( m\right) },Z_{m}\right) \right\} \right) \right] \equiv \left[ \hat{%
V}_{2n}^{\left\{ k_{i}\right\} _{l}}\left( \left\{ \left( \theta ^{\left(
m\right) },Z_{m}\right) \right\} _{m=1,...,n}\right) \right]
\end{equation*}%
and $m$ runs implicitly from $1$ to $n$.

As a consequence, the contribution of a graph made of an arbitrary sequence
of vertices is:%
\begin{eqnarray}
\left[ \mathcal{G}_{0}^{\otimes n}\left( \left( \theta ^{\left( m_{0}\right)
},\theta ^{\left( m_{1}\right) },Z_{m_{0}}\right) \right) \right] \ast &&%
\left[ \hat{V}_{2n}^{\left\{ k_{i}^{1}\right\} _{l_{1}}}\left( \left\{
\left( \theta ^{\left( m_{1}\right) },Z_{m_{1}}\right) \right\} \right) %
\right] \ast \left[ \mathcal{G}_{0}^{\otimes n}\left( \left( \theta ^{\left(
m_{1}\right) },\theta ^{\left( m_{2}\right) },Z_{m_{1}}\right) \right) %
\right]  \label{gx} \\
&&\ast ...\ast \left[ \hat{V}_{2n}^{\left\{ k_{i}^{p}\right\}
_{l_{p}}}\left( \left\{ \left( \theta ^{\left( m_{k-1}\right)
},Z_{m_{k-1}}\right) \right\} \right) \right] \ast \left[ \mathcal{G}%
_{0}^{\otimes n}\left( \left( \theta ^{\left( m_{p-1}\right) },\theta
^{\left( m_{p}\right) },Z_{m_{p}}\right) \right) \right]  \notag
\end{eqnarray}%
with the constraint that $\theta ^{\left( m_{0}\right) }=\theta _{f}^{\left(
m_{0}\right) }$, $\theta ^{\left( m_{p}\right) }=\theta _{i}^{\left(
m_{p}\right) }$, and $\left\{ Z_{m_{i}}\right\} =\left\{ Z_{m}\right\} $ are
fixed. The $2n$ points propagators are defined by the product of individual
propagators:%
\begin{equation*}
\mathcal{G}_{0}^{\otimes n}\left( \left( \theta ^{\left( m\right) },\theta
^{\left( m^{\prime }\right) },Z_{m}\right) \right) \equiv
\dprod\limits_{m=1}^{n}\mathcal{G}_{0}\left( \theta ^{\left( m\right)
},\theta ^{\left( m^{\prime }\right) },Z_{m}\right)
\end{equation*}

The sum over all possible vertices is then:%
\begin{eqnarray}
&&\sum_{p}\frac{1}{p!}\left[ \mathcal{G}_{0}^{\otimes n}\left( \left( \theta
^{\left( m_{0}\right) },\theta ^{\left( m_{1}\right) },Z_{m_{0}}\right)
\right) \right]  \label{sgx} \\
&&\times \left( \sum_{l}\int \sum_{\left\{ k_{1},...,k_{l}\right\} \subset
\left\{ 1,...,n\right\} }\left[ \hat{V}_{2n}^{\left\{ k_{i}\right\}
_{l}}\left( \left\{ \left( \theta ^{\left( m_{l}\right) },Z_{m_{l}}\right)
\right\} \right) \right] \ast \left[ \mathcal{G}_{0}^{\otimes n}\left(
\left( \theta ^{\left( m_{l}\right) },\theta ^{\left( m_{l}^{\prime }\right)
},Z_{k_{i}}\right) \right) \right] \dprod d\theta ^{\left( m_{l}\right)
}\right) ^{p}  \notag
\end{eqnarray}%
The $p!$ arises to avoid counting of equivalent graphs. The power is
understood as successive convolutions.

As a consequence, the sum of graphs rewrites:%
\begin{equation}
\mathcal{G}_{0}^{\otimes n}\ast \exp \left( \sum_{l=1}^{n}\sum_{\left\{
k_{1},...,k_{l}\right\} \subset \left\{ 1,...,n\right\} }\left[ \hat{V}%
_{2n}^{\left\{ k_{i}\right\} _{l}}\ast \mathcal{G}_{0}^{\otimes n}\right]
\right) =\mathcal{G}_{0}^{\otimes n}\ast \exp \left( \left[ \hat{V}_{2n}\ast 
\mathcal{G}_{0}^{\otimes n}\right] \right)  \label{gm}
\end{equation}%
where $\left[ \hat{V}_{2n}^{\left\{ k_{i}\right\} _{l}}\ast \mathcal{G}%
_{0}^{\otimes n}\right] $ is the operator with kernel:%
\begin{equation*}
\left[ \hat{V}_{2n}^{\left\{ k_{i}\right\} _{l}}\left( \left\{ \left( \theta
^{\left( m_{l}\right) },Z_{m_{l}}\right) \right\} \right) \right] \ast \left[
\mathcal{G}_{0}\left( \left( \theta ^{\left( m_{l}\right) },\theta ^{\left(
m_{l}^{\prime }\right) },Z_{k_{i}}\right) \right) \right]
\end{equation*}%
and where we defined:%
\begin{equation*}
\left[ \hat{V}_{2n}\right] =\sum_{l=1}^{n}\sum_{\left\{
k_{1},...,k_{l}\right\} \subset \left\{ 1,...,n\right\} }\left[ \hat{V}%
_{2n}^{\left\{ k_{i}\right\} _{l}}\right]
\end{equation*}%
Alternatively it is also given by:%
\begin{equation}
\exp \left( \sum_{l=1}^{n}\sum_{\left\{ k_{1},...,k_{l}\right\} \subset
\left\{ 1,...,n\right\} }\left[ \mathcal{G}_{0}^{\otimes n}\ast \hat{V}%
_{2n}^{\left\{ k_{i}\right\} _{l}}\right] \right) \ast \mathcal{G}%
_{0}^{\otimes n}=\exp \left( \left[ \hat{V}_{2n}\ast \mathcal{G}%
_{0}^{\otimes n}\right] \right) \ast \mathcal{G}_{0}^{\otimes n}  \label{mg1}
\end{equation}%
We expand this formula in the next paragraph to show the convergence of the
graph expansion. In turn, this proves the convergence of the one-particule
irreducible graphs ($1$PI graphs) series expansion. The series expansion of $%
1$PI graphs compute the effective action. Its precise form will be obtained
by an other method in appendix 3.

\subsection*{2.2 Convergence of the series expansion}

\subsubsection*{2.2.1 Expression of vertices arising in (\protect\ref{gm})}

The series expansion of (\ref{gm}) has to be computed using the Wick theorem
on the terms (\ref{gx}). Such terms are computed by inserting vertices (\ref%
{vh}) between propagators. These ones have the form between $n$ pairs $%
\left( \theta _{1}^{\left( i\right) },\theta _{2}^{\left( i\right)
},Z_{i}\right) _{i=1,...,n}$:%
\begin{equation}
\mathcal{G}_{0}\left( \left( \theta _{1}^{\left( i\right) },\theta
_{2}^{\left( i\right) },Z_{i}\right) \right) =\frac{\exp \left( -\Lambda
_{1}\left( \sum_{i=1}^{n}\theta _{1}^{\left( i\right) }-\sum_{i=1}^{n}\theta
_{2}^{\left( i\right) }\right) \right) }{\Lambda }  \label{ngr}
\end{equation}%
with:%
\begin{eqnarray*}
\Lambda &=&\sqrt{\frac{\pi }{2}}\sqrt{\left( \frac{1}{\sigma ^{2}\bar{X}_{r}}%
\right) ^{2}+\frac{2\alpha }{\sigma ^{2}}} \\
\Lambda _{1} &=&\sqrt{\left( \frac{1}{\sigma ^{2}\bar{X}_{r}}\right) ^{2}+%
\frac{2\alpha }{\sigma ^{2}}}-\frac{1}{\sigma ^{2}\bar{X}_{r}}
\end{eqnarray*}%
As a consequence, the convolution of $m$ propagators leads to a global
factor:%
\begin{equation*}
\frac{\exp \left( -\Lambda _{1}\left( \sum_{i=1}^{n}\theta _{f}^{\left(
i\right) }-\sum_{i=1}^{n}\theta _{i}^{\left( i\right) }\right) \right) }{%
\Lambda ^{m}}
\end{equation*}%
The power $m$ is given by the total number of vertices in the graph, each of
them weighted by its valence. The vertices induce multiplicative factor at
some times $\theta ^{\left( i\right) }$ and integrations are performed over $%
\theta ^{\left( i\right) }$. The presence of derivatives in the vertices
induce terms of the form:%
\begin{equation*}
\frac{1}{2}\overline{\Psi ^{\dagger }\left( \theta ,Z\right) \nabla _{\theta
}V\left( \theta ,Z\right) \Psi \left( \theta ,Z\right) }
\end{equation*}%
where $V\left( \theta ,Z\right) $ is any type of vertex and the upper bar
denotes the contraction through Wick's theorem. It is equivalent to replace
this term by:%
\begin{eqnarray*}
\frac{1}{2}\nabla _{\theta }\left( V\left( \theta ,Z\right) \mathcal{G}%
_{0}\left( \left( \theta ,\theta ,Z\right) \right) \right) &=&\mathcal{G}%
_{0}\left( \left( \theta ,\theta ,Z\right) \right) \nabla _{\theta }V\left(
\theta ,Z\right) \mathcal{-}\Lambda _{1}\mathcal{G}_{0}\left( \left( \theta
,\theta ,Z\right) \right) V\left( \theta ,Z\right) \\
&=&\frac{1}{\Lambda }\left( \nabla _{\theta }V\left( \theta ,Z\right) 
\mathcal{-}\Lambda _{1}V\left( \theta ,Z\right) \right)
\end{eqnarray*}%
given (\ref{ngr}). This means that in a sequence of propagators defining a
graph, the vertex $\nabla _{\theta }V\left( \theta ,Z\right) $ can be
replaced by:%
\begin{eqnarray*}
&&\mathcal{G}_{0}^{-1}\left( \left( \theta ,\theta ,Z\right) \right) \left(
\nabla _{\theta }\left( V\left( \theta ,Z\right) \left( \mathcal{G}%
_{0}\left( \theta _{1},\theta ,Z\right) \right) \right) \right) _{\theta
_{1}=\theta } \\
&\equiv &\mathcal{G}_{0}^{-1}\left( \left( 0,Z\right) \right) \left( \nabla
_{\theta }\left( V\left( \theta ,Z\right) \mathcal{G}_{0}\left(
0^{+},Z\right) \right) \right)
\end{eqnarray*}%
It implies that when a vertex is inserted at the left of a propagator, it
can be replaced by:%
\begin{eqnarray}
&&\hat{V}_{2l}\left( \left\{ \left( \theta ^{\left( k_{i}\right)
},Z_{k_{i}}\right) \right\} _{i=1,...,l}\right)  \label{ldv} \\
&=&\left[ \frac{\delta ^{l}\left[ \int \mathcal{G}_{0}^{-1}\left( \left(
0,Z\right) \right) \Psi ^{\dagger }\left( \theta ,Z\right) \nabla _{\theta
}\omega ^{-1}\left( J,\theta ,Z\right) \mathcal{G}_{0}\left( 0^{+},Z\right)
\Psi \left( \theta ,Z\right) dZd\theta +V\left( \Psi \right) \right] }{%
l!\dprod\limits_{i=1}^{l}\delta \left\vert \Psi \left( \theta ^{\left(
k_{i}\right) },Z_{k_{i}}\right) \right\vert ^{2}}\right] _{\substack{ %
\left\vert \Psi \left( \theta ,Z\right) \right\vert ^{2}  \\ =\mathcal{G}%
_{0}\left( 0,Z\right) }}  \notag \\
&\equiv &\left[ \frac{\delta ^{l}\left[ \int \Psi ^{\dagger }\left( \theta
,Z\right) \nabla _{\theta }\omega ^{-1}\left( J,\theta ,Z\right) \Psi \left(
\theta ,Z\right) dZd\theta +V\left( \Psi \right) \right] _{\mathcal{G}_{0}}}{%
l!\dprod\limits_{i=1}^{l}\delta \left\vert \Psi \left( \theta ^{\left(
k_{i}\right) },Z_{k_{i}}\right) \right\vert ^{2}}\right] _{\left\vert \Psi
\left( \theta ,Z\right) \right\vert ^{2}=\mathcal{G}_{0}\left( 0,Z\right) } 
\notag
\end{eqnarray}

To compute the graphs series, the vertices (\ref{ldv}) have to be expanded
by taking into account the form of the potential $V$.

\subsubsection*{2.2.2 Expanded form of the vertices}

As presented in the text, the potential for maintaining and activating new
connections is chosen to be equal to:%
\begin{eqnarray}
V\left( \Psi \right) &=&-\frac{\zeta _{1}}{2}\int \left( \left\vert \Psi
\left( \theta ,Z\right) \right\vert ^{2}\left\vert \Psi \left( \theta -\frac{%
\left\vert Z-Z^{\prime }\right\vert }{c},Z^{\prime }\right) \right\vert
^{2}\right)  \notag \\
&&+\sum_{n=1}^{\infty }\frac{\zeta _{n}}{2n!}\int \left\vert \Psi \left(
\theta ,Z\right) \right\vert ^{2}\left( \dprod\limits_{i=1}^{n-1}\left\vert
\Psi \left( \theta -\frac{\left\vert Z-Z_{i}\right\vert }{c},Z_{i}\right)
\right\vert ^{2}\right) \\
&=&\sum_{n=2}^{\infty }\frac{1}{2n!}\zeta ^{\left( n\right) }\int \left\vert
\Psi \left( \theta ,Z\right) \right\vert ^{2}\left(
\dprod\limits_{i=1}^{n-1}\left\vert \Psi \left( \theta -\frac{\left\vert
Z-Z_{i}\right\vert }{c},Z_{i}\right) \right\vert ^{2}\right)  \notag
\end{eqnarray}%
with:%
\begin{eqnarray*}
\zeta ^{\left( l\right) } &=&\zeta _{l}\text{, }l>2 \\
\zeta ^{\left( 2\right) } &=&\zeta _{2}-\zeta _{1} \\
\zeta ^{\left( 1\right) } &=&0
\end{eqnarray*}%
The second term represents the limitation in increasing the number of
connections. This amounts to shift the vertices by $+\zeta _{2}$. The factor 
$-\zeta _{1}$ accounts for a minimal number of connections maintained. It
depends on external activity $J$.

The first term modifies the $4$-th vertices by $-\zeta _{1}$. The vertices
involved in the $2n$ points correlation function are given by an expansion
of $V\left( \Psi \right) $ around $\mathcal{G}_{0}\left( 0,Z\right) $. The
first order expansion in $\left\vert \Psi \left( \theta ,Z\right)
\right\vert ^{2}$ modifies the $2$ points propagator by replacing $\alpha $
with:%
\begin{equation*}
\alpha +\sum_{k\geqslant 2}\frac{1}{k!}C_{k}^{1}\frac{\zeta ^{\left(
k\right) }}{\Lambda ^{k-1}}=\sum_{k\geqslant l}\frac{1}{\left( k-1\right) !}%
\frac{\zeta ^{\left( k\right) }}{\Lambda ^{k-1}}
\end{equation*}%
which modifies the values of $\Lambda $ and $\Lambda _{1}$.

For $n\geqslant 2$, the $2n$ vertex is then: 
\begin{eqnarray}
\hat{V}_{2l}\left( \left( \theta ^{\left( i\right) },Z_{i}\right)
_{i=1,...,n}\right) &=&\frac{1}{2\left( l\right) !}\left[ \frac{\delta ^{l}%
\left[ \int \Psi ^{\dagger }\left( \theta ,Z\right) \nabla _{\theta }\omega
^{-1}\left( J,\theta ,Z\right) \Psi \left( \theta ,Z\right) dZd\theta \right]
}{\dprod\limits_{i=1}^{l}\delta \left\vert \Psi \left( \theta ^{\left(
i\right) },Z_{i}\right) \right\vert ^{2}}\right] _{\substack{ \left\vert
\Psi \left( \theta ,Z\right) \right\vert ^{2}  \\ =\mathcal{G}_{0}\left(
0,Z\right) }}  \label{vh} \\
&&-\left[ \frac{\delta ^{l}\left( \sum_{k=l}^{\infty }\frac{1}{2k!}\zeta
^{\left( k\right) }\int \left\vert \Psi \left( \theta ,Z\right) \right\vert
^{2}\left( \dprod\limits_{i=1}^{k-1}\left\vert \Psi \left( \theta -\frac{%
\left\vert Z-Z_{i}\right\vert }{c},Z_{i}\right) \right\vert ^{2}\right)
\right) }{\dprod\limits_{j=1}^{l}\delta \left\vert \Psi \left( \theta
_{j},Z_{j}\right) \right\vert ^{2}}\right] _{\substack{ \left\vert \Psi
\left( \theta _{j},Z_{j}\right) \right\vert ^{2}  \\ =\mathcal{G}_{0}\left(
0,Z_{j}\right) }}  \notag
\end{eqnarray}%
To do so, we decompose (\ref{vh}) in two types of vertices for $n$ given
points $\left( \theta ^{\left( 1\right) },Z_{1}\right) $,...,$\left( \theta
^{\left( n\right) },Z_{n}\right) $:%
\begin{eqnarray*}
&&\hat{V}_{2l}^{\left( 1\right) }\left( \left( \theta ^{\left( i\right)
},Z_{i}\right) ,\left\{ Z_{j},\theta ^{\left( j\right) }\right\} _{j\neq
i}\right) \\
&=&\frac{1}{2\left( l\right) !}\sum_{i=1}^{l}\left[ \frac{\left[ \int \Psi
^{\dagger }\left( \theta ^{\left( i\right) },Z_{i}\right) \delta ^{l-1}\left[
\nabla _{\theta }\omega ^{-1}\left( J,\theta ^{\left( i\right)
},Z_{i}\right) \right] \Psi \left( \theta ^{\left( i\right) },Z_{i}\right) %
\right] _{\mathcal{G}_{0}}}{\dprod\limits_{j=1,j\neq i}^{l}\delta \left\vert
\Psi \left( \theta ^{\left( j\right) },Z_{j}\right) \right\vert ^{2}}\right]
_{\left\vert \Psi \left( \theta _{j},Z_{j}\right) \right\vert ^{2}=\mathcal{G%
}_{0}\left( 0,Z_{j}\right) } \\
&&-\frac{1}{2\left( l\right) !}\sum_{i=1}^{l}\left[ \frac{\delta ^{l-1}\left[
\sum_{k=l}^{\infty }\frac{1}{k!}\zeta ^{\left( k\right) }\left(
\dprod\limits_{i=1}^{k-1}\left\vert \Psi \left( \theta -\frac{\left\vert
Z-Z_{i}\right\vert }{c},Z_{i}\right) \right\vert ^{2}\right) \right] }{%
\dprod\limits_{j=1,j\neq i}^{l}\delta \left\vert \Psi \left( \theta ^{\left(
j\right) },Z_{j}\right) \right\vert ^{2}}\right] _{\left\vert \Psi \left(
\theta _{j},Z_{j}\right) \right\vert ^{2}=\mathcal{G}_{0}\left(
0,Z_{j}\right) }
\end{eqnarray*}%
and:%
\begin{eqnarray}
\hat{V}_{2l}^{\left( 2\right) }\left( \left( \theta ^{\left( i\right)
},Z_{i}\right) _{i=1,...,n}\right) &=&\frac{1}{2\left( l\right) !}\left[ 
\frac{\left[ \int \Psi ^{\dagger }\left( \theta ,Z\right) \nabla _{\theta
}\delta ^{l}\left[ \omega ^{-1}\left( J,\theta ,Z\right) \right] \Psi \left(
\theta ,Z\right) dZd\theta \right] _{\mathcal{G}_{0}}}{\dprod%
\limits_{i=1}^{l}\delta \left\vert \Psi \left( \theta ^{\left( i\right)
},Z_{i}\right) \right\vert ^{2}}\right] _{\left\vert \Psi \left( \theta
,Z\right) \right\vert ^{2}=\mathcal{G}_{0}\left( 0,Z\right) } \\
&&-\left[ \int \mathcal{G}_{0}\left( 0,Z\right) dZd\theta \frac{\delta ^{l}%
\left[ \sum_{k=l}^{\infty }\frac{1}{2k!}\zeta ^{\left( k\right) }\left(
\dprod\limits_{i=1}^{k-1}\left\vert \Psi \left( \theta -\frac{\left\vert
Z-Z_{i}\right\vert }{c},Z_{i}\right) \right\vert ^{2}\right) \right] }{%
\dprod\limits_{j=1}^{l}\delta \left\vert \Psi \left( \theta
_{j},Z_{j}\right) \right\vert ^{2}}\right] _{\substack{ \left\vert \Psi
\left( \theta _{j},Z_{j}\right) \right\vert ^{2}  \\ =\mathcal{G}_{0}\left(
0,Z_{j}\right) }}  \notag
\end{eqnarray}

As explained before, the index $\mathcal{G}_{0}$ denotes the contraction
between the field on the left with the one on the right when the vertex has
a propagator on its right. It is equivalent to introduce $\mathcal{G}%
_{0}\left( 0,Z\right) $ inside the gradient, to remove $\Psi ^{\dagger
}\left( \theta ,Z\right) $ and $\Psi \left( \theta ,Z\right) $ and multiply
by $\mathcal{G}_{0}^{-1}\left( 0,Z\right) $.

To complete the computation of the vertices $\hat{V}_{2l}^{\left( 1\right) }$
and $\hat{V}_{2l}^{\left( 2\right) }$ we can regroup the terms involving the
coefficients of the potential $V$. To do so, we simplify the derivatives of
the potential by writing:%
\begin{eqnarray*}
&\sum_{i=1}^{l}&\left[ \sum_{k=l}^{\infty }\frac{1}{2k!}\zeta ^{\left(
k\right) }\frac{\delta ^{l-1}}{\dprod\limits_{j=1,j\neq i}^{l}\delta
\left\vert \Psi \left( \theta _{j},Z_{j}\right) \right\vert ^{2}}\left(
\dprod\limits_{i=1}^{k-1}\left\vert \Psi \left( \theta -\frac{\left\vert
Z-Z_{i}\right\vert }{c},Z_{i}\right) \right\vert ^{2}\right) \right] 
_{\substack{ \left\vert \Psi \left( \theta _{j},Z_{j}\right) \right\vert
^{2}  \\ =\mathcal{G}_{0}\left( 0,Z_{j}\right) }} \\
&&+\left[ \sum_{k=l}^{\infty }\frac{1}{2k!}\zeta ^{\left( k\right) }\int 
\mathcal{G}_{0}\left( 0,Z\right) dZd\theta \frac{\delta ^{l}\left[ \left(
\dprod\limits_{i=1}^{k-1}\left\vert \Psi \left( \theta -\frac{\left\vert
Z-Z_{i}\right\vert }{c},Z_{i}\right) \right\vert ^{2}\right) \right] }{%
\dprod\limits_{j=1}^{l}\delta \left\vert \Psi \left( \theta
_{j},Z_{j}\right) \right\vert ^{2}}\right] _{\substack{ \left\vert \Psi
\left( \theta _{j},Z_{j}\right) \right\vert ^{2}  \\ =\mathcal{G}_{0}\left(
0,Z_{j}\right) }} \\
&=&\sum_{i=1}^{l}\sum_{k=l}^{\infty }\frac{1}{2k!}C_{k-1}^{l-1}\frac{\zeta
^{\left( k\right) }}{\Lambda ^{k-l}}\dprod\limits_{j=1,j\neq i}^{l}\delta
\left( \theta _{i}-\frac{\left\vert Z-Z_{i}\right\vert }{c}-\theta
_{j}\right) \\
&&+\sum_{k=l+1}^{\infty }\frac{1}{2k!}\zeta ^{\left( k\right) }\int \mathcal{%
G}_{0}\left( \theta ,\theta ,Z\right)
C_{k-1}^{l}\dprod\limits_{i=1}^{l}\delta \left( \theta -\frac{\left\vert
Z-Z_{i}\right\vert }{c}-\theta _{i}\right) \\
&=&\frac{1}{2\left( l-1\right) !}\sum_{k=l}^{\infty }\frac{1}{k\left(
k-l\right) !}\frac{\zeta ^{\left( k\right) }}{\Lambda ^{k-l}}%
\dprod\limits_{j=1,j\neq i}^{l}\delta \left( \theta _{i}-\frac{\left\vert
Z-Z_{i}\right\vert }{c}-\theta _{j}\right) \\
&&+\frac{1}{2l!}\sum_{k=l}^{\infty }\int \frac{1}{k\left( k-l-1\right) !}%
\frac{\zeta ^{\left( k\right) }}{\Lambda ^{k-l-1}}\dprod\limits_{i=1}^{l}%
\delta \left( \theta -\frac{\left\vert Z-Z_{i}\right\vert }{c}-\theta
_{i}\right) \\
&=&\frac{1}{2\left( l-1\right) !}\zeta _{e}^{\left( l\right)
}\dprod\limits_{j=1,j\neq i}^{l}\delta \left( \theta _{i}-\frac{\left\vert
Z-Z_{i}\right\vert }{c}-\theta _{j}\right) +\frac{1}{2l!}\int \zeta
_{e}^{\left( l+1\right) }\dprod\limits_{i=1}^{l}\delta \left( \theta -\frac{%
\left\vert Z-Z_{i}\right\vert }{c}-\theta _{i}\right)
\end{eqnarray*}%
with:%
\begin{equation}
\zeta _{e}^{\left( l\right) }=\sum_{k\geqslant l}\frac{1}{k\left( k-l\right)
!}\frac{\zeta ^{\left( k\right) }}{\Lambda ^{k-l}}  \label{ftc}
\end{equation}%
For $\zeta ^{\left( k\right) }$ slowly varying and $\Lambda >>1$, this is
approximatively equal to $\zeta ^{\left( l\right) }$. We keep the notation $%
\zeta _{e}^{\left( l\right) }\rightarrow \zeta ^{\left( l\right) }$.

As a consequence, $\hat{V}_{2l}^{\left( 1\right) }$ and $\hat{V}%
_{2l}^{\left( 2\right) }$ write: 
\begin{eqnarray}
&&\sum_{i=1}^{n}\hat{V}_{2l}^{\left( 1\right) }\left( \left( \theta ^{\left(
i\right) },Z_{i}\right) ,\left\{ Z_{j},\theta ^{\left( j\right) }\right\}
_{j\neq i}\right)  \label{lv} \\
&=&\sum_{i=1}^{n}\left[ \frac{1}{2\left( l\right) !}\frac{\delta ^{l-1}\left[
\nabla _{\theta }\omega ^{-1}\left( J,\theta ^{\left( i\right)
},Z_{i}\right) \mathcal{G}_{0}\left( 0,Z_{i}\right) \right] }{\mathcal{G}%
_{0}\left( 0,Z_{i}\right) \dprod\limits_{j=1,j\neq i}^{l}\delta \left\vert
\Psi \left( \theta ^{\left( j\right) },Z_{j}\right) \right\vert ^{2}}-\frac{1%
}{2\left( l-1\right) !}\zeta ^{\left( l\right) }\dprod\limits_{j=1,j\neq
i}^{l}\delta \left( \theta _{i}-\frac{\left\vert Z-Z_{i}\right\vert }{c}%
-\theta _{j}\right) \right] _{\substack{ \left\vert \Psi \left( \theta
_{j},Z_{j}\right) \right\vert ^{2}  \\ =\mathcal{G}_{0}\left( 0,Z_{j}\right) 
}}  \notag
\end{eqnarray}

and:%
\begin{eqnarray}
&&\hat{V}_{2l}^{\left( 2\right) }\left( \left( \theta ^{\left( i\right)
},Z_{i}\right) _{i=1,...,n}\right)  \label{vl} \\
&=&\frac{1}{2\left( l\right) !}\left[ \int \nabla _{\theta }\frac{\delta ^{l}%
\left[ \omega ^{-1}\left( J,\theta ,Z\right) \right] }{\dprod%
\limits_{i=1}^{l}\delta \left\vert \Psi \left( \theta ^{\left( i\right)
},Z_{i}\right) \right\vert ^{2}}\mathcal{G}_{0}\left( 0,Z\right) dZd\theta
-\int \zeta ^{\left( l+1\right) }\dprod\limits_{i=1}^{l}\delta \left( \theta
-\frac{\left\vert Z-Z_{i}\right\vert }{c}-\theta _{i}\right) \right] 
_{\substack{ \left\vert \Psi \left( \theta ,Z\right) \right\vert ^{2}  \\ =%
\mathcal{G}_{0}\left( 0,Z\right) }}  \notag
\end{eqnarray}%
\bigskip

with $\zeta _{e}^{\left( l\right) }$ given by (\ref{ftc}). For $\zeta
^{\left( k\right) }$ slowly varying and $\Lambda >>1$, this is
approximatively equal to $\zeta ^{\left( l\right) }$. For the sake of
simplicity we keep the notation $\zeta _{e}^{\left( l\right) }\rightarrow
\zeta ^{\left( l\right) }$.

As explained before, the vertex $\hat{V}_{2l}^{\left( 2\right) }\left(
\left( \theta ^{\left( i\right) },Z_{i}\right) _{i=1,...,n}\right) $ can be
neglected with respect to $\hat{V}_{2l}^{\left( 1\right) }\left( \left(
\theta ^{\left( i\right) },Z_{i}\right) _{i=1,...,n}\right) $\ in first
approximation. We will compute the graphs associated to $\hat{V}%
_{2l}^{\left( 1\right) }\left( \left( \theta ^{\left( i\right)
},Z_{i}\right) _{i=1,...,n}\right) $ in the next paragraph, and will then
compute the contributions due to $\hat{V}_{2l}^{\left( 2\right) }\left(
\left( \theta ^{\left( i\right) },Z_{i}\right) _{i=1,...,n}\right) $ as
corrections.

Ultimately, and for later purpose, we also define:%
\begin{equation*}
\bar{\zeta}_{n}=\sum_{l=2}^{n}\sum_{\left\{ k_{1},...,k_{l-1}\right\}
\subset \left\{ 1,...,n-1\right\} ,k_{j}\neq i}\frac{\zeta ^{\left( l\right)
}}{\Lambda ^{l}}=\sum_{l=1}^{n}C_{n-1}^{l-1}\frac{\zeta ^{\left( l\right) }}{%
\Lambda ^{l}}
\end{equation*}%
For example:%
\begin{equation*}
\bar{\zeta}_{2}=\frac{\zeta ^{\left( 2\right) }}{\Lambda }\text{, }\bar{\zeta%
}_{3}=\frac{\zeta ^{\left( 3\right) }}{\Lambda ^{2}}+3\frac{\zeta ^{\left(
2\right) }}{\Lambda }
\end{equation*}%
If we express $\bar{\zeta}_{n}$ as a function of the initial set of
variables $\zeta ^{\left( l\right) }$, we have: 
\begin{equation*}
\bar{\zeta}_{n}=\sum_{l=1}^{n}C_{n}^{l}\frac{\sum_{k\geqslant l}\frac{1}{%
\left( k-l\right) !}\frac{\zeta ^{\left( k\right) }}{\Lambda ^{k-l}}}{%
\Lambda ^{l-1}}=\sum_{l=1}^{n}C_{n}^{l}\sum_{k\geqslant l}\frac{1}{\left(
k-l\right) !}\frac{\zeta ^{\left( k\right) }}{\Lambda ^{k-1}}
\end{equation*}%
so that: 
\begin{equation*}
\bar{\zeta}_{2}=\sum_{k\geqslant 2}\frac{1}{\left( k-2\right) !}\frac{\zeta
^{\left( k\right) }}{\Lambda ^{k-1}}=\frac{\zeta ^{\left( 2\right) }}{%
\Lambda }+\sum_{k\geqslant 3}\frac{1}{\left( k-2\right) !}\frac{\zeta
^{\left( k\right) }}{\Lambda ^{k-1}}
\end{equation*}%
and:%
\begin{eqnarray*}
\bar{\zeta}_{3} &=&\sum_{k\geqslant 3}\frac{1}{\left( k-3\right) !}\frac{%
\zeta ^{\left( k\right) }}{\Lambda ^{k-1}}+\sum_{k\geqslant 2}\frac{3}{%
\left( k-2\right) !}\frac{\zeta ^{\left( k\right) }}{\Lambda ^{k-1}} \\
&=&3\frac{\zeta ^{\left( 2\right) }}{\Lambda }+3\sum_{k\geqslant 3}\frac{1}{%
\left( k-2\right) !}\frac{\zeta ^{\left( k\right) }}{\Lambda ^{k-1}}%
+\sum_{k\geqslant 3}\frac{1}{\left( k-3\right) !}\frac{\zeta ^{\left(
k\right) }}{\Lambda ^{k-1}} \\
&=&3\frac{\zeta ^{\left( 2\right) }}{\Lambda }+3\sum_{k\geqslant 3}\frac{1}{%
\left( k-2\right) !}\frac{\zeta ^{\left( k\right) }}{\Lambda ^{k-1}}%
+\sum_{k\geqslant 3}\frac{Sup\left( 1,\left( k-3\right) \right) \zeta
^{\left( k\right) }}{\left( k-2\right) !\Lambda ^{k-1}}
\end{eqnarray*}%
We will assume that $\bar{\zeta}_{2}<0$ and $\bar{\zeta}_{n}>0$ for $n>2$.
This is possible under the conditions:%
\begin{equation*}
\sum_{k\geqslant 3}\frac{1}{\left( k-2\right) !}\frac{\zeta ^{\left(
k\right) }}{\Lambda ^{k-1}}<\frac{\left\vert \zeta ^{\left( 2\right)
}\right\vert }{\Lambda }<\sum_{k\geqslant 3}\frac{1}{\left( k-2\right) !}%
\frac{\zeta ^{\left( k\right) }}{\Lambda ^{k-1}}+\sum_{k\geqslant 3}\frac{%
Sup\left( 1,\left( k-2\right) \right) \zeta ^{\left( k\right) }}{3\left(
k-2\right) !\Lambda ^{k-1}}
\end{equation*}%
that are satisfied for a certain range of the parameters, since: 
\begin{equation*}
\sum_{k\geqslant 3}\frac{Sup\left( 1,\left( k-2\right) \right) \zeta
^{\left( k\right) }}{3\left( k-2\right) !\Lambda ^{k-1}}-\sum_{k\geqslant 3}%
\frac{1}{\left( k-2\right) !}\frac{\zeta ^{\left( k\right) }}{\Lambda ^{k-1}}%
=\sum_{k\geqslant 3}\frac{\left( Sup\left( 1,\left( k-2\right) \right)
-3\right) \zeta ^{\left( k\right) }}{3\left( k-2\right) !\Lambda ^{k-1}}
\end{equation*}%
is positive for $\zeta ^{\left( k\right) }$ large enough for $k>5$.

\subsubsection*{2.2.3 Integral form of (\protect\ref{gm})}

The expansion of (\ref{gm}) can be performed using the previous results.
Each vertex of valence $v$ can be attributed a factor $\frac{1}{\Lambda ^{v}}
$. In (\ref{lv}) we also change the variables $\theta ^{\left( j\right)
}=\theta ^{\left( i\right) }-l_{i}$. We also define:%
\begin{eqnarray}
\bar{\Xi}_{1}^{\left( l\right) }\left( Z_{i},\theta ^{\left( i\right)
},\left\{ Z_{j}\right\} _{j\neq i}\right) &=&\frac{\sum_{\substack{ \left\{
k_{1},...,k_{l-1}\right\}  \\ \subset \left\{ 1,...,n-1\right\} }}%
\dprod\limits_{k_{j}}\int_{\frac{\left\vert Z_{i}-Z_{k_{j}}\right\vert }{c}%
}^{\theta ^{\left( i\right) }-\theta _{i}^{\left( k_{j}\right) }}dl_{k_{j}}}{%
2\Lambda ^{l}}  \label{lx} \\
&&\times \left[ \frac{\delta ^{l-1}\left[ \nabla _{\theta }\omega
^{-1}\left( J,\theta ^{\left( i\right) },Z_{i}\right) \mathcal{G}_{0}\left(
0,Z_{i}\right) \right] }{\mathcal{G}_{0}\left( 0,Z_{i}\right)
\dprod\limits_{j=1,j\neq i}^{l}\delta \left\vert \Psi \left( \theta ^{\left(
i\right) }-l_{k_{j}},Z_{k_{j}}\right) \right\vert ^{2}}\right] _{\left\vert
\Psi \left( \theta ,Z\right) \right\vert ^{2}=\mathcal{G}_{0}\left(
0,Z\right) }  \notag
\end{eqnarray}%
\begin{equation}
\hat{\Xi}_{1}^{\left( l\right) }\left( Z_{i},\theta ^{\left( i\right)
},\left\{ Z_{j}\right\} _{j\neq i}\right) =\bar{\Xi}_{1}^{\left( l\right)
}\left( Z_{i},\theta ^{\left( i\right) },\left\{ Z_{j}\right\} _{j\neq
i}\right) -\frac{\zeta ^{\left( l\right) }}{\Lambda ^{l}}  \label{xh}
\end{equation}

\begin{equation}
\bar{\Lambda}_{1}^{\left( l\right) }\left( Z,\theta ,\left\{ Z_{i},\theta
^{\left( i\right) }\right\} _{i=1,...l}\right) =\frac{1}{2\left( l\right)
!\Lambda ^{l}}\left[ \int \nabla _{\theta }\frac{\delta ^{l}\left[ \omega
^{-1}\left( J,\theta ,Z\right) \right] }{\dprod\limits_{i=1}^{l}\delta
\left\vert \Psi \left( \theta ^{\left( i\right) },Z_{i}\right) \right\vert
^{2}}\mathcal{G}_{0}\left( 0,Z\right) dZd\theta \right] _{\left\vert \Psi
\left( \theta ,Z\right) \right\vert ^{2}=\mathcal{G}_{0}\left( 0,Z\right) }
\label{dl}
\end{equation}%
and:%
\begin{equation*}
\hat{\Lambda}_{1}^{\left( l\right) }\left( Z,\theta ,\left\{ Z_{i},\theta
^{\left( i\right) }\right\} _{i=1,...l}\right) =\bar{\Lambda}_{1}^{\left(
l\right) }\left( Z,\theta ,\left\{ Z_{i},\theta ^{\left( i\right) }\right\}
_{i=1,...l}\right) -\frac{\zeta ^{\left( l+1\right) }}{\Lambda ^{l}}
\end{equation*}

The propagators induced by the vertices in (\ref{trct}) have been included
in the definition of $\bar{\Xi}_{1}^{\left( l\right) }\left( Z_{i},\theta
^{\left( i\right) },\left\{ Z_{j}\right\} _{j\neq i}\right) $. The functions
depend implicitely on the border of the timespans $\left[ \theta
_{f}^{\left( j\right) },\theta _{i}^{\left( j\right) }\right] $. Actually,
the integrations $\int_{\frac{\left\vert Z_{i}-Z_{k_{j}}\right\vert }{c}%
}^{\theta ^{\left( i\right) }-\theta _{i}^{\left( k_{j}\right) }}dl_{k_{j}}$
induces the presence of products of Heaviside functions $H\left( \theta
_{f}^{\left( i\right) }-\theta _{i}^{\left( k_{j}\right) }-\frac{\left\vert
Z_{i}-Z_{k_{j}}\right\vert }{c}\right) $.

We\ first compute the full sum of graphs arising from all combination of
vertices $\hat{V}_{2l}^{\left( 1\right) }$ defined in (\ref{lv}) between $n$
initial points and $n$ final points. The vertices $\hat{V}_{2l}^{\left(
2\right) }$ (see (\ref{vl})) will be included later. The vertices $\hat{V}%
_{2l}^{\left( 1\right) }\left( \left( \theta ^{\left( i\right)
},Z_{i}\right) ,\left\{ Z_{j},\theta ^{\left( j\right) }\right\} _{j\neq
i}\right) $ can be associated to an initial point $\left( \theta ^{\left(
i\right) },Z_{i}\right) $ with $l-1$ final points among the $n-1$ others.

The factors associated to the vertices in the expansion of (\ref{gm}) have
been found in the previous paragraph. We aasociate a global factor $\frac{%
\exp \left( -\Lambda _{1}\left( \sum_{i=1}^{n}\theta _{f}^{\left( i\right)
}-\sum_{i=1}^{n}\theta _{i}^{\left( i\right) }\right) \right) }{\Lambda ^{m}}
$ to a sequence of vertices. Then the insertion of vertices at $\left(
\theta ^{\left( i\right) },Z_{i}\right) $, the sum over the final points and
the integration over the $\left\{ Z_{j},\theta ^{\left( j\right) }\right\}
_{j\neq i}$ leads to a factor:%
\begin{equation*}
\frac{1}{\Lambda ^{l}}\sum_{\substack{ \left\{ k_{1},...,k_{l-1}\right\}  \\ %
\subset \left\{ 1,...,n-1\right\} }}\int \hat{V}_{2l}^{\left( 1\right)
}\left( \left( \theta ^{\left( i\right) },Z_{i}\right) ,\left\{ Z_{j},\theta
^{\left( k_{j}\right) }\right\} _{k_{j}\neq i}\right) d\theta ^{\left(
k_{j}\right) }=\hat{\Xi}_{1}^{\left( l\right) }\left( Z_{i},\theta ^{\left(
i\right) },\left\{ Z_{j}\right\} _{j\neq i}\right)
\end{equation*}%
For each line associated to $Z_{i}$, the insertion of $\sum_{l=1}^{n}k_{l}^{%
\left( i\right) }$ vertices where $k_{l}^{\left( i\right) }$ vertices have
valence $l$ implies the integration over $\theta _{i}^{\left( i\right)
}<\theta _{1}^{\left( i\right) }<...\theta _{\sum_{l=1}^{n}k_{l}^{\left(
i\right) }}^{\left( i\right) }<\theta _{f}^{\left( i\right) }$ \ of the
product of terms $\hat{\Xi}_{1}^{\left( l_{q}\right) }\left( Z_{i},\theta
^{\left( i\right) },\left\{ Z_{j}\right\} _{j\neq i}\right) $ for $q=1$ to $%
\sum_{l=1}^{n}k_{l}^{\left( i\right) }$. The number of $l_{q}$ equal to $l$
is $k_{l}^{\left( i\right) }$. Once an order $l_{1},l_{2},...$ is chosen,
there are $\dprod k_{l}^{\left( i\right) }!$ ways to order the vertices
satisfying this order. Then, summing over the various orders $%
l_{1},l_{2},... $ and over the $k_{l}^{\left( i\right) }$ such that $%
\sum_{l=1}^{n}k_{l}^{\left( i\right) }=m$ is fixed, the global factor
associated to the vertices is: 
\begin{equation}
\int_{\theta _{i}^{\left( i\right) }<\theta _{1}^{\left( i\right)
}<...\theta _{m}^{\left( i\right) }<\theta _{f}^{\left( i\right)
}}\dprod\limits_{q=1}^{m}\left( \sum_{l=2}^{n}\hat{\Xi}_{1}^{\left(
l_{q}\right) }\left( Z_{i},\theta _{q}^{\left( i\right) },\left\{
Z_{j}\right\} _{j\neq i}\right) \right) \delta \left( \theta _{1}^{\left(
i\right) }-\theta _{i}^{\left( i\right) }\right) \delta \left( \theta
_{m}^{\left( i\right) }-\theta _{f}^{\left( i\right) }\right) d\theta
_{q}^{\left( i\right) }  \label{wtl}
\end{equation}%
The delta functions accounts for the fact that without external legs, two
vertices are set at the borders of the interval. If we approximate $\sum_{l}%
\hat{\Xi}_{1}^{\left( l\right) }\left( Z_{i},\theta _{q}^{\left( i\right)
},\left\{ Z_{j}\right\} _{j\neq i}\right) $ by its average on the interval $%
\left[ \theta _{i}^{\left( i\right) },\theta _{f}^{\left( i\right) }\right] $%
, that is: 
\begin{equation*}
\frac{\int_{\theta _{i}^{\left( i\right) }}^{\theta _{f}^{\left( i\right)
}}\sum_{l=2}^{n}\hat{\Xi}_{1}^{\left( l\right) }\left( Z_{i},\theta
_{q}^{\left( i\right) },\left\{ Z_{j}\right\} _{j\neq i}\right) d\theta
_{q}^{\left( i\right) }}{\theta _{f}^{\left( i\right) }-\theta _{i}^{\left(
i\right) }}
\end{equation*}%
As a consequence (\ref{wtl}) becomes:%
\begin{eqnarray*}
&&\left( \frac{\int_{\theta _{i}^{\left( i\right) }}^{\theta _{f}^{\left(
i\right) }}\sum_{l=2}^{n}\hat{\Xi}_{1}^{\left( l\right) }\left( Z_{i},\theta
^{\left( i\right) },\left\{ Z_{j}\right\} _{j\neq i}\right) d\theta ^{\left(
i\right) }}{\theta _{f}^{\left( i\right) }-\theta _{i}^{\left( i\right) }}%
\right) ^{m}\int_{\theta _{i}^{\left( i\right) }<\theta _{1}^{\left(
i\right) }<...\theta _{m}^{\left( i\right) }<\theta _{f}^{\left( i\right)
}}\delta \left( \theta _{1}^{\left( i\right) }-\theta _{i}^{\left( i\right)
}\right) \delta \left( \theta _{m}^{\left( i\right) }-\theta _{f}^{\left(
i\right) }\right) \dprod\limits_{q=1}^{m}d\theta _{q}^{\left( i\right) } \\
&=&\frac{1}{m!}\left( \int_{\theta _{i}^{\left( i\right) }}^{\theta
_{f}^{\left( i\right) }}\sum_{l=2}^{n}\hat{\Xi}_{1}^{\left( l\right) }\left(
Z_{i},\theta ^{\left( i\right) },\left\{ Z_{j}\right\} _{j\neq i}\right)
d\theta ^{\left( i\right) }\right) ^{m}
\end{eqnarray*}%
To obtain the expansion (\ref{gm}) due to the vertices at $Z_{i}$, we sum
over $m$ and we multiply with a free propagator on the left which amounts to
introduce a factor $\frac{1}{\Lambda }$ which leads to a contribution:%
\begin{equation*}
\frac{1}{\Lambda }\exp \left( \int_{\theta _{i}^{\left( i\right) }}^{\theta
_{f}^{\left( i\right) }}\sum_{l=2}^{n}\hat{\Xi}_{1}^{\left( l\right) }\left(
Z_{i},\theta ^{\left( i\right) },\left\{ Z_{j}\right\} _{j\neq i}\right)
d\theta ^{\left( i\right) }\right)
\end{equation*}%
The full sum of graphs is then obtained by taking the product over $i$ of
these contributions, introducing the global factor $\exp \left( -\Lambda
_{1}\left( \sum_{i=1}^{n}\theta _{f}^{\left( i\right) }-\sum_{i=1}^{n}\theta
_{i}^{\left( i\right) }\right) \right) $ and suming over $n$. The sum of
graphs is thus:%
\begin{eqnarray*}
&&\frac{\exp \left( -\Lambda _{1}\left( \sum_{i=1}^{n}\theta _{f}^{\left(
i\right) }-\sum_{i=1}^{n}\theta _{i}^{\left( i\right) }\right) \right) }{%
\Lambda ^{n}}\exp \left( \sum_{i}\int_{\theta _{i}^{\left( i\right)
}}^{\theta _{f}^{\left( i\right) }}\sum_{l}\hat{\Xi}_{1}^{\left( l\right)
}\left( Z_{i},\theta ^{\left( i\right) },\left\{ Z_{j}\right\} _{j\neq
i}\right) d\theta ^{\left( i\right) }\right) \\
&=&\frac{\exp \left( -\Lambda _{1}\left( \sum_{i=1}^{n}\theta _{f}^{\left(
i\right) }-\sum_{i=1}^{n}\theta _{i}^{\left( i\right) }\right) \right) }{%
\Lambda ^{n}}\exp \left( \sum_{i}\hat{\Xi}_{1,n}\left( Z_{i},\left\{
Z_{j}\right\} _{j\neq i},\theta _{i}^{\left( i\right) },\theta _{f}^{\left(
i\right) }\right) \right)
\end{eqnarray*}%
with: 
\begin{eqnarray}
\hat{\Xi}_{1,n}\left( Z_{i},\left\{ Z_{j}\right\} _{j\neq i},\theta
_{i}^{\left( i\right) },\theta _{f}^{\left( i\right) }\right)
&=&\int_{\theta _{i}^{\left( i\right) }}^{\theta _{f}^{\left( i\right)
}}\sum_{l=2}^{n}\hat{\Xi}_{1}^{\left( l\right) }\left( Z_{i},\theta ^{\left(
i\right) },\left\{ Z_{j}\right\} _{j\neq i}\right) d\theta ^{\left( i\right)
}  \label{tx} \\
&=&\sum_{l=2}^{n}\sum_{\left\{ k_{1},...,k_{l}\right\} \subset \left\{
1,...,n\right\} ,k_{j}\neq i}\hat{\Xi}_{1}^{\left( l\right) }\left(
Z_{i},\left\{ Z_{k_{j}}\right\} ,\theta _{i}^{\left( i\right) },\theta
_{f}^{\left( i\right) }\right)  \notag
\end{eqnarray}%
or alternatively:

\begin{eqnarray*}
\hat{\Xi}_{1,n}\left( Z_{i},\left\{ Z_{j}\right\} _{j\neq i},\theta
_{i}^{\left( i\right) },\theta _{f}^{\left( i\right) }\right) &=&\bar{\Xi}%
_{1,n}\left( Z_{i},\left\{ Z_{j}\right\} _{j\neq i},\theta _{i}^{\left(
i\right) },\theta _{f}^{\left( i\right) }\right) -\bar{\zeta}_{n}\left(
\theta _{f}^{\left( i\right) }-\theta _{i}^{\left( i\right) }\right) \\
&=&\sum_{l=2}^{n}\sum_{\left\{ k_{1},...,k_{l-1}\right\} \subset \left\{
1,...,n-1\right\} ,k_{j}\neq i}\left( \bar{\Xi}_{1}^{\left( l\right) }\left(
Z_{i},\left\{ Z_{k_{j}}\right\} ,\theta _{i}^{\left( i\right) },\theta
_{f}^{\left( i\right) }\right) -\frac{\zeta ^{\left( l\right) }\left( \theta
_{f}^{\left( i\right) }-\theta _{i}^{\left( i\right) }\right) }{\Lambda ^{l}}%
\right)
\end{eqnarray*}

\subsubsection*{2.2.4 Contributions of $\hat{V}_{2l}^{\left( 2\right)
}\left( \left( \protect\theta ^{\left( i\right) },Z_{i}\right)
_{i=1,...,n}\right) $}

We can include the contributions due to the vertices $\hat{V}_{2l}^{\left(
2\right) }\left( \left( \theta ^{\left( i\right) },Z_{i}\right)
_{i=1,...,n}\right) $ defined in (\ref{mv}) to the sum of graphs. These
vertices are inserted at some times $\theta _{1}<...$ $<\theta _{n}$ and at
each insertion $\theta _{k}$ one has $\theta _{i}^{\left( i\right) }<\theta
_{k}^{\left( i\right) }<\theta _{k}$. As before this leads to an overall
factor:%
\begin{equation}
\exp \left( \sum_{\substack{ \left\{ k_{1},...,k_{l}\right\}  \\ \subset
\left\{ 1,...,n\right\} }}\int \dprod\limits_{i=1}^{l}\int_{\theta
_{i}^{\left( k_{i}\right) }}^{\theta }d\theta ^{\left( k_{i}\right) }\hat{%
\Lambda}_{1}^{\left( l\right) }\left( Z,\theta ,\left\{ Z_{k_{i}},\theta
^{\left( k_{i}\right) }\right\} _{i=1,...l}\right) d\theta \right) =\exp
\left( \hat{\Lambda}_{1,n}\left( \left\{ Z_{i},\theta _{i}^{\left( i\right)
},\theta _{f}^{\left( i\right) }\right\} \right) \right)  \label{ld}
\end{equation}%
As a consequence, the sum (\ref{gm}) of graphs with $2n$ external points
becomes:

\begin{eqnarray}
&&\frac{\exp \left( -\Lambda _{1}\left( \sum_{i=1}^{n}\theta _{f}^{\left(
i\right) }-\sum_{i=1}^{n}\theta _{i}^{\left( i\right) }\right) \right) }{%
\Lambda ^{n}}  \label{xm} \\
&&\times \exp \left( \sum_{i}\hat{\Xi}_{1,n}\left( Z_{i},\left\{
Z_{j}\right\} _{j\neq i},\theta _{i}^{\left( i\right) },\theta _{f}^{\left(
i\right) }\right) \right) \exp \left( \hat{\Lambda}_{1,n}\left( \left\{
Z_{i},\theta _{i}^{\left( i\right) },\theta _{f}^{\left( i\right) }\right\}
\right) \right)  \notag
\end{eqnarray}%
Given (\ref{lx})%
\begin{equation*}
\hat{\Xi}_{1}^{\left( l\right) }\left( Z_{i},\theta ^{\left( i\right)
},\left\{ Z_{j}\right\} _{j\neq i}\right) =\bar{\Xi}_{1}^{\left( l\right)
}\left( Z_{i},\theta ^{\left( i\right) },\left\{ Z_{j}\right\} _{j\neq
i}\right) -\frac{\zeta ^{\left( l\right) }}{\Lambda ^{l}}
\end{equation*}%
and (\ref{tx}):%
\begin{equation*}
\hat{\Xi}_{1,n}\left( Z_{i},\left\{ Z_{j}\right\} _{j\neq i},\theta
_{i}^{\left( i\right) },\theta _{f}^{\left( i\right) }\right)
=\sum_{l=2}^{n}\sum_{\left\{ k_{1},...,k_{l}\right\} \subset \left\{
1,...,n\right\} ,k_{j}\neq i}\hat{\Xi}_{1}^{\left( l\right) }\left(
Z_{i},\left\{ Z_{k_{j}}\right\} ,\theta _{i}^{\left( i\right) },\theta
_{f}^{\left( i\right) }\right)
\end{equation*}%
we can estimate in average the magnitude of $\hat{\Xi}_{1,n}\left(
Z_{i},\left\{ Z_{j}\right\} _{j\neq i},\theta _{i}^{\left( i\right) },\theta
_{f}^{\left( i\right) }\right) $:%
\begin{equation*}
\hat{\Xi}_{1,n}\left( Z_{i},\left\{ Z_{j}\right\} _{j\neq i},\theta
_{i}^{\left( i\right) },\theta _{f}^{\left( i\right) }\right) \simeq
\sum_{l=2}^{n}C_{n}^{l}\left( \frac{\zeta ^{\left( l\right) }}{\Lambda ^{l}}%
\right)
\end{equation*}%
if $\bar{\Xi}_{1}^{\left( l\right) }\left( Z_{i},\theta ^{\left( i\right)
},\left\{ Z_{j}\right\} _{j\neq i}\right) $ and $\zeta ^{\left( l\right) }$
are of the same order. For $\zeta ^{\left( l\right) }$ decreasing faster
than $\exp \left( -l\right) $, for example $\zeta ^{\left( l\right) }\simeq
\exp \left( -l^{\alpha }\right) $ with $\alpha >1$, the sum is converging.
Actually, writing $C_{n}^{l}\simeq \exp \left( n\ln n-\left( n-l\right) \ln
\left( n-l\right) -l\ln l\right) $, we have%
\begin{equation*}
C_{n}^{l}\left( \frac{\zeta ^{\left( l\right) }}{\Lambda ^{l}}\right) \simeq
\exp \left( -n\left( x\ln x+\left( 1-x\right) \ln \left( 1-x\right)
+n^{\alpha -1}x^{\alpha }+x\ln \Lambda \right) \right)
\end{equation*}%
with $x=\frac{l}{n}$. As a consequence:%
\begin{equation*}
\sum_{l=2}^{n}C_{n}^{l}\left( \frac{\zeta ^{\left( l\right) }}{\Lambda ^{l}}%
\right) \simeq \int_{0}^{1}\exp \left( -n\left( x\ln x+\left( 1-x\right) \ln
\left( 1-x\right) +n^{\alpha -1}x^{\alpha }+x\ln \Lambda \right) \right) dx
\end{equation*}%
and the integral converges for $n\rightarrow \infty $. For $\zeta ^{\left(
2\right) }<<1$ we thus have $\hat{\Xi}_{1,n}\left( Z_{i},\left\{
Z_{j}\right\} _{j\neq i},\theta _{i}^{\left( i\right) },\theta _{f}^{\left(
i\right) }\right) <<1$ and for slowly varying parameters, we can replace $%
\hat{\Xi}_{1,n}\left( Z_{i},\left\{ Z_{j}\right\} _{j\neq i},\theta
_{i}^{\left( i\right) },\theta _{f}^{\left( i\right) }\right) $ by its limit 
$\hat{\Xi}_{1,\infty }\left( Z_{i},\left\{ Z_{j}\right\} _{j\neq i},\theta
_{i}^{\left( i\right) },\theta _{f}^{\left( i\right) }\right) $.

\subsubsection*{2.2.5 Convergence of the graph expansion (\protect\ref{gm})}

An estimation of the expansion of graphs of order higher than $2$ is given
by:%
\begin{equation*}
\sum \frac{1}{n!}\int \dprod\limits_{i=1}^{n}\Psi ^{\dagger }\left( \theta
_{i}^{\left( i\right) },Z_{i}\right) \exp \left( \left[ \bar{V}_{2n}\right]
\right) \left[ \bar{V}_{2n}\right] \left[ \bar{V}_{2n}\right]
\dprod\limits_{i=1}^{n}\Psi \left( \theta _{i}^{\left( i\right)
},Z_{i}\right)
\end{equation*}%
where all the graphs are taken into account. Given our previous results,
this is equal to:%
\begin{eqnarray*}
&\simeq &\sum \frac{1}{n!}\int \dprod\limits_{i=1}^{n}\Psi ^{\dagger }\left(
\theta _{i}^{\left( i\right) },Z_{i}\right) \exp \left( \hat{\Xi}_{1,\infty
}\left( Z_{i}\right) \right) \left( \hat{\Xi}_{1,\infty }\left( Z_{i}\right)
\right) ^{2}\dprod\limits_{i=1}^{n}\Psi \left( \theta _{i}^{\left( i\right)
},Z_{i}\right) \\
&\simeq &\hat{\Xi}_{1,\infty }\hat{\Xi}_{1,\infty }\sum \frac{1}{n!}\int
\dprod\limits_{i=1}^{n}\Psi ^{\dagger }\left( \theta _{i}^{\left( i\right)
},Z_{i}\right) \exp \left( \hat{\Xi}_{1,\infty }\left( Z_{i}\right) \right)
\dprod\limits_{i=1}^{n}\Psi \left( \theta _{i}^{\left( i\right)
},Z_{i}\right) \\
&=&\left( \hat{\Xi}_{1,\infty }\right) ^{2}\exp \left( \int \Psi ^{\dagger
}\left( \theta _{i}^{\left( i\right) },Z_{i}\right) \exp \left( \hat{\Xi}%
_{1,\infty }\left( Z_{i}\right) \right) \Psi \left( \theta _{i}^{\left(
i\right) },Z_{i}\right) \right)
\end{eqnarray*}%
where $\hat{\Xi}_{1,\infty }$ is the average of $\hat{\Xi}_{1,\infty }\left(
Z_{i}\right) $ over the thread. The perturbative expansion in $\hat{\Xi}%
_{1,\infty }$ is thus convergent.\bigskip

\section*{Appendix 3. Estimation of the effective action and its minimum}

The previous section showed the convergence of the full graphs series
expansion. To find the effective action we have to restrict the sum to the $%
1 $PI graphs. Once the effective action will be found, we will write the
equation of its minimum and compute the background field.

\subsection*{3.1 Effective action at the lowest order}

We have seen in equation (\ref{fctc}) that the $2$ points Green function are
computed using the action:%
\begin{eqnarray*}
\Gamma _{0}\left( \Psi ^{\dagger },\Psi \right) &\equiv &-\frac{1}{2}\int
\Psi ^{\dagger }\left( \theta ,Z\right) \left( \nabla _{\theta }\frac{\sigma
_{\theta }^{2}}{2}\nabla _{\theta }\right) \Psi \left( \theta ,Z\right) \\
&&+\frac{1}{2}\int \Psi ^{\dagger }\left( \theta ,Z\right) \left[ \frac{%
\delta \left[ \Psi ^{\dagger }\left( \theta ,Z\right) \nabla _{\theta
}\left( \omega ^{-1}\left( J,\theta ,Z,\left\vert \Psi \right\vert
^{2}\right) \Psi \left( \theta ,Z\right) \right) \right] }{\delta \left\vert
\Psi \right\vert ^{2}}\right] _{\substack{ \left\vert \Psi \left( \theta
,Z\right) \right\vert ^{2}  \\ =\mathcal{G}_{0}\left( 0,Z\right) }}\Psi
\left( \theta ,Z\right) \\
&&+\frac{1}{2}\int \Psi ^{\dagger }\left( \theta ,Z\right) \left[ \frac{%
\delta \left[ V\left( \Psi \right) \right] }{\delta \left\vert \Psi
\right\vert ^{2}}\right] _{\substack{ \left\vert \Psi \left( \theta
,Z\right) \right\vert ^{2}  \\ =\mathcal{G}_{0}\left( 0,Z\right) }}\Psi
\left( \theta ,Z\right) \\
&=&\Psi ^{\dagger }\left( \theta ,Z\right) \left[ \frac{\delta \left[
S_{cl}\left( \Psi ^{\dagger },\Psi \right) \right] }{\delta \left\vert \Psi
\right\vert ^{2}}\right] _{\substack{ \left\vert \Psi \left( \theta
,Z\right) \right\vert ^{2}  \\ =\mathcal{G}_{0}\left( 0,Z\right) }}\Psi
\left( \theta ,Z\right)
\end{eqnarray*}%
with:%
\begin{eqnarray}
&&S_{cl}\left( \Psi ^{\dagger },\Psi \right) =-\frac{1}{2}\Psi ^{\dagger
}\left( \theta ,Z\right) \left( \nabla _{\theta }\left( \frac{\sigma
_{\theta }^{2}}{2}\nabla _{\theta }-\omega ^{-1}\left( J,\theta
,Z,\left\vert \Psi \right\vert ^{2}\right) \right) \right) \Psi \left(
\theta ,Z\right) \\
&&+\alpha \int \left\vert \Psi \left( \theta ^{\left( i\right)
},Z_{i}\right) \right\vert ^{2}+V\left( \Psi \right)  \notag
\end{eqnarray}%
and: 
\begin{equation*}
\left[ \frac{\delta \left[ \Psi ^{\dagger }\left( \theta ,Z\right) \nabla
_{\theta }\left( \omega ^{-1}\left( J,\theta ,Z,\left\vert \Psi \right\vert
^{2}\right) \Psi \left( \theta ,Z\right) \right) \right] }{\delta \left\vert
\Psi \right\vert ^{2}}\right] _{\substack{ \left\vert \Psi \left( \theta
,Z\right) \right\vert ^{2}  \\ =\mathcal{G}_{0}\left( 0,Z\right) }}
\end{equation*}
defined as:%
\begin{eqnarray*}
&&\left[ \frac{\delta \left[ \Psi ^{\dagger }\left( \theta ,Z\right) \nabla
_{\theta }\left( \omega ^{-1}\left( J,\theta ,Z,\left\vert \Psi \right\vert
^{2}\right) \Psi \left( \theta ,Z\right) \right) \right] }{\delta \left\vert
\Psi \right\vert ^{2}}\right] _{\substack{ \left\vert \Psi \left( \theta
,Z\right) \right\vert ^{2}  \\ =\mathcal{G}_{0}\left( 0,Z\right) }} \\
&=&\omega ^{-1}\left( \mathcal{G}_{0}\left( 0,Z\right) \right) \nabla
_{\theta }+\left( \nabla _{\theta }\left( \frac{\delta \left[ \omega
^{-1}\left( \bar{J},Z,\mathcal{G}_{0}\right) \right] }{\delta \mathcal{G}%
_{0}\left( 0,Z\right) }\mathcal{G}_{0}\left( \theta ^{\prime },\theta
,Z\right) \right) \right) _{\theta =\theta ^{\prime }} \\
&=&\left( \omega ^{-1}\left( \mathcal{G}_{0}\left( 0,Z\right) \right) +%
\mathcal{G}_{0}\left( 0,Z\right) \frac{\delta \left[ \omega ^{-1}\left( \bar{%
J},Z,\mathcal{G}_{0}\right) \right] }{\delta \mathcal{G}_{0}\left(
0,Z\right) }\right) \nabla _{\theta }+\frac{\delta \left[ \omega ^{-1}\left( 
\bar{J},Z,\mathcal{G}_{0}\right) \right] }{\delta \mathcal{G}_{0}\left(
0,Z\right) }\left( \nabla _{\theta }\mathcal{G}_{0}\left( \theta ^{\prime
},\theta ,Z\right) \right) _{\theta =\theta ^{\prime }}
\end{eqnarray*}

As computed before, $\omega ^{-1}\left( \bar{J},Z,\mathcal{G}_{0}\right) $
satisfies: 
\begin{equation*}
\omega ^{-1}\left( \bar{J},Z,\mathcal{G}_{0}\right) =G\left( \bar{J}+\frac{%
\kappa }{N}\int T\left( Z,Z_{1}\right) W\left( \frac{\omega \left( \bar{J},Z,%
\mathcal{G}_{0}\right) }{\omega \left( \bar{J},Z_{1},\mathcal{G}_{0}\right) }%
\right) \frac{\omega \left( \bar{J},Z_{1},\mathcal{G}_{0}\right) \mathcal{G}%
_{0}\left( Z_{1},0\right) dZ_{1}}{\omega \left( \bar{J},Z,\mathcal{G}%
_{0}\right) }\right)
\end{equation*}

\subsection*{3.2 General formula for the effective action}

The perturbative corrections for the effective action are found by adding
the $1$PI graphs with $2n$ external points with $n\geqslant 2$. The
corrections are ordered by the number of vertices involved in them.

\subsubsection*{3.2.1 First and second order corrections\protect\bigskip}

Using that, for $n\geqslant 2$:%
\begin{equation*}
\left[ \frac{\delta ^{n}\left[ \int \Psi ^{\dagger }\left( \theta ,Z\right)
\nabla _{\theta }\omega ^{-1}\left( J,\theta ,Z\right) \Psi \left( \theta
,Z\right) dZd\theta +V\left( \Psi \right) \right] }{\dprod\limits_{i=1}^{n}%
\delta \left\vert \Psi \left( \theta ^{\left( i\right) },Z_{i}\right)
\right\vert ^{2}}\right] _{\substack{ \left\vert \Psi \left( \theta
,Z\right) \right\vert ^{2}  \\ =\mathcal{G}_{0}\left( 0,Z\right) }}=\left[ 
\frac{\delta ^{n}\left[ S_{cl}\left( \Psi ^{\dagger },\Psi \right) \right] }{%
\dprod\limits_{i=1}^{n}\delta \left\vert \Psi \left( \theta ^{\left(
i\right) },Z_{i}\right) \right\vert ^{2}}\right] _{\substack{ \left\vert
\Psi \left( \theta ,Z\right) \right\vert ^{2}  \\ =\mathcal{G}_{0}\left(
0,Z\right) }}
\end{equation*}%
the lowest order expansion of $1$PI graphs, consists of graphs with $n$
horizontal propagators, connected by one vertex of valence $n$. The sum of
these contributions becomes:%
\begin{eqnarray*}
&&\Gamma _{0}\left( \Psi ^{\dagger },\Psi \right) +\sum_{n=2}^{\infty }\frac{%
1}{n!}\dprod\limits_{i=1}^{n}\Psi ^{\dagger }\left( \theta _{i}^{\left(
i\right) },Z_{i}\right) \left[ \frac{\delta ^{n}\left[ S_{cl}\left( \Psi
^{\dagger },\Psi \right) \right] }{\dprod\limits_{i=1}^{n}\delta \left\vert
\Psi \left( \theta ^{\left( i\right) },Z_{i}\right) \right\vert ^{2}}\right] 
_{\substack{ \left\vert \Psi \left( \theta ,Z\right) \right\vert ^{2}  \\ =%
\mathcal{G}_{0}\left( 0,Z\right) }}\Psi \left( \theta _{i}^{\left( i\right)
},Z_{i}\right) \\
&=&\sum_{n=1}^{\infty }\frac{1}{n!}\dprod\limits_{i=1}^{n}\Psi ^{\dagger
}\left( \theta _{i}^{\left( i\right) },Z_{i}\right) \left[ \frac{\delta ^{n}%
\left[ S_{cl}\left( \Psi ^{\dagger },\Psi \right) \right] }{%
\dprod\limits_{i=1}^{n}\delta \left\vert \Psi \left( \theta ^{\left(
i\right) },Z_{i}\right) \right\vert ^{2}}\right] _{\substack{ \left\vert
\Psi \left( \theta ,Z\right) \right\vert ^{2}  \\ =\mathcal{G}_{0}\left(
0,Z\right) }}\Psi \left( \theta _{i}^{\left( i\right) },Z_{i}\right)
\end{eqnarray*}%
so that, up to the field independent term $\left[ S_{cl}\left( \Psi
^{\dagger },\Psi \right) \right] _{\left\vert \Psi \left( \theta ,Z\right)
\right\vert ^{2}=\mathcal{G}_{0}\left( 0,Z\right) }$, we have the effective
action at the first order in vertices:%
\begin{eqnarray*}
\Gamma _{0}\left( \Psi ^{\dagger },\Psi \right) +\Gamma _{1}\left( \Psi
^{\dagger },\Psi \right) &=&S_{cl}\left( \mathcal{G}_{0}\left( 0,Z\right)
+\left\vert \Psi \right\vert ^{2}\right) \\
&\equiv &-\frac{1}{2}\int \left( \left( \nabla _{\theta }\left( \frac{\sigma
_{\theta }^{2}}{2}\nabla _{\theta }-\omega ^{-1}\left( \left\vert \Psi
\left( \theta ,Z\right) \right\vert ^{2}\right) \right) \right) \left( 
\mathcal{G}_{0}\left( \theta ^{\prime },\theta ,Z\right) +\Psi ^{\dagger
}\left( \theta ^{\prime },Z\right) \Psi \left( \theta ,Z\right) \right)
\right) _{\theta ^{\prime }=\theta } \\
&&+\alpha \int \left( \mathcal{G}_{0}\left( 0,Z_{i}\right) \mathcal{+}%
\left\vert \Psi \left( \theta ^{\left( i\right) },Z_{i}\right) \right\vert
^{2}\right) +\sum_{n\geqslant 2}V_{n}\left( \mathcal{G}_{0}\left(
0,Z_{i}\right) \mathcal{+}\left\vert \Psi \left( \theta ^{\left( i\right)
},Z_{i}\right) \right\vert ^{2}\right)
\end{eqnarray*}%
To find the second order in vertices-number corrections, we need to define
the notion of multiple point. Consider, for $i=1,...,n$, the set $\left(
\left( \theta _{f}^{\left( i\right) },Z_{i}\right) ,\left( \theta
_{i}^{\left( i\right) },Z_{i}\right) \right) $ of $n$ initial points and $n$
final points and any graph connecting the initial and final points . The
index $i\in \left\{ 1,...,n\right\} $ labels a multiple point of valence $k$
of the graph, if the line connecting $\left( \theta _{f}^{\left( i\right)
},Z_{i}\right) $ and$\left( \theta _{i}^{\left( i\right) },Z_{i}\right) $ is
reached by $k$ legs of the vertices defining the graph. A multiple point of
valence $2$ is a double point and a multiple point of valence $l$ is also
refered as a $l$-multiple point.

At the second order in products of vertices the sum of $1$PI graphs is: 
\begin{eqnarray}
&&\int \sum_{n\geqslant 2}\frac{1}{n!}\sum_{\substack{ l_{1}=2,l_{2}=2  \\ %
l_{1}+l_{2}\geqslant n+2}}^{n}\sum_{\substack{ \left\{
k_{1},...,k_{l_{1}}\right\} \cup \left\{ k_{1}^{\prime
},...,k_{l_{2}}^{\prime }\right\}  \\ =\left\{ 1,...,n\right\} }}\left(
\dprod\limits_{k_{d}\in D}\Psi ^{\dagger }\left( \theta _{f}^{\left(
k_{d}\right) },Z_{i}\right) \frac{\exp \left( -\Lambda _{1}\left( \theta
_{f}^{\left( k_{d}\right) }-\theta _{i}^{\left( k_{d}\right) }\right)
\right) }{\Lambda ^{l_{1}+l_{2}}}\right) \dprod\limits_{i=1,i\notin
D}^{n}\Psi ^{\dagger }\left( \theta _{i}^{\left( i\right) },Z_{i}\right) 
\notag \\
&&\times \left( \left[ \frac{\delta ^{l_{1}}\left[ S_{cl}\left( \Psi
^{\dagger },\Psi \right) \right] }{\dprod\limits_{i=1}^{l_{1}}\delta
\left\vert \Psi \left( \theta ^{\left( k_{i}\right) },Z_{k_{i}}\right)
\right\vert ^{2}}\right] _{\substack{ \left\vert \Psi \left( \theta
,Z\right) \right\vert ^{2}  \\ =\mathcal{G}_{0}\left( 0,Z\right) }}\right)
\left( \left[ \frac{\delta ^{l_{2}}\left[ S_{cl}\left( \Psi ^{\dagger },\Psi
\right) \right] }{\dprod\limits_{i=1}^{l_{2}}\delta \left\vert \Psi \left(
\theta ^{\left( k_{i}\right) },Z_{k_{i}}\right) \right\vert ^{2}}\right]
_{\left\vert \Psi \left( \theta ,Z\right) \right\vert ^{2}=\mathcal{G}%
_{0}\left( 0,Z\right) }\right) \dprod\limits_{i=1}^{n}\Psi \left( \theta
_{i}^{\left( i\right) },Z_{i}\right)  \label{scn}
\end{eqnarray}%
with $D$ the set of double points of $\left\{ 1,...,n\right\} $, that is $%
\left\{ k_{1},...,k_{l_{1}}\right\} \cap \left\{ k_{1}^{\prime
},...,k_{l_{2}}^{\prime }\right\} $.

\subsubsection*{3.2.2 Including higher order corrections}

Including all contributions is straightforward and generalizes (\ref{scn}).
The sum of these contributions are: 
\begin{eqnarray*}
&&S_{cl}\left( \mathcal{G}_{0}\left( 0,Z\right) +\left\vert \Psi \right\vert
^{2}\right) \\
&&+\int \sum_{n\geqslant 2,p\geqslant 2}\frac{1}{n!}\sum_{\substack{ %
l_{1}+...l_{p}\geqslant n+2  \\ 2\leqslant l_{m}\leqslant n}}\sum_{\substack{
\underset{j=1...p}{\cup }\left\{ k_{1},...,k_{l_{j}}\right\}  \\ =\left\{
1,...,n\right\} }}\left( \dprod\limits_{k_{d}\in D}\Psi ^{\dagger }\left(
\theta _{f}^{\left( k_{d}\right) },Z_{k_{d}}\right)
\dprod\limits_{i=1,i\notin D}^{n}\Psi ^{\dagger }\left( \theta _{i}^{\left(
i\right) },Z_{i}\right) \right) \\
&&\times \frac{\exp \left( -\Lambda _{1}\left( \theta _{f}^{\left(
k_{d}\right) }-\theta _{i}^{\left( k_{d}\right) }\right) \right) }{\Lambda
^{l_{1}+l_{2}}}\left( \dprod\limits_{j=1}^{p}\left[ \frac{\delta ^{l_{j}}%
\left[ S_{cl}\left( \Psi ^{\dagger },\Psi \right) \right] }{%
\dprod\limits_{i=1}^{l_{j}}\delta \left\vert \Psi \left( \theta ^{\left(
k_{i}\right) },Z_{k_{i}}\right) \right\vert ^{2}}\right] _{\substack{ %
\left\vert \Psi \left( \theta ,Z\right) \right\vert ^{2}  \\ =\mathcal{G}%
_{0}\left( 0,Z\right) }}\right) \dprod\limits_{i=1}^{n}\Psi \left( \theta
_{i}^{\left( i\right) },Z_{i}\right)
\end{eqnarray*}%
where $D$ is the subset of multiple points of $\left\{ 1,...,n\right\} $,
that is the elements belonging at least to two distinct sets $\left\{
k_{1},...,k_{l_{j}}\right\} $. The sum is constrained to $1$PI graphs.

The sum can be regrouped in a different way. The graphs can be gathered in
classes with respects to the number of multiple points and the legs ending
at these points.

To do so, we split the multiple points on the the line $l_{i}$ connecting $%
\left( \theta _{f}^{\left( i\right) },Z_{i}\right) $ and $\left( \theta
_{i}^{\left( i\right) },Z_{i}\right) $ into multiple individual points by
cutting the segments between any two vertices endpoints. The associated
resulting graph belongs to the reduced class. This is the class of graphs in
which the vertices with $n$ legs are connected to $n$ points once, and two
different vertices are connected to different points.

The contribution of a graph in the reduced class with $p$ vertices of
valence $l_{j}$, $j=1,...,p$ is a product:%
\begin{equation*}
\frac{1}{\dprod \left( \natural l_{j}\right) !}\dprod\limits_{j=1}^{p}\left(
\dprod\limits_{i=1}^{l_{j}}\Psi ^{\dag }\left( \theta ^{\left( k_{i}\right)
},Z_{k_{i}}\right) \left[ \frac{1}{l_{j}!}\frac{\delta ^{l_{j}}\left[
S_{cl}\left( \Psi ^{\dagger },\Psi \right) \right] }{\dprod%
\limits_{i=1}^{l_{j}}\delta \left\vert \Psi \left( \theta ^{\left(
k_{i}\right) },Z_{k_{i}}\right) \right\vert ^{2}}\right] _{\substack{ %
\left\vert \Psi \left( \theta ,Z\right) \right\vert ^{2}  \\ =\mathcal{G}%
_{0}\left( 0,Z\right) }}\dprod\limits_{i=1}^{l_{j}}\Psi \left( \theta
^{\left( k_{i}\right) },Z_{k_{i}}\right) \right)
\end{equation*}%
of independent terms, where $\left( \natural l_{j}\right) $ is the number of
vertices of identical valence and the product runs other the set of valences
present in the graph. The graphs in the reduced class can be summed to
produce global contributions of the form:%
\begin{eqnarray*}
&&\frac{1}{p!}\sum \dprod\limits_{j=1}^{p}\left(
\dprod\limits_{i=1}^{l_{j}}\Psi ^{\dag }\left( \theta ^{\left( k_{i}\right)
},Z_{k_{i}}\right) \left[ \sum_{l_{j}}\frac{1}{l_{j}!}\frac{\delta ^{l_{j}}%
\left[ S_{cl}\left( \Psi ^{\dagger },\Psi \right) \right] }{%
\dprod\limits_{i=1}^{l_{j}}\delta \left\vert \Psi \left( \theta ^{\left(
k_{i}\right) },Z_{k_{i}}\right) \right\vert ^{2}}\right] _{\substack{ %
\left\vert \Psi \left( \theta ,Z\right) \right\vert ^{2}  \\ =\mathcal{G}%
_{0}\left( 0,Z\right) }}\dprod\limits_{i=1}^{l_{j}}\Psi \left( \theta
^{\left( k_{i}\right) },Z_{k_{i}}\right) \right) \\
&=&\frac{1}{p!}\dprod\limits_{j=1}^{p}\hat{S}_{cl}\left( \Psi ^{\dagger
},\Psi \right)
\end{eqnarray*}%
where:%
\begin{equation*}
\hat{S}_{cl}\left( \Psi ^{\dagger },\Psi \right) \equiv S_{cl}\left( 
\mathcal{G}_{0}\left( 0,Z\right) +\left\vert \Psi \right\vert ^{2}\right)
\end{equation*}

The factor $\frac{1}{p!}$ arises from the graph expansion of $%
\dprod\limits_{j=1}^{p}\hat{S}_{cl}\left( \Psi ^{\dagger },\Psi \right) $.
Actually, expanding the products yields a multiplicity of the products of
contribution which is $\frac{p!}{\dprod \left( \natural l_{j}\right) !}$.
Dividing by $p!$ thus restores the factor $\frac{1}{\dprod \left( \natural
l_{j}\right) !}$.

Reintroducing the multiple points to compute the graphs in a certain class
amounts to differentiate the factors $\hat{S}_{cl}\left( \Psi ^{\dagger
},\Psi \right) $ with respect to $\left\vert \Psi \right\vert ^{2}$and to
introduce products of fields corresponding to these multiple points.

Actually, let us consider the graphs with $p$ vertices and $%
n=\sum_{i\geqslant 1}l^{i}$ external points with $l^{i}$ points of
multiplicity $i$. We have $l^{i}=\sum_{j=1}^{p}l_{j}^{i}$, where $l_{j}^{i}$
is the number of legs of vertex $j$ connected to a point of multiplicity $i$%
. The factor associated to the repartition of the simple and multiple points
times the global factor $\frac{1}{n!}$ is thus:%
\begin{equation*}
\frac{1}{n!}C_{n}^{l^{1}}C_{n-l^{1}}^{l^{2}}...=\dprod\limits_{i}\frac{1}{%
\left( l^{i}\right) !}
\end{equation*}%
The attribution of the simple points yields the factors $\dprod \frac{1}{%
l_{j}^{1}!}$. \ A factor $\frac{1}{\dprod \left( \natural \sum_{i\geqslant
1}l_{j}^{i}\right) !}$ is associated for identical vertices. The multiple
points are then connected to the vertices in all possible manners compatible
with the $l_{j}^{i}$.

Then, starting with a factor:%
\begin{equation*}
\frac{1}{\dprod \left( \natural l_{j}\right) !}\dprod\limits_{j=1}^{p}\left(
\dprod\limits_{i=1}^{l_{j}}\Psi ^{\dag }\left( \theta ^{\left( k_{i}\right)
},Z_{k_{i}}\right) \left[ \frac{1}{l_{j}!}\frac{\delta ^{l_{j}}\left[
S_{cl}\left( \Psi ^{\dagger },\Psi \right) \right] }{\dprod%
\limits_{i=1}^{l_{j}}\delta \left\vert \Psi \left( \theta ^{\left(
k_{i}\right) },Z_{k_{i}}\right) \right\vert ^{2}}\right] _{\substack{ %
\left\vert \Psi \left( \theta ,Z\right) \right\vert ^{2}  \\ =\mathcal{G}%
_{0}\left( 0,Z\right) }}\dprod\limits_{i=1}^{l_{j}}\Psi \left( \theta
^{\left( k_{i}\right) },Z_{k_{i}}\right) \right)
\end{equation*}%
and applying:%
\begin{equation*}
\frac{1}{2}\int \frac{\delta ^{2}}{\delta \left\vert \Psi \left( \theta
^{\left( 1\right) },Z\right) \right\vert ^{2}\delta \left\vert \Psi \left(
\theta ^{\left( 2\right) },Z\right) \right\vert ^{2}}d\theta ^{\left(
1\right) }d\theta ^{\left( 2\right) }
\end{equation*}%
yields two factors of the form: 
\begin{equation*}
\left( \dprod\limits_{i=1}^{l_{j}-1}\Psi ^{\dag }\left( \theta ^{\left(
k_{i}\right) },Z_{k_{i}}\right) \left[ \frac{1}{\left( l_{j}-1\right) !}%
\frac{\delta ^{l_{j}}\left[ S_{cl}\left( \Psi ^{\dagger },\Psi \right) %
\right] }{\Lambda \dprod\limits_{i=1}^{l_{j}-1}\delta \left\vert \Psi \left(
\theta ^{\left( k_{i}\right) },Z_{k_{i}}\right) \right\vert ^{2}\delta
\left\vert \Psi \left( \theta ^{\left( 1,2\right) },Z\right) \right\vert ^{2}%
}\right] _{\substack{ \left\vert \Psi \left( \theta ,Z\right) \right\vert
^{2}  \\ =\mathcal{G}_{0}\left( 0,Z\right) }}\dprod\limits_{i=1}^{l_{j}-1}%
\Psi \left( \theta ^{\left( k_{i}\right) },Z_{k_{i}}\right) \right)
\end{equation*}%
or one factor:%
\begin{equation*}
\left( \dprod\limits_{i=1}^{l_{j}-2}\Psi ^{\dag }\left( \theta ^{\left(
k_{i}\right) },Z_{k_{i}}\right) \left[ \frac{1}{\left( l_{j}-2\right) !}%
\frac{\delta ^{l_{j}}\left[ S_{cl}\left( \Psi ^{\dagger },\Psi \right) %
\right] }{\Lambda \dprod\limits_{i=1}^{l_{j}-2}\delta \left\vert \Psi \left(
\theta ^{\left( k_{i}\right) },Z_{k_{i}}\right) \right\vert ^{2}\delta
\left\vert \Psi \left( \theta ^{\left( 1\right) },Z\right) \right\vert
^{2}\delta \left\vert \Psi \left( \theta ^{\left( 2\right) },Z\right)
\right\vert ^{2}}\right] _{\substack{ \left\vert \Psi \left( \theta
,Z\right) \right\vert ^{2}  \\ =\mathcal{G}_{0}\left( 0,Z\right) }}%
\dprod\limits_{i=1}^{l_{j}-2}\Psi \left( \theta ^{\left( k_{i}\right)
},Z_{k_{i}}\right) \right)
\end{equation*}%
This corresponds to the introduction of a double point, with the required
factors for simple points, the overall factor $\frac{1}{\dprod \left(
\natural l_{j}\right) !}$ corresponding to $\frac{1}{\dprod \left( \natural
\sum_{i\geqslant 1}l_{j}^{i}\right) !}$ with $l_{j}^{1}=l_{j}$ except for
the two factors case where $l_{j}^{1}=l_{j}-1$ and $l_{j}^{2}=1$, or for the
single factor $l_{j}^{1}=l_{j}-2$ and $l_{j}^{2}=2$.

To account for the global factors $\frac{1}{l^{i}!}$ and the propagators
associated to the multiple points, the operator corresponding to the
introduction of $k$ double points is then: 
\begin{equation*}
\frac{1}{k!}\left( \frac{1}{2}\int \Psi ^{\dagger }\left( \theta
_{f},Z\right) \frac{\exp \left( -\Lambda _{1}\left( \theta _{f}-\theta
_{i}\right) \right) }{\Lambda ^{2}}\Psi \left( \theta _{i},Z\right) \int 
\frac{\delta ^{2}}{\delta \left\vert \Psi \left( \theta ^{\left( 1\right)
},Z\right) \right\vert ^{2}\delta \left\vert \Psi \left( \theta ^{\left(
2\right) },Z\right) \right\vert ^{2}}d\theta ^{\left( 1\right) }d\theta
^{\left( 2\right) }\right) ^{k}
\end{equation*}%
More generally, the introduction of $l$-multiple points are generated by
operators:%
\begin{equation*}
\frac{1}{k!}\left( \frac{1}{l!}\int \Psi ^{\dagger }\left( \theta
_{f},Z\right) \frac{\exp \left( -\Lambda _{1}\left( \theta _{f}-\theta
_{i}\right) \right) }{\Lambda ^{l}}\Psi \left( \theta _{i},Z\right) \int 
\frac{\delta ^{2}}{\dprod\limits_{s=1}^{l}\delta \left\vert \Psi \left(
\theta ^{\left( s\right) },Z\right) \right\vert ^{2}}\dprod%
\limits_{s=1}^{l}d\theta ^{\left( s\right) }\right) ^{k}
\end{equation*}%
So that the whole series generating the multiple points is:%
\begin{eqnarray}
&&\Gamma \left( \Psi ^{\dagger },\Psi \right)  \label{epF} \\
&=&\exp \left( \sum_{l\geqslant 2}\frac{1}{l!}\int \Psi ^{\dagger }\left(
\theta _{f},Z\right) \frac{\exp \left( -\Lambda _{1}\left( \theta
_{f}-\theta _{i}\right) \right) }{\Lambda ^{l}}\Psi \left( \theta
_{i},Z\right) \int \frac{\delta ^{2}}{\Lambda \dprod\limits_{s=1}^{l}\delta
\left\vert \Psi \left( \theta ^{\left( s\right) },Z\right) \right\vert ^{2}}%
\dprod\limits_{s=1}^{l}d\theta ^{\left( s\right) }\right) \sum \frac{1}{p!}%
\dprod\limits_{j=1}^{p}\hat{S}_{cl}\left( \Psi ^{\dagger },\Psi \right) 
\notag \\
&=&\exp \left( \sum_{l\geqslant 2}\frac{1}{l!}\int \Psi ^{\dagger }\left(
\theta _{f},Z\right) \frac{\exp \left( -\Lambda _{1}\left( \theta
_{f}-\theta _{i}\right) \right) }{\Lambda ^{l}}\Psi \left( \theta
_{i},Z\right) \int \frac{\delta ^{2}}{\dprod\limits_{s=1}^{l}\delta
\left\vert \Psi \left( \theta ^{\left( s\right) },Z\right) \right\vert ^{2}}%
\dprod\limits_{s=1}^{l}d\theta ^{\left( s\right) }\right) \exp \left( \hat{S}%
_{cl}\left( \Psi ^{\dagger },\Psi \right) \right)  \notag
\end{eqnarray}%
It is understood that only the $1$PI graphs are kept in the series expansion.

The expansion can also be written in expanded form and yields the effective
action $\Gamma \left( \Psi ^{\dagger },\Psi \right) $:

\begin{eqnarray}
&&\Gamma \left( \Psi ^{\dagger },\Psi \right) =\hat{S}_{cl}\left( \Psi
^{\dagger },\Psi \right) +\sum_{\substack{ j\geqslant 2  \\ m\geqslant 2}}%
\sum _{\substack{ \left( p_{l}^{i}\right) _{m\times j}  \\ %
\sum_{i}p_{l}^{i}\geqslant 2}}\int \left( \dprod\limits_{l=1}^{j}\Psi
^{\dagger }\left( \theta _{f}^{\left( l\right) },Z_{l}\right) \right)
\label{pnxd} \\
&&\times \dprod\limits_{i=1}^{m}\left[ \int_{\dprod\limits_{l=1}^{j}\left[
\theta _{i}^{\left( l\right) },\theta _{f}^{\left( l\right) }\right]
^{p_{l}^{i}}}\frac{\delta ^{\sum_{l}p_{l}^{i}}\left[ \hat{S}_{cl}\left( \Psi
^{\dagger },\Psi \right) \right] }{\dprod\limits_{l=1}^{j}\dprod%
\limits_{k_{l}^{i}=1}^{p_{l}^{i}}\delta \left\vert \Psi \left( \theta
^{\left( k_{l}^{i}\right) },Z_{_{l}}\right) \right\vert ^{2}}%
\dprod\limits_{l=1}^{j}\dprod\limits_{k_{l}^{i}=1}^{p_{l}^{i}}d\theta
^{\left( k_{l}^{i}\right) }\right]  \notag \\
&&\times \frac{\exp \left( -\Lambda _{1}\left( \theta _{f}^{\left( l\right)
}-\theta _{i}^{\left( l\right) }\right) \right) }{m!\dprod\limits_{k}\left(
\sharp _{k}\right) !\Lambda ^{\sum_{i}p_{l}^{i}}}\left(
\dprod\limits_{l=1}^{j}\Psi \left( \theta _{i}^{\left( l\right)
},Z_{l}\right) \right)  \notag
\end{eqnarray}%
with:%
\begin{equation}
\sharp _{k,j,m}\left( \left( p_{l}^{i}\right) \right) =\sum_{l=1}^{j}\delta
_{k,\sum_{i=1}^{m}p_{l}^{i}}  \label{dsK}
\end{equation}%
and where:%
\begin{equation*}
\theta _{i}^{\left( l\right) }<\theta ^{\left( k_{l}^{i}\right) }<\theta
_{f}^{\left( l\right) }
\end{equation*}%
The notation $\left( p_{l}^{i}\right) $ in (\ref{dsK}) stands for the
dependency of $\sharp _{k,j,m}$ in the whole collection of indices $\left(
p_{l}^{i}\right) $, $i=1...m$ and $l=1...j$.

The sums over the indices in (\ref{pnxd}) represent the sum over the
different class of graphs with respect to the multiple points. Actually, to
each of the $m$ copies of $\hat{S}_{cl}\left( \Psi ^{\dagger },\Psi \right) $
we associate a point $P_{i}$ and to each $Z_{_{l}}$ we associate a line $%
L_{l}$. For each derivative of the copy with respect to $\left\vert \Psi
\left( \theta ^{\left( k_{l}^{i}\right) },Z_{_{l}}\right) \right\vert ^{2}$
we draw a segment from $P_{i}$ to $L_{l}$. The number of segments between $%
P_{i}$ and $L_{l}$ is equal to $p_{l}^{i}$. This produces a graph with
multiple points and the corresponding expression computes the sum of graphs
in the class of this multiple-points graph. The sum over the indices are
constrained to produce $1$PI graphs.

In the local approximation, we replace $\frac{\exp \left( -\Lambda
_{1}\left( \theta _{f}^{\left( l\right) }-\theta _{i}^{\left( l\right)
}\right) \right) }{\Lambda ^{2}}$ by $\frac{\delta \left( \theta
_{f}^{\left( l\right) }-\theta _{i}^{\left( l\right) }\right) }{\Lambda
_{1}\Lambda ^{2}}$, and as a consequence the effective action writes: 
\begin{eqnarray*}
\Gamma \left( \Psi ^{\dagger },\Psi \right) &=&\hat{S}_{cl}\left( \Psi
^{\dagger },\Psi \right) \\
&&+\sum_{j\geqslant 2}\int \left[ \left( \dprod\limits_{l=1}^{j}\Psi
^{\dagger }\left( \theta ^{\left( l\right) },Z_{l}\right) \right) \sum 
_{\substack{ m\geqslant 1,\left( p_{l}^{i}\right) _{m\times j}  \\ %
\sum_{i}p_{l}^{i}\geqslant 2}}\dprod\limits_{i=1}^{m}\frac{\left[ \frac{%
\delta ^{\sum_{l}p_{l}^{i}}\left[ \hat{S}_{cl}\left( \Psi ^{\dagger },\Psi
\right) \right] }{\dprod\limits_{l=1}^{j}\prod%
\limits_{k_{l}^{i}=1}^{p_{l}^{i}}\delta \left\vert \Psi \left( \theta
^{\left( l\right) },Z_{_{l}}\right) \right\vert ^{2}}\right] }{%
m!\dprod\limits_{k}\left( \sharp _{k,j,m}\left( \left( p_{l}^{i}\right)
\right) \right) !\Lambda _{1}^{j}\Lambda ^{\sum_{i,l}p_{l}^{i}}}\left(
\dprod\limits_{l=1}^{j}\Psi \left( \theta ^{\left( l\right) },Z_{l}\right)
\right) \right]
\end{eqnarray*}

Note that, for $\frac{\delta ^{\sum_{l}p_{l}^{i}}\left[ \hat{S}_{cl}\left(
\Psi ^{\dagger },\Psi \right) \right] }{\dprod\limits_{l}p_{l}^{i}!\dprod%
\limits_{k_{l}^{i}=1}^{p_{l}^{i}}\delta \left\vert \Psi \left( \theta
^{\left( k_{l}^{i}\right) },Z_{_{l}}\right) \right\vert ^{2}}<<1$ when $%
\sum_{l}p_{l}^{i}$ \ increases, the dominant part of graphs for $j\geqslant
2 $ \ is for $m=j$ and $\sum_{l}p_{l}^{i}=2$. In this case the effective
action rewrites: 
\begin{eqnarray*}
&&\Gamma \left( \Psi ^{\dagger },\Psi \right) =\hat{S}_{cl}\left( \Psi
^{\dagger },\Psi \right) +\sum_{j\geqslant 2}\sum_{\substack{ \cup \left(
l_{1}^{i},l_{2}^{i}\right) _{i=1,...j}  \\ =\left( 1,...,j\right) ^{2}}}\int
\left( \dprod\limits_{l=1}^{j}\Psi ^{\dagger }\left( \theta _{f}^{\left(
l\right) },Z_{l}\right) \right) \\
&&\times \dprod\limits_{i=1}^{j}\left[ \int_{\substack{ \left[ \theta
_{i}^{\left( l_{1}^{i}\right) },\theta _{f}^{\left( l_{1}^{i}\right) }\right]
\\ \times \left[ \theta _{i}^{\left( l_{2}^{i}\right) },\theta _{f}^{\left(
l_{2}^{i}\right) }\right] }}\frac{\delta ^{2}\left[ \hat{S}_{cl}\left( \Psi
^{\dagger },\Psi \right) \right] }{\delta \left\vert \Psi \left( \theta
^{\left( l_{1}^{i}\right) },Z_{_{l_{1}^{i}}}\right) \right\vert ^{2}\delta
\left\vert \Psi \left( \theta ^{\left( l_{2}^{i}\right)
},Z_{_{l_{2}^{i}}}\right) \right\vert ^{2}}d\theta ^{\left( l_{1}^{i}\right)
}d\theta ^{\left( l_{2}^{i}\right) }\right] \\
&&\times \frac{\exp \left( -\Lambda _{1}\left( \theta _{f}^{\left( l\right)
}-\theta _{i}^{\left( l\right) }\right) \right) }{\Lambda ^{2}}\left(
\dprod\limits_{l=1}^{j}\Psi \left( \theta _{i}^{\left( l\right)
},Z_{l}\right) \right)
\end{eqnarray*}%
and it the local approximation, we have:%
\begin{eqnarray*}
&&\Gamma \left( \Psi ^{\dagger },\Psi \right) =\hat{S}_{cl}\left( \Psi
^{\dagger },\Psi \right) \\
&&+\sum_{j\geqslant 2}\int \left( \dprod\limits_{l=1}^{j}\Psi ^{\dagger
}\left( \theta _{f}^{\left( l\right) },Z_{l}\right) \right)
\dprod\limits_{i=1}^{j}\left[ \int_{\substack{ \left[ \theta _{i}^{\left(
i\right) },\theta _{f}^{\left( i+1\right) }\right]  \\ \times \left[ \theta
_{i}^{\left( i\right) },\theta _{f}^{\left( i+1\right) }\right] }}\frac{%
\delta ^{2}\left[ \hat{S}_{cl}\left( \Psi ^{\dagger },\Psi \right) \right] }{%
\delta \left\vert \Psi \left( \theta ^{\left( i\right) },Z_{_{i}}\right)
\right\vert ^{2}\delta \left\vert \Psi \left( \theta ^{\left( i+1\right)
},Z_{_{i+1}}\right) \right\vert ^{2}}d\theta ^{\left( i\right) }d\theta
^{\left( i+1\right) }\right] \\
&&\times \frac{\exp \left( -\Lambda _{1}\left( \theta _{f}^{\left( l\right)
}-\theta _{i}^{\left( l\right) }\right) \right) }{\Lambda ^{2}}\left(
\dprod\limits_{l=1}^{j}\Psi \left( \theta _{i}^{\left( l\right)
},Z_{l}\right) \right)
\end{eqnarray*}

with the convention that $i+1\equiv 1$ for $i=j$.

\subsection*{3.3 Alternative form for the effective action}

For later purposes, (\ref{pnxd}) can be written in developped form. To dos
so, we first give an expanded form for $\dprod\limits_{i=1}^{m}\hat{S}%
_{cl}\left( \Psi ^{\dagger },\Psi \right) $ arising in (\ref{pnxd}).

\subsubsection*{3.3.1 Expanded form for $\protect\dprod\limits_{i=1}^{m}\hat{%
S}_{cl}\left( \Psi ^{\dagger },\Psi \right) $}

We start with:%
\begin{eqnarray*}
\hat{S}_{cl}\left( \Psi ^{\dagger },\Psi \right) &=&-\frac{1}{2}\int \left(
\left( \nabla _{\theta }\frac{\sigma _{\theta }^{2}}{2}\nabla _{\theta
}-\omega ^{-1}\left( \left\vert \Psi \left( \theta ,Z\right) \right\vert
^{2}\right) \right) \left( \mathcal{G}_{0}\left( \theta ^{\prime },\theta
,Z\right) +\Psi ^{\dagger }\left( \theta ^{\prime },Z\right) \Psi \left(
\theta ,Z\right) \right) \right) _{\theta ^{\prime }=\theta } \\
&&+\alpha \int \left( \mathcal{G}_{0}\left( 0,Z_{i}\right) \mathcal{+}%
\left\vert \Psi \left( \theta ^{\left( i\right) },Z_{i}\right) \right\vert
^{2}\right) +\sum_{n\geqslant 2}V_{n}\left( \mathcal{G}_{0}\left(
0,Z_{i}\right) \mathcal{+}\left\vert \Psi \left( \theta ^{\left( i\right)
},Z_{i}\right) \right\vert ^{2}\right)
\end{eqnarray*}%
Given the form of $V_{n}\left( \mathcal{G}_{0}\left( 0,Z_{i}\right) \mathcal{%
+}\left\vert \Psi \left( \theta ^{\left( i\right) },Z_{i}\right) \right\vert
^{2}\right) $ and given that:%
\begin{eqnarray*}
&&-\frac{1}{2}\int \left( \left( \nabla _{\theta }\frac{\sigma _{\theta }^{2}%
}{2}\nabla _{\theta }-\omega ^{-1}\left( \left\vert \Psi \left( \theta
,Z\right) \right\vert ^{2}\right) \right) \left( \mathcal{G}_{0}\left(
\theta ^{\prime },\theta ,Z\right) +\Psi ^{\dagger }\left( \theta ^{\prime
},Z\right) \Psi \left( \theta ,Z\right) \right) \right) _{\theta ^{\prime
}=\theta } \\
&\simeq &-\frac{1}{2}\int \Psi ^{\dagger }\left( \theta ,Z\right) \left(
\nabla _{\theta }\frac{\sigma _{\theta }^{2}}{2}\nabla _{\theta }-\omega
^{-1}\left( \left\vert \Psi \left( \theta ,Z\right) \right\vert ^{2}\right)
\right) \Psi \left( \theta ,Z\right)
\end{eqnarray*}%
we can rewrite $\hat{S}_{cl}\left( \Psi ^{\dagger },\Psi \right) $, up to
the constant $\alpha \int \mathcal{G}_{0}\left( 0,Z_{i}\right) $:%
\begin{eqnarray}
\hat{S}_{cl}\left( \Psi ^{\dagger },\Psi \right) &\simeq &\int \Psi
^{\dagger }\left( \theta ,Z\right) \left( -\frac{1}{2}\left( \nabla _{\theta
}\left( \frac{\sigma _{\theta }^{2}}{2}\nabla _{\theta }-\omega ^{-1}\left(
\left\vert \Psi \left( \theta ,Z\right) \right\vert ^{2}\right) \right)
\right) +\alpha +U\left( \left\vert \Psi \left( \theta ,Z\right) \right\vert
^{2}\right) \right) \Psi \left( \theta ,Z\right)  \notag \\
&\equiv &\int \Psi ^{\dagger }\left( \theta ,Z\right) L\left( \Psi ^{\dagger
}\left( \theta ,Z\right) ,\Psi \left( \theta ,Z\right) \right) \Psi \left(
\theta ,Z\right)  \label{shT}
\end{eqnarray}%
where $U\left( \int \left\vert \Psi \left( \theta -\frac{\left\vert
Z-Z^{\prime }\right\vert }{c},Z^{\prime }\right) \right\vert ^{2}\right) $
is obtained by the series expansion of $U_{0}\left( \mathcal{G}_{0}\left(
0,Z_{i}\right) +\int \left\vert \Psi \left( \theta -\frac{\left\vert
Z-Z^{\prime }\right\vert }{c},Z^{\prime }\right) \right\vert ^{2}\right) $
and collecting its terms of degree $2$ and higher in fields.

Then, in (\ref{pnxd}), The product of $m$ copies of $\hat{S}_{cl}\left( \Psi
^{\dagger },\Psi \right) $ can be reordered as:%
\begin{equation*}
\dprod\limits_{i=1}^{m}\hat{S}_{cl}\left( \Psi ^{\dagger },\Psi \right)
=m!\int_{\theta _{1}<...<\theta _{m}}\Psi ^{\dagger }\left( \theta
_{i},Z_{i}\right) L\left( \Psi ^{\dagger }\left( \theta _{i},Z_{i}\right)
,\Psi \left( \theta _{i},Z_{i}\right) \right) \Psi \left( \theta
_{i},Z_{i}\right)
\end{equation*}%
and (\ref{pnxd}) becomes:

\begin{eqnarray}
&&\Gamma \left( \Psi ^{\dagger },\Psi \right) =\hat{S}_{cl}\left( \Psi
^{\dagger },\Psi \right) +\sum_{\substack{ j\geqslant 2  \\ m\geqslant 2}}%
\sum _{\substack{ \left( p_{l}^{i}\right) _{m\times j}  \\ %
\sum_{i}p_{l}^{i}\geqslant 2}}\int \left( \dprod\limits_{l=1}^{j}\Psi
^{\dagger }\left( \theta _{f}^{\left( l\right) },Z_{l}\right) \right)
\label{prL} \\
&&\times \int_{\theta _{1}<...<\theta _{m}}\dprod\limits_{i=1}^{m}\left[
\int_{\dprod\limits_{l=1}^{j}\left[ \theta _{i}^{\left( l\right) },\theta
_{f}^{\left( l\right) }\right] ^{p_{l}^{i}}}\frac{\delta ^{\sum_{l}p_{l}^{i}}%
\left[ \Psi ^{\dagger }\left( \theta _{i},Z_{i}\right) L\left( \Psi
^{\dagger }\left( \theta _{i},Z_{i}\right) ,\Psi \left( \theta
_{i},Z_{i}\right) \right) \Psi \left( \theta _{i},Z_{i}\right) \right] }{%
\dprod\limits_{l=1}^{j}\dprod\limits_{k_{l}^{i}=1}^{p_{l}^{i}}\delta
\left\vert \Psi \left( \theta ^{\left( k_{l}^{i}\right) },Z_{_{l}}\right)
\right\vert ^{2}}\dprod\limits_{l=1}^{j}\dprod%
\limits_{k_{l}^{i}=1}^{p_{l}^{i}}d\theta ^{\left( k_{l}^{i}\right) }\right]
d\left( \theta _{i},Z_{i}\right)  \notag \\
&&\times \frac{\exp \left( -\Lambda _{1}\left( \theta _{f}^{\left( l\right)
}-\theta _{i}^{\left( l\right) }\right) \right) }{\dprod\limits_{k}\left(
\sharp _{k,j,m}\left( \left( p_{l}^{i}\right) \right) \right) !\Lambda
^{\sum_{i}p_{l}^{i}}}\left( \dprod\limits_{l=1}^{j}\Psi \left( \theta
_{i}^{\left( l\right) },Z_{l}\right) \right)  \notag
\end{eqnarray}

\subsubsection*{3.3.2 Computation of the factor for $i=m$ in (\protect\ref%
{prL})}

To compute:%
\begin{equation}
\frac{\delta ^{\sum_{l}p_{l}^{m}}\left[ \Psi ^{\dagger }\left( \theta
_{m},Z_{m}\right) L\left( \Psi ^{\dagger }\left( \theta _{m},Z_{m}\right)
,\Psi \left( \theta _{m},Z_{m}\right) \right) \Psi \left( \theta
_{m},Z_{m}\right) \right] }{\dprod\limits_{l=1}^{j}\dprod%
\limits_{k_{l}^{i}=1}^{p_{l}^{m}}\delta \left\vert \Psi \left( \theta
^{\left( k_{l}^{i}\right) },Z_{_{l}}\right) \right\vert ^{2}}  \label{thg}
\end{equation}%
in (\ref{prL}), we decompose this term as a sum:%
\begin{eqnarray}
&&\Psi ^{\dagger }\left( \theta _{m},Z_{m}\right) \frac{\delta
^{\sum_{l}p_{l}^{m}}\left[ L\left( \Psi ^{\dagger }\left( \theta
_{m},Z_{m}\right) ,\Psi \left( \theta _{m},Z_{m}\right) \right) \right] }{%
\dprod\limits_{l=1}^{j}\dprod\limits_{k_{l}^{i}=1}^{p_{l}^{m}}\delta
\left\vert \Psi \left( \theta ^{\left( k_{l}^{i}\right) },Z_{_{l}}\right)
\right\vert ^{2}}\Psi \left( \theta _{m},Z_{m}\right)  \label{scD} \\
&&+\sum_{\substack{ p_{l}^{m\prime }\leqslant p_{l}^{m}  \\ %
\sum_{l}p_{l}^{m\prime }=\sum_{l}p_{l}^{m}-1}}\frac{\delta
^{\sum_{l}p_{l}^{m}-1}\left[ L\left( \Psi ^{\dagger }\left( \theta
_{m},Z_{m}\right) ,\Psi \left( \theta _{m},Z_{m}\right) \right) \right] }{%
\dprod\limits_{l=1}^{j}\dprod\limits_{k_{l}^{i}=1}^{p_{l}^{m\prime }}\delta
\left\vert \Psi \left( \theta ^{\left( k_{l}^{i}\right) },Z_{_{l}}\right)
\right\vert ^{2}}  \notag
\end{eqnarray}%
Each configuration $\left( p_{l}^{m^{\prime }}\right) $ is reached by $j$
configuration $\left( p_{l}^{m}\right) $. We perform a change of variable $%
p_{l}^{m^{\prime }}\rightarrow p_{l}^{m}$, and the factor $\frac{1}{%
\dprod\limits_{k}\left( \sharp _{j,m;k}\right) !}$ for each configuration
has to be replaced for each of the $j$ configurations it is issued from.

Now, we consider the second term in (\ref{scD}) multiplied by $\frac{1}{%
\dprod\limits_{k}\left( \sharp _{k,j,m}\left( \left( p_{l}^{i}\right)
\right) \right) !}=\frac{1}{\sharp _{j,m}\left( \left( p_{l}^{i}\right)
\right) }$:%
\begin{eqnarray*}
&&\frac{1}{\dprod\limits_{k}\left( \sharp _{j,m,k}\right) !}\frac{\delta
^{\sum_{l}p_{l}^{m}-1}\left[ L\left( \Psi ^{\dagger }\left( \theta
_{m},Z_{m}\right) ,\Psi \left( \theta _{m},Z_{m}\right) \right) \right] }{%
\dprod\limits_{l=1}^{j}\dprod\limits_{k_{l}^{i}=1}^{p_{l}^{m\prime }}\delta
\left\vert \Psi \left( \theta ^{\left( k_{l}^{i}\right) },Z_{_{l}}\right)
\right\vert ^{2}} \\
&=&\sum_{p=1}^{j}\frac{1}{\dprod\limits_{k}\left( \sum_{l=1}^{j}\delta
_{k,\sum_{i=1}^{m}p_{l}^{i}}\right) !}\frac{\delta ^{\sum_{l}p_{l}^{m}-1}%
\left[ L\left( \Psi ^{\dagger }\left( \theta _{m},Z_{m}\right) ,\Psi \left(
\theta _{m},Z_{m}\right) \right) \right] }{\dprod\limits_{l=1}^{j}\dprod%
\limits_{k_{l}^{i}=1}^{p_{l}^{m\prime }}\delta \left\vert \Psi \left( \theta
^{\left( k_{l}^{i}\right) },Z_{_{l}}\right) \right\vert ^{2}}
\end{eqnarray*}%
In the computation of (\ref{prL}), this term can be replaced by:%
\begin{eqnarray*}
&&\left( \sum_{p=1}^{j}\frac{1}{\dprod\limits_{k}\left( \sum_{l}\delta
_{k,\sum_{i}p_{l}^{i}+\delta _{l,p}}\right) !}\right) \frac{\delta
^{\sum_{l}p_{l}^{m}}\left[ L\left( \Psi ^{\dagger }\left( \theta
_{m},Z_{m}\right) ,\Psi \left( \theta _{m},Z_{m}\right) \right) \right] }{%
\dprod\limits_{l=1}^{j}\dprod\limits_{k_{l}^{i}=1}^{p_{l}^{m}}\delta
\left\vert \Psi \left( \theta ^{\left( k_{l}^{i}\right) },Z_{_{l}}\right)
\right\vert ^{2}} \\
&=&\frac{\delta ^{\sum_{l}p_{l}^{m}}\left[ L\left( \Psi ^{\dagger }\left(
\theta _{m},Z_{m}\right) ,\Psi \left( \theta _{m},Z_{m}\right) \right) %
\right] }{\bar{\sharp}_{j,m}\left( \left( p_{l}^{i}\right) \right)
\dprod\limits_{l=1}^{j}\dprod\limits_{k_{l}^{i}=1}^{p_{l}^{m}}\delta
\left\vert \Psi \left( \theta ^{\left( k_{l}^{i}\right) },Z_{_{l}}\right)
\right\vert ^{2}}
\end{eqnarray*}%
where we define:%
\begin{equation}
\frac{1}{\bar{\sharp}_{j,m}\left( \left( p_{l}^{i}\right) \right) }=\frac{1}{%
\dprod\limits_{k}\left( \bar{\sharp}_{j,m,k}\left( \left( p_{l}^{i}\right)
\right) \right) !}=\sum_{p=1}^{j}\frac{1}{\dprod\limits_{k}\left(
\sum_{l=1}^{j}\delta _{k,\sum_{i=1}^{m}p_{l}^{i}+\delta _{l,p}}\right) !}
\label{dSK}
\end{equation}%
As a consequence (\ref{thg}) can be replaced in (\ref{prL}) by:%
\begin{eqnarray}
&&\frac{\delta ^{\sum_{l}p_{l}^{m}}\left[ \Psi ^{\dagger }\left( \theta
_{m},Z_{m}\right) L\left( \Psi ^{\dagger }\left( \theta _{m},Z_{m}\right)
,\Psi \left( \theta _{m},Z_{m}\right) \right) \Psi \left( \theta
_{m},Z_{m}\right) \right] }{\dprod\limits_{l=1}^{j}\dprod%
\limits_{k_{l}^{i}=1}^{p_{l}^{m}}\delta \left\vert \Psi \left( \theta
^{\left( k_{l}^{i}\right) },Z_{_{l}}\right) \right\vert ^{2}}  \label{sdr} \\
&\rightarrow &\Psi ^{\dagger }\left( \theta _{m},Z_{m}\right) \frac{\delta
^{\sum_{l}p_{l}^{m}}\left[ L\left( \Psi ^{\dagger }\left( \theta
_{m},Z_{m}\right) ,\Psi \left( \theta _{m},Z_{m}\right) \right) \right] }{%
\dprod\limits_{l=1}^{j}\dprod\limits_{k_{l}^{i}=1}^{p_{l}^{m}}\delta
\left\vert \Psi \left( \theta ^{\left( k_{l}^{i}\right) },Z_{_{l}}\right)
\right\vert ^{2}}\Psi \left( \theta _{m},Z_{m}\right)  \notag \\
&&+\frac{\sharp _{j,m}\left( \left( p_{l}^{i}\right) \right) }{\bar{\sharp}%
_{j,m}\left( \left( p_{l}^{i}\right) \right) }\frac{\delta
^{\sum_{l}p_{l}^{m}}\left[ L\left( \Psi ^{\dagger }\left( \theta
_{m},Z_{m}\right) ,\Psi \left( \theta _{m},Z_{m}\right) \right) \right] }{%
\dprod\limits_{l=1}^{j}\dprod\limits_{k_{l}^{i}=1}^{p_{l}^{m}}\delta
\left\vert \Psi \left( \theta ^{\left( k_{l}^{i}\right) },Z_{_{l}}\right)
\right\vert ^{2}}  \notag
\end{eqnarray}%
where:%
\begin{equation}
\sharp _{j,m}\left( \left( p_{l}^{i}\right) \right) =\dprod\limits_{k}\left(
\sharp _{k,j,m}\left( \left( p_{l}^{i}\right) \right) \right) !  \label{dSL}
\end{equation}%
and $\bar{\sharp}_{j,m}\left( \left( p_{l}^{i}\right) \right) $ is defined
by (\ref{dSK}).

\subsubsection*{3.3.3 Reintroducing (\protect\ref{sdr}) in (\protect\ref{prL}%
) and local approximation}

The first contribution in the RHS of (\ref{sdr}):%
\begin{equation}
\Psi ^{\dagger }\left( \theta _{m},Z_{m}\right) \frac{\delta
^{\sum_{l}p_{l}^{m}}\left[ L\left( \Psi ^{\dagger }\left( \theta
_{m},Z_{m}\right) ,\Psi \left( \theta _{m},Z_{m}\right) \right) \right] }{%
\dprod\limits_{l=1}^{j}\dprod\limits_{k_{l}^{i}=1}^{p_{l}^{m}}\delta
\left\vert \Psi \left( \theta ^{\left( k_{l}^{i}\right) },Z_{_{l}}\right)
\right\vert ^{2}}\Psi \left( \theta _{m},Z_{m}\right)  \label{fRT}
\end{equation}%
once reintroduced in the effective action (\ref{prL}), yields:%
\begin{eqnarray}
&&\int_{\theta _{1}<...<\theta _{m}}\Psi ^{\dagger }\left( \theta
_{m},Z_{m}\right) \frac{\delta ^{\sum_{l}p_{l}^{m}}\left[ L\left( \Psi
^{\dagger }\left( \theta _{m},Z_{m}\right) ,\Psi \left( \theta
_{m},Z_{m}\right) \right) \right] }{\dprod\limits_{l=1}^{j}\dprod%
\limits_{k_{l}^{i}=1}^{p_{l}^{m}}\delta \left\vert \Psi \left( \theta
^{\left( k_{l}^{i}\right) },Z_{_{l}}\right) \right\vert ^{2}}\Psi \left(
\theta _{m},Z_{m}\right) \int \left( \dprod\limits_{l=1}^{j}\Psi ^{\dagger
}\left( \theta _{f}^{\left( l\right) },Z_{l}\right) \right)  \label{cbN} \\
&&\times \dprod\limits_{i=1}^{m-1}\left[ \int_{\dprod\limits_{l=1}^{j}\left[
\theta _{i}^{\left( l\right) },\theta _{f}^{\left( l\right) }\right]
^{p_{l}^{i}}}\frac{\delta ^{\sum_{l}p_{l}^{i}}\left[ \Psi ^{\dagger }\left(
\theta _{i},Z_{i}\right) L\left( \Psi ^{\dagger }\left( \theta
_{i},Z_{i}\right) ,\Psi \left( \theta _{i},Z_{i}\right) \right) \Psi \left(
\theta _{i},Z_{i}\right) \right] }{\dprod\limits_{l=1}^{j}\dprod%
\limits_{k_{l}^{i}=1}^{p_{l}^{i}}\delta \left\vert \Psi \left( \theta
^{\left( k_{l}^{i}\right) },Z_{_{l}}\right) \right\vert ^{2}}%
\dprod\limits_{l=1}^{j}\dprod\limits_{k_{l}^{i}=1}^{p_{l}^{i}}d\theta
^{\left( k_{l}^{i}\right) }\right] d\left( \theta _{i},Z_{i}\right)  \notag
\\
&&\times \frac{\exp \left( -\Lambda _{1}\left( \theta _{f}^{\left( l\right)
}-\theta _{i}^{\left( l\right) }\right) \right) }{\dprod\limits_{k}\left(
\sharp _{j,k}\left( \left( p_{l}^{i}\right) \right) \right) !\Lambda
^{\sum_{i}p_{l}^{i}}}\left( \dprod\limits_{l=1}^{j}\Psi \left( \theta
_{i}^{\left( l\right) },Z_{l}\right) \right)  \notag \\
&=&\int_{\theta _{m}}\Psi ^{\dagger }\left( \theta _{m},Z_{m}\right) \frac{%
\delta ^{\sum_{l}p_{l}^{m}}\left[ L\left( \Psi ^{\dagger }\left( \theta
_{m},Z_{m}\right) ,\Psi \left( \theta _{m},Z_{m}\right) \right) \right] }{%
\dprod\limits_{l=1}^{j}\dprod\limits_{k_{l}^{i}=1}^{p_{l}^{m}}\delta
\left\vert \Psi \left( \theta ^{\left( k_{l}^{i}\right) },Z_{_{l}}\right)
\right\vert ^{2}}\Psi \left( \theta _{m},Z_{m}\right)  \notag \\
&&\times \int \left( \dprod\limits_{l=1}^{j}\Psi ^{\dagger }\left( \theta
_{f}^{\left( l\right) },Z_{l}\right) \right) \dprod\limits_{i=1}^{m-1}\left[
\int_{\dprod\limits_{l=1}^{j}\left[ \theta _{i}^{\left( l\right) },\theta
_{f}^{\left( l\right) }\right] ^{p_{l}^{i}}}\frac{\delta ^{\sum_{l}p_{l}^{i}}%
\left[ \hat{S}_{cl,\theta _{m}}\left( \Psi ^{\dagger },\Psi \right) \right] 
}{\dprod\limits_{l=1}^{j}\dprod\limits_{k_{l}^{i}=1}^{p_{l}^{i}}\delta
\left\vert \Psi \left( \theta ^{\left( k_{l}^{i}\right) },Z_{_{l}}\right)
\right\vert ^{2}}\dprod\limits_{l=1}^{j}\dprod%
\limits_{k_{l}^{i}=1}^{p_{l}^{i}}d\theta ^{\left( k_{l}^{i}\right) }\right] 
\notag \\
&&\times \frac{\exp \left( -\Lambda _{1}\left( \theta _{f}^{\left( l\right)
}-\theta _{i}^{\left( l\right) }\right) \right) }{\left( m-1\right)
!\dprod\limits_{k}\left( \sharp _{j,k}\left( \left( p_{l}^{i}\right) \right)
\right) !\Lambda ^{\sum_{i}p_{l}^{i}}}\left( \dprod\limits_{l=1}^{j}\Psi
\left( \theta _{i}^{\left( l\right) },Z_{l}\right) \right)  \notag
\end{eqnarray}%
with:%
\begin{equation*}
\hat{S}_{cl,\theta _{m}}\left( \Psi ^{\dagger },\Psi \right) =\int^{\theta
_{m}}\Psi ^{\dagger }\left( \theta ,Z\right) L\left( \Psi ^{\dagger }\left(
\theta ,Z\right) ,\Psi \left( \theta ,Z\right) \right) \Psi \left( \theta
,Z\right)
\end{equation*}%
The second contribution in (\ref{sdr}) is constrained by that, among the
fields $\Psi ^{\dagger }\left( \theta _{f}^{\left( l\right) },Z_{l}\right) $
one of them is set to $\Psi ^{\dagger }\left( \theta _{m},Z_{m}\right) $.
This leads to the following contribution to (\ref{prL}):%
\begin{eqnarray}
&&\int_{\theta _{m}}\Psi ^{\dagger }\left( \theta _{m},Z_{m}\right) \frac{%
\delta ^{\sum_{l}p_{l}^{m}}\left[ L\left( \Psi ^{\dagger }\left( \theta
_{m},Z_{m}\right) ,\Psi \left( \theta _{m},Z_{m}\right) \right) \right] }{%
\dprod\limits_{l=1}^{j-1}\dprod\limits_{k_{l}^{i}=1}^{p_{l}^{m}}\delta
\left\vert \Psi \left( \theta ^{\left( k_{l}^{i}\right) },Z_{_{l}}\right)
\right\vert ^{2}}\Psi \left( \theta _{i}^{\left( m\right) },Z_{m}\right)
\label{cbT} \\
&&\times \int \left( \dprod\limits_{l=1}^{j-1}\Psi ^{\dagger }\left( \theta
_{f}^{\left( l\right) },Z_{l}\right) \right) \dprod\limits_{i=1}^{m-1}\left[
\int_{\dprod\limits_{l=1}^{j-1}\left[ \theta _{i}^{\left( l\right) },\theta
_{f}^{\left( l\right) }\right] ^{p_{l}^{i}}}\frac{\delta ^{\sum_{l}p_{l}^{i}}%
\left[ \hat{S}_{cl,\theta _{m}}\left( \Psi ^{\dagger },\Psi \right) \right] 
}{\dprod\limits_{l=1}^{j}\dprod\limits_{k_{l}^{i}=1}^{p_{l}^{i}}\delta
\left\vert \Psi \left( \theta ^{\left( k_{l}^{i}\right) },Z_{_{l}}\right)
\right\vert ^{2}}\dprod\limits_{l=1}^{j-1}\dprod%
\limits_{k_{l}^{i}=1}^{p_{l}^{i}}d\theta ^{\left( k_{l}^{i}\right) }\right] 
\notag \\
&&\times \frac{\sharp _{j,m}\left( \left( p_{l}^{i}\right) \right) }{\bar{%
\sharp}_{j,m}\left( \left( p_{l}^{i}\right) \right) }\frac{\exp \left(
-\Lambda _{1}\left( \theta _{f}^{\left( l\right) }-\theta _{i}^{\left(
l\right) }\right) \right) }{\left( m-1\right) !\Lambda ^{\sum_{i}p_{l}^{i}}}%
\left( \dprod\limits_{l=1}^{j-1}\Psi \left( \theta _{i}^{\left( l\right)
},Z_{l}\right) \right)  \notag
\end{eqnarray}%
and the product of derivatives can be decomposed by isolating those
corresponding to $Z_{_{m}}$:%
\begin{eqnarray*}
&&\int_{\dprod\limits_{l=1}^{j-1}\left[ \theta _{i}^{\left( l\right)
},\theta _{f}^{\left( l\right) }\right] ^{p_{l}^{i}}}\frac{\delta
^{\sum_{l}p_{l}^{i}}\left[ \hat{S}_{cl,\theta _{m}}\left( \Psi ^{\dagger
},\Psi \right) \right] }{\dprod\limits_{l=1}^{j}\dprod%
\limits_{k_{l}^{i}=1}^{p_{l}^{i}}\delta \left\vert \Psi \left( \theta
^{\left( k_{l}^{i}\right) },Z_{_{l}}\right) \right\vert ^{2}}%
\dprod\limits_{l=1}^{j-1}\dprod\limits_{k_{l}^{i}=1}^{p_{l}^{i}}d\theta
^{\left( k_{l}^{i}\right) } \\
&=&\int_{\left[ \theta _{i}^{\left( l\right) },\theta _{f}^{\left( l\right) }%
\right] ^{p^{i}}}\frac{\delta ^{p^{i}}}{\dprod\limits_{k_{l}^{i}=1}^{p^{i}}%
\delta \left\vert \Psi \left( \theta ^{\left( k^{i}\right) },Z_{m}\right)
\right\vert ^{2}}d\theta ^{\left( k^{i}\right)
}\int_{\dprod\limits_{l=1}^{j-1}\left[ \theta _{i}^{\left( l\right) },\theta
_{f}^{\left( l\right) }\right] ^{p_{l}^{i}}}\frac{\delta ^{\sum_{l}p_{l}^{i}}%
\left[ \hat{S}_{cl,\theta _{m}}\left( \Psi ^{\dagger },\Psi \right) \right] 
}{\dprod\limits_{l=1}^{j-1}\dprod\limits_{k_{l}^{i}=1}^{p_{l}^{i}}\delta
\left\vert \Psi \left( \theta ^{\left( k_{l}^{i}\right) },Z_{_{l}}\right)
\right\vert ^{2}}\dprod\limits_{l=1}^{j-1}\dprod%
\limits_{k_{l}^{i}=1}^{p_{l}^{i}}d\theta ^{\left( k_{l}^{i}\right) }
\end{eqnarray*}

Gathering the contributions (\ref{cbN}) and (\ref{cbT})\ yields $\Gamma
\left( \Psi ^{\dagger },\Psi \right) $:%
\begin{eqnarray}
&&\Gamma \left( \Psi ^{\dagger },\Psi \right) =\hat{S}_{cl}\left( \Psi
^{\dagger },\Psi \right)  \label{fcV} \\
&&+\sum_{\substack{ j\geqslant 1  \\ m\geqslant 1}}\sum_{\substack{ %
p_{l},\left( p_{l}^{i}\right) _{m\times j}  \\ p_{l}+\sum_{i}p_{l}^{i}%
\geqslant 2}}\int \Psi ^{\dagger }\left( \theta ,Z\right) \frac{\delta
^{\sum_{l}p_{l}}\left[ L\left( \Psi ^{\dagger }\left( \theta ,Z\right) ,\Psi
\left( \theta ,Z\right) \right) \right] }{\dprod\limits_{l=1}^{j}\dprod%
\limits_{k_{l}^{i}=1}^{p_{l}}\delta \left\vert \Psi \left( \theta ^{\left(
k_{l}^{i}\right) },Z_{_{l}}\right) \right\vert ^{2}}a_{j,m}\left( \theta
,\theta _{i}\right) \Psi \left( \theta _{i},Z\right)  \notag \\
&&\times \int \left( \dprod\limits_{l=1}^{j}\Psi ^{\dagger }\left( \theta
_{f}^{\left( l\right) },Z_{l}\right) \right) \dprod\limits_{i=1}^{m}\left[
\int_{\dprod\limits_{l=1}^{j}\left[ \theta _{i}^{\left( l\right) },\theta
_{f}^{\left( l\right) }\right] ^{p_{l}^{i}}}\frac{\delta ^{\sum_{l}p_{l}^{i}}%
\left[ \hat{S}_{cl,\theta }\left( \Psi ^{\dagger },\Psi \right) \right] }{%
\dprod\limits_{l=1}^{j}\dprod\limits_{k_{l}^{i}=1}^{p_{l}^{i}}\delta
\left\vert \Psi \left( \theta ^{\left( k_{l}^{i}\right) },Z_{_{l}}\right)
\right\vert ^{2}}\dprod\limits_{l=1}^{j}\dprod%
\limits_{k_{l}^{i}=1}^{p_{l}^{i}}d\theta ^{\left( k_{l}^{i}\right) }\right] 
\notag \\
&&\times \frac{\exp \left( -\Lambda _{1}\left( \theta _{f}^{\left( l\right)
}-\theta _{i}^{\left( l\right) }\right) \right) }{m!\dprod\limits_{k}\left(
\sharp _{j,k}\left( \left( p_{l}^{i}\right) \right) \right) !\Lambda
^{\sum_{i}p_{l}^{i}}}\left( \dprod\limits_{l=1}^{j}\Psi \left( \theta
_{i}^{\left( l\right) },Z_{l}\right) \right)  \notag
\end{eqnarray}%
with 
\begin{equation*}
a_{1}\left( \theta ,\theta _{i}\right) =\frac{\exp \left( -\Lambda
_{1}\left( \theta -\theta _{i}^{\left( l\right) }\right) \right) }{\Lambda
^{\sum_{i}p_{l}^{i}}}\prod\limits_{i=1}^{m}\int_{\left[ \theta _{i}^{\left(
l\right) },\theta _{f}^{\left( l\right) }\right] ^{p^{i}}}\frac{\delta
^{p^{i}}}{\dprod\limits_{k_{l}^{i}=1}^{p^{i}}\delta \left\vert \Psi \left(
\theta ^{\left( k^{i}\right) },Z_{m}\right) \right\vert ^{2}}d\theta
^{\left( k^{i}\right) }
\end{equation*}%
\ and 
\begin{equation*}
a_{j,m}\left( \theta ,\theta _{i}\right) =\delta \left( \theta -\theta
_{i}\right) +\frac{\sharp _{j+1,m+1}\left( \left( p_{l}\right) ,\left(
p_{l}^{i}\right) \right) }{\bar{\sharp}_{j+1,m+1}\left( \left( p_{l}\right)
,\left( p_{l}^{i}\right) \right) }\frac{\exp \left( -\Lambda _{1}\left(
\theta -\theta _{i}^{\left( l\right) }\right) \right) }{\Lambda
^{\sum_{i}p_{l}^{i}}}\prod\limits_{i=1}^{m}\int_{\left[ \theta _{i}^{\left(
l\right) },\theta _{f}^{\left( l\right) }\right] ^{p^{i}}}\frac{\delta
^{p^{i}}}{\dprod\limits_{k_{l}^{i}=1}^{p^{i}}\delta \left\vert \Psi \left(
\theta ^{\left( k^{i}\right) },Z_{m}\right) \right\vert ^{2}}d\theta
^{\left( k^{i}\right) }
\end{equation*}%
for $j>1$. The derivatives $i=1,,,,m$ implicitly act independently on each
factor:%
\begin{equation*}
\int_{\dprod\limits_{l=1}^{j}\left[ \theta _{i}^{\left( l\right) },\theta
_{f}^{\left( l\right) }\right] ^{p_{l}^{i}}}\frac{\delta ^{\sum_{l}p_{l}^{i}}%
\left[ \hat{S}_{cl,\theta }\left( \Psi ^{\dagger },\Psi \right) \right] }{%
\dprod\limits_{l=1}^{j}\dprod\limits_{k_{l}^{i}=1}^{p_{l}^{i}}\delta
\left\vert \Psi \left( \theta ^{\left( k_{l}^{i}\right) },Z_{_{l}}\right)
\right\vert ^{2}}\dprod\limits_{l=1}^{j}\dprod%
\limits_{k_{l}^{i}=1}^{p_{l}^{i}}d\theta ^{\left( k_{l}^{i}\right) }
\end{equation*}%
in (\ref{fcV}).

The notations $\sharp _{j+1,m+1}\left( \left( p_{l}\right) ,\left(
p_{l}^{i}\right) \right) $ and $\bar{\sharp}_{j+1,m+1}\left( \left(
p_{l}\right) ,\left( p_{l}^{i}\right) \right) $ are defined by (\ref{dSK})
and (\ref{dSL}) in which the multi-indices $\left( p_{l}^{i}\right)
_{l=1,...l}^{i=1,...m}$ are replaced by the collection for $m+1$ and $j+1$
obtained by gathering $\left( p_{l}\right) _{i=1...m}$ and $\left(
p_{l}^{i}\right) _{l=1,...l}^{i=1,...m}$.

In the local approximation, equation (\ref{fcV}) simplifies and writes:%
\begin{eqnarray}
&&\Gamma \left( \Psi ^{\dagger },\Psi \right) =\hat{S}_{cl}\left( \Psi
^{\dagger },\Psi \right)  \label{fCR} \\
&&+\sum_{\substack{ j\geqslant 1  \\ m\geqslant 1}}\sum_{\substack{ %
p_{l},\left( p_{l}^{i}\right) _{m\times j}  \\ p_{l}+\sum_{i}p_{l}^{i}%
\geqslant 2}}\int \int \Psi ^{\dagger }\left( \theta ,Z\right) \frac{\delta
^{\sum_{l}p_{l}}\left[ L\left( \Psi ^{\dagger }\left( \theta ,Z\right) ,\Psi
\left( \theta ,Z\right) \right) \right] }{\dprod\limits_{l=1}^{j}\dprod%
\limits_{k_{l}^{i}=1}^{p_{l}}\delta \left\vert \Psi \left( \theta ^{\left(
l\right) },Z_{_{l}}\right) \right\vert ^{2}}\Psi \left( \theta ,Z\right) 
\notag \\
&&\times \left( \dprod\limits_{l=1}^{j}\Psi ^{\dagger }\left( \theta
^{\left( l\right) },Z_{l}\right) \right) \dprod\limits_{i=1}^{m}\left[ \frac{%
\delta ^{\sum_{l}p_{l}^{i}}\left[ \hat{S}_{cl,\theta }\left( \Psi ^{\dagger
},\Psi \right) \right] }{\dprod\limits_{l=1}^{j}\delta
^{p_{l}^{i}}\left\vert \Psi \left( \theta _{l},Z_{_{l}}\right) \right\vert
^{2}}\right] \frac{\left( \dprod\limits_{l=1}^{j}\Psi \left( \theta ^{\left(
l\right) },Z_{l}\right) \right) }{m!\dprod\limits_{k}\left( \sharp
_{j,k}\right) !\Lambda ^{\sum_{i}p_{l}^{i}}}\dprod\limits_{l=1}^{j}dZ_{l}d%
\theta _{l}  \notag
\end{eqnarray}%
with:%
\begin{equation*}
a_{1,m}=1
\end{equation*}%
\ and 
\begin{equation*}
a_{j,m}=1+\frac{\sharp _{j+1,m+1}\left( \left( p_{l}\right) ,\left(
p_{l}^{i}\right) \right) }{\bar{\sharp}_{j+1,m+1}\left( \left( p_{l}\right)
,\left( p_{l}^{i}\right) \right) }
\end{equation*}%
for $j>1$.

\subsubsection*{3.3.4 Recursive expansion of (\protect\ref{fCR})}

The procedure that led to rewrite (\ref{fRT}), can be applied in (\ref{fCR}%
), to the terms:%
\begin{equation}
\frac{1}{m!}\dprod\limits_{i=1}^{m}\left[ \frac{\delta ^{\sum_{l}p_{l}^{i}}%
\left[ \hat{S}_{cl,\theta }\left( \Psi ^{\dagger },\Psi \right) \right] }{%
\dprod\limits_{l=1}^{j}\delta ^{p_{l}^{i}}\left\vert \Psi \left( \theta
_{l},Z_{_{l}}\right) \right\vert ^{2}}\right]  \label{rPC}
\end{equation}%
Expression (\ref{rPC}) can be expanded recursively. As we did above,
expression (\ref{rPC}) is replaced by an integration over the restricted
domain $\theta _{1}<\theta _{m},...,\theta _{m-1}<\theta _{m}<\theta $:%
\begin{equation*}
\int_{\theta _{1}<\theta _{m},...,\theta _{m-1}<\theta _{m}<\theta
}\dprod\limits_{i=1}^{m}\left[ \frac{\delta ^{\sum_{l}p_{l}^{i}}\left[ \Psi
^{\dagger }\left( \theta _{i},Z_{i}\right) L\left( \Psi ^{\dagger }\left(
\theta _{i},Z_{i}\right) ,\Psi \left( \theta _{i},Z_{i}\right) \right) \Psi
\left( \theta _{i},Z_{i}\right) \right] }{\dprod\limits_{l=1}^{j}\delta
^{p_{l}^{i}}\left\vert \Psi \left( \theta _{l},Z_{_{l}}\right) \right\vert
^{2}}\right] d\left( \theta _{i},Z_{i}\right)
\end{equation*}%
and formula (\ref{sdr}) applies to replace: 
\begin{equation*}
\frac{\delta ^{\sum_{l}p_{l}^{m}}\left[ \Psi ^{\dagger }\left( \theta
_{m},Z_{m}\right) L\left( \Psi ^{\dagger }\left( \theta _{m},Z_{m}\right)
,\Psi \left( \theta _{m},Z_{m}\right) \right) \Psi \left( \theta
_{m},Z_{m}\right) \right] }{\dprod\limits_{l=1}^{j}\delta
^{p_{l}^{i}}\left\vert \Psi \left( \theta _{l},Z_{_{l}}\right) \right\vert
^{2}}
\end{equation*}%
by:%
\begin{equation}
\Psi ^{\dagger }\left( \theta _{m},Z_{m}\right) \frac{\delta
^{\sum_{l}p_{l}^{m}}\left[ L\left( \Psi ^{\dagger }\left( \theta
_{m},Z_{m}\right) ,\Psi \left( \theta _{m},Z_{m}\right) \right) \right] }{%
\dprod\limits_{k_{l}^{i}=1}^{p_{l}^{m}}\delta ^{p_{l}^{m}}\left\vert \Psi
\left( \theta ^{\left( l\right) },Z_{_{l}}\right) \right\vert ^{2}}\Psi
\left( \theta _{m},Z_{m}\right) +\frac{\sharp _{j,m}\left( \left(
p_{l}^{i}\right) \right) }{\bar{\sharp}_{j,m}\left( \left( p_{l}^{i}\right)
\right) }\frac{\delta ^{\sum_{l}p_{l}^{m}}\left[ L\left( \Psi ^{\dagger
}\left( \theta _{m},Z_{m}\right) ,\Psi \left( \theta _{m},Z_{m}\right)
\right) \right] }{\dprod\limits_{k_{l}^{i}=1}^{p_{l}^{m}}\delta
^{p_{l}^{m}}\left\vert \Psi \left( \theta ^{\left( l\right)
},Z_{_{l}}\right) \right\vert ^{2}}  \label{prS}
\end{equation}%
Inserting the first term of expression (\ref{prS}) in (\ref{fCR}) yields the
contribution:%
\begin{eqnarray}
&&\sum_{\substack{ j\geqslant 1  \\ m\geqslant 1}}\sum_{\substack{ %
p_{l},\left( p_{l}^{i}\right) _{m\times j}  \\ p_{l}+\sum_{i}p_{l}^{i}%
\geqslant 2}}\int \Psi ^{\dagger }\left( \theta ,Z\right) \frac{\delta
^{\sum_{l}p_{l}}\left[ L\left( \Psi ^{\dagger }\left( \theta ,Z\right) ,\Psi
\left( \theta ,Z\right) \right) \right] }{\dprod\limits_{l=1}^{j}\dprod%
\limits_{k_{l}^{i}=1}^{p_{l}}\delta \left\vert \Psi \left( \theta ^{\left(
l\right) },Z_{_{l}}\right) \right\vert ^{2}}\Psi \left( \theta ,Z\right)
\label{trN} \\
&&\times \int \Psi ^{\dagger }\left( \theta _{m},Z_{m}\right) \frac{\delta
^{\sum_{l}p_{l}^{m}}\left[ L\left( \Psi ^{\dagger }\left( \theta
_{m},Z_{m}\right) ,\Psi \left( \theta _{m},Z_{m}\right) \right) \right] }{%
\dprod\limits_{l=1}^{j}\delta ^{p_{l}^{m}}\left\vert \Psi \left( \theta
^{\left( l\right) },Z_{_{l}}\right) \right\vert ^{2}}\Psi \left( \theta
_{m},Z_{m}\right)  \notag \\
&&\times \int a_{j,m}\left( \dprod\limits_{l=1}^{j}\Psi ^{\dagger }\left(
\theta ^{\left( l\right) },Z_{l}\right) \right) \dprod\limits_{i=1}^{m-1} 
\left[ \frac{\delta ^{\sum_{l}p_{l}^{i}}\left[ \hat{S}_{cl,\theta }\left(
\Psi ^{\dagger },\Psi \right) \right] }{\dprod\limits_{l=1}^{j}\delta
^{p_{l}^{i}}\left\vert \Psi \left( \theta ^{\left( l\right)
},Z_{_{l}}\right) \right\vert ^{2}}\right] \frac{\left(
\dprod\limits_{l=1}^{j}\Psi \left( \theta _{i}^{\left( l\right)
},Z_{l}\right) \right) }{\left( m-1\right) !\dprod\limits_{k}\left( \sharp
_{j,k}\left( \left( p_{l}^{i}\right) \right) \right) !\Lambda
^{\sum_{i}p_{l}^{i}}}\dprod\limits_{l=1}^{j}dZ_{l}d\theta ^{\left( l\right) }
\notag
\end{eqnarray}

The second contribution of (\ref{prS}), after insertion in (\ref{fcV})
yields:%
\begin{eqnarray}
&&\sum_{\substack{ j\geqslant 1  \\ m\geqslant 1}}\sum_{\substack{ %
p_{l},\left( p_{l}^{i}\right) _{m\times j}  \\ p_{l}+\sum_{i}p_{l}^{i}%
\geqslant 2}}\int \Psi ^{\dagger }\left( \theta ,Z\right) \frac{\delta
^{\sum_{l}p_{l}}\left[ L\left( \Psi ^{\dagger }\left( \theta ,Z\right) ,\Psi
\left( \theta ,Z\right) \right) \right] }{\dprod\limits_{l=1}^{j}\delta
^{p_{l}^{i}}\left\vert \Psi \left( \theta ^{\left( l\right)
},Z_{_{l}}\right) \right\vert ^{2}}\Psi \left( \theta ,Z\right)  \label{tRT}
\\
&&\times \int \Psi ^{\dagger }\left( \theta _{m},Z_{m}\right) \frac{\delta
^{\sum_{l}p_{l}^{m}}\left[ L\left( \Psi ^{\dagger }\left( \theta
_{m},Z_{m}\right) ,\Psi \left( \theta _{m},Z_{m}\right) \right) \right] }{%
\dprod\limits_{l=1}^{j-1}\delta ^{p_{l}^{m}}\left\vert \Psi \left( \theta
^{\left( l\right) },Z_{_{l}}\right) \right\vert ^{2}}\Psi \left( \theta
_{m},Z_{m}\right)  \notag \\
&&\times \int \left( \dprod\limits_{l=1}^{j-1}\Psi ^{\dagger }\left( \theta
^{\left( l\right) },Z_{l}\right) \right) \dprod\limits_{i=1}^{m-1}\left[ 
\frac{\delta ^{\sum_{l}p_{l}^{i}}\left[ \hat{S}_{cl,\theta _{m}}\left( \Psi
^{\dagger },\Psi \right) \right] }{\dprod\limits_{l=1}^{j-1}\delta
^{p_{l}^{i}}\left\vert \Psi \left( \theta ^{\left( l\right)
},Z_{_{l}}\right) \right\vert ^{2}}\right] \times \frac{\sharp _{j,m}\left(
\left( p_{l}^{m}\right) ,\left( p_{l}^{i}\right) \right) }{\bar{\sharp}%
_{j,m}\left( \left( p_{l}^{m}\right) ,\left( p_{l}^{i}\right) \right) }%
a_{j,m}\frac{\left( \dprod\limits_{l=1}^{j-1}\Psi \left( \theta ^{\left(
l\right) },Z_{l}\right) \right) }{\left( m-1\right) !\Lambda
^{\sum_{i}p_{l}^{i}}}\dprod\limits_{l=1}^{j-1}dZ_{l}d\theta ^{\left(
l\right) }  \notag
\end{eqnarray}

Isolating the derivative with respect to $\left\vert \Psi ^{\dagger }\left(
\theta _{m},Z_{m}\right) \right\vert ^{2}$ in the first term of (\ref{tRT})
allows to write:%
\begin{eqnarray}
&&\sum_{\left( p_{l}\right) ^{1\times j}}\int \Psi ^{\dagger }\left( \theta
,Z\right) \frac{\delta ^{\sum_{l=1}^{j}p_{l}}\left[ L\left( \Psi ^{\dagger
}\left( \theta ,Z\right) ,\Psi \left( \theta ,Z\right) \right) \right] }{%
\dprod\limits_{l=1}^{j}\delta ^{p_{l}^{i}}\left\vert \Psi \left( \theta
^{\left( l\right) },Z_{_{l}}\right) \right\vert ^{2}}\Psi \left( \theta
,Z\right)  \label{tRN} \\
&=&\sum_{p_{m},\left( p_{l}\right) ^{1\times \left( j-1\right) }}\left( 
\frac{\delta }{\delta \left\vert \Psi ^{\dagger }\left( \theta
_{m},Z_{m}\right) \right\vert ^{2}}\right) ^{p_{m}}\int \Psi ^{\dagger
}\left( \theta ,Z\right) \frac{\delta ^{\sum_{l=1}^{j-1}p_{l}}\left[ L\left(
\Psi ^{\dagger }\left( \theta ,Z\right) ,\Psi \left( \theta ,Z\right)
\right) \right] }{\dprod\limits_{l=1}^{j-1}\delta ^{p_{l}^{i}}\left\vert
\Psi \left( \theta ^{\left( l\right) },Z_{_{l}}\right) \right\vert ^{2}}\Psi
\left( \theta ,Z\right)  \notag
\end{eqnarray}%
We replace (\ref{tRN}) in (\ref{tRT}) and change of variable $j-1\rightarrow
j$. Then we gather (\ref{trN}) and (\ref{tRT}) to obtain:%
\begin{eqnarray}
\Gamma \left( \Psi ^{\dagger },\Psi \right) &=&\hat{S}_{cl}\left( \Psi
^{\dagger },\Psi \right) +\sum_{\substack{ j\geqslant 1  \\ m\geqslant 1}}%
\sum _{\substack{ p_{l},\left( p_{l}^{i}\right) _{m\times j}  \\ %
p_{l}+\sum_{i}p_{l}^{i}\geqslant 2}}\left( 1+\sum_{p_{m}}\left( \frac{\sharp
_{j+1,m}\left( \left( p_{l}^{m}\right) ,\left( p_{l}^{i}\right) \right) }{%
\bar{\sharp}_{j+1,m}\left( \left( p_{l}^{m}\right) ,\left( p_{l}^{i}\right)
\right) }\left( \frac{\delta }{\delta \left\vert \Psi ^{\dagger }\left(
\theta _{m},Z_{m}\right) \right\vert ^{2}}\right) ^{p_{m}}\right) \right) 
\notag \\
&&\times \int \Psi ^{\dagger }\left( \theta ,Z\right) \frac{\delta
^{\sum_{l=1}^{j}p_{l}}\left[ L\left( \Psi ^{\dagger }\left( \theta ,Z\right)
,\Psi \left( \theta ,Z\right) \right) \right] }{\dprod\limits_{l=1}^{j}%
\delta ^{p_{l}^{i}}\left\vert \Psi \left( \theta ^{\left( l\right)
},Z_{_{l}}\right) \right\vert ^{2}}\Psi \left( \theta ,Z\right)  \label{frT}
\\
&&\times \int \Psi ^{\dagger }\left( \theta _{m},Z_{m}\right) \frac{\delta
^{\sum_{l}p_{l}^{m}}\left[ L\left( \Psi ^{\dagger }\left( \theta
_{m},Z_{m}\right) ,\Psi \left( \theta _{m},Z_{m}\right) \right) \right] }{%
\dprod\limits_{l=1}^{j}\delta ^{p_{l}^{m}}\left\vert \Psi \left( \theta
^{\left( l\right) },Z_{_{l}}\right) \right\vert ^{2}}\Psi \left( \theta
_{m},Z_{m}\right)  \notag \\
&&\times \int a_{j,m}\left( \dprod\limits_{l=1}^{j}\Psi ^{\dagger }\left(
\theta _{f}^{\left( l\right) },Z_{l}\right) \right)
\dprod\limits_{i=1}^{m-1} \left[ \frac{\delta ^{\sum_{l}p_{l}^{i}}\left[ 
\hat{S}_{cl,\theta }\left( \Psi ^{\dagger },\Psi \right) \right] }{%
\dprod\limits_{l=1}^{j}\delta ^{p_{l}^{i}}\left\vert \Psi \left( \theta
^{\left( k_{l}^{i}\right) },Z_{_{l}}\right) \right\vert ^{2}}\right] \frac{%
\left( \dprod\limits_{l=1}^{j}\Psi \left( \theta _{i}^{\left( l\right)
},Z_{l}\right) \right) }{\left( m-1\right) !\Lambda ^{\sum_{i}p_{l}^{i}}}%
dZ_{l}d\theta ^{\left( l\right) }  \notag
\end{eqnarray}%
This relation can be further iterated and we find:%
\begin{eqnarray}
&&\Gamma \left( \Psi ^{\dagger },\Psi \right)  \label{fCT} \\
&=&\hat{S}_{cl}\left( \Psi ^{\dagger },\Psi \right) +\sum_{\substack{ %
j\geqslant 1  \\ m\geqslant 1}}\sum_{\substack{ p_{l},\left(
p_{l}^{i}\right) _{m\times j}  \\ p_{l}+\sum_{i}p_{l}^{i}\geqslant 2}}\int
\int \left( \dprod\limits_{l=1}^{j}\Psi ^{\dagger }\left( \theta ^{\left(
l\right) },Z_{l}\right) \right)  \notag \\
&&\prod\limits_{i=1}^{m}\left\{ \left( \int_{\theta _{i}<\theta _{i+1}}\Psi
^{\dagger }\left( \theta _{i},Z_{i}\right) \frac{\delta ^{\sum_{l}p_{l}^{i}}%
\left[ L\left( \Psi ^{\dagger }\left( \theta _{i},Z_{i}\right) ,\Psi \left(
\theta _{i},Z_{i}\right) \right) \right] }{\dprod\limits_{l=1}^{j}\delta
^{p_{l}^{i}}\left\vert \Psi \left( \theta ^{\left( l\right)
},Z_{_{l}}\right) \right\vert ^{2}}\Psi \left( \theta _{i},Z_{i}\right)
\right) \right.  \notag \\
&&\left. \times \left( 1+\sum_{p_{i}}\left( \frac{\sharp _{j+1,i}\left(
\left( p_{i}\right) ,\left( p_{l}^{i}\right) \right) }{\bar{\sharp}%
_{j+1,i}\left( \left( p_{i}\right) ,\left( p_{l}^{i}\right) \right) }\left( 
\frac{\delta }{\delta \left\vert \Psi ^{\dagger }\left( \theta
_{i},Z_{i}\right) \right\vert ^{2}}\right) ^{p_{i}}\right) \right) \right\} 
\notag \\
&&\times a_{j,m}\left( \int \Psi ^{\dagger }\left( \theta ,Z\right) \frac{%
\delta ^{\sum_{l=1}^{j}p_{l}}\left[ L\left( \Psi ^{\dagger }\left( \theta
,Z\right) ,\Psi \left( \theta ,Z\right) \right) \right] }{%
\dprod\limits_{l=1}^{j}\delta ^{p_{l}^{i}}\left\vert \Psi \left( \theta
^{\left( l\right) },Z_{_{l}}\right) \right\vert ^{2}}\Psi \left( \theta
,Z\right) \right) \frac{\left( \dprod\limits_{l=1}^{j}\Psi \left( \theta
^{\left( l\right) },Z_{l}\right) \right) }{\left( m-1\right) !\Lambda
^{\sum_{i}p_{l}^{i}}}\prod\limits_{i=1}^{m}d\theta
_{i}dZ_{_{l}}\dprod\limits_{l=1}^{j}d\theta ^{\left( l\right) }dZ_{l}  \notag
\end{eqnarray}

\subsubsection*{3.3.5 Limit of slowly time dependent fields}

In the limit of slowly time dependent fields $\Psi \left( \theta
_{i},Z_{i}\right) $ and of a potential slowly varying in field $\Psi \left(
\theta _{i},Z_{i}\right) $, we have:%
\begin{eqnarray*}
&&\int_{\theta _{i}<\theta _{i+1}}\Psi ^{\dagger }\left( \theta
_{i},Z_{i}\right) \frac{\delta ^{\sum_{l}p_{l}^{i}}\left[ L\left( \Psi
^{\dagger }\left( \theta _{i},Z_{i}\right) ,\Psi \left( \theta
_{i},Z_{i}\right) \right) \right] }{\dprod\limits_{l=1}^{j}\delta
^{p_{l}^{i}}\left\vert \Psi \left( \theta ^{\left( l\right)
},Z_{_{l}}\right) \right\vert ^{2}}\Psi \left( \theta _{i},Z_{i}\right) \\
&=&\int_{\theta _{i}<\theta _{i+1}}\Psi ^{\dagger }\left( \theta
_{i},Z_{i}\right) \left\{ \frac{1}{2}\frac{\delta ^{\sum_{l}p_{l}^{i}}\left(
\nabla _{\theta }\omega ^{-1}\left( \theta _{i},Z_{i},\left\vert \Psi
\right\vert ^{2}\right) +U\left( \theta _{i},Z_{i},\left\vert \Psi
\right\vert ^{2}\right) \right) }{\dprod\limits_{l=1}^{j}\delta
^{p_{l}^{i}}\left\vert \Psi \left( \theta ^{\left( l\right)
},Z_{_{l}}\right) \right\vert ^{2}}\right\} \Psi \left( \theta
_{i},Z_{i}\right) \\
&\simeq &\int_{\theta _{i}<\theta _{i+1}}\Psi ^{\dagger }\left( \theta
_{i},Z_{i}\right) \left\{ \frac{1}{2}\frac{\delta ^{\sum_{l}p_{l}^{i}}\left(
\nabla _{\theta }\omega ^{-1}\left( \theta _{i},Z_{i},\left\vert \Psi
\right\vert ^{2}\right) \right) }{\dprod\limits_{l=1}^{j}\delta
^{p_{l}^{i}}\left\vert \Psi \left( \theta ^{\left( l\right)
},Z_{_{l}}\right) \right\vert ^{2}}\right\} \Psi \left( \theta
_{i},Z_{i}\right)
\end{eqnarray*}%
If the field varies slowly, this expression reduces to:%
\begin{equation*}
\int_{\theta _{i}<\theta _{i+1}}\Psi ^{\dagger }\left( \theta
_{i},Z_{i}\right) \frac{\delta ^{\sum_{l}p_{l}^{i}}\left[ L\left( \Psi
^{\dagger }\left( \theta _{i},Z_{i}\right) ,\Psi \left( \theta
_{i},Z_{i}\right) \right) \right] }{\dprod\limits_{l=1}^{j}\delta
^{p_{l}^{i}}\left\vert \Psi \left( \theta ^{\left( l\right)
},Z_{_{l}}\right) \right\vert ^{2}}\Psi \left( \theta _{i},Z_{i}\right)
\simeq \frac{1}{2}\frac{\delta ^{\sum_{l}p_{l}^{i}}\left( \nabla _{\theta
}\omega ^{-1}\left( \theta _{i+1},Z_{i+1},\left\vert \Psi \right\vert
^{2}\right) \right) }{\dprod\limits_{l=1}^{j}\delta ^{p_{l}^{i}}\left\vert
\Psi \left( \theta ^{\left( l\right) },Z_{_{l}}\right) \right\vert ^{2}}%
\left\vert \Psi \left( \theta _{i+1},Z_{i+1}\right) \right\vert ^{2}
\end{equation*}%
and equation (\ref{fCT}) can also be rewritten in the approximation of
slowly varying background fields:%
\begin{eqnarray}
&&\Gamma \left( \Psi ^{\dagger },\Psi \right)  \label{fCV} \\
&=&\hat{S}_{cl}\left( \Psi ^{\dagger },\Psi \right) +\sum_{\substack{ %
j\geqslant 1  \\ m\geqslant 1}}\sum_{\substack{ p_{l},\left(
p_{l}^{i}\right) _{m\times j}  \\ p_{l}+\sum_{i}p_{l}^{i}\geqslant 2}}\int
\int \left( \dprod\limits_{l=1}^{j}\Psi ^{\dagger }\left( \theta ^{\left(
l\right) },Z_{l}\right) \right)  \notag \\
&&\times \prod\limits_{i=1}^{m}\left( \left( \frac{1}{2}\frac{\delta
^{\sum_{l}p_{l}^{i}}\left( \nabla _{\theta }\omega ^{-1}\left( \theta
_{i+1},Z_{i+1},\left\vert \Psi \right\vert ^{2}\right) \right) }{%
\dprod\limits_{l=1}^{j}\delta ^{p_{l}^{i}}\left\vert \Psi \left( \theta
^{\left( l\right) },Z_{_{l}}\right) \right\vert ^{2}}\left\vert \Psi \left(
\theta _{i+1},Z_{i+1}\right) \right\vert ^{2}\right) \right.  \notag \\
&&\left. \times \left( 1+\sum_{p_{i}}\left( \frac{\sharp _{j+1,i}\left(
\left( p_{i}\right) ,\left( p_{l}^{i}\right) \right) }{\bar{\sharp}%
_{j+1,i}\left( \left( p_{i}\right) ,\left( p_{l}^{i}\right) \right) }\left( 
\frac{\delta }{\delta \left\vert \Psi ^{\dagger }\left( \theta
_{i},Z_{i}\right) \right\vert ^{2}}\right) ^{p_{i}}\right) \right) \right)
a_{j,m}  \notag \\
&&\times \left( \int \Psi ^{\dagger }\left( \theta ,Z\right) \frac{1}{2}%
\frac{\delta ^{\sum_{l}p_{l}}\left( \nabla _{\theta }\omega ^{-1}\left(
\theta ,Z,\left\vert \Psi \right\vert ^{2}\right) \right) }{%
\dprod\limits_{l=1}^{j}\delta ^{p_{l}}\left\vert \Psi \left( \theta ^{\left(
l\right) },Z_{_{l}}\right) \right\vert ^{2}}\Psi \left( \theta ,Z\right)
\right) \frac{\left( \dprod\limits_{l=1}^{j}\Psi \left( \theta ^{\left(
l\right) },Z_{l}\right) \right) }{\left( m-1\right) !\Lambda
^{\sum_{i}p_{l}^{i}}}\prod\limits_{i=1}^{m}d\theta
_{i}dZ_{_{l}}\dprod\limits_{l=1}^{j}d\theta ^{\left( l\right) }dZ_{l}  \notag
\end{eqnarray}%
with the convention that $\left( \theta _{m+1},Z_{m+1}\right) =\left( \theta
,Z\right) $.

\subsubsection*{3.3.6 Strong field approximation}

For relatively strong fields, the derivatives $\frac{\delta }{\delta
\left\vert \Psi ^{\dagger }\left( \theta _{i},Z_{i}\right) \right\vert ^{2}}$
are negligible and (\ref{fCV}) reduces to:%
\begin{eqnarray}
&&\Gamma \left( \Psi ^{\dagger },\Psi \right) \\
&=&\hat{S}_{cl}\left( \Psi ^{\dagger },\Psi \right) +\sum_{\substack{ %
j\geqslant 1  \\ m\geqslant 1}}\sum_{\substack{ p_{l},\left(
p_{l}^{i}\right) _{m\times j}  \\ p_{l}+\sum_{i}p_{l}^{i}\geqslant 2}}\int
\int \prod\limits_{i=1}^{m}\left( \int \Psi ^{\dagger }\left( \theta
_{i},Z_{i}\right) \left\{ \frac{1}{2}\frac{\delta ^{\sum_{l}p_{l}^{i}}\left(
\nabla _{\theta }\omega ^{-1}\left( \theta _{i},Z_{i},\left\vert \Psi
\right\vert ^{2}\right) \right) }{\dprod\limits_{l=1}^{j}\delta
^{\sum_{l}p_{l}^{i}}\left\vert \Psi \left( \theta ^{\left( l\right)
},Z_{_{l}}\right) \right\vert ^{2}}\right\} \Psi \left( \theta
_{i},Z_{i}\right) \right)  \notag \\
&&\times \left( a_{j,m}\int \Psi ^{\dagger }\left( \theta ,Z\right) \left\{ 
\frac{1}{2}\frac{\delta ^{\sum_{l}p_{l}}\left( \nabla _{\theta }\omega
^{-1}\left( \theta ,Z,\left\vert \Psi \right\vert ^{2}\right) \right) }{%
\dprod\limits_{l=1}^{j}\delta ^{p_{l}}\left\vert \Psi \left( \theta ^{\left(
l\right) },Z_{_{l}}\right) \right\vert ^{2}}\right\} \Psi \left( \theta
,Z\right) \right)  \notag \\
&&\times \left( \dprod\limits_{l=1}^{j}\left\vert \Psi \left( \theta
^{\left( l\right) },Z_{l}\right) \right\vert ^{2}\right)
\prod\limits_{i=1}^{m}d\theta _{i}dZ_{_{l}}\dprod\limits_{l=1}^{j}d\theta
^{\left( l\right) }dZ_{l}  \notag
\end{eqnarray}

\subsubsection*{3.3.7 Weak field approximation}

For relatively weak fields, the term $\frac{\sharp _{j+1,i}\left( \left(
p_{i}\right) ,\left( p_{l}^{i}\right) \right) }{\bar{\sharp}_{j+1,i}\left(
\left( p_{i}\right) ,\left( p_{l}^{i}\right) \right) }\left( \frac{\delta }{%
\delta \left\vert \Psi ^{\dagger }\left( \theta _{i},Z_{i}\right)
\right\vert ^{2}}\right) $ in:%
\begin{equation*}
1+\sum_{p_{i}}\left( \frac{\sharp _{j+1,i}\left( \left( p_{i}\right) ,\left(
p_{l}^{i}\right) \right) }{\bar{\sharp}_{j+1,i}\left( \left( p_{i}\right)
,\left( p_{l}^{i}\right) \right) }\left( \frac{\delta }{\delta \left\vert
\Psi ^{\dagger }\left( \theta _{i},Z_{i}\right) \right\vert ^{2}}\right)
^{p_{i}}\right)
\end{equation*}%
is dominant and:%
\begin{eqnarray*}
&&\int \Psi ^{\dagger }\left( \theta _{m},Z_{m}\right) \frac{\delta
^{\sum_{l}p_{l}^{m}}\left[ L\left( \Psi ^{\dagger }\left( \theta
_{m},Z_{m}\right) ,\Psi \left( \theta _{m},Z_{m}\right) \right) \right] }{%
\dprod\limits_{l=1}^{j}\dprod\limits_{k_{l}^{i}=1}^{p_{l}^{m}}\delta
\left\vert \Psi \left( \theta ^{\left( k_{l}^{i}\right) },Z_{_{l}}\right)
\right\vert ^{2}}\Psi \left( \theta _{m},Z_{m}\right) \\
&&\times \left( 1+\sum_{p_{m}}\left( \frac{\sharp _{j+1,m}\left( \left(
p_{m}\right) ,\left( p_{l}^{m}\right) \right) }{\bar{\sharp}_{j+1,m}\left(
\left( p_{m}\right) ,\left( p_{l}^{m}\right) \right) }\left( \frac{\delta }{%
\delta \left\vert \Psi ^{\dagger }\left( \theta _{m},Z_{m}\right)
\right\vert ^{2}}\right) ^{p_{m}}\right) \right) \\
&&\times a_{j,m}\left( \int \Psi ^{\dagger }\left( \theta ,Z\right) \frac{1}{%
2}\frac{\delta ^{\sum_{l}p_{l}}\left( \nabla _{\theta }\omega ^{-1}\left(
\theta ,Z,\left\vert \Psi \right\vert ^{2}\right) \right) }{%
\dprod\limits_{l=1}^{j}\delta ^{p_{l}}\left\vert \Psi \left( \theta ^{\left(
l\right) },Z_{_{l}}\right) \right\vert ^{2}}\Psi \left( \theta ,Z\right)
\right) \\
&\rightarrow &\int \Psi ^{\dagger }\left( \theta _{m},Z_{m}\right) \frac{1}{2%
}\frac{\delta ^{\sum_{l}p^{m}}\left( \nabla _{\theta }\omega ^{-1}\left(
\theta _{m},Z_{m},\left\vert \Psi \right\vert ^{2}\right) \right) }{%
\dprod\limits_{l=1}^{j}\delta ^{p_{l}^{m}}\left\vert \Psi \left( \theta
^{\left( l\right) },Z_{_{l}}\right) \right\vert ^{2}}\frac{\sharp
_{j+1,m}\left( \left( p_{m},\left( p_{l}^{m}\right) \right) \right) }{\bar{%
\sharp}_{j+1,m}\left( \left( p_{i},\left( p_{l}^{m}\right) \right) \right) }%
\frac{\delta }{\delta \left\vert \Psi ^{\dagger }\left( \theta
_{m},Z_{m}\right) \right\vert ^{2}} \\
&&a_{j,m}\left( \int \Psi ^{\dagger }\left( \theta ,Z\right) \frac{1}{2}%
\frac{\delta ^{\sum_{l}p_{l}}\left( \nabla _{\theta }\omega ^{-1}\left(
\theta ,Z,\left\vert \Psi \right\vert ^{2}\right) \right) }{%
\dprod\limits_{l=1}^{j}\delta ^{p_{l}}\left\vert \Psi \left( \theta ^{\left(
l\right) },Z_{_{l}}\right) \right\vert ^{2}}\Psi \left( \theta ,Z\right)
\right) \\
&\simeq &\frac{\sharp _{j+1,m}\left( \left( p_{m}\right) ,\left(
p_{l}^{m}\right) \right) }{\bar{\sharp}_{j+1,m}\left( \left( p_{m}\right)
,\left( p_{l}^{m}\right) \right) }a_{j,m}\left( \int \Psi ^{\dagger }\left(
\theta ,Z\right) \frac{1}{2}\frac{\delta ^{\sum_{l}p_{l}^{m}}\left( \nabla
_{\theta }\omega ^{-1}\left( \theta ,Z,\left\vert \Psi \right\vert
^{2}\right) \right) }{\dprod\limits_{l=1}^{j}\delta ^{p_{l}^{m}}\left\vert
\Psi \left( \theta ^{\left( l\right) },Z_{_{l}}\right) \right\vert ^{2}}%
\frac{1}{2}\frac{\delta ^{\sum_{l}p_{l}}\left( \nabla _{\theta }\omega
^{-1}\left( \theta ,Z,\left\vert \Psi \right\vert ^{2}\right) \right) }{%
\dprod\limits_{l=1}^{j}\delta ^{p_{l}}\left\vert \Psi \left( \theta ^{\left(
l\right) },Z_{_{l}}\right) \right\vert ^{2}}\Psi \left( \theta ,Z\right)
\right)
\end{eqnarray*}%
Iterating this relation yields:%
\begin{eqnarray}
&&\Gamma \left( \Psi ^{\dagger },\Psi \right) \\
&=&\hat{S}_{cl}\left( \Psi ^{\dagger },\Psi \right) +\sum_{\substack{ %
j\geqslant 1  \\ m\geqslant 1}}\sum_{\substack{ p_{l},\left(
p_{l}^{i}\right) _{m\times j}  \\ p_{l}+\sum_{i}p_{l}^{i}\geqslant 2}}\left(
\prod\limits_{i=1}^{m}\frac{\sharp _{j+1,m}\left( \left( p_{m}\right)
,\left( p_{l}^{m}\right) \right) }{4\bar{\sharp}_{j+1,m}\left( \left(
p_{m}\right) ,\left( p_{l}^{m}\right) \right) }\right) \frac{a_{j,m}}{2} 
\notag \\
&&\times \int \int \Psi ^{\dagger }\left( \theta ,Z\right)
\prod\limits_{i=1}^{m}\left\{ \frac{\delta ^{\sum_{l}p_{l}^{i}}\left(
\nabla _{\theta }\omega ^{-1}\left( \theta ,Z,\left\vert \Psi \right\vert
^{2}\right) \right) }{\dprod\limits_{l=1}^{j}\delta ^{p_{l}^{i}}\left\vert
\Psi \left( \theta ^{\left( l\right) },Z_{_{l}}\right) \right\vert ^{2}}%
\right\} \frac{\delta ^{\sum_{l}p_{l}}\left( \nabla _{\theta }\omega
^{-1}\left( \theta ,Z,\left\vert \Psi \right\vert ^{2}\right) \right) }{%
\dprod\limits_{l=1}^{j}\delta ^{p_{l}}\left\vert \Psi \left( \theta ^{\left(
l\right) },Z_{_{l}}\right) \right\vert ^{2}}\Psi \left( \theta ,Z\right) 
\notag \\
&&\times \dprod\limits_{l=1}^{j}\left\vert \Psi \left( \theta ^{\left(
l\right) },Z_{l}\right) \right\vert ^{2}d\theta ^{\left( l\right) }dZ_{l} 
\notag
\end{eqnarray}%
and this reduces to:%
\begin{eqnarray}
&&\Gamma \left( \Psi ^{\dagger },\Psi \right)  \label{wkf} \\
&=&\hat{S}_{cl}\left( \Psi ^{\dagger },\Psi \right) +\sum_{\substack{ %
j\geqslant 1  \\ m\geqslant 1}}\sum_{\substack{ p_{l},\left(
p_{l}^{i}\right) _{m\times j}  \\ p_{l}+\sum_{i}p_{l}^{i}\geqslant 2}}\left(
\prod\limits_{i=1}^{m}\frac{\sharp _{j+1,m}\left( \left( p_{m}\right)
,\left( p_{l}^{m}\right) \right) }{4\bar{\sharp}_{j+1,m}\left( \left(
p_{m}\right) ,\left( p_{l}^{m}\right) \right) }\right) \frac{a_{j,m}}{2} 
\notag \\
&&\times \int \int \Psi ^{\dagger }\left( \theta ,Z\right) \frac{\delta
^{\sum_{l,i}p_{l}^{i}+p_{l}}}{\dprod\limits_{l=1}^{j}\delta
^{\sum_{i}p_{l}^{i}+p_{l}}\left\vert \Psi \left( \theta ^{\left( l\right)
},Z_{_{l}}\right) \right\vert ^{2}}\left( \frac{\left( \nabla _{\theta
}\omega ^{-1}\left( \theta ,Z,\left\vert \Psi \right\vert ^{2}\right)
\right) }{\dprod\limits_{l=1}^{j}\delta ^{p_{l}^{i}}\left\vert \Psi \left(
\theta ^{\left( l\right) },Z_{_{l}}\right) \right\vert ^{2}}\right)
^{m+1}\Psi \left( \theta ,Z\right)  \notag \\
&&\times \dprod\limits_{l=1}^{j}\left\vert \Psi \left( \theta ^{\left(
l\right) },Z_{l}\right) \right\vert ^{2}d\theta ^{\left( l\right) }dZ_{l} 
\notag
\end{eqnarray}

\subsection*{3.4 First order condition and non trivial vacuum}

\subsubsection*{3.4.1 Constant potential}

\paragraph{3.4.1.1 First order condition for classical action}

We consider the saddle point equation at the lowest order in perturbation.
Given our assumptions $\bar{\zeta}_{n+1}>0$ for all $n\geqslant 2$, $\zeta
^{\left( n+1\right) }>0$, $\zeta ^{\left( 2\right) }<0$, the potential:%
\begin{equation*}
\alpha \int \left\vert \Psi \left( \theta ^{\left( i\right) },Z_{i}\right)
\right\vert ^{2}dZ_{i}+\sum \frac{\zeta ^{\left( n\right) }}{n!}\left( 
\mathcal{G}_{0}\left( 0,Z\right) +\int \left\vert \Psi \left( \theta
_{i}^{\left( i\right) }-\frac{\left\vert Z_{i}-Z_{j}\right\vert }{c}%
,Z_{j}\right) \right\vert ^{2}dZ_{j}\right) ^{n}
\end{equation*}%
has a minimum for $\alpha <<1$ and for $\left\vert \zeta ^{\left( 2\right)
}\right\vert $ large. This minimum is reached for a value $X_{0}$ of $\int
\left\vert \Psi \left( \theta ^{\left( i\right) },Z_{i}\right) \right\vert
^{2}dZ_{i}$. Up to an irrelevant phase, $\Psi _{0}\left( \theta ^{\left(
i\right) },Z_{i}\right) =\Psi _{0}^{\dagger }\left( \theta ,Z\right) =\sqrt{%
\frac{X_{0}}{V}}$ where $V$ is the volume of the thread.

Moreover the operator $O=\nabla _{\theta }\frac{\sigma _{\theta }^{2}}{2}%
\left( \nabla _{\theta }-\omega ^{-1}\left( J\left( \theta \right) ,\theta
,Z,\mathcal{G}_{0}\right) \right) $ has positive eigenvalues. Developing $%
\Psi \left( \theta ,Z\right) =\sum a_{n}\Psi _{n}\left( \theta ,Z\right) $
where $\Psi _{n}\left( \theta ,Z\right) $ are the eigenstates of $O$, the
definition of $\Psi ^{\dagger }\left( \theta ,Z\right) $ (see \cite{GL1} and 
\cite{GL2}) is given by:%
\begin{equation*}
\sum \bar{a}_{n}\Psi _{n}^{\dagger }\left( \theta ,Z\right)
\end{equation*}%
where $\Psi _{n}^{\dagger }\left( \theta ,Z\right) $ are the eigenstates of
the adjoint operator of $O$. As a consequence 
\begin{equation*}
\int -\frac{1}{2}\Psi ^{\dagger }\left( \theta ,Z\right) \left( \nabla
_{\theta }\left( \frac{\sigma _{\theta }^{2}}{2}\nabla _{\theta }-\omega
^{-1}\left( J\left( \theta \right) ,\theta ,Z,\mathcal{G}_{0}\right) \right)
\right) \Psi \left( \theta ,Z\right)
\end{equation*}%
is positive, and null for constant $\Psi ^{\dagger }\left( \theta ,Z\right) $
and $\Psi \left( \theta ,Z\right) $. \ As a consequence, for $\left\vert
\zeta ^{\left( n\right) }\right\vert >\omega ^{-1}\left( J\left( \theta
\right) ,\theta ,Z,\mathcal{G}_{0}\right) $ the minimum of $\Gamma \left(
\Psi \right) $ is reached for $\Psi \left( \theta ,Z\right) =\Psi _{0}\left(
\theta ,Z\right) +\delta \Psi \left( \theta ,Z\right) $ and $\Psi ^{\dagger
}\left( \theta ,Z\right) =\Psi _{0}^{\dagger }\left( \theta ,Z\right)
+\delta \Psi ^{\dagger }\left( \theta ,Z\right) $ where $\left\vert \delta
\Psi \left( \theta ,Z\right) \right\vert <<\left\vert \Psi _{0}\left( \theta
,Z\right) \right\vert $ and $\left\vert \delta \Psi ^{\dagger }\left( \theta
,Z\right) \right\vert <<\left\vert \Psi _{0}^{\dagger }\left( \theta
,Z\right) \right\vert $. We assume that the successive derivatives of $U$
decrease quickly and we neglect the terms involving $U^{\left( n\right)
}\left( X_{0}\right) $ for $n\geqslant 3$.

Expanding the potential around $\Psi _{0}\left( \theta ,Z\right) $ and
setting $V=1$, yields at the second order:%
\begin{eqnarray*}
\Gamma \left( \Psi ,\Psi ^{\dag }\right) &=&-\frac{1}{2}\int \delta \Psi
^{\dagger }\left( \theta ,Z\right) \left( \nabla _{\theta }\left( \frac{%
\sigma _{\theta }^{2}}{2}\nabla _{\theta }-\omega ^{-1}\left( J\left( \theta
\right) ,\theta ,Z,\mathcal{G}_{0}+\left\vert \Psi \right\vert ^{2}\right)
\right) \right) X_{0} \\
&&-\frac{1}{2}\int \delta \Psi ^{\dagger }\left( \theta ,Z\right) \left(
\nabla _{\theta }\left( \frac{\sigma _{\theta }^{2}}{2}\nabla _{\theta
}-\omega ^{-1}\left( J\left( \theta \right) ,\theta ,Z,\mathcal{G}%
_{0}+\left\vert \Psi \right\vert ^{2}\right) \right) \right) \delta \Psi
\left( \theta ,Z\right) \\
&&+\frac{1}{2}\int \delta \Psi ^{\dagger }\left( \theta ,Z\right) U^{\prime
\prime }\left( X_{0}\right) \delta \Psi \left( \theta ,Z\right)
\end{eqnarray*}%
with $\left\vert \Psi \right\vert ^{2}=X_{0}+\sqrt{X_{0}}\left( \delta
\left( \Psi ^{\dagger }+\delta \Psi \right) \right) $. This leads to the
first order condition for $\delta \Psi \left( \theta _{1},Z_{1}\right) $:%
\begin{eqnarray*}
0 &=&\frac{1}{2}\delta \Psi ^{\dagger }\left( \theta ,Z\right) \left(
-\nabla _{\theta }\left( \frac{\sigma _{\theta }^{2}}{2}\nabla _{\theta
}-\omega ^{-1}\left( J\left( \theta \right) ,\theta ,Z,\mathcal{G}%
_{0}+X_{0}\right) \right) +U^{\prime \prime }\left( X_{0}\right) \right) \\
&&-\frac{1}{2}\int \delta \Psi ^{\dagger }\left( \theta _{1},Z_{1}\right) 
\sqrt{X_{0}}\left( \nabla _{\theta }\frac{\delta \omega ^{-1}\left( J\left(
\theta _{1}\right) ,\theta _{1},Z_{1},\mathcal{G}_{0}+X_{0}\right) }{\delta
\left\vert \Psi \left( \theta ,Z\right) \right\vert ^{2}}\right)
X_{0}d\theta _{1}dZ_{1}
\end{eqnarray*}%
with solution $\delta \Psi ^{\dagger }\left( \theta ,Z\right) =0$. This
implies that the first order condition for $\delta \Psi ^{\dag }\left(
\theta ,Z\right) $ becomes: 
\begin{eqnarray}
&&0=-\frac{1}{2}\left( \nabla _{\theta }\left( \frac{\sigma _{\theta }^{2}}{2%
}\nabla _{\theta }-\omega ^{-1}\left( J\left( \theta \right) ,\theta ,Z,%
\mathcal{G}_{0}+\left\vert \Psi \right\vert ^{2}\right) \right) \right) X_{0}
\label{sDP} \\
&&-\frac{1}{2}\left( \nabla _{\theta }\left( \frac{\sigma _{\theta }^{2}}{2}%
\nabla _{\theta }-\omega ^{-1}\left( J\left( \theta \right) ,\theta ,Z,%
\mathcal{G}_{0}+\left\vert \Psi \right\vert ^{2}\right) \right) \right)
\delta \Psi \left( \theta ,Z\right)  \notag \\
&&+\frac{1}{2}U^{\prime \prime }\left( X_{0}\right) \delta \Psi \left(
\theta ,Z\right)  \notag
\end{eqnarray}%
Equation (\ref{sDP}) also rewrites:%
\begin{equation}
\left( -\left( \nabla _{\theta }\left( \frac{\sigma _{\theta }^{2}}{2}\nabla
_{\theta }-\omega ^{-1}\left( J\left( \theta \right) ,\theta ,Z,\mathcal{G}%
_{0}+\left\vert \Psi \right\vert ^{2}\right) \right) \right) +U^{\prime
\prime }\left( X_{0}\right) \right) \left( \delta \Psi \left( \theta
,Z\right) +X_{0}\right) =U^{\prime \prime }\left( X_{0}\right) X_{0}
\label{sDT}
\end{equation}%
Equation (\ref{sDT}) can be used to write $\delta \Psi \left( \theta
,Z\right) $ as a function of $\omega ^{-1}\left( J\left( \theta \right)
,\theta ,Z,\mathcal{G}_{0}+\left\vert \Psi \right\vert ^{2}\right) $:%
\begin{eqnarray}
\delta \Psi \left( \theta ,Z\right) &=&\left( \frac{\left( \nabla _{\theta
}\left( \frac{\sigma _{\theta }^{2}}{2}\nabla _{\theta }-\omega ^{-1}\left(
J\left( \theta \right) ,\theta ,Z,\mathcal{G}_{0}+\left\vert \Psi
\right\vert ^{2}\right) \right) \right) }{U^{\prime \prime }\left(
X_{0}\right) -\left( \nabla _{\theta }\left( \frac{\sigma _{\theta }^{2}}{2}%
\nabla _{\theta }-\omega ^{-1}\left( J\left( \theta \right) ,\theta ,Z,%
\mathcal{G}_{0}+\left\vert \Psi \right\vert ^{2}\right) \right) \right) }%
\right) X_{0}  \label{psv} \\
&=&-\frac{\nabla _{\theta }\left( \omega ^{-1}\left( J\left( \theta \right)
,\theta ,Z,\mathcal{G}_{0}+\left\vert \Psi \right\vert ^{2}\right) \right) }{%
U^{\prime \prime }\left( X_{0}\right) -\left( \nabla _{\theta }\left( \frac{%
\sigma _{\theta }^{2}}{2}\nabla _{\theta }-\omega ^{-1}\left( J\left( \theta
\right) ,\theta ,Z,\mathcal{G}_{0}+\left\vert \Psi \right\vert ^{2}\right)
\right) \right) }X_{0}  \notag
\end{eqnarray}%
In first approximation, for $U^{\prime \prime }\left( X_{0}\right) >>1$ and $%
\sigma _{\theta }^{2}<<1$, this yields: 
\begin{eqnarray}
\delta \Psi \left( \theta ,Z\right) &\simeq &-\frac{\nabla _{\theta }\omega
^{-1}\left( J\left( \theta \right) ,\theta ,Z,\mathcal{G}_{0}+\left\vert
\Psi \right\vert ^{2}\right) }{U^{\prime \prime }\left( X_{0}\right) +\nabla
_{\theta }\omega ^{-1}\left( J\left( \theta \right) ,\theta ,Z,\mathcal{G}%
_{0}+\left\vert \Psi \right\vert ^{2}\right) }X_{0}  \label{psG} \\
&\simeq &-\frac{\nabla _{\theta }\omega ^{-1}\left( J\left( \theta \right)
,\theta ,Z,\mathcal{G}_{0}+\left\vert \Psi \right\vert ^{2}\right) }{%
U^{\prime \prime }\left( X_{0}\right) }X_{0}  \notag
\end{eqnarray}%
Note that for a slowly background field $\Psi _{0}\left( \theta ,Z\right) $,
equation (\ref{psG}) becomes:%
\begin{equation}
\delta \Psi \left( \theta ,Z\right) \simeq -\frac{\nabla _{\theta }\omega
^{-1}\left( J\left( \theta \right) ,\theta ,Z,\mathcal{G}_{0}+\left\vert
\Psi \right\vert ^{2}\right) }{U^{\prime \prime }\left( X_{0}\right) }\Psi
_{0}\left( \theta ,Z\right)  \label{pSG}
\end{equation}

\paragraph*{3.4.1.2 Solution of classical action's first order condition}

To solve equations (\ref{psG})\ and (\ref{pSG}), the dependency of $\omega
^{-1}\left( J\left( \theta \right) ,\theta ,Z,\mathcal{G}_{0}+\left\vert
\Psi \right\vert ^{2}\right) $ in $\left\vert \Psi \right\vert ^{2}$ has to
be explicited Note that in first approximation, the solution of (\ref{pSG})
is:%
\begin{eqnarray*}
\delta \Psi \left( \theta ,Z\right) &\simeq &-\frac{\nabla _{\theta }\omega
^{-1}\left( J\left( \theta \right) ,\theta ,Z,\mathcal{G}_{0}+\left\vert
\Psi \right\vert ^{2}\right) }{U^{\prime \prime }\left( X_{0}\right) }\Psi
_{0}\left( \theta ,Z\right) \\
&=&\frac{\nabla _{\theta }\omega \left( J\left( \theta \right) ,\theta ,Z,%
\mathcal{G}_{0}+\left\vert \Psi _{0}\right\vert ^{2}\right) }{U^{\prime
\prime }\left( X_{0}\right) \omega ^{2}\left( J\left( \theta \right) ,\theta
,Z,\mathcal{G}_{0}+\left\vert \Psi _{0}\right\vert ^{2}\right) }\Psi
_{0}\left( \theta ,Z\right)
\end{eqnarray*}%
and this approximation is sufficient as a first approximation.

However, to find a more precise expression for $\delta \Psi \left( \theta
,Z\right) $, we use (\ref{fqt}) that defines $\omega ^{-1}\left( J\left(
\theta \right) ,\theta ,Z,\mathcal{G}_{0}+\left\vert \Psi \right\vert
^{2}\right) $ at the classical order:%
\begin{eqnarray}
&&\omega ^{-1}\left( J,\theta ,Z,\left\vert \Psi \right\vert ^{2}\right)
\label{ftq} \\
&=&G\left( J\left( \theta ,Z\right) +\int \frac{\kappa }{N}\frac{\omega
\left( J,\theta -\frac{\left\vert Z-Z_{1}\right\vert }{c},Z_{1},\Psi \right)
T\left( Z,\theta ,Z_{1},\theta -\frac{\left\vert Z-Z_{1}\right\vert }{c}%
\right) }{\omega \left( J,\theta ,Z,\left\vert \Psi \right\vert ^{2}\right) }%
\right.  \notag \\
&&\times \left. \left( \mathcal{G}_{0}\left( Z_{1}\right) +\left\vert \Psi
\left( \theta -\frac{\left\vert Z-Z_{1}\right\vert }{c},Z_{1}\right)
\right\vert ^{2}\right) dZ_{1}\right)  \notag
\end{eqnarray}%
Using (\ref{ftq}), the defining equation (\ref{psG}) for $\delta \Psi \left(
\theta ,Z\right) $ becomes:%
\begin{eqnarray*}
&&G^{-1}\left( -\frac{U^{\prime \prime }\left( X_{0}\right) }{X_{0}}%
\int^{\theta }\delta \Psi \left( \theta ,Z\right) \right) \\
&=&\int \frac{\kappa }{N}\frac{\omega \left( J,\theta -\frac{\left\vert
Z-Z_{1}\right\vert }{c},Z_{1},\Psi \right) }{\omega \left( J,\theta
,Z,\left\vert \Psi \right\vert ^{2}\right) }T\left( Z,\theta ,Z_{1},\theta -%
\frac{\left\vert Z-Z_{1}\right\vert }{c}\right) \left( \mathcal{G}_{0}\left(
Z_{1}\right) +\left\vert \Psi \left( \theta -\frac{\left\vert
Z-Z_{1}\right\vert }{c},Z_{1}\right) \right\vert ^{2}\right) dZ_{1}
\end{eqnarray*}%
This equation can be rewritten in the local approximation:%
\begin{equation}
G^{-1}\left( -\frac{U^{\prime \prime }\left( X_{0}\right) }{X_{0}}%
\int^{\theta }\delta \Psi \left( \theta ,Z\right) \right) \simeq J\left(
\theta ,Z\right) +\frac{\left( -\Gamma \nabla _{\theta }+\Gamma ^{\prime
}\nabla _{Z}^{2}\right) \left( \omega \left( J,\theta ,Z\right) \left( 
\mathcal{G}_{0}\left( Z\right) +\left\vert \Psi \left( \theta ,Z\right)
\right\vert ^{2}\right) \right) }{\omega \left( J,\theta ,Z,\left\vert \Psi
\right\vert ^{2}\right) }  \label{pSM}
\end{equation}%
where $\Gamma $ and $\Gamma ^{\prime }$ are defined by: 
\begin{eqnarray*}
\Gamma &=&\int \frac{\kappa }{N}\frac{\left\vert Z-Z_{1}\right\vert }{c}%
T\left( Z,\theta ,Z_{1},\theta -\frac{\left\vert Z-Z_{1}\right\vert }{c}%
\right) dZ_{1} \\
\Gamma ^{\prime } &=&\int \frac{\kappa }{N}\left\vert Z-Z_{1}\right\vert
^{2}T\left( Z,\theta ,Z_{1},\theta -\frac{\left\vert Z-Z_{1}\right\vert }{c}%
\right) dZ_{1}
\end{eqnarray*}%
At the lowest order in derivatives, equation (\ref{pSM}) becomes:%
\begin{eqnarray}
G^{-1}\left( -\frac{U^{\prime \prime }\left( X_{0}\right) }{X_{0}}%
\int^{\theta }\delta \Psi \left( \theta ,Z\right) \right) &\simeq &J\left(
\theta ,Z\right) -\frac{\Gamma \nabla _{\theta }\omega \left( J,\theta
,Z\right) \left( \mathcal{G}_{0}\left( Z_{1}\right) +\left\vert \Psi \left(
\theta ,Z\right) \right\vert ^{2}\right) }{\omega \left( J,\theta
,Z,\left\vert \Psi \right\vert ^{2}\right) }  \label{rtm} \\
&=&J\left( \theta ,Z\right) -\Gamma \nabla _{\theta }\left\vert \Psi \left(
\theta ,Z\right) \right\vert ^{2}+\Gamma \frac{\delta \Psi \left( \theta
,Z\right) }{\int^{\theta }\delta \Psi \left( \theta ,Z\right) }\left( 
\mathcal{G}_{0}\left( Z\right) +\left\vert \Psi \left( \theta ,Z\right)
\right\vert ^{2}\right)  \notag \\
&\simeq &J\left( \theta ,Z\right) -\Gamma \sqrt{X_{0}}\nabla _{\theta
}\delta \Psi \left( \theta ,Z\right) +\Gamma \frac{\mathcal{G}_{0}\left(
Z\right) +X_{0}+\sqrt{X_{0}}\delta \Psi \left( \theta ,Z\right) }{%
\int^{\theta }\delta \Psi \left( \theta ,Z\right) }\delta \Psi \left( \theta
,Z\right)  \notag \\
&\simeq &J\left( \theta ,Z\right) +\Gamma \frac{\mathcal{G}_{0}\left(
Z_{1}\right) +X_{0}}{\int^{\theta }\delta \Psi \left( \theta ,Z\right) }%
\delta \Psi \left( \theta ,Z\right)  \notag
\end{eqnarray}%
We set:%
\begin{equation*}
Y=\ln \left( \int \delta \Psi \left( \theta ,Z\right) \right)
\end{equation*}%
and (\ref{rtm}) writes: 
\begin{eqnarray}
G^{-1}\left( -\frac{U^{\prime \prime }\left( X_{0}\right) }{X_{0}}\exp
Y\right) &=&J\left( \theta ,Z\right) +\Gamma \left( \mathcal{G}_{0}\left(
Z_{1}\right) +X_{0}\right) \nabla _{\theta }Y  \label{qrt} \\
&\simeq &\left\langle J\right\rangle \left( Z\right) +\Gamma \left( \mathcal{%
G}_{0}\left( Z_{1}\right) +X_{0}\right) \nabla _{\theta }Y  \notag
\end{eqnarray}%
where $\left\langle J\right\rangle \left( Z\right) $ is the current averaged
over time. The solution of (\ref{qrt}) is: 
\begin{equation*}
\int \delta \Psi \left( \theta ,Z\right) =\exp \left( Y\right) =\exp \left(
H^{-1}\left( \frac{\theta }{\Gamma \left( \mathcal{G}_{0}\left( Z_{1}\right)
+X_{0}\right) }+d\right) \right)
\end{equation*}%
with: 
\begin{equation*}
H\left( Y\right) =\int \frac{dY}{G^{-1}\left( -\frac{U^{\prime \prime
}\left( X_{0}\right) }{X_{0}}\exp Y\right) -\left\langle J\right\rangle
\left( Z\right) }
\end{equation*}%
and:%
\begin{eqnarray}
\delta \Psi \left( \theta ,Z\right) &=&\left( G^{-1}\left( -\frac{U^{\prime
\prime }\left( X_{0}\right) }{X_{0}}\exp \left( H^{-1}\left( \frac{\theta }{%
\Gamma \left( \mathcal{G}_{0}\left( Z_{1}\right) +\sqrt{X_{0}}\right) }%
+d\right) \right) \right) -\left\langle J\right\rangle \left( Z\right)
\right)  \label{nls} \\
&&\times \exp \left( H^{-1}\left( \frac{\theta }{\Gamma \left( \mathcal{G}%
_{0}\left( Z_{1}\right) +\sqrt{X_{0}}\right) }+d\right) \right)  \notag
\end{eqnarray}%
The constant $d$ is chosen so that $\lim_{\theta \rightarrow \infty }\delta
\Psi \left( \theta ,Z\right) =0$. For slowly varyng currents, $\left\langle
J\right\rangle \left( Z\right) $ can replaced by $J\left( \theta ,Z\right) $
in the formula.

\paragraph{Higher order corrections}

We use the series expansion (\ref{clS}) to compute higher order corrections
to the background field equation.%
\begin{equation}
\hat{S}_{cl}\left( \Psi ^{\dagger },\Psi \right) +A
\end{equation}%
with:%
\begin{eqnarray}
&&A=-\sum_{j\geqslant 1}\int \sum_{\substack{ m\geqslant 2,\left(
p_{l}^{i}\right) _{m\times j}  \\ \sum_{i}p_{l}^{i}\geqslant 2}}%
\dprod\limits_{l=1}^{j}d\theta ^{\left( l\right) }dZ_{l}\left( \frac{\omega
^{-1}\left( J,\theta ^{\left( l\right) },Z_{_{l}}\right) }{\left\vert \Psi
\left( \theta ^{\left( l\right) },Z_{_{l}}\right) \right\vert ^{2}}\right)
^{\sum_{i}p_{l}^{i}}\left\vert \Psi \left( \theta ^{\left( l\right)
},Z_{l}\right) \right\vert ^{2} \\
&&\times \frac{\dprod\limits_{i=1}^{m}\int c\exp \left( -c\left( \theta
_{i}-\theta _{i,j}\right) -\alpha \left( \sum_{l=1,p_{l}^{i}\neq
0}^{j}\left( \left( c\left( \theta ^{\left( l-1,i\right) }-\theta ^{\left(
l,i\right) }\right) \right) ^{2}-\left\vert Z_{l-1}^{\left( i\right)
}-Z_{l}^{\left( i\right) }\right\vert ^{2}\right) \right) \right) \left\vert
\Psi \left( \theta _{i},Z_{i}\right) \right\vert ^{2}d\theta _{i}dZ_{i}}{%
\left( -2\right) ^{m}D^{\sum_{i,l}p_{l}^{i}}m!\dprod\limits_{k}\left( \sharp
_{k}\right) !\Lambda _{1}^{j}\Lambda ^{\sum_{i,l}p_{l}^{i}}}  \notag
\end{eqnarray}%
The corrective term to $\frac{\delta \hat{S}_{cl}\left( \Psi ^{\dagger
},\Psi \right) }{\delta \Psi }$ is $\frac{\delta A}{\delta \Psi }$ evaluated
at $\sqrt{X_{0}}+\delta \Psi $. At first order in $\delta \Psi $\ it is
given by:%
\begin{eqnarray*}
&&\frac{\delta A}{\delta \Psi }=-\sum_{j\geqslant 1}\int \sum_{\substack{ %
m\geqslant 2,\left( p_{l}^{i}\right) _{m\times j}  \\ \sum_{i}p_{l}^{i}%
\geqslant 2}}\left[ \frac{\dprod\limits_{i=1}^{m}\int \left[ c\exp \left(
-cl_{j}-\alpha \left( \sum_{r=1}^{n-1}\left( \left( c\left(
l_{r}-l_{r+1}\right) \right) ^{2}-\left\vert Z_{r}-Z_{r+1}\right\vert
^{2}\right) \right) \right) X_{0}\right] }{X_{0}^{2}2^{m}D^{%
\sum_{i,l}p_{l}^{i}}m!\dprod\limits_{k}\left( \sharp _{k}\right) !\Lambda
_{1}^{j}\Lambda ^{\sum_{i,l}p_{l}^{i}}}\right. \\
&&\left. \times \left( m+\sum_{i}p_{l}^{i}\right) \left(
m+\sum_{i}p_{l}^{i}-1\right) X_{0}^{j-2}\dprod\limits_{l=1}^{j}\left( \frac{%
\omega _{0}^{-1}\left( J,Z_{i}\right) }{X_{0}}\right)
^{p_{l}^{i}}dl_{i}dZ_{i}\right] \times
\dprod\limits_{l=1}^{j}dl_{l}dZ_{l}\times \sqrt{X_{0}}\delta \Psi
\end{eqnarray*}%
where we have replaced $\omega ^{-1}\left( J,\theta _{i}-l_{l},Z_{i}\right) $
by it's static solution approximation $\omega _{0}^{-1}\left( J,Z_{i}\right) 
$.

The integral may be computed and yield a constant $-\frac{1}{2}C\left(
X_{0}\right) $ times $\delta \Psi $. Equation (\ref{sDP}) is thus replaced
by: 
\begin{eqnarray}
&&0=-\frac{1}{2}\left( \nabla _{\theta }\left( \frac{\sigma _{\theta }^{2}}{2%
}\nabla _{\theta }-\omega ^{-1}\left( J\left( \theta \right) ,\theta ,Z,%
\mathcal{G}_{0}+\left\vert \Psi \right\vert ^{2}\right) \right) \right) X_{0}
\\
&&-\frac{1}{2}\left( \nabla _{\theta }\left( \frac{\sigma _{\theta }^{2}}{2}%
\nabla _{\theta }-\omega ^{-1}\left( J\left( \theta \right) ,\theta ,Z,%
\mathcal{G}_{0}+\left\vert \Psi \right\vert ^{2}\right) \right) \right)
\delta \Psi \left( \theta ,Z\right)  \notag \\
&&+\frac{1}{2}\left( U^{\prime \prime }\left( X_{0}\right) -C\left(
X_{0}\right) \right) \delta \Psi \left( \theta ,Z\right)  \notag
\end{eqnarray}%
As a consequence, equation (\ref{psv}) is thus transformed into:%
\begin{eqnarray}
\delta \Psi \left( \theta ,Z\right) &\simeq &-\frac{\nabla _{\theta }\omega
^{-1}\left( J\left( \theta \right) ,\theta ,Z,\mathcal{G}_{0}+\left\vert
\Psi \right\vert ^{2}\right) }{U^{\prime \prime }\left( X_{0}\right)
-C\left( X_{0}\right) +\nabla _{\theta }\omega ^{-1}\left( J\left( \theta
\right) ,\theta ,Z,\mathcal{G}_{0}+\left\vert \Psi \right\vert ^{2}\right) }%
X_{0} \\
&\simeq &-\frac{\nabla _{\theta }\omega ^{-1}\left( J\left( \theta \right)
,\theta ,Z,\mathcal{G}_{0}+\left\vert \Psi \right\vert ^{2}\right) }{%
U^{\prime \prime }\left( X_{0}\right) -C\left( X_{0}\right) }X_{0}  \notag
\end{eqnarray}%
Solution (\ref{nls}) is thus still valid, with $U^{\prime \prime }\left(
X_{0}\right) $\ shifted: $U^{\prime \prime }\left( X_{0}\right) \rightarrow $
$U^{\prime \prime }\left( X_{0}\right) -C\left( X_{0}\right) $.

\subsubsection{Time dependent potential}

When the system is interacting with external signals, the average number of
activations may be shiftted and the assumption of a constant minimum $\sqrt{%
\frac{X_{0}}{V}}$ of the potential has to be modified. We consider a time
dependent potential with minimum $\Psi _{0}\left( \theta ,Z\right) $.
Expanding the the effective action at the second order around $\Psi
_{0}\left( \theta ,Z\right) $ yields:%
\begin{eqnarray*}
\Gamma \left( \Psi \right) &=&-\frac{1}{2}\int \delta \Psi ^{\dagger }\left(
\theta ,Z\right) \left( \nabla _{\theta }\left( \frac{\sigma _{\theta }^{2}}{%
2}\nabla _{\theta }-\omega ^{-1}\left( J\left( \theta \right) ,\theta ,Z,%
\mathcal{G}_{0}+\left\vert \Psi \right\vert ^{2}\right) \right) \right) \Psi
_{0}\left( \theta ,Z\right) \\
&&-\frac{1}{2}\int \Psi _{0}^{\dag }\left( \theta ,Z\right) \left( \nabla
_{\theta }\left( \frac{\sigma _{\theta }^{2}}{2}\nabla _{\theta }-\omega
^{-1}\left( J\left( \theta \right) ,\theta ,Z,\mathcal{G}_{0}+\left\vert
\Psi \right\vert ^{2}\right) \right) \right) \delta \Psi \left( \theta
,Z\right) \\
&&-\frac{1}{2}\int \delta \Psi ^{\dagger }\left( \theta ,Z\right) \left(
\nabla _{\theta }\left( \frac{\sigma _{\theta }^{2}}{2}\nabla _{\theta
}-\omega ^{-1}\left( J\left( \theta \right) ,\theta ,Z,\mathcal{G}%
_{0}+\left\vert \Psi \right\vert ^{2}\right) \right) \right) \delta \Psi
\left( \theta ,Z\right) \\
&&+\frac{1}{2}\int \delta \Psi ^{\dagger }\left( \theta ,Z\right) U^{\prime
\prime }\left( X_{0}\right) \delta \Psi \left( \theta ,Z\right)
\end{eqnarray*}%
The second term can still be neglected for relatively slow variations of $%
\Psi _{0}^{\dag }\left( \theta ,Z\right) $. The solutions are thus similar
to the previous paragraph:%
\begin{equation*}
\delta \Psi ^{\dagger }\left( \theta ,Z\right) =0
\end{equation*}%
and:%
\begin{eqnarray*}
\delta \Psi \left( \theta ,Z\right) &=&\left( \frac{\left( \nabla _{\theta
}\left( \frac{\sigma _{\theta }^{2}}{2}\nabla _{\theta }-\omega ^{-1}\left(
J\left( \theta \right) ,\theta ,Z,\mathcal{G}_{0}+\left\vert \Psi
\right\vert ^{2}\right) \right) \right) }{U^{\prime \prime }\left(
X_{0}\right) -\left( \nabla _{\theta }\left( \frac{\sigma _{\theta }^{2}}{2}%
\nabla _{\theta }-\omega ^{-1}\left( J\left( \theta \right) ,\theta ,Z,%
\mathcal{G}_{0}+\left\vert \Psi \right\vert ^{2}\right) \right) \right) }%
\right) \Psi _{0}\left( \theta ,Z\right) \\
&\simeq &-\frac{\nabla _{\theta }\left( \omega ^{-1}\left( J\left( \theta
\right) ,\theta ,Z,\mathcal{G}_{0}+\left\vert \Psi \right\vert ^{2}\right)
\right) }{U^{\prime \prime }\left( X_{0}\right) -\left( \nabla _{\theta
}\left( \frac{\sigma _{\theta }^{2}}{2}\nabla _{\theta }-\omega ^{-1}\left(
J\left( \theta \right) ,\theta ,Z,\mathcal{G}_{0}+\left\vert \Psi
\right\vert ^{2}\right) \right) \right) }\Psi _{0}\left( \theta ,Z\right) \\
&\simeq &-\frac{\nabla _{\theta }\omega ^{-1}\left( J\left( \theta \right)
,\theta ,Z,\mathcal{G}_{0}+\left\vert \Psi \right\vert ^{2}\right) }{%
U^{\prime \prime }\left( X_{0}\right) +\nabla _{\theta }\omega ^{-1}\left(
J\left( \theta \right) ,\theta ,Z,\mathcal{G}_{0}+\left\vert \Psi
\right\vert ^{2}\right) }\Psi _{0}\left( \theta ,Z\right)
\end{eqnarray*}

\section*{Appendix 4. Correlation functions and corrections to frequencies
equations}

\subsection*{\protect\bigskip 4.1 Two points correlation functions}

\subsubsection*{4.1.1 Second order effective vertex}

The two points Green function is the inverse of the second derivative of the
effective action $\Gamma \left( \Psi ^{\dagger },\Psi \right) $, defined by:%
\begin{equation*}
\Gamma _{1,1}\left( \left( \theta _{f},Z_{f}\right) ,\left( \theta
_{i},Z_{i}\right) \right) =\frac{\delta ^{2}\Gamma \left( \Psi ^{\dagger
},\Psi \right) }{\delta \Psi ^{\dagger }\left( \theta _{f},Z_{f}\right)
\delta \Psi \left( \theta _{i},Z_{i}\right) }
\end{equation*}%
where, in first approximation:%
\begin{eqnarray*}
&&\Gamma _{1,1}\left( \left( \theta _{f},Z_{f}\right) ,\left( \theta
_{i},Z_{i}\right) \right) \\
&=&-\nabla _{\theta }\left( \frac{\sigma _{\theta }^{2}}{2}\nabla _{\theta
}-\omega ^{-1}\left( J\left( \theta \right) ,\theta ,Z,\mathcal{G}%
_{0}+\left\vert \Psi \right\vert ^{2}\right) \right) +\hat{\Gamma}%
_{1,1}\left( \left( \theta _{f},Z_{f}\right) ,\left( \theta
_{i},Z_{i}\right) \right)
\end{eqnarray*}%
with $\hat{\Gamma}_{1,1}\left( \left( \theta _{f},Z_{f}\right) ,\left(
\theta _{i},Z_{i}\right) \right) $ given by the second derivatives of (\ref%
{ffc}). We decompose this vertex in two parts:%
\begin{equation*}
\hat{\Gamma}_{1,1}\left( \left( \theta _{f},Z_{f}\right) ,\left( \theta
_{i},Z_{i}\right) \right) =\hat{\Gamma}_{1,1}^{\left( 1\right) }\left(
\left( \theta _{f},Z_{f}\right) ,\left( \theta _{i},Z_{i}\right) \right) +%
\hat{\Gamma}_{1,1}^{\left( 2\right) }\left( \left( \theta _{f},Z_{f}\right)
,\left( \theta _{i},Z_{i}\right) \right)
\end{equation*}%
where:%
\begin{eqnarray}
&&\hat{\Gamma}_{1,1}^{\left( 1\right) }\left( \left( \theta
_{f},Z_{f}\right) ,\left( \theta _{i},Z_{i}\right) ,\Psi ^{\dagger },\Psi
\right) \\
&=&\sum_{\substack{ j\geqslant 2  \\ m\geqslant 2}}\sum_{\substack{ \left(
p_{l}^{i}\right) _{m\times j}  \\ \sum_{i}p_{l}^{i}\geqslant 2}}%
\sum_{l_{f}^{\prime }=1,l_{i}^{\prime }=1}^{j}\int \frac{\left(
\dprod\limits_{l=1,l\neq l_{f}^{\prime }}^{j}\Psi ^{\dagger }\left( \theta
_{f}^{\left( l\right) },Z_{l}\right) \right) }{m!\dprod\limits_{k}\left(
\sharp _{k}\right) !}\times \frac{\dprod\limits_{l=1}^{j}\exp \left(
-\Lambda _{1}\left( \theta _{f}^{\left( l\right) }-\theta _{i}^{\left(
l\right) }\right) \right) }{\Lambda ^{\sum_{i,l}p_{l}^{i}}}  \notag \\
&&\times \dprod\limits_{i=1}^{m}\left[ \int_{\dprod\limits_{l=1}^{j}\left[
\theta _{i}^{\left( l\right) },\theta _{f}^{\left( l\right) }\right]
^{p_{l}^{i}}}\frac{\delta ^{\sum_{l}p_{l}^{i}}\left[ \hat{S}_{cl}\left( \Psi
^{\dagger },\Psi \right) \right] }{\dprod\limits_{l=1}^{j}\dprod%
\limits_{k_{l}^{i}=1}^{p_{l}^{i}}\delta \left\vert \Psi \left( \theta
^{\left( k_{l}^{i}\right) },Z_{_{l}}\right) \right\vert ^{2}}%
\dprod\limits_{l=1}^{j}\dprod\limits_{k_{l}^{i}=1}^{p_{l}^{i}}d\theta
^{\left( k_{l}^{i}\right) }\right] _{\substack{ \left( \theta ^{\left(
l_{f}^{\prime }\right) },Z_{l_{f}^{\prime }}\right) =\left( \theta
_{f},Z_{f}\right)  \\ \left( \theta ^{\left( l_{i}^{\prime }\right)
},Z_{l_{i}^{\prime }}\right) =\left( \theta _{i},Z_{i}\right) }}\times
\left( \dprod\limits_{l=1,l\neq l_{i}^{\prime }}^{j}\Psi \left( \theta
_{i}^{\left( l\right) },Z_{l}\right) \right)  \notag
\end{eqnarray}%
and:%
\begin{eqnarray}
&&\hat{\Gamma}_{1,1}^{\left( 1\right) }\left( \left( \theta
_{f},Z_{f}\right) ,\left( \theta _{i},Z_{i}\right) ,\Psi ^{\dagger },\Psi
\right) \\
&=&\sum_{\substack{ j\geqslant 2  \\ m\geqslant 2}}\sum_{\substack{ \left(
p_{l}^{i}\right) _{m\times j}  \\ \sum_{i}p_{l}^{i}\geqslant 2}}\int \Psi
^{\dagger }\left( \theta _{f},Z_{f}\right) \left(
\dprod\limits_{l=1}^{j}\Psi ^{\dagger }\left( \theta _{f}^{\left( l\right)
},Z_{l}\right) \right)  \notag \\
&&\times \frac{\delta ^{2}}{\delta \left\vert \Psi \left( \theta
_{f},Z_{f}\right) \right\vert ^{2}\delta \left\vert \Psi \left( \theta
_{i},Z_{i}\right) \right\vert ^{2}}\dprod\limits_{i=1}^{m}\left[ \underset{%
\prod\limits_{l}\left[ \theta _{i}^{\left( l\right) },\theta _{f}^{\left(
l\right) }\right] ^{p_{l}^{i}}}{\int }\frac{\delta ^{\sum_{l}p_{l}^{i}}\left[
\hat{S}_{cl}\left( \Psi ^{\dagger },\Psi \right) \right] }{%
\dprod\limits_{l=1}^{j}\dprod\limits_{k_{l}^{i}=1}^{p_{l}^{i}}\delta
\left\vert \Psi \left( \theta ^{\left( k_{l}^{i}\right) },Z_{_{l}}\right)
\right\vert ^{2}}\dprod\limits_{l=1}^{j}\dprod%
\limits_{k_{l}^{i}=1}^{p_{l}^{i}}d\theta ^{\left( k_{l}^{i}\right) }\right] 
\notag \\
&&\times \frac{\dprod\limits_{l=1}^{j}\exp \left( -\Lambda _{1}\left( \theta
_{f}^{\left( l\right) }-\theta _{i}^{\left( l\right) }\right) \right) }{%
m!\dprod\limits_{k}\left( \sharp _{k}\right) !\Lambda ^{\sum_{i,l}p_{l}^{i}}}%
\left( \dprod\limits_{l=1}^{j}\Psi \left( \theta _{i}^{\left( l\right)
},Z_{l}\right) \right) \Psi \left( \theta _{i},Z_{i}\right)  \notag
\end{eqnarray}%
The second contribution corresponds to the influence of the propagation over
the whole system and can be neglected, as said in the text.

To find an approximate series expansion of $\hat{\Gamma}_{1,1}\left( \left(
\theta _{f},Z_{f}\right) ,\left( \theta _{i},Z_{i}\right) \right) $, we work
with the local series expansion (\ref{mmg}) of $\hat{\Gamma}_{1,1}\left(
\Psi ^{\dagger },\Psi \right) $. Using (\ref{clS}) alongside with the
associated notations, we can rewrite (\ref{mmg}) as:%
\begin{eqnarray}
&&\hat{\Gamma}_{1,1}\left( \left( \theta ,Z_{f}\right) ,\left( \theta
,Z_{i}\right) ,\Psi ^{\dagger },\Psi \right) \\
&=&\sum_{\substack{ j\geqslant 2  \\ m\geqslant 2}}\sum_{\substack{ \left(
p_{l}^{i}\right) _{m\times j}  \\ \sum_{i}p_{l}^{i}\geqslant 2}}%
\sum_{l_{f}^{\prime }=1,l_{i}^{\prime }=1}^{j}\int \left[ \left(
\dprod\limits_{l=1,l\neq l_{f}^{\prime }}^{j}\Psi ^{\dagger }\left( \theta
^{\left( l\right) },Z_{l}\right) \right) \times \right.  \notag \\
&&\frac{\dprod\limits_{i=1}^{m}\int c\exp \left( -c\left( \theta _{i}-\theta
_{i,j}\right) -\alpha \left( \sum_{l=1,p_{l}^{i}\neq 0}^{j}\left( \left(
c\left( \theta ^{\left( l-1,i\right) }-\theta ^{\left( l,i\right) }\right)
\right) ^{2}-\left\vert Z_{l-1}^{\left( i\right) }-Z_{l}^{\left( i\right)
}\right\vert ^{2}\right) \right) \right) \left\vert \Psi \left( \theta
_{i},Z_{i}\right) \right\vert ^{2}d\theta _{i}dZ_{i}}{\left( -2\right)
^{m}D^{\sum_{i,l}p_{l}^{i}}m!\dprod\limits_{k}\left( \sharp _{k}\right)
!\Lambda _{1}^{j}\Lambda ^{\sum_{i,l}p_{l}^{i}}}  \notag \\
&&\left. \left( \frac{\omega ^{-1}\left( J,\theta ^{\left( l\right)
},Z_{_{l}}\right) }{\left\vert \Psi \left( \theta ^{\left( l\right)
},Z_{_{l}}\right) \right\vert ^{2}}\right) ^{\sum_{i}p_{l}^{i}}\left(
\dprod\limits_{l=1,l\neq l_{i}^{\prime }}^{j}\Psi \left( \theta ^{\left(
l\right) },Z_{l}\right) d\theta ^{\left( l\right) }dZ_{l}\right) \right] _{ 
_{\substack{ \left( \theta ^{\left( l^{\prime }\right) },Z_{l_{f}^{\prime
}}\right) =\left( \theta _{i},Z_{f}\right)  \\ \left( \theta ^{\left(
l^{\prime }\right) },Z_{l_{i}^{\prime }}\right) =\left( \theta
_{f},Z_{i}\right) }}}  \notag
\end{eqnarray}

For $\Lambda >>1$, the dominant contributions is obtained for each point
connected by two vertices of valence $2$. As a consequence $%
\dprod\limits_{k}\left( \sharp _{k}\right) !=j$. Moreover there are $j\left(
j-1\right) $ possibilities to choose the external points (dismissing the
same point) and $m\left( m-1\right) $ possibilities to differentiate with
respect to $\left\vert \Psi \left( \theta _{i},Z_{i}\right) \right\vert ^{2}$
and $\left\vert \Psi \left( \theta _{f},Z_{f}\right) \right\vert ^{2}$. For
internal points that are integrated on, we can replace the sums over $\left(
\theta ^{\left( l\right) },Z_{l}\right) $ by the values of the quantities
evaluated at their average $\left( \bar{\theta},\bar{Z}\right) $ and we have:%
\begin{eqnarray}
&&\hat{\Gamma}_{1,1}\left( \left( \theta ,Z_{f}\right) ,\left( \theta
,Z_{i}\right) ,\Psi ^{\dagger },\Psi \right) \\
&\simeq &\sum_{\substack{ j\geqslant 2  \\ m\geqslant 2}}\int \left[ d\bar{%
\theta}d\bar{Z}\left( \left\vert \Psi \left( \bar{\theta},\bar{Z}\right)
\right\vert ^{2}\right) ^{j-2}\left( \frac{\omega ^{-1}\left( \bar{\theta},%
\bar{Z}\right) }{\left\vert \Psi \left( \bar{\theta},\bar{Z}\right)
\right\vert ^{2}}\right) ^{j-2}\Psi ^{\dag }\left( \theta _{i},Z_{i}\right)
\Psi \left( \theta _{f},Z_{f}\right) \frac{\omega ^{-1}\left( \theta
_{i},Z_{i}\right) }{\left\vert \Psi \left( \theta _{i},Z_{i}\right)
\right\vert ^{2}}\frac{\omega ^{-1}\left( \theta _{f},Z_{f}\right) }{%
\left\vert \Psi \left( \theta _{f},Z_{f}\right) \right\vert ^{2}}\right. 
\notag \\
&&\times \frac{\left( \int c\exp \left( -c\left( \theta -\bar{\theta}\right)
-\alpha \left( \left( c\left( \theta -\bar{\theta}\right) \right)
^{2}-\left\vert Z-\bar{Z}\right\vert ^{2}\right) \right) \left\vert \Psi
\left( \theta ,Z\right) \right\vert ^{2}d\theta dZ\right) ^{m-2}}{\left(
-2\right) ^{m}D^{m}\left( m-2\right) !\left( j-2\right) !\Lambda
_{1}^{j}\Lambda ^{2j}}  \notag \\
&&\times \frac{\left( \int c\exp \left( -c\left( \theta -\theta _{i}\right)
-\alpha \left( \left( c\left( \theta -\theta _{i}\right) \right)
^{2}-\left\vert Z-Z_{i}\right\vert ^{2}\right) \right) \left\vert \Psi
\left( \theta ,Z\right) \right\vert ^{2}d\theta dZ\right) }{\left( -2\right)
D} \\
&&\times \frac{\left( \int c\exp \left( -c\left( \theta -\theta _{f}\right)
-\alpha \left( \left( c\left( \theta -\theta _{f}\right) \right)
^{2}-\left\vert Z-Z_{f}\right\vert ^{2}\right) \right) \left\vert \Psi
\left( \theta ,Z\right) \right\vert ^{2}d\theta dZ\right) }{\left( -2\right)
D}
\end{eqnarray}%
We can replace the exponentials:%
\begin{equation*}
\int c\exp \left( -c\left( \theta -\bar{\theta}\right) -\alpha \left( \left(
c\left( \theta -\bar{\theta}\right) \right) ^{2}-\left\vert Z-\bar{Z}%
\right\vert ^{2}\right) \right) \left\vert \Psi \left( \theta
_{i},Z_{i}\right) \right\vert ^{2}d\theta _{i}dZ_{i}
\end{equation*}%
by its dominant contribution: 
\begin{equation*}
\int c\exp \left( -c\left( \theta -\bar{\theta}\right) \right) \left\vert
\Psi \left( \theta _{i},\bar{Z}+c\left( \theta -\bar{\theta}\right) \mathbf{e%
}\right) \right\vert ^{2}d\theta d\mathbf{e}
\end{equation*}%
where $\mathbf{e}$ is a unit vector. As a consequence:%
\begin{eqnarray*}
&&\hat{\Gamma}_{1,1}\left( \left( \theta ,Z_{f}\right) ,\left( \theta
,Z_{i}\right) ,\Psi ^{\dagger },\Psi \right) \\
&\simeq &\frac{\omega ^{-1}\left( \theta _{i},Z_{i}\right) }{\Psi \left(
\theta _{i},Z_{i}\right) }\frac{\omega ^{-1}\left( \theta _{f},Z_{f}\right) 
}{\Psi ^{\dag }\left( \theta _{f},Z_{f}\right) }\exp \left( \int \frac{%
\omega ^{-1}\left( \bar{\theta},\bar{Z}\right) }{\Lambda _{1}}d\bar{\theta}d%
\bar{Z}\right) \\
&&\times \exp \left( -\int \frac{c}{2D\Lambda }\exp \left( -c\left( \theta -%
\bar{\theta}\right) \right) \left\vert \Psi \left( \theta ,\bar{Z}+c\left(
\theta -\bar{\theta}\right) \mathbf{e}\right) \right\vert ^{2}d\theta d%
\mathbf{e}\right) \\
&&\times \left( \int \frac{c}{2D\Lambda }\exp \left( -c\left( \theta -\theta
_{i}\right) \right) \left\vert \Psi \left( \theta ,Z_{i}+c\left( \theta
-\theta _{i}\right) \mathbf{e}\right) \right\vert ^{2}d\theta d\mathbf{e}%
\right) \\
&&\times \left( \int \frac{c}{2D\Lambda }\exp \left( -c\left( \theta -\theta
_{f}\right) \right) \left\vert \Psi \left( \theta ,Z_{f}+c\left( \theta
-\theta _{f}\right) \mathbf{e}\right) \right\vert ^{2}d\theta d\mathbf{e}%
\right) \\
&\simeq &\frac{\omega ^{-1}\left( \theta _{i},Z_{i}\right) }{\Psi \left(
\theta _{i},Z_{i}\right) }\frac{\omega ^{-1}\left( \theta _{f},Z_{f}\right) 
}{\Psi ^{\dag }\left( \theta _{f},Z_{f}\right) }\exp \left( \int \frac{%
\omega ^{-1}\left( \bar{\theta},\bar{Z}\right) }{\Lambda _{1}}d\bar{\theta}d%
\bar{Z}\right) \times \exp \left( -\int \frac{1}{2D\Lambda }\left\vert \Psi
\left( \bar{\theta},\bar{Z}\right) \right\vert ^{2}d\bar{\theta}d\bar{Z}%
\right) \\
&&\times \frac{1}{2D\Lambda }\left\vert \Psi \left( \theta _{i},Z_{i}\right)
\right\vert ^{2}\times \frac{1}{2D\Lambda }\left\vert \Psi \left( \theta
_{f},Z_{f}\right) \right\vert ^{2} \\
&=&\omega ^{-1}\left( \theta _{i},Z_{i}\right) \omega ^{-1}\left( \theta
_{f},Z_{f}\right) \Psi \left( \theta _{f},Z_{f}\right) \Psi ^{\dag }\left(
\theta _{i},Z_{i}\right) C\left( \bar{\omega},\Psi \right)
\end{eqnarray*}%
We thus have:%
\begin{equation*}
\hat{\Gamma}_{1,1}\left( \left( \theta ,Z_{f}\right) ,\left( \theta
,Z_{i}\right) ,\Psi ^{\dagger },\Psi \right) =\omega ^{-1}\left( \theta
_{i},Z_{i}\right) \omega ^{-1}\left( \theta _{f},Z_{f}\right) \Psi \left(
\theta _{f},Z_{f}\right) \Psi ^{\dag }\left( \theta _{i},Z_{i}\right)
C\left( \bar{\omega},\Psi \right)
\end{equation*}

where:%
\begin{equation*}
C\left( \bar{\omega},\Psi \right) =\frac{1}{2D\Lambda }\frac{1}{2D\Lambda }%
\exp \left( \int \frac{\omega ^{-1}\left( \bar{\theta},\bar{Z}\right) }{%
\Lambda _{1}}d\bar{\theta}d\bar{Z}\right) \times \exp \left( -\int \frac{c}{%
2D\Lambda }\left\vert \Psi \left( \bar{\theta},\bar{Z}\right) \right\vert
^{2}d\bar{\theta}d\bar{Z}\right)
\end{equation*}

\subsubsection*{4.1.2 2 points correlation function}

Inverting $\Gamma _{1,1}\left( \theta _{f},\theta _{i}\right) $ yields the
two-points Green function:

\begin{equation*}
G_{2}\left( \theta _{f},\theta _{i}\right) =\mathcal{G}\left( \theta
_{f},\theta _{i}\right) +\mathcal{G\ast }\sum_{n\geqslant 2}\left( -1\right)
^{n-1}\left( \hat{\Gamma}_{1,1}\left( \left( \theta _{f},Z_{f}\right)
,\left( \theta _{i},Z_{i}\right) ,\Psi ^{\dagger },\Psi \right) \ast 
\mathcal{G}\right) ^{n}
\end{equation*}

where:

\begin{equation}
-\nabla _{\theta }\left( \frac{\sigma _{\theta }^{2}}{2}\nabla _{\theta
}-\omega ^{-1}\left( J\left( \theta \right) ,\theta ,Z,\mathcal{G}%
_{0}+\left\vert \Psi \right\vert ^{2}\right) \right) \mathcal{G}\left(
\theta _{f},\theta _{i},Z_{f},Z_{i}\right) =\mathcal{\delta }\left( \theta
_{f}-\theta _{i}\right) \delta \left( Z_{f}-Z_{i}\right)  \label{grn}
\end{equation}%
For relatively slow variations in frequencies, we can use (\ref{rdrzr}) by
replacing $\omega ^{-1}\left( J,\theta ,Z,\mathcal{G}_{0}\right) $ by its
average on the interval $\left[ \theta ,\theta ^{\prime }\right] $: 
\begin{equation*}
\omega ^{-1}\left( J,\theta ,Z,\mathcal{G}_{0}\right) =\frac{1}{\bar{X}_{r}}%
\rightarrow \left\langle \omega ^{-1}\left( J\left( \theta \right) ,\theta
,Z,\mathcal{G}_{0}+\left\vert \Psi \right\vert ^{2}\right) \right\rangle _{%
\left[ \theta ,\theta ^{\prime }\right] }\equiv \bar{\omega}^{-1}\left(
J\left( \theta \right) ,\theta ,Z,\mathcal{G}_{0}+\left\vert \Psi
\right\vert ^{2}\right)
\end{equation*}%
in (\ref{mgv}): 
\begin{equation*}
\mathcal{G}\left( \theta _{f},\theta _{i},Z_{f},Z_{i}\right) =\delta \left(
Z_{f}-Z_{i}\right) \frac{\exp \left( -\left( \sqrt{\left( \frac{\bar{\omega}%
^{-1}\left( J\left( \theta \right) ,\theta ,Z,\mathcal{G}_{0}+\left\vert
\Psi \right\vert ^{2}\right) }{\sigma ^{2}}\right) ^{2}+\frac{2\alpha }{%
\sigma ^{2}}}-\frac{\bar{\omega}^{-1}\left( J\left( \theta \right) ,\theta
,Z,\mathcal{G}_{0}+\left\vert \Psi \right\vert ^{2}\right) }{\sigma ^{2}}%
\right) \left( \theta -\theta ^{\prime }\right) \right) }{\sqrt{\frac{\pi }{2%
}}\sqrt{\left( \frac{\bar{\omega}^{-1}\left( J\left( \theta \right) ,\theta
,Z,\mathcal{G}_{0}+\left\vert \Psi \right\vert ^{2}\right) }{\sigma ^{2}}%
\right) ^{2}+\frac{2\alpha }{\sigma ^{2}}}}H\left( \theta -\theta ^{\prime
}\right)
\end{equation*}%
As a consequence, the solution of (\ref{grn}) is a series expansion:

\begin{eqnarray*}
G_{2}\left( \theta _{f},\theta _{i},Z_{f},Z_{i}\right) &=&\mathcal{G}\left(
\theta _{f},\theta _{i},Z_{f},Z_{i}\right) +\mathcal{G\ast }\sum_{n\geqslant
2}\left( -1\right) ^{n-1}\left( \hat{\Gamma}_{1,1}\ast \mathcal{G}\right)
^{n} \\
&=&\mathcal{G}\left( \theta _{f},\theta _{i},Z_{f},Z_{i}\right) +\int \left(
\prod\limits_{k=1}^{n}d\theta _{k}dZ_{k}\right) \mathcal{G}\left( \theta
_{f},\theta _{n},Z_{f},Z_{f}\right) \frac{\Psi \left( \theta
_{n},Z_{f}\right) }{\omega \left( \theta _{n},Z_{f}\right) }\times
\sum_{n\geqslant 2}\left( -1\right) ^{n-1} \\
&&\times \prod\limits_{k=1}^{n-1}\left( \left\vert \Psi \left( \theta
_{k},Z_{k}\right) \right\vert ^{2}\left( \omega ^{-1}\left( \theta
_{k},Z_{k}\right) \right) ^{2}\mathcal{G}\left( \theta _{k+1},\theta
_{k},Z_{k},Z_{k}\right) \right) \frac{\Psi ^{\dag }\left( \theta
_{1},Z_{i}\right) }{\omega \left( \theta _{1},Z_{i}\right) }\mathcal{G}%
\left( \theta _{1},\theta _{i},Z_{i},Z_{i}\right)
\end{eqnarray*}

\subsection*{4.2 $\left( k,n\right) $ points correlation functions}

The $\left( k,n\right) $ points correlation functions, are found using the
standard techniques. We first derive the $\left( k,n\right) $ effective
vertex, the associated connected correlation function, from which the $%
\left( k,n\right) $ correlation function is derived.

\subsubsection*{4.2.1 $\left( k,n\right) $ effective vertices\protect\bigskip%
}

The $\left( k,n\right) $-th effective vertex are defined by:%
\begin{equation*}
\Gamma _{k,n}\left( \left( \theta _{f}^{\left( l\right) },Z_{l}\right)
_{l=1,..,k},\left( \theta _{i}^{\left( l\right) },Z_{l}\right)
_{l=1,..,n}\right) =\frac{\delta ^{k+n}\Gamma \left( \Psi ^{\dagger },\Psi
\right) }{\delta ^{k}\left( \Psi ^{\dagger }\left( \theta _{f}^{\left(
l\right) },Z_{l}\right) \right) _{l=1,..,k}\delta ^{n}\left( \Psi \left(
\theta _{i}^{\left( l\right) },Z_{l}\right) \right) _{l=1,..,n}}
\end{equation*}%
Neglecting the derivatives corresponding to the impact of propagation
between $\theta _{i}$ and $\theta _{f}$, we find:

\begin{eqnarray*}
&&\Gamma _{k,n}\left( \left( \theta _{f}^{\left( r\right) },Z_{r}\right)
_{r=1,..,k},\left( \theta _{i}^{\left( s\right) },Z_{s}\right)
_{s=1,..,n}\right) \\
&=&\sum_{\substack{ j\geqslant \max \left( k,n\right)  \\ m\geqslant 2}}\sum 
_{\substack{ \left( p_{l}^{i}\right) _{m\times j}  \\ \sum_{i}p_{l}^{i}%
\geqslant 2}}\sum_{\substack{ \left( l_{1}^{\prime }...,l_{k}^{\prime
}\right)  \\ \times \left( l_{1}^{\prime \prime }...,l_{n}^{\prime \prime
}\right)  \\ \subset \left( 1,...,j\right) ^{2}}}^{j}\int \left(
\dprod\limits_{l=1,l\notin \left( l_{i}^{\prime }\right) }^{j}\Psi ^{\dagger
}\left( \theta _{f}^{\left( l\right) },Z_{l}\right) \right) \\
&&\times \dprod\limits_{i=1}^{m}\left[ \int_{\dprod\limits_{l=1}^{j}\left[
\theta _{i}^{\left( l\right) },\theta _{f}^{\left( l\right) }\right]
^{p_{l}^{i}}}\frac{\delta ^{\sum_{l}p_{l}^{i}}\left[ \hat{S}_{cl}\left( \Psi
^{\dagger },\Psi \right) \right] }{\dprod\limits_{l=1}^{j}\dprod%
\limits_{k_{l}^{i}=1}^{p_{l}^{i}}\delta \left\vert \Psi \left( \theta
^{\left( k_{l}^{i}\right) },Z_{_{l}}\right) \right\vert ^{2}}%
\dprod\limits_{l=1}^{j}\dprod\limits_{k_{l}^{i}=1}^{p_{l}^{i}}d\theta
^{\left( k_{l}^{i}\right) }\right] _{_{\substack{ \left( \theta ^{\left(
l_{i}^{\prime }\right) },Z_{l_{i}^{\prime }}\right) =\left( \theta
_{f}^{\left( r\right) },Z_{r}\right)  \\ \left( \theta ^{\left(
l_{i}^{\prime \prime }\right) },Z_{l_{i}^{\prime \prime }}\right) =\left(
\theta _{i}^{\left( s\right) },Z_{s}\right) }}} \\
&&\times \frac{\exp \left( -\Lambda _{1}\left( \theta _{f}^{\left( l\right)
}-\theta _{i}^{\left( l\right) }\right) \right) }{m!\dprod\limits_{k}\left(
\sharp _{k}\right) !\Lambda ^{\sum_{i}p_{l}^{i}}}\left(
\dprod\limits_{l=1,l\notin \left( l_{i}^{\prime \prime }\right) }^{j}\Psi
\left( \theta _{i}^{\left( l\right) },Z_{l}\right) \right)
\end{eqnarray*}%
Under the approximation that $\frac{\delta ^{p}\left[ \hat{S}_{cl}\left(
\Psi ^{\dagger },\Psi \right) \right] }{\dprod\limits_{i=1}^{p}\delta
\left\vert \Psi \left( \theta ^{\left( k_{i}\right) },Z_{_{k_{i}}}\right)
\right\vert ^{2}}$ is decreasing with $p$, the previous expression reduces
to:%
\begin{eqnarray}
&&\Gamma _{k,n}\left( \left( \theta _{f}^{\left( r\right) },Z_{r}\right)
_{r=1,..,k},\left( \theta _{i}^{\left( s\right) },Z_{s}\right)
_{s=1,..,n}\right)  \label{xpr} \\
&=&\sum_{\substack{ j=\max \left( k,n\right)  \\ m\geqslant 2}}\sum 
_{\substack{ \cup L_{i}  \\ =2\left\{ 1,...,j\right\} }}\sum_{\substack{ %
\left( l_{1}^{\prime }...,l_{k}^{\prime }\right)  \\ \times \left(
l_{1}^{\prime \prime }...,l_{n}^{\prime \prime }\right)  \\ \subset \left(
1,...,j\right) ^{2}}}^{j}\int \frac{\left( \dprod\limits_{l=1,l\notin \left(
l_{i}^{\prime }\right) }^{j}\Psi ^{\dagger }\left( \theta _{f}^{\left(
l\right) },Z_{l}\right) \right) }{\Lambda ^{2j}}  \notag \\
&&\times \underset{\prod\limits_{l=1}^{j}\left[ \theta _{i}^{\left(
l\right) },\theta _{f}^{\left( l\right) }\right] ^{2}}{\int }\left[
\dprod\limits_{i=1}^{m}\frac{\delta ^{\sharp L_{i}}\left[ \hat{S}_{cl}\left(
\Psi ^{\dagger },\Psi \right) \right] }{\dprod\limits_{l_{i}\in L_{i}}\delta
\left\vert \Psi \left( \theta ^{\left( l_{i}\right) },Z_{_{l_{i}}}\right)
\right\vert ^{2}}\dprod\limits_{l_{i}\in L_{i}}d\theta ^{\left( l_{i}\right)
}\right] _{_{\substack{ \left( \theta ^{\left( l_{i}^{\prime }\right)
},Z_{l_{i}^{\prime }}\right) =\left( \theta _{f}^{\left( r\right)
},Z_{r}\right)  \\ \left( \theta ^{\left( l_{i}^{\prime \prime }\right)
},Z_{l_{i}^{\prime \prime }}\right) =\left( \theta _{i}^{\left( s\right)
},Z_{s}\right) }}}  \notag \\
&&\times \frac{\exp \left( -\Lambda _{1}\left( \theta _{f}^{\left( l\right)
}-\theta _{i}^{\left( l\right) }\right) \right) }{m!\dprod\limits_{k}\left(
\sharp _{k}\right) !}\left( \dprod\limits_{l=1,l\notin \left( l_{i}^{\prime
\prime }\right) }^{j}\Psi \left( \theta _{i}^{\left( l\right) },Z_{l}\right)
\right)  \notag
\end{eqnarray}%
In first approximation:%
\begin{equation*}
\frac{\delta ^{\sharp L_{i}}\left[ \hat{S}_{cl}\left( \Psi ^{\dagger },\Psi
\right) \right] }{\dprod\limits_{l_{i}\in L_{i}}\delta \left\vert \Psi
\left( \theta ^{\left( l_{i}\right) },Z_{_{l_{i}}}\right) \right\vert ^{2}}%
\simeq \dprod\limits_{l_{i}\in L_{i}}\nabla _{\theta _{i}}\omega ^{-1}\left(
J\left( \theta ^{\left( l_{i}\right) }\right) ,\theta ^{\left( l_{i}\right)
},Z_{_{l_{i}}},\mathcal{G}_{0}+\left\vert \Psi \right\vert ^{2}\right)
\end{equation*}

where $2\left\{ 1,...,j\right\} $ denotes two copies of $\left\{
1,...,j\right\} $ and (\ref{ngv}) gives:%
\begin{eqnarray}
&&\nabla _{\theta _{i}}\omega ^{-1}\left( \theta ^{\left( i\right) },Z\right)
\label{rnv} \\
&=&\nabla _{\theta _{i}}\frac{G^{\prime }\left[ J,\omega ,\theta ,Z,\Psi %
\right] }{F^{\prime }\left[ J,\omega ,\theta ,Z,\Psi \right] }\int
\sum_{k=0}^{\infty }\frac{\exp \left( -\alpha \left( \theta _{2}-\theta
_{1}-\sum_{l=0}^{k}\frac{\left\vert Z_{l}-Z_{l+1}\right\vert }{c}\right)
\right) }{A^{k+1}}  \notag \\
&&\times \left( \dprod\limits_{l=1}^{k}\int \left( \left\vert \Psi \left(
\theta -l_{l},Z_{l}\right) \right\vert ^{2}-\frac{\left( \hat{T}_{1}\omega
_{0}\left\vert \Psi \right\vert ^{2}\right) \left( \theta
-l_{l},Z_{l}\right) }{\left( \omega _{0}+\hat{T}\omega _{0}\left\vert \Psi
\right\vert ^{2}\right) \left( \theta -l_{l},Z_{l}\right) }\right)
dZ_{l}dl_{l}\right) \omega _{0}\left( J,\theta -l_{k},Z_{k}\right)
\left\vert \Psi \left( \theta -l_{k},Z_{k}\right) \right\vert ^{2}  \notag
\end{eqnarray}

\subsubsection*{4.2.2 \ $\left( k,n\right) $ connected correlation functions}

The link between connected correlation functions and effective vertices is
obtained by the recursive relation:%
\begin{eqnarray*}
G_{k,n}^{\left( c\right) }\left( \left( \theta _{f}^{\left( r\right)
},Z_{r}\right) _{r=1,..,k},\left( \theta _{i}^{\left( s\right)
},Z_{s}\right) _{s=1,..,n}\right) &=&-\Gamma _{k,n}\left( \left( \theta
_{f}^{\left( r\right) },Z_{r}\right) _{r=1,..,k},\left( \theta _{i}^{\left(
s\right) },Z_{s}\right) _{s=1,..,n}\right) \\
&&+\sum_{l}\left( -1\right) ^{l}\sum_{_{\substack{ \sum k_{i}=k+l  \\ \sum
n_{i}=n+l}}}\sum_{G}\left[ \Gamma _{k_{1},n_{1}}\ast _{\mathcal{G}}...%
\mathcal{\ast }_{\mathcal{G}}\Gamma _{k_{l},n_{l}}\right] _{G}
\end{eqnarray*}%
where the sum over $G$ denotes the graphs with trivial fundamental group
obtained by drawing $l$ vertices labelled $\Gamma _{k_{i},n_{i}}$. The
vertex $\Gamma _{k_{i},n_{i}}$ has two types of valences, denoted in and
out, of order $\left( k_{1},n_{1}\right) $. The $\Gamma _{k_{i},n_{i}}$ are
connected by segments issued from a valence of type out to a valence of type
in. Only $\left( k,l\right) $ valences labelled $\left( \theta _{f}^{\left(
r\right) },Z_{r}\right) _{r=1,..,k},\left( \theta _{i}^{\left( s\right)
},Z_{s}\right) _{s=1,..,n}$ are left free.

The expression $\left[ \Gamma _{k_{1},n_{1}}\ast _{\mathcal{G}}...\mathcal{%
\ast }_{\mathcal{G}}\Gamma _{k_{l},n_{l}}\right] _{G}$ is computed for each
graph $G$ by associating a propagator $\mathcal{G}$ to each leg of the
graph: and convoluting all expressions that are connected. For slowly
varying $\hat{S}_{cl}\left( \Psi ^{\dagger },\Psi \right) $ in fields, $%
\left\vert \frac{\delta ^{p}\left[ \hat{S}_{cl}\left( \Psi ^{\dagger },\Psi
\right) \right] }{\dprod\limits_{i=1}^{p}\delta \left\vert \Psi \left(
\theta ^{\left( k_{i}\right) },Z_{_{k_{i}}}\right) \right\vert ^{2}}%
\right\vert <<1$ and in first approximation: 
\begin{equation*}
G_{k,n}^{c}\left( \left( \theta _{f}^{\left( r\right) },Z_{r}\right)
_{r=1,..,k},\left( \theta _{i}^{\left( s\right) },Z_{s}\right)
_{s=1,..,n}\right) \simeq -\Gamma _{k,n}\left( \left( \theta _{f}^{\left(
r\right) },Z_{r}\right) _{r=1,..,k},\left( \theta _{i}^{\left( s\right)
},Z_{s}\right) _{s=1,..,n}\right)
\end{equation*}

\subsubsection*{4.2.3 $\left( k,n\right) $\ correlation functions}

The Green functions can ultimately be computed in the usual way. They are
defined by:%
\begin{eqnarray*}
&&G_{k,n}\left( \left( \theta _{f}^{\left( l\right) },Z_{l}\right)
_{l=1,..,k},\left( \theta _{i}^{\left( l\right) },Z_{l}\right)
_{l=1,..,n}\right) \\
&=&\sum_{i=1,j=1}^{k,n}\sum_{P_{i}\left( k\right) ,P_{j}\left( n\right)
}\dprod\limits_{\substack{ r\in P_{i}\left( k\right)  \\ s\in P_{j}\left(
n\right) }}G_{k_{r},n_{s}}^{\left( c\right) }\left( \left( \theta
_{f}^{\left( l,r\right) },Z_{l,r}\right) _{l=1,..,k_{r}},\left( \theta
_{i}^{\left( l,s\right) },Z_{l,s}\right) _{l=1,..,n_{s}}\right)
\end{eqnarray*}%
where $P_{i}\left( k\right) $ and $P_{j}\left( n\right) $ denote the
partitions of $k$ and $n$ in $i$ and $j$ subsets and: 
\begin{equation*}
\cup _{r}\left( \theta _{f}^{\left( l,r\right) },Z_{l,r}\right)
_{l=1,..,k_{r}}=\left( \theta _{f}^{\left( l\right) },Z_{l}\right)
_{l=1,..,k}
\end{equation*}
\begin{equation*}
\cup _{s}\left( \theta _{i}^{\left( l,s\right) },Z_{l,s}\right)
_{l=1,..,n_{s}}=\left( \theta _{i}^{\left( l\right) },Z_{l}\right)
_{l=1,..,n}
\end{equation*}%
In first approximation, the Green function can thus be written:%
\begin{eqnarray*}
&&G_{k,n}\left( \left( \theta _{f}^{\left( l\right) },Z_{l}\right)
_{l=1,..,k},\left( \theta _{i}^{\left( l\right) },Z_{l}\right)
_{l=1,..,n}\right) \\
&=&\sum_{\sigma _{k},\sigma _{n}}\sum_{u=0}^{\inf \left( k,n\right)
}\prod\limits_{j=0}^{u}G_{2}\left( \left( \theta _{f}^{\left( j\right)
},Z_{f}\right) ,\left( \theta _{i}^{\left( i\right) },Z_{i}\right) \right) \\
&&\times \sum_{i=1,j=1}^{k-u,n-u}\left( -1\right) ^{i+j}\sum_{\substack{ %
P_{i}\left( k-u\right)  \\ P_{j}\left( n-u\right) }}\dprod\limits_{\substack{
r\in P_{i}\left( k-u\right)  \\ s\in P_{j}\left( n-u\right) }}\Gamma
_{k_{r},n_{s}}\left( \left( \theta _{f}^{\left( l,r,u\right)
},Z_{l,r,u}\right) _{l=1,..,k_{r}},\left( \theta _{i}^{\left( l,s,u\right)
},Z_{l,s,u}\right) _{l=1,..,n_{s}}\right)
\end{eqnarray*}%
where $\cup _{r}\left( \theta _{f}^{\left( l,r,u\right) },Z_{l,r,u}\right)
_{l=1,..,k_{r}}=\left( \theta _{f}^{\left( l\right) },Z_{l}\right)
_{l=u+1,..,k}$ and $\cup _{s}\left( \theta _{i}^{\left( l,s,u\right)
},Z_{l,s,u}\right) _{l=u+1,..,n_{s}}=\left( \theta _{i}^{\left( l\right)
},Z_{l}\right) _{l=u+1,..,n}$ as ordered sets and the sum over $\sigma _{k}$
and $\sigma _{n}$ is over all permutations of the $\left( \theta
_{f}^{\left( l\right) },Z_{l}\right) _{l=1,..,k}$ and $\left( \theta
_{i}^{\left( l\right) },Z_{l}\right) _{l=1,..,n}$ respectively.

\subsection*{4.3 Corrections to (\protect\ref{pw})}

\subsubsection*{4.3.1 Series expansion}

The corrective terms to $\omega ^{-1}\left( J\left( \theta \right) ,\theta
,Z,\mathcal{G}_{0}+\left\vert \Psi \right\vert ^{2}\right) $ are obtained by
isolating $\Omega \left( \theta ,Z\right) $ in (\ref{rfq}):%
\begin{equation*}
\Gamma \left( \Psi ^{\dagger },\Psi \right) \simeq \int \Psi ^{\dagger
}\left( \theta ,Z\right) \left( -\nabla _{\theta }\left( \frac{\sigma
_{\theta }^{2}}{2}\nabla _{\theta }-\omega ^{-1}\left( J\left( \theta
\right) ,\theta ,Z,\mathcal{G}_{0}+\left\vert \Psi \right\vert ^{2}\right)
\right) \delta \left( \theta _{f}-\theta _{i}\right) +\Omega \left( \theta
,Z\right) \right) \Psi \left( \theta ,Z\right)
\end{equation*}%
and by integration over $\theta $ (see (\ref{fft})). As explained in the
text, we study the weak field case in the local approximation, so that we
use formula (\ref{wkf}) for the effective action. Combining (\ref{wkf}) with
(\ref{fft}) we obtain $\omega _{e}^{-1}\left( J\left( \theta \right) ,\theta
,Z,\mathcal{G}_{0}+\left\vert \Psi \right\vert ^{2}\right) $: 
\begin{equation}
\omega _{e}^{-1}\left( J\left( \theta \right) ,\theta ,Z,\mathcal{G}%
_{0}+\left\vert \Psi \right\vert ^{2}\right) =\omega ^{-1}\left( J\left(
\theta \right) ,\theta ,Z,\mathcal{G}_{0}+\left\vert \Psi \right\vert
^{2}\right) +Z  \label{fwK}
\end{equation}%
where:%
\begin{eqnarray}
Z &=&\int^{\theta }d\theta \sum_{\substack{ j\geqslant 1  \\ m\geqslant 1}}%
\sum_{\substack{ p_{l},\left( p_{l}^{i}\right) _{m\times j}  \\ %
p_{l}+\sum_{i}p_{l}^{i}\geqslant 2}}\left( \prod\limits_{i=1}^{m}\frac{%
\sharp _{j+1,m}\left( \left( p_{m},\left( p_{l}^{m}\right) \right) \right) }{%
4\bar{\sharp}_{j+1,m}\left( \left( p_{i},\left( p_{l}^{m}\right) \right)
\right) }\right) \frac{a_{j,m}}{2} \\
&&\times \int \prod\limits_{i=1}^{m}\left\{ \frac{\delta
^{\sum_{l}p_{l}^{i}}\left( \nabla _{\theta }\omega ^{-1}\left( \theta
,Z,\left\vert \Psi \right\vert ^{2}\right) \right) }{\dprod\limits_{l=1}^{j}%
\delta ^{p_{l}^{i}}\left\vert \Psi \left( \theta ^{\left( l\right)
},Z_{_{l}}\right) \right\vert ^{2}}\right\} \frac{\delta
^{\sum_{l}p_{l}}\left( \nabla _{\theta }\omega ^{-1}\left( \theta
,Z,\left\vert \Psi \right\vert ^{2}\right) \right) }{\dprod\limits_{l=1}^{j}%
\delta ^{p_{l}}\left\vert \Psi \left( \theta ^{\left( l\right)
},Z_{_{l}}\right) \right\vert ^{2}}\dprod\limits_{l=1}^{j}\left\vert \Psi
\left( \theta ^{\left( l\right) },Z_{l}\right) \right\vert ^{2}d\theta
^{\left( l\right) }dZ_{l}  \notag
\end{eqnarray}%
Equation (\ref{fwK}) is a series in $\nabla _{\theta }\omega ^{-1}\left(
J\left( \theta \right) ,\theta ,Z,\mathcal{G}_{0}+\left\vert \Psi
\right\vert ^{2}\right) $. We provide below a detailed computation of the
lowest order terms.

To conclude, note that the series can be divided in two different parts and
writes:%
\begin{eqnarray*}
Z &=&\left[ \frac{1}{n!}\left( \sum_{l\geqslant 2}\frac{1}{l!}\int \Psi
^{\dagger }\left( \theta _{i},Z_{i}\right) \Psi \left( \theta
_{i},Z_{i}\right) \frac{\delta ^{l}}{\Lambda ^{l}\delta ^{l}\left\vert \Psi
\left( \theta _{i},Z_{i}\right) \right\vert ^{2}}d\theta _{i}\right)
^{n}\right. \\
&&\times \left. \sum \frac{1}{p!}\left( \nabla _{\theta }\omega ^{-1}\left(
\theta ,Z,\left\vert \Psi \right\vert ^{2}\right) +\Psi ^{\dag }\left(
\theta ,Z\right) \nabla _{\theta }\omega ^{-1}\left( \theta ,Z,\left\vert
\Psi \right\vert ^{2}\right) \Psi \left( \theta ,Z\right) \right) ^{p}\right]
_{\Psi \left( \theta ,Z\right) =\Psi ^{\dag }\left( \theta ,Z\right) =0}
\end{eqnarray*}%
and that this can also be written by using the exponential form given in (%
\ref{epF}) in the local approximation:%
\begin{eqnarray}
Z &=&\left[ \exp \left( \sum_{l\geqslant 2}\frac{1}{l!}\int \Psi ^{\dagger
}\left( \theta _{i},Z_{i}\right) \Psi \left( \theta _{i},Z_{i}\right) \frac{%
\delta ^{l}}{\Lambda ^{l}\delta ^{l}\left\vert \Psi \left( \theta
_{i},Z_{i}\right) \right\vert ^{2}}d\theta _{i}\right) \right. \\
&&\left. \exp \left( \nabla _{\theta }\omega ^{-1}\left( \theta
,Z,\left\vert \Psi \right\vert ^{2}\right) +\Psi ^{\dag }\left( \theta
,Z\right) \nabla _{\theta }\omega ^{-1}\left( \theta ,Z,\left\vert \Psi
\right\vert ^{2}\right) \Psi \left( \theta ,Z\right) \right) \right] _{\Psi
\left( \theta ,Z\right) =\Psi ^{\dag }\left( \theta ,Z\right) =0}  \notag
\end{eqnarray}

\subsubsection*{4.3.2 Lowest order corrections}

Given that $p_{l}+\sum_{i}p_{l}^{i}\geqslant 2$, the lowest order correction
in (\ref{fwk}) is for $m=1$,$j=1$ , $p_{l}=p_{1}^{1}=1$:%
\begin{eqnarray}
&&\omega _{e}^{-1}\left( J\left( \theta \right) ,\theta ,Z,\mathcal{G}%
_{0}+\left\vert \Psi \right\vert ^{2}\right)  \label{fcR} \\
&=&\omega ^{-1}\left( J\left( \theta \right) ,\theta ,Z,\mathcal{G}%
_{0}+\left\vert \Psi \right\vert ^{2}\right)  \notag \\
&&+\frac{1}{4}\int \int^{\theta }\frac{\delta \left( \nabla _{\theta }\omega
^{-1}\left( J\left( \theta \right) ,\theta ,Z,\mathcal{G}_{0}+\left\vert
\Psi \right\vert ^{2}\right) \right) }{\delta \left\vert \Psi \left( \theta
^{\left( l\right) },Z_{_{l}}\right) \right\vert ^{2}}\times \frac{\delta
\left( \nabla _{\theta }\omega ^{-1}\left( J\left( \theta \right) ,\theta ,Z,%
\mathcal{G}_{0}+\left\vert \Psi \right\vert ^{2}\right) \right) }{\delta
\left\vert \Psi \left( \theta ^{\left( l\right) },Z_{_{l}}\right)
\right\vert ^{2}}\left\vert \Psi \left( \theta ^{\left( l\right)
},Z_{_{l}}\right) \right\vert ^{2}  \notag
\end{eqnarray}%
We rewrite the second term in the RHS of (\ref{fcR}):%
\begin{equation*}
Z\equiv \int \int^{\theta }\frac{\delta \left( \nabla _{\theta }\omega
^{-1}\left( J\left( \theta \right) ,\theta ,Z,\mathcal{G}_{0}+\left\vert
\Psi \right\vert ^{2}\right) \right) }{\delta \left\vert \Psi \left( \theta
^{\left( l\right) },Z_{_{l}}\right) \right\vert ^{2}}\frac{\delta \left(
\nabla _{\theta }\omega ^{-1}\left( J\left( \theta \right) ,\theta ,Z,%
\mathcal{G}_{0}+\left\vert \Psi \right\vert ^{2}\right) \right) }{\delta
\left\vert \Psi \left( \theta ^{\left( l\right) },Z_{_{l}}\right)
\right\vert ^{2}}\left\vert \Psi \left( \theta ^{\left( l\right)
},Z_{_{l}}\right) \right\vert ^{2}
\end{equation*}%
using an integration by part for the variable $\theta $:%
\begin{eqnarray*}
&&Z=\int \frac{\delta \left( \omega ^{-1}\left( J\left( \theta \right)
,\theta ,Z,\mathcal{G}_{0}+\left\vert \Psi \right\vert ^{2}\right) \right) }{%
\delta \left\vert \Psi \left( \theta ^{\left( l\right) },Z_{_{l}}\right)
\right\vert ^{2}}\frac{\delta \left( \nabla _{\theta }\omega ^{-1}\left(
J\left( \theta \right) ,\theta ,Z,\mathcal{G}_{0}+\left\vert \Psi
\right\vert ^{2}\right) \right) }{\delta \left\vert \Psi \left( \theta
^{\left( l\right) },Z_{_{l}}\right) \right\vert ^{2}}\left\vert \Psi \left(
\theta ^{\left( l\right) },Z_{_{l}}\right) \right\vert ^{2} \\
&&-\int \int^{\theta }\frac{\delta \left( \omega ^{-1}\left( J\left( \theta
\right) ,\theta ,Z,\mathcal{G}_{0}+\left\vert \Psi \right\vert ^{2}\right)
\right) }{\delta \left\vert \Psi \left( \theta ^{\left( l\right)
},Z_{_{l}}\right) \right\vert ^{2}}\frac{\delta \left( \nabla _{\theta
}^{2}\omega ^{-1}\left( J\left( \theta \right) ,\theta ,Z,\mathcal{G}%
_{0}+\left\vert \Psi \right\vert ^{2}\right) \right) }{\delta \left\vert
\Psi \left( \theta ^{\left( l\right) },Z_{_{l}}\right) \right\vert ^{2}}%
\left\vert \Psi \left( \theta ^{\left( l\right) },Z_{_{l}}\right)
\right\vert ^{2}
\end{eqnarray*}%
We then use the local frequency equation (\ref{pw}) to rewrite $\nabla
_{\theta }^{2}\omega ^{-1}\left( J\left( \theta \right) ,\theta ,Z,\mathcal{G%
}_{0}+\left\vert \Psi \right\vert ^{2}\right) $ in the last term. In first
approximation:%
\begin{equation*}
\nabla _{\theta }^{2}\omega ^{-1}\left( J\left( \theta \right) ,\theta ,Z,%
\mathcal{G}_{0}+\left\vert \Psi \right\vert ^{2}\right) \simeq \frac{f\Omega
\left( \theta ,Z\right) -\left( \frac{\hat{f}_{1}}{\omega \left( \theta
,Z\right) }+N_{1}\right) \nabla _{\theta }\Omega \left( \theta ,Z\right)
-c^{2}\hat{f}_{3}\nabla _{Z}^{2}\Omega \left( \theta ,Z\right) }{\omega
_{0}^{2}\left( \theta ,Z\right) \left( \frac{\hat{f}_{3}}{\omega \left(
\theta ,Z\right) }-N_{2}\right) }
\end{equation*}%
We will neglect the term $c^{2}\hat{f}_{3}\nabla _{Z}^{2}\Omega \left(
\theta ,Z\right) $ assuming that $\Omega \left( \theta ,Z\right) $ varies
slowly for a given time $\theta $ across space. Consequently, we have: 
\begin{eqnarray*}
Z &=&\int \frac{\delta \left( \omega ^{-1}\left( J\left( \theta \right)
,\theta ,Z,\mathcal{G}_{0}+\left\vert \Psi \right\vert ^{2}\right) \right) }{%
\delta \left\vert \Psi \left( \theta ^{\left( l\right) },Z_{_{l}}\right)
\right\vert ^{2}}\frac{\delta \left( \nabla _{\theta }\omega ^{-1}\left(
J\left( \theta \right) ,\theta ,Z,\mathcal{G}_{0}+\left\vert \Psi
\right\vert ^{2}\right) \right) }{\delta \left\vert \Psi \left( \theta
^{\left( l\right) },Z_{_{l}}\right) \right\vert ^{2}}\left\vert \Psi \left(
\theta ^{\left( l\right) },Z_{_{l}}\right) \right\vert ^{2} \\
&&+\int \int^{\theta }\frac{\delta \left( \omega ^{-1}\left( J\left( \theta
\right) ,\theta ,Z,\mathcal{G}_{0}+\left\vert \Psi \right\vert ^{2}\right)
\right) }{\delta \left\vert \Psi \left( \theta ^{\left( l\right)
},Z_{_{l}}\right) \right\vert ^{2}} \\
&&\times \frac{\delta }{\delta \left\vert \Psi \left( \theta ^{\left(
l\right) },Z_{_{l}}\right) \right\vert ^{2}}\frac{f\Omega \left( \theta
,Z\right) -\left( \frac{\hat{f}_{1}}{\omega \left( \theta ,Z\right) }%
+N_{1}\right) \nabla _{\theta }\Omega \left( \theta ,Z\right) -c^{2}\hat{f}%
_{3}\nabla _{Z}^{2}\Omega \left( \theta ,Z\right) }{\omega _{0}^{2}\left(
\theta ,Z\right) \left( \frac{\hat{f}_{3}}{\omega \left( \theta ,Z\right) }%
-N_{2}\right) }\left\vert \Psi \left( \theta ^{\left( l\right)
},Z_{_{l}}\right) \right\vert ^{2}
\end{eqnarray*}%
Regrouping the terms in the previous expression yields:

\begin{eqnarray*}
Z &\simeq &\int \frac{\delta \left( \omega ^{-1}\left( J\left( \theta
\right) ,\theta ,Z,\mathcal{G}_{0}+\left\vert \Psi \right\vert ^{2}\right)
\right) }{\delta \left\vert \Psi \left( \theta ^{\left( l\right)
},Z_{_{l}}\right) \right\vert ^{2}}\frac{\delta \left( \nabla _{\theta
}\omega ^{-1}\left( J\left( \theta \right) ,\theta ,Z,\mathcal{G}%
_{0}+\left\vert \Psi \right\vert ^{2}\right) \right) }{\delta \left\vert
\Psi \left( \theta ^{\left( l\right) },Z_{_{l}}\right) \right\vert ^{2}}%
\left\vert \Psi \left( \theta ^{\left( l\right) },Z_{_{l}}\right)
\right\vert ^{2} \\
&&+\frac{f}{\left( \frac{\hat{f}_{3}}{\omega \left( \theta ,Z\right) }%
-N_{2}\right) }\int \int^{\theta }\frac{\delta \left( \omega ^{-1}\left(
J\left( \theta \right) ,\theta ,Z,\mathcal{G}_{0}+\left\vert \Psi
\right\vert ^{2}\right) \right) }{\delta \left\vert \Psi \left( \theta
^{\left( l\right) },Z_{_{l}}\right) \right\vert ^{2}}\frac{\delta \omega
\left( J\left( \theta \right) ,\theta ,Z,\mathcal{G}_{0}+\left\vert \Psi
\right\vert ^{2}\right) }{\omega _{0}^{2}\left( \theta ,Z\right) \delta
\left\vert \Psi \left( \theta ^{\left( l\right) },Z_{_{l}}\right)
\right\vert ^{2}}\left\vert \Psi \left( \theta ^{\left( l\right)
},Z_{_{l}}\right) \right\vert ^{2} \\
&&-\frac{\frac{\hat{f}_{1}}{\omega \left( \theta ,Z\right) }+N_{1}}{\frac{%
\hat{f}_{3}}{\omega \left( \theta ,Z\right) }-N_{2}}\int \int^{\theta }\frac{%
\delta \left( \omega ^{-1}\left( J\left( \theta \right) ,\theta ,Z,\mathcal{G%
}_{0}+\left\vert \Psi \right\vert ^{2}\right) \right) }{\omega
_{0}^{2}\left( \theta ,Z\right) \delta \left\vert \Psi \left( \theta
^{\left( l\right) },Z_{_{l}}\right) \right\vert ^{2}}\frac{\delta \nabla
_{\theta }\omega \left( J\left( \theta \right) ,\theta ,Z,\mathcal{G}%
_{0}+\left\vert \Psi \right\vert ^{2}\right) }{\delta \left\vert \Psi \left(
\theta ^{\left( l\right) },Z_{_{l}}\right) \right\vert ^{2}}\left\vert \Psi
\left( \theta ^{\left( l\right) },Z_{_{l}}\right) \right\vert ^{2} \\
&\simeq &\int \frac{\delta \left( \omega ^{-1}\left( J\left( \theta \right)
,\theta ,Z,\mathcal{G}_{0}+\left\vert \Psi \right\vert ^{2}\right) \right) }{%
\delta \left\vert \Psi \left( \theta ^{\left( l\right) },Z_{_{l}}\right)
\right\vert ^{2}}\frac{\delta \left( \nabla _{\theta }\omega ^{-1}\left(
J\left( \theta \right) ,\theta ,Z,\mathcal{G}_{0}+\left\vert \Psi
\right\vert ^{2}\right) \right) }{\delta \left\vert \Psi \left( \theta
^{\left( l\right) },Z_{_{l}}\right) \right\vert ^{2}}\left\vert \Psi \left(
\theta ^{\left( l\right) },Z_{_{l}}\right) \right\vert ^{2} \\
&&-\frac{f}{\left( \frac{\hat{f}_{3}}{\omega \left( \theta ,Z\right) }%
-N_{2}\right) }\int \int^{\theta }\left( \frac{\delta \left( \omega
^{-1}\left( J\left( \theta \right) ,\theta ,Z,\mathcal{G}_{0}+\left\vert
\Psi \right\vert ^{2}\right) \right) }{\delta \left\vert \Psi \left( \theta
^{\left( l\right) },Z_{_{l}}\right) \right\vert ^{2}}\right) ^{2}\left\vert
\Psi \left( \theta ^{\left( l\right) },Z_{_{l}}\right) \right\vert ^{2} \\
&&+\frac{\frac{\hat{f}_{1}}{\omega \left( \theta ,Z\right) }+N_{1}}{\frac{%
\hat{f}_{3}}{\omega \left( \theta ,Z\right) }-N_{2}}\int \int^{\theta }\frac{%
\delta \left( \omega ^{-1}\left( J\left( \theta \right) ,\theta ,Z,\mathcal{G%
}_{0}+\left\vert \Psi \right\vert ^{2}\right) \right) }{\delta \left\vert
\Psi \left( \theta ^{\left( l\right) },Z_{_{l}}\right) \right\vert ^{2}}%
\frac{\delta \nabla _{\theta }\omega ^{-1}\left( J\left( \theta \right)
,\theta ,Z,\mathcal{G}_{0}+\left\vert \Psi \right\vert ^{2}\right) }{\delta
\left\vert \Psi \left( \theta ^{\left( l\right) },Z_{_{l}}\right)
\right\vert ^{2}}\left\vert \Psi \left( \theta ^{\left( l\right)
},Z_{_{l}}\right) \right\vert ^{2}
\end{eqnarray*}%
and this rewrites ultimately as:%
\begin{eqnarray*}
&&\int \int^{\theta }\frac{\delta \left( \nabla _{\theta }\omega ^{-1}\left(
J\left( \theta \right) ,\theta ,Z,\mathcal{G}_{0}+\left\vert \Psi
\right\vert ^{2}\right) \right) }{\delta \left\vert \Psi \left( \theta
^{\left( l\right) },Z_{_{l}}\right) \right\vert ^{2}}\frac{\delta \left(
\nabla _{\theta }\omega ^{-1}\left( J\left( \theta \right) ,\theta ,Z,%
\mathcal{G}_{0}+\left\vert \Psi \right\vert ^{2}\right) \right) }{\delta
\left\vert \Psi \left( \theta ^{\left( l\right) },Z_{_{l}}\right)
\right\vert ^{2}}\left\vert \Psi \left( \theta ^{\left( l\right)
},Z_{_{l}}\right) \right\vert ^{2} \\
&\simeq &\int \frac{1}{2}\nabla _{\theta }\left( \frac{\delta \left( \omega
^{-1}\left( J\left( \theta \right) ,\theta ,Z,\mathcal{G}_{0}+\left\vert
\Psi \right\vert ^{2}\right) \right) }{\delta \left\vert \Psi \left( \theta
^{\left( l\right) },Z_{_{l}}\right) \right\vert ^{2}}\right) ^{2} \\
&&-\frac{f}{\left( \frac{\hat{f}_{3}}{\omega \left( \theta ,Z\right) }%
-N_{2}\right) }\int \int^{\theta }\left( \frac{\delta \left( \omega
^{-1}\left( J\left( \theta \right) ,\theta ,Z,\mathcal{G}_{0}+\left\vert
\Psi \right\vert ^{2}\right) \right) }{\delta \left\vert \Psi \left( \theta
^{\left( l\right) },Z_{_{l}}\right) \right\vert ^{2}}\right) ^{2}\left\vert
\Psi \left( \theta ^{\left( l\right) },Z_{_{l}}\right) \right\vert ^{2} \\
&&+\frac{\frac{\hat{f}_{1}}{\omega \left( \theta ,Z\right) }+N_{1}}{2\left( 
\frac{\hat{f}_{3}}{\omega \left( \theta ,Z\right) }-N_{2}\right) }\int
\left( \frac{\delta \left( \omega ^{-1}\left( J\left( \theta \right) ,\theta
,Z,\mathcal{G}_{0}+\left\vert \Psi \right\vert ^{2}\right) \right) }{\delta
\left\vert \Psi \left( \theta ^{\left( l\right) },Z_{_{l}}\right)
\right\vert ^{2}}\right) ^{2}\left\vert \Psi \left( \theta ^{\left( l\right)
},Z_{_{l}}\right) \right\vert ^{2}
\end{eqnarray*}

Which is the result stated in the text.

\section*{Appendix 5 Estimation of $\frac{\protect\delta \protect\omega %
^{-1}\left( J,\protect\theta ,Z\right) }{\protect\delta \left\vert \Psi
\left( \protect\theta -l_{1},Z_{1}\right) \right\vert ^{2}}$}

To compute the effective action, the vacuum, the Green functions and to find
non local solutions of (\ref{fft}) we will need to compute $\omega
^{-1}\left( J,\theta ,Z\right) $ and its derivatives $\frac{\delta
^{n}\omega ^{-1}\left( J,\theta ,Z\right) }{\dprod\limits_{i=1}^{n}\delta
\left\vert \Psi \left( \theta -l_{i},Z_{i}\right) \right\vert ^{2}}$
appearing in (\ref{psn}). In Appendix $6$ we will show that, in first
approximation, the computation relies on the case $n=1$.

The first order derivatives $\frac{\delta \omega ^{-1}\left( J,\theta
,Z\right) }{\delta \left\vert \Psi \left( \theta -l_{1},Z_{1}\right)
\right\vert ^{2}}$ can be computed recursively. To do so, we will need to
approximate the results around some static solution. We define $\bar{\omega}$
as solution of:%
\begin{eqnarray}
\bar{\omega}^{-1}\left( J,Z\right) &=&G\left( \bar{J}\left( Z\right) +\int 
\frac{\kappa }{N}\frac{\bar{\omega}\left( \bar{J},Z_{1}\right) }{\bar{\omega}%
\left( \bar{J},Z\right) }\mathcal{G}_{0}\left( 0,0,Z_{1}\right) T\left(
Z,Z_{1}\right) dZ_{1}\right)  \label{gbr} \\
&=&G\left( \bar{J}\left( Z\right) +\int \frac{\kappa }{N}\frac{\bar{\omega}%
\left( \bar{J},Z_{1}\right) }{\bar{\omega}\left( \bar{J},Z\right) \sqrt{%
\frac{\pi }{2}\left( \frac{1}{\sigma ^{2}\bar{X}_{r}}\right) ^{2}+\frac{2\pi
\alpha }{\sigma ^{2}}}}T\left( Z,Z_{1}\right) dZ_{1}\right)  \notag
\end{eqnarray}%
where $\bar{J}\left( Z\right) $ is the average of $J\left( \theta ,Z\right) $
over the full timespan. We also define:%
\begin{equation}
G_{0}^{\prime }\left( J,Z\right) \equiv G^{\prime }\left( \bar{J}\left(
Z\right) +\int \frac{\kappa }{N}\frac{\bar{\omega}\left( \bar{J}%
,Z_{1}\right) }{\bar{\omega}\left( J,\theta ,Z\right) }G_{0Z_{1}}\left(
0,0\right) T\left( Z,Z_{1}\right) dZ_{1}\right)  \label{frp}
\end{equation}%
These quantities will be useful below.

\subsection*{5.1 Computation of the first order derivatives in (\protect\ref%
{psn})}

\subsubsection*{5.1.1 General formula}

Using the recursive definition of $\omega ^{-1}\left( J,\theta ,Z\right) $: 
\begin{equation}
\omega ^{-1}\left( J,\theta ,Z\right) =G\left( J\left( \theta ,Z\right)
+\int \frac{\kappa }{N}\frac{\omega \left( J,\theta -\frac{\left\vert
Z-Z_{1}\right\vert }{c},Z_{1}\right) }{\omega \left( J,\theta ,Z\right) }%
\bar{W}\left( \frac{\omega \left( \theta ,Z\right) }{\omega \left( \theta -%
\frac{\left\vert Z-Z_{1}\right\vert }{c},Z_{1}\right) }\right) \left\vert
\Psi \left( \theta -\frac{\left\vert Z-Z_{1}\right\vert }{c},Z_{1}\right)
\right\vert ^{2}T\left( Z,Z_{1}\right) dZ_{1}\right)  \label{btr}
\end{equation}%
with:%
\begin{equation*}
\bar{W}\left( \frac{\omega \left( J,\theta -\frac{\left\vert
Z-Z_{1}\right\vert }{c},Z_{1}\right) }{\omega \left( J,\theta ,Z\right) }%
\right) =\frac{\omega \left( J,\theta -\frac{\left\vert Z-Z_{1}\right\vert }{%
c},Z_{1}\right) }{\omega \left( J,\theta ,Z\right) }W\left( \frac{\omega
\left( \theta ,Z\right) }{\omega \left( \theta -\frac{\left\vert
Z-Z_{1}\right\vert }{c},Z_{1}\right) }\right)
\end{equation*}%
we first compute $\frac{\delta \omega ^{-1}\left( J,\theta ,Z\right) }{%
\delta \left\vert \Psi \left( \theta -l_{1},Z_{1}\right) \right\vert ^{2}}$:%
\begin{eqnarray}
&&\frac{\delta \omega ^{-1}\left( J,\theta ,Z\right) }{\delta \left\vert
\Psi \left( \theta -l_{1},Z_{1}\right) \right\vert ^{2}}  \label{vrtft} \\
&=&\frac{\delta G\left( J\left( \theta ,Z\right) +\int \frac{\kappa }{N}\bar{%
W}\left( \frac{\omega \left( J,\theta -\frac{\left\vert Z-Z_{1}\right\vert }{%
c},Z_{1}\right) }{\omega \left( J,\theta ,Z\right) }\right) \left\vert \Psi
\left( \theta -\frac{\left\vert Z-Z^{\prime }\right\vert }{c},Z^{\prime
}\right) \right\vert ^{2}T\left( Z,Z^{\prime }\right) dZ^{\prime }\right) }{%
\delta \left\vert \Psi \left( \theta -l_{1},Z_{1}\right) \right\vert ^{2}} 
\notag
\end{eqnarray}%
Expanding the right hand side and regrouping $\frac{\delta \omega
^{-1}\left( J,\theta ,Z\right) }{\delta \left\vert \Psi \left( \theta
-l_{1},Z_{1}\right) \right\vert ^{2}}$ on the left yields: 
\begin{eqnarray}
&&\frac{\delta \omega ^{-1}\left( J,\theta ,Z\right) }{\delta \left\vert
\Psi \left( \theta -l_{1},Z_{1}\right) \right\vert ^{2}}  \notag \\
&=&\frac{\frac{\kappa }{N}\bar{W}\left( \frac{\omega \left( J,\theta -\frac{%
\left\vert Z-Z_{1}\right\vert }{c},Z_{1}\right) }{\omega \left( J,\theta
,Z\right) }\right) T\left( Z,Z_{1}\right) G^{\prime }\left[ J,\omega ,\theta
,Z,\Psi \right] \delta \left( l_{1}-\frac{\left\vert Z-Z_{1}\right\vert }{c}%
\right) }{1-\left( \int \frac{\kappa }{N}\omega \left( J,\theta -\frac{%
\left\vert Z-Z^{\prime }\right\vert }{c},Z^{\prime }\right) \bar{W}^{\prime
}\left( \frac{\omega \left( J,\theta -\frac{\left\vert Z-Z^{\prime
}\right\vert }{c},Z^{\prime }\right) }{\omega \left( J,\theta ,Z\right) }%
\right) \left\vert \Psi \left( \theta -\frac{\left\vert Z-Z^{\prime
}\right\vert }{c},Z^{\prime }\right) \right\vert ^{2}T\left( Z,Z^{\prime
}\right) dZ^{\prime }\right) G^{\prime }\left[ J,\omega ,\theta ,Z,\Psi %
\right] }  \notag \\
&&+\frac{\frac{1}{\omega \left( J,\theta ,Z\right) }\int \frac{\kappa }{N}%
\frac{\delta \omega \left( J,\theta -\frac{\left\vert Z-Z^{\prime
}\right\vert }{c},Z^{\prime }\right) }{\delta \left\vert \Psi \left( \theta -%
\frac{\left\vert Z-Z_{1}\right\vert }{c},Z_{1}\right) \right\vert ^{2}}\bar{W%
}^{\prime }\left( \frac{\omega \left( J,\theta -\frac{\left\vert
Z-Z_{1}\right\vert }{c},Z_{1}\right) }{\omega \left( J,\theta ,Z\right) }%
\right) \left\vert \Psi \left( \theta -\frac{\left\vert Z-Z^{\prime
}\right\vert }{c},Z^{\prime }\right) \right\vert ^{2}T\left( Z,Z^{\prime
}\right) dZ^{\prime }G^{\prime }\left[ J,\omega ,\theta ,Z,\Psi \right] }{%
1-\left( \int \frac{\kappa }{N}\omega \left( J,\theta -\frac{\left\vert
Z-Z^{\prime }\right\vert }{c},Z^{\prime }\right) \bar{W}^{\prime }\left( 
\frac{\omega \left( J,\theta -\frac{\left\vert Z-Z^{\prime }\right\vert }{c}%
,Z^{\prime }\right) }{\omega \left( J,\theta ,Z\right) }\right) \left\vert
\Psi \left( \theta -\frac{\left\vert Z-Z^{\prime }\right\vert }{c},Z^{\prime
}\right) \right\vert ^{2}T\left( Z,Z^{\prime }\right) dZ^{\prime }\right)
G^{\prime }\left[ J,\omega ,\theta ,Z,\Psi \right] }  \notag \\
&=&\omega \left( J,\theta -l_{1},Z_{1}\right) \hat{T}_{1}\left( \theta
,Z,Z_{1},\omega ,\Psi \right) \delta \left( l_{1}-\frac{\left\vert
Z-Z_{1}\right\vert }{c}\right)  \notag \\
&&+\int \frac{\delta \omega \left( J,\theta -\frac{\left\vert Z-Z^{\prime
}\right\vert }{c},Z^{\prime }\right) }{\delta \left\vert \Psi \left( \theta
-l_{1},Z_{1}\right) \right\vert ^{2}}\left\vert \Psi \left( \theta -\frac{%
\left\vert Z-Z^{\prime }\right\vert }{c},Z^{\prime }\right) \right\vert ^{2}%
\hat{T}_{1}\left( \theta ,Z,Z^{\prime },\omega ,\Psi \right) dZ^{\prime }
\label{cr}
\end{eqnarray}%
where we defined:%
\begin{eqnarray}
&&\hat{T}_{1}\left( \theta ,Z,Z_{1},\omega ,\Psi \right) =\frac{1}{\omega
\left( J,\theta ,Z\right) }  \label{vrtbb} \\
&&\times \frac{\frac{\kappa }{N}T\left( Z,Z_{1}\right) \bar{W}^{\prime
}\left( \frac{\omega \left( J,\theta -\frac{\left\vert Z-Z_{1}\right\vert }{c%
},Z_{1}\right) }{\omega \left( J,\theta ,Z\right) }\right) G^{\prime }\left[
J,\omega ,\theta ,Z,\Psi \right] }{1-\left( \int \frac{\kappa }{N}\omega
\left( J,\theta -\frac{\left\vert Z-Z^{\prime }\right\vert }{c},Z^{\prime
}\right) \bar{W}^{\prime }\left( \frac{\omega \left( J,\theta -\frac{%
\left\vert Z-Z^{\prime }\right\vert }{c},Z^{\prime }\right) }{\omega \left(
J,\theta ,Z\right) }\right) \left\vert \Psi \left( \theta -\frac{\left\vert
Z-Z^{\prime }\right\vert }{c},Z^{\prime }\right) \right\vert ^{2}T\left(
Z,Z^{\prime }\right) dZ^{\prime }\right) G^{\prime }\left[ J,\omega ,\theta
,Z,\Psi \right] }  \notag
\end{eqnarray}%
Equation (\ref{vrtft}) shows that we also need $\frac{\delta \omega \left(
J,\theta ,Z\right) }{\delta \left\vert \Psi \left( \theta
-l_{1},Z_{1}\right) \right\vert ^{2}}$ to compute $\frac{\delta \omega
^{-1}\left( J,\theta ,Z\right) }{\delta \left\vert \Psi \left( \theta
-l_{1},Z_{1}\right) \right\vert ^{2}}$. This is obtained by: 
\begin{eqnarray}
\frac{\delta \omega \left( J,\theta ,Z\right) }{\delta \left\vert \Psi
\left( \theta -l_{1},Z_{1}\right) \right\vert ^{2}} &=&\frac{\delta F\left(
J\left( \theta ,Z\right) +\int \frac{\kappa }{N}\bar{W}\left( \frac{\omega
\left( J,\theta -\frac{\left\vert Z-Z^{\prime }\right\vert }{c},Z^{\prime
}\right) }{\omega \left( J,\theta ,Z\right) }\right) \left\vert \Psi \left(
\theta -\frac{\left\vert Z-Z^{\prime }\right\vert }{c},Z^{\prime }\right)
\right\vert ^{2}T\left( Z,Z^{\prime }\right) dZ^{\prime }\right) }{\delta
\left\vert \Psi \left( \theta -l_{1},Z_{1}\right) \right\vert ^{2}}  \notag
\\
&=&\omega \left( J,\theta -l_{1},Z_{1}\right) \hat{T}\left( \theta
,Z,Z_{1},\omega ,\Psi \right) \delta \left( l_{1}-\frac{\left\vert
Z-Z_{1}\right\vert }{c}\right)  \notag \\
&&+\int \frac{\delta \omega \left( J,\theta -\frac{\left\vert Z-Z^{\prime
}\right\vert }{c},Z^{\prime }\right) }{\delta \left\vert \Psi \left( \theta
-l_{1},Z_{1}\right) \right\vert ^{2}}\left\vert \Psi \left( \theta -\frac{%
\left\vert Z-Z^{\prime }\right\vert }{c},Z^{\prime }\right) \right\vert ^{2}%
\hat{T}\left( \theta ,Z,Z^{\prime },\omega ,\Psi \right) dZ^{\prime }
\label{vrttt}
\end{eqnarray}%
with:%
\begin{eqnarray}
&&\hat{T}\left( \theta ,Z,Z_{1}\omega ,\Psi \right)  \label{vrtftbs} \\
&=&\frac{\frac{\kappa }{N}\omega \left( J,\theta ,Z\right) T\left(
Z,Z_{1}\right) \bar{W}^{\prime }\left( \frac{\omega \left( J,\theta -\frac{%
\left\vert Z-Z_{1}\right\vert }{c},Z_{1}\right) }{\omega \left( J,\theta
,Z\right) }\right) F^{\prime }\left[ J,\omega ,\theta ,Z,\Psi \right] }{%
\omega ^{2}\left( J,\theta ,Z\right) +\left( \int \frac{\kappa }{N}\omega
\left( J,\theta -\frac{\left\vert Z-Z^{\prime }\right\vert }{c},Z^{\prime
}\right) \bar{W}^{\prime }\left( \frac{\omega \left( J,\theta -\frac{%
\left\vert Z-Z^{\prime }\right\vert }{c},Z^{\prime }\right) }{\omega \left(
J,\theta ,Z\right) }\right) \left\vert \Psi \left( \theta -\frac{\left\vert
Z-Z^{\prime }\right\vert }{c},Z^{\prime }\right) \right\vert ^{2}T\left(
Z,Z^{\prime }\right) dZ^{\prime }\right) F^{\prime }\left[ J,\omega ,\theta
,Z,\Psi \right] }  \notag
\end{eqnarray}%
Equation (\ref{vrttt}) and (\ref{vrtftbs}) define $\frac{\delta \omega
\left( J,\theta ,Z\right) }{\delta \left\vert \Psi \left( \theta
-l_{1},Z_{1}\right) \right\vert ^{2}}$\ recursively. Actually, writing: 
\begin{eqnarray*}
&&\frac{\delta \omega \left( J,\theta -\frac{\left\vert Z-Z^{\prime
}\right\vert }{c},Z^{\prime }\right) }{\delta \left\vert \Psi \left( \theta
-l_{1},Z_{1}\right) \right\vert ^{2}} \\
&=&\int \omega \left( J,\theta -\frac{\left\vert Z-Z^{\prime }\right\vert }{c%
}-\frac{\left\vert Z^{\prime }-Z^{\prime \prime }\right\vert }{c},Z^{\prime
\prime }\right) \hat{T}\left( \theta -\frac{\left\vert Z-Z^{\prime
}\right\vert }{c},Z^{\prime },Z^{\prime \prime },\omega ,\Psi \right) \delta
\left( \frac{\left\vert Z-Z^{\prime }\right\vert }{c}+\frac{\left\vert
Z^{\prime }-Z^{\prime \prime }\right\vert }{c}-l_{1}\right) dZ^{\prime
\prime } \\
&&+\int \frac{\delta \omega \left( J,\theta -\frac{\left\vert Z-Z^{\prime
}\right\vert }{c}-\frac{\left\vert Z^{\prime }-Z^{\prime \prime }\right\vert 
}{c},Z^{\prime \prime }\right) }{\delta \left\vert \Psi \left( \theta
-l_{1},Z_{1}\right) \right\vert ^{2}}\left\vert \Psi \left( \theta -\frac{%
\left\vert Z-Z^{\prime }\right\vert }{c}-\frac{\left\vert Z^{\prime
}-Z^{\prime \prime }\right\vert }{c},Z^{\prime \prime }\right) \right\vert
^{2}\hat{T}\left( \theta -\frac{\left\vert Z-Z^{\prime }\right\vert }{c}%
,Z^{\prime },Z^{\prime \prime },\omega ,\Psi \right) dZ^{\prime \prime }
\end{eqnarray*}%
we have:

\begin{eqnarray*}
&&\frac{\delta \omega \left( J,\theta ,Z\right) }{\delta \left\vert \Psi
\left( \theta -l_{1},Z_{1}\right) \right\vert ^{2}} \\
&=&\int \omega \left( J,\theta -\frac{\left\vert Z-Z^{\prime }\right\vert }{c%
},Z^{\prime }\right) \hat{T}\left( \theta ,Z,Z_{1},\omega ,\Psi \right)
\delta \left( \frac{\left\vert Z-Z^{\prime }\right\vert }{c}-l_{1}\right)
dZ^{\prime } \\
&&+\int \omega \left( J,\theta -\frac{\left\vert Z-Z^{\prime }\right\vert }{c%
}-\frac{\left\vert Z^{\prime }-Z^{\prime \prime }\right\vert }{c},Z^{\prime
\prime }\right) \hat{T}\left( \theta -\frac{\left\vert Z-Z^{\prime
}\right\vert }{c},Z^{\prime },Z^{\prime \prime },\omega ,\Psi \right) \\
&&\times \left\vert \Psi \left( \theta -\frac{\left\vert Z-Z^{\prime
}\right\vert }{c},Z^{\prime }\right) \right\vert ^{2}\hat{T}\left( \theta
,Z,Z^{\prime },\omega ,\Psi \right) \delta \left( \frac{\left\vert
Z-Z^{\prime }\right\vert }{c}+\frac{\left\vert Z^{\prime }-Z^{\prime \prime
}\right\vert }{c}-l_{1}\right) dZ^{\prime }dZ^{\prime \prime } \\
&&+\int \frac{\delta \omega \left( J,\theta -\frac{\left\vert Z-Z^{\prime
}\right\vert }{c}-\frac{\left\vert Z^{\prime }-Z^{\prime \prime }\right\vert 
}{c},Z^{\prime \prime }\right) }{\delta \left\vert \Psi \left( \theta
-l_{1},Z_{1}\right) \right\vert ^{2}}\left\vert \Psi \left( \theta -\frac{%
\left\vert Z-Z^{\prime }\right\vert }{c}-\frac{\left\vert Z^{\prime
}-Z^{\prime \prime }\right\vert }{c},Z^{\prime \prime }\right) \right\vert
^{2} \\
&&\times \hat{T}\left( \theta -\frac{\left\vert Z-Z^{\prime }\right\vert }{c}%
,Z^{\prime },Z^{\prime \prime },\omega ,\Psi \right) \left\vert \Psi \left(
\theta -\frac{\left\vert Z-Z^{\prime }\right\vert }{c},Z^{\prime }\right)
\right\vert ^{2}\hat{T}\left( \theta ,Z,Z^{\prime },\omega ,\Psi \right)
dZ^{\prime }dZ^{\prime \prime }
\end{eqnarray*}%
By a redefinition of $\hat{T}$ and $\hat{T}_{1}$:%
\begin{eqnarray*}
\hat{T}\left( \theta ,Z,Z^{\prime },\omega ,\Psi \right) \left\vert \Psi
\left( \theta -\frac{\left\vert Z-Z^{\prime }\right\vert }{c},Z^{\prime
}\right) \right\vert ^{2} &\rightarrow &\hat{T}\left( \theta ,Z,Z^{\prime
},\omega ,\Psi \right) \\
\hat{T}_{1}\left( \theta ,Z,Z^{\prime },\omega ,\Psi \right) \left\vert \Psi
\left( \theta -\frac{\left\vert Z-Z^{\prime }\right\vert }{c},Z^{\prime
}\right) \right\vert ^{2} &\rightarrow &\hat{T}_{1}\left( \theta
,Z,Z^{\prime },\omega ,\Psi \right)
\end{eqnarray*}%
which yields the series expansion:%
\begin{eqnarray}
\frac{\delta \omega \left( J,\theta ,Z\right) }{\delta \left\vert \Psi
\left( \theta -l_{1},Z_{1}\right) \right\vert ^{2}} &=&\sum_{n=1}^{\infty }%
\frac{1}{\left\vert \Psi \left( \theta -l_{1},Z_{1}\right) \right\vert ^{2}}%
\int \omega \left( J,\theta -\sum_{l=1}^{n}\frac{\left\vert Z^{\left(
l-1\right) }-Z^{\left( l\right) }\right\vert }{c},Z_{1}\right)  \label{xpng}
\\
&&\times \dprod\limits_{l=1}^{n}\hat{T}\left( \theta -\sum_{j=1}^{l-1}\frac{%
\left\vert Z^{\left( j-1\right) }-Z^{\left( j\right) }\right\vert }{c}%
,Z^{\left( l-1\right) },Z^{\left( l\right) },\omega ,\Psi \right) \delta
\left( l_{1}-\sum_{l=1}^{n}\frac{\left\vert Z^{\left( l-1\right) }-Z^{\left(
l\right) }\right\vert }{c}\right) \dprod\limits_{l=1}^{n-1}dZ^{\left(
l\right) }  \notag
\end{eqnarray}%
and:%
\begin{eqnarray}
\frac{\delta \omega ^{-1}\left( J,\theta ,Z\right) }{\delta \left\vert \Psi
\left( \theta -l_{1},Z_{1}\right) \right\vert ^{2}} &=&\sum_{n=1}^{\infty }%
\frac{1}{\left\vert \Psi \left( \theta -l_{1},Z_{1}\right) \right\vert ^{2}}%
\int \omega \left( J,\theta -\sum_{l=1}^{n}\frac{\left\vert Z^{\left(
l-1\right) }-Z^{\left( l\right) }\right\vert }{c},Z_{1}\right) \hat{T}%
_{1}\left( \theta ,Z,Z^{\left( 1\right) },\omega ,\Psi \right)  \label{drvt}
\\
&&\times \dprod\limits_{l=2}^{n}\hat{T}\left( \theta -\sum_{j=1}^{l-1}\frac{%
\left\vert Z^{\left( j-1\right) }-Z^{\left( j\right) }\right\vert }{c}%
,Z^{\left( l-1\right) },Z^{\left( l\right) },\omega ,\Psi \right) \delta
\left( l_{1}-\sum_{l=1}^{n}\frac{\left\vert Z^{\left( l-1\right) }-Z^{\left(
l\right) }\right\vert }{c}\right) \dprod\limits_{l=1}^{n-1}dZ^{\left(
l\right) }  \notag
\end{eqnarray}%
with the convention that $Z^{\left( 0\right) }=Z$ and $Z^{\left( n\right)
}=Z_{1}$.

We can write (\ref{drvt}) in a more symetric way. Defining:%
\begin{equation*}
\check{T}\left( \theta ,Z,Z^{\left( 1\right) },\omega ,\Psi \right) =-\omega
^{2}\left( J,\theta ,Z\right) \hat{T}_{1}\left( \theta ,Z,Z^{\left( 1\right)
},\omega ,\Psi \right)
\end{equation*}%
Relation (\ref{cr}) writes:%
\begin{eqnarray*}
\frac{\delta \omega ^{-1}\left( J,\theta ,Z\right) }{\delta \left\vert \Psi
\left( \theta -l_{1},Z_{1}\right) \right\vert ^{2}} &=&\omega ^{-1}\left(
J,\theta -l_{1},Z_{1}\right) \check{T}\left( \theta ,Z,Z_{1},\omega ,\Psi
\right) \delta \left( l_{1}-\frac{\left\vert Z-Z_{1}\right\vert }{c}\right)
\\
&&+\int \frac{\delta \omega ^{-1}\left( J,\theta -\frac{\left\vert
Z-Z^{\prime }\right\vert }{c},Z^{\prime }\right) }{\delta \left\vert \Psi
\left( \theta -l_{1},Z_{1}\right) \right\vert ^{2}}\left\vert \Psi \left(
\theta -\frac{\left\vert Z-Z^{\prime }\right\vert }{c},Z^{\prime }\right)
\right\vert ^{2}\check{T}\left( \theta ,Z,Z^{\prime },\omega ,\Psi \right)
dZ^{\prime }
\end{eqnarray*}%
and we have:%
\begin{eqnarray}
\frac{\delta \omega ^{-1}\left( J,\theta ,Z\right) }{\delta \left\vert \Psi
\left( \theta -l_{1},Z_{1}\right) \right\vert ^{2}} &=&\sum_{n=1}^{\infty }%
\frac{1}{\left\vert \Psi \left( \theta -l_{1},Z_{1}\right) \right\vert ^{2}}%
\int \omega ^{-1}\left( J,\theta -\sum_{l=1}^{n}\frac{\left\vert Z^{\left(
l-1\right) }-Z^{\left( l\right) }\right\vert }{c},Z_{1}\right)  \label{dvtr}
\\
&&\times \dprod\limits_{l=1}^{n}\check{T}\left( \theta -\sum_{j=1}^{l-1}%
\frac{\left\vert Z^{\left( j-1\right) }-Z^{\left( j\right) }\right\vert }{c}%
,Z^{\left( l-1\right) },Z^{\left( l\right) },\omega ,\Psi \right) \delta
\left( l_{1}-\sum_{l=1}^{n}\frac{\left\vert Z^{\left( l-1\right) }-Z^{\left(
l\right) }\right\vert }{c}\right) \dprod\limits_{l=1}^{n-1}dZ^{\left(
l\right) }  \notag
\end{eqnarray}

\subsubsection*{5.1.2 Static approximation\protect\bigskip}

We now use the static approximations (\ref{gbr}) and (\ref{frp}). Actually,
the values of $\hat{T}_{1}\left( \theta ,Z,Z_{1}\omega ,\Psi \right) $ and $%
\hat{T}\left( \theta ,Z,Z_{1}\omega ,\Psi \right) $ can be estimated for $%
\bar{\omega}^{-1}\left( \bar{J},Z\right) $. Moreover, in the limit of small
fluctuations, $\bar{\omega}^{-1}\left( \bar{J},Z\right) $, $F^{\prime }\left[
J,\bar{\omega},Z,\Psi \right] $ and $G^{\prime }\left[ J,\bar{\omega},Z,\Psi %
\right] $ can be approximated by their average over $Z$, denoted $\bar{\omega%
}^{-1}$, $\bar{F}^{\prime }$and $\bar{G}^{\prime }$. We also have:%
\begin{equation*}
\frac{\bar{\omega}\left( J,Z^{\prime }\right) }{\bar{\omega}\left(
J,Z\right) }\simeq 1
\end{equation*}%
We also replace $\left\vert \Psi \right\vert ^{2}$ by $\frac{1}{\sqrt{\frac{%
\pi }{2}\left( \frac{1}{\sigma ^{2}\bar{X}_{r}}\right) ^{2}+\frac{2\pi
\alpha }{\sigma ^{2}}}}$. Moreover for $\bar{\omega}$, both $\hat{T}_{1}$
and $\hat{T}$ can be considered independent of $\theta $: 
\begin{eqnarray*}
\hat{T}_{1}\left( \theta ,Z,Z_{1}\bar{\omega},\Psi \right) &\simeq &\hat{T}%
_{1}\left( Z,Z_{1},\bar{\omega}\right) \\
&=&\frac{1}{\sqrt{\frac{\pi }{2}\left( \frac{1}{\sigma ^{2}\bar{X}_{r}}%
\right) ^{2}+\frac{2\pi \alpha }{\sigma ^{2}}}}\frac{\frac{\kappa }{N}\bar{%
\omega}^{-1}T\left( Z,Z_{1}\right) \bar{G}^{\prime }}{1-\frac{\bar{G}%
^{\prime }\bar{\omega}\int \frac{\kappa }{N}T\left( Z,Z^{\prime }\right)
dZ^{\prime }}{\sqrt{\frac{\pi }{2}\left( \frac{1}{\sigma ^{2}\bar{X}_{r}}%
\right) ^{2}+\frac{2\pi \alpha }{\sigma ^{2}}}}}
\end{eqnarray*}%
\begin{eqnarray*}
\hat{T}\left( \theta ,Z,Z_{1}\omega ,\Psi \right) &\simeq &\hat{T}\left(
Z,Z_{1},\bar{\omega}\right) \\
&=&\frac{1}{\sqrt{\frac{\pi }{2}\left( \frac{1}{\sigma ^{2}\bar{X}_{r}}%
\right) ^{2}+\frac{2\pi \alpha }{\sigma ^{2}}}}\frac{\frac{\kappa }{N}%
T\left( Z,Z_{1}\right) \bar{F}^{\prime }}{\bar{\omega}+\frac{\bar{F}^{\prime
}\int \frac{\kappa }{N}T\left( Z,Z^{\prime }\right) dZ^{\prime }}{\sqrt{%
\frac{\pi }{2}\left( \frac{1}{\sigma ^{2}\bar{X}_{r}}\right) ^{2}+\frac{2\pi
\alpha }{\sigma ^{2}}}}}
\end{eqnarray*}%
as a consequence $\hat{T}_{1}\left( Z,Z_{1},\bar{\omega}\right) $ and $\hat{T%
}\left( Z,Z_{1},\bar{\omega}\right) $ are functions of $\left\vert
Z-Z_{1}\right\vert $ denoted $\hat{T}_{1}\left( \left\vert
Z-Z_{1}\right\vert \right) $. \ As a consequence (\ref{drvt}) can be
estimated by:%
\begin{eqnarray}
\frac{\delta \omega ^{-1}\left( J,\theta ,Z\right) }{\delta \left\vert \Psi
\left( \theta -l_{1},Z_{1}\right) \right\vert ^{2}} &=&\sum_{n=1}^{\infty }%
\frac{1}{\left\vert \Psi \left( \theta -l_{1},Z_{1}\right) \right\vert ^{2}}%
\int \omega \left( J,\theta -l_{1},Z_{1}\right) \hat{T}_{1}\left( \left\vert
Z-Z^{\left( 1\right) }\right\vert \right)  \label{vrtftbscc} \\
&&\times \dprod\limits_{l=2}^{n}\hat{T}\left( \left\vert Z^{\left(
l-1\right) }-Z^{\left( l\right) }\right\vert \right) \delta \left(
l_{1}-\sum_{l=1}^{n}\frac{\left\vert Z^{\left( l-1\right) }-Z^{\left(
l\right) }\right\vert }{c}\right)  \notag \\
&&\times \delta \left( Z-Z_{1}-\sum_{l=1}^{n}\left( Z^{\left( l-1\right)
}-Z^{\left( l\right) }\right) \right) \dprod\limits_{l=1}^{n-1}dZ^{\left(
l\right) }  \notag
\end{eqnarray}%
and (\ref{xpng}) is:%
\begin{eqnarray}
\frac{\delta \omega \left( J,\theta ,Z\right) }{\delta \left\vert \Psi
\left( \theta -l_{1},Z_{1}\right) \right\vert ^{2}} &=&\sum_{n=1}^{\infty }%
\frac{1}{\left\vert \Psi \left( \theta -l_{1},Z_{1}\right) \right\vert ^{2}}%
\int \omega \left( J,\theta -\sum_{l=1}^{n}\frac{\left\vert Z^{\left(
l-1\right) }-Z^{\left( l\right) }\right\vert }{c},Z_{1}\right)
\dprod\limits_{l=1}^{n}\hat{T}\left( \left\vert Z^{\left( l-1\right)
}-Z^{\left( l\right) }\right\vert \right)  \notag \\
&&\times \delta \left( l_{1}-\sum_{l=1}^{n}\frac{\left\vert Z^{\left(
l-1\right) }-Z^{\left( l\right) }\right\vert }{c}\right) \times \delta
\left( Z-Z_{1}-\sum_{l=1}^{n}\left( Z^{\left( l-1\right) }-Z^{\left(
l\right) }\right) \right) \dprod\limits_{l=1}^{n-1}dZ^{\left( l\right) }
\label{xpgk}
\end{eqnarray}

\subsection*{5.2 Estimation of (\protect\ref{vrtftbscc}) and (\protect\ref%
{xpng}) close to the permanent regime}

The series (\ref{vrtftbscc}) can be computed by using the Fourier transform
of the Dirac functions:%
\begin{eqnarray}
\left\vert \Psi \left( \theta -l_{1},Z_{1}\right) \right\vert ^{2}\frac{%
\delta \omega ^{-1}\left( J,\theta ,Z\right) }{\delta \left\vert \Psi \left(
\theta -l_{1},Z_{1}\right) \right\vert ^{2}} &=&\sum_{n=1}^{\infty }\int
\omega \left( J,\theta -l_{1},Z_{1}\right) \times \hat{T}_{1}\left(
\left\vert Z-Z^{\left( 1\right) }\right\vert \right)  \label{gr} \\
&&\times \dprod\limits_{l=2}^{n}\hat{T}\left( \left\vert Z^{\left(
l-1\right) }-Z^{\left( l\right) }\right\vert \right) \exp \left( i\lambda
\left( cl_{1}-\sum_{l=1}^{n}\left\vert Z^{\left( l-1\right) }-Z^{\left(
l\right) }\right\vert \right) \right)  \notag \\
&&\times \exp \left( i\lambda _{1}.\left( Z-Z_{1}-\sum_{l=1}^{n}\left(
Z^{\left( l-1\right) }-Z^{\left( l\right) }\right) \right) \right) d\lambda
d\lambda _{1}  \notag \\
&&\times \dprod\limits_{l=1}^{n}\left\vert Z^{\left( l-1\right) }-Z^{\left(
l\right) }\right\vert ^{2}d\left\vert Z^{\left( l-1\right) }-Z^{\left(
l\right) }\right\vert dv_{l}  \notag
\end{eqnarray}%
where the unit vectors $v_{l}$ are defined such that:%
\begin{equation*}
Z^{\left( l-1\right) }-Z^{\left( l\right) }=v_{l}\left\vert Z^{\left(
l-1\right) }-Z^{\left( l\right) }\right\vert
\end{equation*}%
We also define: 
\begin{eqnarray*}
\lambda _{1}.\left( Z-Z_{1}\right) &=&\left\vert \lambda _{1}\right\vert
\left\vert Z-Z_{1}\right\vert \cos \left( \theta _{1}\right) \\
\lambda _{1}.v_{l} &=&\left\vert \lambda _{1}\right\vert \cos \left( \theta
_{l}\right)
\end{eqnarray*}%
The angles $\theta _{l}$ are computed in the plane $\left( \lambda
_{1},Z-Z_{1}\right) $ between the projection of $v_{l}$ and $Z-Z_{1}$.

Before computing the integrals in (\ref{gr}) for arbitrary transfer
functions, we develop the particular case of an exponential transfer
function.

\subsubsection*{5.2.1 Exponential transfer function}

We first choose:%
\begin{eqnarray}
\hat{T}\left( \left\vert Z^{\left( l-1\right) }-Z^{\left( l\right)
}\right\vert \right) &=&C\frac{\exp \left( -c\left\vert Z^{\left( l-1\right)
}-Z^{\left( l\right) }\right\vert \right) }{\left\vert Z^{\left( l-1\right)
}-Z^{\left( l\right) }\right\vert }  \label{srt} \\
\hat{T}\left( \left\vert Z^{\left( l-1\right) }-Z^{\left( l\right)
}\right\vert \right) &\simeq &\frac{A_{1}}{A}\hat{T}\left( \left\vert
Z^{\left( l-1\right) }-Z^{\left( l\right) }\right\vert \right)  \notag
\end{eqnarray}%
and discard the factor $\frac{A_{1}}{A}$ that will be reintroduced in the
end of the computation.

Using that $\sum_{l=1}^{n}\left( Z^{\left( l-1\right) }-Z^{\left( l\right)
}\right) =cl_{1}$, the right hand side of (\ref{gr}) becomes:

\begin{eqnarray*}
&&\exp \left( -cl_{1}\right) \times \sum_{n=1}^{\infty }\int \exp \left(
i\lambda \left( cl_{1}-\sum_{l=1}^{n}\left\vert Z^{\left( l-1\right)
}-Z^{\left( l\right) }\right\vert \right) \right) \\
&&\times \exp \left( i\lambda _{1}.\left( Z-Z_{1}-\sum_{l=1}^{n}\left(
Z^{\left( l-1\right) }-Z^{\left( l\right) }\right) \right) \right) d\lambda
d\lambda _{1}\dprod\limits_{l=1}^{n}C\left\vert Z^{\left( l-1\right)
}-Z^{\left( l\right) }\right\vert d\left\vert Z^{\left( l-1\right)
}-Z^{\left( l\right) }\right\vert dv_{l}
\end{eqnarray*}%
that can be written in terms of the angles as:%
\begin{eqnarray}
&&\exp \left( -cl_{1}\right) \times \sum_{n=1}^{\infty }\int \exp \left(
i\lambda cl_{1}+i\left\vert \lambda _{1}\right\vert \left\vert
Z-Z_{1}\right\vert \cos \left( \theta _{1}\right) \right)  \label{sh} \\
&&\times \exp \left( -i\sum_{l=1}^{n}\left( \lambda +\left\vert \lambda
_{1}\right\vert \cos \left( \theta _{l}\right) \right) \left\vert Z^{\left(
l-1\right) }-Z^{\left( l\right) }\right\vert \right) d\lambda d\lambda
_{1}\dprod\limits_{l=1}^{n}C\left\vert Z^{\left( l-1\right) }-Z^{\left(
l\right) }\right\vert d\left\vert Z^{\left( l-1\right) }-Z^{\left( l\right)
}\right\vert dv_{l}  \notag
\end{eqnarray}%
The integration over $\theta _{l}$ is:%
\begin{eqnarray*}
&&\pi \int_{0}^{\pi }\exp \left( -i\left( \lambda +\left\vert \lambda
_{1}\right\vert \cos \left( \theta _{l}\right) \right) \left\vert Z^{\left(
l-1\right) }-Z^{\left( l\right) }\right\vert \right) \sin \left( \theta
_{l}\right) d\theta _{l} \\
&=&-\frac{\pi i}{\left\vert \lambda _{1}\right\vert \left\vert Z^{\left(
l-1\right) }-Z^{\left( l\right) }\right\vert }\left( \exp \left( -i\left(
\lambda -\left\vert \lambda _{1}\right\vert \right) \left\vert Z^{\left(
l-1\right) }-Z^{\left( l\right) }\right\vert \right) -\exp \left( -i\left(
\lambda +\left\vert \lambda _{1}\right\vert \right) \left\vert Z^{\left(
l-1\right) }-Z^{\left( l\right) }\right\vert \right) \right) \\
&=&\frac{\pi i}{\left\vert \lambda _{1}\right\vert \left\vert Z^{\left(
l-1\right) }-Z^{\left( l\right) }\right\vert }\left( \exp \left( -i\left(
\lambda +\left\vert \lambda _{1}\right\vert \right) \left\vert Z^{\left(
l-1\right) }-Z^{\left( l\right) }\right\vert \right) -\exp \left( -i\left(
\lambda -\left\vert \lambda _{1}\right\vert \right) \left\vert Z^{\left(
l-1\right) }-Z^{\left( l\right) }\right\vert \right) \right)
\end{eqnarray*}%
and (\ref{sh}) rewrites:%
\begin{eqnarray*}
&&\exp \left( -cl_{1}\right) \times \sum_{n=1}^{\infty }\int \frac{-\pi i}{%
\left\vert \lambda _{1}\right\vert \left\vert Z-Z_{1}\right\vert }\left(
\exp \left( i\lambda cl_{1}+i\left\vert \lambda _{1}\right\vert \left\vert
Z-Z_{1}\right\vert \right) -\exp \left( i\lambda cl_{1}-i\left\vert \lambda
_{1}\right\vert \left\vert Z-Z_{1}\right\vert \right) \right) \\
&&\times \dprod\limits_{l=1}^{n}C\frac{\pi i}{\left\vert \lambda
_{1}\right\vert }\left( \exp \left( -i\left( \lambda +\left\vert \lambda
_{1}\right\vert \right) \left\vert Z^{\left( l-1\right) }-Z^{\left( l\right)
}\right\vert \right) -\exp \left( -i\left( \lambda -\left\vert \lambda
_{1}\right\vert \right) \left\vert Z^{\left( l-1\right) }-Z^{\left( l\right)
}\right\vert \right) \right) d\left\vert Z^{\left( l-1\right) }-Z^{\left(
l\right) }\right\vert d\lambda \left\vert \lambda _{1}\right\vert
^{2}d\left\vert \lambda _{1}\right\vert
\end{eqnarray*}%
We can then perform the integrals over the norms $\left\vert Z^{\left(
l-1\right) }-Z^{\left( l\right) }\right\vert $, which yields:

\begin{eqnarray*}
&&\exp \left( -cl_{1}\right) \times \sum_{n=1}^{\infty }\int \frac{-\pi i}{%
\left\vert \lambda _{1}\right\vert \left\vert Z-Z_{1}\right\vert }\left(
\exp \left( i\lambda cl_{1}+i\left\vert \lambda _{1}\right\vert \left\vert
Z-Z_{1}\right\vert \right) -\exp \left( i\lambda cl_{1}-i\left\vert \lambda
_{1}\right\vert \left\vert Z-Z_{1}\right\vert \right) \right) \\
&&\times \dprod\limits_{l=1}^{n}C\frac{\pi }{\left\vert \lambda
_{1}\right\vert }\left( \frac{1}{\lambda +\left\vert \lambda _{1}\right\vert
-i\varepsilon }-\frac{1}{\lambda -\left\vert \lambda _{1}\right\vert
-i\varepsilon }\right) d\lambda \left\vert \lambda _{1}\right\vert
^{2}d\left\vert \lambda _{1}\right\vert
\end{eqnarray*}%
Performing the sum yields then the following expression for (\ref{sh}):

\begin{eqnarray*}
&&\exp \left( -cl_{1}\right) \times \int \frac{-\pi i}{\left\vert \lambda
_{1}\right\vert \left\vert Z-Z_{1}\right\vert }\left( \exp \left( i\lambda
cl_{1}+i\left\vert \lambda _{1}\right\vert \left\vert Z-Z_{1}\right\vert
\right) -\exp \left( i\lambda cl_{1}-i\left\vert \lambda _{1}\right\vert
\left\vert Z-Z_{1}\right\vert \right) \right) \\
&&\times \frac{-C\frac{2\pi }{\left( \lambda +\left\vert \lambda
_{1}\right\vert -i\varepsilon \right) \left( \lambda -\left\vert \lambda
_{1}\right\vert -i\varepsilon \right) }}{1+C\frac{2\pi }{\left( \lambda
+\left\vert \lambda _{1}\right\vert -i\varepsilon \right) \left( \lambda
-\left\vert \lambda _{1}\right\vert -i\varepsilon \right) }}d\lambda
\left\vert \lambda _{1}\right\vert ^{2}d\left\vert \lambda _{1}\right\vert \\
&=&\exp \left( -cl_{1}\right) \times \int \frac{-\pi i}{\left\vert \lambda
_{1}\right\vert \left\vert Z-Z_{1}\right\vert }\left( \exp \left( i\lambda
cl_{1}+i\left\vert \lambda _{1}\right\vert \left\vert Z-Z_{1}\right\vert
\right) -\exp \left( i\lambda cl_{1}-i\left\vert \lambda _{1}\right\vert
\left\vert Z-Z_{1}\right\vert \right) \right) \\
&&\times \frac{-2\pi C}{\left( \lambda +\left\vert \lambda _{1}\right\vert
-i\varepsilon \right) \left( \lambda -\left\vert \lambda _{1}\right\vert
-i\varepsilon \right) +2\pi C}d\lambda \left\vert \lambda _{1}\right\vert
^{2}d\left\vert \lambda _{1}\right\vert
\end{eqnarray*}

Ultimately, the previous formula can be reduced to a single expression, by
performing the change of variable $x=-$ $\left\vert \lambda _{1}\right\vert $
in the term with $\exp \left( i\lambda cl_{1}-i\left\vert \lambda
_{1}\right\vert \left\vert Z-Z_{1}\right\vert \right) $ in factor. We obtain:%
\begin{equation*}
\exp \left( -cl_{1}\right) \times \int \frac{-\pi i}{\left\vert
Z-Z_{1}\right\vert }\exp \left( i\lambda cl_{1}+i\lambda _{1}\left\vert
Z-Z_{1}\right\vert \right) \frac{-2\pi C\lambda _{1}}{\left( \lambda
+\lambda _{1}-i\varepsilon \right) \left( \lambda -\lambda _{1}-i\varepsilon
\right) +2\pi C}d\lambda d\lambda _{1}
\end{equation*}%
where the integral over $\lambda _{1}$ is now performed with $\lambda
_{1}\in 
\mathbb{R}
$. This integral is computed by the residue theorem, where the residues
satisfy:%
\begin{equation*}
\lambda _{1}^{2}=\left( \lambda -i\varepsilon \right) ^{2}+2\pi C
\end{equation*}%
leading to write (\ref{sh}) as: 
\begin{eqnarray}
&&\exp \left( -cl_{1}\right) \times \int \frac{-\pi i}{\left\vert
Z-Z_{1}\right\vert }\exp \left( i\lambda cl_{1}+i\sqrt{\left( \lambda
-i\varepsilon \right) ^{2}+2\pi C}\left\vert Z-Z_{1}\right\vert \right)
d\lambda  \label{hs} \\
&&+\exp \left( -cl_{1}\right) \times \int \frac{-\pi i}{\left\vert
Z-Z_{1}\right\vert }\exp \left( i\lambda cl_{1}-i\sqrt{\left( \lambda
-i\varepsilon \right) ^{2}+2\pi C}\left\vert Z-Z_{1}\right\vert \right)
d\lambda  \notag
\end{eqnarray}%
We then perform the change of variable:%
\begin{eqnarray*}
x &=&\lambda +\sqrt{\lambda ^{2}+2\pi C} \\
dx &=&\left( 1+\frac{\lambda }{\sqrt{\lambda ^{2}+2\pi C}}\right) d\lambda \\
&=&\frac{x}{\sqrt{\lambda ^{2}+2\pi C}}d\lambda =\frac{2x^{2}}{x^{2}+2\pi C}%
d\lambda
\end{eqnarray*}%
and rewrite the exponents in (\ref{hs}) as:%
\begin{eqnarray*}
\lambda cl_{1}+\sqrt{\left( \lambda -i\varepsilon \right) ^{2}+2\pi C}%
\left\vert Z-Z_{1}\right\vert &=&\frac{cl_{1}+\left\vert Z-Z_{1}\right\vert 
}{2}\left( \lambda +\sqrt{\lambda ^{2}+2\pi C}\right) \\
&&+\frac{cl_{1}-\left\vert Z-Z_{1}\right\vert }{2}\left( \lambda -\sqrt{%
\lambda ^{2}+2\pi C}\right) \\
&=&\frac{cl_{1}+\left\vert Z-Z_{1}\right\vert }{2}\left( \lambda +\sqrt{%
\lambda ^{2}+2\pi C}\right) -\frac{cl_{1}-\left\vert Z-Z_{1}\right\vert }{2}%
\frac{2\pi C}{\lambda +\sqrt{\lambda ^{2}+2\pi C}} \\
&=&\frac{cl_{1}+\left\vert Z-Z_{1}\right\vert }{2}x-\frac{cl_{1}-\left\vert
Z-Z_{1}\right\vert }{2}\frac{2\pi C}{x}
\end{eqnarray*}%
and:%
\begin{equation*}
\lambda cl_{1}-\sqrt{\left( \lambda -i\varepsilon \right) ^{2}+2\pi C}%
\left\vert Z-Z_{1}\right\vert =\frac{cl_{1}-\left\vert Z-Z_{1}\right\vert }{2%
}x-\frac{cl_{1}+\left\vert Z-Z_{1}\right\vert }{2}\frac{2\pi C}{x}
\end{equation*}%
As a consequence, expression (\ref{hs}) becomes: 
\begin{eqnarray*}
&&\exp \left( -cl_{1}\right) \times \int \frac{-\pi i}{\left\vert
Z-Z_{1}\right\vert }\exp \left( i\left( \frac{cl_{1}+\left\vert
Z-Z_{1}\right\vert }{2}x-\frac{cl_{1}-\left\vert Z-Z_{1}\right\vert }{2}%
\frac{2\pi C}{x}\right) \right) dx \\
&&+\exp \left( -cl_{1}\right) \times \int \frac{-\pi i}{\left\vert
Z-Z_{1}\right\vert }\exp \left( i\left( \frac{cl_{1}-\left\vert
Z-Z_{1}\right\vert }{2}x-\frac{cl_{1}+\left\vert Z-Z_{1}\right\vert }{2}%
\frac{2\pi C}{x}\right) \right) dx \\
&&+2\pi C\exp \left( -cl_{1}\right) \times \int \frac{-\pi i}{\left\vert
Z-Z_{1}\right\vert }\exp \left( i\left( \frac{cl_{1}+\left\vert
Z-Z_{1}\right\vert }{2}x-\frac{cl_{1}-\left\vert Z-Z_{1}\right\vert }{2}%
\frac{2\pi C}{x}\right) \right) \frac{1}{x^{2}}dx \\
&&+2\pi C\times \int \frac{-\pi i}{\left\vert Z-Z_{1}\right\vert }\exp
\left( i\left( \frac{cl_{1}-\left\vert Z-Z_{1}\right\vert }{2}x-\frac{%
cl_{1}+\left\vert Z-Z_{1}\right\vert }{2}\frac{2\pi C}{x}\right) \right) 
\frac{1}{x^{2}}dx
\end{eqnarray*}%
Performing the change of variable $y=\frac{1}{x}$ in the two last
expressions yields:%
\begin{eqnarray*}
&&\exp \left( -cl_{1}\right) \left( 1+2\pi C\right) \times \left( \int \frac{%
-\pi i}{\left\vert Z-Z_{1}\right\vert }\exp \left( i\left( \frac{%
cl_{1}+\left\vert Z-Z_{1}\right\vert }{2}x-\frac{cl_{1}-\left\vert
Z-Z_{1}\right\vert }{2}\frac{2\pi C}{x}\right) \right) dx\right. \\
&&\left. +\int \frac{-\pi i}{\left\vert Z-Z_{1}\right\vert }\exp \left(
i\left( \frac{cl_{1}-\left\vert Z-Z_{1}\right\vert }{2}x-\frac{%
cl_{1}+\left\vert Z-Z_{1}\right\vert }{2}\frac{2\pi C}{x}\right) \right)
dx\right)
\end{eqnarray*}%
and by analytic continuation $x\rightarrow ix$, this becomes:%
\begin{eqnarray*}
&&\exp \left( -cl_{1}\right) \left( 1+2\pi C\right) \times \left( \int \frac{%
\pi }{\left\vert Z-Z_{1}\right\vert }\exp \left( -\left( \frac{%
cl_{1}+\left\vert Z-Z_{1}\right\vert }{2}x-\frac{cl_{1}-\left\vert
Z-Z_{1}\right\vert }{2}\frac{2\pi C}{x}\right) \right) dx\right. \\
&&\left. +\int \frac{\pi }{\left\vert Z-Z_{1}\right\vert }\exp \left(
-\left( \frac{cl_{1}-\left\vert Z-Z_{1}\right\vert }{2}x-\frac{%
cl_{1}+\left\vert Z-Z_{1}\right\vert }{2}\frac{2\pi C}{x}\right) \right)
dx\right)
\end{eqnarray*}%
Ultimately, reintroducing the constraint $H\left( cl_{1}-\left\vert
Z-Z_{1}\right\vert \right) $ and the factor $\frac{A_{1}}{A}$, (\ref{gr})
writes:%
\begin{eqnarray}
\left\vert \Psi \left( \theta -l_{1},Z_{1}\right) \right\vert ^{2}\frac{%
\delta \omega ^{-1}\left( J,\theta ,Z\right) }{\delta \left\vert \Psi \left(
\theta -l_{1},Z_{1}\right) \right\vert ^{2}} &=&\left( 1+2\pi C\right) \frac{%
A_{1}}{A}\frac{\exp \left( -cl_{1}\right) }{\left\vert Z-Z_{1}\right\vert }%
\left( \sqrt{\frac{cl_{1}-\left\vert Z-Z_{1}\right\vert }{cl_{1}+\left\vert
Z-Z_{1}\right\vert }}+\sqrt{\frac{cl_{1}+\left\vert Z-Z_{1}\right\vert }{%
cl_{1}-\left\vert Z-Z_{1}\right\vert }}\right)  \notag \\
&&\times K_{1}\left( \frac{cl_{1}-\left\vert Z-Z_{1}\right\vert }{2}2\pi C%
\frac{cl_{1}+\left\vert Z-Z_{1}\right\vert }{2}\right) \omega \left(
J,\theta -l_{1},Z_{1}\right)  \notag \\
&=&\left( 1+2\pi C\right) \frac{A_{1}}{A}\frac{\exp \left( -cl_{1}\right) }{%
\left\vert Z-Z_{1}\right\vert }\left( \sqrt{\frac{cl_{1}-\left\vert
Z-Z_{1}\right\vert }{cl_{1}+\left\vert Z-Z_{1}\right\vert }}+\sqrt{\frac{%
cl_{1}+\left\vert Z-Z_{1}\right\vert }{cl_{1}-\left\vert Z-Z_{1}\right\vert }%
}\right)  \notag \\
&&\times K_{1}\left( \pi C\frac{\left( cl_{1}\right) ^{2}-\left\vert
Z-Z_{1}\right\vert ^{2}}{2}\right) \omega \left( J,\theta -l_{1},Z_{1}\right)
\label{cdl}
\end{eqnarray}%
In first approximation, the right hand side of (\ref{cdl}) is:%
\begin{eqnarray}
&&\frac{\exp \left( -cl_{1}\right) \left( cl_{1}+\left\vert
Z-Z_{1}\right\vert \right) }{B\left\vert Z-Z_{1}\right\vert }\exp \left(
-\pi C\frac{\left( cl_{1}\right) ^{2}-\left\vert Z-Z_{1}\right\vert ^{2}}{2}%
\right) \omega \left( J,\theta -l_{1},Z_{1}\right)  \label{rtt} \\
&\sim &\frac{\exp \left( -cl_{1}\right) }{B}\exp \left( -\pi Ccl_{1}\frac{%
cl_{1}-\left\vert Z-Z_{1}\right\vert }{2}\right) H\left( cl_{1}-\left\vert
Z-Z_{1}\right\vert \right) \omega \left( J,\theta -l_{1},Z_{1}\right)  \notag
\end{eqnarray}%
for $cl_{1}>>\left\vert Z-Z_{1}\right\vert $. This can also be replaced by a
simplest form:%
\begin{equation}
\left\vert \Psi \left( \theta -l_{1},Z_{1}\right) \right\vert ^{2}\frac{%
\delta \omega ^{-1}\left( J,\theta ,Z\right) }{\delta \left\vert \Psi \left(
\theta -l_{1},Z_{1}\right) \right\vert ^{2}}\simeq \frac{\exp \left(
-cl_{1}-\alpha \left( \left( cl_{1}\right) ^{2}-\left\vert
Z-Z_{1}\right\vert ^{2}\right) \right) }{B}H\left( cl_{1}-\left\vert
Z-Z_{1}\right\vert \right) \omega \left( J,\theta -l_{1},Z_{1}\right)
\label{rrt}
\end{equation}%
where $B$ and $\alpha $ are constants.

Using (\ref{dvtr}), the same computation can be performed by replacing $\hat{%
T}$ with $\check{T}$ and we obtain: 
\begin{equation}
\left\vert \Psi \left( \theta -l_{1},Z_{1}\right) \right\vert ^{2}\frac{%
\delta \omega ^{-1}\left( J,\theta ,Z\right) }{\delta \left\vert \Psi \left(
\theta -l_{1},Z_{1}\right) \right\vert ^{2}}\simeq \frac{\exp \left(
-cl_{1}-\alpha \left( \left( cl_{1}\right) ^{2}-\left\vert
Z-Z_{1}\right\vert ^{2}\right) \right) }{D}H\left( cl_{1}-\left\vert
Z-Z_{1}\right\vert \right) \omega ^{-1}\left( J,\theta -l_{1},Z_{1}\right)
\label{trr}
\end{equation}%
with $D$ a constant.

\subsubsection*{\protect\bigskip 5.2.2 General formula}

For an arbitrary transfer function:%
\begin{equation*}
\hat{T}\left( \left\vert Z^{\left( l-1\right) }-Z^{\left( l\right)
}\right\vert \right) =C\exp \left( -c\left\vert Z^{\left( l-1\right)
}-Z^{\left( l\right) }\right\vert \right) f\left( \left\vert Z^{\left(
l-1\right) }-Z^{\left( l\right) }\right\vert \right)
\end{equation*}%
we can factor $C\exp \left( -cl\right) $ as in the previous paragraph. It
amounts to replace: 
\begin{equation*}
\hat{T}\left( \left\vert Z^{\left( l-1\right) }-Z^{\left( l\right)
}\right\vert \right) \rightarrow f\left( \left\vert Z^{\left( l-1\right)
}-Z^{\left( l\right) }\right\vert \right)
\end{equation*}%
We rewrite (\ref{gr}) as:%
\begin{eqnarray}
&&\left\vert \Psi \left( \theta -l_{1},Z_{1}\right) \right\vert ^{2}\frac{%
\delta \omega ^{-1}\left( J,\theta ,Z\right) }{\delta \left\vert \Psi \left(
\theta -l_{1},Z_{1}\right) \right\vert ^{2}}  \label{Srs} \\
&=&\sum_{n=1}^{\infty }\int \omega \left( J,\theta -l_{1},Z_{1}\right)
\times \mathit{T}_{1}^{\prime \prime }\left( \lambda +\lambda
_{1}.v_{1}\right) dv_{1}\dprod\limits_{l=2}^{n}\int \mathit{T}^{\prime
\prime }\left( \lambda +\lambda _{1}.v_{l}\right) dv_{l}\exp \left( i\lambda
cl_{1}+i\lambda _{1}.\left( Z-Z_{1}\right) \right) d\lambda d\lambda _{1} 
\notag \\
&=&\delta \left( \left\vert Z_{1}-Z\right\vert -cl_{1}\right) \hat{T}%
_{1}\left( \left\vert Z-Z^{\left( 1\right) }\right\vert \right) \omega
\left( J,\theta -l_{1},Z_{1}\right)  \notag \\
&&+\left( -1\right) ^{n}\int \omega \left( J,\theta -l_{1},Z_{1}\right)
\times \frac{\mathit{T}_{1}^{\prime \prime }\left( \lambda +\lambda
_{1}.v_{1}\right) }{2}dv_{1}\dprod\limits_{l=2}^{n}\int \frac{\mathit{T}%
^{\prime \prime }\left( \lambda +\lambda _{1}.v_{l}\right) }{2}dv_{l}\exp
\left( i\lambda cl_{1}+i\lambda _{1}.\left( Z-Z_{1}\right) \right) d\lambda
d\lambda _{1}  \notag
\end{eqnarray}%
With the convention that for $n=1$, the product $\dprod\limits_{l=2}^{n}$ is
set to be equal to $1$. The functions $\mathit{T}_{1}$ and $\mathit{T}$ are
the fourier transform of $\hat{T}_{1}H$\ and $\hat{T}H$\ respectively, and $%
H $ is the heaviside function. Remark that the first term of (\ref{Srs})
expresses the Dirac function $\delta \left( \left\vert Z_{1}-Z\right\vert
-cl_{1}\right) $ as a Fourier transform:%
\begin{eqnarray*}
&&\exp \left( i\lambda \left( cl_{1}-\sum_{l=1}^{n}\left\vert Z^{\left(
0\right) }-Z^{\left( 1\right) }\right\vert \right) \right) \\
&&\times \exp \left( i\lambda _{1}.\left( Z-Z_{1}-\sum_{l=1}^{n}\left(
Z^{\left( 0\right) }-Z^{\left( 1\right) }\right) \right) \right) d\lambda
d\lambda _{1}\left\vert Z^{\left( 0\right) }-Z^{\left( 1\right) }\right\vert
^{2}d\left\vert Z^{\left( 0\right) }-Z^{\left( 1\right) }\right\vert dv_{l}
\end{eqnarray*}%
Some terms of (\ref{Srs}) can be written in a useful form for the sequel: 
\begin{eqnarray}
\frac{1}{2}\int \mathit{T}^{\prime \prime }\left( \lambda +\lambda
_{1}.v_{l}\right) dv_{l} &=&\pi \int_{0}^{\pi }\mathit{T}^{\prime \prime
}\left( \lambda +\left\vert \lambda _{1}\right\vert \cos \left( \theta
_{l}\right) \right) \sin \left( \theta _{l}\right) d\theta _{l}  \notag \\
&=&\pi \int_{-1}^{1}\mathit{T}^{\prime \prime }\left( \lambda +\left\vert
\lambda _{1}\right\vert u\right) du  \notag \\
&=&\frac{2\pi \left( \mathit{T}^{\prime }\left( \lambda +\left\vert \lambda
_{1}\right\vert \right) -\mathit{T}^{\prime }\left( \lambda -\left\vert
\lambda _{1}\right\vert \right) \right) }{2\left\vert \lambda
_{1}\right\vert }  \notag \\
&\equiv &\mathit{\bar{T}}\left( \lambda ,\left\vert \lambda _{1}\right\vert
\right)  \label{brT}
\end{eqnarray}%
\begin{eqnarray}
\int \mathit{T}_{1}^{\prime \prime }\left( \lambda +\lambda
_{1}.v_{l}\right) dv_{l} &=&\frac{2\pi \left( \mathit{T}_{1}^{\prime }\left(
\lambda +\left\vert \lambda _{1}\right\vert \right) -\mathit{T}_{1}^{\prime
}\left( \lambda -\left\vert \lambda _{1}\right\vert \right) \right) }{%
2\left\vert \lambda _{1}\right\vert }  \notag \\
&\equiv &\mathit{\bar{T}}_{1}\left( \lambda ,\left\vert \lambda
_{1}\right\vert \right)  \label{rbT}
\end{eqnarray}%
\begin{eqnarray}
\exp \left( i\lambda _{1}.\left( Z-Z_{1}\right) \right) d\lambda _{1}
&=&\exp \left( i\cos \left( \theta _{1}\right) \left\vert \lambda
_{1}\right\vert \left\vert Z-Z_{1}\right\vert \right) \sin \left( \theta
_{1}\right) \left\vert \lambda _{1}\right\vert ^{2}d\left\vert \lambda
_{1}\right\vert d\theta _{1}  \label{rF} \\
&=&\exp \left( iu\left\vert \lambda _{1}\right\vert \left\vert
Z-Z_{1}\right\vert \right) \left\vert \lambda _{1}\right\vert
^{2}d\left\vert \lambda _{1}\right\vert du  \notag
\end{eqnarray}%
Remark that the functions of $x$:%
\begin{equation*}
\mathit{\bar{T}}\left( \lambda ,x\right) =\frac{2\pi \left( \mathit{T}%
^{\prime }\left( \lambda +x\right) -\mathit{T}^{\prime }\left( \lambda
-x\right) \right) }{2x}\text{ and }\mathit{\bar{T}}_{1}\left( \lambda
,x\right) =\frac{2\pi \left( \mathit{T}_{1}^{\prime }\left( \lambda
+x\right) -\mathit{T}_{1}^{\prime }\left( \lambda -x\right) \right) }{2x}
\end{equation*}%
are even.

\subsubsection*{5.2.3 Estimation of (\protect\ref{Srs})}

Using (\ref{brT}), (\ref{rbT}) and (\ref{rF}), equation (\ref{Srs}) becomes:%
\begin{eqnarray}
&&\left\vert \Psi \left( \theta -l_{1},Z_{1}\right) \right\vert ^{2}\frac{%
\delta \omega ^{-1}\left( J,\theta ,Z\right) }{\delta \left\vert \Psi \left(
\theta -l_{1},Z_{1}\right) \right\vert ^{2}}  \notag \\
&=&\sum_{n=1}^{\infty }\left( -1\right) ^{n}\int \omega \left( J,\theta
-l_{1},Z_{1}\right) \times \mathit{T}_{1}\left( \lambda +\lambda
_{1}.v_{1}\right) dv_{1}\dprod\limits_{l=2}^{n}\int \mathit{T}\left( \lambda
+\lambda _{1}.v_{l}\right) dv_{l}\exp \left( i\lambda cl_{1}+i\lambda
_{1}.\left( Z-Z_{1}\right) \right) d\lambda d\lambda _{1}  \notag \\
&=&-\int \omega \left( J,\theta -l_{1},Z_{1}\right) \times \frac{\mathit{%
\bar{T}}_{1}\left( \lambda ,\left\vert \lambda _{1}\right\vert \right) }{1+%
\mathit{\bar{T}}\left( \lambda ,\left\vert \lambda _{1}\right\vert \right) }%
\exp \left( i\lambda cl_{1}\right) \int_{-1}^{1}\exp \left( iu\left\vert
\lambda _{1}\right\vert \left\vert Z-Z_{1}\right\vert \right) \left\vert
\lambda _{1}\right\vert ^{2}d\left\vert \lambda _{1}\right\vert dud\lambda 
\notag \\
&=&-\int \omega \left( J,\theta -l_{1},Z_{1}\right) \times \frac{\mathit{%
\bar{T}}_{1}\left( \lambda ,\left\vert \lambda _{1}\right\vert \right) }{1+%
\mathit{\bar{T}}\left( \lambda ,\left\vert \lambda _{1}\right\vert \right) }%
\exp \left( i\lambda cl_{1}\right) \left( 2\frac{\sin \left( \left\vert
\lambda _{1}\right\vert \left\vert Z-Z_{1}\right\vert \right) }{\left\vert
Z-Z_{1}\right\vert }\left\vert \lambda _{1}\right\vert \right) d\left\vert
\lambda _{1}\right\vert d\lambda  \label{tmn}
\end{eqnarray}%
We remark that for even functions $f$, the following identity holds: 
\begin{eqnarray*}
&&\int_{0}^{+\infty }f\left( \left\vert \lambda _{1}\right\vert \right) 2%
\frac{\sin \left( \left\vert \lambda _{1}\right\vert \left\vert
Z-Z_{1}\right\vert \right) }{\left\vert Z-Z_{1}\right\vert }\left\vert
\lambda _{1}\right\vert d\left\vert \lambda _{1}\right\vert \\
&=&\int_{0}^{+\infty }f\left( x\right) \frac{\exp \left( ix\left\vert
Z-Z_{1}\right\vert \right) -\exp \left( -ix\left\vert Z-Z_{1}\right\vert
\right) }{i\left\vert Z-Z_{1}\right\vert }xdx \\
&=&\int_{0}^{+\infty }f\left( x\right) \frac{\exp \left( ix\left\vert
Z-Z_{1}\right\vert \right) }{i\left\vert Z-Z_{1}\right\vert }%
xdx+\int_{-\infty }^{0}f\left( -x\right) \frac{\exp \left( ix\left\vert
Z-Z_{1}\right\vert \right) }{i\left\vert Z-Z_{1}\right\vert }xdx \\
&=&\int_{-\infty }^{+\infty }f\left( x\right) \frac{\exp \left( ix\left\vert
Z-Z_{1}\right\vert \right) }{i\left\vert Z-Z_{1}\right\vert }xdx
\end{eqnarray*}%
so that (\ref{tmn}) becomes:%
\begin{eqnarray}
&&\left\vert \Psi \left( \theta -l_{1},Z_{1}\right) \right\vert ^{2}\frac{%
\delta \omega ^{-1}\left( J,\theta ,Z\right) }{\delta \left\vert \Psi \left(
\theta -l_{1},Z_{1}\right) \right\vert ^{2}}  \label{nmt} \\
&=&-\int \omega \left( J,\theta -l_{1},Z_{1}\right) \times \frac{\mathit{%
\bar{T}}_{1}\left( \lambda ,\lambda _{1}\right) }{1+\mathit{\bar{T}}\left(
\lambda ,\lambda _{1}\right) }\frac{\lambda _{1}}{i\left\vert
Z-Z_{1}\right\vert }\exp \left( i\lambda cl_{1}+i\lambda _{1}\left\vert
Z-Z_{1}\right\vert \right) d\lambda _{1}d\lambda  \notag \\
&=&-\int \omega \left( J,\theta -l_{1},Z_{1}\right) \times \frac{\pi \left( 
\mathit{T}_{1}^{\prime }\left( \lambda +\lambda _{1}\right) -\mathit{T}%
_{1}^{\prime }\left( \lambda -\lambda _{1}\right) \right) }{\lambda _{1}+\pi
\left( \mathit{T}^{\prime }\left( \lambda +\lambda _{1}\right) -\mathit{T}%
^{\prime }\left( \lambda -\lambda _{1}\right) \right) }\frac{\lambda _{1}}{%
i\left\vert Z-Z_{1}\right\vert }  \notag \\
&&\times \exp \left( i\lambda cl_{1}+i\lambda _{1}\left\vert
Z-Z_{1}\right\vert \right) d\lambda _{1}d\lambda  \notag \\
&=&-\int \omega \left( J,\theta -l_{1},Z_{1}\right) \times \frac{\pi \left( 
\mathit{T}_{1}^{\prime }\left( \lambda +\lambda _{1}\right) -\mathit{T}%
_{1}^{\prime }\left( \lambda -\lambda _{1}\right) \right) }{\lambda _{1}+\pi
\left( \mathit{T}^{\prime }\left( \lambda +\lambda _{1}\right) -\mathit{T}%
^{\prime }\left( \lambda -\lambda _{1}\right) \right) }\frac{\lambda _{1}}{%
i\left\vert Z-Z_{1}\right\vert }  \notag \\
&&\times \exp \left( iu\left( cl_{1}+\left\vert Z-Z_{1}\right\vert \right)
\right) \times \exp \left( iv\left( cl_{1}-\left\vert Z-Z_{1}\right\vert
\right) \right) d\lambda _{1}d\lambda  \notag
\end{eqnarray}%
As in the previous paragraph, we also simplify (\ref{nmt}) by writing $%
\mathit{T}_{1}$ as a function of $\mathit{T}$: 
\begin{equation*}
\mathit{T}_{1}^{\prime }\left( \lambda +\lambda _{1}\right) -\mathit{T}%
_{1}^{\prime }\left( \lambda -\lambda _{1}\right) =\frac{A_{1}}{A}\left( 
\mathit{T}^{\prime }\left( \lambda +\lambda _{1}\right) -\mathit{T}^{\prime
}\left( \lambda -\lambda _{1}\right) \right)
\end{equation*}%
and by setting:%
\begin{eqnarray*}
u &=&\frac{\lambda +\lambda _{1}}{2} \\
v &=&\frac{\lambda +\lambda _{1}}{2}
\end{eqnarray*}%
so that we are lead to:%
\begin{eqnarray}
\left\vert \Psi \left( \theta -l_{1},Z_{1}\right) \right\vert ^{2}\frac{%
\delta \omega ^{-1}\left( J,\theta ,Z\right) }{\delta \left\vert \Psi \left(
\theta -l_{1},Z_{1}\right) \right\vert ^{2}} &=&-\int \omega \left( J,\theta
-l_{1},Z_{1}\right) \times \frac{A_{1}}{A}\frac{\pi \left( \mathit{T}%
_{1}^{\prime }\left( \lambda +\lambda _{1}\right) -\mathit{T}_{1}^{\prime
}\left( \lambda -\lambda _{1}\right) \right) }{\lambda _{1}+\pi \left( 
\mathit{T}^{\prime }\left( \lambda +\lambda _{1}\right) -\mathit{T}^{\prime
}\left( \lambda -\lambda _{1}\right) \right) }\frac{\lambda _{1}}{%
i\left\vert Z-Z_{1}\right\vert }  \notag \\
&&\times \exp \left( iu\left( cl_{1}+\left\vert Z-Z_{1}\right\vert \right)
\right) \times \exp \left( iv\left( cl_{1}-\left\vert Z-Z_{1}\right\vert
\right) \right) d\lambda _{1}d\lambda  \label{tnm}
\end{eqnarray}%
Remark that the particular case of the exponential transfer function is
encompassed in (\ref{nmt}). Actually, if we choose:%
\begin{equation*}
\hat{T}\left( \left\vert Z^{\left( l-1\right) }-Z^{\left( l\right)
}\right\vert \right) =C\frac{\exp \left( -c\left\vert Z^{\left( l-1\right)
}-Z^{\left( l\right) }\right\vert \right) }{\left\vert Z^{\left( l-1\right)
}-Z^{\left( l\right) }\right\vert }
\end{equation*}%
we have:%
\begin{equation*}
\dprod\limits_{l=1}^{n}\hat{T}\left( \left\vert Z^{\left( l-1\right)
}-Z^{\left( l\right) }\right\vert \right) =\exp \left( -cl_{1}\right)
\dprod\limits_{l=1}^{n}\frac{C}{\left\vert Z^{\left( l-1\right) }-Z^{\left(
l\right) }\right\vert }
\end{equation*}%
For such a choice, we have formally: $\mathit{T}=-iC\int \left( FH\right) $
where $H$ is the heaviside function. As a consequence:%
\begin{equation*}
\mathit{T}^{\prime }\left( \lambda \right) =CFH=-\frac{C}{\lambda
+i\varepsilon }
\end{equation*}%
and (\ref{tnm}) is equivalent to the expressions of appendix 1.3.2.1.

In the general case, we write $\lambda _{1}^{\left( r\right) }$, $r=1,...$
the solutions to the pole equation of (\ref{tnm}):%
\begin{equation*}
\lambda _{1}+\pi \left( \mathit{T}^{\prime }\left( \lambda +\lambda
_{1}\right) -\mathit{T}^{\prime }\left( \lambda -\lambda _{1}\right) \right)
=0
\end{equation*}%
For regular functions $\mathit{T}^{\prime }\left( \lambda +\lambda
_{1}\right) $ such that for $\lambda \rightarrow \infty $:%
\begin{equation*}
\mathit{T}^{\prime }\left( \lambda +\lambda _{1}\right) \simeq \frac{g\left(
\lambda +\lambda _{1}\right) }{\left( \lambda +\lambda _{1}\right) ^{l}}
\end{equation*}%
\begin{equation*}
\int \frac{1}{\left( \lambda -s\right) ^{l}}\left\vert \Psi \left( s\right)
\right\vert
\end{equation*}%
with $l>0$ given and $g$ bounded, the poles equation implies that for $%
\lambda \rightarrow \infty $:%
\begin{equation*}
\lambda _{1}\simeq \pm \lambda
\end{equation*}%
and as a consequence, we can write: 
\begin{equation}
\lambda _{1}^{\left( r\right) }=\sqrt{\lambda ^{2}+h_{r}\left( \lambda
\right) }  \label{psl}
\end{equation}%
where $h_{r}\left( \lambda \right) $ is bounded.

We can compute the values of the residues at each pole by the first order
expansion of $1+\pi \frac{\mathit{T}^{\prime }\left( \lambda +\lambda
_{1}\right) -\mathit{T}^{\prime }\left( \lambda -\lambda _{1}\right) }{%
\lambda _{1}}$:%
\begin{eqnarray*}
&&1+\pi \frac{\mathit{T}^{\prime }\left( \lambda +\lambda _{1}\right) -%
\mathit{T}^{\prime }\left( \lambda -\lambda _{1}\right) }{\lambda _{1}} \\
&\simeq &\pi \left( \frac{\mathit{T}^{\prime }\left( \lambda +\lambda
_{1}\right) -\mathit{T}^{\prime }\left( \lambda -\lambda _{1}\right) }{%
\lambda _{1}}-\frac{\mathit{T}^{\prime }\left( \lambda +\lambda _{1}^{\left(
r\right) }\right) -\mathit{T}^{\prime }\left( \lambda -\lambda _{1}^{\left(
r\right) }\right) }{\lambda _{1}^{\left( r\right) }}\right) \\
&\simeq &\pi \left( \frac{\frac{1}{\pi }+\mathit{T}^{\prime \prime }\left(
\lambda +\lambda _{1}^{\left( r\right) }\right) +\mathit{T}^{\prime \prime
}\left( \lambda -\lambda _{1}^{\left( r\right) }\right) }{\lambda
_{1}^{\left( r\right) }}\right) \\
&\simeq &\pi \left( \frac{\mathit{T}^{\prime \prime }\left( \lambda +\lambda
_{1}^{\left( r\right) }\right) +\mathit{T}^{\prime \prime }\left( \lambda
-\lambda _{1}^{\left( r\right) }\right) -\frac{\mathit{T}^{\prime }\left(
\lambda +\lambda _{1}^{\left( r\right) }\right) -\mathit{T}^{\prime }\left(
\lambda -\lambda _{1}^{\left( r\right) }\right) }{\lambda _{1}^{\left(
r\right) }}}{\lambda _{1}^{\left( r\right) }}\right)
\end{eqnarray*}%
For regular functions $\mathit{T}^{\prime }\left( \lambda +\lambda
_{1}\right) $, this can be expanded as:%
\begin{equation*}
2\pi \lambda _{1}^{\left( r\right) }\left( \sum_{k\geqslant 1}\frac{\mathit{T%
}^{\left( 2k+2\right) }\left( \lambda \right) }{\left( 2k\right) !}\left(
\lambda _{1}^{\left( r\right) }\right) ^{2k-2}-\sum_{k\geqslant 1}\frac{%
\mathit{T}^{\left( 2k+2\right) }\left( \lambda \right) }{\left( 2k+1\right) !%
}\left( \lambda _{1}^{\left( r\right) }\right) ^{2k-2}\right)
\end{equation*}%
and for relatively slowly varying functions, this reduces to:%
\begin{equation}
1+\pi \frac{\mathit{T}^{\prime }\left( \lambda +\lambda _{1}\right) -\mathit{%
T}^{\prime }\left( \lambda -\lambda _{1}\right) }{\lambda _{1}}\simeq 2\pi
\lambda _{1}^{\left( r\right) }\frac{\mathit{T}^{\left( 4\right) }\left(
\lambda \right) }{3}  \label{spl}
\end{equation}%
and the residue theorem implies to replace:%
\begin{eqnarray}
&&\frac{\pi \left( \mathit{T}_{1}^{\prime }\left( \lambda +\lambda
_{1}\right) -\mathit{T}_{1}^{\prime }\left( \lambda -\lambda _{1}\right)
\right) }{\lambda _{1}+\pi \left( \mathit{T}^{\prime }\left( \lambda
+\lambda _{1}\right) -\mathit{T}^{\prime }\left( \lambda -\lambda
_{1}\right) \right) }\frac{\lambda _{1}}{i\left\vert Z-Z_{1}\right\vert }
\label{dsr} \\
&\rightarrow &-\frac{i}{\pi }\frac{3}{\left\vert Z-Z_{1}\right\vert \mathit{T%
}^{\left( 4\right) }\left( \lambda \right) }  \notag
\end{eqnarray}%
in (\ref{tnm}). Using (\ref{psl}) and (\ref{dsr}) in (\ref{tnm}) leads to:%
\begin{eqnarray*}
\left\vert \Psi \left( \theta -l_{1},Z_{1}\right) \right\vert ^{2}\frac{%
\delta \omega ^{-1}\left( J,\theta ,Z\right) }{\delta \left\vert \Psi \left(
\theta -l_{1},Z_{1}\right) \right\vert ^{2}} &\simeq &\sum_{r}\frac{i}{\pi }%
\frac{1}{\left\vert Z-Z_{1}\right\vert }\int \omega \left( J,\theta
-l_{1},Z_{1}\right) \\
&&\times \frac{3}{\mathit{T}^{\left( 4\right) }\left( \lambda \right) }\exp
\left( iu\left( cl_{1}+\left\vert Z-Z_{1}\right\vert \right) \right) \times
\exp \left( iv\left( cl_{1}-\left\vert Z-Z_{1}\right\vert \right) \right)
d\lambda \\
&=&\sum_{r}\frac{i}{\pi }\frac{1}{\left\vert Z-Z_{1}\right\vert }\int \omega
\left( J,\theta -l_{1},Z_{1}\right) \\
&&\times \frac{3}{\mathit{T}^{\left( 4\right) }\left( \lambda \right) }\exp
\left( iu\left( cl_{1}+\left\vert Z-Z_{1}\right\vert \right) \right) \times
\exp \left( iv\left( cl_{1}-\left\vert Z-Z_{1}\right\vert \right) \right)
d\lambda
\end{eqnarray*}%
where:%
\begin{eqnarray*}
u &=&\frac{\lambda +\lambda _{1}^{\left( r\right) }}{2}=\frac{\lambda
+f^{\left( r\right) }\left( \lambda \right) }{2} \\
v &=&\frac{\lambda +\lambda _{1}^{\left( r\right) }}{2}=\frac{\lambda
-f^{\left( r\right) }\left( \lambda \right) }{2}
\end{eqnarray*}%
As a consequence:%
\begin{eqnarray*}
v &=&\lambda -\sqrt{\lambda ^{2}+h_{r}\left( \lambda \right) } \\
&=&-\frac{h_{r}\left( \lambda \right) }{\lambda +\sqrt{\lambda
^{2}+h_{r}\left( \lambda \right) }} \\
&=&-\frac{h_{r}\left( \lambda \right) }{u}
\end{eqnarray*}%
For $h_{r}\left( \lambda \right) $ varying slowly, we can replace $%
h_{r}\left( \lambda \right) $ by its average $\bar{h}_{r}$, and we have:%
\begin{equation*}
v=-\frac{\bar{h}_{r}}{u}
\end{equation*}%
Replacing $\mathit{T}^{\left( 4\right) }\left( \lambda \right) $ by its
average $\mathit{\bar{T}}^{\left( 4\right) }$, we find:%
\begin{eqnarray*}
\left\vert \Psi \left( \theta -l_{1},Z_{1}\right) \right\vert ^{2}\frac{%
\delta \omega ^{-1}\left( J,\theta ,Z\right) }{\delta \left\vert \Psi \left(
\theta -l_{1},Z_{1}\right) \right\vert ^{2}} &\simeq &\sum_{r}\frac{i}{\pi }%
\frac{1}{\left\vert Z-Z_{1}\right\vert }\int \omega \left( J,\theta
-l_{1},Z_{1}\right) \\
&&\times \frac{3}{\mathit{\bar{T}}^{\left( 4\right) }}\exp \left( iu\left(
cl_{1}+\left\vert Z-Z_{1}\right\vert \right) \right) \times \exp \left( -i%
\frac{\bar{h}_{r}}{u}\left( cl_{1}-\left\vert Z-Z_{1}\right\vert \right)
\right) d\lambda
\end{eqnarray*}%
We can then apply the results of the previous paragraph for each $r$, and
has a consequence, we obtain:%
\begin{eqnarray}
\left\vert \Psi \left( \theta -l_{1},Z_{1}\right) \right\vert ^{2}\frac{%
\delta \omega ^{-1}\left( J,\theta ,Z\right) }{\delta \left\vert \Psi \left(
\theta -l_{1},Z_{1}\right) \right\vert ^{2}} &\simeq &\sum_{r}\left( 1+\bar{h%
}_{r}\right) \frac{3}{\mathit{\bar{T}}^{\left( 4\right) }}\omega \left(
J,\theta -l_{1},Z_{1}\right) \frac{\exp \left( -cl_{1}\right) }{\left\vert
Z-Z_{1}\right\vert }  \label{rgc} \\
&&\times \left( \sqrt{\frac{cl_{1}-\left\vert Z-Z_{1}\right\vert }{%
cl_{1}+\left\vert Z-Z_{1}\right\vert }}+\sqrt{\frac{cl_{1}+\left\vert
Z-Z_{1}\right\vert }{cl_{1}-\left\vert Z-Z_{1}\right\vert }}\right)
K_{1}\left( \bar{h}_{r}\frac{\left( cl_{1}\right) ^{2}-\left\vert
Z-Z_{1}\right\vert ^{2}}{4}\right)  \notag
\end{eqnarray}%
that becomes in first approximation:%
\begin{equation*}
\left\vert \Psi \left( \theta -l_{1},Z_{1}\right) \right\vert ^{2}\frac{%
\delta \omega ^{-1}\left( J,\theta ,Z\right) }{\delta \left\vert \Psi \left(
\theta -l_{1},Z_{1}\right) \right\vert ^{2}}\simeq \sum_{r}\frac{\exp \left(
-cl_{1}-\alpha _{r}\left( \left( cl_{1}\right) ^{2}-\left\vert
Z-Z_{1}\right\vert ^{2}\right) \right) }{B_{r}}H\left( cl_{1}-\left\vert
Z-Z_{1}\right\vert \right) \omega \left( J,\theta -l_{1},Z_{1}\right)
\end{equation*}%
where the $B_{r}$ are constant coefficients and $\alpha _{r}=\frac{\bar{h}%
_{r}}{4}$. As for (\ref{trr}), this also writes:%
\begin{equation}
\left\vert \Psi \left( \theta -l_{1},Z_{1}\right) \right\vert ^{2}\frac{%
\delta \omega ^{-1}\left( J,\theta ,Z\right) }{\delta \left\vert \Psi \left(
\theta -l_{1},Z_{1}\right) \right\vert ^{2}}\simeq \sum_{r}\frac{\exp \left(
-cl_{1}-\alpha _{r}\left( \left( cl_{1}\right) ^{2}-\left\vert
Z-Z_{1}\right\vert ^{2}\right) \right) }{D_{r}}H\left( cl_{1}-\left\vert
Z-Z_{1}\right\vert \right) \omega \left( J,\theta -l_{1},Z_{1}\right)
\label{ttr}
\end{equation}%
for some constants $D_{r}$.

\subsubsection*{5.2.4 Application: Gaussian transfer function}

We apply the previous method to the case of a Gaussian transfer function.
The results will be similar to the exponential case, confirming that the
results obtained in appendix 5.2.1 are quite general and can be used
generally in first approximation

\paragraph*{\protect\bigskip 5.2.4.1 Estimation of the poles}

We can refine (\ref{rgc}) by computing more precisely the poles in (\ref{tnm}%
). To do so, we perform the change of variable:%
\begin{eqnarray*}
u &=&\lambda +\lambda _{1} \\
v &=&\lambda -\lambda _{1}
\end{eqnarray*}%
before computing the poles, and equation (\ref{tnm}) becomes:%
\begin{eqnarray}
\left\vert \Psi \left( \theta -l_{1},Z_{1}\right) \right\vert ^{2}\frac{%
\delta \omega ^{-1}\left( J,\theta ,Z\right) }{\delta \left\vert \Psi \left(
\theta -l_{1},Z_{1}\right) \right\vert ^{2}} &=&-\frac{A_{1}}{2iA\left\vert
Z-Z_{1}\right\vert }\int \omega \left( J,\theta -l_{1},Z_{1}\right) \times 
\frac{\pi \left( \mathit{T}^{\prime }\left( u\right) -\mathit{T}^{\prime
}\left( v\right) \right) }{1+2\pi \frac{\left( \mathit{T}^{\prime }\left(
u\right) -\mathit{T}^{\prime }\left( v\right) \right) }{u-v}}  \label{mtn} \\
&&\times \exp \left( i\frac{u}{2}\left( cl_{1}+\left\vert Z-Z_{1}\right\vert
\right) +i\frac{v}{2}\left( cl_{1}-\left\vert Z-Z_{1}\right\vert \right)
\right) dudv  \notag
\end{eqnarray}%
\bigskip We first estimate the $v$ integral using the residues theorem. The
poles are solutions of:%
\begin{equation*}
u-v+2\pi \left( \mathit{T}^{\prime }\left( u\right) -\mathit{T}^{\prime
}\left( v\right) \right) =0
\end{equation*}%
we write the solutions $v_{k}\left( u\right) $ with $k\geqslant 1$. As a
consequence (\ref{mtn}) rewrites:%
\begin{eqnarray*}
\left\vert \Psi \left( \theta -l_{1},Z_{1}\right) \right\vert ^{2}\frac{%
\delta \omega ^{-1}\left( J,\theta ,Z\right) }{\delta \left\vert \Psi \left(
\theta -l_{1},Z_{1}\right) \right\vert ^{2}} &=&-\sum_{k\geqslant 1}\frac{%
A_{1}i\pi }{2iA\left\vert Z-Z_{1}\right\vert }\int \omega \left( J,\theta
-l_{1},Z_{1}\right) \times \frac{-2\pi \left( v_{k}\left( u\right) -u\right)
^{2}}{1+2\pi \mathit{T}^{\prime \prime }\left( v_{k}\left( u\right) \right) }
\\
&&\times \exp \left( i\frac{u}{2}\left( cl_{1}+\left\vert Z-Z_{1}\right\vert
\right) +i\frac{v_{k}\left( u\right) }{2}\left( cl_{1}-\left\vert
Z-Z_{1}\right\vert \right) \right) du
\end{eqnarray*}

To compute (\ref{mtn}), we study its two components independently:%
\begin{eqnarray}
&&-\frac{A_{1}}{iA\left\vert Z-Z_{1}\right\vert }\int \omega \left( J,\theta
-l_{1},Z_{1}\right) \times \frac{\pi \mathit{T}^{\prime }\left( u\right) }{%
1+2\pi \frac{\left( \mathit{T}^{\prime }\left( u\right) -\mathit{T}^{\prime
}\left( v\right) \right) }{u-v}}  \label{fts} \\
&&\times \exp \left( i\frac{u}{2}\left( cl_{1}+\left\vert Z-Z_{1}\right\vert
\right) +i\frac{v}{2}\left( cl_{1}-\left\vert Z-Z_{1}\right\vert \right)
\right) dudv  \notag
\end{eqnarray}%
and: 
\begin{eqnarray}
&&\frac{A_{1}}{iA\left\vert Z-Z_{1}\right\vert }\int \omega \left( J,\theta
-l_{1},Z_{1}\right) \times \frac{\pi \mathit{T}^{\prime }\left( v\right) }{%
1+2\pi \frac{\left( \mathit{T}^{\prime }\left( u\right) -\mathit{T}^{\prime
}\left( v\right) \right) }{u-v}}  \label{sdc} \\
&&\times \exp \left( i\frac{u}{2}\left( cl_{1}+\left\vert Z-Z_{1}\right\vert
\right) +i\frac{v}{2}\left( cl_{1}-\left\vert Z-Z_{1}\right\vert \right)
\right) dudv  \notag
\end{eqnarray}%
In the integral (\ref{fts}), we first estimate the $v$ integral using the
residues theorem. The poles are solutions of:%
\begin{equation*}
1+2\pi \frac{\mathit{T}^{\prime }\left( u\right) -\mathit{T}^{\prime }\left(
v\right) }{u-v}=0
\end{equation*}%
That is: 
\begin{equation}
v+2\pi \mathit{T}^{\prime }\left( v\right) =u+2\pi \mathit{T}^{\prime
}\left( u\right)  \label{lP}
\end{equation}%
with $v\neq u$.

Now, we consider the following gaussian form for the transfer functions:%
\begin{equation}
\mathit{T}\left( \lambda \right) =A\exp \left( -\eta \frac{\lambda ^{2}}{4}%
\right) \left( 1-\func{erf}\left( i\sqrt{\eta }\lambda \right) \right)
\label{lT}
\end{equation}%
and its derivative satisfies:%
\begin{equation*}
\mathit{T}^{\prime }\left( \lambda \right) =-\eta \frac{\lambda }{2}\mathit{T%
}\left( \lambda \right) -i\sqrt{\eta }
\end{equation*}%
As a consequence of these two identities, the solutions of (\ref{lP}) are
given by:%
\begin{equation}
v\left( 1-\pi \eta \mathit{T}\left( v\right) \right) =z  \label{pl}
\end{equation}%
with:%
\begin{equation*}
z=u\left( 1-\pi \eta \mathit{T}\left( u\right) \right)
\end{equation*}%
To solve (\ref{pl}) it will be useful to expand $\mathit{T}\left( \lambda
\right) $\ as a series expansion. In first approximation, one has (see
Abramovitz stegun):%
\begin{equation*}
\func{Im}\func{erf}\left( i\sqrt{\eta }\lambda \right) \simeq \frac{2}{\pi }%
\sum_{k=1}^{+\infty }\frac{\exp \left( -k^{2}\right) \sinh k\sqrt{\eta }%
\lambda }{k}\simeq \sqrt{\frac{\eta }{\pi }}\lambda
\end{equation*}%
and: 
\begin{equation*}
\func{Im}\mathit{T}\left( \lambda \right) \simeq A\sqrt{\eta }\left( \frac{1%
}{2\sqrt{\pi }}\exp \left( -\eta \frac{\lambda ^{2}}{4}\right) \eta \lambda
^{2}\right) >0
\end{equation*}%
as a consequence $\func{Im}z>0$ and asymptotically, equation (\ref{pl})
reduces to:%
\begin{equation*}
v\left( \pi \eta \mathit{T}\left( v\right) \right) =-z
\end{equation*}%
that is:%
\begin{equation*}
\left( A\pi \eta v\right) ^{2}\exp \left( -\eta \frac{v^{2}}{2}\right)
\left( 1-\func{erf}\left( i\sqrt{\eta }v\right) \right) ^{2}=z^{2}
\end{equation*}%
for $\eta <<1$%
\begin{equation*}
\left( \frac{A\pi \eta v}{2}\right) ^{2}\exp \left( -\eta \frac{v^{2}}{2}%
\right) =z^{2}
\end{equation*}%
and the poles arising in (\ref{fts}) are given by the Whittaker functions $%
W_{k}$:%
\begin{equation*}
v=\sqrt{-\frac{2}{\eta }W_{k}\left( \frac{-2z^{2}}{\left( A\pi \right)
^{2}\eta }\right) }
\end{equation*}%
for $k>0$. They are approximatively equal to:%
\begin{equation*}
v\simeq \pm i\sqrt{\frac{2}{\eta }\left( \ln \left( \frac{2u^{2}}{\left(
A\pi \right) ^{2}\eta }\right) +i\left( 2k+1\right) \pi \right) }
\end{equation*}%
The terms involved in (\ref{fts}) can thus be evaluated at the poles. First,
for $\left( A\pi \right) ^{2}<<1$: 
\begin{eqnarray*}
\pi \eta \left\vert \mathit{T}\left( v\right) \right\vert &\simeq &A\pi \eta
\left\vert \exp \left( -\eta \frac{v^{2}}{4}\right) \right\vert \\
&\simeq &A\pi \eta \exp \left( \ln \left( \frac{\sqrt{2u^{2}}}{\left( A\pi
\right) \sqrt{\eta }}\right) \right) =\sqrt{2\eta }u
\end{eqnarray*}%
Asymptotically, for $\sqrt{2\eta }u>>1$, this formula justifies our previous
approximation $v\left( 1-\pi \eta \mathit{T}\left( v\right) \right) \simeq
-v\pi \eta \mathit{T}\left( v\right) $. For $\sqrt{2\eta }u<<1$, the
solution is $v=u$ and there is no pole. Second, we have:%
\begin{eqnarray*}
&&\left( 1+2\pi \frac{\left( \mathit{T}^{\prime }\left( u\right) -\mathit{T}%
^{\prime }\left( v\right) \right) }{u-v}\right) ^{\prime } \\
&=&-2\pi \frac{\mathit{T}^{\prime \prime }\left( v\right) }{u-v}+2\pi \frac{%
\left( \mathit{T}^{\prime }\left( u\right) -\mathit{T}^{\prime }\left(
v\right) \right) }{\left( u-v\right) ^{2}} \\
&=&-\frac{1+2\pi \mathit{T}^{\prime \prime }\left( v\right) }{u-v}
\end{eqnarray*}%
and (\ref{fts}) becomes:%
\begin{eqnarray*}
\sum_{k\neq 0} &&\frac{2\pi A_{1}}{A\left\vert Z-Z_{1}\right\vert }\int
\omega \left( J,\theta -l_{1},Z_{1}\right) \times \frac{\left( u-v\right) 
\mathit{T}^{\prime }\left( u\right) }{2\left( 1+\mathit{T}^{\prime \prime
}\left( v\right) \right) } \\
&&\times \exp \left( i\frac{u}{2}\left( cl_{1}+\left\vert Z-Z_{1}\right\vert
\right) -\sqrt{\frac{2}{\eta }\left( \ln \left( \frac{2u^{2}}{\left( A\pi
\right) ^{2}\eta }\right) +i\left( 2k+1\right) \pi \right) }\left\vert
cl_{1}-\left\vert Z-Z_{1}\right\vert \right\vert \right) du
\end{eqnarray*}%
Note that for $\left( A\pi \right) ^{2}\eta <<1$, we recover $\delta \left(
cl_{1}-\left\vert Z-Z_{1}\right\vert \right) $ as needed in the lowest order
approximation.

The second integral (\ref{sdc}) is obtained by inverting the role of $u$ and 
$v$. It yields:%
\begin{eqnarray*}
-\sum_{k\neq 0} &&\frac{2\pi A_{1}}{A\left\vert Z-Z_{1}\right\vert }\int
\omega \left( J,\theta -l_{1},Z_{1}\right) \times \frac{\left( u-v\right) 
\mathit{T}^{\prime }\left( v\right) }{2\left( 1+\mathit{T}^{\prime \prime
}\left( u\right) \right) } \\
&&\times \exp \left( -\sqrt{\frac{2}{\eta }\left( \ln \left( \frac{2u^{2}}{%
\left( A\pi \right) ^{2}\eta }\right) +i\left( 2k+1\right) \pi \right) }%
\left( cl_{1}+\left\vert Z-Z_{1}\right\vert \right) +i\frac{v}{2}\left(
cl_{1}-\left\vert Z-Z_{1}\right\vert \right) \right) du
\end{eqnarray*}%
and this can be neglected, since $cl_{1}+\left\vert Z-Z_{1}\right\vert >0$
and for $\left( A\pi \right) ^{2}\eta <<1$ this becomes $\delta \left(
cl_{1}+\left\vert Z-Z_{1}\right\vert \right) =0$.

Gathering the results for (\ref{fts}) and (\ref{sdc}), we are left with: 
\begin{eqnarray}
\left\vert \Psi \left( \theta -l_{1},Z_{1}\right) \right\vert ^{2} &&\frac{%
\delta \omega ^{-1}\left( J,\theta ,Z\right) }{\delta \left\vert \Psi \left(
\theta -l_{1},Z_{1}\right) \right\vert ^{2}}  \label{mrt} \\
=\sum_{k\neq 0} &&\frac{2\pi A_{1}}{A\left\vert Z-Z_{1}\right\vert }\int
\omega \left( J,\theta -l_{1},Z_{1}\right) \times \frac{\left( u-v\right) 
\mathit{T}^{\prime }\left( u\right) }{2\left( 1+\mathit{T}^{\prime \prime
}\left( v\right) \right) }  \notag \\
&&\times \exp \left( i\frac{u}{2}\left( cl_{1}+\left\vert Z-Z_{1}\right\vert
\right) -\sqrt{\frac{2}{\eta }\left( \ln \left( \frac{2u^{2}}{\left( A\pi
\right) ^{2}\eta }\right) +i\left( 2k+1\right) \pi \right) }\left\vert
cl_{1}-\left\vert Z-Z_{1}\right\vert \right\vert \right) du  \notag
\end{eqnarray}%
Ulitmately, some simplifications can be performed on (\ref{mrt}). Actually,
we have the following identities for $\mathit{T}$:

\begin{equation*}
\mathit{T}^{\prime \prime }\left( \lambda \right) =-\frac{\nu }{2}\mathit{T}%
\left( \lambda \right) +\left( \nu \frac{\lambda }{2}\right) ^{2}\mathit{T}%
\left( \lambda \right) +Ai\left( \sqrt{\nu }\right) ^{3}\frac{\lambda }{2}
\end{equation*}%
\begin{eqnarray*}
v\left( 1-\pi \eta \mathit{T}\left( v\right) \right) &=&u\left( 1-\pi \eta 
\mathit{T}\left( u\right) \right) \simeq u \\
\pi \eta \mathit{T}\left( v\right) &\simeq &v-u
\end{eqnarray*}%
and this two equations imply that, for $A\left( \sqrt{\eta }\right) ^{3}<<1$:%
\begin{eqnarray}
&&1+2\pi \mathit{T}^{\prime \prime }\left( v\right)  \label{spm} \\
&=&1-\pi \eta \mathit{T}\left( v\right) +2\pi \left( \eta \frac{v}{2}\right)
^{2}\mathit{T}\left( v\right) +2\pi iA\left( \sqrt{\eta }\right) ^{3}\frac{v%
}{2}  \notag \\
&\simeq &\left( v-u\right) \left( -1+\pi \eta v\right) \simeq i\left(
v-u\right) \sqrt{\frac{2}{\eta }}C  \notag
\end{eqnarray}%
where $C=\sqrt{\left( \ln \left( \frac{2u^{2}}{\left( A\pi \right) ^{2}\eta }%
\right) +i\left( 2k+1\right) \pi \right) }$. A consequence of (\ref{spm}) is
that:%
\begin{equation*}
-\frac{1+2\pi \mathit{T}^{\prime \prime }\left( v\right) }{u-v}\simeq \frac{1%
}{i\pi \sqrt{2\eta }C}
\end{equation*}%
Moreover, for $\left( A\pi \right) ^{2}\eta <<1$, the function $\mathit{T}%
^{\prime }\left( u\right) $ can be replaced by the multiplication by $i\frac{%
cl_{1}+\left\vert Z-Z_{1}\right\vert }{2}$. We are thus led to rewrite (\ref%
{mrt}):%
\begin{eqnarray}
&&\left\vert \Psi \left( \theta -l_{1},Z_{1}\right) \right\vert ^{2}\frac{%
\delta \omega ^{-1}\left( J,\theta ,Z\right) }{\delta \left\vert \Psi \left(
\theta -l_{1},Z_{1}\right) \right\vert ^{2}}  \label{mtr} \\
&=&\frac{A_{1}}{A}\sum_{k\neq 0}\frac{\left( cl_{1}+\left\vert
Z-Z_{1}\right\vert \right) }{2\left\vert Z-Z_{1}\right\vert }\int \omega
\left( J,\theta -l_{1},Z_{1}\right) \times \frac{\mathit{T}\left( u\right) }{%
\sqrt{2\eta }C}  \notag \\
&&\times \exp \left( i\frac{u}{2}\left( cl_{1}+\left\vert Z-Z_{1}\right\vert
\right) -\sqrt{\frac{2}{\eta }\left( \ln \left( \frac{2u^{2}}{\left( A\pi
\right) ^{2}\eta }\right) +i\left( 2k+1\right) \pi \right) }\left\vert
cl_{1}-\left\vert Z-Z_{1}\right\vert \right\vert \right) du  \notag \\
&\equiv &\frac{A_{1}}{A}\Xi \left( \left\vert Z_{1}-Z\right\vert ,l_{1},\bar{%
\omega}\right) \omega \left( J,\theta -l_{1},Z_{1}\right)  \notag
\end{eqnarray}%
Remark that, for $\left( A\pi \right) ^{2}\eta <<1$: 
\begin{eqnarray*}
&&\sqrt{\frac{2}{\eta }\left( \ln \left( \frac{2u^{2}}{\left( A\pi \right)
^{2}\eta }\right) +i\left( 2k+1\right) \pi \right) } \\
&&\times \exp \left( -\sqrt{\frac{2}{\eta }\left( \ln \left( \frac{2u^{2}}{%
\left( A\pi \right) ^{2}\eta }\right) +i\left( 2k+1\right) \pi \right) }%
\left\vert cl_{1}-\left\vert Z-Z_{1}\right\vert \right\vert \right) \\
&\simeq &\delta \left( cl_{1}-\left\vert Z-Z_{1}\right\vert \right)
\end{eqnarray*}%
so that one recovers the first order term.

For $\left( A\pi \right) ^{2}\eta <<1$, \ $\Xi \left( \left\vert
Z_{1}-Z\right\vert ,l_{1},\bar{\omega}\right) $ is a function of $\left\vert
Z_{1}-Z\right\vert $ written $\Xi \left( \left\vert Z_{1}-Z\right\vert ,\bar{%
\omega}\right) $.

Finally, the sum in (\ref{mtr}) can be estimated in the following way:

\begin{eqnarray*}
&&\sum_{k\neq 0}\exp \left( -\sqrt{\frac{2}{\eta }\left( \ln \left( \frac{%
2u^{2}}{\left( A\pi \right) ^{2}\eta }\right) +i\left( 2k+1\right) \pi
\right) }\left\vert cl_{1}-\left\vert Z-Z_{1}\right\vert \right\vert \right)
\\
&=&\sum_{k\neq 0}\exp \left( -\sqrt{\frac{2}{\eta }\ln \left( \frac{2u^{2}}{%
\left( A\pi \right) ^{2}\eta }\right) }\sqrt{1+i\frac{\left( 2k+1\right) }{%
\ln \left( \frac{2u^{2}}{\left( A\pi \right) ^{2}\eta }\right) }\pi }%
\left\vert cl_{1}-\left\vert Z-Z_{1}\right\vert \right\vert \right) \\
&\simeq &\frac{C^{2}}{\pi }\func{Re}\int_{0}^{1}\exp \left( -C\sqrt{\frac{2}{%
\eta }}\sqrt{1+ix}\left\vert cl_{1}-\left\vert Z-Z_{1}\right\vert
\right\vert \right) dx \\
&=&\frac{C^{2}}{\pi }\func{Re}\int_{0}^{1}\exp \left( -C\sqrt{\frac{2}{\eta }%
}\left( 1+x^{2}\right) ^{\frac{1}{4}}\exp \left( \frac{i}{2}\arctan \left(
x\right) \right) \left\vert cl_{1}-\left\vert Z-Z_{1}\right\vert \right\vert
\right) dx
\end{eqnarray*}%
with:%
\begin{equation*}
C=\sqrt{\ln \left( \frac{2u^{2}}{\left( A\pi \right) ^{2}\eta }\right) }
\end{equation*}%
The upper bound of the integral is set to $1$, in agreement with our
approximation $\ln \left( \frac{2u^{2}}{\left( A\pi \right) ^{2}\eta }%
\right) >>1$. It amounts to neglect the poles for $k>>1$, whose
contributions are decreasing quicly with $k$ as given by oscillatory
integrals of frequencies proportional to $k$.

By a change of variable, the last integral is also given by: 
\begin{equation*}
\frac{2C^{2}}{\pi }\func{Re}\int_{0}^{1}\exp \left( -C\sqrt{\frac{2}{\eta }}%
\left( \sqrt{1+v^{2}}+iv\right) \left\vert cl_{1}-\left\vert
Z-Z_{1}\right\vert \right\vert \right) \left( \sqrt{1+v^{2}}+\frac{v^{2}}{%
\sqrt{1+v^{2}}}\right) dv
\end{equation*}%
and we are left with the estimation for the first vertex:%
\begin{eqnarray}
&&\left\vert \Psi \left( \theta -l_{1},Z_{1}\right) \right\vert ^{2}\frac{%
\delta \omega ^{-1}\left( J,\theta ,Z\right) }{\delta \left\vert \Psi \left(
\theta -l_{1},Z_{1}\right) \right\vert ^{2}}  \label{vtr} \\
&=&\frac{A_{1}}{A}\sum_{k\neq 0}\frac{\left( cl_{1}+\left\vert
Z-Z_{1}\right\vert \right) }{2\left\vert Z-Z_{1}\right\vert }\int \omega
\left( J,\theta -l_{1},Z_{1}\right) \times \frac{\mathit{T}\left( u\right) }{%
\sqrt{2\eta }C}du  \notag \\
&&\frac{2C^{2}}{\pi }\func{Re}\int_{0}^{1}\exp \left( -C\sqrt{\frac{2}{\eta }%
}\left( \sqrt{1+v^{2}}+iv\right) \left\vert cl_{1}-\left\vert
Z-Z_{1}\right\vert \right\vert \right) \left( \sqrt{1+v^{2}}+\frac{v^{2}}{%
\sqrt{1+v^{2}}}\right) dv  \notag
\end{eqnarray}

\paragraph*{5.2.4.2 Gaussian approximation}

We can estimate the integral $\int_{0}^{1}dv$ in (\ref{vtr}) by integrating
between $0$ and $+\infty $.

\begin{eqnarray*}
&&\frac{2C^{2}}{\pi }\func{Re}\int_{0}^{+\infty }\exp \left( -C\sqrt{\frac{2%
}{\eta }}\left( \sqrt{1+v^{2}}+iv\right) \left\vert cl_{1}-\left\vert
Z-Z_{1}\right\vert \right\vert \right) \left( 2\sqrt{1+v^{2}}-\frac{1}{\sqrt{%
1+v^{2}}}\right) dv \\
&=&\frac{2C^{2}}{\pi }\func{Re}\int_{0}^{+\infty }\exp \left( -C\sqrt{\frac{2%
}{\eta }}\left( \sqrt{1+v^{2}}a+ivb\right) \right) \left( 2\sqrt{1+v^{2}}-%
\frac{1}{\sqrt{1+v^{2}}}\right) dv
\end{eqnarray*}%
with $a=b=\left\vert cl_{1}-\left\vert Z-Z_{1}\right\vert \right\vert $. The
last integral can be rewritten:%
\begin{eqnarray}
&&\left( -\frac{2\partial _{a}}{C\sqrt{\frac{2}{\eta }}}+C\sqrt{\frac{2}{%
\eta }}\int da\right) \frac{2C^{2}}{\pi }\func{Re}\int_{0}^{+\infty }\exp
\left( -C\sqrt{\frac{2}{\eta }}\left( \sqrt{1+v^{2}}a+ivb\right) \right) dv 
\notag \\
&=&\left( -\frac{2\partial _{a}}{C\sqrt{\frac{2}{\eta }}}+C\sqrt{\frac{2}{%
\eta }}\int da\right) \frac{2C^{2}}{\pi b}\func{Re}\int_{0}^{+\infty }\exp
\left( -C\sqrt{\frac{2}{\eta }}\left( \sqrt{b^{2}+v^{2}}\frac{a}{b}%
+iv\right) \right) dv  \notag \\
&\simeq &\left( -\frac{2\partial _{a}}{C\sqrt{\frac{2}{\eta }}}+C\sqrt{\frac{%
2}{\eta }}\int da\right) \frac{2C^{2}}{\pi b}\func{Re}\int_{0}^{+\infty
}\exp \left( -C\sqrt{\frac{2}{\eta }}\left( -\left( \left( \frac{a}{b}%
+i\right) v+\frac{ab}{2v}\right) \right) \right) dv  \label{ldt}
\end{eqnarray}%
for $a\simeq b<<1$. We use that:%
\begin{eqnarray}
&&\left( -\frac{2\partial _{a}}{C\sqrt{\frac{2}{\eta }}}+C\sqrt{\frac{2}{%
\eta }}\int da\right) \frac{2C^{2}}{\pi }\func{Re}\int_{0}^{+\infty }\exp
\left( -C\sqrt{\frac{2}{\eta }}\left( -\left( \left( a+ib\right) v+\frac{a}{%
2v}\right) \right) \right) dv  \notag \\
&=&\left( -\frac{2\partial _{a}}{C\sqrt{\frac{2}{\eta }}}+C\sqrt{\frac{2}{%
\eta }}\int da\right) \frac{2C^{2}}{\pi }\func{Re}\sqrt{\frac{2a}{a+ib}}%
K_{1}\left( 2C\sqrt{\frac{a\left( a+ib\right) }{\eta }}\right)  \label{tdl}
\end{eqnarray}%
where $K_{1}$ is a modified Bessel function, and that the following identity
holds for $K_{1}$: 
\begin{equation*}
\sqrt{\frac{2a}{a+ib}}K_{1}\left( 2C\sqrt{\frac{a\left( a+ib\right) }{\eta }}%
\right) \simeq \sqrt{\frac{2a}{a+ib}}\sqrt{\frac{\pi }{4C\sqrt{\frac{a\left(
a+ib\right) }{\eta }}}}\exp \left( -2C\sqrt{\frac{a\left( a+ib\right) }{\eta 
}}\right)
\end{equation*}%
for $C>>1$. Then computing the integral $\int da$ in (\ref{tdl}) yields:%
\begin{eqnarray*}
&&C\sqrt{\frac{2}{\eta }}\int da\left( \frac{\sqrt{\frac{2a}{a+ib}}\sqrt{%
\frac{\pi }{4C\sqrt{\frac{a\left( a+ib\right) }{\eta }}}}}{-\frac{C}{\eta }%
\frac{2a+ib}{\sqrt{\frac{a}{\eta }\left( a+ib\right) }}}\left( -\frac{C}{%
\eta }\frac{2a+ib}{\sqrt{\frac{a}{\eta }\left( a+ib\right) }}\right) \exp
\left( -2C\sqrt{\frac{a\left( a+ib\right) }{\eta }}\right) \right) \\
&\simeq &C\sqrt{\frac{2}{\eta }}\frac{\sqrt{\frac{2a}{a+ib}}\sqrt{\frac{\pi 
}{4C\sqrt{\frac{a\left( a+ib\right) }{\eta }}}}}{-\frac{C}{\eta }\frac{2a+ib%
}{\sqrt{\frac{a}{\eta }\left( a+ib\right) }}}\exp \left( -2C\sqrt{\frac{%
a\left( a+ib\right) }{\eta }}\right) \\
&=&-2\sqrt{\pi }a\frac{\sqrt{\frac{1}{4C\sqrt{\frac{a}{\eta }\left(
a+ib\right) }}}}{2a+ib}\exp \left( -2C\sqrt{\frac{a\left( a+ib\right) }{\eta 
}}\right)
\end{eqnarray*}%
for $C>>1$. For $a=b$, this identity reduces to:%
\begin{equation}
-\sqrt{\pi }\allowbreak \frac{\frac{2}{5}-\frac{1}{5}i}{\sqrt[4]{\left(
1+i\right) }}\sqrt{\frac{\sqrt{\eta }}{Ca}}\exp \left( -2\frac{Ca}{\sqrt{%
\eta }}\sqrt{1+i}\right)  \label{ct}
\end{equation}%
The derivative arising in (\ref{tdl}) can be estimated by:%
\begin{eqnarray}
&&-\frac{2\partial _{a}}{C\sqrt{\frac{2}{\eta }}}\left( \sqrt{\frac{2a}{a+ib}%
}\sqrt{\frac{\pi }{4C\sqrt{\frac{a\left( a+ib\right) }{\eta }}}}\exp \left(
-2C\sqrt{\frac{a\left( a+ib\right) }{\eta }}\right) \right)  \notag \\
&=&-\frac{1}{8}\sqrt{\left( \frac{1}{2}-\frac{1}{2}i\right) }\frac{\sqrt{\pi 
}}{C^{2}a^{3}\eta }\frac{\exp \left( -2\sqrt{\left( 1+i\right) }C\sqrt{\frac{%
a^{2}}{\eta }}\right) }{\left( \frac{1}{\eta }\right) ^{\frac{3}{2}}\sqrt{%
\frac{1}{C\sqrt{\frac{a^{2}}{\eta }}}}}\allowbreak \left( \left( 1+2i\right)
\left( \left( 1+i\right) \right) ^{\frac{3}{4}}\eta \sqrt{\frac{a^{2}}{\eta }%
}-\left( 12-4i\right) \sqrt[4]{\left( 1+i\right) }Ca^{2}\right)  \notag \\
&\simeq &\frac{1}{8}\sqrt{\left( \frac{1}{2}-\frac{1}{2}i\right) }\frac{%
\sqrt{\pi }}{C^{2}a^{3}\eta }\frac{\exp \left( -2\sqrt{\left( 1+i\right) }C%
\sqrt{\frac{a^{2}}{\eta }}\right) }{\left( \frac{1}{\eta }\right) ^{\frac{3}{%
2}}\sqrt{\frac{1}{C\sqrt{\frac{a^{2}}{\eta }}}}}\allowbreak \left( \left(
12-4i\right) \sqrt[4]{\left( 1+i\right) }Ca^{2}\right)  \notag \\
&=&\frac{\sqrt{\pi }}{8}\sqrt{\left( \frac{1}{2}-\frac{1}{2}i\right) }\frac{%
\exp \left( -2\sqrt{\frac{\left( 1+i\right) }{\eta }}Ca\right) }{\sqrt{\frac{%
Ca}{\sqrt{\eta }}}}\allowbreak \left( \left( 12-4i\right) \sqrt[4]{\left(
1+i\right) }\right)  \label{tc}
\end{eqnarray}%
Gathering (\ref{ct}) and (\ref{tc}), we find that for $C>>1$, For $%
a=b=\left\vert cl_{1}-\left\vert Z-Z_{1}\right\vert \right\vert $, we find
for (\ref{tdl}):

\begin{eqnarray*}
&&C^{2}\frac{\sqrt{130}}{5\sqrt{\pi }}\left( \frac{1}{\sqrt{2}}\right) ^{%
\frac{1}{4}}\frac{\exp \left( -\frac{2^{\frac{9}{8}}\cos \left( \frac{\pi }{8%
}\right) C}{\sqrt{\eta }}\left\vert cl_{1}-\left\vert Z-Z_{1}\right\vert
\right\vert \right) }{\sqrt{\frac{C\left\vert cl_{1}-\left\vert
Z-Z_{1}\right\vert \right\vert }{\sqrt{\eta }}}}\cos \left( \frac{2^{\frac{9%
}{8}}\cos \left( \frac{\pi }{8}\right) C}{\sqrt{\eta }}\left\vert
cl_{1}-\left\vert Z-Z_{1}\right\vert \right\vert \right) \\
&=&C^{2}\frac{\sqrt{65\sqrt{2}}}{5\sqrt{\pi }}\frac{\exp \left( -\frac{2^{%
\frac{3}{8}}\sqrt{\sqrt{2}+1}C}{\sqrt{\eta }}\left\vert cl_{1}-\left\vert
Z-Z_{1}\right\vert \right\vert \right) }{\sqrt{\frac{C\left\vert
cl_{1}-\left\vert Z-Z_{1}\right\vert \right\vert }{\sqrt{\eta }}}}\cos
\left( \frac{2^{\frac{3}{8}}\sqrt{\sqrt{2}+1}C}{\sqrt{\eta }}\left\vert
cl_{1}-\left\vert Z-Z_{1}\right\vert \right\vert \right)
\end{eqnarray*}%
In the sequel, for $\left( A\pi \right) ^{2}\eta <<1$, we approximate: 
\begin{equation*}
C=\sqrt{\ln \left( \frac{2u^{2}}{\left( A\pi \right) ^{2}\eta }\right) }%
\simeq \sqrt{\ln \left( \frac{2}{\left( A\pi \right) ^{2}\eta }\right) }
\end{equation*}%
Finally, the integral over $u$ in (\ref{vtr}) is:%
\begin{equation*}
\int \frac{\mathit{T}\left( u\right) }{\sqrt{2\eta }C}\exp \left( i\frac{u}{2%
}\left( cl_{1}+\left\vert Z-Z_{1}\right\vert \right) \right) du=\frac{1}{%
\sqrt{2\eta }C}\hat{T}\left( \frac{cl_{1}+\left\vert Z-Z_{1}\right\vert }{2}%
\right) du
\end{equation*}%
so that, using that 
\begin{equation*}
\hat{T}_{1}\left( \frac{cl_{1}+\left\vert Z-Z_{1}\right\vert }{2}\right) =%
\frac{A_{1}}{A}\hat{T}\left( \frac{cl_{1}+\left\vert Z-Z_{1}\right\vert }{2}%
\right)
\end{equation*}%
The result for (\ref{vtr}) is: 
\begin{eqnarray*}
&&\frac{\delta \omega ^{-1}\left( J,\theta ,Z\right) }{\delta \left\vert
\Psi \left( \theta -l_{1},Z_{1}\right) \right\vert ^{2}}\simeq \frac{\sqrt{65%
}}{5\sqrt{\pi }2^{\frac{3}{8}}\left( \sqrt{2}+1\right) ^{\frac{1}{4}}}D\frac{%
\exp \left( -D\left\vert cl_{1}-\left\vert Z-Z_{1}\right\vert \right\vert
\right) }{\sqrt{\frac{C\left\vert cl_{1}-\left\vert Z-Z_{1}\right\vert
\right\vert }{\sqrt{\eta }}}}\cos \left( D\left\vert cl_{1}-\left\vert
Z-Z_{1}\right\vert \right\vert \right) \\
&&\times \frac{\left( cl_{1}+\left\vert Z-Z_{1}\right\vert \right) }{%
2\left\vert Z-Z_{1}\right\vert }\hat{T}_{1}\left( \frac{cl_{1}+\left\vert
Z-Z_{1}\right\vert }{2}\right)
\end{eqnarray*}%
where:%
\begin{equation*}
D=\frac{2^{\frac{3}{8}}\sqrt{\sqrt{2}+1}C}{\sqrt{\eta }}
\end{equation*}%
We also write this result in a more compact form: 
\begin{equation}
\frac{\delta \omega ^{-1}\left( J,\theta ,Z\right) }{\delta \left\vert \Psi
\left( \theta -l_{1},Z_{1}\right) \right\vert ^{2}}=\frac{A_{1}}{A}\Xi
\left( \left\vert Z_{1}-Z\right\vert ,l_{1},\bar{\omega}\right) \omega
\left( J,\theta -l_{1},Z_{1}\right)  \label{drv}
\end{equation}%
for $\frac{\left\vert Z_{1}-Z\right\vert }{cl_{1}}<1$, and $0$ otherwise,
with:%
\begin{eqnarray}
\Xi \left( \left\vert Z_{1}-Z\right\vert ,l_{1},\bar{\omega}\right) &=&\frac{%
\sqrt{65}}{5\sqrt{\pi }2^{\frac{3}{8}}\left( \sqrt{2}+1\right) ^{\frac{1}{4}}%
}\frac{D\exp \left( -D\left\vert cl_{1}-\left\vert Z-Z_{1}\right\vert
\right\vert \right) }{\sqrt{D\left\vert cl_{1}-\left\vert Z-Z_{1}\right\vert
\right\vert }}\cos \left( D\left\vert cl_{1}-\left\vert Z-Z_{1}\right\vert
\right\vert \right)  \label{thtt} \\
&&\times \frac{\left( cl_{1}+\left\vert Z-Z_{1}\right\vert \right) }{%
2\left\vert Z-Z_{1}\right\vert }\hat{T}\left( \frac{cl_{1}+\left\vert
Z-Z_{1}\right\vert }{2}\right)  \notag
\end{eqnarray}%
Similarly:%
\begin{equation}
\frac{\delta \omega \left( J,\theta ,Z\right) }{\delta \left\vert \Psi
\left( \theta -l_{1},Z_{1}\right) \right\vert ^{2}}=\Xi \left( \left\vert
Z_{1}-Z\right\vert ,l_{1},\bar{\omega}\right) \omega \left( J,\theta
-l_{1},Z_{1}\right)  \label{drvbs}
\end{equation}%
The appearance of the $\cos \left( D\left\vert cl_{1}-\left\vert
Z-Z_{1}\right\vert \right\vert \right) $ in (\ref{thtt}) is a consequence of
our approximation computing the integral between $0$ and $+\infty $. This
approximation breaks down when the $\cos $ function becomes negative. As a
consequence, for $D\left\vert cl_{1}-\left\vert Z-Z_{1}\right\vert
\right\vert >\frac{\pi }{2}$, we can set $\Xi \left( \left\vert
Z_{1}-Z\right\vert ,l_{1},\bar{\omega}\right) \simeq 0$.

As stated in the beginning of this paragraph, formula (\ref{thtt}) is
similar to the case of an exponential transfer function.

\section*{Appendix 6 Non local expansion for $\protect\omega \left( \protect%
\theta ,Z\right) $}

\subsection*{6.1 n-th derivatives of at $\left\vert \Psi \right\vert ^{2}=0$}

\subsubsection*{6.1.1 General formula}

Based on the results of Appendix 5, we can now compute $\omega \left(
J,\theta ,Z\right) $, $\omega ^{-1}\left( J,\theta ,Z\right) $ and their
derivatives $\frac{\delta ^{n}\omega \left( J,\theta ,Z\right) }{%
\dprod\limits_{i=1}^{n}\delta \left\vert \Psi \left( \theta
-l_{i},Z_{i}\right) \right\vert ^{2}}$ and $\frac{\delta ^{n}\omega
^{-1}\left( J,\theta ,Z\right) }{\dprod\limits_{i=1}^{n}\delta \left\vert
\Psi \left( \theta -l_{i},Z_{i}\right) \right\vert ^{2}}$. It allows to
compute the expansion of the effective action, and also to study the
solutions of (\ref{fft}) without the locality assumption.

\paragraph*{6.1.1.1 Series expansion for the first order derivative of $%
\protect\omega \left( \protect\theta ,Z\right) $}

Recall that $\omega \left( \theta ,Z\right) $ is solution of\ (\ref{fqt}):

\begin{eqnarray}
\omega \left( \theta ,Z\right) &=&F\left( J\left( \theta \right) +\frac{%
\kappa }{N}\int T\left( Z,Z_{1}\right) \frac{\omega \left( \theta -\frac{%
\left\vert Z-Z_{1}\right\vert }{c},Z_{1}\right) }{\omega \left( \theta
,Z\right) }\right.  \label{qtf} \\
&&\times \left. W\left( \frac{\omega \left( \theta ,Z\right) }{\omega \left(
\theta -\frac{\left\vert Z-Z_{1}\right\vert }{c},Z_{1}\right) }\right)
\left( \mathcal{\bar{G}}_{0}\left( 0,Z_{1}\right) +\left\vert \Psi \left(
\theta -\frac{\left\vert Z-Z_{1}\right\vert }{c},Z_{1}\right) \right\vert
^{2}\right) dZ_{1}\right)  \notag
\end{eqnarray}%
\begin{equation*}
\mathcal{\bar{G}}_{0}\left( 0,Z_{1}\right) =\mathcal{G}_{0}\left(
0,Z_{1}\right) +X_{0}
\end{equation*}%
and $\omega ^{-1}\left( \theta ,Z\right) $ is solution of:%
\begin{eqnarray*}
\omega ^{-1}\left( \theta ,Z\right) &=&G\left( J\left( \theta \right) +\frac{%
\kappa }{N}\int T\left( Z,Z_{1}\right) \frac{\omega \left( \theta -\frac{%
\left\vert Z-Z_{1}\right\vert }{c},Z_{1}\right) }{\omega \left( \theta
,Z\right) }\right. \\
&&\times \left. W\left( \frac{\omega \left( \theta ,Z\right) }{\omega \left(
\theta -\frac{\left\vert Z-Z_{1}\right\vert }{c},Z_{1}\right) }\right)
\left( \mathcal{\bar{G}}_{0}\left( 0,Z_{1}\right) +\left\vert \Psi \left(
\theta -\frac{\left\vert Z-Z_{1}\right\vert }{c},Z_{1}\right) \right\vert
^{2}\right) dZ_{1}\right)
\end{eqnarray*}%
To find the internal dynamics of the system we will consider $J\left( \theta
\right) =J$, a constant external current, usually $J=0$. We use a series
expansion in $\left\vert \Psi \left( \theta ^{\left( j\right) },Z_{1}\right)
\right\vert ^{2}$ of the right hand side of (\ref{qtf}) and write:%
\begin{eqnarray}
\omega \left( \theta ^{\left( i\right) },Z\right) &=&\omega \left( \theta
^{\left( i\right) },Z\right) _{\left\vert \Psi \right\vert ^{2}=0}
\label{snp} \\
&&+\int \sum_{n=1}^{\infty }\left( \frac{\delta ^{n}\omega \left( J,\theta
,Z\right) }{\dprod\limits_{i=1}^{n}\delta \left\vert \Psi \left( \theta
-l_{i},Z_{i}\right) \right\vert ^{2}}\right) _{\left\vert \Psi \right\vert
^{2}=0}\dprod\limits_{i=1}^{n}\left\vert \Psi \left( \theta
-l_{i},Z_{i}\right) \right\vert ^{2}  \notag
\end{eqnarray}%
The first term (\ref{snp}), i.e. $\omega \left( \theta ^{\left( i\right)
},Z\right) _{\left\vert \Psi \right\vert ^{2}=0}$, is a solution of:%
\begin{equation*}
F\left( J+\frac{\kappa }{N}\int T\left( Z,Z_{1}\right) \frac{\omega \left(
\theta -\frac{\left\vert Z-Z_{1}\right\vert }{c},Z_{1}\right) }{\omega
\left( \theta ,Z\right) }W\left( \frac{\omega \left( \theta ,Z\right) }{%
\omega \left( \theta -\frac{\left\vert Z-Z_{1}\right\vert }{c},Z_{1}\right) }%
\right) \left( \mathcal{\bar{G}}_{0}\left( 0,Z_{1}\right) \right)
dZ_{1}\right)
\end{equation*}%
One solution is the static frequency (\ref{frs}) solution of:%
\begin{eqnarray*}
\omega \left( J,Z\right) &=&F\left( J+\frac{\kappa }{N}\int T\left(
Z,Z_{1}\right) \frac{\omega \left( Z_{1}\right) }{\omega \left( Z\right) }%
W\left( \frac{\omega \left( Z\right) }{\omega \left( Z_{1}\right) }\right) 
\mathcal{\bar{G}}_{0}\left( 0,Z_{i}\right) dZ_{1}\right) \\
&\equiv &F\left[ J,\omega ,Z\right]
\end{eqnarray*}%
but any time dependent solution for $\left\vert \Psi \right\vert ^{2}=0$ is
also possible. This arises for non constant current $J\left( \theta \right) $%
. Equation (\ref{snp}) is the expansion of $\omega \left( \theta ^{\left(
i\right) },Z\right) $ around this solution, the dynamics depending on $%
\left\vert \Psi \left( \theta ^{\left( j\right) },Z_{1}\right) \right\vert
^{2}$. We set:%
\begin{equation*}
\omega \left( \theta ^{\left( i\right) },Z\right) _{\left\vert \Psi
\right\vert ^{2}=0}=\omega _{0}\left( J,Z\right)
\end{equation*}%
The first derivative $\frac{\delta \omega \left( J,\theta ,Z\right) }{\delta
\left\vert \Psi \left( \theta -l_{1},Z_{1}\right) \right\vert ^{2}}$ \ in (%
\ref{snp})\ has been computed in Appendix 5. It is given by:%
\begin{eqnarray}
\frac{\delta \omega \left( J,\theta ,Z\right) }{\delta \left\vert \Psi
\left( \theta -l_{1},Z_{1}\right) \right\vert ^{2}} &=&\sum_{n=1}^{\infty
}\int \frac{\omega \left( J,\theta -\sum_{l=1}^{n}\frac{\left\vert Z^{\left(
l-1\right) }-Z^{\left( l\right) }\right\vert }{c},Z_{1}\right) }{\left( 
\mathcal{\bar{G}}_{0}\left( 0,Z_{1}\right) +\left\vert \Psi \left( \theta
-l_{1},Z_{1}\right) \right\vert ^{2}\right) }  \label{rdv} \\
&&\times \dprod\limits_{l=1}^{n}\hat{T}\left( \theta -\sum_{j=1}^{l-1}\frac{%
\left\vert Z^{\left( j-1\right) }-Z^{\left( j\right) }\right\vert }{c}%
,Z^{\left( l-1\right) },Z^{\left( l\right) },\omega ,\Psi \right) \delta
\left( l_{1}-\sum_{l=1}^{n}\frac{\left\vert Z^{\left( l-1\right) }-Z^{\left(
l\right) }\right\vert }{c}\right) \dprod\limits_{l=1}^{n-1}dZ^{\left(
l\right) }  \notag
\end{eqnarray}%
where:%
\begin{eqnarray}
&&\hat{T}\left( \theta ,Z,Z_{1}\omega ,\Psi \right)  \label{vdR} \\
&=&\frac{\frac{\kappa }{N}\omega \left( J,\theta ,Z\right) T\left(
Z,Z_{1}\right) \bar{W}^{\prime }\left( \frac{\omega \left( J,\theta -\frac{%
\left\vert Z-Z_{1}\right\vert }{c},Z_{1}\right) }{\omega \left( J,\theta
,Z\right) }\right) F^{\prime }\left[ J,\omega ,\theta ,Z,\Psi \right] \left( 
\mathcal{\bar{G}}_{0}\left( 0,Z_{1}\right) +\left\vert \Psi \left( \theta -%
\frac{\left\vert Z-Z_{1}\right\vert }{c},Z_{1}\right) \right\vert
^{2}\right) }{\omega ^{2}\left( J,\theta ,Z\right) +\left( \int \frac{\kappa 
}{N}\omega \left( J,\theta -\frac{\left\vert Z-Z^{\prime }\right\vert }{c}%
,Z^{\prime }\right) \bar{W}^{\prime }\left( \frac{\omega \left( J,\theta -%
\frac{\left\vert Z-Z^{\prime }\right\vert }{c},Z^{\prime }\right) }{\omega
\left( J,\theta ,Z\right) }\right) \left( \mathcal{\bar{G}}_{0}\left(
0,Z^{\prime }\right) +\left\vert \Psi \left( \theta -\frac{\left\vert
Z-Z^{\prime }\right\vert }{c},Z^{\prime }\right) \right\vert ^{2}\right)
T\left( Z,Z^{\prime }\right) dZ^{\prime }\right) F^{\prime }\left[ J,\omega
,\theta ,Z,\Psi \right] }  \notag
\end{eqnarray}%
with the convention that $Z^{\left( 0\right) }=Z$ and $Z^{\left( n\right)
}=Z_{1}$. The derivative (\ref{rdv}) was then evaluated in Appendix 5 using
combinations of $K_{1}$ functions, but for the purpose of the computation of
the successive derivatives of $\omega \left( J,\theta ,Z\right) $, we will
work, temporarily, with the general formula (\ref{rdv}). Equation (\ref{rdv}%
) yield recursively $\frac{\delta \omega \left( J,\theta ,Z\right) }{\delta
\left\vert \Psi \left( \theta -l_{1},Z_{1}\right) \right\vert ^{2}}$ in
terms of past frequencies. Applied to the case $\left\vert \Psi \right\vert
^{2}=0$, the factor (\ref{vdR}) simplifies:%
\begin{eqnarray}
\hat{T}\left( \theta ,Z,Z_{1},\omega _{0}\right) &\equiv &\hat{T}\left(
\theta ,Z,Z_{1}\omega _{0},0\right)  \label{rvd} \\
&=&\frac{\frac{\kappa }{N}\omega _{0}\left( J,\theta ,Z\right) \bar{W}%
^{\prime }\left( \frac{\omega _{0}\left( Z\right) }{\omega _{0}\left(
Z_{1}\right) }\right) T\left( Z,Z_{1}\right) F^{\prime }\left[ J,\omega
,\theta ,Z,\Psi \right] \mathcal{\bar{G}}_{0}\left( 0,Z_{1}\right) }{\omega
_{0}^{2}\left( J,\theta ,Z\right) +\left( \int \frac{\kappa }{N}\omega
_{0}\left( J,Z^{\prime }\right) \bar{W}^{\prime }\left( \frac{\omega
_{0}\left( Z\right) }{\omega _{0}\left( Z_{1}\right) }\right) \left( 
\mathcal{\bar{G}}_{0}\left( 0,Z^{\prime }\right) \right) T\left( Z,Z^{\prime
}\right) dZ^{\prime }\right) F^{\prime }\left[ J,\omega ,\theta ,Z,\Psi %
\right] }  \notag
\end{eqnarray}%
or in first aproximation:%
\begin{eqnarray}
\hat{T}\left( \theta ,Z,Z_{1},\omega _{0},\Psi \right) &\equiv &\hat{T}%
\left( Z,Z_{1},\omega _{0}\right)  \label{rvD} \\
&\simeq &\frac{\frac{\kappa }{N}T\left( Z,Z_{1}\right) F^{\prime }\left[
J,\omega _{0},\theta ,Z\right] \mathcal{\bar{G}}_{0}\left( 0,Z_{1}\right) }{%
\omega _{0}\left( J,Z\right) }  \notag
\end{eqnarray}%
and (\ref{rdv}) becomes:%
\begin{eqnarray}
\left( \frac{\delta \omega \left( J,\theta ,Z\right) }{\delta \left\vert
\Psi \left( \theta -l_{1},Z_{1}\right) \right\vert ^{2}}\right) _{\left\vert
\Psi \right\vert ^{2}=0} &=&\sum_{n=1}^{\infty }\int \frac{\omega _{0}\left(
J,\theta -\sum_{l=1}^{n}\frac{\left\vert Z^{\left( l-1\right) }-Z^{\left(
l\right) }\right\vert }{c},Z_{1}\right) }{\mathcal{\bar{G}}_{0}\left(
0,Z_{1}\right) }  \label{zdv} \\
&&\times \dprod\limits_{l=1}^{n}\hat{T}\left( \theta -\sum_{j=1}^{l-1}\frac{%
\left\vert Z^{\left( j-1\right) }-Z^{\left( j\right) }\right\vert }{c}%
,Z^{\left( l-1\right) },Z^{\left( l\right) },\omega _{0},0\right)  \notag \\
&&\times \delta \left( l_{1}-\sum_{l=1}^{n}\frac{\left\vert Z^{\left(
l-1\right) }-Z^{\left( l\right) }\right\vert }{c}\right)
\dprod\limits_{l=1}^{n-1}dZ^{\left( l\right) }  \notag
\end{eqnarray}

\paragraph*{6.1.1.2 Graphical representation of the successive derivatives}

The $n$-th term in (\ref{zdv}) can be understood graphically as a sum over
the set of broken paths with $n$ segments, each path linking $Z^{\left(
l-1\right) }$ and $Z^{\left( l\right) }$ during a timespan of $\frac{%
\left\vert Z^{\left( l-1\right) }-Z^{\left( l\right) }\right\vert }{c}$. To
each point of the segment, we associate the factor: 
\begin{equation}
\hat{T}\left( \theta -\sum_{j=1}^{l-1}\frac{\left\vert Z^{\left( j-1\right)
}-Z^{\left( j\right) }\right\vert }{c},Z^{\left( l-1\right) },Z^{\left(
l\right) },\omega _{0},\Psi \right) \simeq \frac{\frac{\kappa }{N}T\left(
Z^{\left( l-1\right) },Z^{\left( l\right) }\right) F^{\prime }\left[
J,\omega _{0},\theta -\sum_{j=1}^{l-1}\frac{\left\vert Z^{\left( j-1\right)
}-Z^{\left( j\right) }\right\vert }{c},Z^{\left( l-1\right) }\right] 
\mathcal{\bar{G}}_{0}\left( 0,Z^{\left( l\right) }\right) }{\omega
_{0}\left( J,\theta -\sum_{j=1}^{l-1}\frac{\left\vert Z^{\left( j-1\right)
}-Z^{\left( j\right) }\right\vert }{c},Z^{\left( l-1\right) }\right) }
\label{tf}
\end{equation}%
Ultimately, the product of factor is multiplied by the frequency at the last
point:%
\begin{equation}
\frac{\omega _{0}\left( J,\theta -\sum_{l=1}^{n}\frac{\left\vert Z^{\left(
l-1\right) }-Z^{\left( l\right) }\right\vert }{c},Z_{1}\right) }{\mathcal{%
\bar{G}}_{0}\left( 0,Z_{1}\right) }  \label{cf}
\end{equation}%
and by $\left\vert \Psi \left( \theta -l_{1},Z_{1}\right) \right\vert ^{2}$.
The integrals over the points $Z^{\left( l\right) }$ and the sum over $n$,
the length of the broken paths, yield the first order contribution to the
expansion (\ref{snp}).

The next terms in the expansion of (\ref{snp}) are the derivatives $\left( 
\frac{\delta ^{n}\omega \left( J,\theta ,Z\right) }{\dprod\limits_{i=1}^{n}%
\delta \left\vert \Psi \left( \theta -l_{i},Z_{i}\right) \right\vert ^{2}}%
\right) _{\left\vert \Psi \right\vert ^{2}=0}$ which are obtained by
successive derivations of (\ref{rdv}) and (\ref{vdR}) by $\left\vert \Psi
\left( \theta -l_{2},Z_{2}\right) \right\vert ^{2}$ and evaluated at $%
\left\vert \Psi \right\vert ^{2}=0$. The $l_{i}$ are ordered such that $%
l_{1}<...<l_{n}$. These derivatives are obtained by differentiating either:%
\begin{equation*}
\omega \left( J,\theta -\sum_{l=1}^{n}\frac{\left\vert Z^{\left( l-1\right)
}-Z^{\left( l\right) }\right\vert }{c},Z_{n}\right)
\end{equation*}%
or the successive factors:%
\begin{equation*}
\dprod\limits_{l=1}^{n}\hat{T}\left( \theta -\sum_{j=1}^{l-1}\frac{%
\left\vert Z^{\left( j-1\right) }-Z^{\left( j\right) }\right\vert }{c}%
,Z^{\left( l-1\right) },Z^{\left( l\right) },\omega ,\Psi \right)
\end{equation*}%
The first possibility amounts to write $\frac{\delta \omega \left( J,\theta
-l_{1},Z_{1}\right) }{\delta \left\vert \Psi \left( \theta
-l_{2},Z_{2}\right) \right\vert ^{2}}$ using (\ref{rdv}). Graphically it
amounts to write broken lines from $Z_{1}$ to $Z_{2}$ and associate to each
broken line the factor (\ref{tf}), (\ref{cf}) and $\left\vert \Psi \left(
\theta -l_{2},Z_{2}\right) \right\vert ^{2}$.

The second possibility is obtained by computing for each $l$:%
\begin{equation}
\frac{\delta \hat{T}\left( \theta -\sum_{j=1}^{l-1}\frac{\left\vert
Z^{\left( j-1\right) }-Z^{\left( j\right) }\right\vert }{c},Z^{\left(
l-1\right) },Z^{\left( l\right) },\omega ,\Psi \right) }{\delta \left\vert
\Psi \left( \theta -l_{2},Z_{2}\right) \right\vert ^{2}}  \label{rds}
\end{equation}%
Which can be written as:%
\begin{eqnarray*}
&&\frac{\delta \hat{T}\left( \theta -\sum_{j=1}^{l-1}\frac{\left\vert
Z^{\left( j-1\right) }-Z^{\left( j\right) }\right\vert }{c},Z^{\left(
l-1\right) },Z^{\left( l\right) },\omega ,\Psi \right) }{\delta \left\vert
\Psi \left( \theta -l_{2},Z_{2}\right) \right\vert ^{2}} \\
&=&\int d\Delta dZ^{\prime }\frac{\delta \hat{T}\left( \theta
-\sum_{j=1}^{l-1}\frac{\left\vert Z^{\left( j-1\right) }-Z^{\left( j\right)
}\right\vert }{c},Z^{\left( l-1\right) },Z^{\left( l\right) },\omega ,\Psi
\right) }{\delta \omega \left( J,\theta -\sum_{j=1}^{l-1}\frac{\left\vert
Z^{\left( j-1\right) }-Z^{\left( j\right) }\right\vert }{c}-\Delta
,Z^{\prime }\right) }\frac{\delta \omega \left( J,\theta -\sum_{j=1}^{l-1}%
\frac{\left\vert Z^{\left( j-1\right) }-Z^{\left( j\right) }\right\vert }{c}%
-\Delta ,Z^{\prime }\right) }{\delta \left\vert \Psi \left( \theta
-l_{2},Z_{2}\right) \right\vert ^{2}}
\end{eqnarray*}%
This derivative can be described graphically by assigning to some point $%
Z^{\left( l\right) }$ of the initial line the factor:%
\begin{equation*}
\frac{\delta \hat{T}\left( \theta -\sum_{j=1}^{l-1}\frac{\left\vert
Z^{\left( j-1\right) }-Z^{\left( j\right) }\right\vert }{c},Z^{\left(
l-1\right) },Z^{\left( l\right) },\omega ,\Psi \right) }{\delta \omega
\left( J,\theta -\sum_{j=1}^{l-1}\frac{\left\vert Z^{\left( j-1\right)
}-Z^{\left( j\right) }\right\vert }{c}-\Delta ,Z^{\prime }\right) }
\end{equation*}%
issuing a new succession of segments representing $\frac{\delta \omega
\left( J,\theta -\sum_{j=1}^{l-1}\frac{\left\vert Z^{\left( j-1\right)
}-Z^{\left( j\right) }\right\vert }{c}-\Delta ,Z^{\prime }\right) }{\delta
\left\vert \Psi \left( \theta -l_{2},Z_{2}\right) \right\vert ^{2}}$ and
then summing over $\Delta $ and $Z^{\prime }$. In first approximation, we
can set $\Delta =0$ and $Z^{\prime }$, so that the factor is: 
\begin{equation*}
\left( \frac{\delta \hat{T}\left( \theta -\sum_{j=1}^{l-1}\frac{\left\vert
Z^{\left( j-1\right) }-Z^{\left( j\right) }\right\vert }{c},Z^{\left(
l-1\right) },Z^{\left( l\right) },\omega ,\Psi \right) }{\delta \omega
\left( J,\theta -\sum_{j=1}^{l-1}\frac{\left\vert Z^{\left( j-1\right)
}-Z^{\left( j\right) }\right\vert }{c},Z^{\left( l-1\right) }\right) }%
\right) _{\left\vert \Psi \right\vert ^{2}=0}
\end{equation*}%
and the new succession of segments represents $\frac{\delta \omega \left(
J,\theta -\sum_{j=1}^{l-1}\frac{\left\vert Z^{\left( j-1\right) }-Z^{\left(
j\right) }\right\vert }{c},Z^{\left( l-1\right) }\right) }{\delta \left\vert
\Psi \left( \theta -l_{2},Z_{2}\right) \right\vert ^{2}}$.

More generally, differentiating successively $\hat{T}\left( \theta
,Z,Z_{1}\omega ,\left\vert \Psi \right\vert ^{2}\right) $, corresponds to
insert the vertices: 
\begin{equation*}
\frac{\delta ^{k}\hat{T}\left( \theta -\sum_{j=1}^{l-1}\frac{\left\vert
Z^{\left( j-1\right) }-Z^{\left( j\right) }\right\vert }{c},Z^{\left(
l-1\right) },Z^{\left( l\right) },\omega ,\Psi \right) }{\dprod%
\limits_{i=1}^{k}\delta \omega \left( J,\theta -\sum_{j=1}^{l-1}\frac{%
\left\vert Z^{\left( j-1\right) }-Z^{\left( j\right) }\right\vert }{c}%
-\Delta _{l},Z_{l}\right) }\simeq \frac{\delta ^{k}\left( \frac{\frac{\kappa 
}{N}T\left( Z,Z^{\left( l\right) }\right) F^{\prime }\left[ J,\theta ,\omega
_{0},Z^{\left( l\right) }\right] \mathcal{\bar{G}}_{0}\left( 0,Z^{\left(
l\right) }\right) }{\omega _{0}\left( J,\theta ,Z^{\left( l\right) }\right) }%
\right) }{\delta ^{k}\omega _{0}\left( J,\theta ,Z^{\left( l\right) }\right) 
}
\end{equation*}%
with $k$ new segments representing $\frac{\delta \omega \left( J,\theta
-\sum_{j=1}^{l-1}\frac{\left\vert Z^{\left( j-1\right) }-Z^{\left( j\right)
}\right\vert }{c}-\Delta _{l},Z_{l}\right) }{\delta \left\vert \Psi \left(
\theta -l_{l},Z_{l}\right) \right\vert ^{2}}$.

Gathering the two possibilities forementionned and iterating this procedures
yields a graphical representation for:%
\begin{equation}
\left( \frac{\delta ^{n}\omega \left( J,\theta ,Z\right) }{%
\dprod\limits_{i=1}^{n}\delta \left\vert \Psi \left( \theta
-l_{i},Z_{i}\right) \right\vert ^{2}}\right) _{\left\vert \Psi \right\vert
^{2}=0}\dprod\limits_{i=1}^{n}\left\vert \Psi \left( \theta
-l_{i},Z_{i}\right) \right\vert ^{2}  \label{sl}
\end{equation}%
We associate the squared field $\left\vert \Psi \left( \theta
-l_{i},Z_{i}\right) \right\vert ^{2}$ to each point $Z_{i}$ . For $m=1,...,n$%
, we draw $m$ lines. At least one of them is starting from $Z$. These lines
are composed of an arbitrary number of segments and all the points $Z_{i}$
are crossed by one line. Each line ends at a point $Z_{i}$. The starting
points of the lines have to branch either at $Z$, either at some point of an
other line. There are $m$ branching points of valence $k$ including the
starting point at $Z$ Apart from $Z$ the branching points have valence $%
3,...,n-1$. To each line $i$ of length $L_{i}$, we associate the factor:%
\begin{eqnarray}
F\left( line_{i}\right) &=&\dprod\limits_{l=1}^{L_{i}}\frac{\frac{\kappa }{N}%
T\left( Z^{\left( l-1\right) },Z^{\left( l\right) }\right) F^{\prime }\left[
J,\omega _{0},\theta -\sum_{j=1}^{l-1}\frac{\left\vert Z^{\left( j-1\right)
}-Z^{\left( j\right) }\right\vert }{c},Z^{\left( l-1\right) }\right] 
\mathcal{\bar{G}}_{0}\left( 0,Z^{\left( l\right) }\right) }{\omega
_{0}\left( J,\theta -\sum_{j=1}^{l-1}\frac{\left\vert Z^{\left( j-1\right)
}-Z^{\left( j\right) }\right\vert }{c},Z^{\left( l-1\right) }\right) }
\label{lf} \\
&&\times \frac{\omega _{0}\left( J,\theta -\sum_{l=1}^{L_{i}}\frac{%
\left\vert Z^{\left( l-1\right) }-Z^{\left( l\right) }\right\vert }{c}%
,Z_{i}\right) }{\mathcal{\bar{G}}_{0}\left( 0,Z_{i}\right) }  \notag \\
&=&\dprod\limits_{l=1}^{L_{i}}\hat{T}\left( \theta -\sum_{j=1}^{l-1}\frac{%
\left\vert Z^{\left( j-1\right) }-Z^{\left( j\right) }\right\vert }{c}%
,Z^{\left( l-1\right) },Z^{\left( l\right) },\omega _{0},\Psi \right) \frac{%
\omega _{0}\left( J,\theta -\sum_{l=1}^{L_{i}}\frac{\left\vert Z^{\left(
l-1\right) }-Z^{\left( l\right) }\right\vert }{c},Z_{i}\right) }{\mathcal{%
\bar{G}}_{0}\left( 0,Z_{i}\right) }  \notag
\end{eqnarray}%
and to each branching point $\left( X,\theta \right) =B$ of valence $k+2$
arising in the expansion, we associate the factor:%
\begin{equation}
F\left( \left( X,\theta \right) \right) =\frac{\delta ^{k}\left( \frac{\frac{%
\kappa }{N}T\left( Z,Z^{\left( l\right) }\right) F^{\prime }\left[ J,\theta
,\omega _{0},Z^{\left( l\right) }\right] \mathcal{\bar{G}}_{0}\left(
0,Z^{\left( l\right) }\right) }{\omega _{0}\left( J,\theta ,Z^{\left(
l\right) }\right) }\right) }{\delta ^{k}\omega _{0}\left( J,\theta
,Z^{\left( l\right) }\right) }  \label{rc}
\end{equation}%
and (\ref{sl}) writes:%
\begin{eqnarray}
&&\left( \frac{\delta ^{n}\omega \left( J,\theta ,Z\right) }{%
\dprod\limits_{i=1}^{n}\delta \left\vert \Psi \left( \theta
-l_{i},Z_{i}\right) \right\vert ^{2}}\right) _{\left\vert \Psi \right\vert
^{2}=0}\dprod\limits_{i=1}^{n}\left\vert \Psi \left( \theta
-l_{i},Z_{i}\right) \right\vert ^{2}  \notag \\
&=&\left( \sum_{m=1}^{n}\sum_{i=1}^{m}\sum_{\left(
line_{1},...,line_{m}\right) }\dprod\limits_{i}F\left( line_{i}\right)
\dprod\limits_{B}F\left( B\right) \right) \dprod\limits_{i=1}^{n}\left\vert
\Psi \left( \theta -l_{i},Z_{i}\right) \right\vert ^{2}  \label{rdt}
\end{eqnarray}%
The integral over the branch points is implicit. The factor $F\left(
B\right) $ for a branch point $B$ is defined in (\ref{rc}) The graphical
representation is generic. While integrating over the set of lines, the
degenerate case of lines that share some segments is taken into account.

\subsubsection*{6.1.2 Approximate expression}

The results of the section 5 can then be used with (\ref{rdt}) to compute:%
\begin{equation*}
\frac{\delta ^{n}\omega \left( J,\theta ,Z\right) }{\dprod\limits_{i=1}^{n}%
\delta \left\vert \Psi \left( \theta -l_{i},Z_{i}\right) \right\vert ^{2}}
\end{equation*}%
in the approximation of the dominant contribution. To each line from a
branching point $\theta -l_{j}^{\prime },Z_{j}^{\prime }$ to $\theta
-l_{i},Z_{i}$ (the branching point can be one of the $\theta -l_{i},Z_{i}$)
we associate a factor of the type, as in (\ref{rrt}): 
\begin{equation*}
\frac{\exp \left( -c\left( l_{i}-l_{j}^{\prime }\right) -\gamma \left(
c\left( l_{i}-l_{j}^{\prime }\right) -\left\vert Z_{j}^{\prime
}-Z_{i}\right\vert \right) \right) }{B}H\left( cl_{1}-\left\vert
Z-Z_{1}\right\vert \right)
\end{equation*}%
The dominant contribution is obtained when the set $\left\{ l_{j}^{\prime
},Z_{j}^{\prime }\right\} $ is equal to $\left\{ l_{j},Z_{j}\right\} $ and
the product over the branching points yields a contribution whose form is:%
\begin{eqnarray}
\frac{\delta ^{n}\omega \left( J,\theta ,Z\right) }{\dprod\limits_{i=1}^{n}%
\delta \left\vert \Psi \left( \theta -l_{i},Z_{i}\right) \right\vert ^{2}}
&\simeq &\frac{\exp \left( -cl_{n}-\gamma \left( \sum_{i=1}^{n-1}\left(
\left( c\left( l_{i}-l_{i+1}\right) \right) ^{2}-\left\vert
Z_{i}-Z_{i+1}\right\vert ^{2}\right) \right) \right) }{B^{n}}  \label{drnv}
\\
&&\times H\left( cl_{n}-\sum_{i=1}^{n-1}\left\vert Z_{i}-Z_{i+1}\right\vert
\right) \dprod\limits_{i=1}^{n}\frac{\omega _{0}\left( J,\theta
-l_{i},Z_{i}\right) }{\mathcal{\bar{G}}_{0}\left( 0,Z_{i}\right) }  \notag
\end{eqnarray}%
with $Z_{1}=Z$ and $l_{n}>...>l_{1}$ and $B$ a constant coefficient (see (%
\ref{trr})).

Formula (\ref{dvtr}) shows that the previous computations are also valid for
the derivatives of $\omega ^{-1}\left( J,\theta ,Z\right) $. We thus obtain
the generalization of (\ref{trr}): 
\begin{eqnarray}
\frac{\delta ^{n}\omega ^{-1}\left( J,\theta ,Z\right) }{\dprod%
\limits_{i=1}^{n}\delta \left\vert \Psi \left( \theta -l_{i},Z_{i}\right)
\right\vert ^{2}} &\simeq &\frac{\exp \left( -cl_{n}-\alpha \left(
\sum_{i=1}^{n-1}\left( \left( c\left( l_{i}-l_{i+1}\right) \right)
^{2}-\left\vert Z_{i}-Z_{i+1}\right\vert ^{2}\right) \right) \right) }{D^{n}}
\label{dnv} \\
&&\times H\left( cl_{n}-\sum_{i=1}^{n-1}\left\vert Z_{i}-Z_{i+1}\right\vert
\right) \dprod\limits_{i=1}^{n}\frac{\omega _{0}^{-1}\left( J,\theta
-l_{i},Z_{i}\right) }{\mathcal{\bar{G}}_{0}\left( 0,Z_{i}\right) }  \notag
\end{eqnarray}%
The only difference is the appearance of different coefficients $\alpha $
and $D$\ in the expression.

\subsection*{6.2 Equation for $\protect\omega \left( \protect\theta %
,Z\right) $}

\subsubsection*{6.2.1 Reordering the graphical sum (\protect\ref{rdt})}

We now sum the series expansion (\ref{snp}):%
\begin{eqnarray}
\omega \left( \theta ^{\left( i\right) },Z\right) &=&\omega \left( \theta
^{\left( i\right) },Z\right) _{\left\vert \Psi \right\vert ^{2}=0} \\
&&+\int \sum_{n=1}^{\infty }\left( \frac{\delta ^{n}\omega \left( J,\theta
,Z\right) }{\dprod\limits_{i=1}^{n}\delta \left\vert \Psi \left( \theta
-l_{i},Z_{i}\right) \right\vert ^{2}}\right) _{\left\vert \Psi \right\vert
^{2}=0}\dprod\limits_{i=1}^{n}\left\vert \Psi \left( \theta
-l_{i},Z_{i}\right) \right\vert ^{2}  \notag
\end{eqnarray}
by reordering the sums in the RHS of (\ref{rdt}).

To do so, we first compute the sum over the lines between $\left( Z,\theta
\right) $ and $\left( Z_{1},\theta _{1}\right) $ and of given length $%
L_{i}=n $ of the product of factors $\hat{T}\left( \theta -\sum_{j=1}^{l-1}%
\frac{\left\vert Z^{\left( j-1\right) }-Z^{\left( j\right) }\right\vert }{c}%
,Z^{\left( l-1\right) },Z^{\left( l\right) },\omega _{0},\Psi \right) $ in $%
F\left( line_{i}\right) $ (see (\ref{lf}) for the definition of $F\left(
line_{i}\right) $). This sum is computed in (\ref{zdv}). We call the result $%
G_{0}^{\left( n\right) }\left( \left( Z,\theta \right) ,\left( Z_{1},\theta
_{1}\right) \right) $, so that: 
\begin{eqnarray*}
&&G_{0}^{\left( n\right) }\left( \left( Z,\theta \right) ,\left(
Z_{1},\theta _{1}\right) \right) =\int \dprod\limits_{l=1}^{n}\hat{T}\left(
\theta -\sum_{j=1}^{l-1}\frac{\left\vert Z^{\left( j-1\right) }-Z^{\left(
j\right) }\right\vert }{c},Z^{\left( l-1\right) },Z^{\left( l\right)
},\omega _{0}\right) \\
&&\times \delta \left( \left( \theta -\theta _{1}\right) -\sum_{l=1}^{n}%
\frac{\left\vert Z^{\left( l-1\right) }-Z^{\left( l\right) }\right\vert }{c}%
\right) \dprod\limits_{l=1}^{n-1}dZ^{\left( l\right) } \\
&=&\int \dprod\limits_{l=1}^{n}\hat{T}\left( \theta -\sum_{j=1}^{l-1}\frac{%
\left\vert Z^{\left( j-1\right) }-Z^{\left( j\right) }\right\vert }{c}%
,Z^{\left( l-1\right) },Z^{\left( l\right) },\omega _{0}\right) \delta
\left( \left( \theta ^{\left( l\right) }-\theta ^{\left( l-1\right) }\right)
-\frac{\left\vert Z^{\left( l-1\right) }-Z^{\left( l\right) }\right\vert }{c}%
\right) \dprod\limits_{l=1}^{n-1}dZ^{\left( l\right) }d\theta _{l}
\end{eqnarray*}

with $\left( Z^{\left( 0\right) },\theta ^{\left( 0\right) }\right) =\left(
Z,\theta \right) $ and $\left( Z^{\left( n\right) },\theta ^{\left( n\right)
}\right) =\left( Z_{1},\theta _{1}\right) $.

Then, we sum over the length $n$ of the lines and the factor associated to
the sum of lines, written $G_{0}\left( \left( Z,\theta \right) ,\left(
Z_{1},\theta _{1}\right) \right) $, is:%
\begin{eqnarray*}
G_{0}\left( \left( Z,\theta \right) ,\left( Z_{1},\theta _{1}\right) \right)
&=&\sum_{n=1}^{\infty }\int \dprod\limits_{l=1}^{n}\hat{T}\left( \theta
-\sum_{j=1}^{l-1}\frac{\left\vert Z^{\left( j-1\right) }-Z^{\left( j\right)
}\right\vert }{c},Z^{\left( l-1\right) },Z^{\left( l\right) },\omega
_{0}\right) \\
&&\times \delta \left( \left( \theta ^{\left( l\right) }-\theta ^{\left(
l-1\right) }\right) -\frac{\left\vert Z^{\left( l-1\right) }-Z^{\left(
l\right) }\right\vert }{c}\right) \dprod\limits_{l=1}^{n-1}dZ^{\left(
l\right) }d\theta _{l}
\end{eqnarray*}%
The function $G_{0}\left( \left( Z,\theta \right) ,\left( Z_{1},\theta
_{1}\right) \right) $ is a series expansion that can be summed:%
\begin{equation}
G_{0}\left( \left( Z,\theta \right) ,\left( Z_{1},\theta _{1}\right) \right)
=\hat{T}\left( 1-\hat{T}\right) ^{-1}\left( \left( Z,\theta \right) ,\left(
Z_{1},\theta _{1}\right) \right)  \label{prT}
\end{equation}%
with:%
\begin{equation*}
\hat{T}\left( \left( Z^{\left( l-1\right) },\theta ^{\left( l-1\right)
}\right) ,\left( Z^{\left( l\right) },\theta ^{\left( l\right) }\right)
\right) =\hat{T}\left( \theta -\sum_{j=1}^{l-1}\frac{\left\vert Z^{\left(
j-1\right) }-Z^{\left( j\right) }\right\vert }{c},Z^{\left( l-1\right)
},Z^{\left( l\right) },\omega _{0}\right) \delta \left( \left( \theta
^{\left( l\right) }-\theta ^{\left( l-1\right) }\right) -\frac{\left\vert
Z^{\left( l-1\right) }-Z^{\left( l\right) }\right\vert }{c}\right)
\end{equation*}%
As a consequence, equation (\ref{rdt}) can be rewritten as a sum over the
branch points.:%
\begin{eqnarray}
\omega \left( \theta ^{\left( i\right) },Z\right) -\omega \left( \theta
^{\left( i\right) },Z\right) _{\left\vert \Psi \right\vert ^{2}=0} &=&\int
\sum_{n}\left( \frac{\delta ^{n}\omega \left( J,\theta ,Z\right) }{%
\dprod\limits_{i=1}^{n}\delta \left\vert \Psi \left( \theta
-l_{i},Z_{i}\right) \right\vert ^{2}}\right) _{\left\vert \Psi \right\vert
^{2}=0}\dprod\limits_{i=1}^{n}\left\vert \Psi \left( \theta
-l_{i},Z_{i}\right) \right\vert ^{2}dl_{i}dZ_{i}  \label{xf} \\
&=&\left( \sum_{m=1}^{n}\sum_{i=1}^{m}\sum_{B}\sum_{\left( \overline{line_{1}%
},...,\overline{line_{m}}\right) }\dprod\limits_{i}G_{0}\left( \overline{%
line_{i}}\right) \dprod\limits_{B}F\left( B\right) \right)
\dprod\limits_{i=1}^{n}\left\vert \Psi \left( \theta -l_{i},Z_{i}\right)
\right\vert ^{2}  \notag
\end{eqnarray}
The sum $\sum_{\left( \overline{line_{1}},...,\overline{line_{m}}\right) }$
is over the finite set of $m$\ segments connecting two branch points and
respecting the constraint given above (\ref{lf}). If $\overline{line_{i}}$
connects two branch points $\left( \left( X_{1},\theta _{1}\right) ,\left(
X_{2},\theta _{2}\right) \right) $, then $G_{0}\left( \overline{line_{i}}%
\right) $ is equal to $G_{0}\left( \left( X_{1},\theta _{1}\right) ,\left(
X_{2},\theta _{2}\right) \right) $. At each branch point we insert $\frac{%
\left\vert \Psi \left( \theta -l_{k},Z_{k}\right) \right\vert ^{2}}{\mathcal{%
\bar{G}}_{0}\left( 0,Z_{k}\right) }$ and for a terminal point $\frac{\omega
_{0}\left( J,\theta -l_{k},Z_{k}\right) \left\vert \Psi \left( \theta
-l_{k},Z_{k}\right) \right\vert ^{2}}{\mathcal{\bar{G}}_{0}\left(
0,Z_{k}\right) }$. We will normalize $\left\vert \Psi \right\vert ^{2}$ by $%
\mathcal{\bar{G}}_{0}$, so that $\left\vert \Psi \left( \theta
-l_{k},Z_{k}\right) \right\vert ^{2}$ will stand for $\frac{\left\vert \Psi
\left( \theta -l_{k},Z_{k}\right) \right\vert ^{2}}{\mathcal{\bar{G}}%
_{0}\left( 0,Z_{k}\right) }$.

Now the sums in (\ref{xf}) can be reordered in the following way. We
consider the lines from $\left( \theta ,Z\right) $ to a final point, and sum
over the branch points of valence $2$ crossed by these lines, that is points
crossed or reached only by this line. We then sum the contributions over all
these lines. For instance, if a line crosses only one branch point, the
associated contribution will include two propagators $G_{0}=\hat{T}\left( 1-%
\hat{T}\right) ^{-1}$, one between the initial point and the branch point,
one between the branch point and the final point plus the factors inserted
at each point. Summing over all possible branch points crossed by a line
yields the factor associated to the overall sum of single lines crossing the
points $Z_{k}$:%
\begin{eqnarray}
&&\hat{T}\left( 1-\hat{T}\right) ^{-1}\sum_{n\geqslant 0}\int
\dprod\limits_{l=1}^{n-1}\left\{ \int \left( \left\vert \Psi \left( \theta
-l_{l},Z_{l}\right) \right\vert ^{2}dZ_{l}dl_{l}\right) \hat{T}\left( 1-\hat{%
T}\right) ^{-1}\right\} \left\vert \Psi \left( \theta -l_{n},Z_{n}\right)
\right\vert ^{2}\frac{\omega _{0}\left( J,\theta -l_{n},Z_{n}\right) }{%
\mathcal{\bar{G}}_{0}\left( 0,Z_{n}\right) }  \notag \\
&=&\hat{T}\left( 1-\hat{T}\right) ^{-1}\frac{1}{1-\left\vert \Psi \left(
\theta ,Z\right) \right\vert ^{2}\hat{T}\left( 1-\hat{T}\right) ^{-1}}%
\left\vert \Psi \left( \theta -l_{n},Z_{n}\right) \right\vert ^{2}\frac{%
\omega _{0}\left( J,\theta -l_{n},Z_{n}\right) }{\mathcal{\bar{G}}_{0}\left(
0,Z_{n}\right) }  \notag \\
&=&\hat{T}\frac{1}{1-\left( 1+\left\vert \Psi \right\vert ^{2}\right) \hat{T}%
}\left\vert \Psi \left( \theta -l_{n},Z_{n}\right) \right\vert ^{2}\frac{%
\omega _{0}\left( J,\theta -l_{n},Z_{n}\right) }{\mathcal{\bar{G}}_{0}\left(
0,Z_{n}\right) }  \label{rl}
\end{eqnarray}%
with $Z_{0}=X_{1}$ and $Z_{k+1}=X_{2}$ and $\dprod\limits_{l=1}^{0}$ is set
to $1$. The $l_{i}$ are ranked such that: $l_{1}<...<l_{k}$ We sum over all
contributions of field insertions between $\left( X_{1},\theta _{1}\right) $
and $\left( X_{2},\theta _{2}\right) $ and integrate over the intermediate
points. The factor $\left\vert \Psi \right\vert ^{2}$ is seen as the
operator multiplication by $\left\vert \Psi \left( \theta ,Z\right)
\right\vert ^{2}$ at the point $\left( \theta ,Z\right) $.

The sum (\ref{rl}) over the single lines is the Green function of the
operator $1-\left( 1+\left\vert \Psi \right\vert ^{2}\right) \hat{T}$ with $%
\hat{T}$ and $\left\vert \Psi \left( \theta -l_{n},Z_{n}\right) \right\vert
^{2}\omega _{0}\left( J,\theta -l_{n},Z_{n}\right) $ inserted at the
starting and ending points. This quantity can be seen as a block $\left[
\left( X_{1},\theta _{1}\right) ,\left( X_{2},\theta _{2}\right) \right] $.

\subsubsection*{6.2.2 Path integral formulation}

Then, the series (\ref{xf}) can ultimately be rewritten as a sum over the
number $m$ of branch points $\left( X_{i},\theta _{i}\right) $ with valence $%
k_{i}>2$: we draw all connected graphs whose vertices are the branch points $%
\left( X_{1},\theta _{1}\right) ...\left( X_{m},\theta _{m}\right) $. We
attach $k_{i}$ blocks to the vertex $\left( X_{i},\theta _{i}\right) $, the
endpoint of one of them and the starting point of the others are fixed by
the vertex. To each vertex, the factor $F\left( \left( X_{i},\theta
_{i}\right) \right) $ defined in (\ref{rc}) is associated. The extremities
of the blocks that are not fixed are free and integrated over, except one of
them which is equal to $\left( Z,\theta \right) $. Then the series (\ref{ft}%
) is the sum over $m$ and over all types of graphs with $m$ vertices.

Note that the sum of graph can be computed without ordering in time the
fields. It amounts to replace (\ref{xf}) by: 
\begin{equation*}
\omega \left( \theta ^{\left( i\right) },Z\right) -\omega \left( \theta
^{\left( i\right) },Z\right) _{\left\vert \Psi \right\vert ^{2}=0}=\int 
\frac{1}{n!}\sum_{n}\left( \frac{\delta ^{n}\omega \left( J,\theta ,Z\right) 
}{\dprod\limits_{i=1}^{n}\delta \left\vert \Psi \left( \theta
-l_{i},Z_{i}\right) \right\vert ^{2}}\right) _{\left\vert \Psi \right\vert
^{2}=0}\dprod\limits_{i=1}^{n}\left\vert \Psi \left( \theta
_{i},Z_{i}\right) \right\vert ^{2}d\theta _{i}dZ_{i}
\end{equation*}%
As a consequence, the symetry factor of equivalent graphs factored by $%
\dprod\limits_{i=1}^{n}\left\vert \Psi \left( \theta _{i},Z_{i}\right)
\right\vert ^{2}$ and integrated over $\dprod\limits_{i=1}^{n}d\theta
_{i}dZ_{i}$ is:%
\begin{equation*}
\frac{1}{n!}\frac{n!}{\prod\limits_{V}k_{V}!}
\end{equation*}%
where the product is over the vertices of valence $k_{V}$ of the graph. The
factor $n!$ comes from the exchange between the vertices $%
\dprod\limits_{i=1}^{n}\left\vert \Psi \left( \theta _{i},Z_{i}\right)
\right\vert ^{2}$ \ The $k_{V}!$ accounts for the exchange of the $k_{V}$
vertices among the same graph.

The sum of lines connected by vertices can then be computed using the Green
function $\frac{1}{1-\left( 1+\left\vert \Psi \right\vert ^{2}\right) \hat{T}%
}$ connecting the vertices of all possible valences.

As a consequence, the generating function for the graphs is equal to the
partition function for an auxiliary complex field $\Lambda \left( X,\theta
\right) $ with free Green function equal to $\frac{1}{1-\left( 1+\left\vert
\Psi \right\vert ^{2}\right) \hat{T}}$ and interaction terms generating the
various graphs with arbitrar vertices. The free part of the action for $%
\Lambda \left( X,\theta \right) $ is thus:%
\begin{equation*}
\int \Lambda \left( X,\theta \right) \left( 1-\left( 1+\left\vert \Psi
\right\vert ^{2}\right) \hat{T}\right) \Lambda ^{\dag }\left( X,\theta
\right) d\left( X,\theta \right)
\end{equation*}%
and the interaction terms have to induce the graphs with factor (\ref{rc}).
The $k+2$ valence vertex, with $k\geqslant 1$ is thus described by a term
involving (\ref{rc}) and writes:%
\begin{eqnarray*}
&&\int \Lambda \left( Z^{\left( 1\right) },\theta ^{\left( 1\right) }\right) 
\frac{\delta ^{k}\left( \hat{T}\left( \theta ^{\left( 1\right) }-\frac{%
\left\vert Z^{\left( 1\right) }-Z^{\left( 2\right) }\right\vert }{c}%
,Z^{\left( 1\right) },Z^{\left( 2\right) },\omega _{0}\right) \right) }{%
k!\dprod\limits_{l=3}^{k+2}\delta ^{k}\omega _{0}\left( J,\theta ^{\left(
l\right) },Z^{\left( l\right) }\right) }\Lambda ^{\dag }\left( Z^{\left(
2\right) },\theta ^{\left( 1\right) }-\frac{\left\vert Z^{\left( 1\right)
}-Z^{\left( 2\right) }\right\vert }{c}\right) \\
&&\times \dprod\limits_{l=3}^{k+2}\hat{T}\left( \left( Z^{\left( 1\right)
},\theta ^{\left( 1\right) }\right) ,\left( \theta ^{\left( l\right)
},Z^{\left( l\right) }\right) \right) \left( \Lambda ^{\dag }\left( \theta
^{\left( l\right) },Z^{\left( l\right) }\right) \right)
\dprod\limits_{l=1}^{k+2}d\left( \theta ^{\left( l\right) },Z^{\left(
l\right) }\right) \\
&=&\int \Lambda \left( Z^{\left( 1\right) },\theta ^{\left( 1\right)
}\right) \frac{\delta ^{k}\left( \hat{T}\left( \theta ^{\left( 1\right) }-%
\frac{\left\vert Z^{\left( 1\right) }-Z^{\left( 2\right) }\right\vert }{c}%
,Z^{\left( 1\right) },Z^{\left( 2\right) },\omega _{0}\right) \right) }{%
k!\dprod\limits_{l=3}^{k+2}\delta ^{k}\omega _{0}\left( J,\theta ^{\left(
l\right) },Z^{\left( l\right) }\right) }\Lambda ^{\dag }\left( Z^{\left(
2\right) },\theta ^{\left( 1\right) }-\frac{\left\vert Z^{\left( 1\right)
}-Z^{\left( 2\right) }\right\vert }{c}\right) \\
&&\times \dprod\limits_{l=3}^{k+2}\hat{T}\left( \theta ^{\left( 1\right) }-%
\frac{\left\vert Z^{\left( 1\right) }-Z^{\left( l\right) }\right\vert }{c}%
,Z^{\left( 1\right) },Z^{\left( l\right) },\omega _{0}\right) \Lambda ^{\dag
}\left( \theta ^{\left( l\right) },Z^{\left( l\right) }\right) d\theta
^{\left( 1\right) }\dprod\limits_{l=1}^{k+2}dZ^{\left( l\right) }
\end{eqnarray*}%
Having found the free part of the action and the required vertices, the sum
of all graphs (\ref{xf}) yields, for $\frac{\left\vert \Psi \left( J,\theta
_{i},Z_{i}\right) \right\vert ^{2}}{\mathcal{\bar{G}}_{0}\left(
0,Z_{i}\right) }\rightarrow \left\vert \Psi \left( J,\theta
_{i},Z_{i}\right) \right\vert ^{2}$:%
\begin{eqnarray}
&&\omega _{0}\left( J,\theta ,Z\right) +\sum_{n=1}^{\infty }\frac{1}{n!}%
\frac{\int \hat{T}\Lambda ^{\dag }\left( Z,\theta \right) \int
\dprod\limits_{i=1}^{n}\omega _{0}\left( J,\theta _{i},Z_{i}\right)
\left\vert \Psi \left( J,\theta _{i},Z_{i}\right) \right\vert ^{2}\Lambda
\left( Z_{i},\theta _{i}\right) d\left( Z_{i},\theta _{i}\right) \exp \left(
-S\left( \Lambda \right) \right) \mathcal{D}\Lambda }{\exp \left( -S\left(
\Lambda \right) \right) \mathcal{D}\Lambda }  \notag \\
&=&\omega _{0}\left( J,\theta ,Z\right) +\frac{\int \hat{T}\Lambda ^{\dag
}\left( Z,\theta \right) \exp \left( -S\left( \Lambda \right) +\int \Lambda
\left( X,\theta \right) \omega _{0}\left( J,\theta ,Z\right) \left\vert \Psi
\left( J,\theta ,Z\right) \right\vert ^{2}d\left( X,\theta \right) \right) 
\mathcal{D}\Lambda }{\int \exp \left( -S\left( \Lambda \right) \right) 
\mathcal{D}\Lambda }  \label{PX}
\end{eqnarray}%
with:%
\begin{eqnarray*}
S\left( \Lambda \right) &=&\int \Lambda \left( X,\theta \right) \left(
1-\left( 1+\left\vert \Psi \right\vert ^{2}\right) \hat{T}\right) \Lambda
^{\dag }\left( X,\theta \right) d\left( X,\theta \right) \\
&&-\int \Lambda \left( Z^{\left( 1\right) },\theta ^{\left( 1\right)
}\right) \sum_{k}\frac{\delta ^{k}\left( \hat{T}\left( \theta ^{\left(
1\right) }-\frac{\left\vert Z^{\left( 1\right) }-Z^{\left( 2\right)
}\right\vert }{c},Z^{\left( 1\right) },Z^{\left( 2\right) },\omega
_{0}\right) \right) }{k!\dprod\limits_{l=3}^{k+2}\delta ^{k}\omega
_{0}\left( J,\theta ^{\left( l\right) },Z^{\left( l\right) }\right) }\Lambda
^{\dag }\left( Z^{\left( 2\right) },\theta ^{\left( 1\right) }-\frac{%
\left\vert Z^{\left( 1\right) }-Z^{\left( 2\right) }\right\vert }{c}\right)
\\
&&\times \dprod\limits_{l=3}^{k+2}\hat{T}\left( \theta ^{\left( 1\right) }-%
\frac{\left\vert Z^{\left( 1\right) }-Z^{\left( l\right) }\right\vert }{c}%
,Z^{\left( 1\right) },Z^{\left( l\right) },\omega _{0}\right) \Lambda ^{\dag
}\left( \theta ^{\left( l\right) },Z^{\left( l\right) }\right) d\theta
^{\left( 1\right) }\dprod\limits_{l=1}^{k+2}dZ^{\left( l\right) } \\
&=&\int \Lambda \left( Z,\theta \right) \left( 1-\left\vert \Psi \right\vert
^{2}\hat{T}\right) \Lambda ^{\dag }\left( Z,\theta \right) d\left( Z,\theta
\right) -\int \Lambda \left( Z,\theta \right) \hat{T}\left( \theta -\frac{%
\left\vert Z-Z^{\left( 1\right) }\right\vert }{c},Z,Z^{\left( 1\right)
},\omega _{0}+\hat{T}\Lambda ^{\dag }\right) \\
&&\times \Lambda ^{\dag }\left( Z^{\left( 1\right) },\theta -\frac{%
\left\vert Z-Z^{\left( 1\right) }\right\vert }{c}\right) dZdZ^{\left(
1\right) }d\theta
\end{eqnarray*}%
where:%
\begin{eqnarray*}
&&\hat{T}\left( \theta -\frac{\left\vert Z^{\left( 1\right) }-Z\right\vert }{%
c},Z^{\left( 1\right) },Z,\omega _{0}+\hat{T}\Lambda ^{\dag }\right) \\
&=&\hat{T}\left( \theta -\frac{\left\vert Z^{\left( 1\right) }-Z\right\vert 
}{c},Z^{\left( 1\right) },Z,\omega _{0}\left( Z,\theta \right) +\int \hat{T}%
\left( \theta -\frac{\left\vert Z-Z^{\left( 1\right) }\right\vert }{c}%
,Z^{\left( 1\right) },Z,\omega _{0}\right) \Lambda ^{\dag }\left( Z^{\left(
1\right) },\theta -\frac{\left\vert Z-Z^{\left( 1\right) }\right\vert }{c}%
\right) dZ^{\left( 1\right) }\right)
\end{eqnarray*}

\subsubsection*{6.2.3 Saddle point approximation}

The saddle point approximation yields the equations for $\Lambda ^{\dag
}\left( Z,\theta \right) $ and $\Lambda \left( Z,\theta \right) $: 
\begin{equation}
\left( \left( 1-\left\vert \Psi \right\vert ^{2}\hat{T}\right) \Lambda
^{\dag }\right) \left( Z,\theta \right) -\left( \hat{T}_{\omega _{0}+\hat{T}%
\Lambda ^{\dag }}\Lambda ^{\dag }\right) \left( Z,\theta \right) -\omega
_{0}\left\vert \Psi \right\vert ^{2}=0  \label{dsn}
\end{equation}%
\begin{equation*}
\Lambda \left( Z,\theta \right) =0
\end{equation*}%
Using that:%
\begin{equation*}
\hat{T}\left( \omega _{0}+\hat{T}\Lambda ^{\dag }\right) \simeq \frac{\omega
_{0}}{\omega _{0}+\hat{T}\Lambda ^{\dag }}\hat{T}
\end{equation*}%
equation (\ref{dsn}) writes:

\begin{equation}
\left( 1-\left\vert \Psi \right\vert ^{2}\hat{T}\right) \Lambda ^{\dag }-%
\frac{\omega _{0}}{\omega _{0}+\hat{T}\Lambda ^{\dag }}\hat{T}\Lambda ^{\dag
}-\omega _{0}\left\vert \Psi \right\vert ^{2}=0  \label{dpn}
\end{equation}%
This can be rewritten as an equation for $\omega $. Actually, using (\ref{PX}%
), under the saddle point approximation:%
\begin{equation}
\omega \left( J,\theta ,Z\right) =\omega _{0}\left( J,\theta ,Z\right) +\hat{%
T}\Lambda ^{\dag }\left( Z,\theta \right)  \label{dst}
\end{equation}%
and:%
\begin{equation*}
\hat{T}\Lambda ^{\dag }\left( Z,\theta \right) =\omega \left( J,\theta
,Z\right) -\omega _{0}\left( J,\theta ,Z\right) \equiv \Omega \left(
J,\theta ,Z\right)
\end{equation*}%
so that (\ref{dsn}) writes:%
\begin{equation}
\Lambda ^{\dag }-\left( \Omega +\omega _{0}\right) \left\vert \Psi
\right\vert ^{2}-\frac{\omega _{0}}{\omega _{0}+\Omega }\Omega =0
\label{snd}
\end{equation}%
Applying the operator $\hat{T}$ on the left leads to: 
\begin{equation}
\Omega -\hat{T}\left( \Omega +\omega _{0}\right) \left\vert \Psi \right\vert
^{2}-\hat{T}\frac{\omega _{0}\Omega }{\omega _{0}+\Omega }=0  \label{mqn}
\end{equation}%
Then using the expression for the background field:%
\begin{eqnarray*}
\Psi \left( \theta ,Z\right) &=&\frac{\nabla _{\theta }\omega \left( J\left(
\theta \right) ,\theta ,Z,\mathcal{G}_{0}+\left\vert \Psi _{0}\right\vert
^{2}\right) }{U^{\prime \prime }\left( X_{0}\right) \omega ^{2}\left(
J\left( \theta \right) ,\theta ,Z,\mathcal{G}_{0}+\left\vert \Psi
_{0}\right\vert ^{2}\right) }\Psi _{0}\left( \theta ,Z\right) \\
&=&\frac{\nabla _{\theta }\omega }{U^{\prime \prime }\left( X_{0}\right)
\omega ^{2}}\Psi _{0}\left( \theta ,Z\right) \\
&=&\frac{X_{0}\nabla _{\theta }\Omega }{U^{\prime \prime }\left(
X_{0}\right) \left( \omega _{0}+\Omega \right) ^{2}}
\end{eqnarray*}%
and:%
\begin{equation*}
\left\vert \Psi \right\vert ^{2}=\frac{X_{0}^{2}}{U^{\prime \prime }\left(
X_{0}\right) }\frac{\nabla _{\theta }\Omega }{\left( \omega _{0}+\Omega
\right) ^{2}}
\end{equation*}%
Equation (\ref{mqn}) becomes:%
\begin{equation*}
\Omega -\hat{T}\frac{X_{0}^{2}}{U^{\prime \prime }\left( X_{0}\right) }\frac{%
\nabla _{\theta }\Omega }{\left( \omega _{0}+\Omega \right) ^{2}}\Omega -%
\hat{T}\frac{\omega _{0}\Omega }{\omega _{0}+\Omega }-\hat{T}\frac{X_{0}^{2}%
}{U^{\prime \prime }\left( X_{0}\right) }\frac{\omega _{0}\nabla _{\theta
}\Omega }{\left( \omega _{0}+\Omega \right) ^{2}}=0
\end{equation*}%
that is:%
\begin{equation}
\Omega -\hat{T}\frac{\frac{X_{0}^{2}}{U^{\prime \prime }\left( X_{0}\right) }%
\nabla _{\theta }\Omega +\omega _{0}\Omega }{\omega _{0}+\Omega }=0
\label{dqt}
\end{equation}%
Remark that this equation is still valid for any background field related to 
$\omega $ by a relation of the type:%
\begin{equation}
\left\vert \Psi \right\vert ^{2}=f\left( \omega ,\nabla _{\theta }^{l}\omega
\right)  \label{spg}
\end{equation}%
which yields:%
\begin{equation*}
\Omega -\hat{T}\omega f\left( \omega ,\nabla _{\theta }^{l}\omega \right) -%
\hat{T}\frac{\omega _{0}\left( \omega -\omega _{0}\right) }{\omega }=0
\end{equation*}

The second order expansion in derivatives of the right hand side of (\ref%
{dqt}) yields a second order linear differential equation similar to the
type of equation derived in the text.

\subsubsection*{6.2.4 Solution as function of external field and series
expansion\protect\bigskip}

We can also write a series expansion for the solution of (\ref{dsn}) in
terms of propagation functions for an external field $\left\vert \Psi
\right\vert ^{2}$. We write (\ref{dsn}) as an operator equation:%
\begin{equation*}
\left( 1-\left\vert \Psi \right\vert ^{2}\hat{T}-\hat{T}_{\omega _{0}+\hat{T}%
\Lambda ^{\dag }}\right) \Lambda ^{\dag }=\omega _{0}\left\vert \Psi
\right\vert ^{2}
\end{equation*}%
which leads to:%
\begin{eqnarray}
\hat{T}\Lambda ^{\dag } &=&\hat{T}\frac{1}{1-\left\vert \Psi \right\vert ^{2}%
\hat{T}-\hat{T}_{\omega _{0}+\hat{T}\Lambda ^{\dag }}}\omega _{0}\left\vert
\Psi \right\vert ^{2}  \notag \\
&=&\hat{T}\frac{1}{1-\left( 1+\left\vert \Psi \right\vert ^{2}\right) \hat{T}%
-\left( \hat{T}_{\omega _{0}+\hat{T}\Lambda ^{\dag }}-\hat{T}\right) }\omega
_{0}\left\vert \Psi \right\vert ^{2}  \notag \\
&\simeq &\frac{\hat{T}}{1-\left( 1+\left\vert \Psi \right\vert ^{2}\right) 
\hat{T}}\frac{1}{1-\left( \hat{T}_{\omega _{0}+\hat{T}\Lambda ^{\dag }}-\hat{%
T}\right) \frac{1}{1-\left( 1+\left\vert \Psi \right\vert ^{2}\right) \hat{T}%
}}\omega _{0}\left\vert \Psi \right\vert ^{2}  \notag \\
&=&A\frac{1}{1-\left( \hat{T}_{\omega _{0}+\hat{T}\Lambda ^{\dag }}-\hat{T}%
\right) \hat{T}^{-1}A}\omega _{0}\left\vert \Psi \right\vert ^{2}
\label{mrf}
\end{eqnarray}%
where:%
\begin{equation}
A=\frac{\hat{T}}{1-\left( 1+\left\vert \Psi \right\vert ^{2}\right) \hat{T}}
\label{mfd}
\end{equation}

\subsubsection*{6.4.1 Recursive expansion of (\protect\ref{mrf}), first
approximation}

Equation (\ref{mrf}) can be solved recursively, by expanding $\hat{T}%
_{\omega _{0}+\hat{T}\Lambda ^{\dag }}-\hat{T}$ order by order:%
\begin{equation*}
\hat{T}\Lambda ^{\dag }=A\frac{1}{1-\left( \hat{T}_{\omega _{0}+A\frac{1}{%
1-\left( \hat{T}_{\omega _{0}+\hat{T}\Lambda ^{\dag }}-\hat{T}\right) \hat{T}%
^{-1}A}\omega _{0}\left\vert \Psi \right\vert ^{2}}-\hat{T}\right) \hat{T}%
^{-1}A}\omega _{0}\left\vert \Psi \right\vert ^{2}
\end{equation*}%
and so on. In first approximation,the series expansion for\ $\hat{T}\Lambda
^{\dag }$\ is:%
\begin{eqnarray*}
\hat{T}\Lambda ^{\dag } &=&A\frac{1}{1-\left( \hat{T}_{\omega _{0}+A\omega
_{0}\left\vert \Psi \right\vert ^{2}}-\hat{T}\right) \hat{T}^{-1}A}\omega
_{0}\left\vert \Psi \right\vert ^{2} \\
&\simeq &A\frac{1}{1-\frac{\omega _{0}}{\omega _{0}+A\omega _{0}\left\vert
\Psi \right\vert ^{2}}A}\omega _{0}\left\vert \Psi \right\vert ^{2}
\end{eqnarray*}%
or written in expanded form:%
\begin{equation}
\omega =\omega _{0}+\sum_{n\geqslant 0}\left( A\frac{\omega _{0}}{\omega
_{0}+A\omega _{0}\left\vert \Psi \right\vert ^{2}}\right) ^{n}A\omega
_{0}\left\vert \Psi \right\vert ^{2}  \label{mGG}
\end{equation}%
We use (\ref{mfd}) to write $A$ as:%
\begin{equation*}
A=\frac{\hat{T}}{1-\left( 1+\left\vert \Psi \right\vert ^{2}\right) \hat{T}}=%
\frac{1}{1+\left\vert \Psi \right\vert ^{2}}\frac{\left( 1+\left\vert \Psi
\right\vert ^{2}\right) \hat{T}}{1-\left( 1+\left\vert \Psi \right\vert
^{2}\right) \hat{T}}
\end{equation*}%
Operator $A$ is defined by successive convolutions, it can thus be
approximated by:%
\begin{equation*}
A\simeq \frac{1}{1+\left\langle \left\vert \Psi \right\vert
^{2}\right\rangle }\frac{\left( 1+\left\langle \left\vert \Psi \right\vert
^{2}\right\rangle \right) \hat{T}}{1-\left( 1+\left\langle \left\vert \Psi
\right\vert ^{2}\right\rangle \right) \hat{T}}
\end{equation*}%
where $\frac{\left( 1+\left\langle \left\vert \Psi \right\vert
^{2}\right\rangle \right) \hat{T}}{1-\left( 1+\left\langle \left\vert \Psi
\right\vert ^{2}\right\rangle \right) \hat{T}}$ can be computed as $\frac{%
\hat{T}}{1-\hat{T}}$ \ Given formulas (\ref{srt}), (\ref{rtt}) and (\ref{trr}%
), it amounts to replace the constant $C$ \ by $\left( 1+\left\langle
\left\vert \Psi \right\vert ^{2}\right\rangle \right) C$ in the expansion of 
$\frac{\hat{T}}{1-\hat{T}}$. This modifies formula (\ref{trr}) by
introducing a dependency in field inside the exponential function. We thus
have:%
\begin{equation}
A\left( \left( Z,\theta \right) ,\left( Z_{1},\theta -l_{1}\right) \right)
\simeq \frac{\exp \left( -cl_{1}-\alpha \left( 1+\left\langle \left\vert
\Psi \right\vert ^{2}\right\rangle \right) \left( \left( cl_{1}\right)
^{2}-\left\vert Z-Z_{1}\right\vert ^{2}\right) \right) }{\left(
1+\left\langle \left\vert \Psi \right\vert ^{2}\right\rangle \right) B}%
H\left( cl_{1}-\left\vert Z-Z_{1}\right\vert \right)  \label{npr}
\end{equation}%
Inserting formula (\ref{npr}) in the expression (\ref{mGG}) for $\omega $
leads to:

\begin{eqnarray}
\omega \left( Z,\theta \right) &=&\omega _{0}\left( J,\theta ,Z\right)
\label{frm} \\
&&+\int \sum_{k=0}^{\infty }\frac{\exp \left( -c\sum_{i=0}^{k}l_{i}-\alpha
\left( 1+\left\langle \left\vert \Psi \right\vert ^{2}\right\rangle \right)
\left( \sum_{i=0}^{k}\left( cl_{i}\right) ^{2}-\sum_{l=0}^{k-1}\frac{%
\left\vert Z_{i}-Z_{i+1}\right\vert }{c}\right) \right) }{B^{k+1}}  \notag \\
&&\times \dprod\limits_{i=1}^{k}\left( \frac{\omega _{0}\left( \theta
-l_{i},Z_{i}\right) }{\omega _{0}\left( \theta -l_{i},Z_{i}\right) +A\omega
_{0}\left\vert \Psi \right\vert ^{2}\left( \theta -l_{i},Z_{i}\right) }%
\right) \frac{\omega _{0}\left( J,\theta -l_{k},Z_{k}\right) }{\left(
1+\left\langle \left\vert \Psi \right\vert ^{2}\right\rangle \right) }%
\left\vert \Psi \left( \theta -l_{k},Z_{k}\right) \right\vert
^{2}dZ_{i}dl_{i}  \notag
\end{eqnarray}%
\bigskip and:%
\begin{eqnarray}
\omega ^{-1}\left( Z,\theta \right) &=&\omega _{0}^{-1}\left( J,\theta
,Z\right) \\
&&+\frac{G^{\prime }\left[ J,\omega ,\theta ,Z,\Psi \right] }{F^{\prime }%
\left[ J,\omega ,\theta ,Z,\Psi \right] }\int \sum_{k=0}^{\infty }\frac{\exp
\left( -c\sum_{i=0}^{k}l_{i}-\alpha \left( 1+\left\langle \left\vert \Psi
\right\vert ^{2}\right\rangle \right) \left( \sum_{i=0}^{k}\left(
cl_{i}\right) ^{2}-\sum_{l=0}^{k-1}\frac{\left\vert Z_{i}-Z_{i+1}\right\vert 
}{c}\right) \right) }{D^{k+1}}  \notag \\
&&\times \dprod\limits_{i=1}^{k}\left( \frac{\omega _{0}\left( \theta
-l_{i},Z_{i}\right) }{\omega _{0}\left( \theta -l_{i},Z_{i}\right) +A\omega
_{0}\left\vert \Psi \right\vert ^{2}\left( \theta -l_{i},Z_{i}\right) }%
\right) \frac{\omega _{0}^{-1}\left( J,\theta -l_{k},Z_{k}\right) }{\left(
1+\left\langle \left\vert \Psi \right\vert ^{2}\right\rangle \right) }%
\left\vert \Psi \left( \theta -l_{k},Z_{k}\right) \right\vert
^{2}dZ_{i}dl_{i}  \notag
\end{eqnarray}

\subsubsection*{6.4.2 Recursive expansion of (\protect\ref{mrf}), full
series expansion}

Formula (\ref{frm}) can be refined by including the full series expansion of
(\ref{mrf}). In a first step we can approximate:%
\begin{eqnarray*}
&&\frac{1}{1-\left( 1+\left\vert \Psi \right\vert ^{2}\right) \hat{T}}\frac{1%
}{1-\left( \hat{T}_{\omega _{0}+\hat{T}\Lambda ^{\dag }}-\hat{T}\right) 
\frac{1}{1-\left( 1+\left\vert \Psi \right\vert ^{2}\right) \hat{T}}} \\
&\simeq &\frac{1}{1-\left( 1+\left\vert \Psi \right\vert ^{2}\right) \hat{T}}%
\frac{1}{1-\frac{\frac{\delta \hat{T}}{\delta \omega _{0}}}{\hat{T}}\hat{T}%
\Lambda ^{\dag }\frac{\hat{T}}{1-\left( 1+\left\vert \Psi \right\vert
^{2}\right) \hat{T}}}
\end{eqnarray*}%
so that:%
\begin{eqnarray*}
\hat{T}\Lambda ^{\dag } &=&A\frac{1}{1-\frac{\frac{\delta \hat{T}}{\delta
\omega _{0}}}{\hat{T}}\hat{T}\Lambda ^{\dag }A}\omega _{0}\left\vert \Psi
\right\vert ^{2} \\
&=&A\sum_{n\geqslant 0}\left( \frac{\frac{\delta \hat{T}}{\delta \omega _{0}}%
}{\hat{T}}\hat{T}\Lambda ^{\dag }A\right) ^{n}\omega _{0}\left\vert \Psi
\right\vert ^{2} \\
&=&A\sum_{k\geqslant 0}\left( \left( \frac{\delta \hat{T}}{\delta \omega _{0}%
}\left( A\sum_{l\geqslant 0}\left( \left( \frac{\delta \hat{T}}{\delta
\omega _{0}}\hat{T}\Lambda ^{\dag }\right) A\right) ^{l}\omega
_{0}\left\vert \Psi \right\vert ^{2}\right) \right) A\right) ^{k}\omega
_{0}\left\vert \Psi \right\vert ^{2}=...
\end{eqnarray*}%
The iteration of the previous relation amounts to the successive action of
the operator:%
\begin{eqnarray}
&&O=\int d\left( Z,\theta \right) \frac{\frac{\delta \hat{T}}{\delta \omega
_{0}}}{\hat{T}}\left( Z,\theta \right) \left( \int A\left( \left( Z,\theta
\right) ,\left( Z^{\prime },\theta ^{\prime }\right) \right) \omega
_{0}\left\vert \Psi \right\vert ^{2}\left( Z^{\prime },\theta ^{\prime
}\right) d\left( Z^{\prime },\theta ^{\prime }\right) \right)  \label{rpt} \\
&&\times \int d\left( Z,\theta \right) _{1}A\left( \left( Z,\theta \right)
,\left( Z,\theta \right) _{1}\right) \omega _{0}\left\vert \Psi \right\vert
^{2}\left( Z_{1},\theta _{1}\right) \frac{\delta }{\delta \omega
_{0}\left\vert \Psi \right\vert ^{2}\left( Z,\theta \right) }  \notag
\end{eqnarray}%
where $\frac{\delta }{\delta \omega _{0}\left\vert \Psi \right\vert
^{2}\left( Z,\theta \right) }$ does not act on $A\left( \left( Z,\theta
\right) ,\left( Z,\theta \right) _{1}\right) $. As a consequence: 
\begin{equation*}
\hat{T}\Lambda ^{\dag }=\sum_{n\geqslant 0}O^{n}\omega _{0}A\left\vert \Psi
\right\vert ^{2}=\frac{1}{1-O}\omega _{0}A\left\vert \Psi \right\vert ^{2}
\end{equation*}%
with $\frac{1}{n!}$ accounting for the permutations between the powers of $%
\left\vert \Psi \right\vert ^{2}$. Setting $\left( Z,\theta \right) =X$\ the
series expansion of $\hat{T}\Lambda ^{\dag }$ can be obtained as:

\begin{eqnarray}
\hat{T}\Lambda ^{\dag }
&=&\sum_{l_{1},...,l_{n},\sum\limits_{i=1}^{n}l_{i}=n-1}\sum_{\substack{ %
f\in \left\{ 1,...,n\right\} ^{\left\{ 1,...,n\right\} },  \\ f\left(
k\right) \notin \left\{
\sum\limits_{i=1}^{k-1}l_{i}+1,...\sum\limits_{i=1}^{k}l_{i}\right\} }}%
A\left( X,\hat{X}_{1}\right) ...A\left( \hat{X}_{l_{1}},X_{0}\right) A\left(
X_{0},X_{1}^{\prime }\right) \omega _{0}\left\vert \Psi \right\vert
^{2}\left( X_{1}^{\prime }\right) dX_{1}^{\prime }  \notag \\
&&\times \prod\limits_{k=1}^{n}\frac{1}{\sharp k!}\times
\prod\limits_{k=2}^{n}A\left( X_{k},\hat{X}_{\sum%
\limits_{i=1}^{k-1}l_{i}+1}\right) ...A\left( X_{l_{1}},\hat{X}%
_{\sum\limits_{i=1}^{k}l_{i}}\right) A\left( \hat{X}_{\sum%
\limits_{i=1}^{k}l_{i}},X_{k}^{\prime }\right) \omega _{0}\left\vert \Psi
\right\vert ^{2}\left( X_{k}^{\prime }\right) dX_{k}^{\prime }  \notag \\
&&\times \prod\limits_{k=2}^{n}\frac{\frac{\delta \hat{T}}{\delta \omega
_{0}}}{\hat{T}}\left( X_{k}\right) \delta \left( X_{k}-\hat{X}_{f\left(
k\right) }\right)  \label{ms}
\end{eqnarray}%
with the convention that for $l_{i}=0$ the successive products of
convolution reduces to:%
\begin{equation*}
A\left( X_{k},X_{k}^{\prime }\right) \omega _{0}\left\vert \Psi \right\vert
^{2}\left( X^{\prime }\right) dX^{\prime }
\end{equation*}%
The factors $\sharp k$ are defined by: 
\begin{equation*}
\sharp k=\sum_{l_{i}=1}^{n}\delta _{l_{i},k}
\end{equation*}%
The series expansion can be written in a more compact way:%
\begin{eqnarray}
&&\hat{T}\Lambda ^{\dag
}=\sum_{n}\sum_{l_{1},...,l_{n},\sum\limits_{i=1}^{n}l_{i}=n-1}\sum 
_{\substack{ f\in \left\{ 1,...,n\right\} ^{\left\{ 1,...,n\right\} },  \\ %
f\left( k\right) \notin \left\{
\sum\limits_{i=1}^{k-1}l_{i}+1,...\sum\limits_{i=1}^{k}l_{i}\right\} }}%
A^{l_{1}}\left( X,\hat{X}_{1},...,\hat{X}_{l_{1}},X_{0},X_{1}^{\prime
}\right) \omega _{0}\left\vert \Psi \right\vert ^{2}\left( X_{1}^{\prime
}\right) dX_{1}^{\prime }  \label{rps} \\
&&\times \prod\limits_{k=1}^{n}\frac{d\hat{X}_{k}}{\sharp k!}\times
\prod\limits_{k=2}^{n}A^{l_{k}}\left( X_{k},\hat{X}_{\sum%
\limits_{i=1}^{k-1}l_{i}+1},...,\hat{X}_{\sum%
\limits_{i=1}^{k}l_{i}},X_{k}^{\prime }\right) \omega _{0}\left\vert \Psi
\right\vert ^{2}\left( X_{k}^{\prime }\right) dX_{k}^{\prime }\times
\prod\limits_{k=2}^{n}\frac{\frac{\delta \hat{T}}{\delta \omega _{0}}}{\hat{%
T}}\left( X_{k}\right) \delta \left( X_{k}-\hat{X}_{f\left( k\right) }\right)
\notag
\end{eqnarray}%
with:%
\begin{equation*}
A^{l_{k}}\left( X_{k},\hat{X}_{\sum\limits_{i=1}^{k-1}l_{i}+1},...,\hat{X}%
_{\sum\limits_{i=1}^{k}l_{i}},X_{k}^{\prime }\right) =A\left( X_{k},\hat{X}%
_{\sum\limits_{i=1}^{k-1}l_{i}+1}\right) ...A\left( X_{l_{1}},\hat{X}%
_{\sum\limits_{i=1}^{k}l_{i}}\right) A\left( \hat{X}_{\sum%
\limits_{i=1}^{k}l_{i}},X_{k}^{\prime }\right)
\end{equation*}%
In a second step, all the derivatives $\frac{\frac{\delta ^{l}\hat{T}}{%
\delta ^{l}\omega _{0}}}{\hat{T}}$ can be included and yield the series
expansion: 
\begin{eqnarray}
&&\hat{T}\Lambda ^{\dag
}=\sum_{n,p}\sum_{\sum\limits_{i=1}^{p}r_{i}=n-1}\sum_{l_{1},...,l_{n},%
\sum\limits_{i=1}^{n}l_{i}=p}\sum_{\substack{ f\in \left\{ 1,...,p\right\}
^{\left\{ 1,...,n\right\} },  \\ f\left( k\right) \notin \left\{
\sum\limits_{i=1}^{k-1}l_{i}+1,...\sum\limits_{i=1}^{k}l_{i}\right\} }}%
A^{l_{1}}\left( X,\hat{X}_{1},...,\hat{X}_{l_{1}},X_{0},X_{1}^{\prime
}\right) \omega _{0}\left\vert \Psi \right\vert ^{2}\left( X_{1}^{\prime
}\right) dX_{1}^{\prime }  \notag \\
&&\times \prod\limits_{k=1}^{p}\frac{d\hat{X}_{k}}{\sharp k!}\times
\prod\limits_{k=2}^{n}A^{l_{k}}\left( X_{k},\hat{X}_{\sum%
\limits_{i=1}^{k-1}l_{i}+1},...,\hat{X}_{\sum%
\limits_{i=1}^{k}l_{i}},X_{k}^{\prime }\right) \omega _{0}\left\vert \Psi
\right\vert ^{2}\left( X_{k}^{\prime }\right) dX_{k}^{\prime }  \notag \\
&&\times \prod\limits_{i}^{p}\frac{\frac{\delta ^{r_{i}}\hat{T}}{\delta
^{r_{i}}\omega _{0}\left( \hat{X}_{i}\right) }}{\hat{T}}\prod%
\limits_{k=2}^{n}\delta \left( X_{k}-\hat{X}_{f\left( k\right) }\right)
\label{rsf}
\end{eqnarray}

The propagators $A^{l_{k}}\left( X_{k},\hat{X}_{\sum%
\limits_{i=1}^{k-1}l_{i}+1},...,\hat{X}_{\sum%
\limits_{i=1}^{k}l_{i}},X_{k}^{\prime }\right) $ can be computed using (\ref%
{npr}):%
\begin{equation*}
A=\frac{\hat{T}}{1-\left( 1+\left\vert \Psi \right\vert ^{2}\right) \hat{T}}=%
\frac{1}{1+\left\vert \Psi \right\vert ^{2}}\frac{\left( 1+\left\vert \Psi
\right\vert ^{2}\right) \hat{T}}{1-\left( 1+\left\vert \Psi \right\vert
^{2}\right) \hat{T}}
\end{equation*}%
Operator $A$ being defined by successive convolutions, it can be
approximated by:%
\begin{equation*}
\frac{1}{1+\left\langle \left\vert \Psi \right\vert ^{2}\right\rangle }\frac{%
\left( 1+\left\langle \left\vert \Psi \right\vert ^{2}\right\rangle \right) 
\hat{T}}{1-\left( 1+\left\langle \left\vert \Psi \right\vert
^{2}\right\rangle \right) \hat{T}}
\end{equation*}%
and $\frac{\left( 1+\left\langle \left\vert \Psi \right\vert
^{2}\right\rangle \right) \hat{T}}{1-\left( 1+\left\langle \left\vert \Psi
\right\vert ^{2}\right\rangle \right) \hat{T}}$ can be computed as $\frac{%
\hat{T}}{1-\hat{T}}$ \ Given formulas (\ref{srt}), (\ref{rtt}) and (\ref{trr}%
), it amounts to replace the constant $C$ \ by $\left( 1+\left\langle
\left\vert \Psi \right\vert ^{2}\right\rangle \right) C$. This modifies (\ref%
{trr}) by introducing a dependency in field inside the exponential function:%
\begin{equation}
A\left( \left( Z,\theta \right) ,\left( Z_{1},\theta -l_{1}\right) \right)
\simeq \frac{\exp \left( -cl_{1}-\alpha \left( 1+\left\langle \left\vert
\Psi \right\vert ^{2}\right\rangle \right) \left( \left( cl_{1}\right)
^{2}-\left\vert Z-Z_{1}\right\vert ^{2}\right) \right) }{B}H\left(
cl_{1}-\left\vert Z-Z_{1}\right\vert \right)  \label{fpr}
\end{equation}%
In (\ref{rps}) the vertices $\frac{\frac{\delta \hat{T}}{\delta \omega _{0}}%
}{\hat{T}}\left( X_{k}\right) $ and the $\omega _{0}$ arising in factor can
be approximated by the average $\frac{1}{\left\langle \omega
_{0}\right\rangle }$ and $\left\langle \omega _{0}\right\rangle $, so that (%
\ref{rps}) writes: 
\begin{eqnarray}
&&\Omega =\hat{T}\Lambda ^{\dag
}=\sum_{n}\sum_{l_{1},...,l_{n},\sum\limits_{i=1}^{n}l_{i}=n-1}\sum 
_{\substack{ f\in \left\{ 1,...,n\right\} ^{\left\{ 1,...,n\right\} },  \\ %
f\left( k\right) \notin \left\{
\sum\limits_{i=1}^{k-1}l_{i}+1,...\sum\limits_{i=1}^{k}l_{i}\right\} }}%
A^{l_{1}}\left( X,\hat{X}_{1},...,\hat{X}_{l_{1}},X_{0},X_{1}^{\prime
}\right) \left\vert \Psi \right\vert ^{2}\left( X_{1}^{\prime }\right)
dX_{1}^{\prime }  \notag \\
&&\times \prod\limits_{k=1}^{n}\frac{d\hat{X}_{k}}{\sharp k!}\times
\prod\limits_{k=2}^{n}A^{l_{k}}\left( X_{k},\hat{X}_{\sum%
\limits_{i=1}^{k-1}l_{i}+1},...,\hat{X}_{\sum%
\limits_{i=1}^{k}l_{i}},X_{k}^{\prime }\right) \left\vert \Psi \right\vert
^{2}\left( X_{k}^{\prime }\right) dX_{k}^{\prime }\times
\prod\limits_{k=2}^{n}\delta \left( X_{k}-\hat{X}_{f\left( k\right) }\right)
\label{sm}
\end{eqnarray}

\subsubsection*{6.2.5 Corrections to the saddle point}

The corrections to the saddle point approximation are obtained by expanding $%
S\left( \Lambda ,\Lambda _{0}^{\dag }+\Lambda ^{\dag }\right) $ around the
solution of (\ref{dsn}) $\Lambda _{0}^{\dag }$. Given that:

\begin{eqnarray*}
&&S\left( \Lambda ,\Lambda _{0}^{\dag }+\Lambda ^{\dag }\right) \simeq \int
\Lambda \left( Z,\theta \right) \left( 1-\left\vert \Psi \right\vert ^{2}%
\hat{T}\right) \Lambda ^{\dag }\left( Z,\theta \right) d\left( Z,\theta
\right) \\
&&-\int \Lambda \left( Z,\theta \right) \hat{T}\left( \theta -\frac{%
\left\vert Z-Z^{\left( 1\right) }\right\vert }{c},Z,Z^{\left( 1\right)
},\omega _{0}+\hat{T}\Lambda _{0}^{\dag }\right) \Lambda ^{\dag }\left(
Z^{\left( 1\right) },\theta -\frac{\left\vert Z-Z^{\left( 1\right)
}\right\vert }{c}\right) dZdZ^{\left( 1\right) }d\theta \\
&&-\int \Lambda \left( Z,\theta \right) \left( \hat{T}\left( \theta -\frac{%
\left\vert Z^{\left( 1\right) }-Z\right\vert }{c},Z^{\left( 1\right)
},Z,\omega _{0}+\hat{T}\left( \Lambda _{0}^{\dag }+\Lambda ^{\dag }\right)
\right) -\hat{T}\left( \theta -\frac{\left\vert Z^{\left( 1\right)
}-Z\right\vert }{c},Z^{\left( 1\right) },Z,\omega _{0}+\hat{T}\Lambda
_{0}^{\dag }\right) \right) \\
&&\times \Lambda _{0}^{\dag }\left( Z^{\left( 1\right) },\theta -\frac{%
\left\vert Z-Z^{\left( 1\right) }\right\vert }{c}\right) dZdZ^{\left(
1\right) }d\theta
\end{eqnarray*}

It yields the second order corrections to $S$:%
\begin{eqnarray*}
&&\hat{S}\left( \Lambda ^{\dag },\Lambda \right) \\
&=&S\left( \Lambda ,\Lambda _{0}^{\dag }+\Lambda ^{\dag }\right) -S\left(
0,\Lambda _{0}^{\dag }\right) \\
&\simeq &\int \Lambda \left( Z,\theta \right) \left( 1-\left\vert \Psi
\right\vert ^{2}\hat{T}\right) \Lambda ^{\dag }\left( Z,\theta \right)
d\left( Z,\theta \right) \\
&&-\int \Lambda \left( Z,\theta \right) \hat{T}\left( \theta -\frac{%
\left\vert Z-Z^{\left( 1\right) }\right\vert }{c},Z,Z^{\left( 1\right)
},\omega _{0}+\hat{T}\Lambda _{0}^{\dag }\right) \Lambda ^{\dag }\left(
Z^{\left( 1\right) },\theta -\frac{\left\vert Z-Z^{\left( 1\right)
}\right\vert }{c}\right) dZdZ^{\left( 1\right) }d\theta \\
&&-\int \Lambda \left( Z,\theta \right) \left( \frac{\delta }{\delta \omega
_{0}\left( Z,\theta \right) }\hat{T}\left( \theta -\frac{\left\vert
Z^{\left( 1\right) }-Z\right\vert }{c},Z^{\left( 1\right) },Z,\omega _{0}+%
\hat{T}\Lambda _{0}\right) \right) \left( \hat{T}\Lambda ^{\dag }\right)
\left( Z,\theta \right) \\
&&\times \Lambda _{0}^{\dag }\left( Z^{\left( 1\right) },\theta -\frac{%
\left\vert Z-Z^{\left( 1\right) }\right\vert }{c}\right) dZdZ^{\left(
1\right) }d\theta \\
&&-\frac{1}{2}\int \Lambda \left( Z,\theta \right) \left( \frac{\delta ^{2}}{%
\delta \omega _{0}^{2}\left( Z,\theta \right) }\hat{T}\left( \theta -\frac{%
\left\vert Z^{\left( 1\right) }-Z\right\vert }{c},Z^{\left( 1\right)
},Z,\omega _{0}+\hat{T}\Lambda _{0}\right) \right) \\
&&\times \left( \left( \hat{T}\Lambda ^{\dag }\right) \left( Z,\theta
\right) \right) ^{2}\Lambda _{0}^{\dag }\left( Z^{\left( 1\right) },\theta -%
\frac{\left\vert Z-Z^{\left( 1\right) }\right\vert }{c}\right) dZdZ^{\left(
1\right) }d\theta
\end{eqnarray*}%
and the quadratic expansion of $\hat{S}$ is:%
\begin{equation*}
\hat{S}\left( \Lambda ^{\dag },\Lambda \right) \simeq \int \Lambda \left(
Z,\theta \right) \left( 1-\left\vert \Psi \right\vert ^{2}\hat{T}_{\omega
_{0}}-\hat{T}_{\omega _{0}+\hat{T}\Lambda _{0}^{\dag }}-\frac{\delta \hat{T}%
_{\omega _{0}+\hat{T}\Lambda _{0}}}{\delta \omega _{0}\left( Z,\theta
\right) }\Lambda _{0}^{\dag }\hat{T}_{\omega _{0}}\right) \Lambda ^{\dag
}\left( Z^{\left( 1\right) },\theta -\frac{\left\vert Z-Z^{\left( 1\right)
}\right\vert }{c}\right) dZdZ^{\left( 1\right) }d\theta
\end{equation*}%
The corrections to $\Lambda _{0}^{\dag }$ \ are computed using this second
order expansion of:%
\begin{equation*}
\int \hat{T}\Lambda ^{\dag }\left( Z,\theta \right) \exp \left( -S\left(
\Lambda \right) +\int \Lambda \left( X,\theta \right) \omega _{0}\left(
J,\theta ,Z\right) \left\vert \Psi \left( J,\theta ,Z\right) \right\vert
^{2}d\left( X,\theta \right) \right) \mathcal{D}\Lambda
\end{equation*}%
around $\Lambda =0$ and $\Lambda ^{\dag }=\Lambda _{0}^{\dag }$ as well as
similar expansion of $\int \exp \left( -S\left( \Lambda \right) \right) 
\mathcal{D}\Lambda $ around its minimum $\Lambda =\Lambda ^{\dag }=0$. Given
that all correlations function with a different number of $\Lambda ^{\dag }$
and $\Lambda $ is null, including the first order corrections yield the
following expression for $\omega $:%
\begin{eqnarray}
\omega &=&\omega _{0}  \label{cqm} \\
&&+\frac{1}{\int \exp \left( -S\left( \Lambda \right) \right) \mathcal{D}%
\Lambda }\hat{T}\Lambda _{0}^{\dag }\int \exp \left( -\int \Lambda \left(
Z,\theta \right) \left( 1-\left( 1+\left\vert \Psi \right\vert ^{2}\right) 
\hat{T}_{\omega _{0}}+\left( 1-\hat{T}_{\omega _{0}+\hat{T}\Lambda
_{0}^{\dag }}-\frac{\delta \hat{T}_{\omega _{0}+\hat{T}\Lambda _{0}}}{\delta
\omega _{0}\left( Z,\theta \right) }\Lambda _{0}^{\dag }\hat{T}_{\omega
_{0}}\right) \right) \right.  \notag \\
&&\times \left. \Lambda ^{\dag }\left( Z^{\left( 1\right) },\theta -\frac{%
\left\vert Z-Z^{\left( 1\right) }\right\vert }{c}\right) dZdZ^{\left(
1\right) }d\theta \right) \mathcal{D}\Lambda  \notag
\end{eqnarray}%
The factor $\int \exp \left( -S\left( \Lambda \right) \right) \mathcal{D}%
\Lambda $ is computed by expanding around $\Lambda _{0}=\Lambda _{0}^{\dag
}=0$:%
\begin{eqnarray*}
\int \exp \left( -S\left( \Lambda \right) \right) \mathcal{D}\Lambda &\simeq
&\int \exp \left( -\int \Lambda \left( X,\theta \right) \left( 1-\left(
1+\left\vert \Psi \right\vert ^{2}\right) \hat{T}\right) \Lambda ^{\dag
}\left( X,\theta \right) d\left( X,\theta \right) \right) \mathcal{D}\Lambda
\\
&=&\det \left( 1-\left( 1+\left\vert \Psi \right\vert ^{2}\right) \hat{T}%
\right) ^{-1}
\end{eqnarray*}%
and the equation (\ref{dst}) is replaced by:%
\begin{eqnarray}
\omega &\simeq &\omega _{0}+\hat{T}\Lambda _{0}^{\dag }\frac{\left( \det
\left( 1-\left( 1+\left\vert \Psi \right\vert ^{2}\right) \hat{T}+\left( 1-%
\hat{T}_{\omega _{0}+\hat{T}\Lambda _{0}^{\dag }}-\frac{\delta \hat{T}%
_{\omega _{0}+\hat{T}\Lambda _{0}}}{\delta \omega _{0}\left( Z,\theta
\right) }\Lambda _{0}^{\dag }\hat{T}_{\omega _{0}}\right) \right) \right)
^{-1}}{\det \left( 1-\left( 1+\left\vert \Psi \right\vert ^{2}\right) \hat{T}%
\right) ^{-1}}  \label{dts} \\
&=&\omega _{0}+\hat{T}\Lambda _{0}^{\dag }\left( \det \left( 1+\left( 1-\hat{%
T}_{\omega _{0}+\hat{T}\Lambda _{0}^{\dag }}-\frac{\delta \hat{T}_{\omega
_{0}+\hat{T}\Lambda _{0}}}{\delta \omega _{0}\left( Z,\theta \right) }%
\Lambda _{0}^{\dag }\hat{T}_{\omega _{0}}\right) \left( 1-\left(
1+\left\vert \Psi \right\vert ^{2}\right) \hat{T}\right) ^{-1}\right)
\right) ^{-1}  \notag
\end{eqnarray}%
Using that $\hat{T}_{\omega _{0}+\hat{T}\Lambda _{0}^{\dag }}$\ is of order $%
\frac{1}{\omega _{0}+\hat{T}\Lambda ^{\dag }}$ and:%
\begin{equation*}
\hat{T}\left( \omega _{0}+\hat{T}\Lambda ^{\dag }\right) \simeq \frac{\omega
_{0}}{\omega _{0}+\hat{T}\Lambda ^{\dag }}\hat{T}
\end{equation*}%
we have:%
\begin{equation*}
\frac{\delta \hat{T}_{\omega _{0}+\hat{T}\Lambda _{0}}}{\delta \omega
_{0}\left( Z,\theta \right) }\simeq -\frac{\omega _{0}}{\left( \omega _{0}+%
\hat{T}\Lambda ^{\dag }\right) ^{2}}\hat{T}
\end{equation*}%
and the factor $-\hat{T}_{\omega _{0}+\hat{T}\Lambda _{0}^{\dag }}-\frac{%
\delta \hat{T}_{\omega _{0}+\hat{T}\Lambda _{0}}}{\delta \omega _{0}\left(
Z,\theta \right) }\Lambda _{0}^{\dag }\hat{T}_{\omega _{0}}$ arising in (\ref%
{cqm}) writes:%
\begin{eqnarray*}
-\hat{T}_{\omega _{0}+\hat{T}\Lambda _{0}^{\dag }}-\frac{\delta \hat{T}%
_{\omega _{0}+\hat{T}\Lambda _{0}}}{\delta \omega _{0}\left( Z,\theta
\right) }\Lambda _{0}^{\dag }\hat{T} &=&-\frac{\omega _{0}}{\omega _{0}+\hat{%
T}\Lambda _{0}^{\dag }}\hat{T}+\frac{\omega _{0}\hat{T}\Lambda _{0}^{\dag }}{%
\left( \omega _{0}+\hat{T}\Lambda _{0}^{\dag }\right) ^{2}}\hat{T} \\
&=&\frac{\omega _{0}^{2}}{\left( \omega _{0}+\hat{T}\Lambda _{0}^{\dag
}\right) ^{2}}\hat{T}
\end{eqnarray*}%
and (\ref{dts}) becomes:%
\begin{equation}
\omega =\omega _{0}+\hat{T}\Lambda _{0}^{\dag }\left( \det \left( 1+\left( 1+%
\frac{\omega _{0}\left( 2\omega _{0}+\hat{T}\Lambda _{0}^{\dag }\right) }{%
\left( \omega _{0}+\hat{T}\Lambda _{0}^{\dag }\right) ^{2}}\hat{T}\left(
1+\left\vert \Psi \right\vert ^{2}\right) \hat{T}\right) ^{-1}\right)
\right) ^{-1}  \label{dns}
\end{equation}%
Equation (\ref{dnt}) is considered along with the defining equation (\ref%
{dpn}) for $\Lambda _{0}^{\dag }$: 
\begin{subequations}
\begin{equation}
\Lambda _{0}^{\dag }-\left( \left\vert \Psi \right\vert ^{2}+\frac{\omega
_{0}}{\omega _{0}+\hat{T}\Lambda _{0}^{\dag }}\right) \hat{T}\Lambda
_{0}^{\dag }-\omega _{0}\left\vert \Psi \right\vert ^{2}=0  \label{dbp}
\end{equation}%
and the relation (\ref{spg}) between background field and frequency: 
\end{subequations}
\begin{equation*}
\left\vert \Psi \right\vert ^{2}=f\left( \omega ,\nabla _{\theta }^{l}\omega
\right)
\end{equation*}%
We set again $\Omega =$ $\omega -\omega _{0}$ and we have:%
\begin{equation*}
\hat{T}\Lambda _{0}^{\dag }\simeq \Omega \det \left( 1+\left( 1+\frac{\omega
_{0}\left( 2\omega _{0}+\Omega \right) }{\left( \omega _{0}+\Omega \right)
^{2}}\hat{T}\left( 1+f\left( \omega ,\nabla _{\theta }^{l}\omega \right)
\right) \hat{T}\right) ^{-1}\right)
\end{equation*}%
so that applying $\hat{T}$ to (\ref{dbp}) yields:%
\begin{eqnarray*}
&&\Omega \det \left( 1+\left( 1+\frac{\omega _{0}\left( 2\omega _{0}+\Omega
\right) }{\left( \omega _{0}+\Omega \right) ^{2}}\hat{T}\left( 1+f\left(
\omega ,\nabla _{\theta }^{l}\omega \right) \right) \hat{T}\right)
^{-1}\right) \\
&=&\hat{T}\left( \left( f\left( \omega ,\nabla _{\theta }^{l}\omega \right) +%
\frac{\omega _{0}}{\omega _{0}+\Omega \det \left( 1+\left( 1+\frac{\omega
_{0}\left( 2\omega _{0}+\Omega \right) }{\left( \omega _{0}+\Omega \right)
^{2}}\hat{T}\left( 1+f\left( \omega ,\nabla _{\theta }^{l}\omega \right)
\right) \hat{T}\right) ^{-1}\right) }\right) \right. \\
&&\left. \times \Omega \det \left( 1+\left( 1+\frac{\omega _{0}\left(
2\omega _{0}+\Omega \right) }{\left( \omega _{0}+\Omega \right) ^{2}}\hat{T}%
\left( 1+f\left( \omega ,\nabla _{\theta }^{l}\omega \right) \right) \hat{T}%
\right) ^{-1}\right) +\omega _{0}f\left( \omega ,\nabla _{\theta }^{l}\omega
\right) \right)
\end{eqnarray*}

\subsubsection*{6.2.6 Time dependent background field}

For non constant background fields, the link between $\Psi \left( \theta
,Z\right) $ and $\omega ^{-1}\left( J\left( \theta \right) ,\theta ,Z,%
\mathcal{G}_{0}+\left\vert \Psi \right\vert ^{2}\right) $ is:%
\begin{equation*}
\Psi =\delta \Psi +\Psi _{0}
\end{equation*}%
and thus: 
\begin{eqnarray*}
\Psi \left( \theta ,Z\right) &=&\left( \frac{\left( \nabla _{\theta }\left( 
\frac{\sigma _{\theta }^{2}}{2}\nabla _{\theta }-\omega ^{-1}\left( J\left(
\theta \right) ,\theta ,Z,\mathcal{G}_{0}+\left\vert \Psi \right\vert
^{2}\right) \right) \right) }{U^{\prime \prime }\left( X_{0}\right) -\left(
\nabla _{\theta }\left( \frac{\sigma _{\theta }^{2}}{2}\nabla _{\theta
}-\omega ^{-1}\left( J\left( \theta \right) ,\theta ,Z,\mathcal{G}%
_{0}+\left\vert \Psi \right\vert ^{2}\right) \right) \right) }\right) \Psi
_{0}\left( \theta ,Z\right) +\Psi _{0}\left( \theta ,Z\right) \\
&\simeq &\left( \frac{\nabla _{\theta }\omega \left( J\left( \theta \right)
,\theta ,Z,\mathcal{G}_{0}+\left\vert \Psi \right\vert ^{2}\right) }{\omega
^{2}\left( J\left( \theta \right) ,\theta ,Z,\mathcal{G}_{0}+\left\vert \Psi
\right\vert ^{2}\right) U^{\prime \prime }\left( X_{0}\right) -\nabla
_{\theta }\omega \left( J\left( \theta \right) ,\theta ,Z,\mathcal{G}%
_{0}+\left\vert \Psi \right\vert ^{2}\right) }\right) \Psi _{0}\left( \theta
,Z\right) +\Psi _{0}\left( \theta ,Z\right)
\end{eqnarray*}%
\begin{equation*}
\Psi ^{\dag }\left( \theta ,Z\right) =\Psi _{0}^{\dag }\left( \theta
,Z\right)
\end{equation*}%
\begin{equation*}
\left\vert \Psi \left( \theta ,Z\right) \right\vert ^{2}\simeq \left\vert
\Psi _{0}\left( \theta ,Z\right) \right\vert ^{2}+\Psi _{0}^{\dag }\left(
\theta ,Z\right) \left( \frac{\nabla _{\theta }\omega \left( J\left( \theta
\right) ,\theta ,Z,\mathcal{G}_{0}+\left\vert \Psi \right\vert ^{2}\right) }{%
\omega ^{2}\left( J\left( \theta \right) ,\theta ,Z,\mathcal{G}%
_{0}+\left\vert \Psi \right\vert ^{2}\right) U^{\prime \prime }\left(
X_{0}\right) -\nabla _{\theta }\omega \left( J\left( \theta \right) ,\theta
,Z,\mathcal{G}_{0}+\left\vert \Psi \right\vert ^{2}\right) }\right) \Psi
_{0}\left( \theta ,Z\right)
\end{equation*}%
Equation (\ref{mqn}) leads to:%
\begin{equation*}
\Omega -\hat{T}\left( \Omega +\omega _{0}\right) \left( \left\vert \Psi
_{0}\left( \theta ,Z\right) \right\vert ^{2}+\Psi _{0}^{\dag }\left( \theta
,Z\right) \frac{\nabla _{\theta }\left( \omega _{0}+\Omega \right) }{\left(
\omega _{0}+\Omega \right) ^{2}U^{\prime \prime }\left( X_{0}\right) -\nabla
_{\theta }\left( \omega _{0}+\Omega \right) }\Psi _{0}\left( \theta
,Z\right) \right) -\hat{T}\frac{\omega _{0}\Omega }{\omega _{0}+\Omega }=0
\end{equation*}%
At the lowest order, this becomes:%
\begin{equation*}
\Omega -\hat{T}\left( \Omega +\omega _{0}\right) \left\vert \Psi _{0}\left(
\theta ,Z\right) \right\vert ^{2}-\hat{T}\frac{\omega _{0}\Omega }{\omega
_{0}+\Omega }=0
\end{equation*}%
\begin{equation*}
\Omega -\hat{T}\Omega \left\vert \Psi _{0}\left( \theta ,Z\right)
\right\vert ^{2}-\hat{T}\frac{\omega _{0}\Omega }{\omega _{0}+\Omega }=\hat{T%
}\left( \omega _{0}\left\vert \Psi _{0}\left( \theta ,Z\right) \right\vert
^{2}\right)
\end{equation*}%
for $\Omega <<\omega _{0}$, this becomes:%
\begin{equation*}
\left( 1-\hat{T}\left( 1+\left\vert \Psi _{0}\left( \theta ,Z\right)
\right\vert ^{2}\right) \right) \Omega =\hat{T}\left( \omega _{0}\left\vert
\Psi _{0}\left( \theta ,Z\right) \right\vert ^{2}\right)
\end{equation*}%
with solution:%
\begin{eqnarray}
\Omega &=&\frac{1}{1-\hat{T}\left( 1+\left\vert \Psi _{0}\left( \theta
,Z\right) \right\vert ^{2}\right) }\hat{T}\left( \omega _{0}\left\vert \Psi
_{0}\left( \theta ,Z\right) \right\vert ^{2}\right)  \label{slG} \\
&=&\frac{1}{1-\frac{-\hat{T}\left\vert \Psi _{0}\left( \theta ,Z\right)
\right\vert ^{2}}{1-T}}\frac{\hat{T}}{1-\hat{T}}\omega _{0}\left\vert \Psi
_{0}\left( \theta ,Z\right) \right\vert ^{2}  \notag
\end{eqnarray}%
The operator $\frac{\hat{T}}{1-\hat{T}}$ summing over all lines between two
given points has been estimated in Appendix 5. Between two points $\left(
Z_{1},\theta _{1}\right) $ and $\left( Z_{2},\theta _{2}\right) $ the factor
associated to the sum of paths is of order: 
\begin{equation}
\frac{\hat{T}}{1-\hat{T}}\left( \left( Z_{1},\theta _{1}\right) ,\left(
Z_{2},\theta _{2}\right) \right) =\frac{\exp \left( -c\left( \theta
_{2}-\theta _{1}\right) -\alpha \left( c^{2}\left( \theta _{2}-\theta
_{1}\right) ^{2}-\left\vert Z_{2}-Z_{1}\right\vert ^{2}\right) \right) }{A}%
H\left( \theta _{2}-\theta _{1}-\frac{\left\vert Z_{2}-Z_{1}\right\vert }{c}%
\right)
\end{equation}%
with $\alpha $ and $A$ are some parameters and $H$ is the heaviside
function. The insertion of the field in (\ref{slG}) yields:

\begin{eqnarray}
\Omega &=&\int \sum_{k=0}^{\infty }\frac{\exp \left( -\alpha \left( \left( 1+%
\frac{c}{\alpha }\right) \left( \theta _{2}-\theta _{1}\right)
-\sum_{l=0}^{k}\frac{\left\vert Z_{l}-Z_{l+1}\right\vert }{c}\right) \right) 
}{A^{k+1}} \\
&&\times \left( \dprod\limits_{l=1}^{k}\int \left\vert \Psi \left( \theta
-l_{l},Z_{l}\right) \right\vert ^{2}dZ_{l}dl_{l}\right) \omega _{0}\left(
J,\theta -l_{k},Z_{k}\right) \left\vert \Psi \left( \theta
-l_{k},Z_{k}\right) \right\vert ^{2}  \notag
\end{eqnarray}%
This value of $\Omega $ represents the fluctuations in frequencies due to
the time dependency in potential. In first approximation, this combines with

\subsection*{6.3 Extension: Excitatory vs inhibitory interaction}

\subsubsection*{6.3.1 Series expansion for the frequencies}

The method of section 6.2 can be extended straightforwarly in the case of
two types of interactions. We will derive a path integral description for
the frequencies.

We consider $n$ populations, each caracterized by their frequencies $%
i=1,...,n$. They interact positively or negatively. Each population is
defined by a field $\Psi _{i}$ and freqncies $\omega _{i}\left( \theta
,Z\right) $. Equations for frequencies are defined by:

\begin{eqnarray}
\omega _{i}\left( \theta ,Z\right) &=&F_{i}\left( J\left( \theta \right) +%
\frac{\kappa }{N}\int T\left( Z,Z_{1}\right) \frac{\omega _{j}\left( \theta -%
\frac{\left\vert Z-Z_{1}\right\vert }{c},Z_{1}\right) }{\omega _{i}\left(
\theta ,Z\right) }G^{ij}\right.  \label{fml} \\
&&\times \left. W\left( \frac{\omega _{i}\left( \theta ,Z\right) }{\omega
_{j}\left( \theta -\frac{\left\vert Z-Z_{1}\right\vert }{c},Z_{1}\right) }%
\right) \left( \mathcal{\bar{G}}_{0j}\left( 0,Z_{1}\right) +\left\vert \Psi
_{j}\left( \theta -\frac{\left\vert Z-Z_{1}\right\vert }{c},Z_{1}\right)
\right\vert ^{2}\right) dZ_{1}\right)  \notag
\end{eqnarray}%
For example, if $i,j=1,2$, a matrix $g$ of the form:%
\begin{equation*}
G=\left( 
\begin{array}{cc}
1 & -g \\ 
-g & 1%
\end{array}%
\right)
\end{equation*}%
represents inhibitory interactions between the two populations. More
generally, the matrix $G$ is $n\times n$ with coefficients in the interval $%
\left[ -1,1\right] $. The sum over indices is understood for $j$. The
resolution of (\ref{fml}) follows the same principle as for (\ref{qtf}),
with a $n$ components vector of frequencies $\omega \left( J,\theta
,Z\right) $. Writing the series expansion for $\omega \left( J,\theta
,Z\right) $: 
\begin{eqnarray}
\omega \left( J,\theta ,Z\right) &=&\sum_{r}\left( \frac{\delta ^{r}\omega
\left( J,\theta ,Z\right) }{\dprod\limits_{i=1}^{r}\delta \left\vert \Psi
\left( \theta -l_{i},Z_{i}\right) \right\vert ^{2}}\right) _{\left\vert \Psi
\right\vert ^{2}=0}\dprod\limits_{i=1}^{r}\left\vert \Psi \left( \theta
-l_{i},Z_{i}\right) \right\vert ^{2}  \label{mgV} \\
&=&\left( \sum_{m=1}^{n}\sum_{i=1}^{m}\sum_{\left(
line_{1},...,line_{m}\right) }\dprod\limits_{i}F\left( line_{i}\right)
\dprod\limits_{B}F\left( B\right) \right) \dprod\limits_{i=1}^{n}\left\vert
\Psi \left( \theta -l_{i},Z_{i}\right) \right\vert ^{2}  \notag
\end{eqnarray}%
where:%
\begin{equation}
\left( \frac{\delta ^{r}\omega \left( J,\theta ,Z\right) }{%
\dprod\limits_{i=1}^{r}\delta \left\vert \Psi \left( \theta
-l_{i},Z_{i}\right) \right\vert ^{2}}\right) _{\left\vert \Psi \right\vert
^{2}=0}  \label{drT}
\end{equation}%
and:%
\begin{equation*}
\left\vert \Psi \left( \theta -l_{i},Z_{i}\right) \right\vert ^{2}
\end{equation*}%
are considered as $\left( 1,r\right) $ and $\left( 1,0\right) $\ tensors
respectively, the expansion of the first order derivative is similar to (\ref%
{xpng}) and is given by:%
\begin{eqnarray*}
\left( \frac{\delta \omega \left( J,\theta ,Z\right) }{\delta \left\vert
\Psi \left( \theta -l_{1},Z_{1}\right) \right\vert ^{2}}\right) _{\left\vert
\Psi \right\vert ^{2}=0} &=&\sum_{n=1}^{\infty }\int \dprod\limits_{l=1}^{n}%
\hat{T}\left( \theta -\sum_{j=1}^{l-1}\frac{\left\vert Z^{\left( j-1\right)
}-Z^{\left( j\right) }\right\vert }{c},Z^{\left( l-1\right) },Z^{\left(
l\right) },\omega _{0},0\right) \\
&&\times \Omega _{0}\left( J,\theta -\sum_{l=1}^{n}\frac{\left\vert
Z^{\left( l-1\right) }-Z^{\left( l\right) }\right\vert }{c},Z_{1}\right)
\times \delta \left( l_{1}-\sum_{l=1}^{n}\frac{\left\vert Z^{\left(
l-1\right) }-Z^{\left( l\right) }\right\vert }{c}\right)
\dprod\limits_{l=1}^{n-1}dZ^{\left( l\right) }
\end{eqnarray*}%
with $\omega _{0}$ a $n$ component vector describing a solution for $%
\left\vert \Psi \right\vert ^{2}=0$ and $\Omega _{0}\left( J,\theta
-\sum_{l=1}^{n}\frac{\left\vert Z^{\left( l-1\right) }-Z^{\left( l\right)
}\right\vert }{c},Z_{1}\right) $ is a diagonal matrix with components $%
\omega _{0i}\left( J,\theta -\sum_{l=1}^{n}\frac{\left\vert Z^{\left(
l-1\right) }-Z^{\left( l\right) }\right\vert }{c},Z_{1}\right) $.

For practical purposes, we also define the diagonal matrix $D\left(
\left\vert \Psi \right\vert ^{2}\right) $ with $\left\vert \Psi
_{i}\right\vert ^{2}$ on the diagonal. More generally, for any expression $%
H\left( \omega _{0i},\left\vert \Psi _{i}\right\vert ^{2}\right) $, we
define $D\left( H\left( \omega _{0},\left\vert \Psi \right\vert ^{2}\right)
\right) $ the diagonal matrix with components $H\left( \omega
_{0i},\left\vert \Psi _{i}\right\vert ^{2}\right) $.

The quantity $\Omega \left( J,\theta ,Z\right) $ $\left\vert \Psi
\right\vert ^{2}$ is a vector with components $\omega _{i}\left( J,\theta
,Z\right) $ $\left\vert \Psi _{i}\right\vert ^{2}$. The expressions $\left( 
\frac{\delta \omega \left( J,\theta ,Z\right) }{\delta \left\vert \Psi
\left( \theta -l_{1},Z_{1}\right) \right\vert ^{2}}\right) _{\left\vert \Psi
\right\vert ^{2}=0}$ and $\hat{T}\left( \theta -\sum_{j=1}^{l-1}\frac{%
\left\vert Z^{\left( j-1\right) }-Z^{\left( j\right) }\right\vert }{c}%
,Z^{\left( l-1\right) },Z^{\left( l\right) },\omega _{0},0\right) $ are $%
n\times n$ matrices:%
\begin{equation*}
\left( \left( \frac{\delta \omega \left( J,\theta ,Z\right) }{\delta
\left\vert \Psi \left( \theta -l_{1},Z_{1}\right) \right\vert ^{2}}\right)
_{\left\vert \Psi \right\vert ^{2}=0}\right) _{ij}=\left( \frac{\delta
\omega _{i}\left( J,\theta ,Z\right) }{\delta \left\vert \Psi _{j}\left(
\theta -l_{1},Z_{1}\right) \right\vert ^{2}}\right) _{\left\vert \Psi
\right\vert ^{2}=0}
\end{equation*}%
and:%
\begin{eqnarray*}
&&\hat{T}_{ij}\left( \theta ,Z,Z_{1}\omega ,\Psi \right) \\
&=&\frac{G^{ij}\frac{\kappa }{N}\omega _{i}\left( J,\theta ,Z\right) T\left(
Z,Z_{1}\right) F_{i}^{\prime }\left[ J,\omega ,\theta ,Z,\Psi \right] }{%
\omega _{i}^{2}\left( J,\theta ,Z\right) +G^{ij}\left( \int \frac{\kappa }{N}%
\omega _{j}\left( J,\theta -\frac{\left\vert Z-Z^{\prime }\right\vert }{c}%
,Z^{\prime }\right) \left( \mathcal{\bar{G}}_{0j}\left( 0,Z^{\prime }\right)
+\left\vert \Psi _{j}\left( \theta -\frac{\left\vert Z-Z^{\prime
}\right\vert }{c},Z^{\prime }\right) \right\vert ^{2}\right) T\left(
Z,Z^{\prime }\right) dZ^{\prime }\right) F^{\prime }\left[ J,\omega ,\theta
,Z,\Psi \right] }
\end{eqnarray*}%
The successive derivatives (\ref{drT}) in (\ref{mgV}) are similar to (\ref%
{rdt}) along with (\ref{lf}) and (\ref{rc}):%
\begin{equation}
\left( \frac{\delta ^{n}\omega \left( J,\theta ,Z\right) }{%
\dprod\limits_{i=1}^{n}\delta \left\vert \Psi \left( \theta
-l_{i},Z_{i}\right) \right\vert ^{2}}\right) _{\left\vert \Psi \right\vert
^{2}=0}=\left( \sum_{m=1}^{n}\sum_{i=1}^{m}\sum_{\left(
line_{1},...,line_{m}\right) }\dprod\limits_{i}F\left( line_{i}\right)
\dprod\limits_{B}F\left( B\right) \right)  \label{srM}
\end{equation}%
where $F\left( line_{i}\right) $ is $n\times n$, in other words a $\left(
1,1\right) $ tensor, given by:

\begin{equation}
\left[ F\left( line_{i}\right) \right] _{ab}=\left[ \dprod%
\limits_{l=1}^{L_{i}}\hat{T}\left( \theta -\sum_{j=1}^{l-1}\frac{\left\vert
Z^{\left( j-1\right) }-Z^{\left( j\right) }\right\vert }{c},Z^{\left(
l-1\right) },Z^{\left( l\right) },\omega _{0},\Psi \right) \right] _{ab}%
\frac{\omega _{0b}\left( J,\theta -\sum_{l=1}^{L_{i}}\frac{\left\vert
Z^{\left( l-1\right) }-Z^{\left( l\right) }\right\vert }{c},Z_{i}\right) }{%
\mathcal{\bar{G}}_{0}\left( 0,Z_{i}\right) }
\end{equation}%
To each branching point $\left( X,\theta \right) =B$ of valence $k+2$
arising in the expansion, we associate the $\left( 1,k+1\right) $ tensor:%
\begin{equation}
\left[ F\left( \left( X,\theta \right) \right) \right] _{abc_{1},...,c_{k}}=%
\frac{\delta ^{k}\left( \frac{\frac{\kappa }{N}Tab\left( Z,Z^{\left(
l\right) }\right) F^{\prime }\left[ J,\theta ,\omega _{0},Z^{\left( l\right)
}\right] \mathcal{\bar{G}}_{0}\left( 0,Z^{\left( l\right) }\right) }{\omega
_{0a}\left( J,\theta ,Z^{\left( l\right) }\right) }\right) }{\delta \omega
_{0c_{1}}\left( J,\theta ,Z^{\left( l\right) }\right) ...\delta \omega
_{0c_{k}}\left( J,\theta ,Z^{\left( l\right) }\right) }
\end{equation}

We attach $1$ line, coming in, and $k+1$ lines, coming out, to each
branching point. As a consequence, the contraction of a branching point of
valence $k+2$ and $k+2$ lines yields a $\left( 1,k+1\right) $ tensor. The
factor associated to the sum of single lines (\ref{rl}) crossing the points $%
Z_{k}$ generalizes straightforwardly and is given by:%
\begin{eqnarray}
&&\hat{T}\left( 1-\hat{T}\right) ^{-1}\dprod\limits_{l=1}^{n-1}\left\{
\left( D\left( \left\vert \Psi \left( \theta -l_{l},Z_{l}\right) \right\vert
^{2}\right) dZ_{l}dl_{l}\right) \hat{T}\left( 1-\hat{T}\right) ^{-1}\right\}
D\left( \left\vert \Psi \left( \theta -l_{n},Z_{n}\right) \right\vert
^{2}\omega _{0}\left( J,\theta -l_{n},Z_{n}\right) \right)  \notag \\
&=&\hat{T}\left( 1-\hat{T}\right) ^{-1}\frac{1}{1-D\left( \left\vert \Psi
\left( \theta ,Z\right) \right\vert ^{2}\right) \hat{T}\left( 1-\hat{T}%
\right) ^{-1}}D\left( \left\vert \Psi \left( \theta -l_{n},Z_{n}\right)
\right\vert ^{2}\omega _{0}\left( J,\theta -l_{n},Z_{n}\right) \right) 
\notag \\
&=&\hat{T}\frac{1}{1-\left( 1+D\left( \left\vert \Psi \right\vert
^{2}\right) \right) \hat{T}}D\left( \left\vert \Psi \left( \theta
-l_{n},Z_{n}\right) \right\vert ^{2}\omega _{0}\left( J,\theta
-l_{n},Z_{n}\right) \right)
\end{eqnarray}

\subsubsection*{6.3.2 Path integral form for the frequencies}

Then, as in section 6.2, the series expansion (\ref{srM}) can be reordered
to compute $\omega \left( J,\theta ,Z\right) $ as a path integral for the
action of an auxiliary field $\left( \Lambda ,\Lambda ^{\dag }\right) $ with 
$n$ components. The result is the same as in section 6.2. The action $%
S\left( \Lambda \right) $\ is:%
\begin{eqnarray*}
S\left( \Lambda \right) &=&\int \Lambda \left( Z,\theta \right) \left(
1-D\left( \left\vert \Psi \right\vert ^{2}\right) \hat{T}\right) \Lambda
^{\dag }\left( Z,\theta \right) d\left( Z,\theta \right) \\
&&-\int \Lambda \left( Z,\theta \right) \hat{T}\left( \theta -\frac{%
\left\vert Z^{\left( 1\right) }-Z\right\vert }{c},Z^{\left( 1\right)
},Z,\omega _{0}+\hat{T}\Lambda ^{\dag }\right) \Lambda ^{\dag }\left(
Z^{\left( 1\right) },\theta -\frac{\left\vert Z-Z^{\left( 1\right)
}\right\vert }{c}\right) dZdZ^{\left( 1\right) }d\theta ^{\left( 1\right) }
\end{eqnarray*}%
where $\Lambda \left( Z,\theta \right) $ is a two components vector, and $%
\Lambda ^{\dag }\left( Z,\theta \right) $ is the hermitian conjugate. The
frequency vector is thus given by the integral:%
\begin{equation}
\omega \left( J,\theta ,Z\right) =\omega _{0}\left( J,\theta ,Z\right) +%
\frac{\int \hat{T}\Lambda ^{\dag }\left( Z,\theta \right) \exp \left(
-S\left( \Lambda \right) +\int \Lambda \left( X,\theta \right) D\left(
\left\vert \Psi \right\vert ^{2}\omega _{0}\left( J,\theta ,Z\right) \right)
d\left( X,\theta \right) \right) \mathcal{D}\Lambda }{\exp \left( -S\left(
\Lambda \right) \right) \mathcal{D}\Lambda }  \label{frQ}
\end{equation}%
\ 

\subsubsection*{6.3.3 Saddle path approximation}

The solution of (\ref{frQ}) is obtained in first approximation by
considering that $\Lambda ^{\dag }$ satisfies the saddle point
approximation: 
\begin{equation}
\left( \left( 1-D\left( \left\vert \Psi \right\vert ^{2}\right) \hat{T}%
\right) \Lambda ^{\dag }\right) \left( Z,\theta \right) -\left( \hat{T}%
_{\omega _{0}+\hat{T}\left( \omega _{0}\left\vert \Psi \right\vert
^{2}\right) }\Lambda ^{\dag }\right) \left( Z,\theta \right) -D\left(
\left\vert \Psi \right\vert ^{2}\omega _{0}\right) =0  \label{sdP}
\end{equation}%
In the sequel, we will define the vector:%
\begin{equation*}
\frac{\omega \left( J,\theta ,Z\right) }{\omega \left( J,\theta ,Z\right) +%
\hat{T}\left( \left( \left\vert \Psi \right\vert ^{2}\omega _{0}\right)
\right) }
\end{equation*}%
as the vector with components:%
\begin{equation*}
\frac{\omega _{i}\left( J,\theta ,Z\right) }{\omega _{i}\left( J,\theta
,Z\right) +\left( \hat{T}\left( \left( \left\vert \Psi \right\vert
^{2}\omega _{0}\right) \right) \right) _{i}}
\end{equation*}%
More generally, we will define for any $F$, the vector $VF\left( \omega
,\Psi ,...\right) $ with components $F\left( \omega _{i},\Psi
_{i},...\right) $.

Using that:%
\begin{equation*}
\hat{T}_{\omega _{0}+\hat{T}\left( \omega _{0}\left\vert \Psi \right\vert
^{2}\right) }\simeq D\left( \frac{\omega \left( J,\theta ,Z\right) }{\omega
\left( J,\theta ,Z\right) +\hat{T}\left( \left( \left\vert \Psi \right\vert
^{2}\omega _{0}\right) \right) }\right) \hat{T}_{\omega _{0}}
\end{equation*}%
the saddle point equation (\ref{sdP}) becomes: 
\begin{equation}
\left( \left( 1-D\left( \left\vert \Psi \right\vert ^{2}\right) \hat{T}%
\right) \Lambda ^{\dag }\right) \left( Z,\theta \right) -D\left( \frac{%
\omega _{0}}{\omega _{0}+\hat{T}\Lambda ^{\dag }}\right) \hat{T}\Lambda
^{\dag }\left( Z,\theta \right) -D\left( \omega _{0}\right) \left\vert \Psi
\right\vert ^{2}\simeq 0  \label{vsr}
\end{equation}%
Equation (\ref{vsr}) can be used in two different ways: first by writing a
non-local equation for $\omega \left( J,\theta ,Z\right) $ and second by
solving recursively (\ref{vsr}) for an externaly-shaped background field.

\subsubsection*{6.3.4 Non local equation for $\protect\omega \left( J,%
\protect\theta ,Z\right) $}

As for the basic case, (\ref{vsr}) can be rewritten as an equation for $%
\omega $. Actually, under the saddle point approximation:%
\begin{equation*}
\omega \left( J,\theta ,Z\right) =\omega _{0}\left( J,\theta ,Z\right) +\hat{%
T}\Lambda ^{\dag }\left( Z,\theta \right)
\end{equation*}%
and:%
\begin{equation*}
\hat{T}\Lambda ^{\dag }\left( Z,\theta \right) =\omega \left( J,\theta
,Z\right) -\omega _{0}\left( J,\theta ,Z\right) \equiv \Omega \left(
J,\theta ,Z\right)
\end{equation*}%
so that (\ref{dsn}) writes:%
\begin{equation*}
\Lambda ^{\dag }-\left( \Omega +\omega _{0}\right) \left\vert \Psi
\right\vert ^{2}-\frac{\omega _{0}}{\omega _{0}+\Omega }\Omega =0
\end{equation*}%
Applying the operator $\hat{T}$ on the left leads to: 
\begin{equation}
\Omega -\hat{T}\left( \Omega +\omega _{0}\right) \left\vert \Psi \right\vert
^{2}-\hat{T}\left( \frac{\omega _{0}}{\omega _{0}+\Omega }\right) \Omega =0
\label{dTN}
\end{equation}%
where $\left( \Omega +\omega _{0}\right) \left\vert \Psi \right\vert ^{2}$
and $\left( \frac{\omega _{0}}{\omega _{0}+\Omega }\right) \Omega $ are
defined as the vectors with components $\left( \Omega +\omega _{0}\right)
_{i}\left\vert \Psi \right\vert _{i}^{2}$ and $\left( \frac{\omega _{i0}}{%
\omega _{i0}+\Omega _{i}}\right) \Omega _{i}$ respectively.

Then we can generalize the expression (\ref{psv}) defining the background
field with several component $\Psi _{i}\left( \theta ,Z\right) $. We assume
a stabilization potential $U_{i}$ for each component, with minimum $X_{i0}$.
Using the notation $V$ defined after equation (\ref{sdP}), the expression
for the vector background field becomes:%
\begin{eqnarray*}
\Psi \left( \theta ,Z\right) &=&V\left( \frac{\nabla _{\theta }\omega \left(
J\left( \theta \right) ,\theta ,Z,\mathcal{G}_{0}+\left\vert \Psi
_{0}\right\vert ^{2}\right) }{U^{\prime \prime }\left( X_{0}\right) \omega
^{2}\left( J\left( \theta \right) ,\theta ,Z,\mathcal{G}_{0}+\left\vert \Psi
_{0}\right\vert ^{2}\right) }\Psi _{0}\left( \theta ,Z\right) \right) \\
&=&V\left( \frac{\nabla _{\theta }\omega }{U^{\prime \prime }\left(
X_{0}\right) \omega ^{2}}\Psi _{0}\left( \theta ,Z\right) \right) \\
&=&V\left( \frac{X_{0}}{U^{\prime \prime }\left( X_{0}\right) \left( \omega
_{0}+\Omega \right) ^{2}}\nabla _{\theta }\Omega \right)
\end{eqnarray*}%
and the vector of squared norms is:%
\begin{equation*}
\left\vert \Psi \right\vert ^{2}=V\left( \frac{X_{0}^{2}}{\left( \omega
_{0}+\Omega \right) ^{2}U^{\prime \prime }\left( X_{0}\right) }\nabla
_{\theta }\Omega \right)
\end{equation*}%
Equation (\ref{dqt}) is then replaced by:%
\begin{equation}
\Omega -\hat{T}V\left( \frac{\frac{X_{0}^{2}}{U^{\prime \prime }\left(
X_{0}\right) }\nabla _{\theta }\Omega +\omega _{0}\Omega }{\omega
_{0}+\Omega }\right) =0  \label{qnt}
\end{equation}

\subsubsection*{6.3.5 Recursive solution of (\protect\ref{vsr})}

Alternatively (\ref{vsr}) can be solved recursively for a given bachground
field. As in the one component field case, we find in first approximation:%
\begin{eqnarray}
\hat{T}\Lambda ^{\dag } &=&A\frac{1}{1-\left( \hat{T}_{\omega _{0}+A\omega
_{0}\left\vert \Psi \right\vert ^{2}}-\hat{T}\right) \hat{T}^{-1}A}\omega
_{0}\left\vert \Psi \right\vert ^{2}  \label{qvn} \\
&\simeq &A\frac{1}{1-D\left( \frac{\omega _{0}}{\omega _{0}+A\omega
_{0}\left\vert \Psi \right\vert ^{2}}\right) A}\omega _{0}\left\vert \Psi
\right\vert ^{2}  \notag
\end{eqnarray}%
with:%
\begin{eqnarray}
A &=&\frac{\hat{T}}{1-\left( 1+D\left( \left\vert \Psi \right\vert
^{2}\right) \right) \hat{T}}=\frac{1}{\left( 1+D\left( \left\vert \Psi
\right\vert ^{2}\right) \right) }\frac{\left( 1+D\left( \left\vert \Psi
\right\vert ^{2}\right) \right) \hat{T}}{1-\left( 1+D\left( \left\vert \Psi
\right\vert ^{2}\right) \right) \hat{T}}  \label{frC} \\
&=&\frac{\hat{T}}{1-\hat{T}-D\left( \left\vert \Psi \right\vert ^{2}\right) 
\hat{T}}  \notag \\
&=&\frac{\hat{T}}{1-\hat{T}}\sum_{n\geqslant 0}\left( D\left( \left\vert
\Psi \right\vert ^{2}\right) \frac{\hat{T}}{1-\hat{T}}\right) ^{n}  \notag
\end{eqnarray}%
and the generalization of (\ref{fpr}) is obtained by diagonalization of $%
\hat{T}$.

\paragraph*{6.3.5.1 Case $n=2$}

To obtain explicit formula, we consider that $n=2$, that is, there are two
type of cells.

Writing: 
\begin{equation*}
\left( 1+D\left( \left\vert \Psi \right\vert ^{2}\right) \right) \hat{T}%
=\left( 
\begin{array}{cc}
\hat{T}_{1}\left( \left( 1+\left( \left\vert \Psi _{1}\right\vert
^{2}\right) \right) \omega _{01}\right) & -g\hat{T}_{2}\left( \left(
1+\left( \left\vert \Psi _{2}\right\vert ^{2}\right) \right) \omega
_{02}\right) \\ 
-g\hat{T}_{1}\left( \left( 1+\left( \left\vert \Psi _{1}\right\vert
^{2}\right) \right) \omega _{01}\right) & \hat{T}_{2}\left( \left( 1+\left(
\left\vert \Psi _{2}\right\vert ^{2}\right) \right) \omega _{02}\right)%
\end{array}%
\right)
\end{equation*}%
and assuming $\omega _{01}$ and $\omega _{02}$ changing slowly in time, we
have:%
\begin{equation*}
\left( 1+D\left( \left\vert \Psi \right\vert ^{2}\right) \right) \hat{T}=U%
\hat{T}_{D}U^{-1}
\end{equation*}%
$\allowbreak \allowbreak $%
\begin{eqnarray*}
\hat{T}_{D} &=&\left( 
\begin{array}{cc}
\frac{1}{2}\left( \hat{T}_{1}+\hat{T}_{2}-\sqrt{4g^{2}\hat{T}_{1}\hat{T}%
_{2}+\left( \hat{T}_{1}-\hat{T}_{2}\right) ^{2}}\right) & 0 \\ 
0 & \frac{1}{2}\left( \hat{T}_{1}+\hat{T}_{2}+\sqrt{4g^{2}\hat{T}_{1}\hat{T}%
_{2}+\left( \hat{T}_{1}-\hat{T}_{2}\right) ^{2}}\right)%
\end{array}%
\right) \\
U &=&\left( 
\begin{array}{cc}
-\frac{1}{2g}\left( \hat{T}_{1}-\hat{T}_{2}-\sqrt{4g^{2}\hat{T}_{1}\hat{T}%
_{2}+\left( \hat{T}_{1}-\hat{T}_{2}\right) ^{2}}\right) & \hat{T}_{2} \\ 
\hat{T}_{1} & \frac{1}{2g}\left( \hat{T}_{1}-\hat{T}_{2}-\sqrt{4g^{2}\hat{T}%
_{1}\hat{T}_{2}+\left( \hat{T}_{1}-\hat{T}_{2}\right) ^{2}}\right)%
\end{array}%
\right)
\end{eqnarray*}%
As a consequence:%
\begin{equation*}
\hat{T}=UD\left( \frac{\exp \left( -cl_{1}-\alpha \left( \hat{T}_{D}\right)
\left( \left( cl_{1}\right) ^{2}-\left\vert Z-Z_{1}\right\vert ^{2}\right)
\right) }{B\left( \hat{T}_{D}\right) }H\left( cl_{1}-\left\vert
Z-Z_{1}\right\vert \right) \right) U^{-1}
\end{equation*}%
with $\alpha \left( \hat{T}\right) $ and $B\left( \hat{T}\right) $ are
vectors. That is, given our conventions:%
\begin{equation*}
\hat{T}=U\left( 
\begin{array}{cc}
\frac{\exp \left( -cl_{1}-\alpha _{1}\left( \hat{T}_{D}\right) \left( \left(
cl_{1}\right) ^{2}-\left\vert Z-Z_{1}\right\vert ^{2}\right) \right) }{%
B_{1}\left( \hat{T}\right) } & 0 \\ 
0 & \frac{\exp \left( -cl_{1}-\alpha _{2}\left( \hat{T}_{D}\right) \left(
\left( cl_{1}\right) ^{2}-\left\vert Z-Z_{1}\right\vert ^{2}\right) \right) 
}{B_{2}\left( \hat{T}\right) }%
\end{array}%
\right) U^{-1}H\left( cl_{1}-\left\vert Z-Z_{1}\right\vert \right)
\end{equation*}%
For transfers functions $T_{i}\left( Z,Z_{1}\right) $ that are proportional $%
T_{i}\left( Z,Z_{1}\right) =C_{i}T_{0}\left( Z,Z_{1}\right) $, the change of
basis yields the diagonalized transfer function:%
\begin{equation*}
T_{D}\left( Z,Z_{1}\right) =\left( 
\begin{array}{cc}
\frac{1}{2}\left( C_{1}+C_{2}-\sqrt{4g_{1}^{2}C_{1}C_{2}+\left(
C_{1}-C_{2}\right) ^{2}}\right) & 0 \\ 
0 & \frac{1}{2}\left( C_{1}+C_{2}+\sqrt{4g_{1}^{2}C_{1}C_{2}+\left(
C_{1}-C_{2}\right) ^{2}}\right)%
\end{array}%
\right) T_{0}\left( Z,Z_{1}\right)
\end{equation*}

Appendix $5.2$ shows that $\alpha _{i}\left( \hat{T}\right) $ and $%
B_{i}\left( \hat{T}\right) $ are proportional to the averages of $\hat{T}%
_{iD}$ and $1+\hat{T}_{iD}$, more precisely:%
\begin{eqnarray*}
D\left( \alpha \left( \hat{T}\right) \right) &\propto &\left( 
\begin{array}{cc}
\frac{1}{2}\left( C_{1}+C_{2}-\sqrt{4g_{1}^{2}C_{1}C_{2}+\left(
C_{1}-C_{2}\right) ^{2}}\right) & 0 \\ 
0 & \frac{1}{2}\left( C_{1}+C_{2}+\sqrt{4g_{1}^{2}C_{1}C_{2}+\left(
C_{1}-C_{2}\right) ^{2}}\right)%
\end{array}%
\right) \\
D\left( B\left( \hat{T}\right) \right) &\propto &\left( 
\begin{array}{cc}
\frac{1}{2}\left( C_{1}+C_{2}-\sqrt{4g_{1}^{2}C_{1}C_{2}+\left(
C_{1}-C_{2}\right) ^{2}}\right) & 0 \\ 
0 & \frac{1}{2}\left( C_{1}+C_{2}+\sqrt{4g_{1}^{2}C_{1}C_{2}+\left(
C_{1}-C_{2}\right) ^{2}}\right)%
\end{array}%
\right)
\end{eqnarray*}%
As a consequence, by multiplication with $U$ and $U^{-1}$, we find that:%
\begin{equation}
\frac{\left( 1+D\left( \left\vert \Psi \right\vert ^{2}\right) \right) \hat{T%
}}{1-\left( 1+D\left( \left\vert \Psi \right\vert ^{2}\right) \right) \hat{T}%
}=\frac{\exp \left( -cl_{1}-\left( 1+D\left( \left\langle \left\vert \Psi
\right\vert ^{2}\right\rangle \right) \right) \Lambda \left( \left(
cl_{1}\right) ^{2}-\left\vert Z-Z_{1}\right\vert ^{2}\right) \right) }{B}%
H\left( cl_{1}-\left\vert Z-Z_{1}\right\vert \right)  \label{fRC}
\end{equation}%
with: 
\begin{eqnarray*}
\Lambda &=&\left( 
\begin{array}{cc}
C_{1} & -gC_{2} \\ 
-gC_{1} & C_{2}%
\end{array}%
\right) \\
B &=&1+2\pi \left( 1+D\left( \left\langle \left\vert \Psi \right\vert
^{2}\right\rangle \right) \right) \Lambda
\end{eqnarray*}%
where the constants $C_{1}$ and $C_{2}$ are as in Appendix 4.3 to define $%
\hat{T}_{1}$ and $\hat{T}_{2}$.

\paragraph*{6.3.5.2 General case}

The formula of the previous paragraph generalize to a system with $n$
interacting components, and with have the generalization of (\ref{fRC}):

\begin{eqnarray}
A\left( \left( Z,\theta \right) ,\left( Z_{1},\theta -l_{1}\right) \right)
&\simeq &D\left( \frac{1}{\left( 1+D\left( \left\langle \left\vert \Psi
\right\vert ^{2}\right\rangle \right) \right) }\right)  \label{rfp} \\
&&\times \left( \frac{\exp \left( -cl_{1}-\left( 1+D\left( \left\langle
\left\vert \Psi \right\vert ^{2}\right\rangle \right) \right) \Lambda \left(
\left( cl_{1}\right) ^{2}-\left\vert Z-Z_{1}\right\vert ^{2}\right) \right) 
}{B}H\left( cl_{1}-\left\vert Z-Z_{1}\right\vert \right) \right)  \notag
\end{eqnarray}

As a consequence, the expansion of (\ref{qvn}) is:%
\begin{eqnarray}
\omega \left( Z,\theta \right) &=&\omega _{0}\left( J,\theta ,Z\right) +\int
\sum_{k=0}^{\infty }\dprod\limits_{i=0}^{k-1}\frac{\exp \left(
-cl_{i}-\left( 1+D\left( \left\langle \left\vert \Psi \right\vert
^{2}\right\rangle \right) \right) \Lambda \left( \left( cl_{i}\right) ^{2}-%
\frac{\left\vert Z_{i}-Z_{i+1}\right\vert }{c}\right) \right) }{B}  \notag \\
&&\times D\left( \frac{\omega _{0}\left( \theta -l_{i},Z_{i}\right) }{\omega
_{0}\left( \theta -l_{i},Z_{i}\right) +A\omega _{0}\left\vert \Psi
\right\vert ^{2}\left( \theta -l_{i},Z_{i}\right) }\frac{\omega _{0}\left(
J,\theta -l_{k},Z_{k}\right) }{\left( 1+D\left( \left\langle \left\vert \Psi
\right\vert ^{2}\right\rangle \right) \right) }\right)  \notag \\
&&\times \frac{\exp \left( -cl_{k}-\left( 1+D\left( \left\langle \left\vert
\Psi \right\vert ^{2}\right\rangle \right) \right) \Lambda \left( \left(
cl_{i}\right) ^{2}-\frac{\left\vert Z_{k-1}-Z_{k}\right\vert }{c}\right)
\right) }{B}\left\vert \Psi \left( \theta -l_{k},Z_{k}\right) \right\vert
^{2}dZ_{i}dl_{i}
\end{eqnarray}%
\begin{eqnarray}
\omega ^{-1}\left( Z,\theta \right) &=&\omega _{0}^{-1}\left( J,\theta
,Z\right)  \notag \\
&&+D\left( \frac{G^{\prime }\left[ J,\omega ,\theta ,Z,\Psi \right] }{%
F^{\prime }\left[ J,\omega ,\theta ,Z,\Psi \right] }\right) \int
\sum_{k=0}^{\infty }\dprod\limits_{i=0}^{k-1}\frac{\exp \left(
-cl_{i}-\left( 1+D\left( \left\langle \left\vert \Psi \right\vert
^{2}\right\rangle \right) \right) \Lambda \left( \left( cl_{i}\right) ^{2}-%
\frac{\left\vert Z_{i}-Z_{i+1}\right\vert }{c}\right) \right) }{D}  \notag \\
&&\times D\left( \frac{\omega _{0}\left( \theta -l_{i},Z_{i}\right) }{\omega
_{0}\left( \theta -l_{i},Z_{i}\right) +A\omega _{0}\left\vert \Psi
\right\vert ^{2}\left( \theta -l_{i},Z_{i}\right) }\frac{\omega _{0}\left(
J,\theta -l_{k},Z_{k}\right) }{\left( 1+D\left( \left\langle \left\vert \Psi
\right\vert ^{2}\right\rangle \right) \right) }\right)  \notag \\
&&\times \frac{\exp \left( -cl_{k}-\left( 1+D\left( \left\langle \left\vert
\Psi \right\vert ^{2}\right\rangle \right) \right) \Lambda \left( \left(
cl_{i}\right) ^{2}-\frac{\left\vert Z_{k-1}-Z_{k}\right\vert }{c}\right)
\right) }{B}\left\vert \Psi \left( \theta -l_{k},Z_{k}\right) \right\vert
^{2}dZ_{i}dl_{i}
\end{eqnarray}

and the full series expansion is obtained as in the previous paragraph:%
\begin{eqnarray}
&&\hat{T}\Lambda ^{\dag
}=\sum_{n,p}\sum_{\sum\limits_{i=1}^{p}r_{i}=n-1}\sum_{l_{1},...,l_{n},%
\sum\limits_{i=1}^{n}l_{i}=p}\sum_{\substack{ f\in \left\{ 1,...,p\right\}
^{\left\{ 1,...,n\right\} },  \\ f\left( k\right) \notin \left\{
\sum\limits_{i=1}^{k-1}l_{i}+1,...\sum\limits_{i=1}^{k}l_{i}\right\} }}%
A^{l_{1}}\left( X,\hat{X}_{1},...,\hat{X}_{l_{1}},X_{0},X_{1}^{\prime
}\right) \omega _{0}\left\vert \Psi \right\vert ^{2}\left( X_{1}^{\prime
}\right) dX_{1}^{\prime }  \notag \\
&&\times \prod\limits_{k=1}^{p}\frac{d\hat{X}_{k}}{\sharp k!}\times
\prod\limits_{k=2}^{n}A^{l_{k}}\left( X_{k},\hat{X}_{\sum%
\limits_{i=1}^{k-1}l_{i}+1},...,\hat{X}_{\sum%
\limits_{i=1}^{k}l_{i}},X_{k}^{\prime }\right) \omega _{0}\left\vert \Psi
\right\vert ^{2}\left( X_{k}^{\prime }\right) dX_{k}^{\prime }  \notag \\
&&\times \prod\limits_{i}^{p}\frac{\frac{\delta ^{r_{i}}\hat{T}}{\delta
^{r_{i}}\omega _{0}\left( \hat{X}_{i}\right) }}{\hat{T}}\prod%
\limits_{k=2}^{n}\delta \left( X_{k}-\hat{X}_{f\left( k\right) }\right)
\end{eqnarray}%
\bigskip

The corrections due to the fluctuations around the saddle point can be
derived as in the previous paragraph, but the computations will be omitted
here.

\section*{Appendix 7. Dynamic equations for connectivity functions}

\subsection*{7.1 General formula}

We adapt the description of (\cite{IFR}) to our context. The transfer
function $T$ from $i$ to $j$ satisfies the following equation: 
\begin{eqnarray}
&&\nabla _{\theta ^{\left( i\right) }\left( n_{i}\right) }T\left( \left(
Z_{i},\theta ^{\left( i\right) }\left( n_{i}\right) ,\omega _{i}\left(
n_{i}\right) \right) ,\left( Z_{j},\theta ^{\left( j\right) }\left(
n_{j}\right) ,\omega _{j}\left( n_{j}\right) \right) \right)  \label{tvl} \\
&=&-\frac{1}{\tau }T\left( \left( Z_{i},\theta ^{\left( i\right) }\left(
n_{i}\right) ,\omega _{i}\left( n_{i}\right) \right) ,\left( Z_{j},\theta
^{\left( j\right) }\left( n_{j}\right) ,\omega _{j}\left( n_{j}\right)
\right) \right)  \notag \\
&&+\lambda \left( \hat{T}\left( \left( Z_{i},\theta ^{\left( i\right)
}\left( n_{i}\right) ,\omega _{i}\left( n_{i}\right) \right) ,\left(
Z_{j},\theta ^{\left( j\right) }\left( n_{j}\right) ,\omega _{j}\left(
n_{j}\right) \right) \right) \right) \delta \left( \theta ^{\left( i\right)
}\left( n_{i}\right) -\theta ^{\left( j\right) }\left( n_{j}\right) -\frac{%
\left\vert Z_{i}-Z_{j}\right\vert }{c}\right)  \notag
\end{eqnarray}%
where $\hat{T}$ measures the variation of $T$ due to the signals send from $%
j $ to $i$ and the signals emitted by $i$. It satisfies the following
equation: 
\begin{eqnarray}
&&\nabla _{\theta ^{\left( i\right) }\left( n_{i}\right) }\hat{T}\left(
\left( Z_{i},\theta ^{\left( i\right) }\left( n_{i}\right) ,\omega
_{i}\left( n_{i}\right) \right) ,\left( Z_{j},\theta ^{\left( j\right)
}\left( n_{j}\right) ,\omega _{j}\left( n_{j}\right) \right) \right)
\label{thvl} \\
&=&\rho \delta \left( \theta ^{\left( i\right) }\left( n_{i}\right) -\theta
^{\left( j\right) }\left( n_{j}\right) -\frac{\left\vert
Z_{i}-Z_{j}\right\vert }{c}\right)  \notag \\
&&\times \left\{ \left( h\left( Z,Z_{1}\right) -\hat{T}\left( \left(
Z_{i},\theta ^{\left( i\right) }\left( n_{i}\right) ,\omega _{i}\left(
n_{i}\right) \right) ,\left( Z_{j},\theta ^{\left( j\right) }\left(
n_{j}\right) ,\omega _{j}\left( n_{j}\right) \right) \right) \right) C\left(
\theta ^{\left( i\right) }\left( n-1\right) \right) h_{C}\left( \omega
_{i}\left( n_{i}\right) \right) \right.  \notag \\
&&\left. -D\left( \theta ^{\left( i\right) }\left( n-1\right) \right) \hat{T}%
\left( \left( Z_{i},\theta ^{\left( i\right) }\left( n_{i}\right) ,\omega
_{i}\left( n_{i}\right) \right) ,\left( Z_{j},\theta ^{\left( j\right)
}\left( n_{j}\right) ,\omega _{j}\left( n_{j}\right) \right) \right)
h_{D}\left( \omega _{j}\left( n_{j}\right) \right) \right\}  \notag
\end{eqnarray}%
where $h_{C}$ and $h_{D}$\ are increasing functions. We depart slightly from
(\cite{IFR}) by the introduction of the function $h\left( Z,Z_{1}\right) $
(they chose $h\left( Z,Z_{1}\right) =1$), to implement some loss due to the
distance. We may chose for example:%
\begin{equation*}
h\left( Z,Z_{1}\right) =\exp \left( -\frac{\left\vert Z_{i}-Z_{j}\right\vert 
}{\nu c}\right)
\end{equation*}%
where $\nu $ is a parameter. Equation (\ref{thvl}) involves two dynamic
factors $C\left( \theta ^{\left( i\right) }\left( n-1\right) \right) $ and $%
D\left( \theta _{i}\left( n-1\right) \right) $. The factor $C\left( \theta
^{\left( i\right) }\left( n-1\right) \right) $ describes the accumulation of
input spikes. It is solution of the differential equation:%
\begin{eqnarray}
\nabla _{\theta ^{\left( i\right) }\left( n-1\right) }C\left( \theta
^{\left( i\right) }\left( n-1\right) \right) &=&-\frac{C\left( \theta
^{\left( i\right) }\left( n-1\right) \right) }{\tau _{C}}  \label{qnC} \\
&&+\alpha _{C}\left( 1-C\left( \theta ^{\left( i\right) }\left( n-1\right)
\right) \right) \omega _{j}\left( Z_{j},\theta ^{\left( i\right) }\left(
n-1\right) -\frac{\left\vert Z_{i}-Z_{j}\right\vert }{c}\right)  \notag
\end{eqnarray}%
In the continuous approximation, the solution of (\ref{qnC}) is:%
\begin{eqnarray*}
C\left( \theta ^{\left( i\right) }\left( n-1\right) \right) &=&\int \exp
\left( -\left( \frac{\left( \theta ^{\left( i\right) }\left( n-1\right)
-\theta ^{\left( i\right) \prime }\right) }{\tau _{C}}+\alpha
_{C}\int_{\theta ^{\left( i\right) \prime }}^{\theta ^{\left( i\right)
}\left( n-1\right) }\omega _{j}\left( Z_{j},\theta ^{\prime }-\frac{%
\left\vert Z_{i}-Z_{j}\right\vert }{c}\right) d\theta ^{\prime }\right)
\right) \\
&&\times \omega _{j}\left( Z_{j},\theta ^{\left( i\right) \prime }-\frac{%
\left\vert Z_{i}-Z_{j}\right\vert }{c}\right) d\theta ^{\left( i\right)
\prime }
\end{eqnarray*}%
If a static equilibrium $\omega _{0}\left( Z_{j}\right) $ exists, expanding
around this equilibrium leads to approximate the integral: 
\begin{equation*}
\int_{\theta _{i}^{\prime }}^{\theta _{i}}\omega _{j}\left( Z_{j},\theta
^{\prime }-\frac{\left\vert Z_{i}-Z_{j}\right\vert }{c}\right) d\theta
^{\prime }
\end{equation*}%
by the quantity:%
\begin{equation*}
\omega _{0}\left( Z_{j}\right) \left( \theta ^{\left( i\right) }\left(
n-1\right) -\theta ^{\left( i\right) \prime }\right)
\end{equation*}%
so that:%
\begin{eqnarray}
C\left( \theta ^{\left( i\right) }\left( n-1\right) \right) &=&\int \exp
\left( -\left( \frac{1}{\tau _{C}}+\alpha _{C}\omega _{0}\left( Z_{j}\right)
\right) \left( \theta ^{\left( i\right) }\left( n-1\right) -\theta ^{\left(
i\right) \prime }\right) \right)  \label{slc} \\
&&\times \left( C_{0}+\omega _{j}\left( Z_{j},\theta ^{\left( i\right)
\prime }-\frac{\left\vert Z_{i}-Z_{j}\right\vert }{c}\right) \right) d\theta
_{i}^{\prime }  \notag
\end{eqnarray}%
The term $D\left( \theta _{i}\left( n-1\right) \right) $ is proportional to
the accumulation of output spikes and is solution of:%
\begin{equation}
\nabla _{\theta ^{\left( i\right) }\left( n-1\right) }D\left( \theta
^{\left( i\right) }\left( n-1\right) \right) =-\frac{D\left( \theta ^{\left(
i\right) }\left( n-1\right) \right) }{\tau _{D}}+\alpha _{D}\left( 1-D\left(
\theta ^{\left( i\right) }\left( n-1\right) \right) \right) \omega
_{i}\left( Z_{i}\right)  \label{qnD}
\end{equation}%
In the continuous approximation, the solution of (\ref{qnD}) is:%
\begin{equation}
D\left( \theta ^{\left( i\right) }\left( n-1\right) \right) =\int \exp
\left( -\left( \frac{1}{\tau _{D}}+\alpha _{D}\omega _{0}\left( Z_{i}\right)
\right) \left( \theta ^{\left( i\right) }\left( n-1\right) -\theta ^{\left(
i\right) \prime }\right) \right) \left( D_{0}+\omega _{i}\left( Z_{i},\theta
^{\left( i\right) \prime }\right) \right) d\theta ^{\left( i\right) \prime }
\label{sld}
\end{equation}%
As a consequence, the dynamics for transfer functions is a set of two
equations:

\begin{eqnarray}
&&\nabla _{\theta ^{\left( i\right) }\left( n_{i}\right) }T\left( \left(
Z_{i},\theta ^{\left( i\right) }\left( n_{i}\right) ,\omega _{i}\left(
n_{i}\right) \right) ,\left( Z_{j},\theta ^{\left( j\right) }\left(
n_{j}\right) ,\omega _{j}\left( n_{j}\right) \right) \right)  \label{rtq} \\
&=&-\frac{1}{\tau }T\left( \left( Z_{i},\theta ^{\left( i\right) }\left(
n_{i}\right) ,\omega _{i}\left( n_{i}\right) \right) ,\left( Z_{j},\theta
^{\left( j\right) }\left( n_{j}\right) ,\omega _{j}\left( n_{j}\right)
\right) \right)  \notag \\
&&+\lambda \left( \hat{T}\left( \left( Z_{i},\theta ^{\left( i\right)
}\left( n_{i}\right) ,\omega _{i}\left( n_{i}\right) \right) ,\left(
Z_{j},\theta ^{\left( j\right) }\left( n_{j}\right) ,\omega _{j}\left(
n_{j}\right) \right) \right) \right) \delta \left( \theta ^{\left( i\right)
}\left( n_{i}\right) -\theta ^{\left( j\right) }\left( n_{j}\right) -\frac{%
\left\vert Z_{i}-Z_{j}\right\vert }{c}\right)  \notag
\end{eqnarray}%
and:%
\begin{eqnarray}
&&\nabla _{\theta ^{\left( i\right) }\left( n_{i}\right) }\hat{T}\left(
\left( Z_{i},\theta ^{\left( i\right) }\left( n_{i}\right) ,\omega
_{i}\left( n_{i}\right) \right) ,\left( Z_{j},\theta ^{\left( j\right)
}\left( n_{j}\right) ,\omega _{j}\left( n_{j}\right) \right) \right)
\label{rthq} \\
&=&\rho \delta \left( \theta ^{\left( i\right) }\left( n_{i}\right) -\theta
^{\left( j\right) }\left( n_{j}\right) -\frac{\left\vert
Z_{i}-Z_{j}\right\vert }{c}\right)  \notag \\
&&\times \left\{ \left( h\left( Z,Z_{1}\right) -\hat{T}\left( \left(
Z_{i},\theta ^{\left( i\right) }\left( n_{i}\right) ,\omega _{i}\left(
n_{i}\right) \right) ,\left( Z_{j},\theta ^{\left( j\right) }\left(
n_{j}\right) ,\omega _{j}\left( n_{j}\right) \right) \right) \right) C\left(
\theta ^{\left( i\right) }\left( n-1\right) \right) h_{C}\left( \omega
_{i}\left( n_{i}\right) \right) \right.  \notag \\
&&\left. -D\left( \theta ^{\left( i\right) }\left( n-1\right) \right) \hat{T}%
\left( \left( Z_{i},\theta ^{\left( i\right) }\left( n_{i}\right) ,\omega
_{i}\left( n_{i}\right) \right) ,\left( Z_{j},\theta ^{\left( j\right)
}\left( n_{j}\right) ,\omega _{j}\left( n_{j}\right) \right) \right)
h_{D}\left( \omega _{j}\left( n_{j}\right) \right) \right\}  \notag
\end{eqnarray}%
with $C\left( \theta ^{\left( i\right) }\left( n-1\right) \right) $ and $%
D\left( \theta ^{\left( i\right) }\left( n-1\right) \right) $ given by (\ref%
{slc}) and (\ref{sld}).

The field translation of (\ref{rtq}) and (\ref{rthq}) is obtained by
including the following potential terms in the action for the field:%
\begin{eqnarray}
&&\int \left( \nabla _{\theta }T\left( \left( Z,\theta ,\omega \right)
,\left( Z_{1},\theta _{1},\omega _{1}\right) \right) +\frac{T\left( \left(
Z,\theta ,\omega \right) ,\left( Z_{1},\theta _{1},\omega _{1}\right)
\right) }{\tau }\right.  \label{tpt} \\
&&\left. -\lambda \left( \hat{T}\left( \left( Z,\theta ,\omega \right)
,\left( Z_{1},\theta _{1},\omega _{1}\right) \right) \right) \delta \left(
\theta -\theta _{1}-\frac{\left\vert Z-Z_{1}\right\vert }{c}\right) \right) 
\notag \\
&&\times \left\vert \Psi \left( \theta ,Z,\omega \right) \right\vert
^{2}\left\vert \Psi \left( \theta _{1},Z_{1},\omega _{1}\right) \right\vert
^{2}  \notag
\end{eqnarray}%
corresponding to (\ref{rtq}) and: 
\begin{eqnarray}
&&\int \left( \nabla _{\theta }\hat{T}\left( \left( Z,\theta ,\omega \right)
,\left( Z_{1},\theta _{1},\omega _{1}\right) \right) -\rho \delta \left(
\theta ^{\left( i\right) }\left( n_{i}\right) -\theta ^{\left( j\right)
}\left( n_{j}\right) -\frac{\left\vert Z-Z_{1}\right\vert }{c}\right) \right.
\label{tpth} \\
&&\left. \times \left\{ \left( h\left( Z,Z_{1}\right) -\hat{T}\left( \left(
Z,\theta ,\omega \right) ,\left( Z_{1},\theta _{1},\omega _{1}\right)
\right) \right) C\left( \theta ,Z,Z_{1}\right) h_{C}\left( \omega \right)
-D\left( \theta ,Z\right) \hat{T}\left( \left( Z,\theta ,\omega \right)
,\left( Z_{1},\theta _{1},\omega _{1}\right) \right) h_{D}\left( \omega
_{1}\right) \right\} \right)  \notag \\
&&\times \left\vert \Psi \left( \theta ,Z,\omega \right) \right\vert
^{2}\left\vert \Psi \left( \theta _{1},Z_{1},\omega _{1}\right) \right\vert
^{2}  \notag
\end{eqnarray}%
for (\ref{rthq}), with $C\left( \theta ,Z,Z_{1}\right) $ and $D\left( \theta
,Z\right) $\ are defined as:%
\begin{equation*}
C\left( \theta ,Z,Z_{1}\right) =\int^{\theta }\exp \left( -\left( \frac{1}{%
\tau _{C}}+\alpha _{C}\omega _{0}\left( Z_{1}\right) \right) \left( \theta
-\theta ^{\prime }\right) \right) \left( C_{0}+\omega \left( Z_{1},\theta
^{\prime }-\frac{\left\vert Z-Z_{1}\right\vert }{c}\right) \right) d\theta
^{\prime }
\end{equation*}%
\begin{equation*}
D\left( \theta ,Z\right) =\int^{\theta }\exp \left( -\left( \frac{1}{\tau
_{D}}+\alpha _{D}\omega _{0}\left( Z\right) \right) \left( \theta -\theta
^{\prime }\right) \right) \left( D_{0}+\omega \left( Z,\theta ^{\prime
}\right) \right) d\theta ^{\prime }
\end{equation*}%
For 
\begin{eqnarray*}
\tau _{C}\left( Z_{1}\right) &=&\frac{1}{\tau _{C}}+\alpha _{C}\omega
_{0}\left( Z_{1}\right) <1 \\
\tau _{D}\left( Z\right) &=&\frac{1}{\tau _{D}}+\alpha _{D}\omega _{0}\left(
Z\right) <1
\end{eqnarray*}%
and if the transfer function adapts slowly with respect to $\omega \left(
Z,\theta \right) $, we can simplify the expressions for $C\left( \theta
,Z,Z_{1}\right) $ and $D\left( \theta ,Z\right) $:

\begin{eqnarray*}
C\left( \theta ,Z,Z_{1}\right) &\simeq &C\left( Z_{1}\right) =\frac{%
C_{0}+\omega _{0}\left( Z_{1}\right) }{\tau _{C}\left( Z_{1}\right) } \\
D\left( \theta ,Z\right) &\simeq &D\left( Z\right) =\frac{D_{0}+\omega
_{0}\left( Z\right) }{\tau _{D}\left( Z\right) }
\end{eqnarray*}%
After projection on the dependent frequency states the transfer functions
become functions $T\left( \left( Z,\theta \right) ,\left( Z_{1},\theta
_{1}\right) \right) $ and $\hat{T}\left( \left( Z,\theta \right) ,\left(
Z_{1},\theta _{1}\right) \right) $ respectively. Moreover, we can simplify
the action by finding the configurations for $T\left( \left( Z,\theta
\right) ,\left( Z_{1},\theta _{1}\right) \right) $ and $\hat{T}\left( \left(
Z,\theta \right) ,\left( Z_{1},\theta _{1}\right) \right) $ that minimize
the potential terms (\ref{tpt}), (\ref{tpth}). It corresponds to set: 
\begin{eqnarray}
0 &=&\nabla _{\theta }T\left( \left( Z,\theta ,\omega \right) ,\left(
Z_{1},\theta _{1},\omega _{1}\right) \right) +\frac{T\left( \left( Z,\theta
,\omega \right) ,\left( Z_{1},\theta _{1},\omega _{1}\right) \right) }{\tau }
\label{flq} \\
&&-\lambda \left( \hat{T}\left( \left( Z,\theta ,\omega \right) ,\left(
Z_{1},\theta _{1},\omega _{1}\right) \right) \right) \delta \left( \theta
-\theta _{1}-\frac{\left\vert Z-Z_{1}\right\vert }{c}\right)  \notag
\end{eqnarray}%
and:%
\begin{eqnarray}
0 &=&\left( \nabla _{\theta }\hat{T}\left( \left( Z,\theta ,\omega \right)
,\left( Z_{1},\theta _{1},\omega _{1}\right) \right) -\rho \delta \left(
\theta ^{\left( i\right) }\left( n_{i}\right) -\theta ^{\left( j\right)
}\left( n_{j}\right) -\frac{\left\vert Z-Z_{1}\right\vert }{c}\right) \right.
\label{flh} \\
&&\times \left\{ \left( h\left( Z,Z_{1}\right) -\hat{T}\left( \left(
Z,\theta ,\omega \right) ,\left( Z_{1},\theta _{1},\omega _{1}\right)
\right) \right) C\left( \theta ,Z,Z_{1}\right) h_{C}\left( \omega \right)
-D\left( \theta ,Z\right) \hat{T}\left( \left( Z,\theta ,\omega \right)
,\left( Z_{1},\theta _{1},\omega _{1}\right) \right) h_{D}\left( \omega
_{1}\right) \right\}  \notag
\end{eqnarray}%
We look for solutions of the form:%
\begin{eqnarray*}
T\left( Z,\theta ,Z_{1},\theta -\frac{\left\vert Z-Z_{1}\right\vert }{c}%
\right) &\equiv &T\left( Z,\theta ,Z_{1}\right) \\
\hat{T}\left( Z,\theta ,Z_{1},\theta -\frac{\left\vert Z-Z_{1}\right\vert }{c%
}\right) &\equiv &\hat{T}\left( Z,\theta ,Z_{1}\right)
\end{eqnarray*}%
so that $T\left( Z,\theta ,Z_{1}\right) $ and $\hat{T}\left( Z,\theta
,Z_{1}\right) $ satisfy:

\begin{equation}
\nabla _{\theta }T\left( Z,\theta ,Z_{1}\right) +\left( \frac{T\left(
Z,\theta ,Z_{1}\right) }{\tau }-\lambda \hat{T}\left( Z,\theta ,Z_{1}\right)
\right) =0  \label{qnth}
\end{equation}

\begin{eqnarray}
&&\nabla _{\theta }\hat{T}\left( Z,\theta ,Z_{1}\right)  \label{qnthbs} \\
&=&\rho \left( \left( h\left( Z,Z_{1}\right) -\hat{T}\left( Z,\theta
,Z_{1}\right) \right) C\left( Z_{1}\right) h_{C}\left( \omega \left(
Z,\theta \right) \right) -\hat{T}\left( Z,\theta ,Z_{1}\right) D\left(
Z\right) h_{D}\left( \omega \left( Z_{1},\theta -\frac{\left\vert
Z-Z_{1}\right\vert }{c}\right) \right) \right)  \notag
\end{eqnarray}%
Using (\ref{qnth}), we replace $\hat{T}\left( Z,\theta ,Z_{1}\right) $ in (%
\ref{qnthbs}):%
\begin{equation*}
\hat{T}\left( Z,\theta ,Z_{1}\right) =\frac{\nabla _{\theta }T\left(
Z,\theta ,Z_{1}\right) }{\lambda }+\frac{T\left( Z,\theta ,Z_{1}\right) }{%
\lambda \tau }
\end{equation*}%
and we arrive to the differential equation satisfied by $T\left( Z,\theta
,Z_{1}\right) $: 
\begin{equation}
\frac{\nabla _{\theta }^{2}T\left( Z,\theta ,Z_{1}\right) }{\lambda }%
+U_{1}\left( \omega \right) \nabla _{\theta }T\left( Z,\theta ,Z_{1}\right)
+U_{2}\left( \omega \right) T\left( Z,\theta ,Z_{1}\right) =\rho C\left(
Z_{1}\right) h\left( Z,Z_{1}\right) h_{C}\left( \omega \left( Z,\theta
\right) \right)  \label{qntrsfct}
\end{equation}%
where:%
\begin{eqnarray*}
U_{1}\left( \omega \right) &=&\left( \frac{1}{\lambda \tau }+\frac{\rho }{%
\lambda }\left( C\left( Z_{1}\right) h_{C}\left( \omega \left( Z,\theta
\right) \right) +D\left( Z\right) h_{D}\left( \omega \left( Z_{1},\theta -%
\frac{\left\vert Z-Z_{1}\right\vert }{c}\right) \right) \right) \right) \\
U_{2}\left( \omega \right) &=&\frac{\rho }{\lambda \tau }\left( C\left(
Z_{1}\right) h_{C}\left( \omega \left( Z,\theta \right) \right) +D\left(
Z\right) h_{D}\left( \omega \left( Z_{1},\theta -\frac{\left\vert
Z-Z_{1}\right\vert }{c}\right) \right) \right)
\end{eqnarray*}%
If we consider that the transfer function varies slowly compared to the
oscillations of the thread, we can approximate (\ref{qntrsfct}) by a quite
static equation:%
\begin{equation*}
U_{2}\left( \omega \right) T\left( Z,\theta ,Z_{1}\right) =\rho C\left(
Z_{1}\right) h\left( Z,Z_{1}\right) h_{C}\left( \omega \left( Z,\theta
\right) \right)
\end{equation*}%
whose solution is:%
\begin{eqnarray}
T\left( Z,\theta ,Z_{1}\right) &=&\frac{\lambda \tau C\left( Z_{1}\right)
h\left( Z,Z_{1}\right) h_{C}\left( \omega \left( Z,\theta \right) \right) }{%
C\left( Z_{1}\right) h_{C}\left( \omega \left( Z,\theta \right) \right)
+D\left( Z\right) h_{D}\left( \omega \left( Z_{1},\theta -\frac{\left\vert
Z-Z_{1}\right\vert }{c}\right) \right) }  \label{sttc} \\
&\simeq &\frac{\lambda \tau h\left( Z,Z_{1}\right) }{1+\frac{D\left(
Z\right) }{C\left( Z_{1}\right) }\frac{h_{D}\left( \omega \left(
Z_{1},\theta -\frac{\left\vert Z-Z_{1}\right\vert }{c}\right) \right) }{%
h_{C}\left( \omega \left( Z,\theta \right) \right) }}  \notag
\end{eqnarray}%
Thus $T\left( Z,\theta ,Z_{1}\right) $ is a decreasing function of $\omega
\left( Z_{1},\theta -\frac{\left\vert Z-Z_{1}\right\vert }{c}\right) $ and
an increasing function of $\omega \left( Z,\theta \right) $, as hypothesized
in the text. The fully static solution associated to (\ref{sttc}) is:%
\begin{equation*}
T_{0}\left( Z,Z_{1}\right) =\frac{\lambda \tau h\left( Z,Z_{1}\right) }{1+%
\frac{D\left( Z\right) }{C\left( Z_{1}\right) }\frac{h_{D}\left( \omega
_{0}\left( Z_{1}\right) \right) }{h_{C}\left( \omega _{0}\left( Z\right)
\right) }}
\end{equation*}

\subsection*{7.2 Linearized dynamics}

We conclude this section by giving the linearized version of (\ref{qntrsfct}%
) around the static solution $(\omega _{0}\left( Z\right) ,T_{0}\left(
Z,Z_{1}\right) )$. It is:

\begin{eqnarray}
0 &=&\frac{\nabla _{\theta }^{2}T\left( Z,\theta ,Z_{1}\right) }{\lambda }%
+U_{1}\left( \omega _{0}\right) \nabla _{\theta }T\left( Z,\theta
,Z_{1}\right) +U_{2}\left( \omega _{0}\right) T\left( Z,\theta ,Z_{1}\right)
\label{lrn} \\
&&-\rho C\left( Z_{1}\right) \left( 1-\frac{T_{0}\left( Z,Z_{1}\right) }{%
\lambda \tau }\right) h_{C}^{\prime }\left( \omega _{0}\left( Z\right)
\right) \Omega \left( Z,\theta \right)  \notag \\
&&+\frac{\rho T_{0}\left( Z,Z_{1}\right) }{\lambda \tau }\left( D\left(
Z\right) h_{D}^{\prime }\left( \omega _{0}\left( Z_{1}\right) \right) \Omega
\left( Z_{1},\theta -\frac{\left\vert Z-Z_{1}\right\vert }{c}\right) \right)
\notag
\end{eqnarray}%
where:%
\begin{eqnarray*}
U_{1}\left( \omega _{0}\right) &=&\frac{1}{\lambda \tau }+\frac{\rho }{%
\lambda }\left( C\left( Z_{1}\right) h_{C}\left( \omega _{0}\left( Z\right)
\right) +D\left( Z\right) h_{D}\left( \omega _{0}\left( Z_{1}\right) \right)
\right) \\
U_{2}\left( \omega _{0}\right) &=&\frac{\rho }{\lambda \tau }\left( C\left(
Z_{1}\right) h_{C}\left( \omega _{0}\left( Z\right) \right) +D\left(
Z\right) h_{D}\left( \omega _{0}\left( Z_{1}\right) \right) \right) \\
T_{0}\left( Z,Z_{1}\right) &=&\frac{\lambda \tau h\left( Z,Z_{1}\right) }{1+%
\frac{D\left( Z\right) }{C\left( Z_{1}\right) }\frac{h_{D}\left( \omega
_{0}\left( Z_{1}\right) \right) }{h_{C}\left( \omega _{0}\left( Z\right)
\right) }} \\
T_{0}\left( Z,Z_{1}\right) &=&\frac{T_{0}\left( Z,Z_{1}\right) }{h\left(
Z,Z_{1}\right) } \\
T\left( Z,\theta ,Z_{1}\right) &=&\frac{T\left( Z,\theta ,Z_{1}\right)
-T_{0}\left( Z,Z_{1}\right) }{h\left( Z,Z_{1}\right) } \\
\Omega \left( Z,\theta \right) &=&\omega \left( Z,\theta \right) -\omega
_{0}\left( Z\right)
\end{eqnarray*}%
for $\omega _{0}\left( Z\right) \equiv \omega _{0}$, this reduces to:%
\begin{eqnarray}
&&\frac{\nabla _{\theta }^{2}T\left( Z,\theta ,Z_{1}\right) }{\lambda }%
+U_{1}\left( \omega _{0}\right) \nabla _{\theta }T\left( Z,\theta
,Z_{1}\right) +U_{2}\left( \omega \right) T\left( Z,\theta ,Z_{1}\right)
\label{lrnct} \\
&=&-\frac{\rho T_{0}\left( Z,Z_{1}\right) D\left( Z\right) h_{D}^{\prime
}\left( \omega _{0}\right) \Omega \left( Z_{1},\theta -\frac{\left\vert
Z-Z_{1}\right\vert }{c}\right) }{\lambda \tau }+\rho C\left( Z_{1}\right)
\left( 1-\frac{T_{0}\left( Z,Z_{1}\right) }{\lambda \tau }\right)
h_{C}^{\prime }\left( \omega _{0}\right) \Omega \left( Z,\theta \right) 
\notag
\end{eqnarray}%
where:%
\begin{eqnarray*}
U_{1}\left( \omega \right) &=&\frac{1}{\lambda \tau }+\frac{\rho }{\lambda }%
\left( Ch_{C}\left( \omega _{0}\right) +Dh_{D}\left( \omega _{0}\right)
\right) \\
U_{2}\left( \omega \right) &=&\frac{\rho }{\lambda \tau }\left( Ch_{C}\left(
\omega _{0}\right) +Dh_{D}\left( \omega _{0}\right) \right) \\
T_{0}\left( Z,Z_{1}\right) &=&\frac{\lambda \tau h\left( Z,Z_{1}\right) }{1+%
\frac{D}{C}\frac{h_{D}\left( \omega _{0}\right) }{h_{C}\left( \omega
_{0}\right) }} \\
C &=&\frac{C_{0}+\omega _{0}}{\tau _{C}} \\
D &=&\frac{D_{0}+\omega _{0}}{\tau _{D}}
\end{eqnarray*}%
which can also be written, up to the second order in derivatives:%
\begin{eqnarray}
&&\frac{\nabla _{\theta }^{2}T\left( Z,\theta ,Z_{1}\right) }{\lambda }%
+U_{1}\left( \omega _{0}\right) \nabla _{\theta }T\left( Z,\theta
,Z_{1}\right) +U_{2}\left( \omega \right) T\left( Z,\theta ,Z_{1}\right)
\label{lrncp} \\
&=&\left( \rho C\left( Z_{1}\right) h_{C}^{\prime }\left( \omega _{0}\right)
-\frac{\rho T_{0}\left( Z,Z_{1}\right) \left( D\left( Z\right) h_{D}^{\prime
}\left( \omega _{0}\right) +C\left( Z_{1}\right) h_{C}^{\prime }\left(
\omega _{0}\right) \right) }{\lambda \tau }\right) \Omega \left( Z,\theta
\right)  \notag \\
&&+\frac{\rho T_{0}\left( Z,Z_{1}\right) D\left( Z\right) h_{D}^{\prime
}\left( \omega _{0}\right) \left( \frac{\left\vert Z-Z_{1}\right\vert }{c}%
\nabla _{\theta }\Omega \left( Z,\theta \right) -\frac{\left( Z-Z_{1}\right)
^{2}}{2c^{2}}\nabla _{\theta }^{2}\Omega \left( Z,\theta \right) -\frac{%
\left( Z-Z_{1}\right) ^{2}}{2}\nabla _{Z}^{2}\Omega \left( Z,\theta \right)
\right) }{\lambda \tau }  \notag
\end{eqnarray}%
Then, to separate the dependences in time and position, we define:%
\begin{eqnarray*}
T\left( Z,\theta \right) &=&\int h\left( Z,Z_{1}\right) \frac{T\left(
Z,\theta ,Z_{1}\right) }{\sqrt{\frac{\pi }{8}\left( \frac{1}{\bar{X}_{r}}%
\right) ^{2}+\frac{\pi }{2}\alpha }} \\
\bar{C}\left( Z\right) &=&\frac{1}{\sqrt{\frac{\pi }{8}\left( \frac{1}{\bar{X%
}_{r}}\right) ^{2}+\frac{\pi }{2}\alpha }}\int h\left( Z,Z_{1}\right)
C\left( Z_{1}\right) \\
\bar{C}_{0}\left( Z\right) &=&\frac{1}{\sqrt{\frac{\pi }{8}\left( \frac{1}{%
\bar{X}_{r}}\right) ^{2}+\frac{\pi }{2}\alpha }}\int h\left( Z,Z_{1}\right)
C\left( Z_{1}\right) T_{0}\left( Z,Z_{1}\right) \\
T_{0}\left( Z\right) &=&\frac{1}{\sqrt{\frac{\pi }{8}\left( \frac{1}{\bar{X}%
_{r}}\right) ^{2}+\frac{\pi }{2}\alpha }}\int h\left( Z,Z_{1}\right)
T_{0}\left( Z,Z_{1}\right)
\end{eqnarray*}%
and $T\left( Z,\theta \right) $ satisfies: 
\begin{eqnarray*}
&&\frac{\nabla _{\theta }^{2}T\left( Z,\theta \right) }{\lambda }%
+U_{1}\left( \omega _{0}\right) \nabla _{\theta }T\left( Z,\theta \right)
+U_{2}\left( \omega \right) T\left( Z,\theta \right) \\
&=&\left( \rho \bar{C}\left( Z\right) h_{C}^{\prime }\left( \omega
_{0}\right) -\frac{\rho \left( D\left( Z\right) T_{0}\left( Z\right)
h_{D}^{\prime }\left( \omega _{0}\right) +\bar{C}_{0}\left( Z\right)
h_{C}^{\prime }\left( \omega _{0}\right) \right) }{\lambda \tau }\right)
\Omega \left( Z,\theta \right) \\
&&+\frac{\rho D\left( Z\right) h_{D}^{\prime }\left( \omega _{0}\right)
\left( \Gamma _{1}\nabla _{\theta }\Omega \left( Z,\theta \right) -\left(
\Gamma _{1}\nabla _{\theta }^{2}\Omega \left( Z,\theta \right) +c^{2}\Gamma
_{2}\nabla _{Z}^{2}\Omega \left( Z,\theta \right) \right) \right) }{\lambda
\tau }
\end{eqnarray*}%
\pagebreak

\end{document}